\documentclass[useAMS,usenatbib]{mn2e}
\usepackage{amssymb}
\usepackage{amsmath}
\usepackage{multirow}
\usepackage{epsfig}

\def\Journal#1#2#3#4{{#1} {\bf #2}, (#3) #4}
\def\cir#1{{\GCN} #1}
\def\rep#1{{\GCR} #1}

\def\etal{{\it et al.}}
\def\AA{\em A.\& A.}

\def\APJ{\em ApJ.}
\def\APL{\em ApJ.Lett.}

\def\GCN{\em GCN Circ.}
\def\GCR{\em GCN Rep.}

\def\MRA{\em MNRAS}
\def\MRAl{\em MNRASL}

\def\NAT{\em Nature}
\def\NCA{\em Nuovo Cimento}

\def\POF{\em Phys. of Fluids}

\def\PTR{{\em Phil.Trans.Roy.Soc.Lond.} A}

\def\SSR{\em Space Sci. Rev.}

\def\be{\begin{equation}}
\def\ee{\end{equation}}
\def\bea{\begin{eqnarray}}
\def\eea{\end{eqnarray}}
\def\bes{\begin{equation*}}
\def\ees{\end{equation*}}
\def\beas{\begin{eqnarray*}}
\def\eeas{\end{eqnarray*}}

\def\swift{{\it Swift }}

\def\secextract{3 of Paper I}

\title{A Systematic Description of Shocks in Gamma Ray Bursts II: Simulation}

\author[H. Ziaeepour]{Houri~Ziaeepour
\thanks{Email: hz@mssl.ucl.ac.uk}\\ 
Mullard Space Science Laboratory, Holmbury St Mary, Dorking, 
Surrey RH5 6NT, UK}

\begin{document}

\date{Accepted $\ldots$; Received $\ldots$; in original form Dec. 2008}

\pagerange{\pageref{firstpage}--\pageref{lastpage}} \pubyear{2008}

\maketitle

\label{firstpage}
\begin{abstract}
In Paper I we presented a detailed formulation of the relativistic shocks and 
synchrotron emission in the context of Gamma-Ray Burst (GRB) physics. To see 
how well this model reproduces the observed characteristics of the GRBs and 
their afterglows, here we present the results of some simulations based on 
this model. They are meant to reproduce the prompt and afterglow emission 
in some intervals of time during a burst. We show that this goal is achieved 
for both short and long GRBs and their afterglows, at least for part of the 
parameter space. Moreover, these results are the evidence of the physical 
relevance of the two phenomenological models we have suggested in Paper I for 
the evolution of the {\it active region}, the synchrotron emitting region in 
a shock. The dynamical active region model seems to reproduce the observed 
characteristics of prompt emissions and late afterglow better than the 
quasi-steady model which is more suitable for the onset of afterglows. 
Therefore these simulations confirm the arguments presented in Paper I about 
the behaviour of these models based on their physical properties. 
\end{abstract}


\section {Introduction}\label{sec:intro}
Gamma ray bursts show a variety of behaviours both in their early/prompt 
emissions in all energy bands and in their late emissions in lower energies - 
usually called the {\it afterglow}. There have been a large 
number of simulations to reproduce some of the aspects of GRB emission. 
Some of them are based on the phenomenological formulation of the 
evolution of a relativistic shock and its synchrotron 
emission~\citep{simul,simul0,dopplerlag1,simul1,simul2}. Their main 
goal has been to show that the idea of shock origin for the GRBs is 
essentially correct. Others are based on heavy numerical magnetohydrodynamics 
simulations of relativistic shocks and permit to reproduce some of the 
properties of GRBs from fundamental 
principles~\citep{simulshock,simulspec,simulspec0,simulexternal,simuljet,simuljet0}. But they present a small volume of the parameter space. 
None of these formulations and related simulations can be directly and 
systematically applied to real bursts data to estimate their parameters. 

With the simplified shock model explained in Paper I, we expect that the 
analytical approximation of the solutions of the kinematics and dynamics 
equations permit to simulate relativistic shocks and their synchrotron 
emission more consistently and realistically than previous simple 
formulations. Moreover, 
the relative simplicity of the formalism and the availability of analytical 
expressions for observables such as flux and lags should facilitate the 
extraction of the parameters and the estimation of the evolution of important 
physical quantities. In its turn, this knowledge will help us to better 
understand the properties of the central engines of GRBs and their 
surroundings. 

A real GRB originates from one or more colliding shells. In Paper I we 
mentioned that simulations of the Fermi processes~\citep{fermiaccspec} show 
that the energy distribution of electrons and fields as well as the evolution 
of the synchrotron emission~\citep{simuljet,simuljet0} are complex and simple 
distributions such as power-law or even a power-law with exponential cutoff 
are not good approximations for physical parameters. Nonetheless, within the 
precision of the available data, they can be sufficiently close 
approximations in a short time interval. Therefore, we can model an entire 
burst by dividing it to separate regimes. However, we must first explore 
the parameter space and understand how various parameters affect the 
synchrotron emission that is our only observable. This task is 
not simple because the model has a large number of parameters. We leave the 
systematic exploration of the parameter space to a future work. Here our goal 
is to see if at least for some set of parameters this model can reproduce the 
observed features of the GRB emission. 

We consider a few set of parameters, determine various physical quantities 
explained in Paper I and their evolution, such as $\beta' (r')$,  
$\omega'_m (r')$ (The definition of parameters are summarized in 
Table \ref{tab:param} and details can found in Paper I.), and observables 
such as light curves and spectrum. We remind that these simulations meant to 
correspond to a time interval in the life of a GRB and {\it do not present the 
whole emission}. For determination of the light curves and spectrum we have 
used equation (60-Paper I) and we have neglected the terms proportional to 
${\mathcal G}$ which are the contribution of high latitude emission and their 
Doppler shift.

\section {Simulations}\label{sec:simul}
\begin{figure*}
\begin{center}
\begin{tabular}{cccc}
1)\includegraphics[height=4cm,angle=-90]{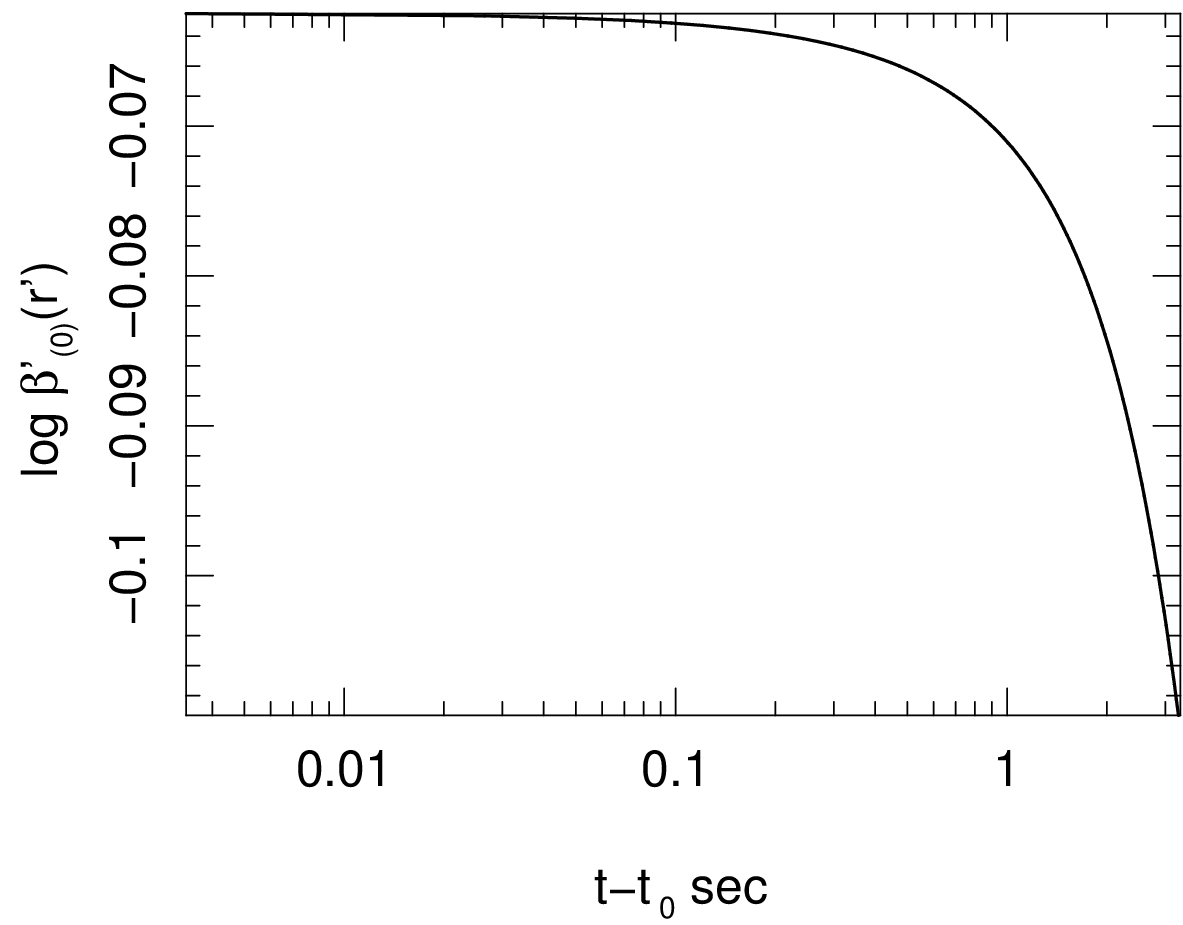} &
2)\includegraphics[height=4cm,angle=-90]{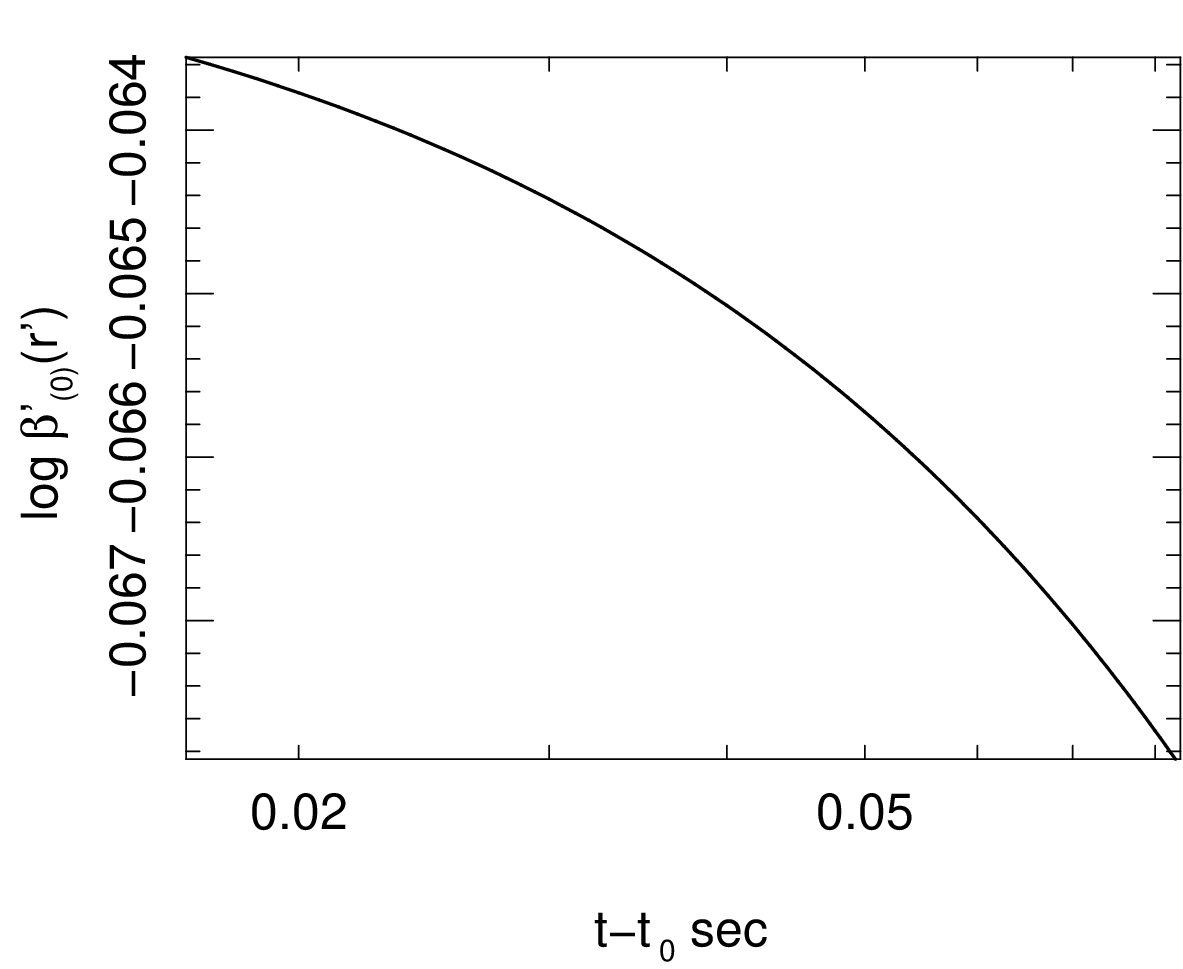} & 
3)\includegraphics[height=4cm,angle=-90]{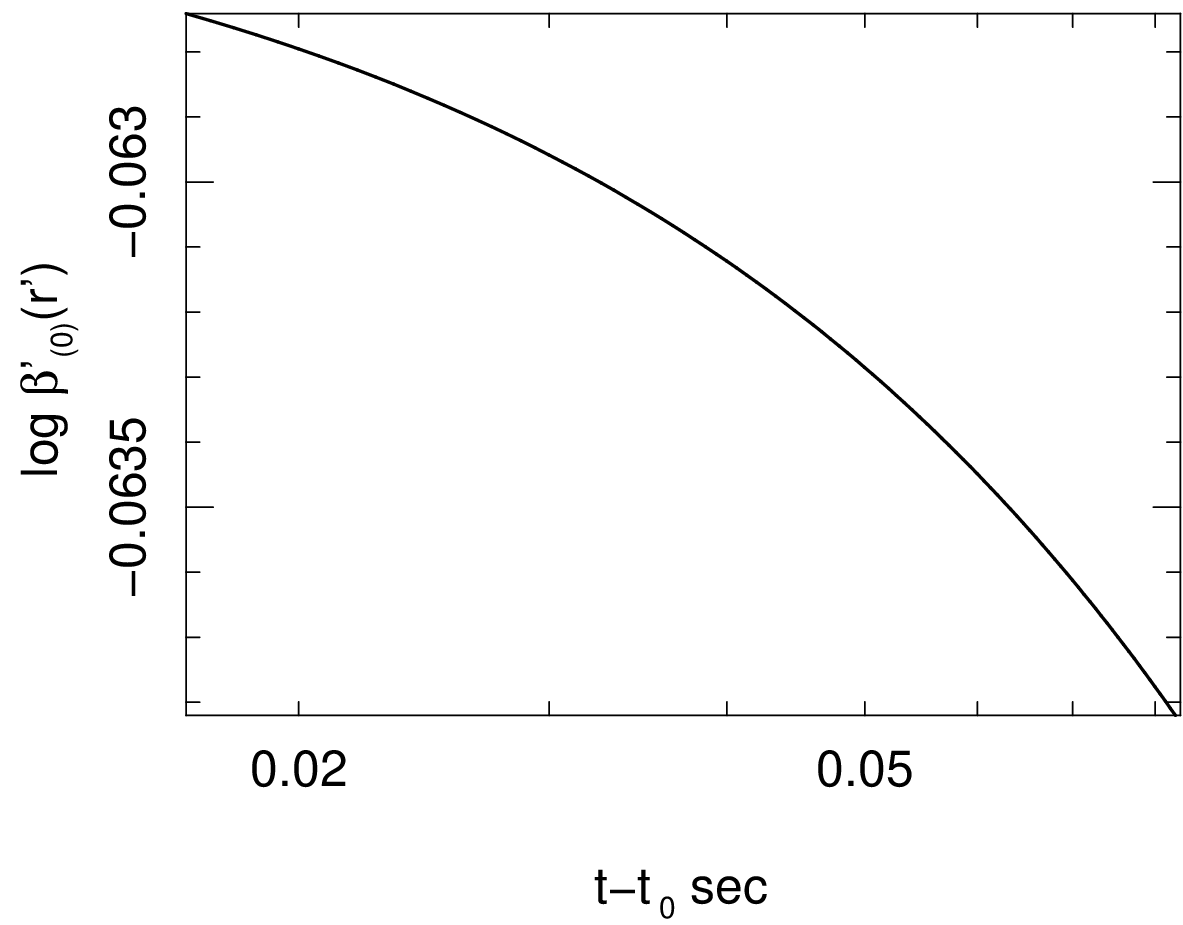} &
4)\includegraphics[height=4cm,angle=-90]{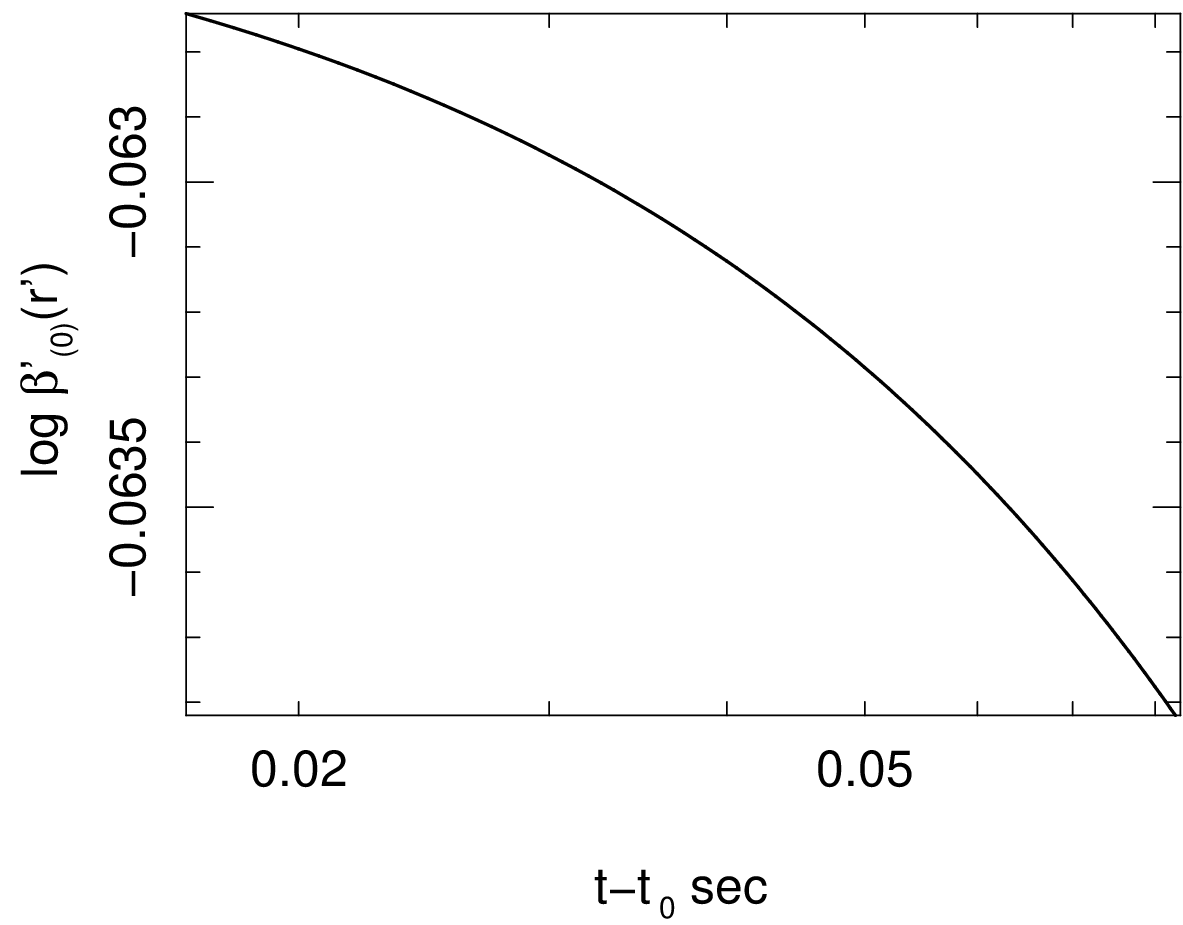} \\ 
5)\includegraphics[height=4cm,angle=-90]{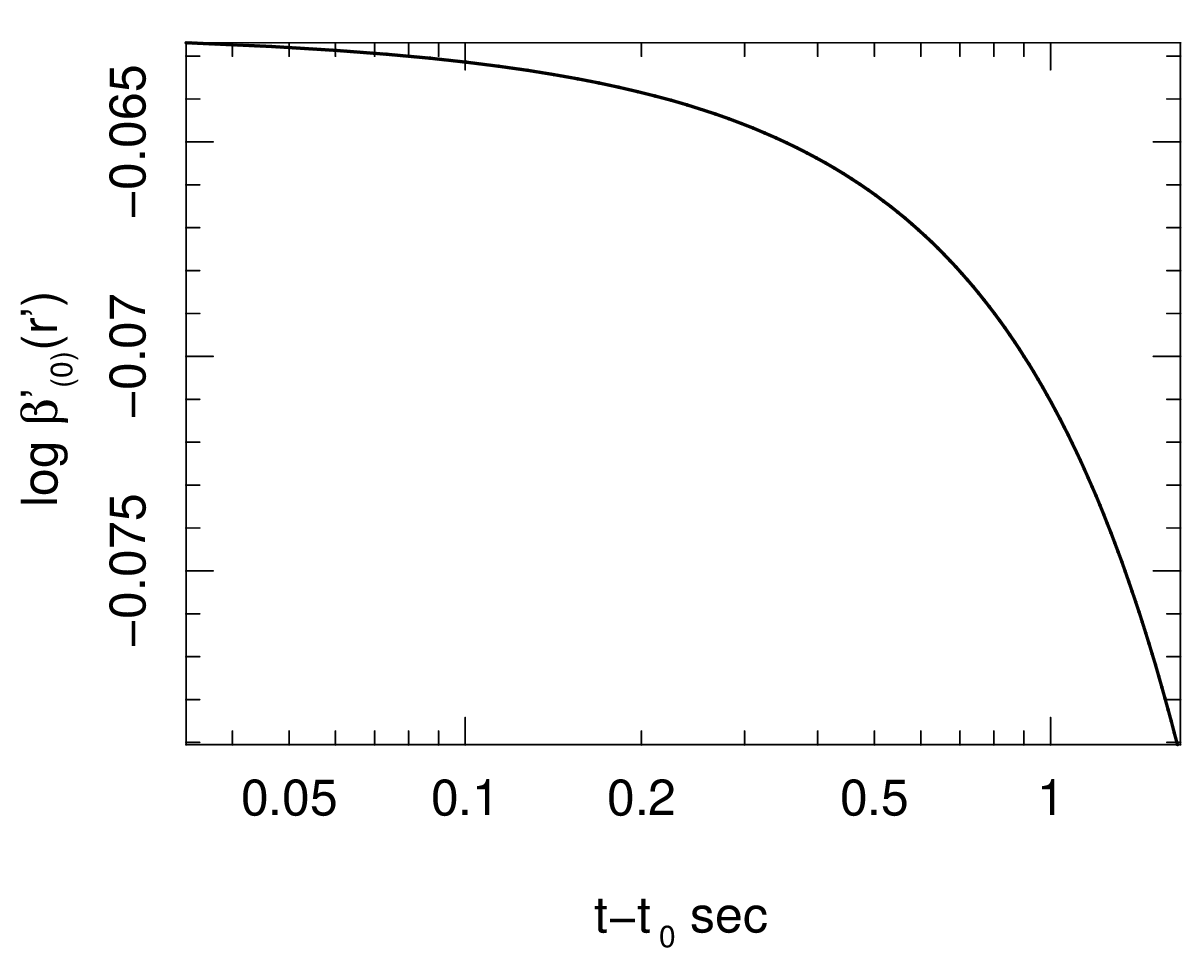} &
6)\includegraphics[height=4cm,angle=-90]{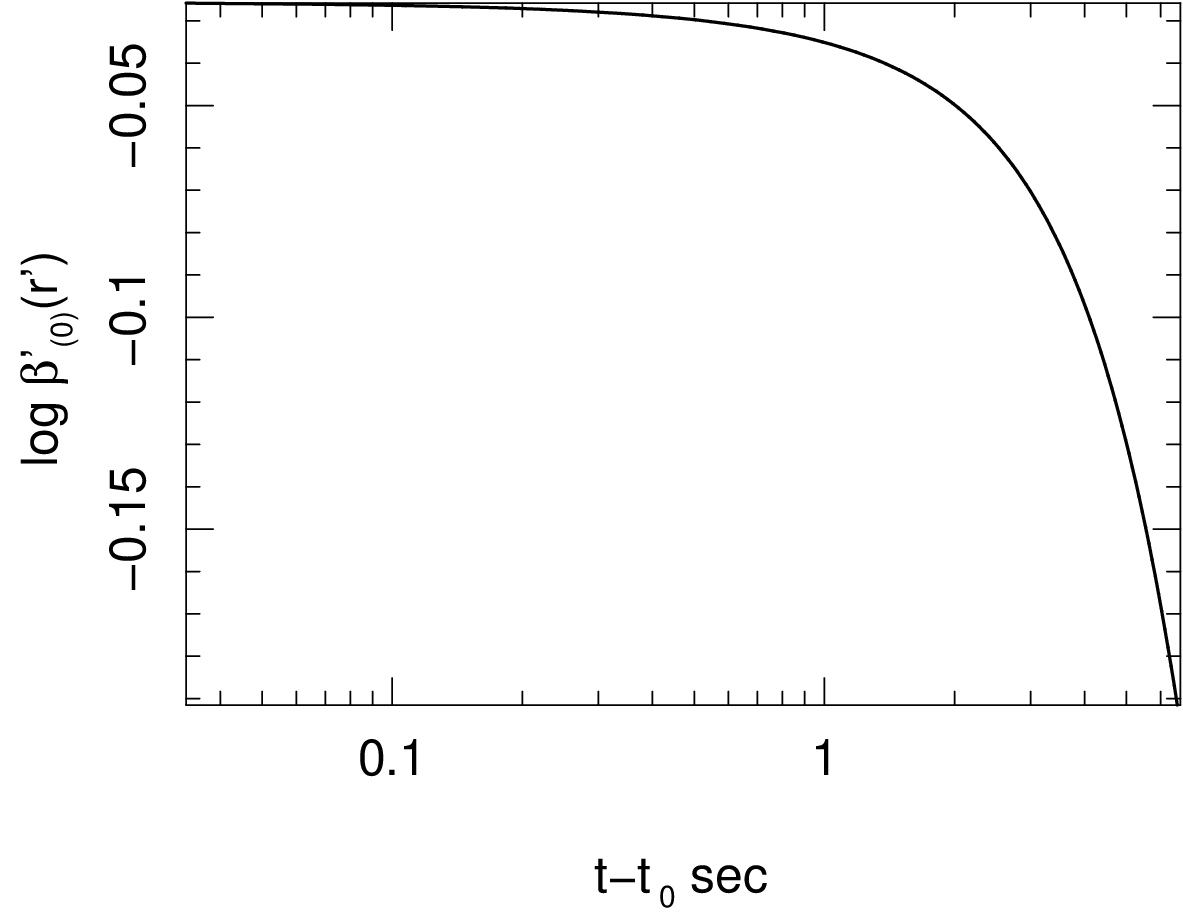} &
7)\includegraphics[height=4cm,angle=-90]{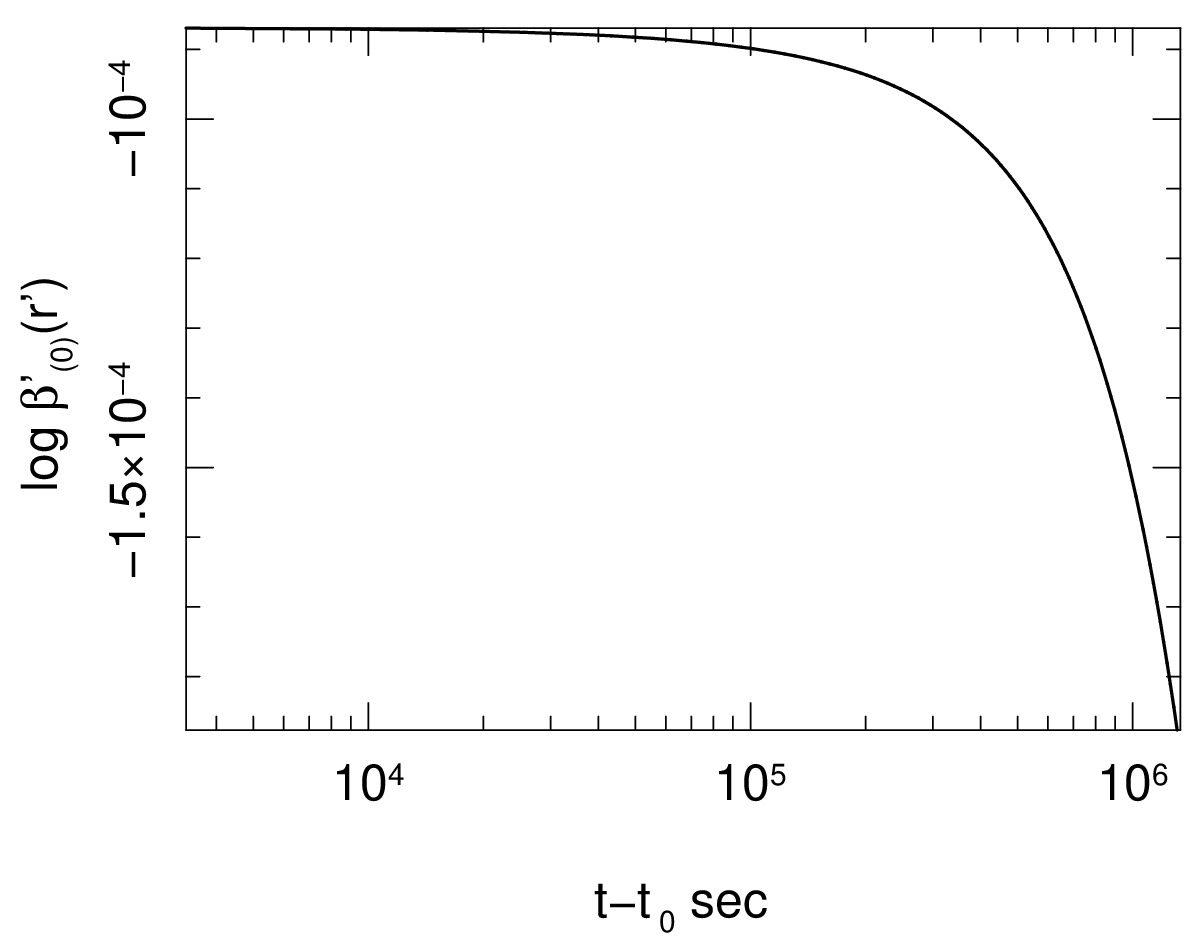} &
8)\includegraphics[height=4cm,angle=-90]{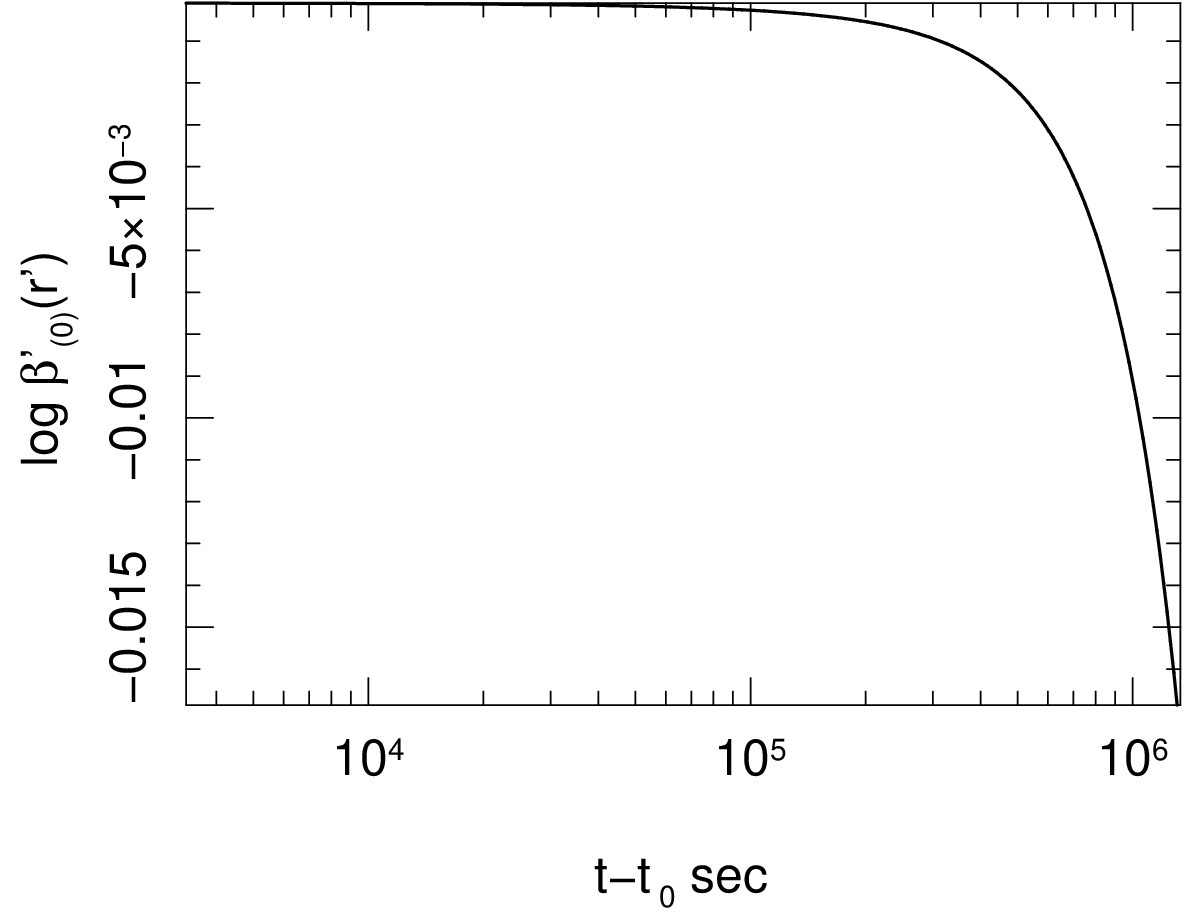} \\
9)\includegraphics[height=4cm,angle=-90]{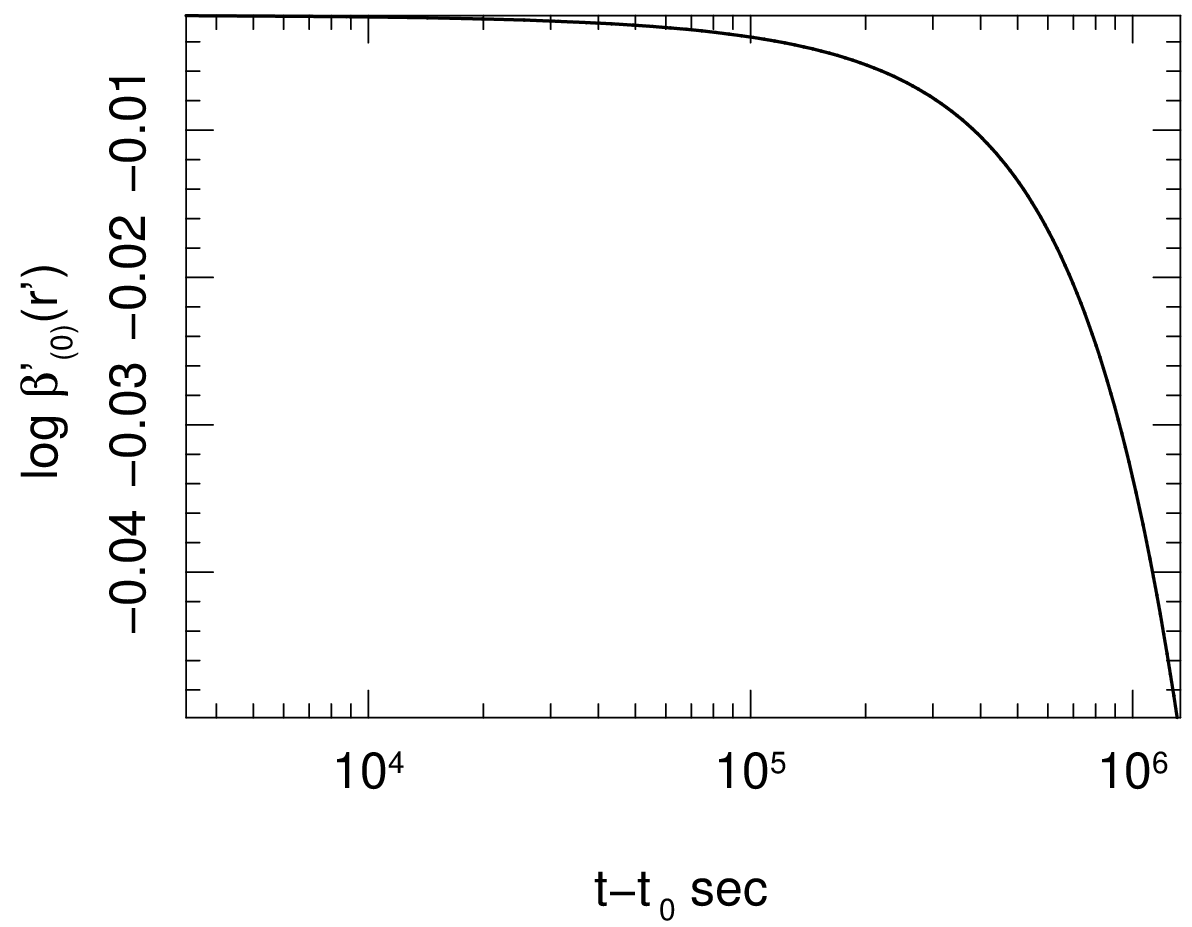} &
10)\includegraphics[height=4cm,angle=-90]{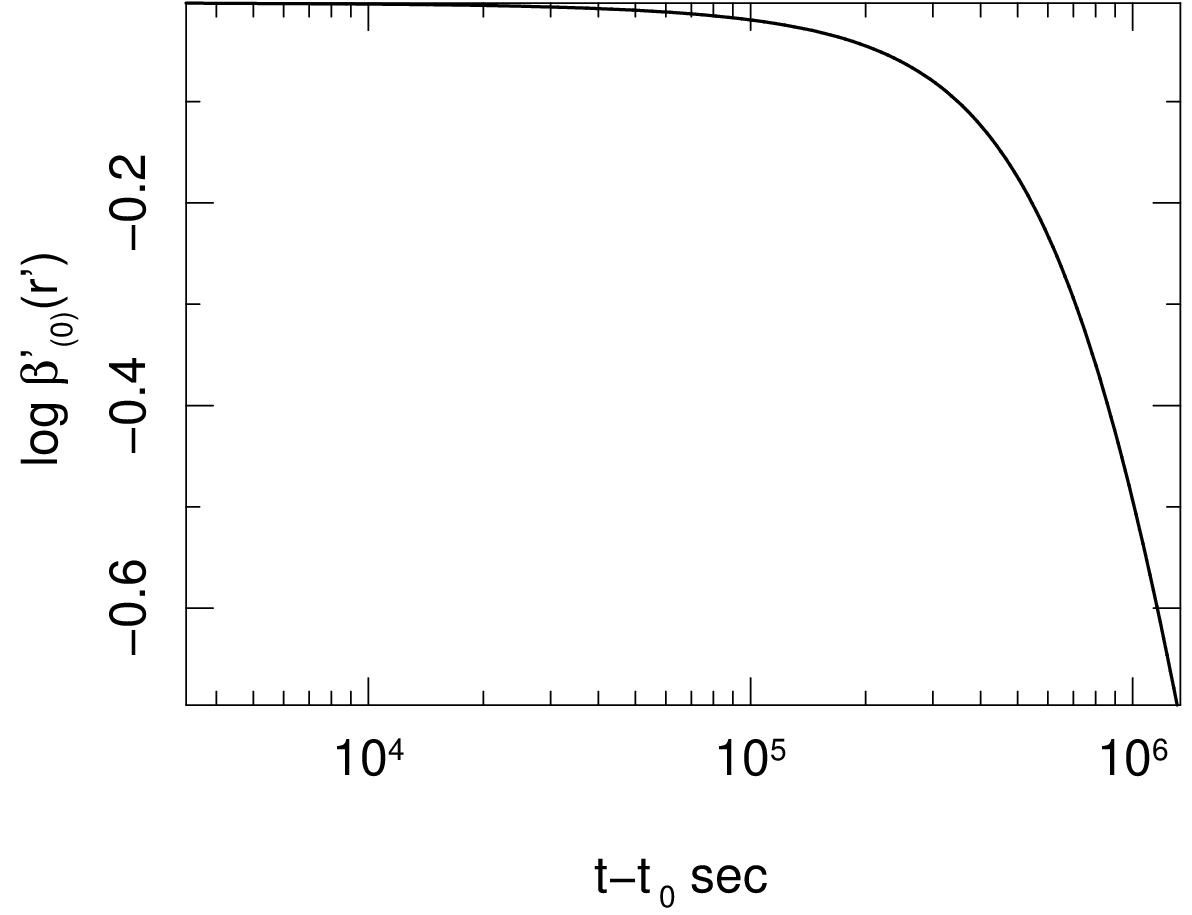} &
11)\includegraphics[height=4cm,angle=-90]{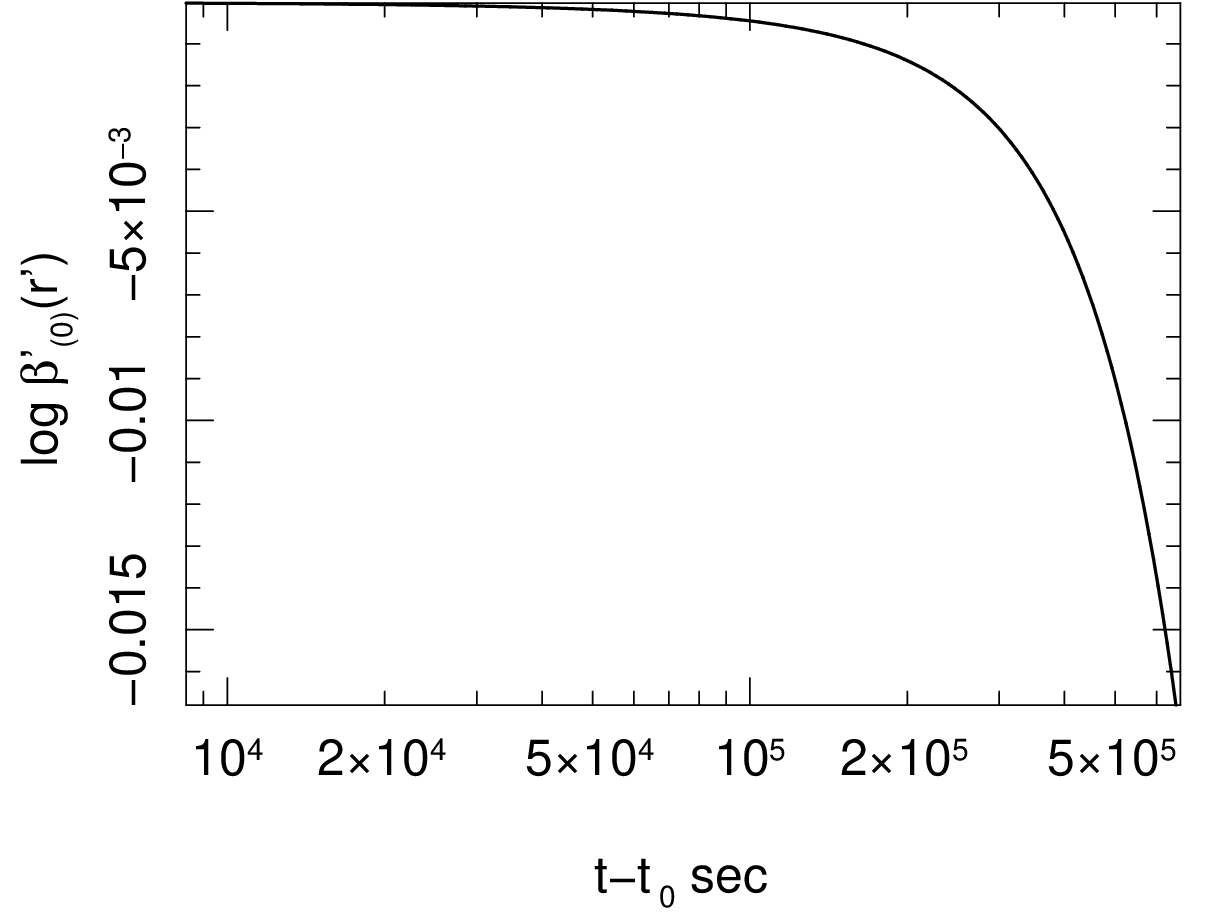} &
12)\includegraphics[height=4cm,angle=-90]{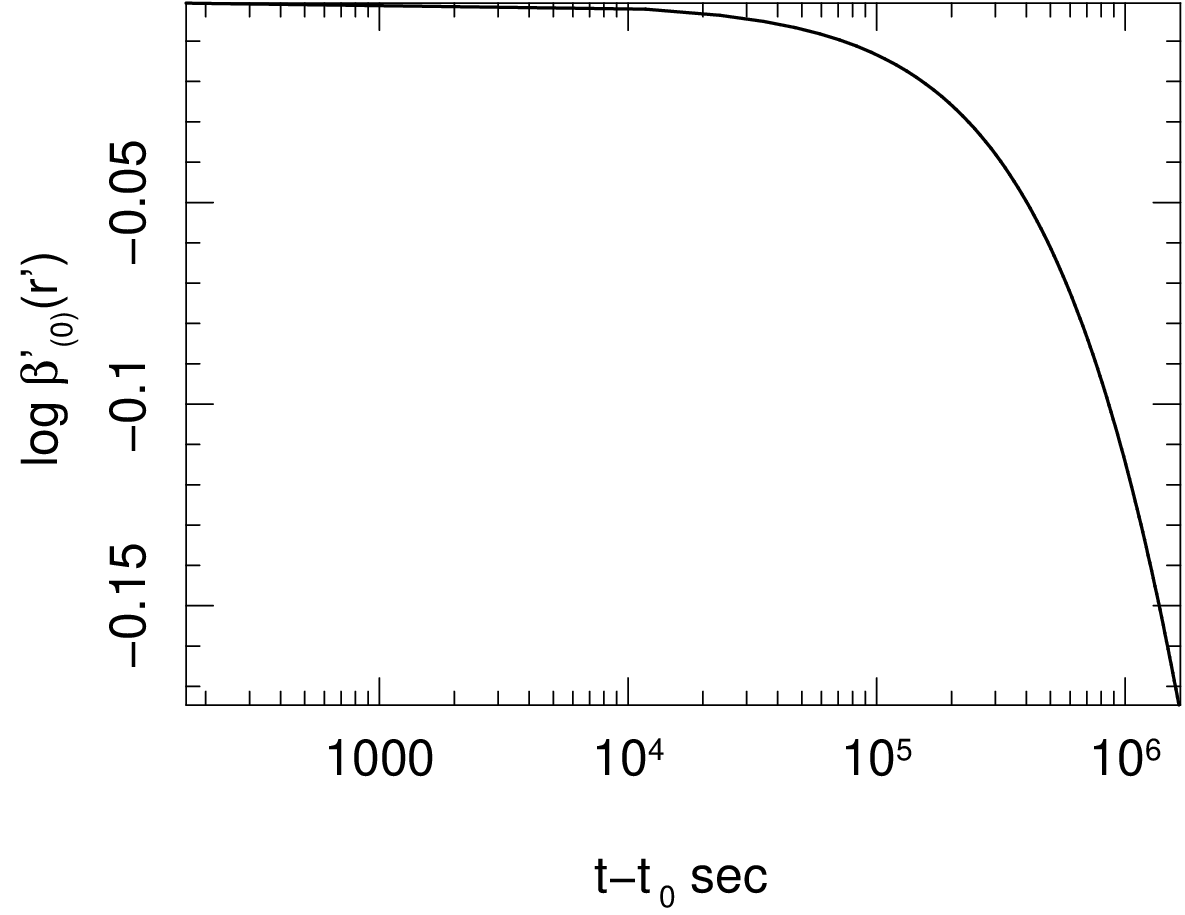} \\
13)\includegraphics[height=4cm,angle=-90]{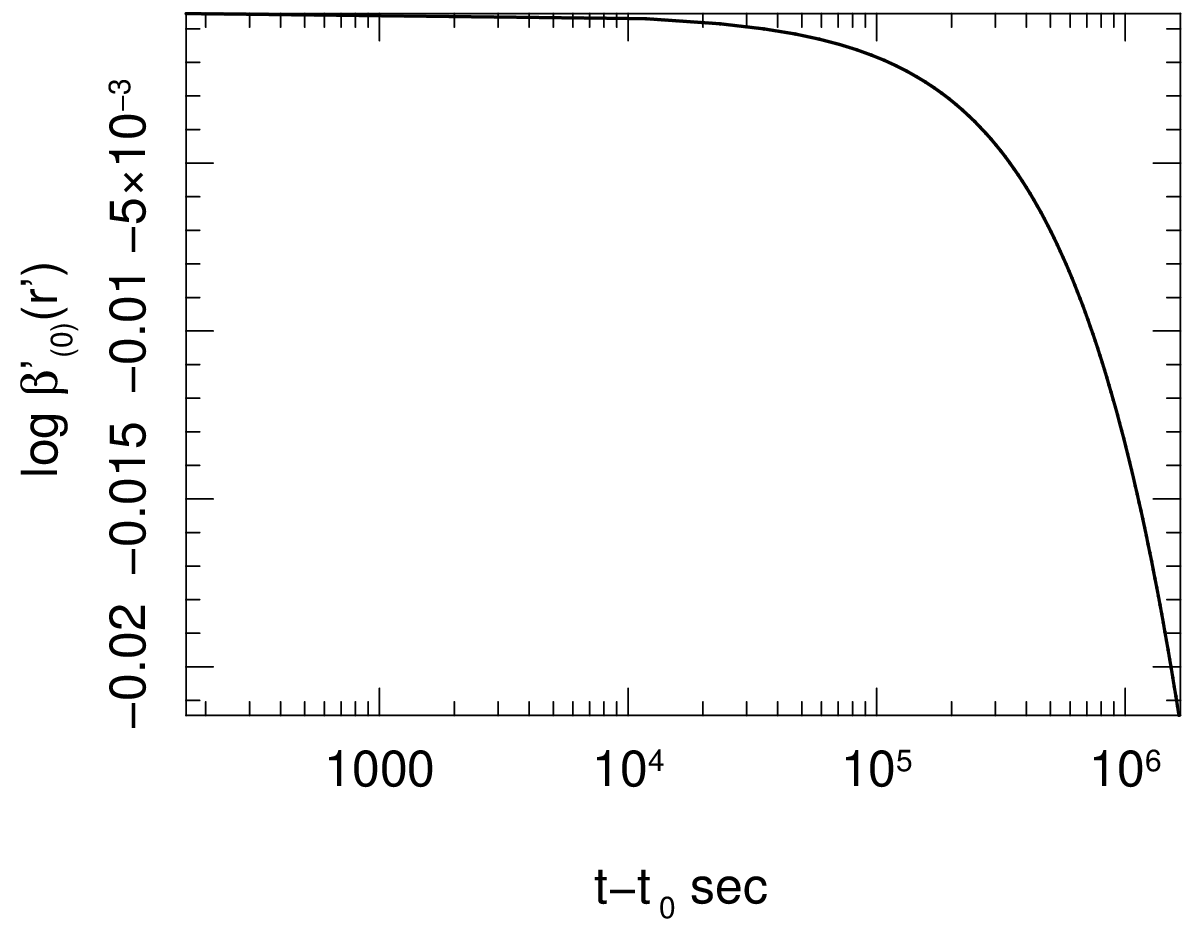} &
14)\includegraphics[height=4cm,angle=-90]{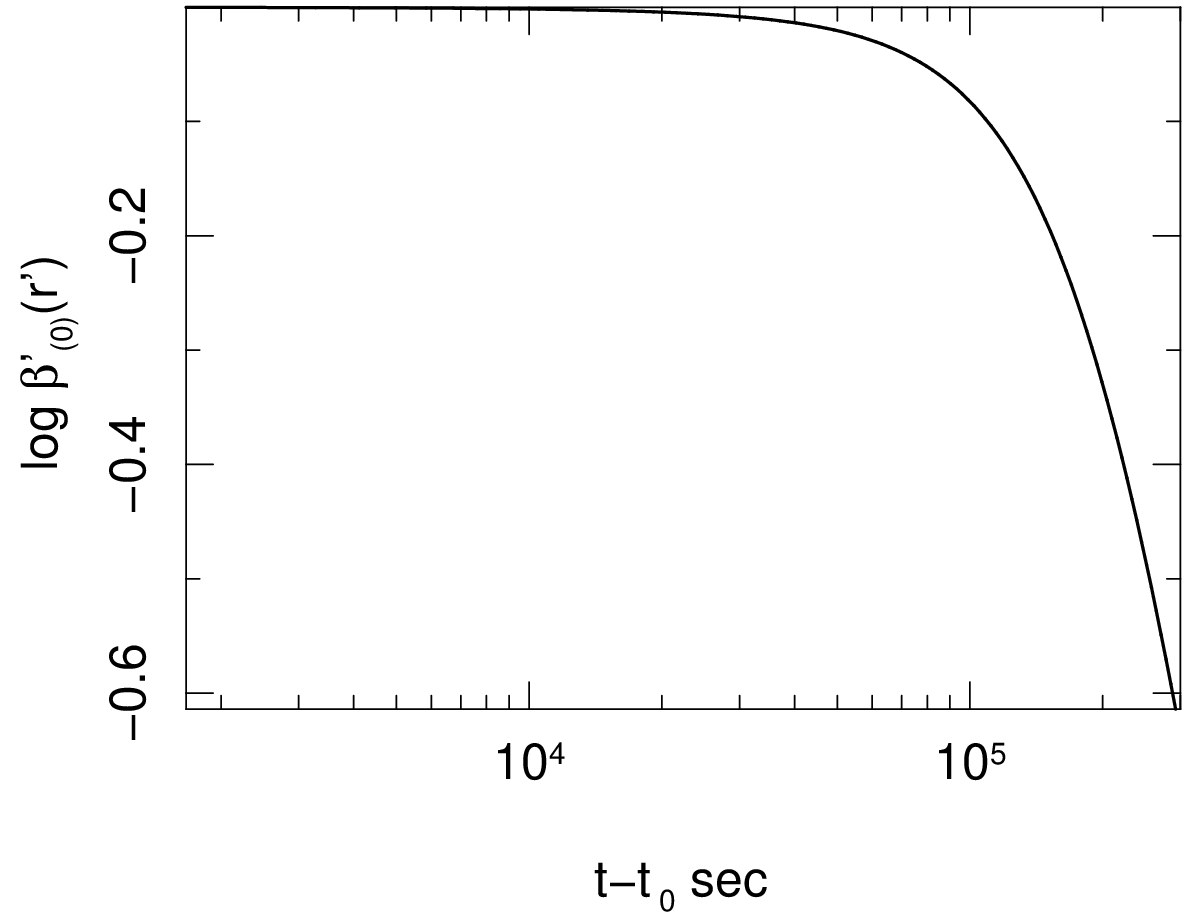} &
15)\includegraphics[height=4cm,angle=-90]{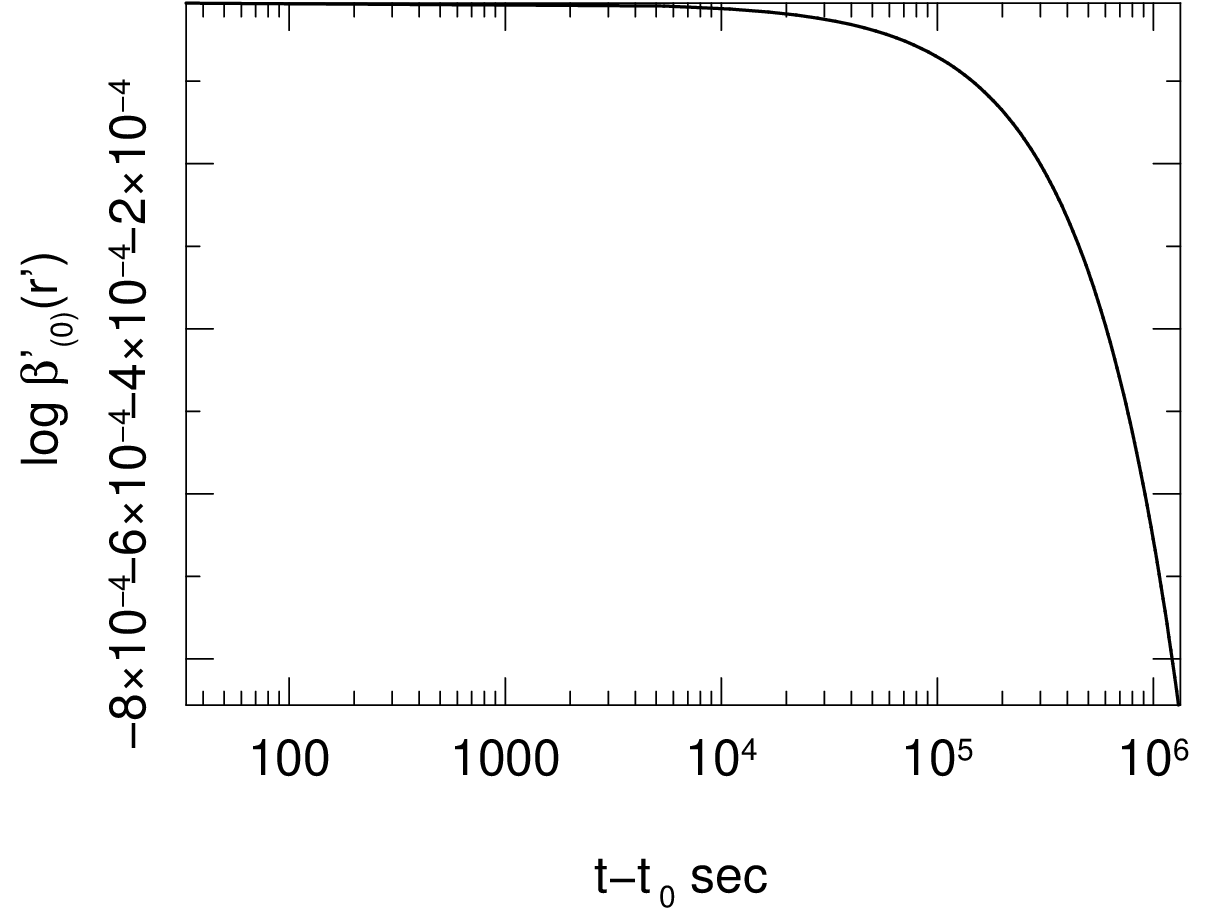} &
16)\includegraphics[height=4cm,angle=-90]{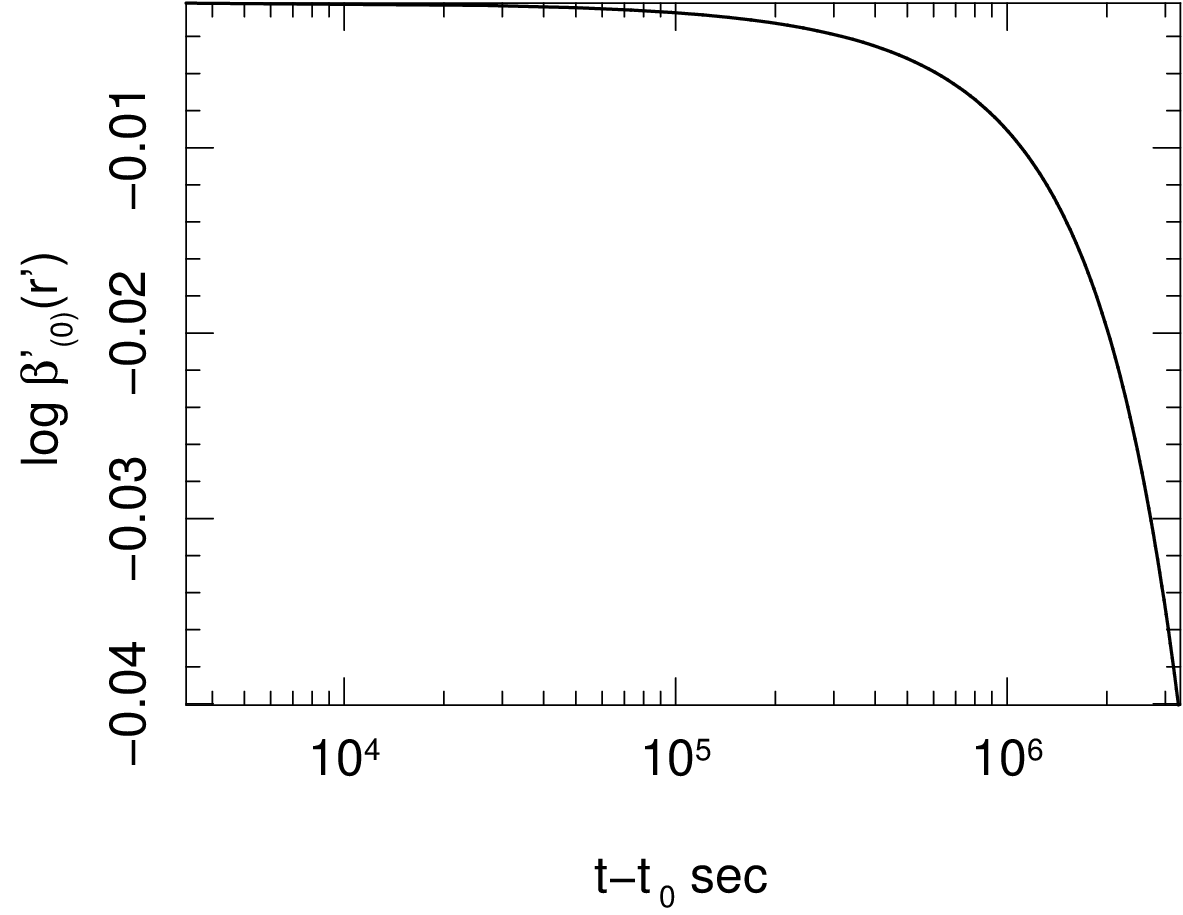} \\
17)\includegraphics[height=4cm,angle=-90]{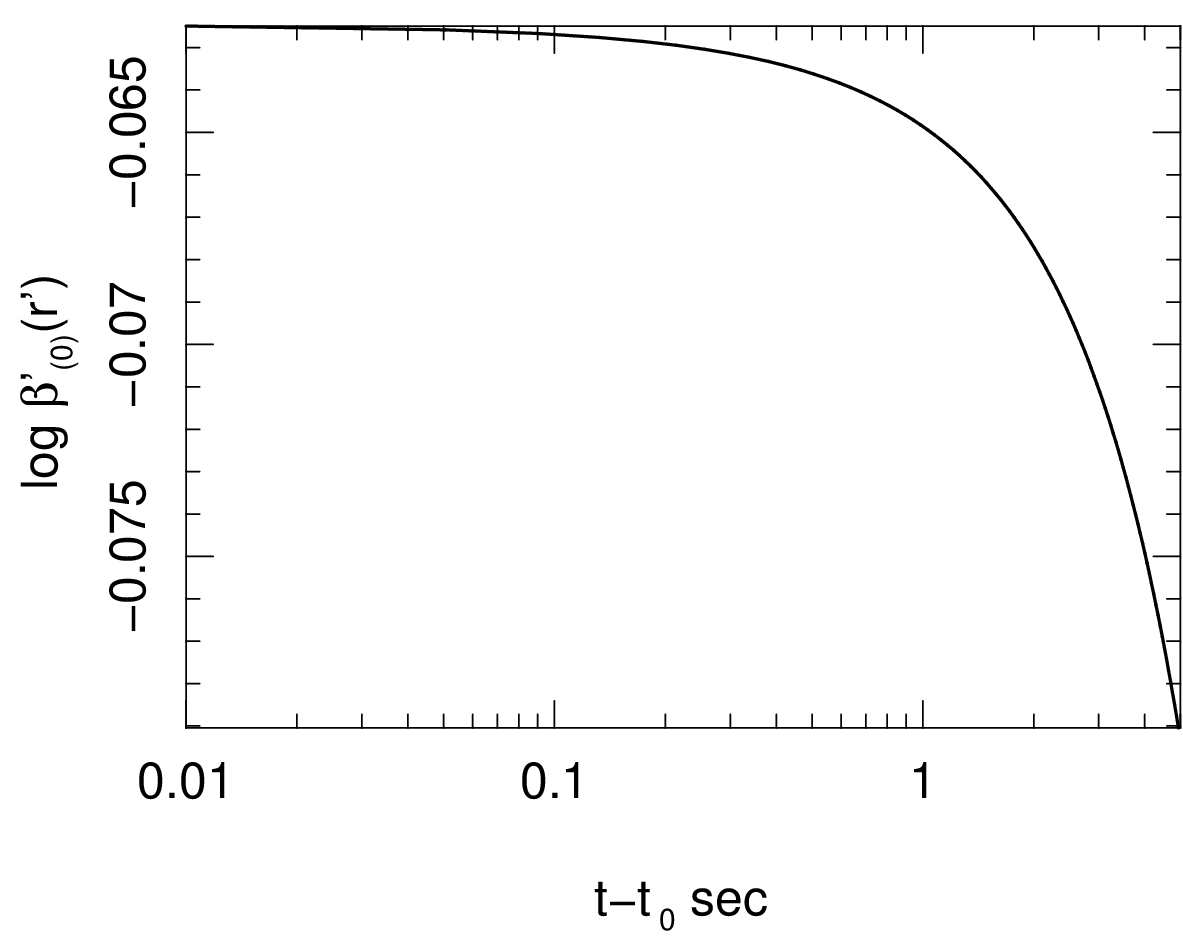} &
18)\includegraphics[height=4cm,angle=-90]{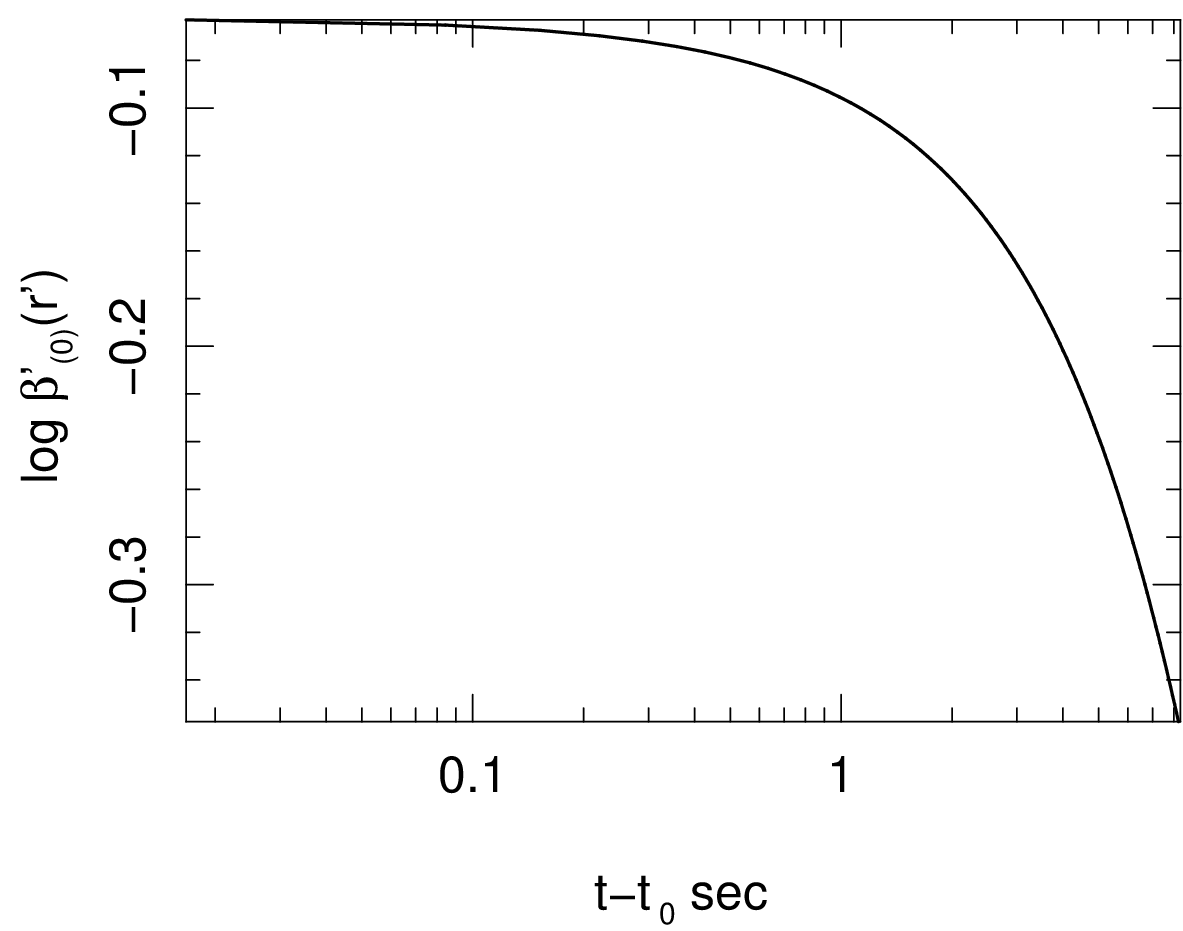} &
19)\includegraphics[height=4cm,angle=-90]{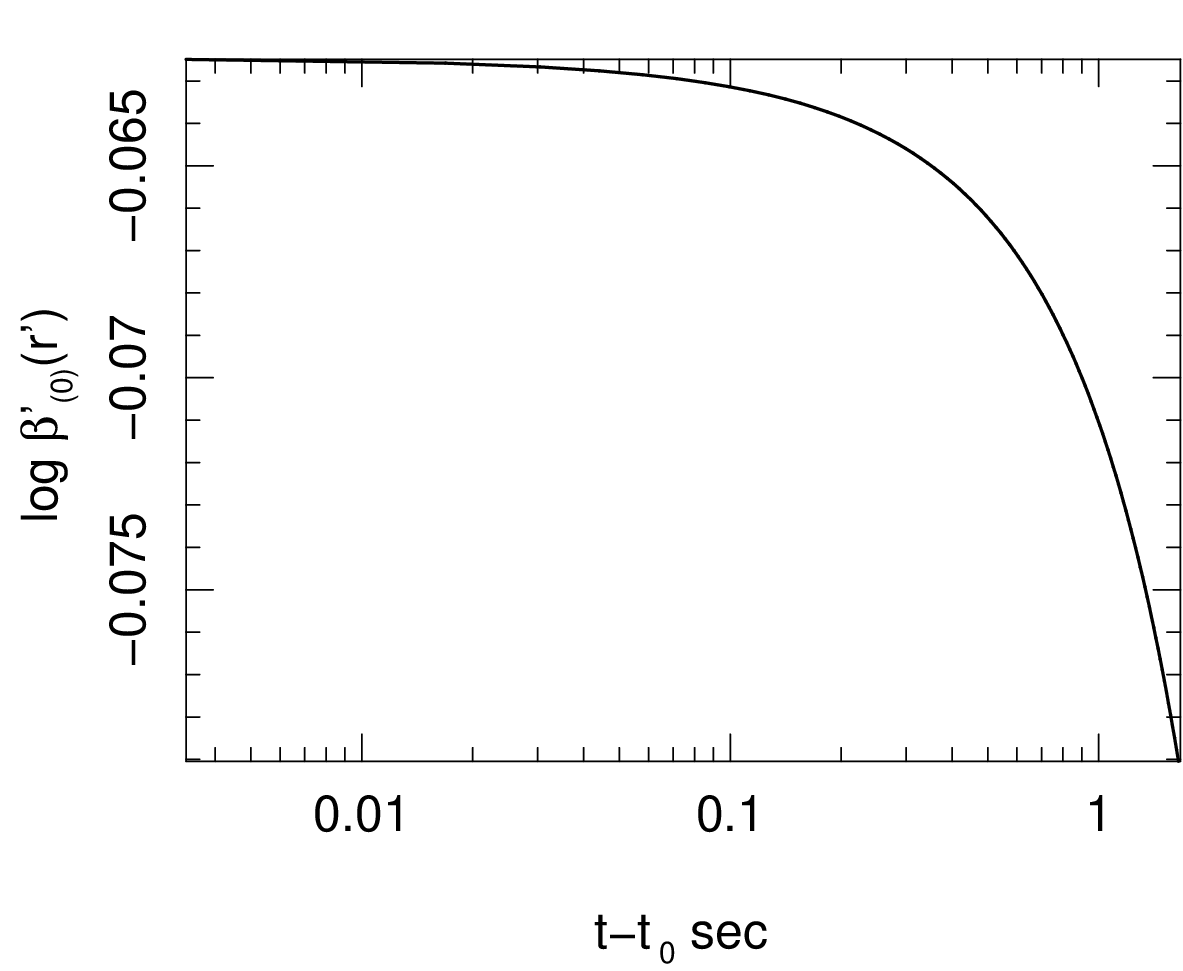} &
20)\includegraphics[height=4cm,angle=-90]{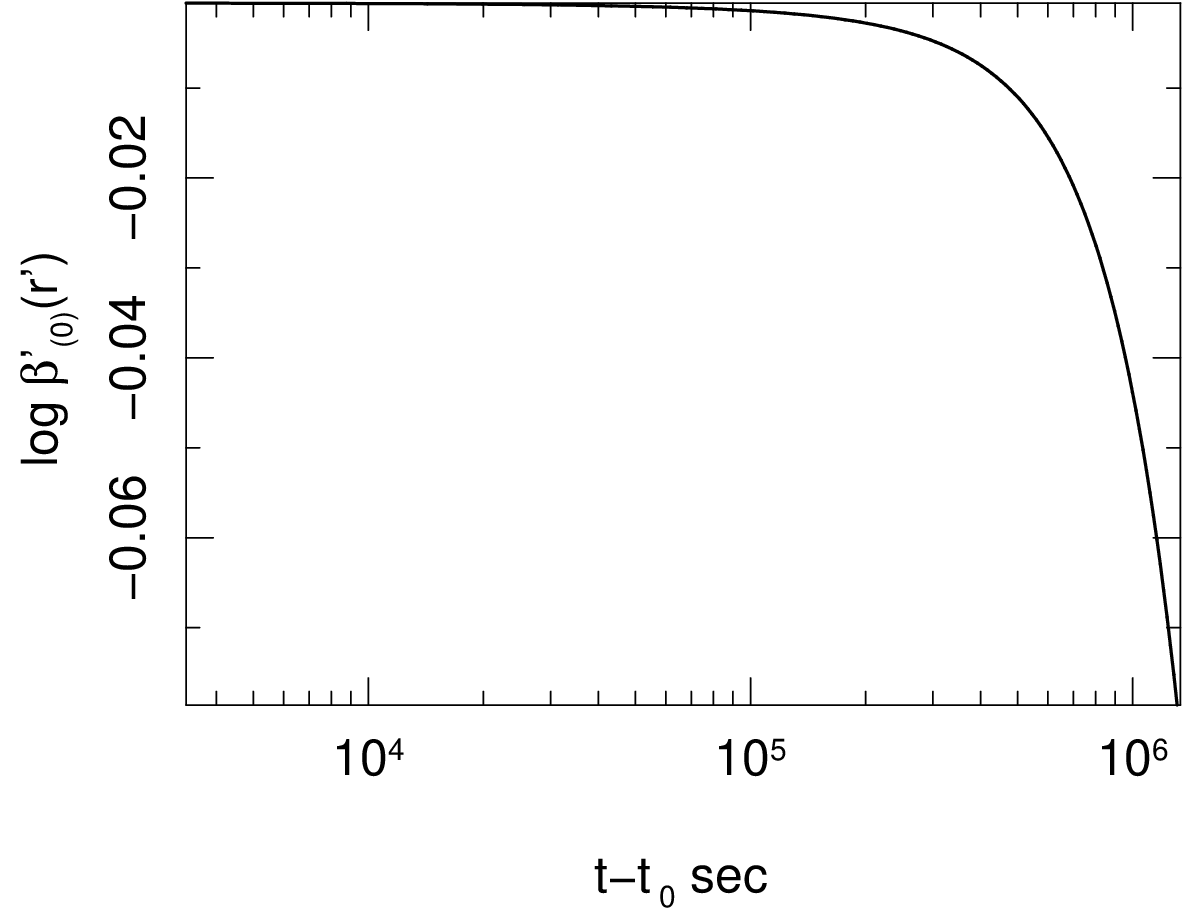} \\
21)\includegraphics[height=4cm,angle=-90]{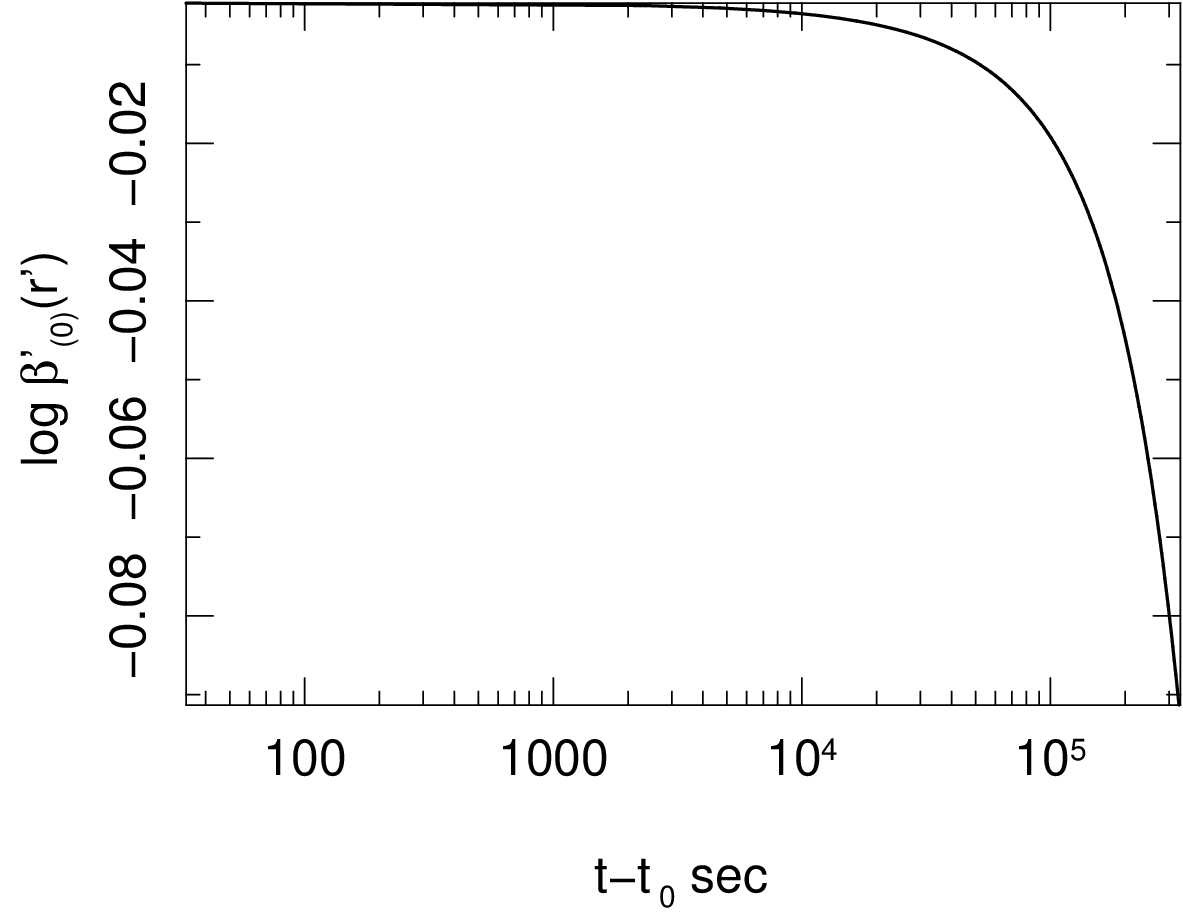} &
22)\includegraphics[height=4cm,angle=-90]{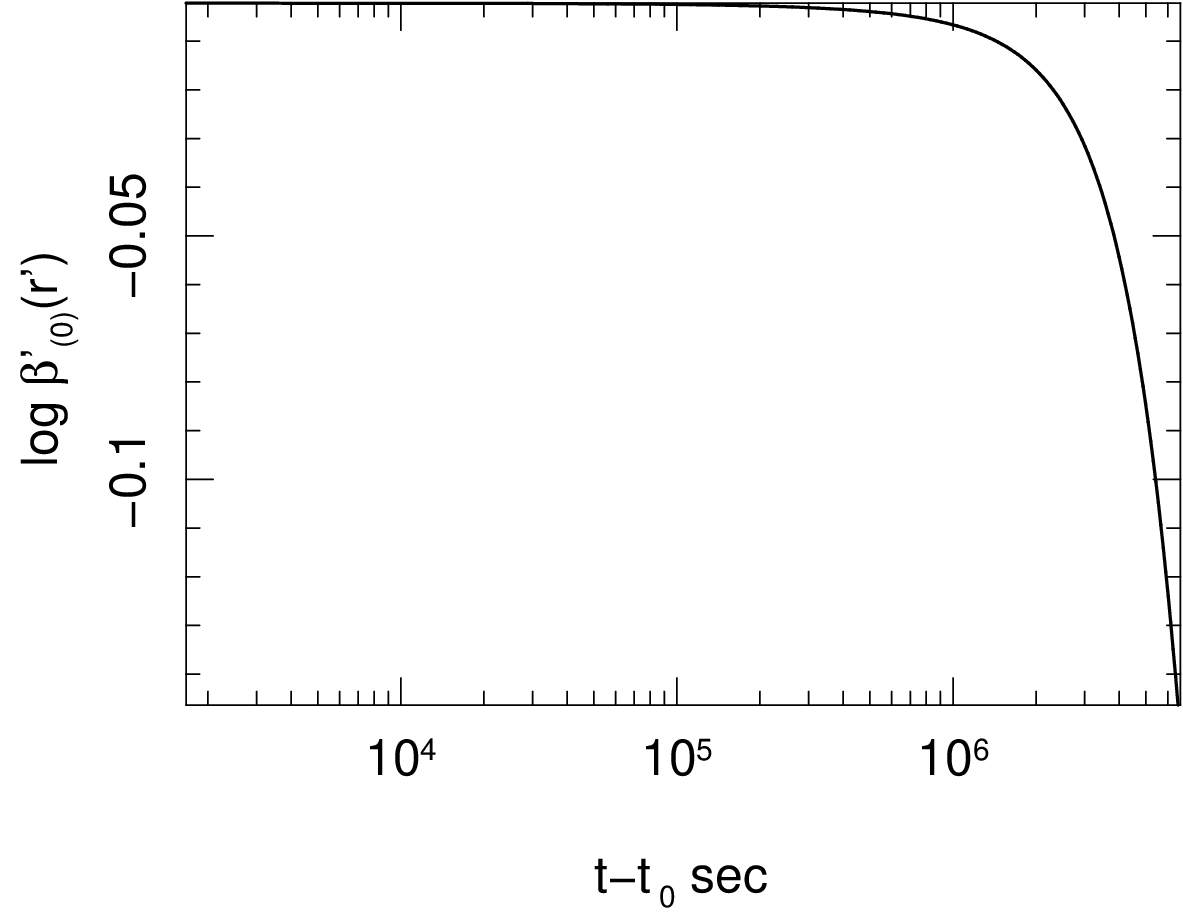} &
23)\includegraphics[height=4cm,angle=-90]{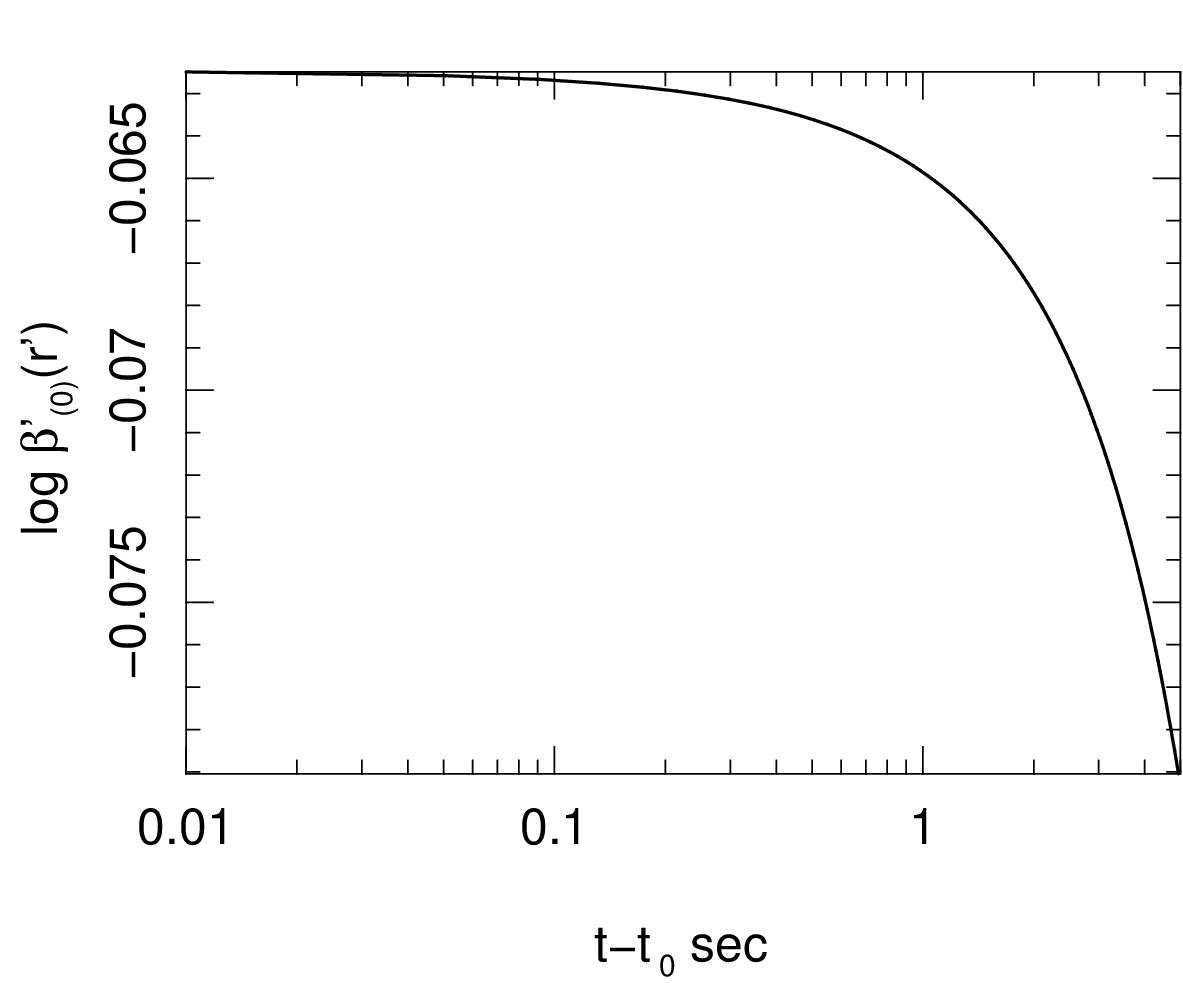} &
24)\includegraphics[height=4cm,angle=-90]{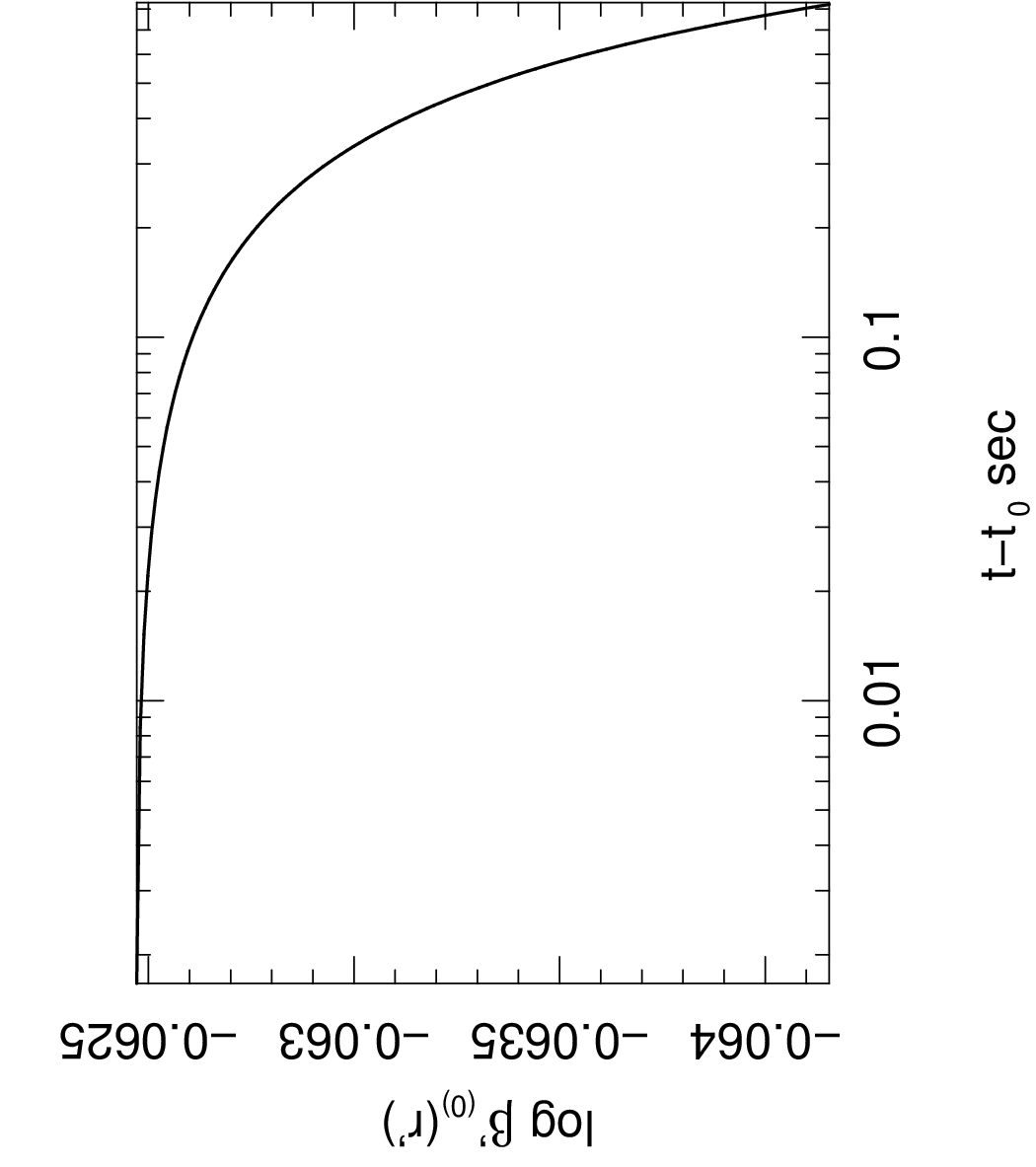} 
\end {tabular}
\end {center}
\caption{Evolution of $\beta'_0$ with time/radius in observer reference frame. 
The number in the corner of each plot refers to the model number in Table 
\ref{tab:param}.
\label{fig:betazero}}
\end{figure*}

\begin{figure*}
\begin{center}
\begin {tabular}{cccc} 
1)\includegraphics[height=4cm,angle=-90]{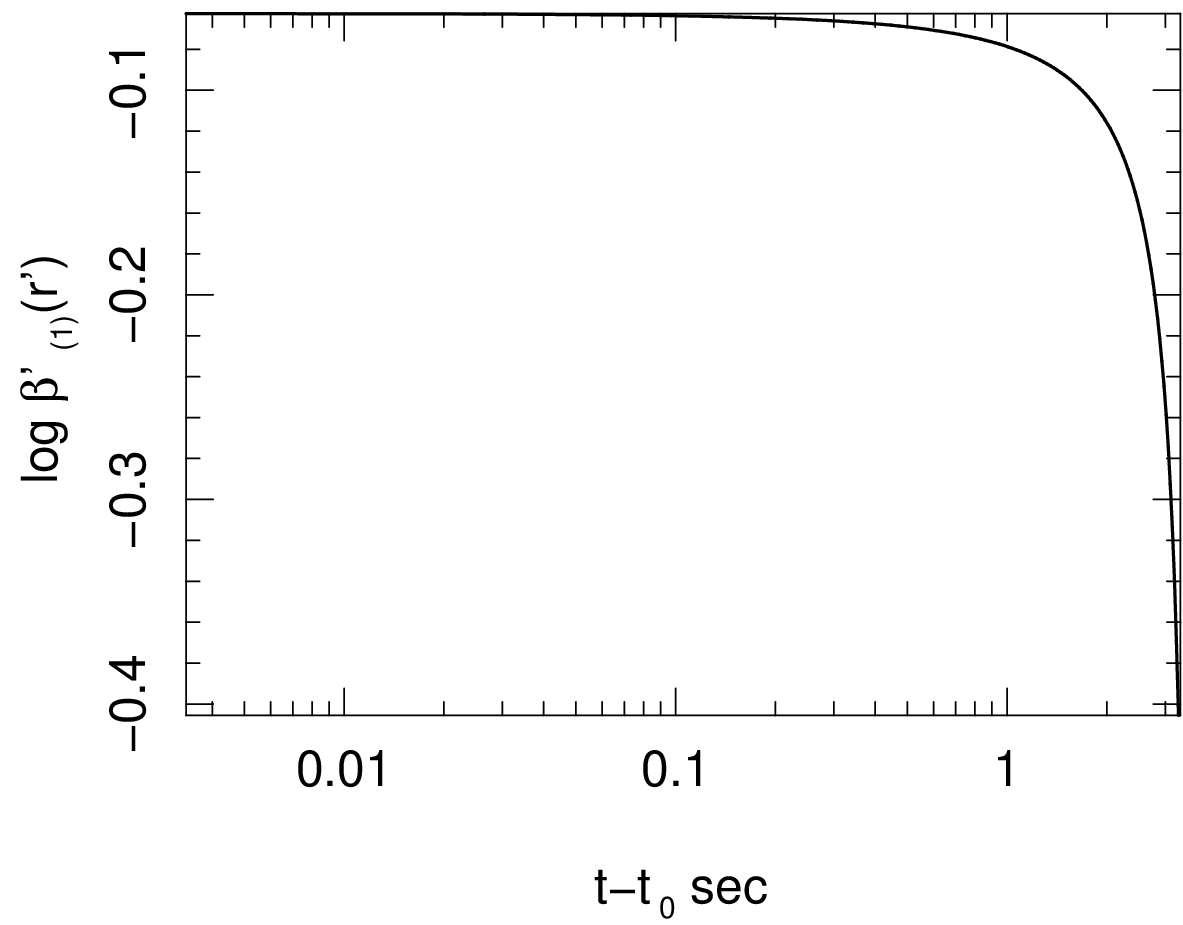} &
2)\includegraphics[height=4cm,angle=-90]{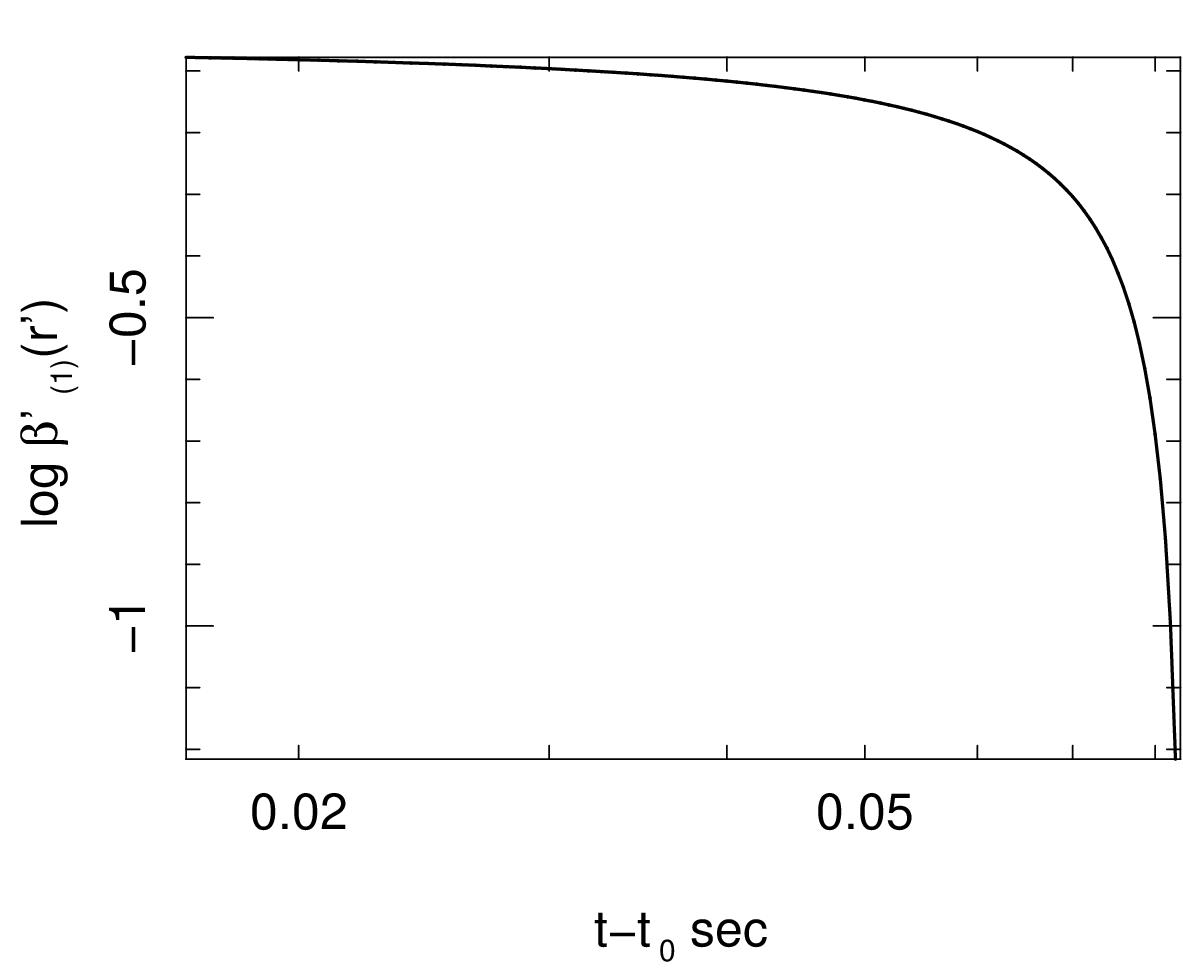} & 
3)\includegraphics[height=4cm,angle=-90]{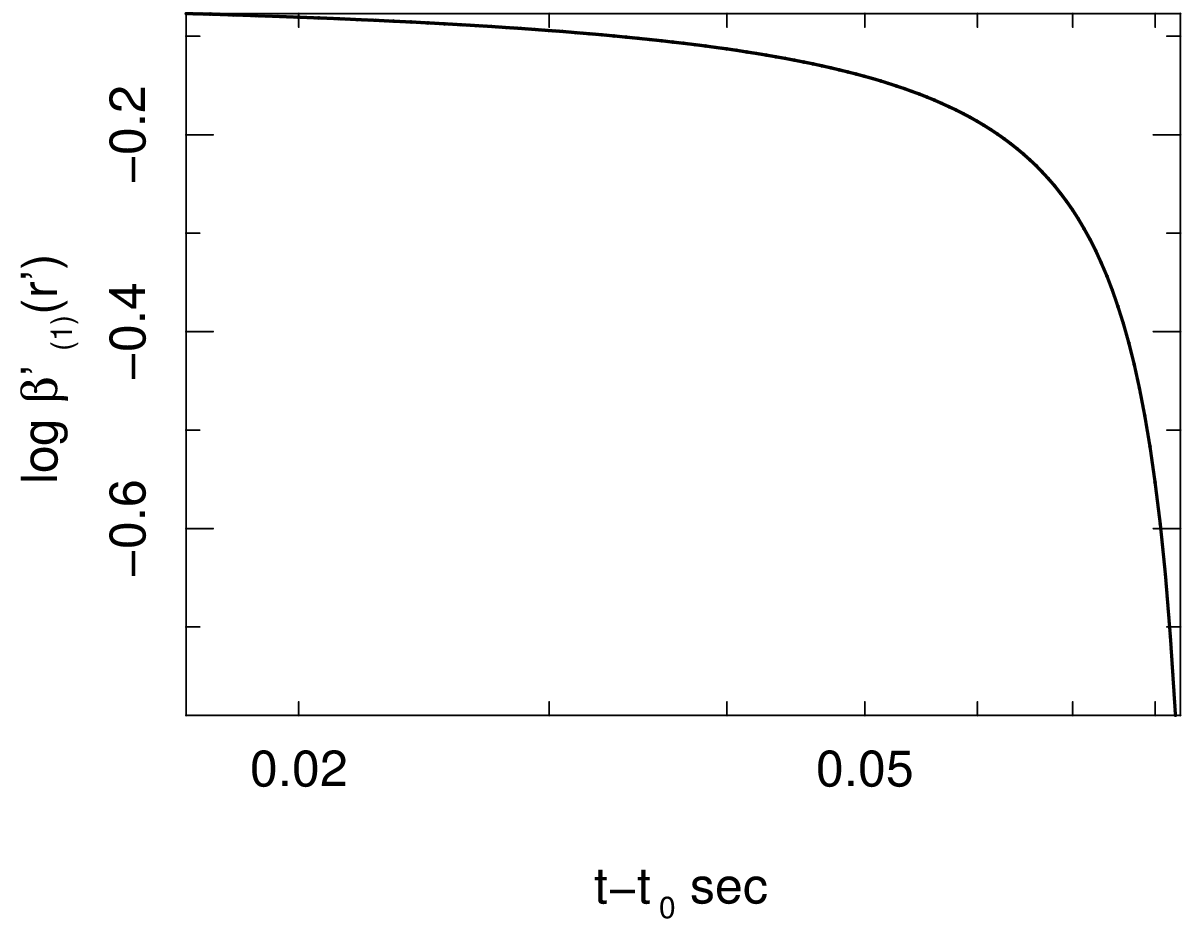} &
4)\includegraphics[height=4cm,angle=-90]{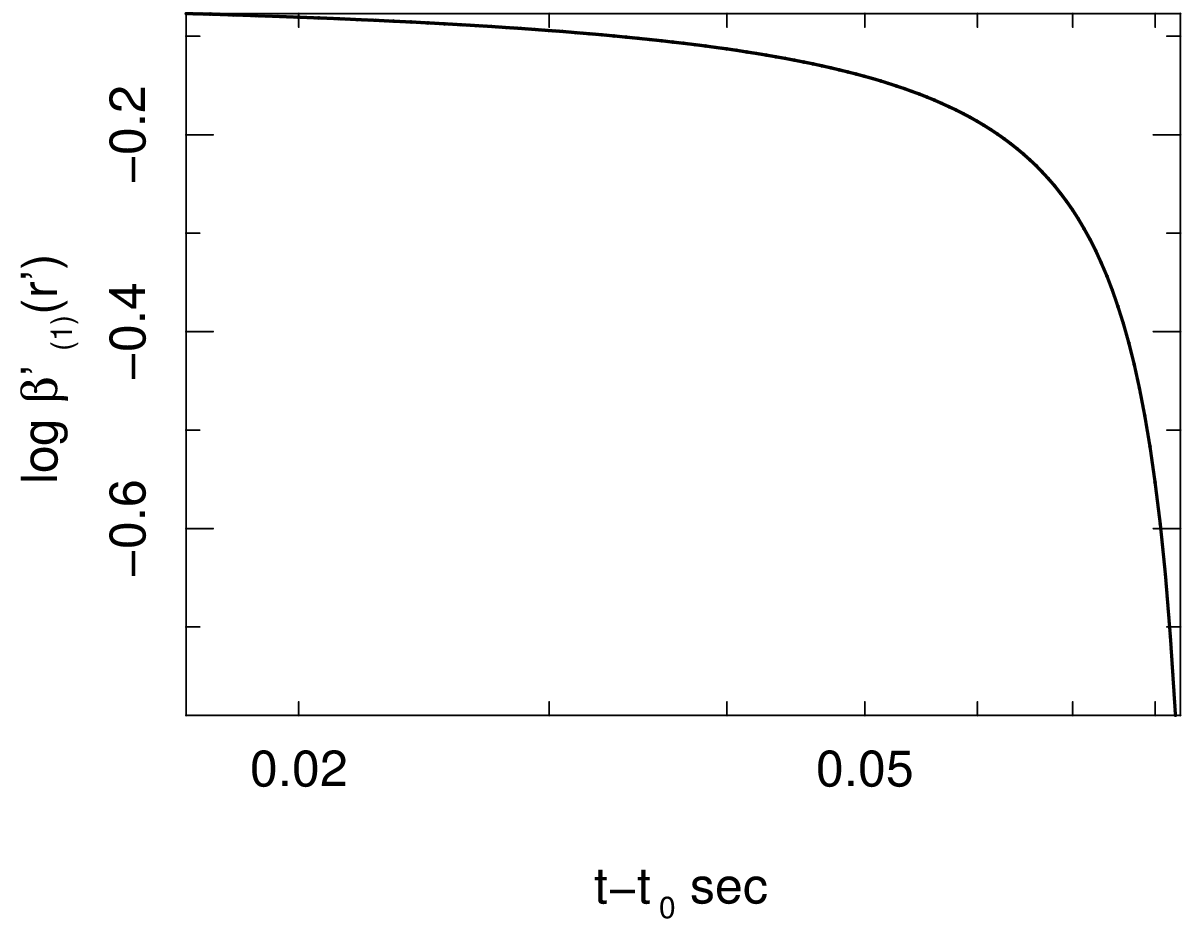} \\ 
5)\includegraphics[height=4cm,angle=-90]{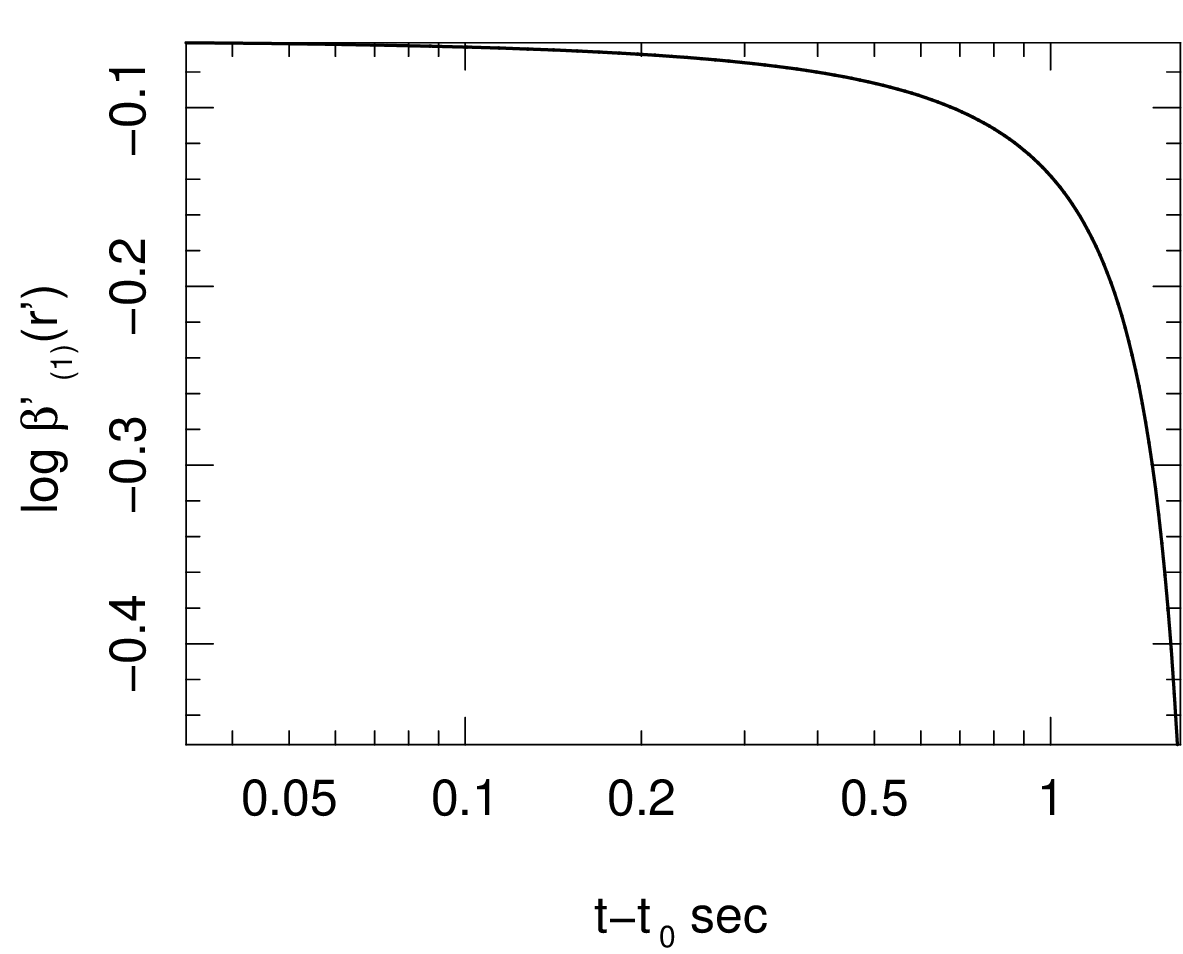} &
6)\includegraphics[height=4cm,angle=-90]{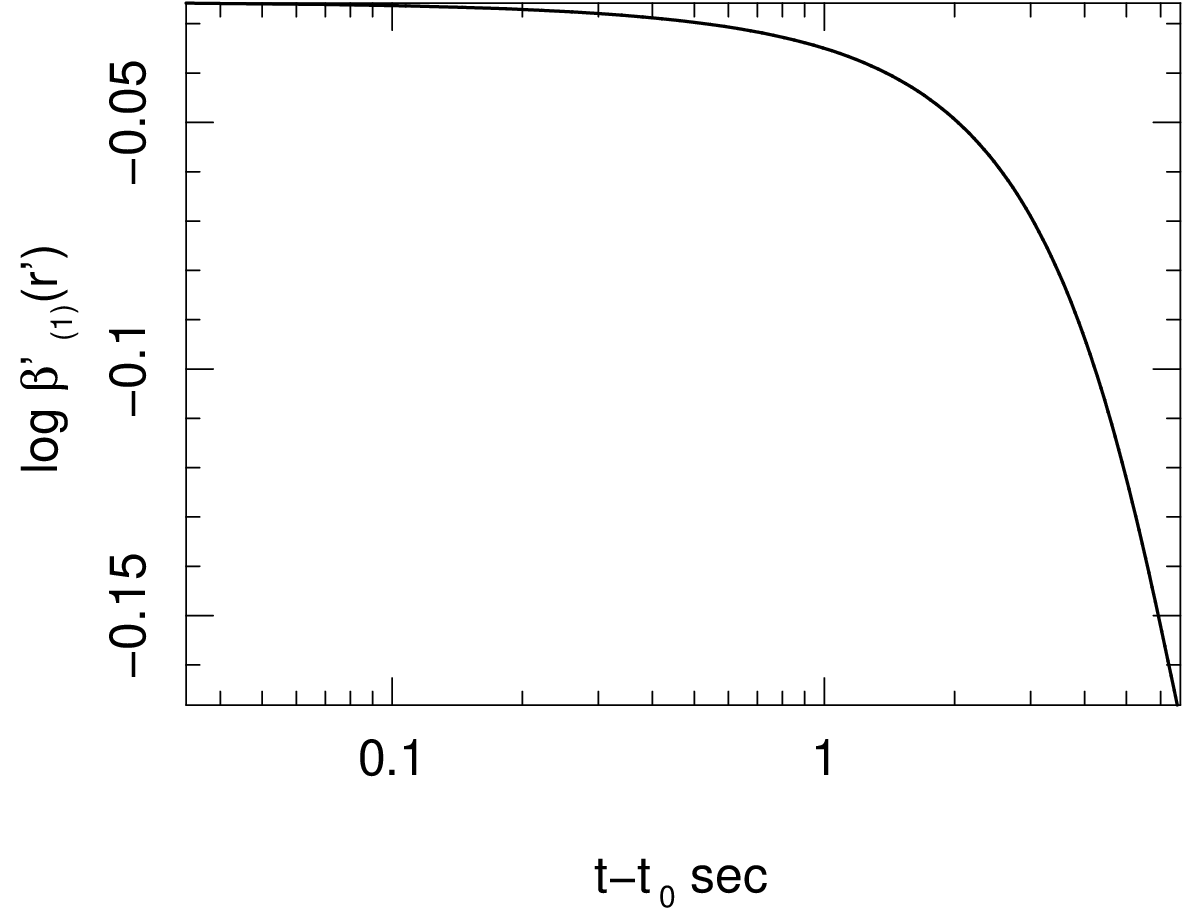} & 
7)\includegraphics[height=4cm,angle=-90]{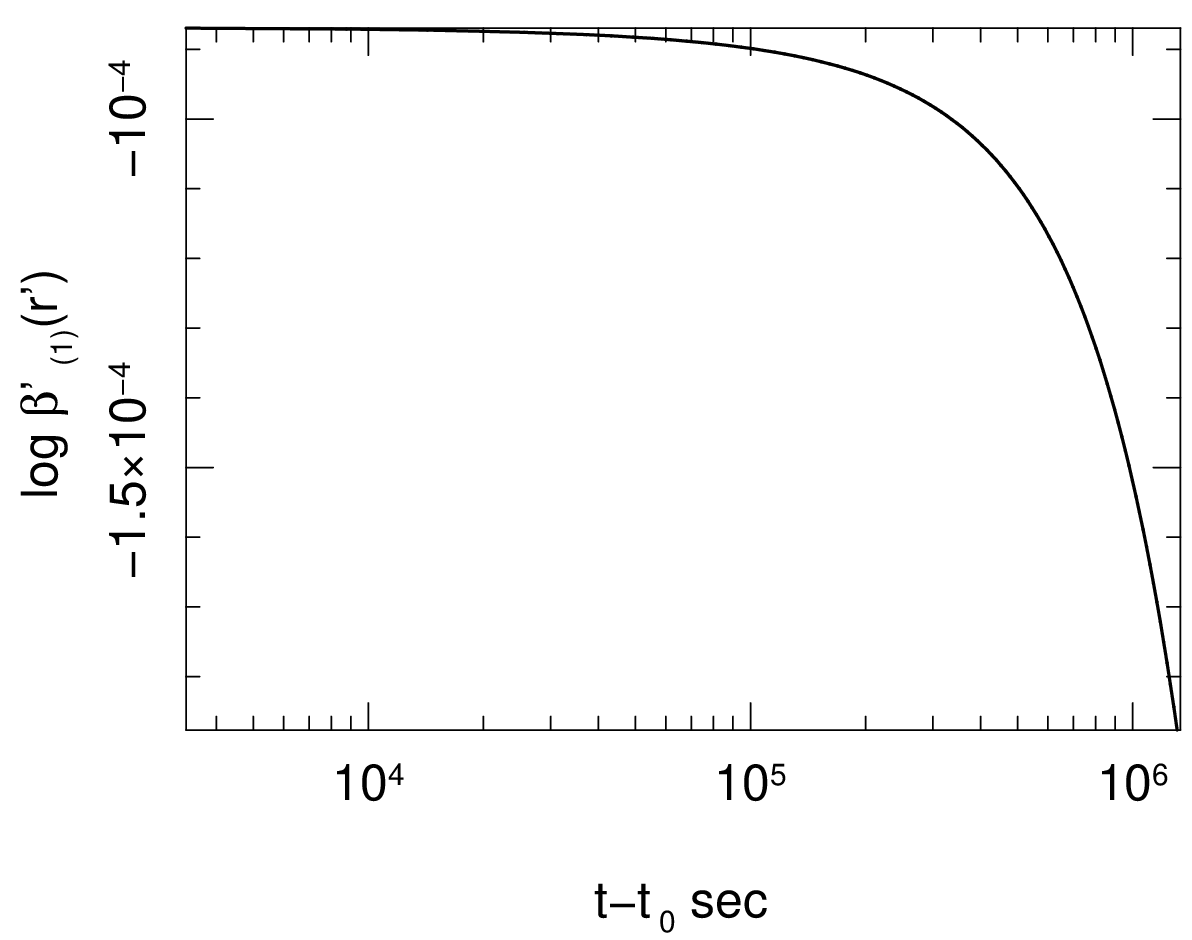} &
8)\includegraphics[height=4cm,angle=-90]{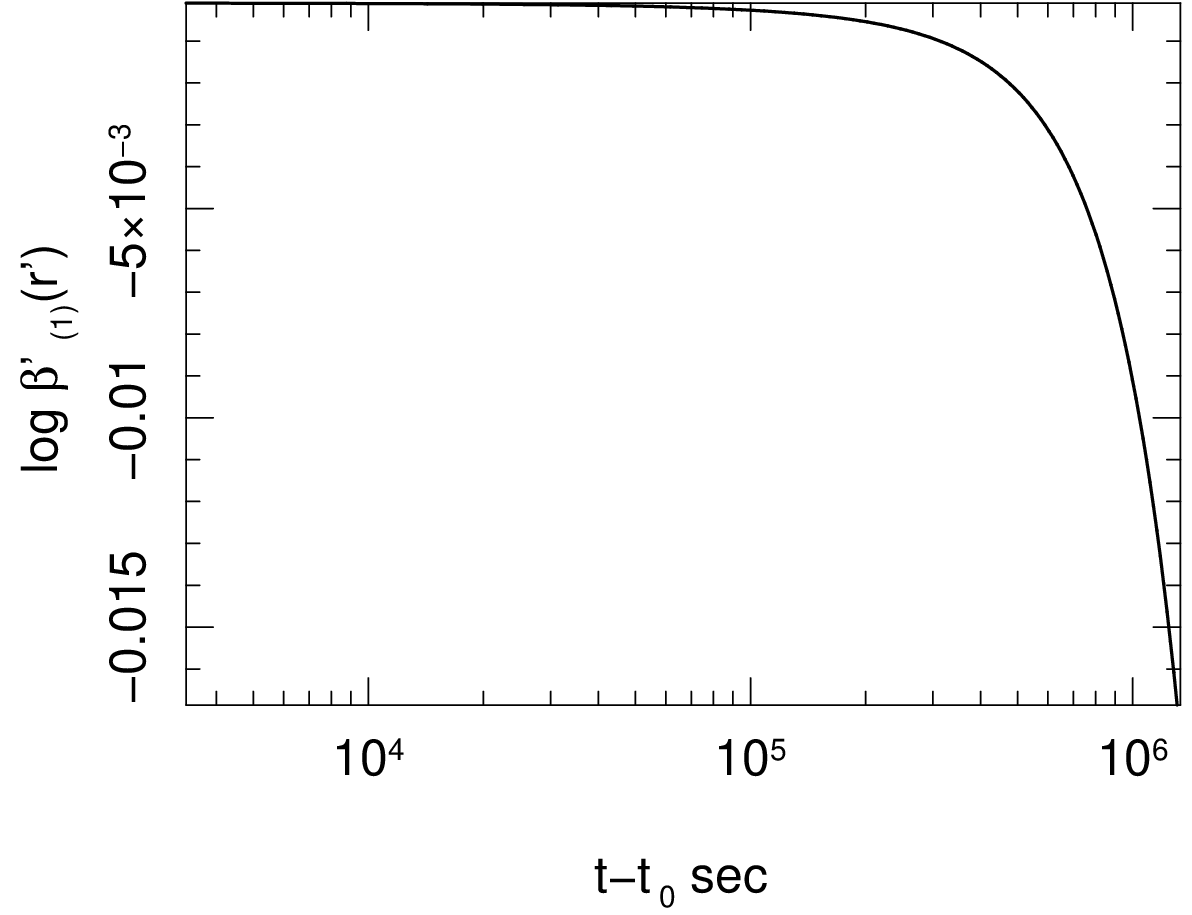} \\
9)\includegraphics[height=4cm,angle=-90]{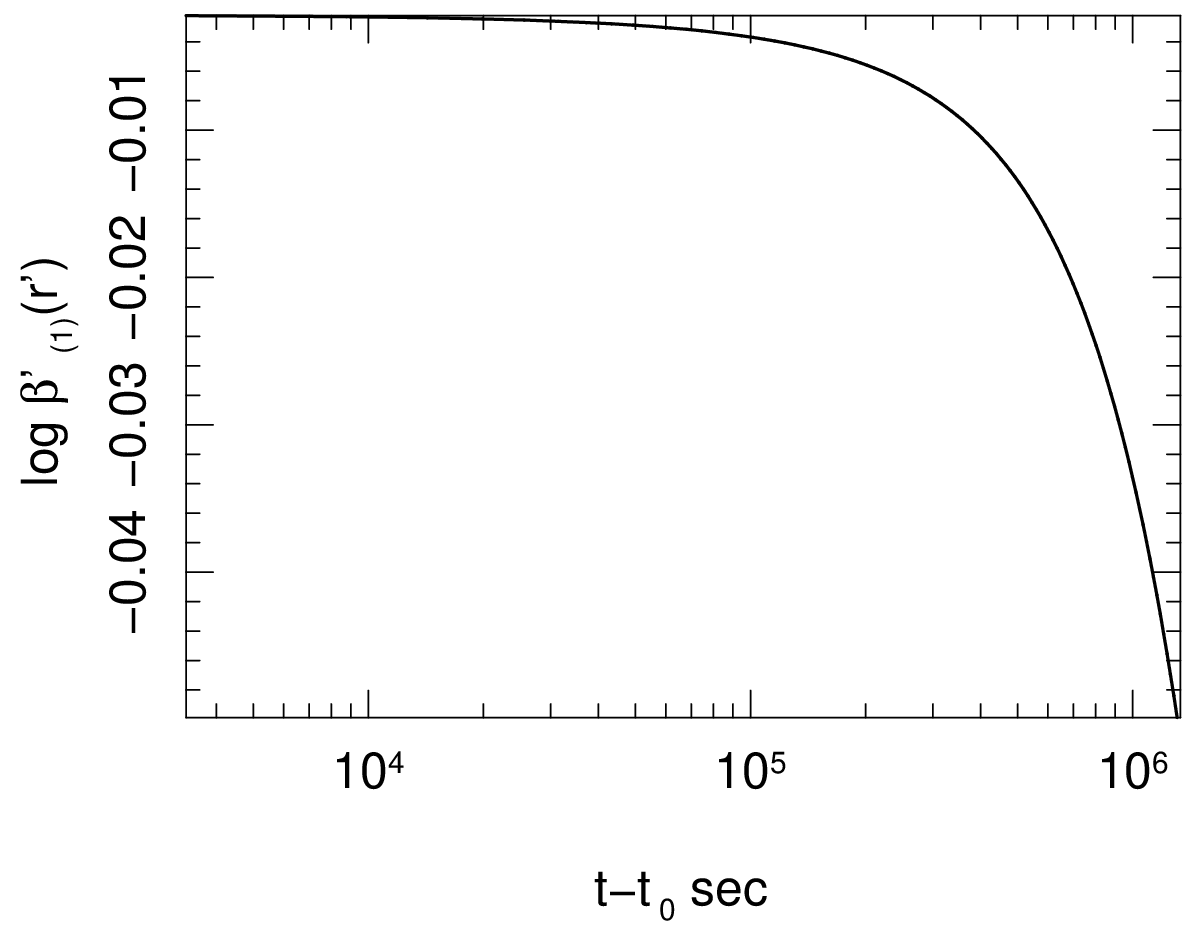} &
10)\includegraphics[height=4cm,angle=-90]{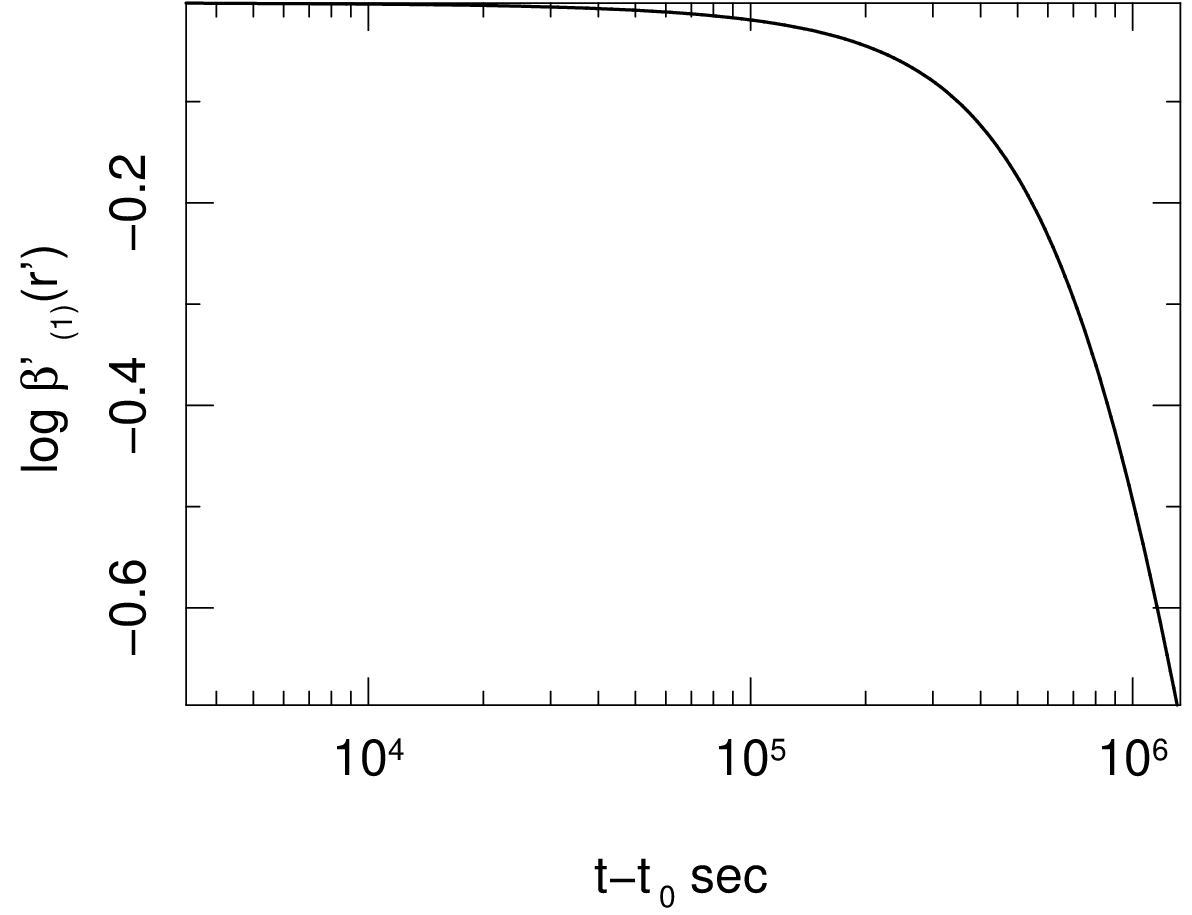} &
11)\includegraphics[height=4cm,angle=-90]{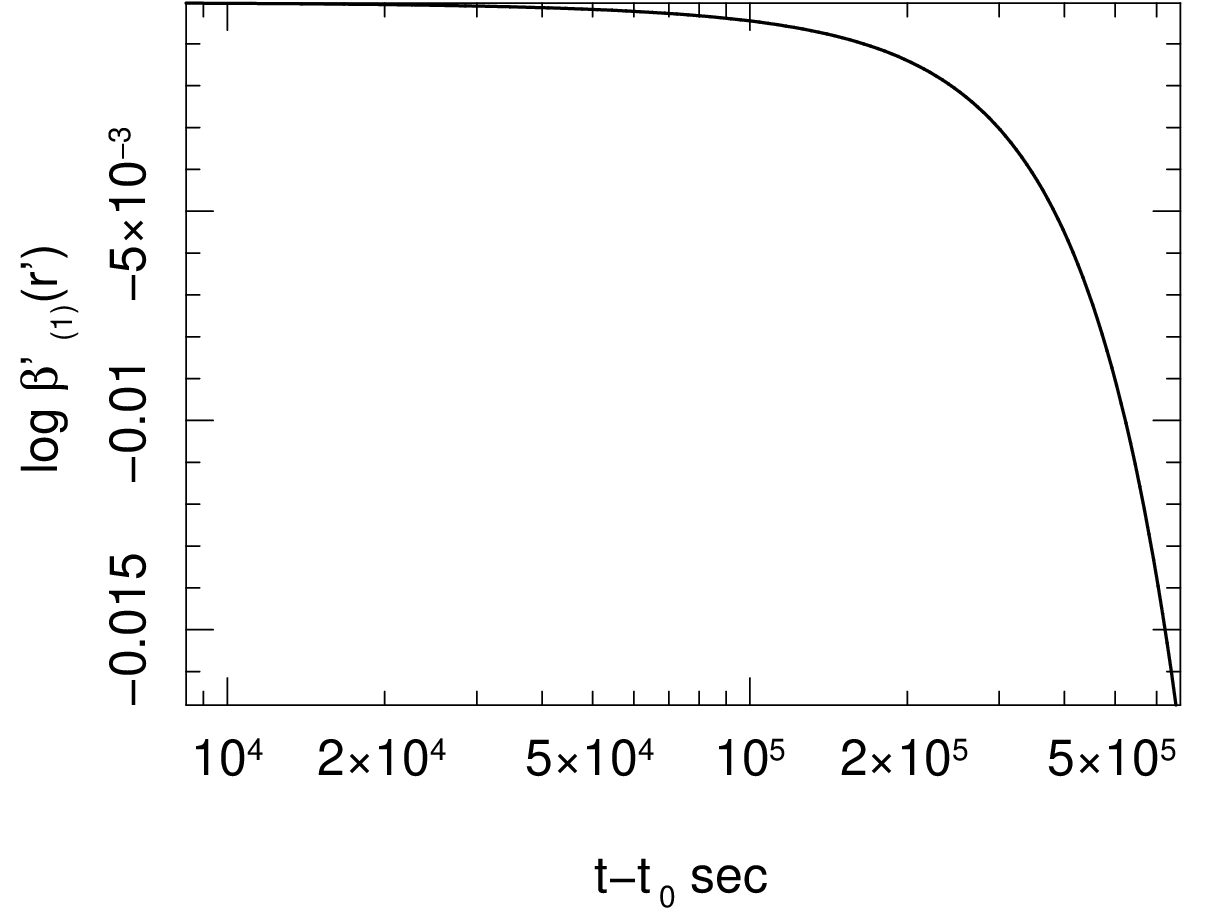} &
12)\includegraphics[height=4cm,angle=-90]{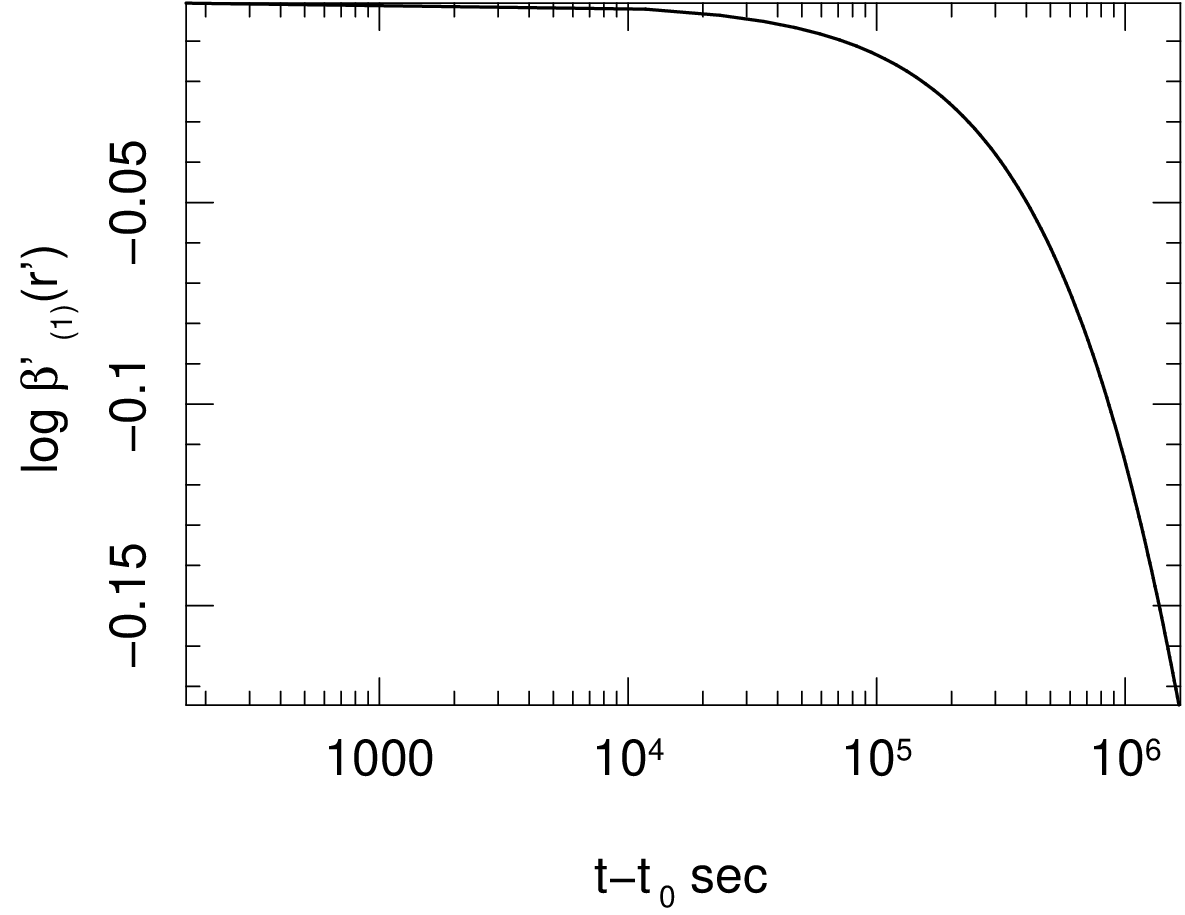} \\
13)\includegraphics[height=4cm,angle=-90]{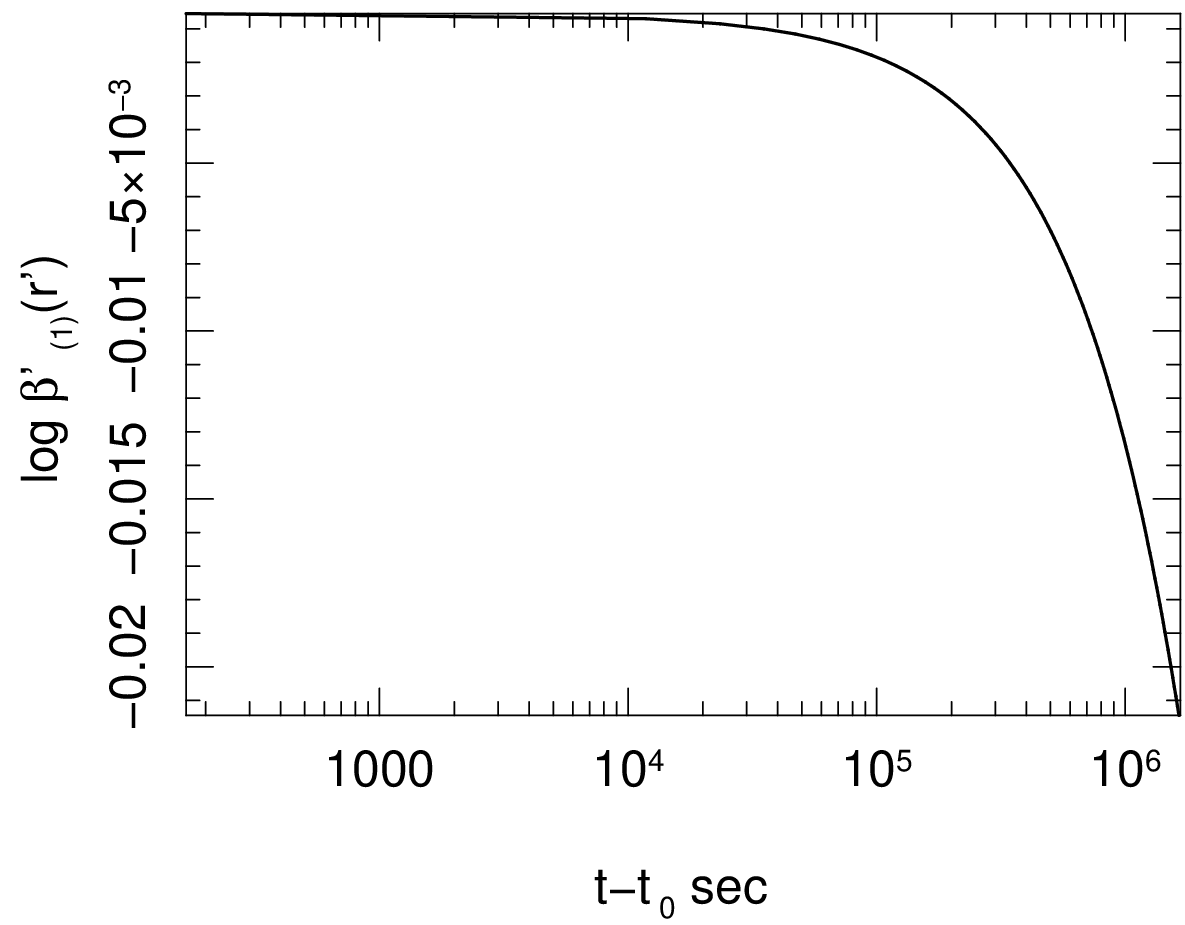} &
14)\includegraphics[height=4cm,angle=-90]{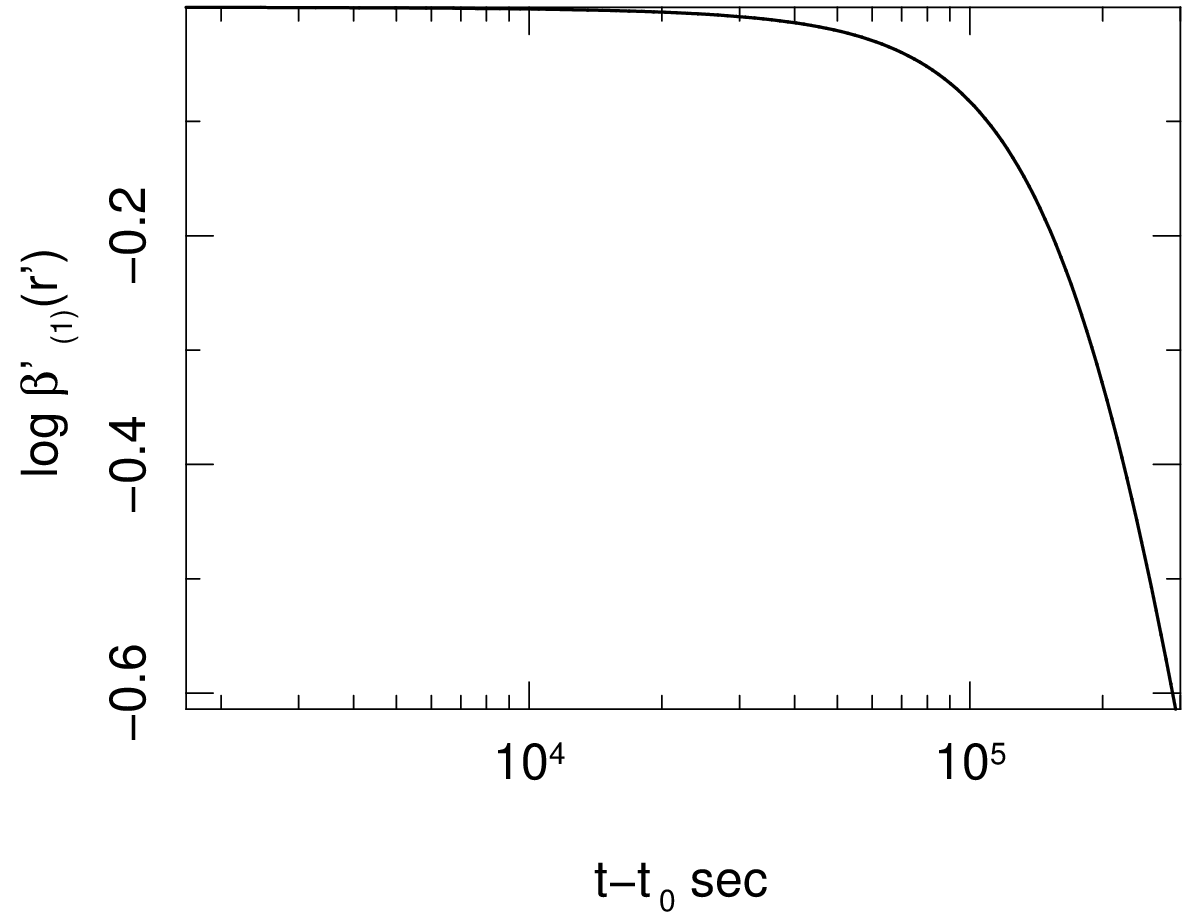} &
15)\includegraphics[height=4cm,angle=-90]{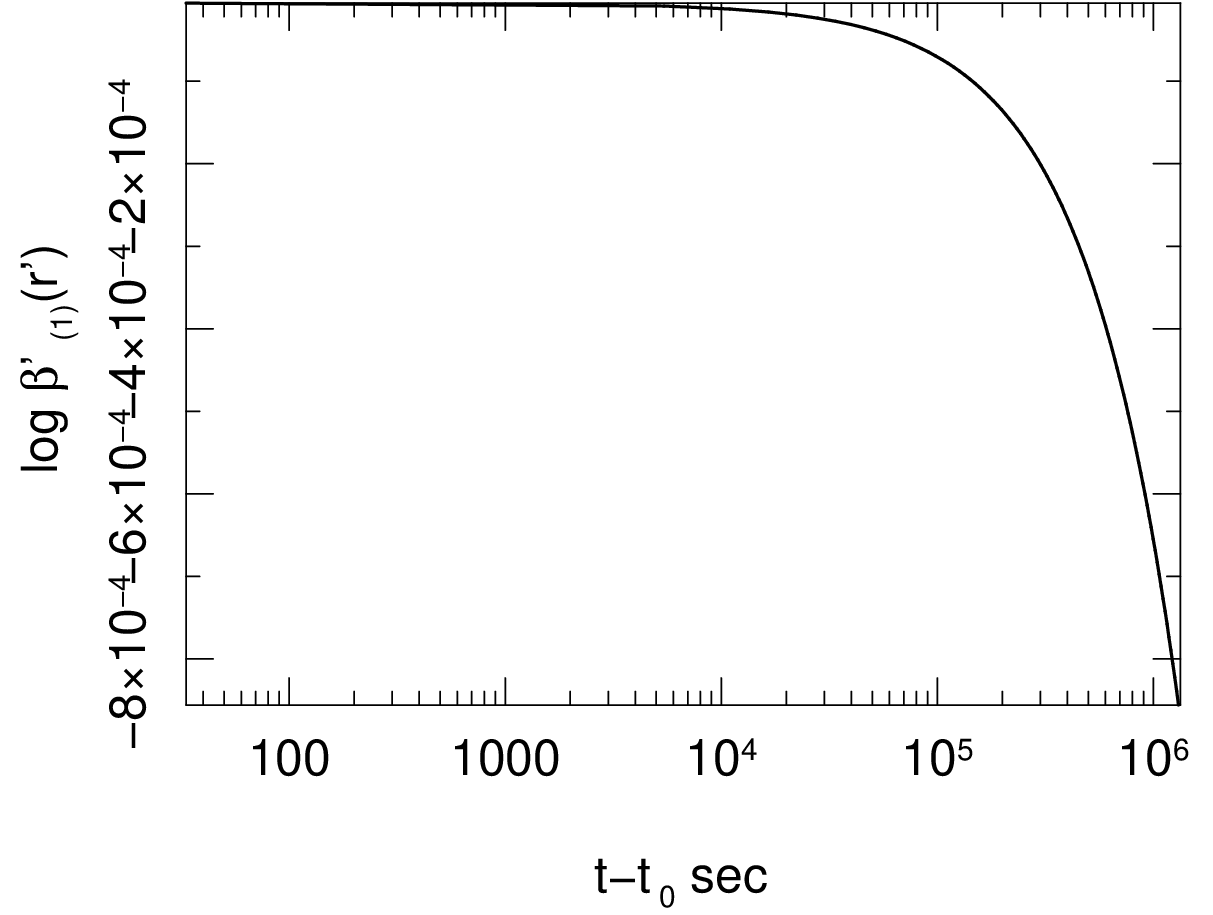} &
16)\includegraphics[height=4cm,angle=-90]{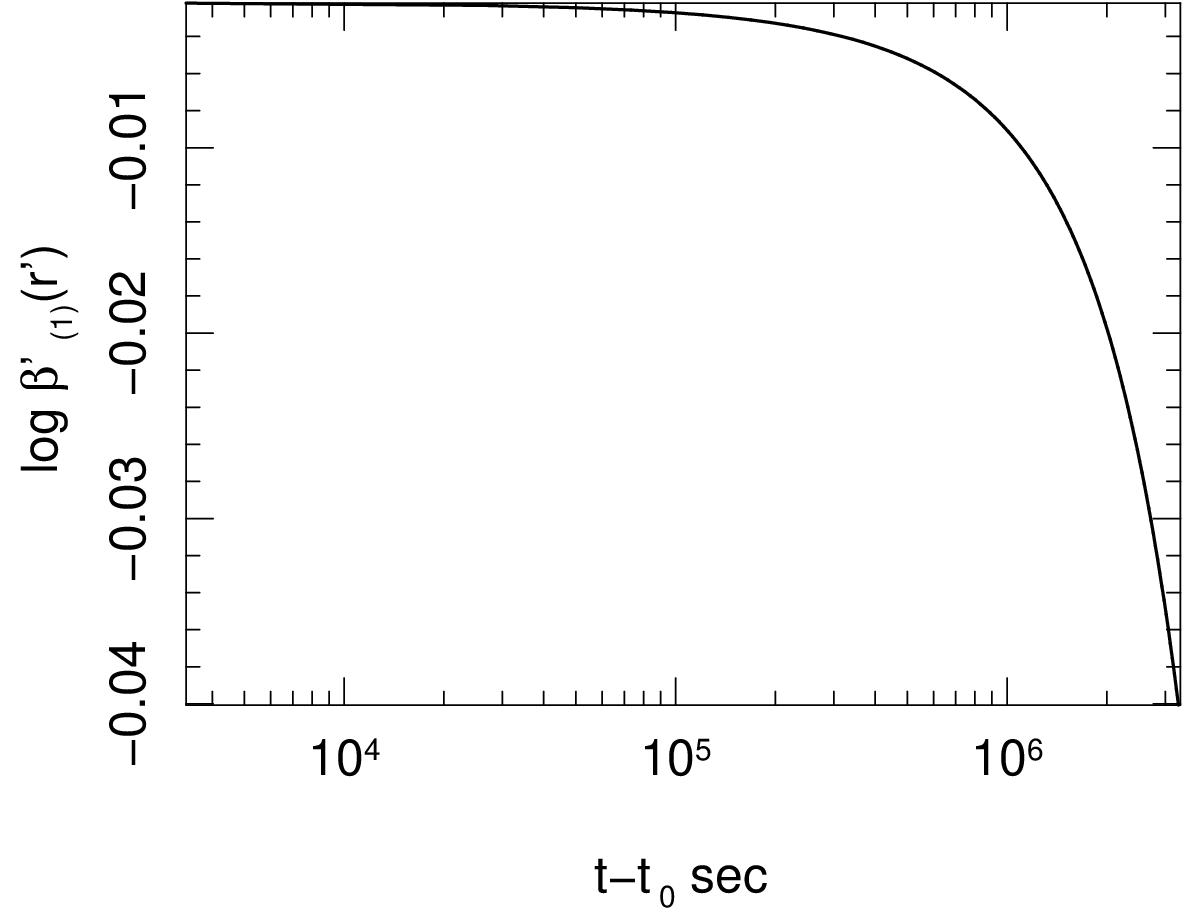} \\
17)\includegraphics[height=4cm,angle=-90]{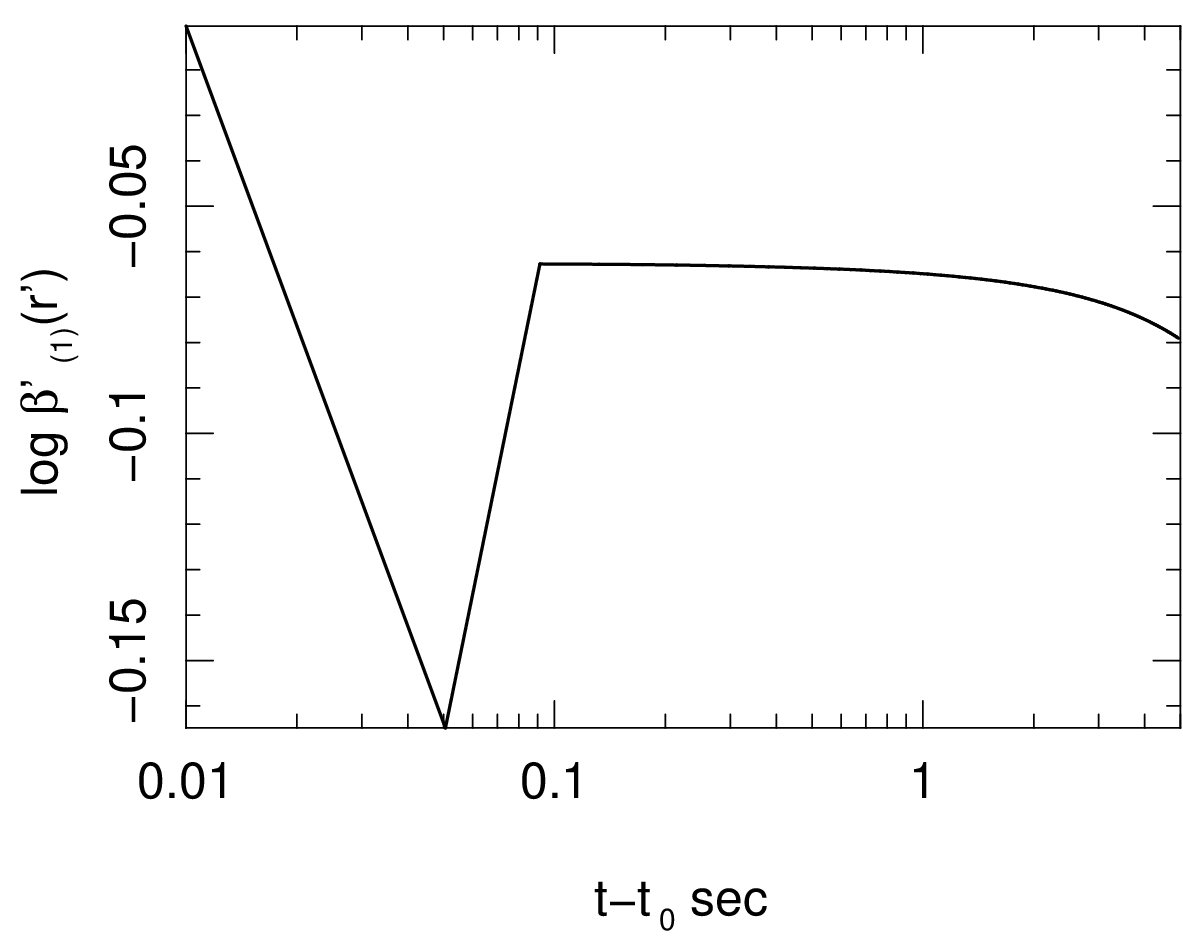} &
18)\includegraphics[height=4cm,angle=-90]{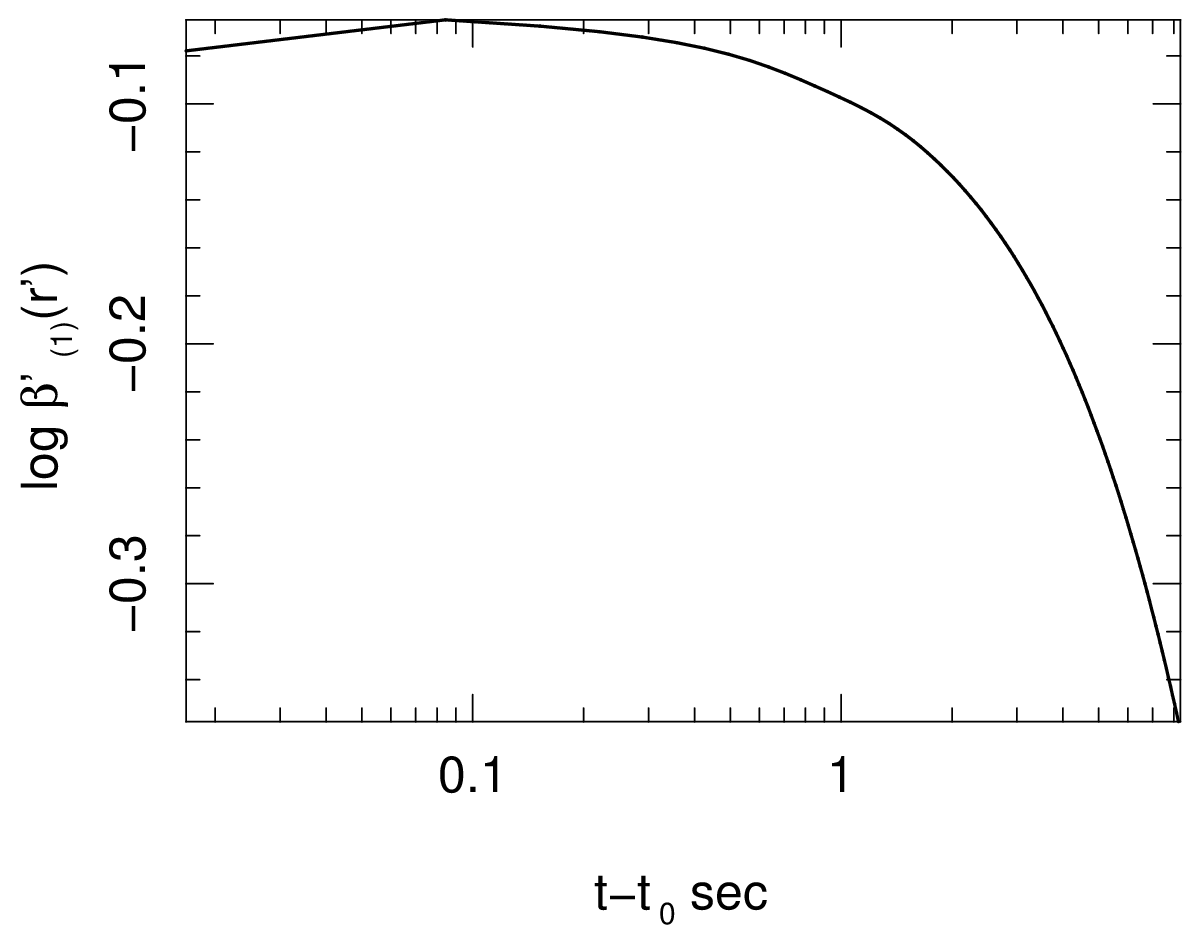} &
19)\includegraphics[height=4cm,angle=-90]{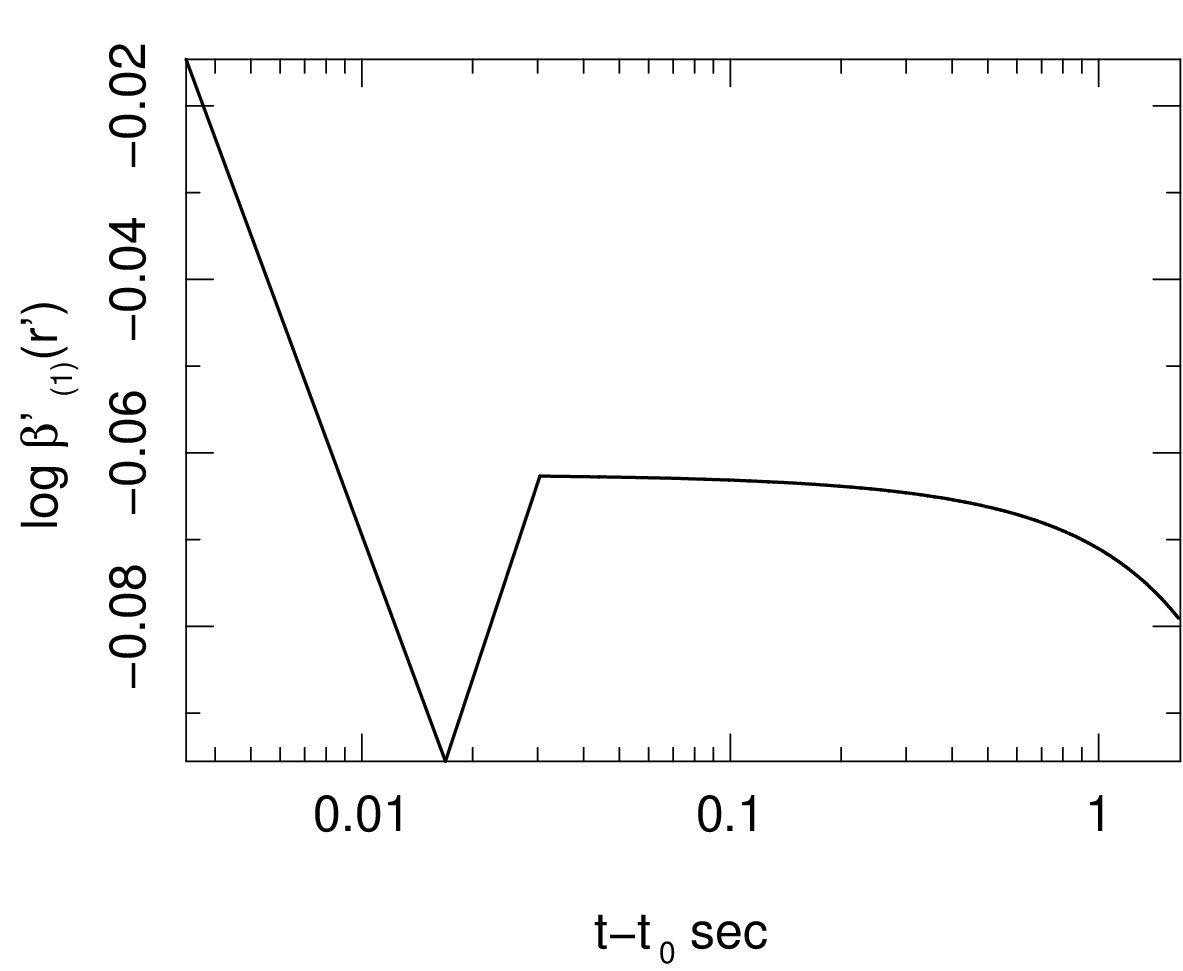} &
20)\includegraphics[height=4cm,angle=-90]{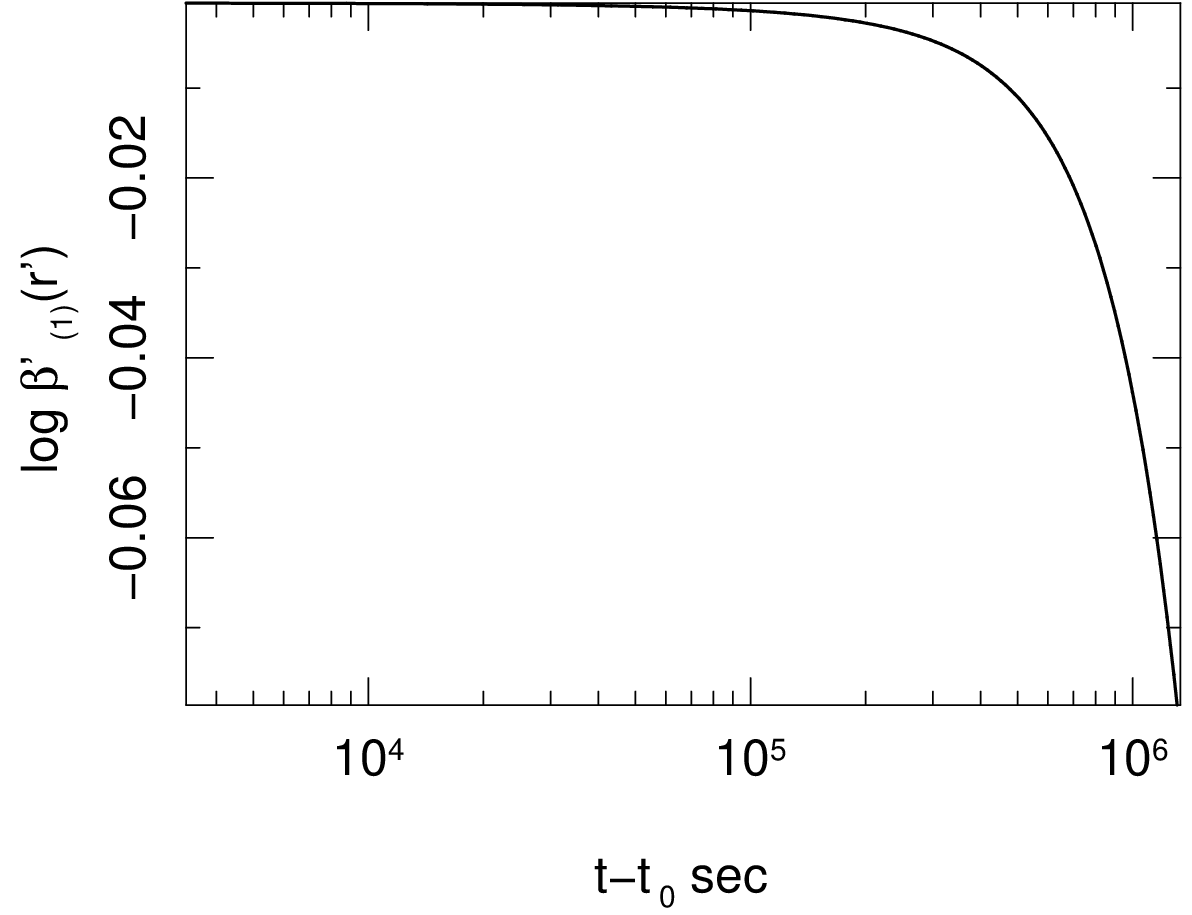} \\
21)\includegraphics[height=4cm,angle=-90]{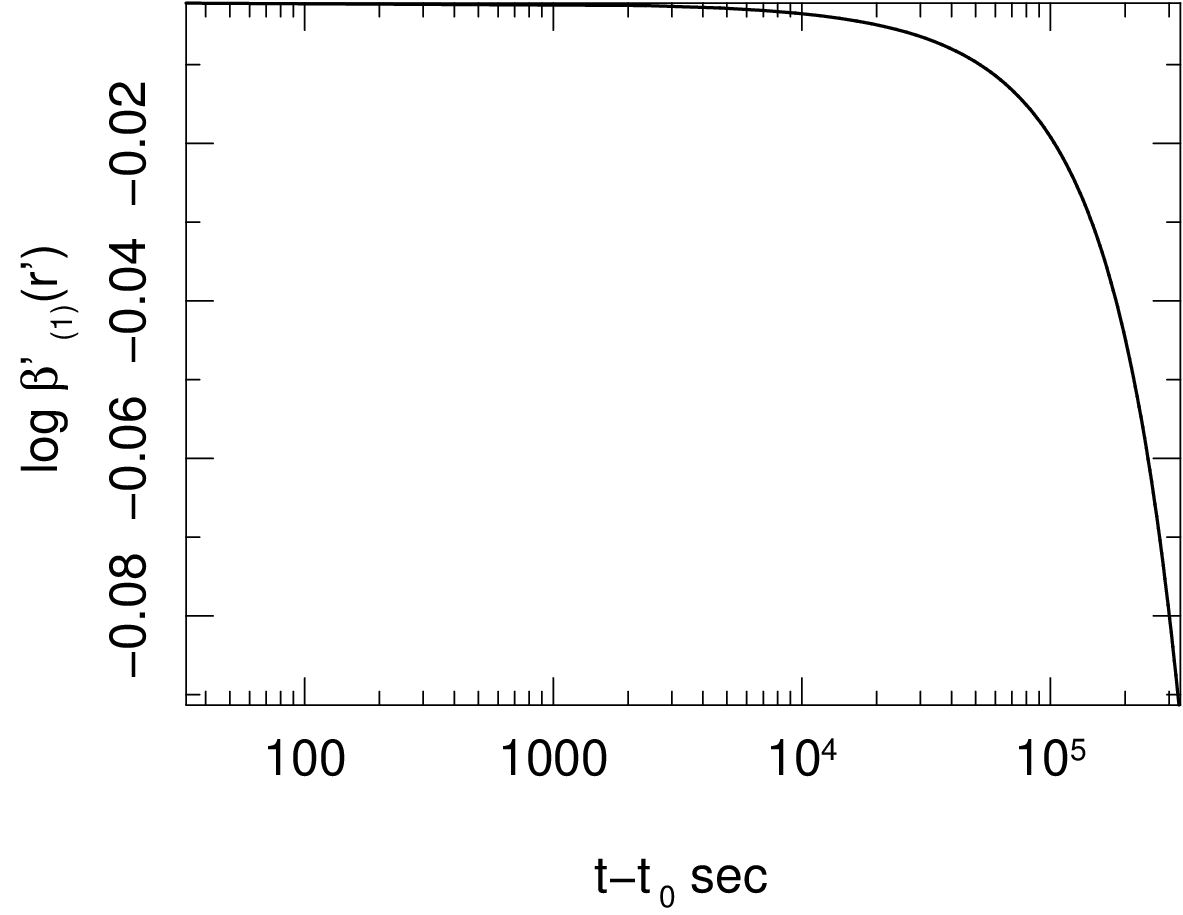} &
22)\includegraphics[height=4cm,angle=-90]{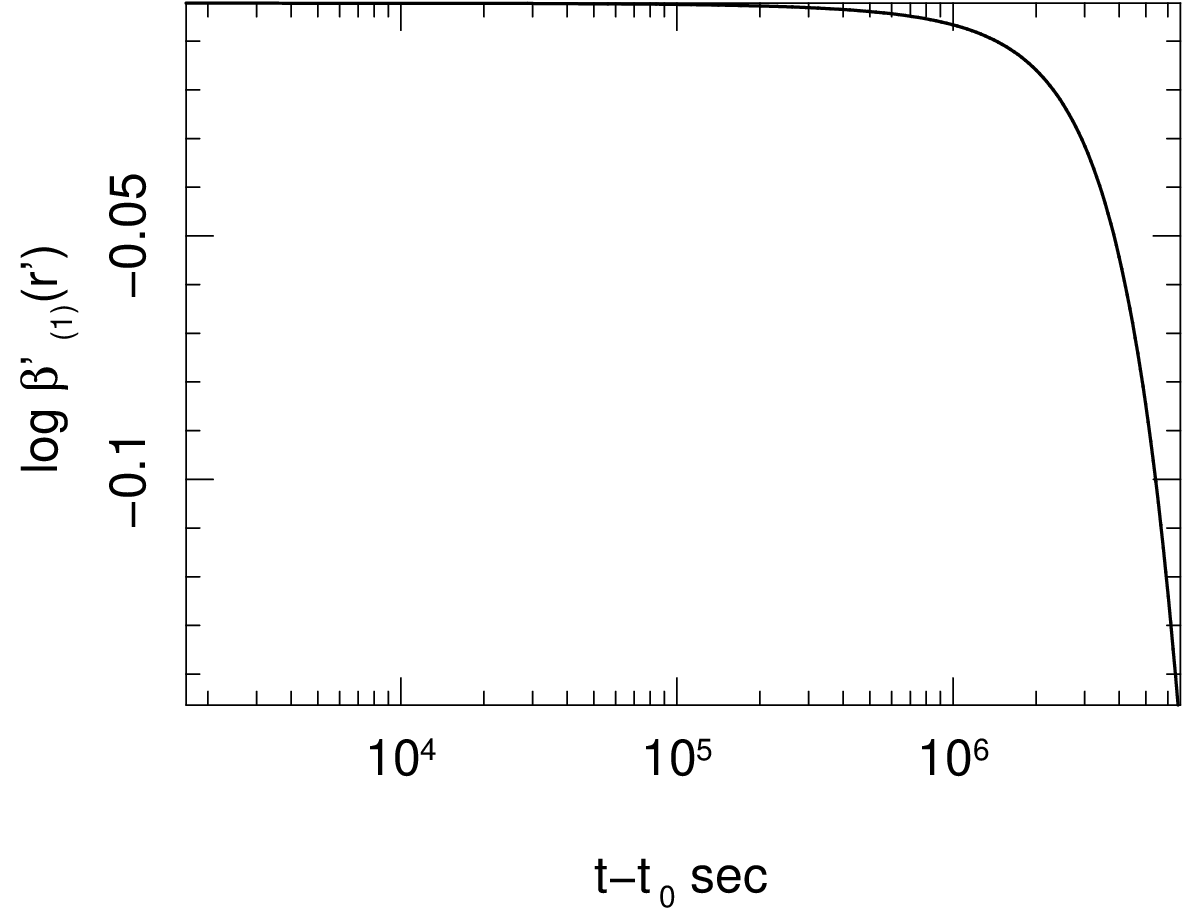} &
23)\includegraphics[height=4cm,angle=-90]{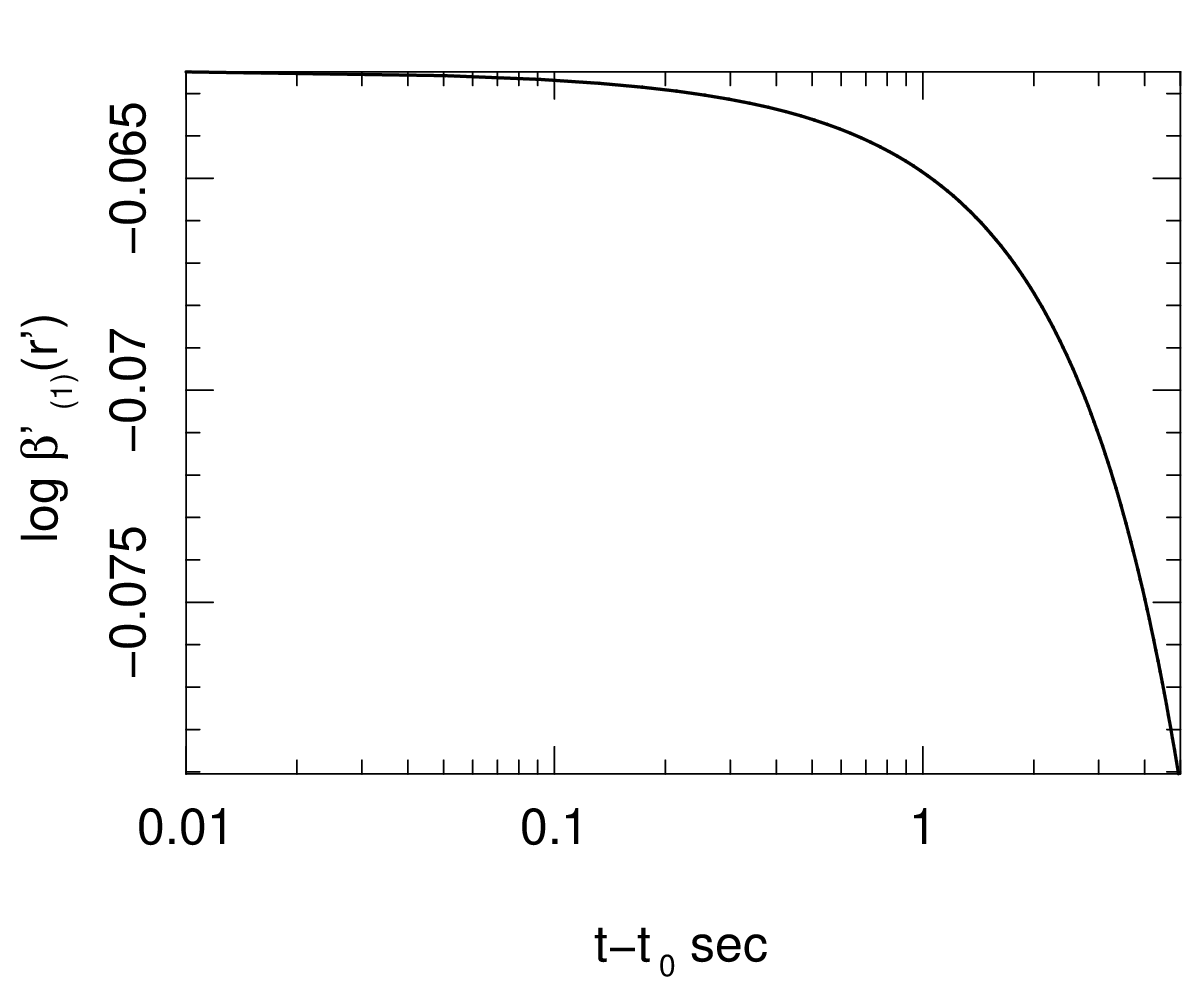} &
24)\includegraphics[height=4cm,angle=-90]{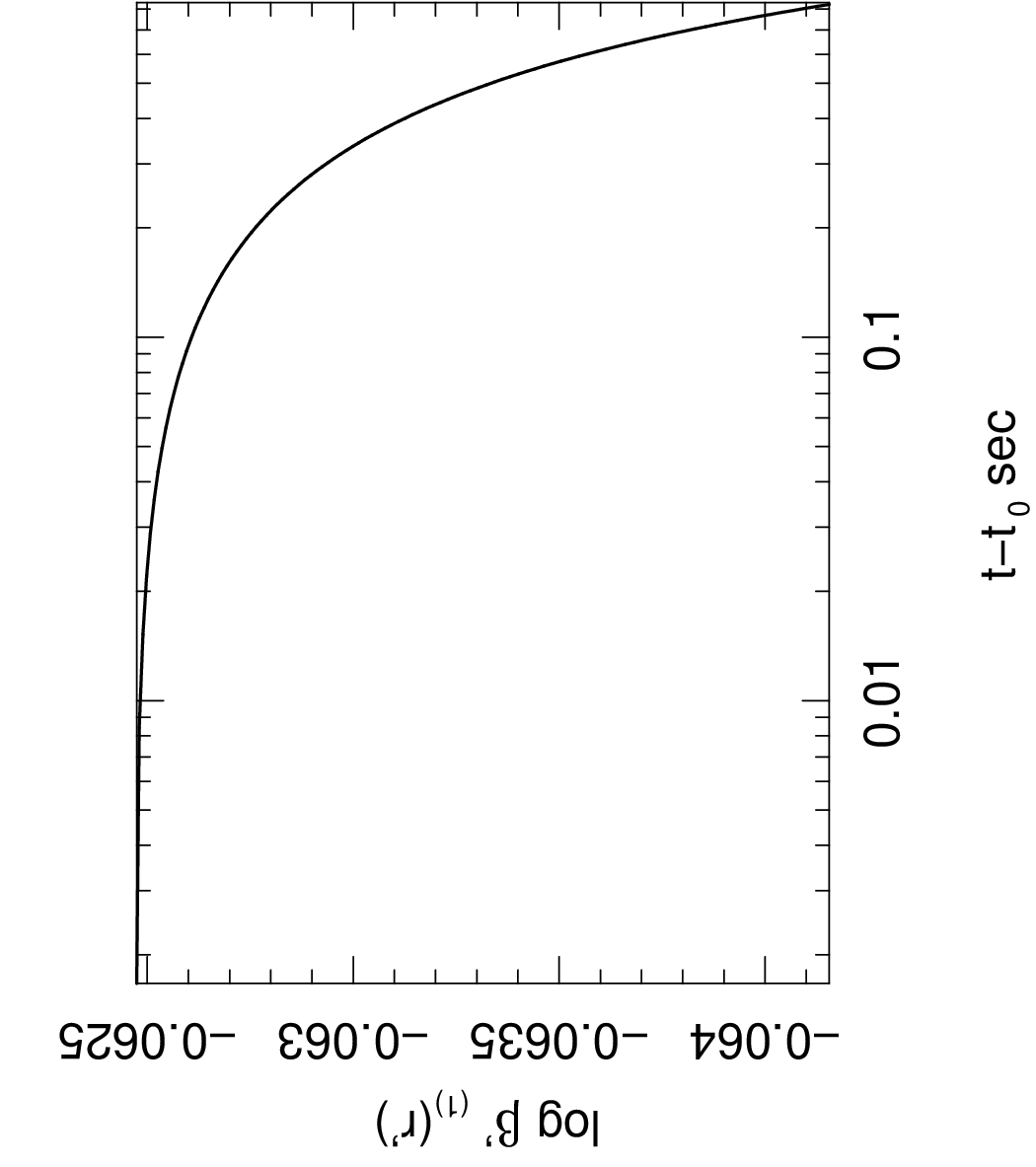} 
\end {tabular}
\end {center}
\caption{Evolution of $\beta'_1$ with time/radius. Definitions of curves are 
the same as Fig. \ref{fig:betazero}. The difference with $\beta'_0$ shows for 
each model the importance radiative correction in the evolution of the 
kinematics.
\label{fig:betaone}}
\end{figure*}

\begin{figure*}
\begin{center}
\begin {tabular}{cccc}
1)\includegraphics[height=4cm,angle=-90]{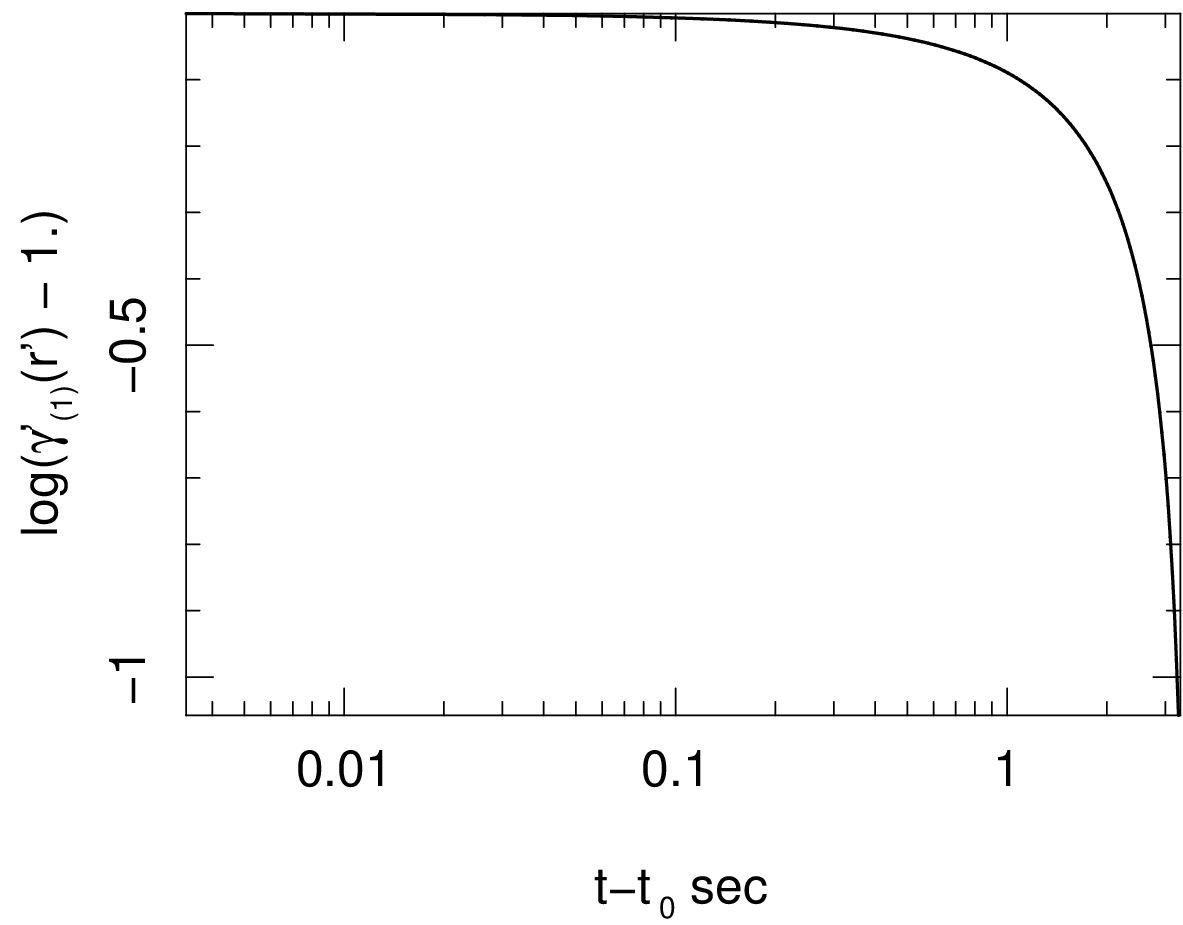} &
2)\includegraphics[height=4cm,angle=-90]{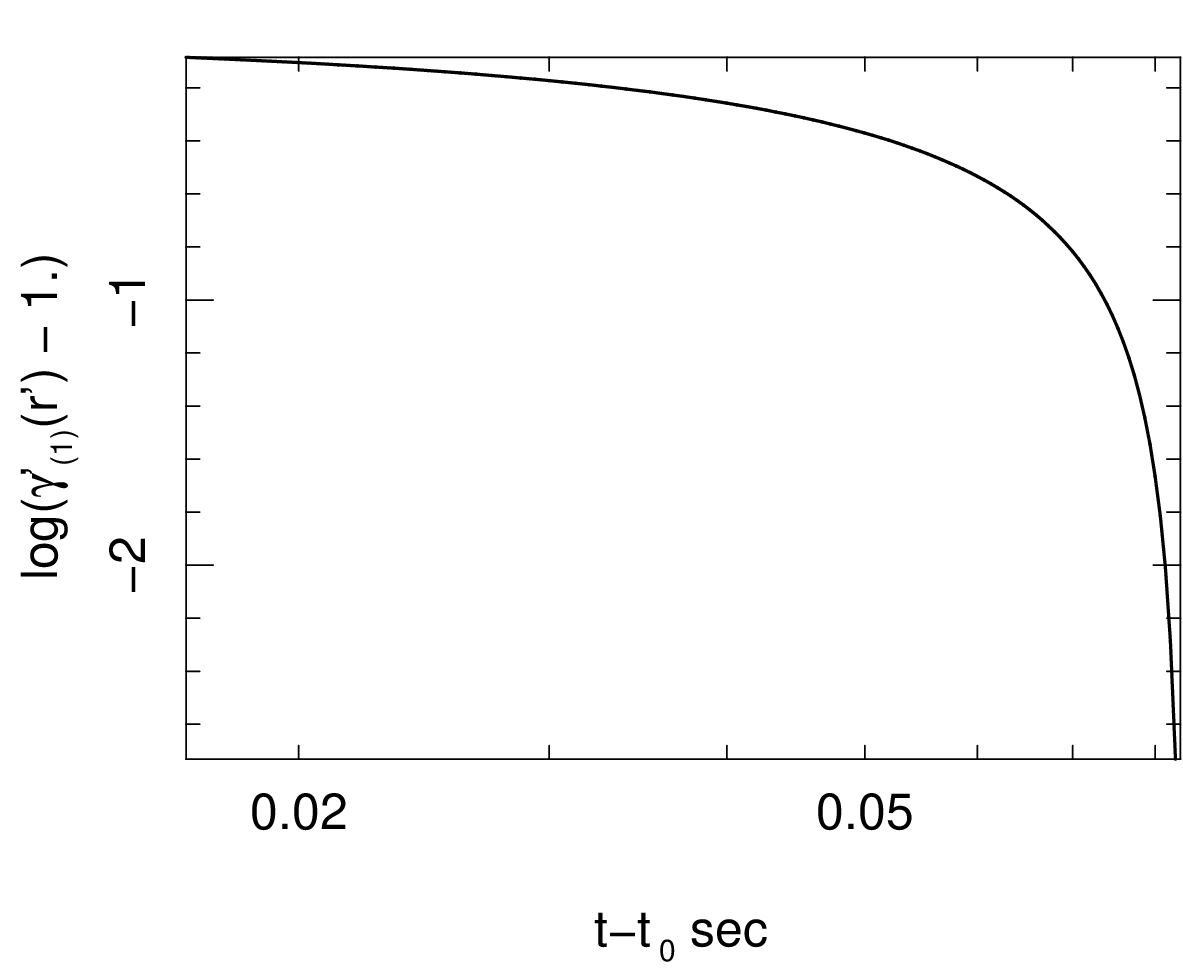} & 
3)\includegraphics[height=4cm,angle=-90]{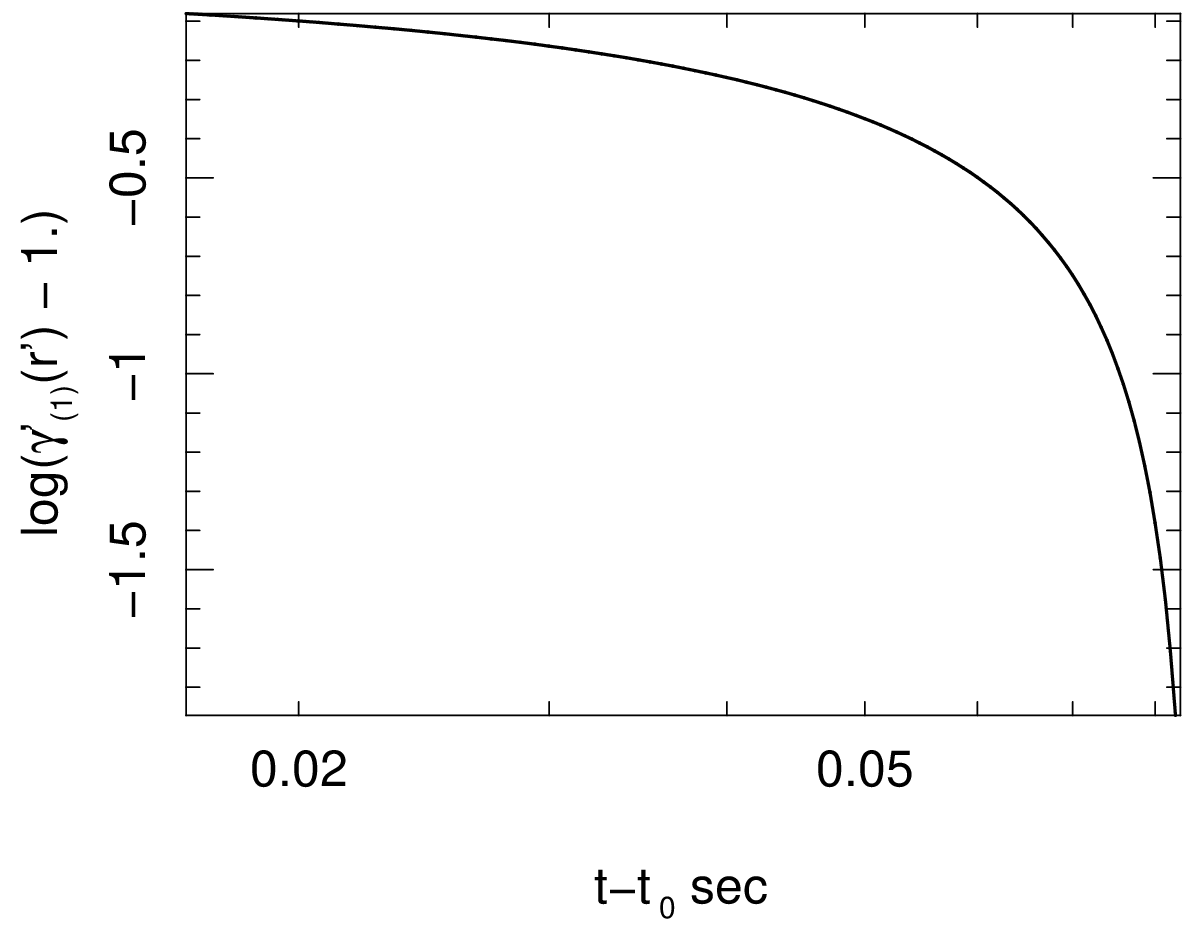} &
4)\includegraphics[height=4cm,angle=-90]{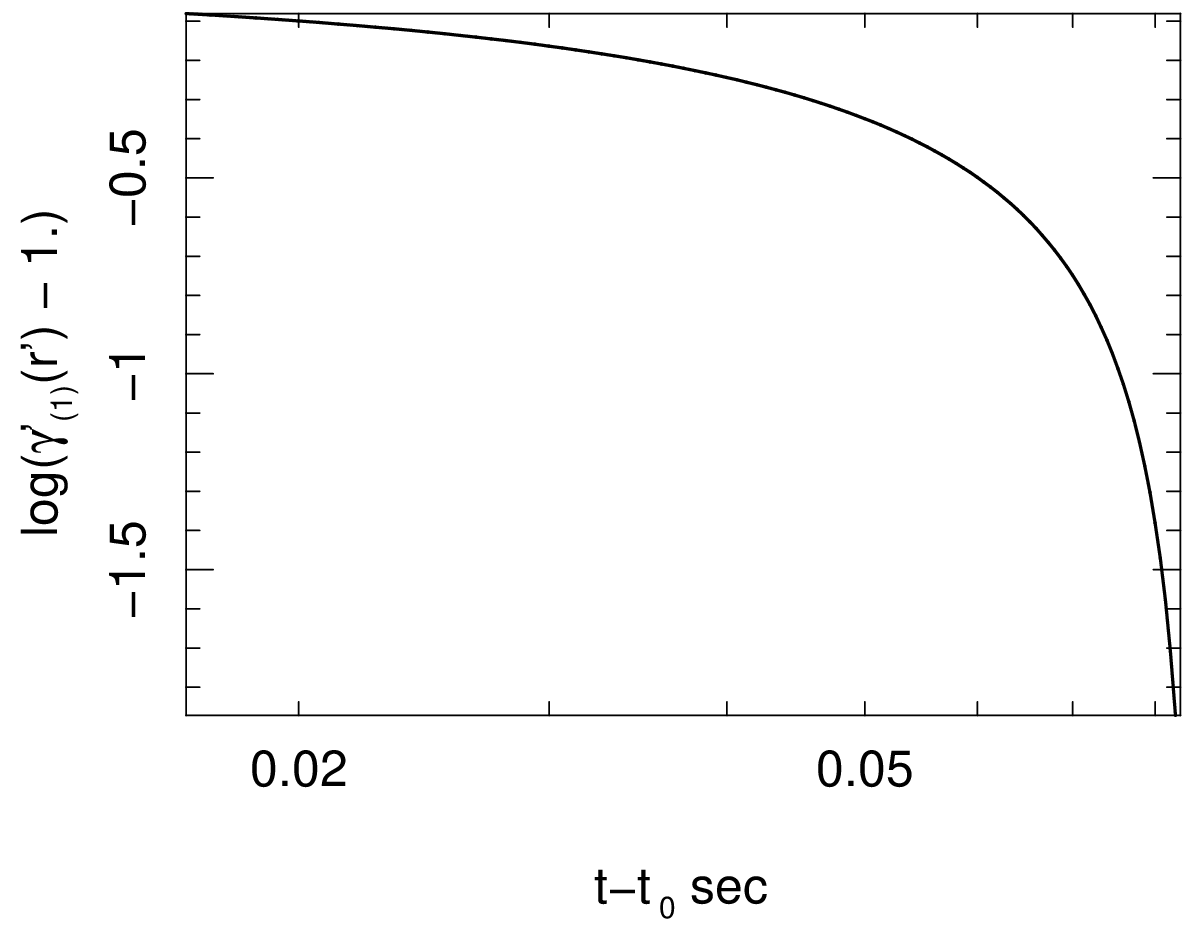} \\ 
5)\includegraphics[height=4cm,angle=-90]{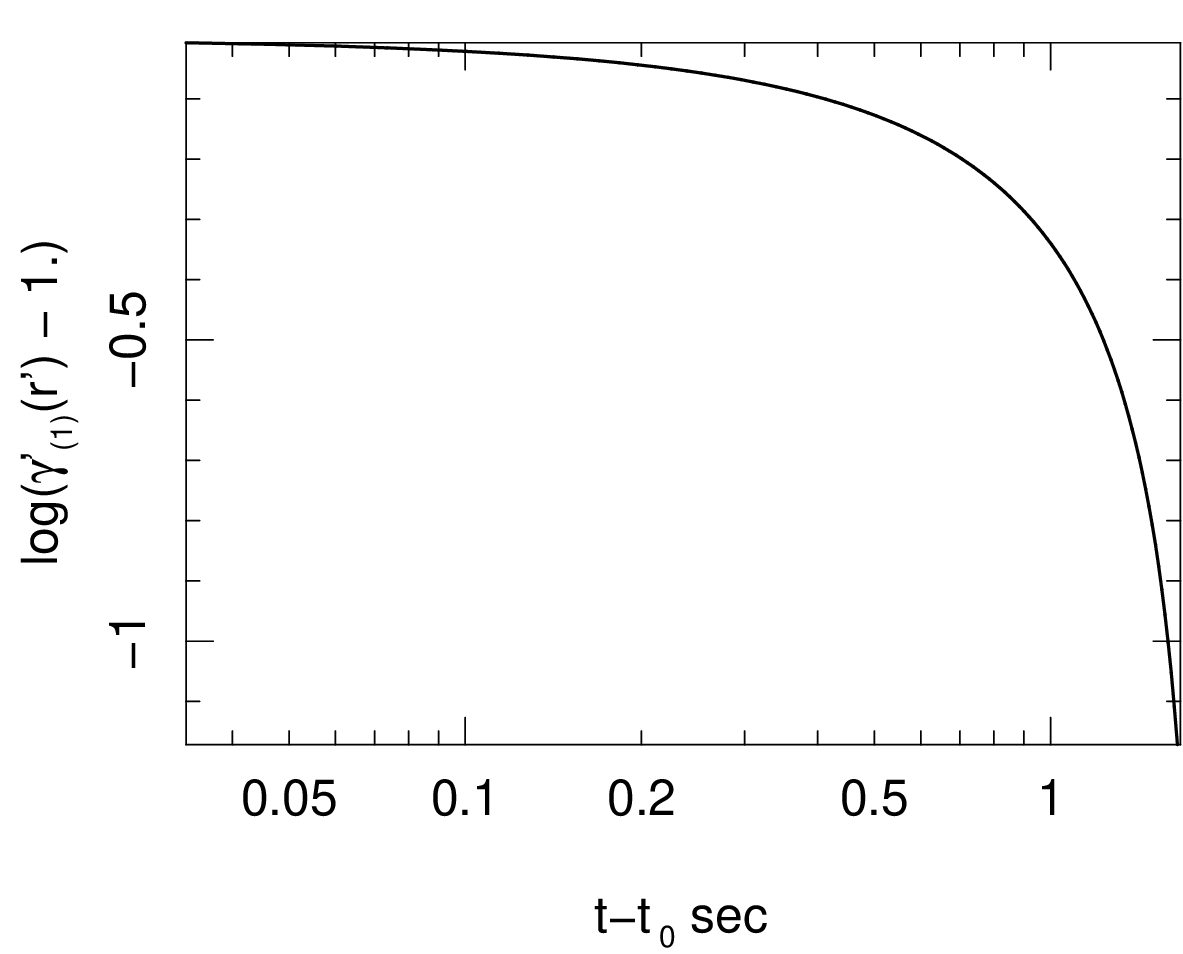} &
6)\includegraphics[height=4cm,angle=-90]{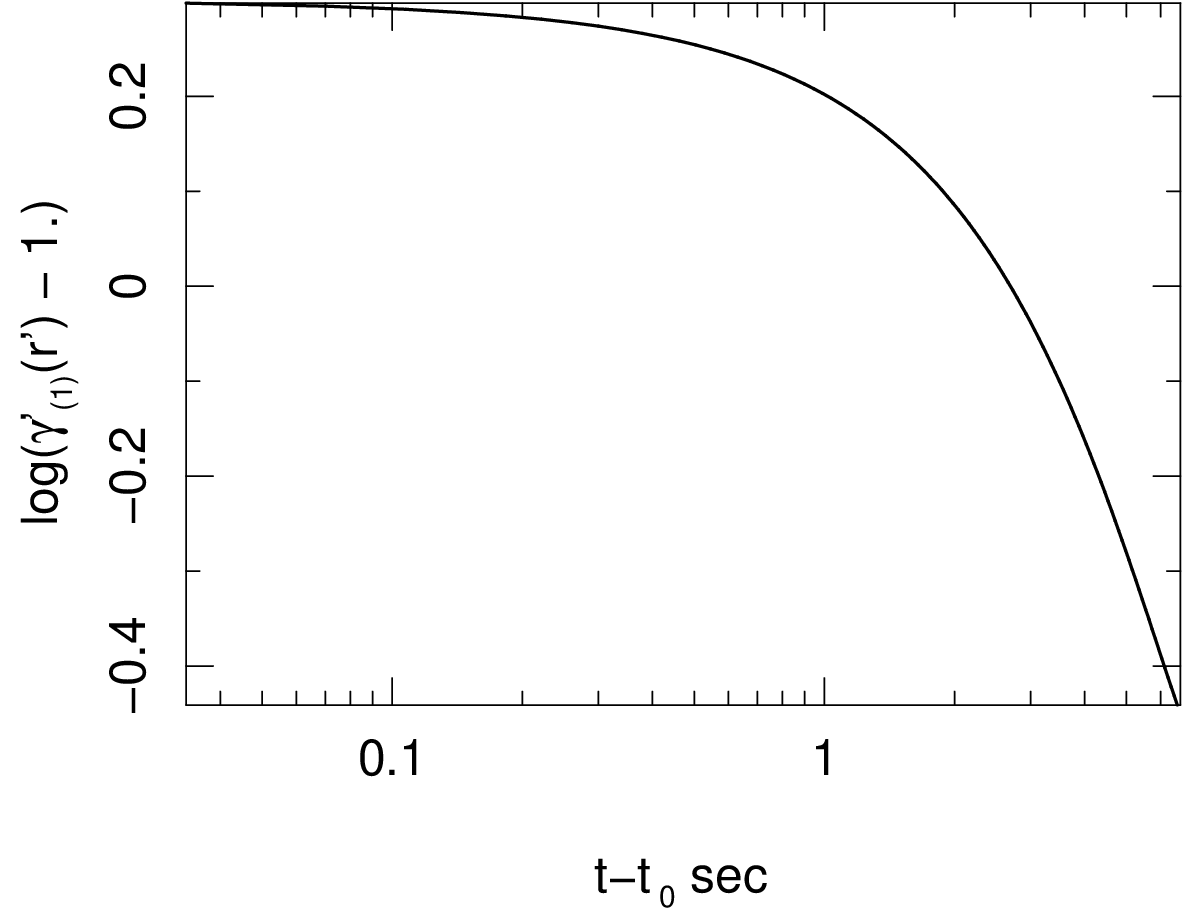} & 
7)\includegraphics[height=4cm,angle=-90]{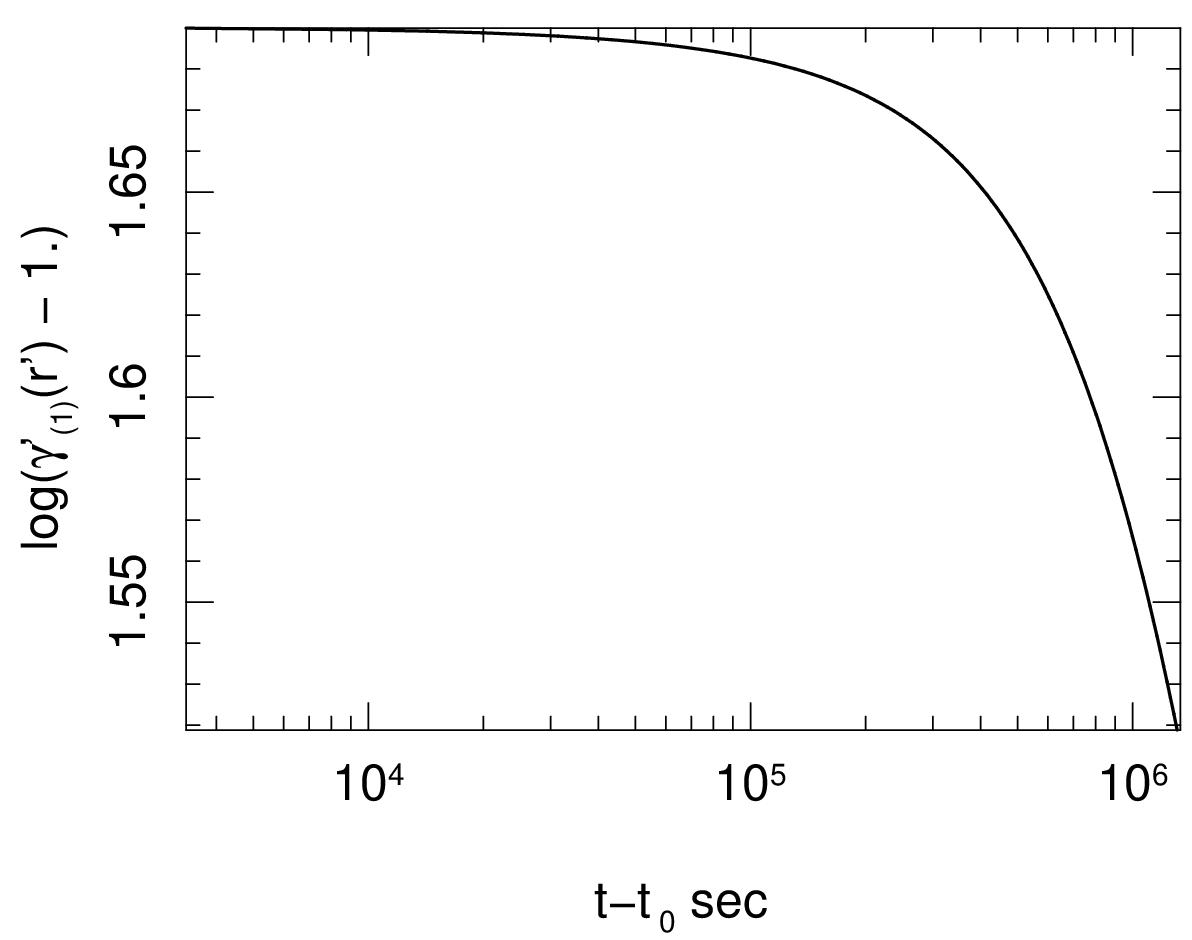} &
8)\includegraphics[height=4cm,angle=-90]{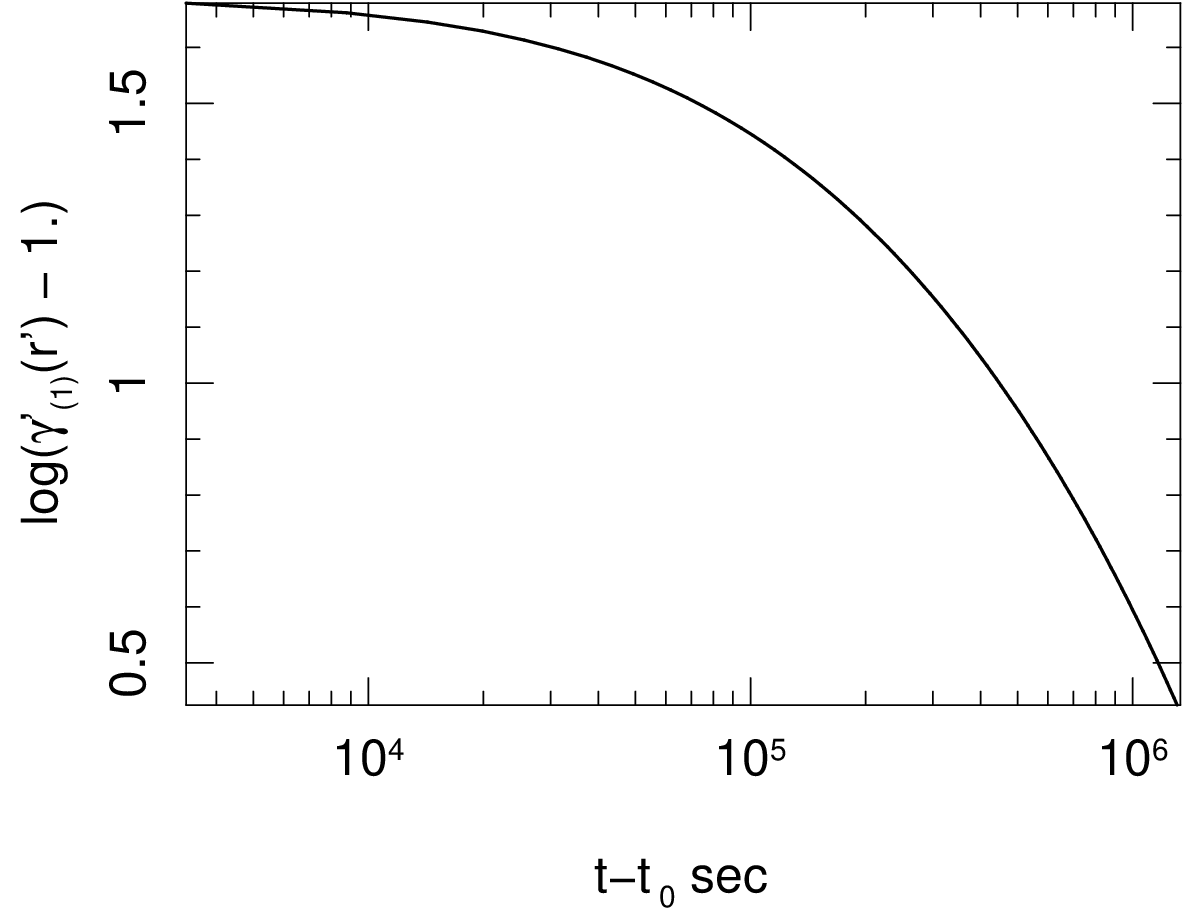} \\
9)\includegraphics[height=4cm,angle=-90]{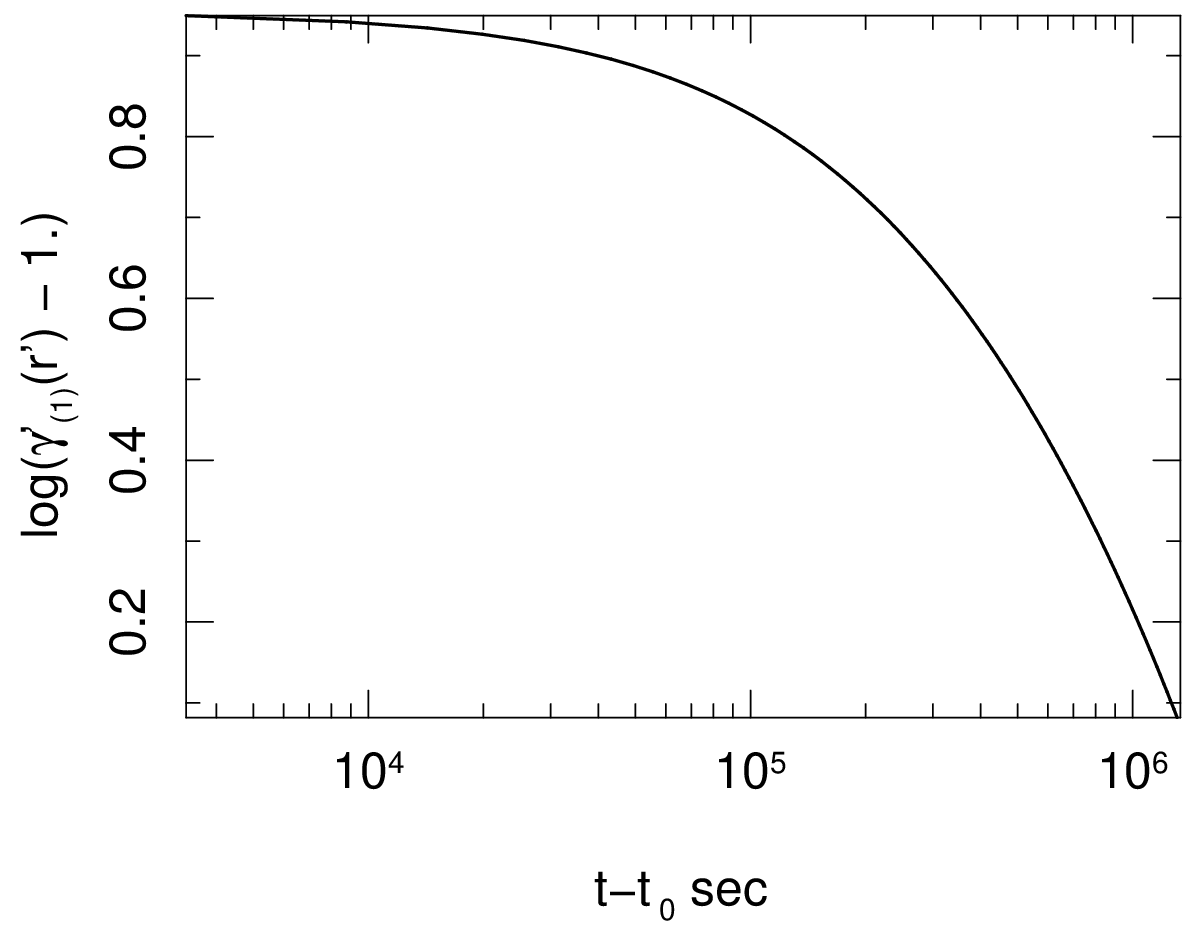} &
10)\includegraphics[height=4cm,angle=-90]{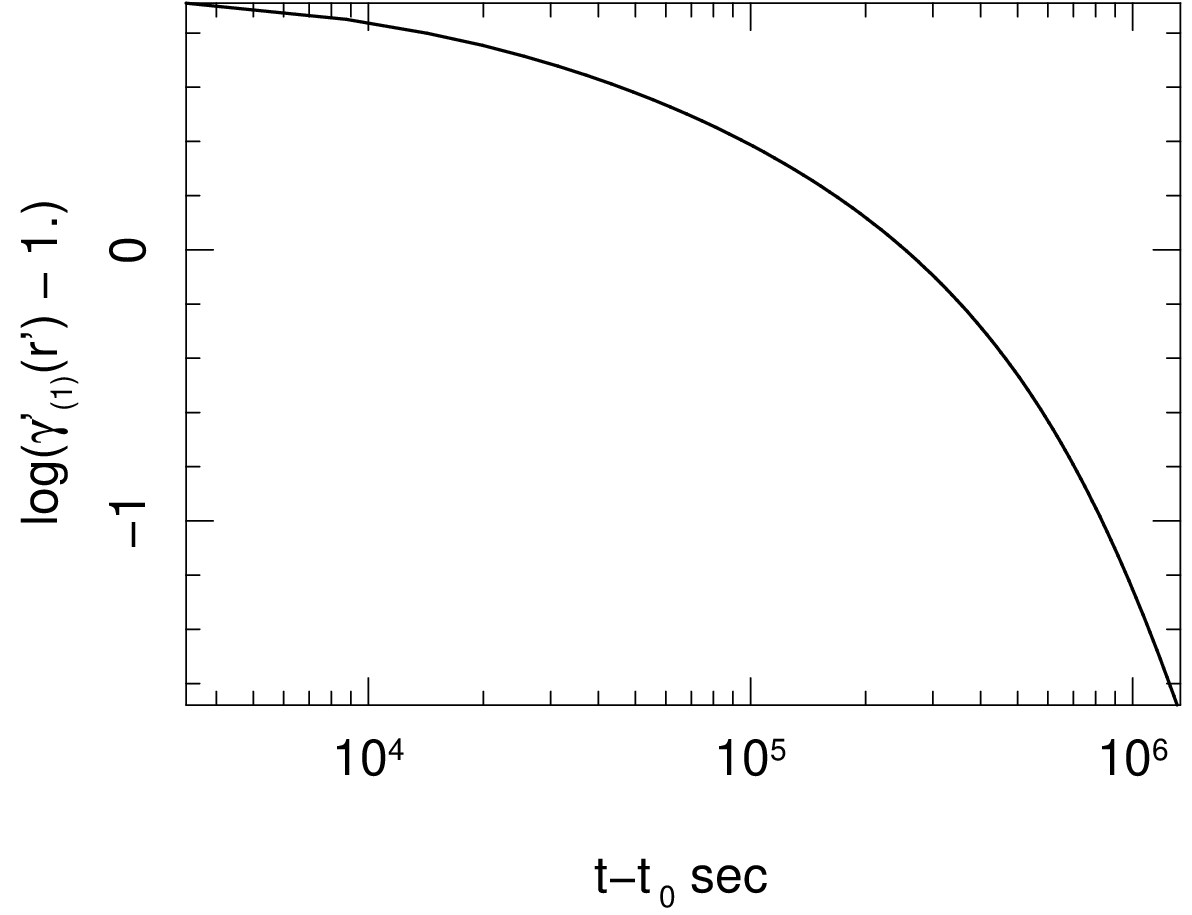} &
11)\includegraphics[height=4cm,angle=-90]{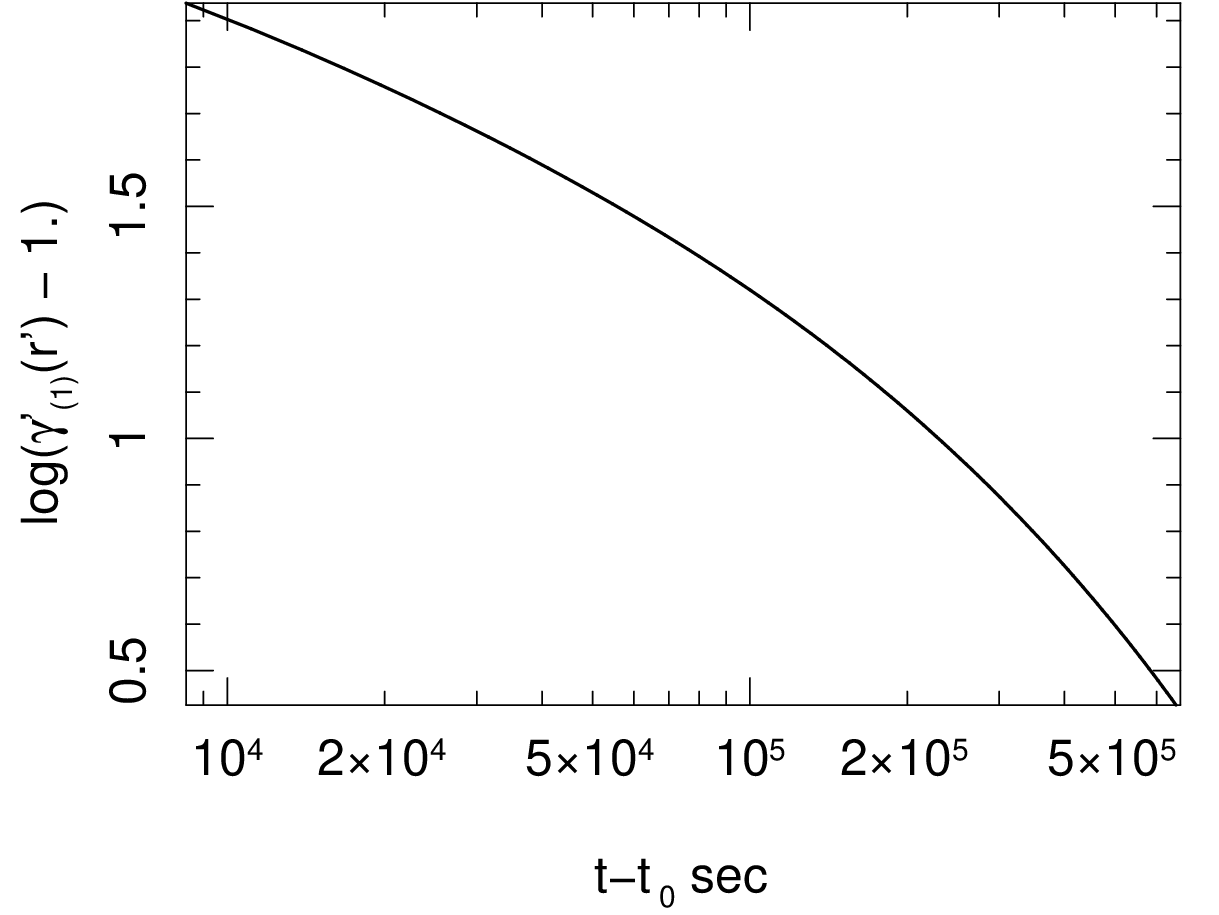} &
12)\includegraphics[height=4cm,angle=-90]{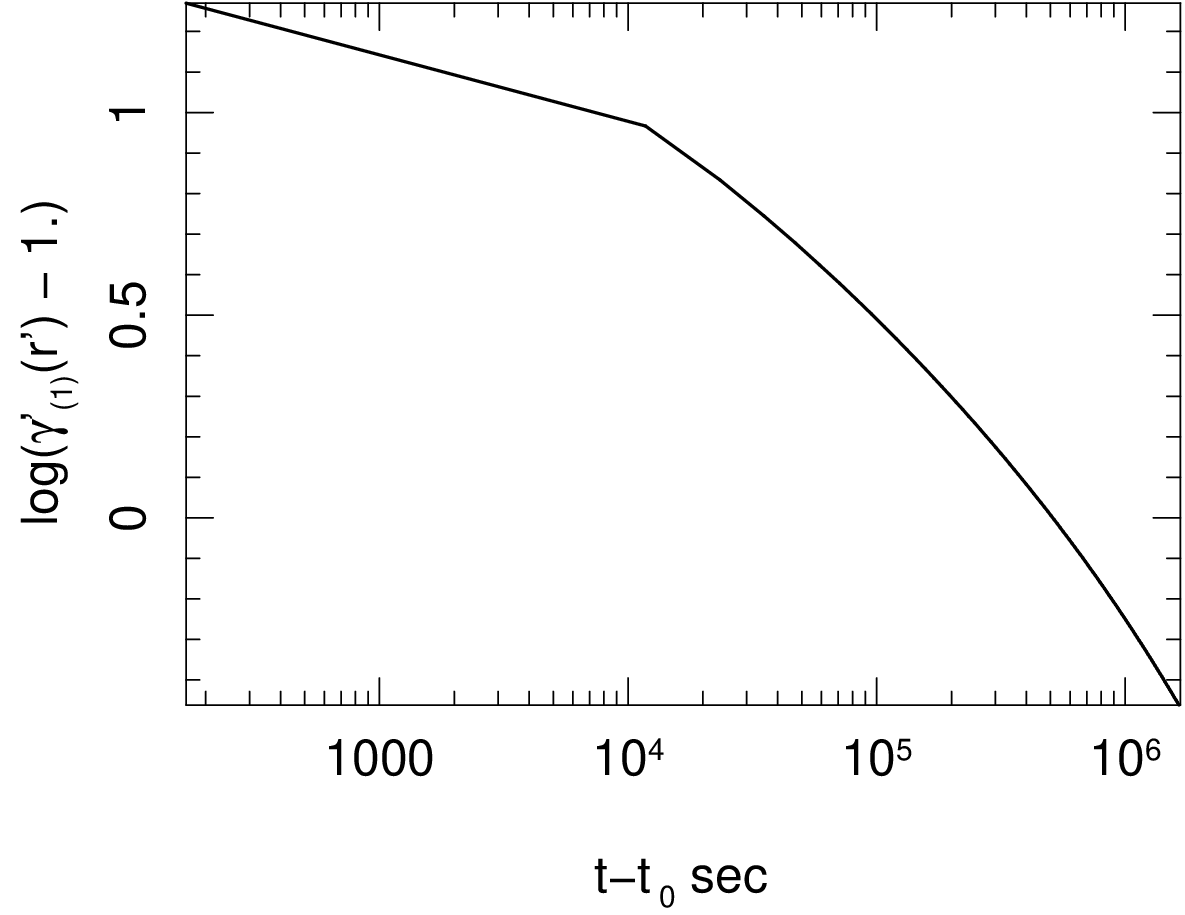} \\
13)\includegraphics[height=4cm,angle=-90]{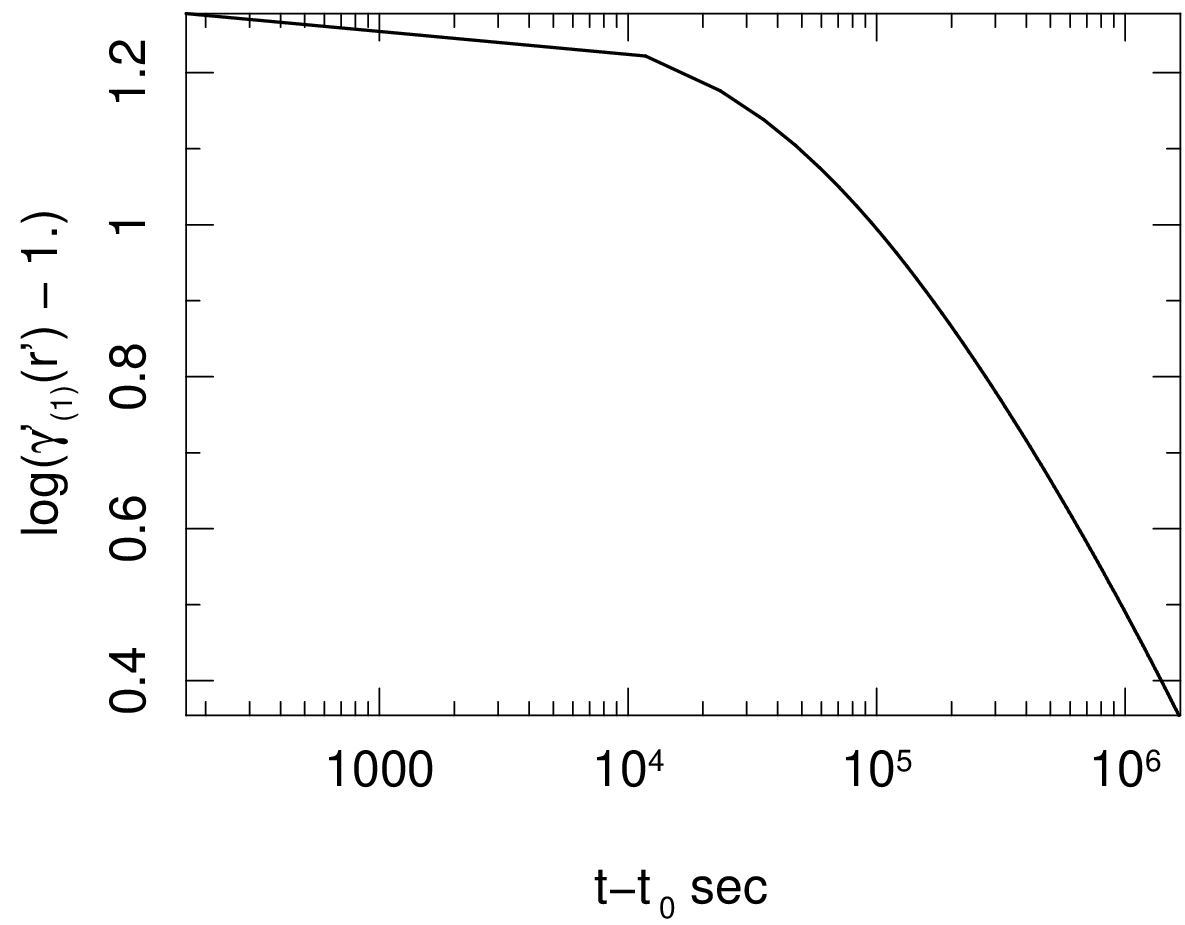} &
14)\includegraphics[height=4cm,angle=-90]{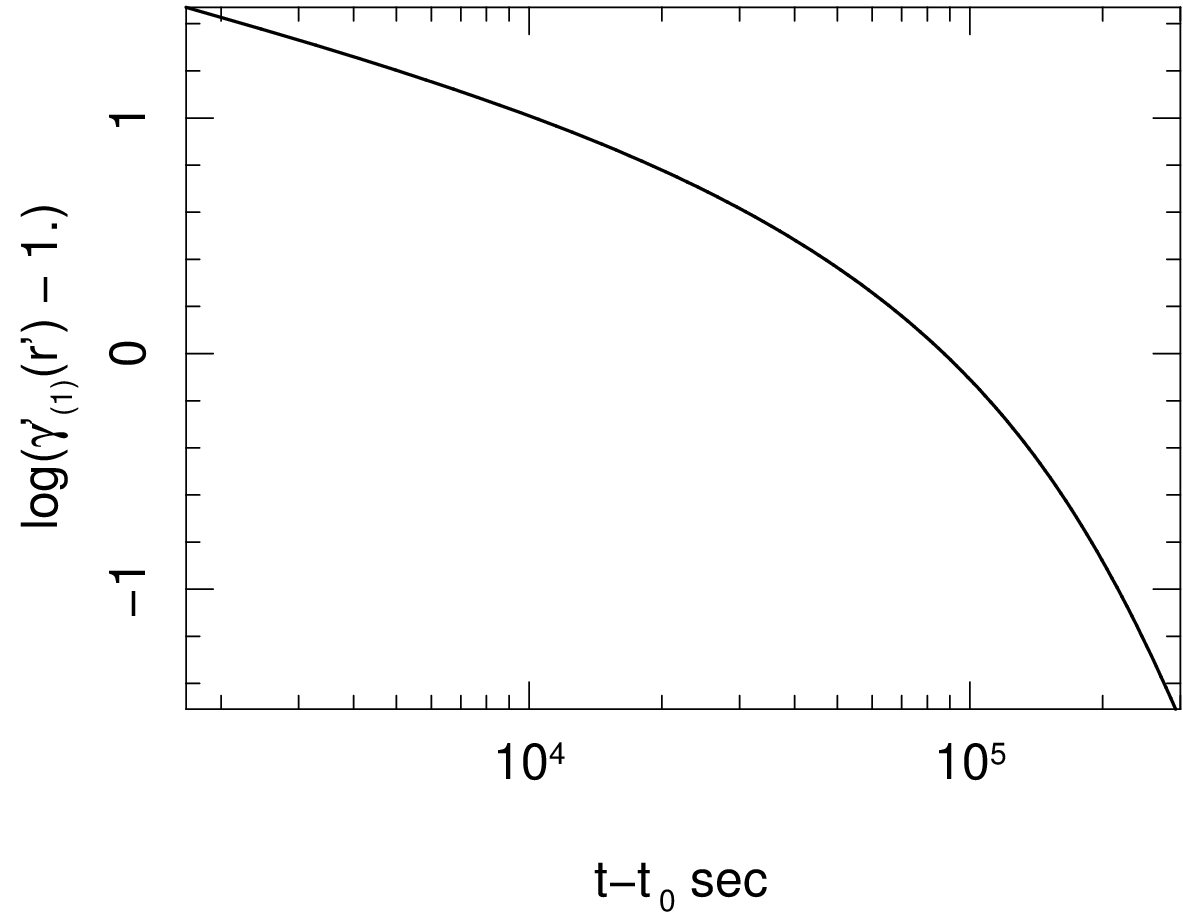} &
15)\includegraphics[height=4cm,angle=-90]{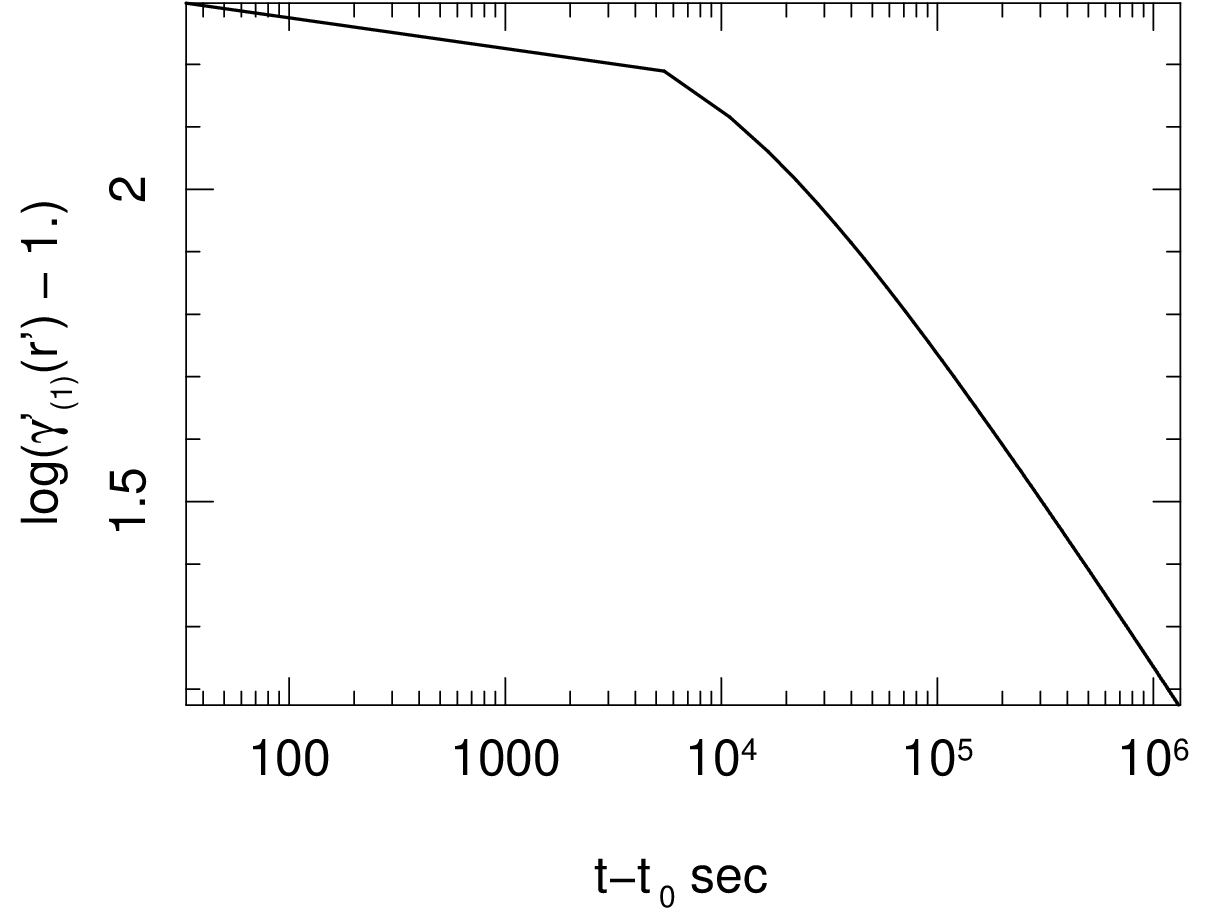} &
16)\includegraphics[height=4cm,angle=-90]{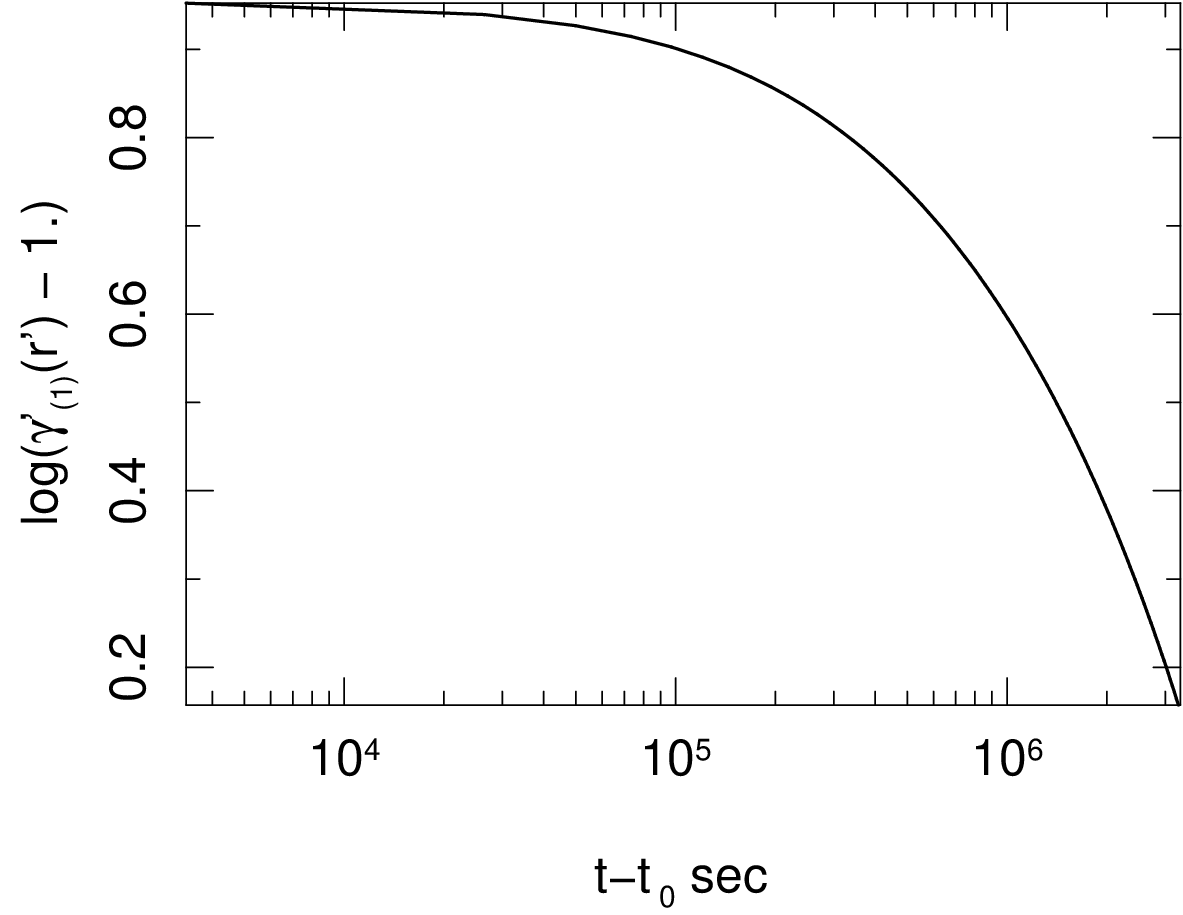} \\
17)\includegraphics[height=4cm,angle=-90]{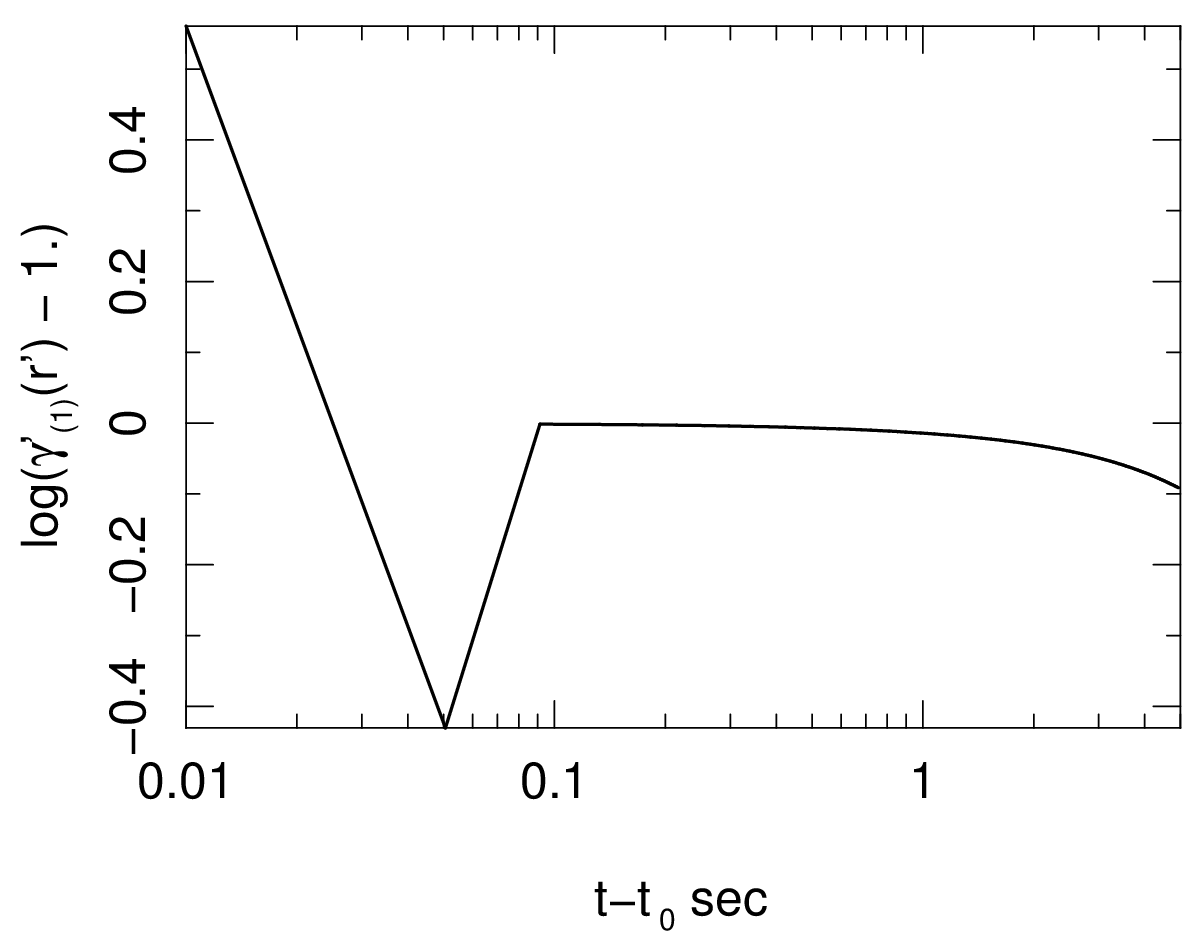} &
18)\includegraphics[height=4cm,angle=-90]{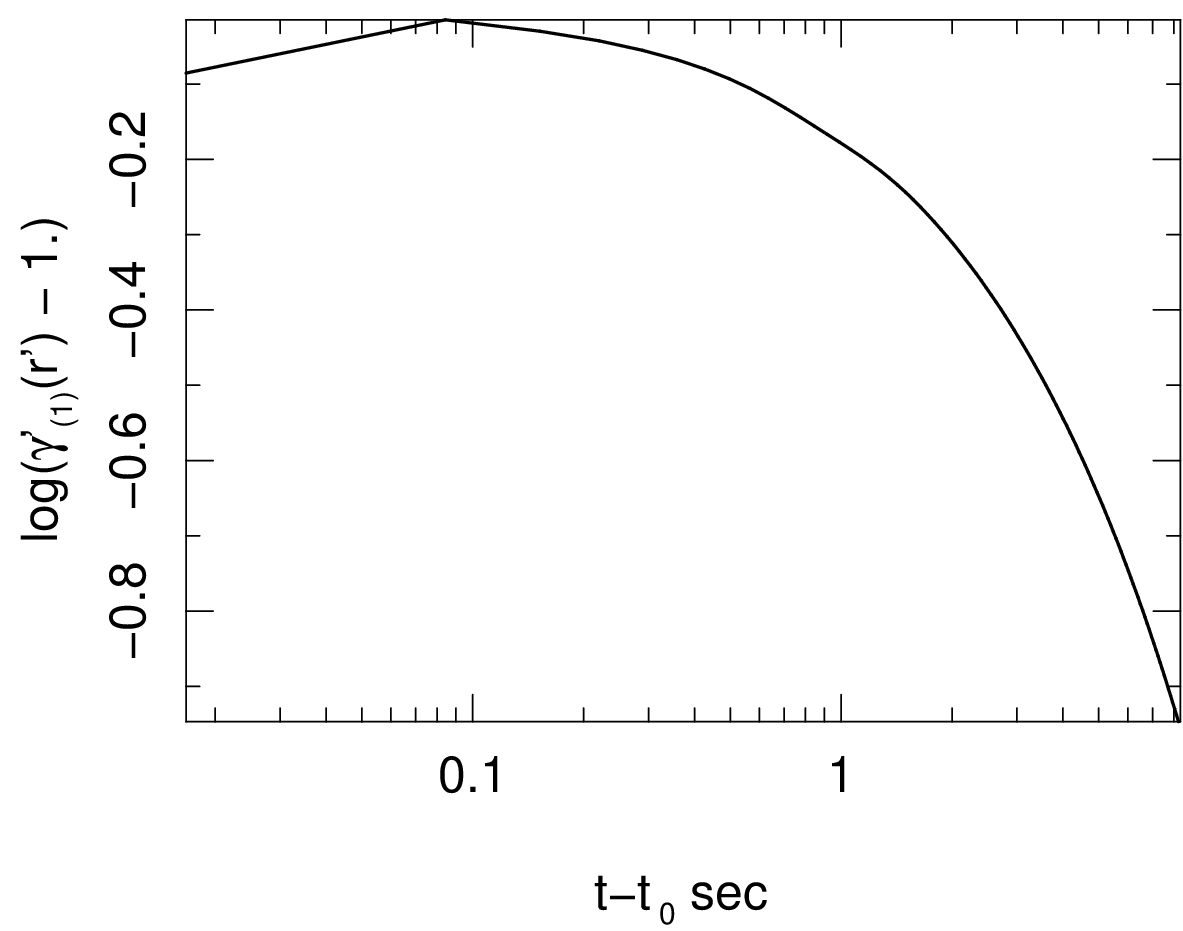} &
19)\includegraphics[height=4cm,angle=-90]{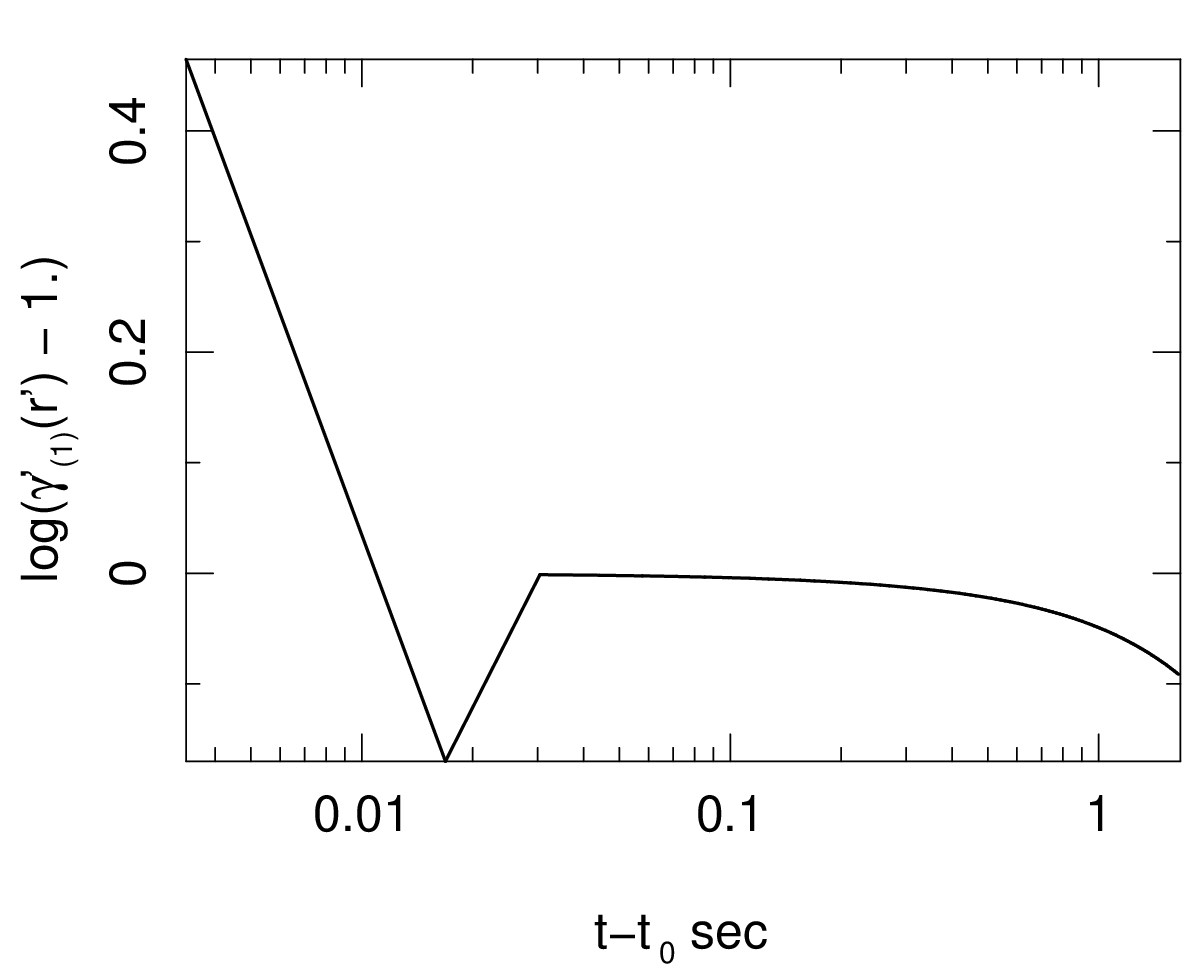} &
20)\includegraphics[height=4cm,angle=-90]{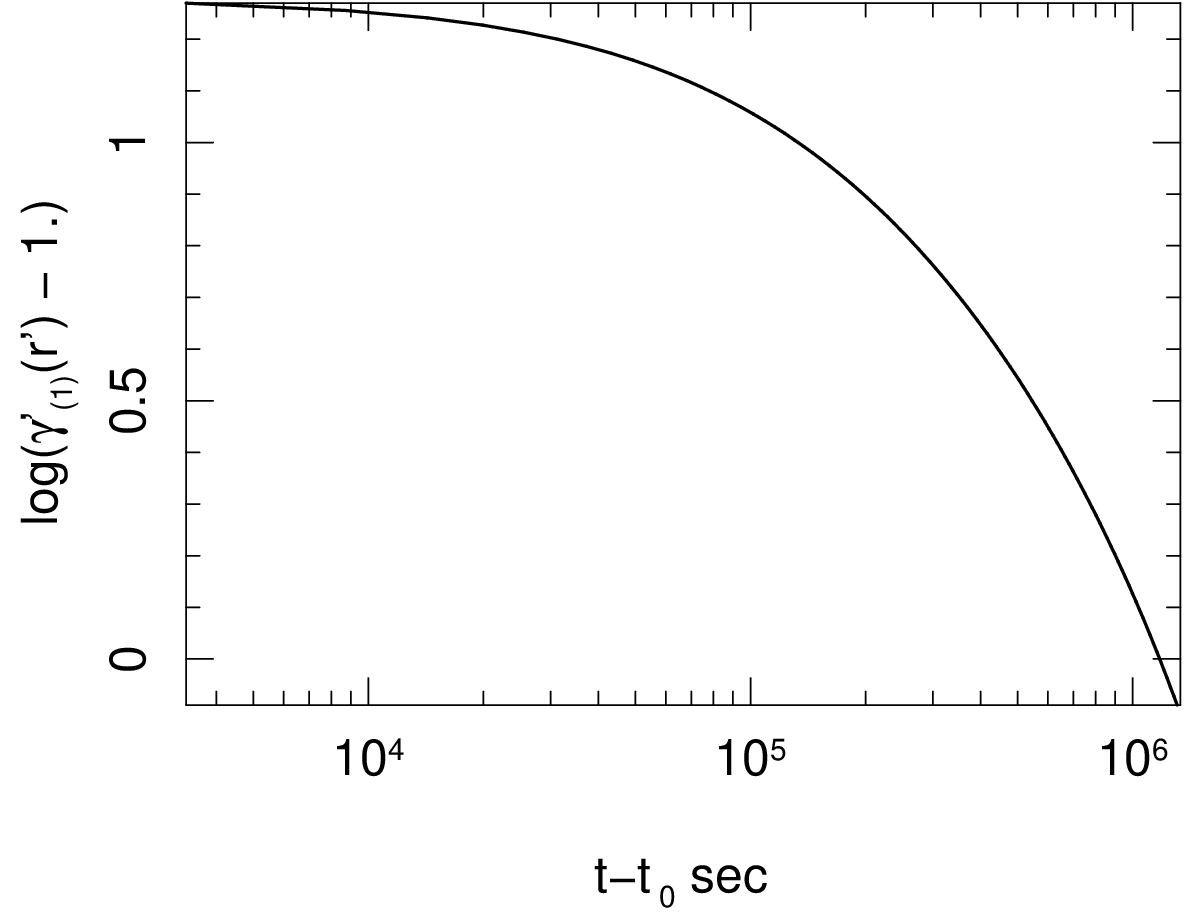} \\
21)\includegraphics[height=4cm,angle=-90]{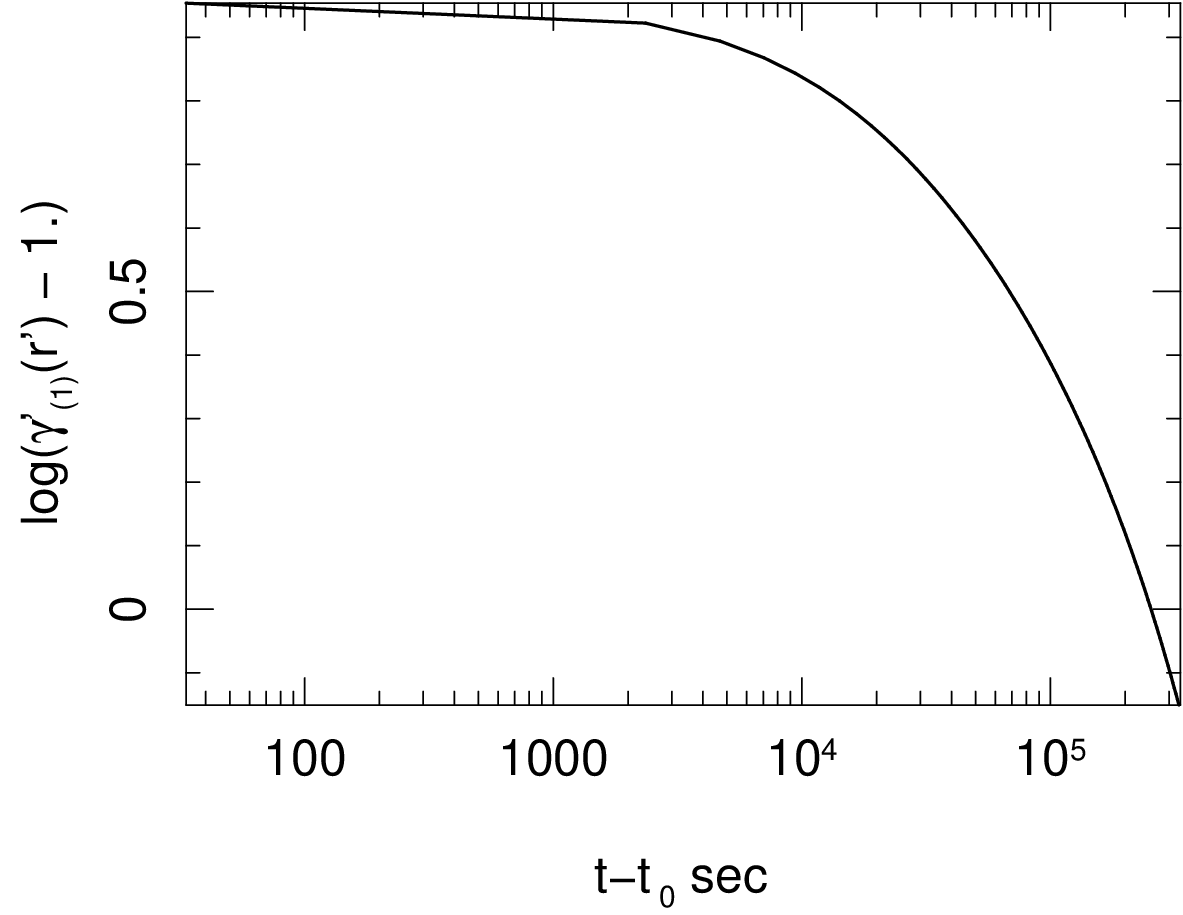} &
22)\includegraphics[height=4cm,angle=-90]{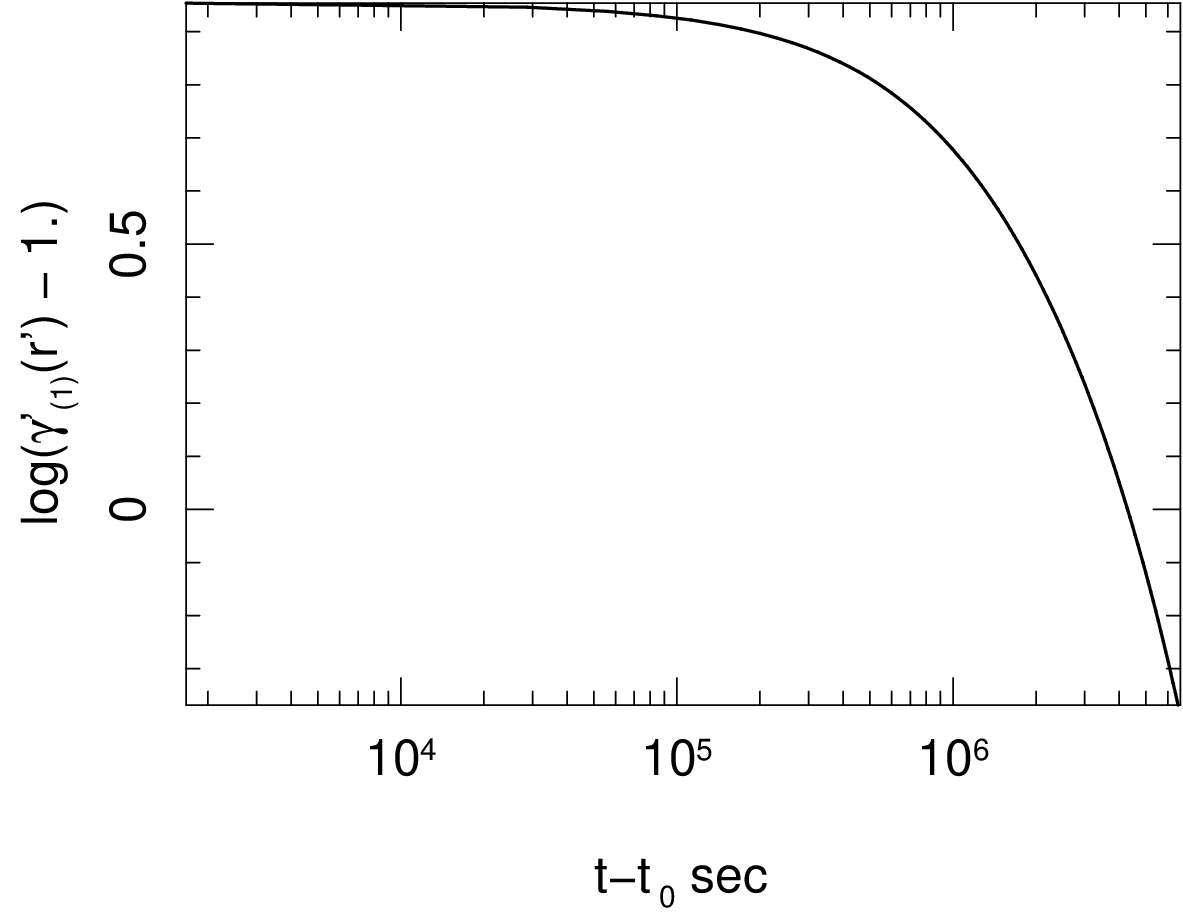} &
23)\includegraphics[height=4cm,angle=-90]{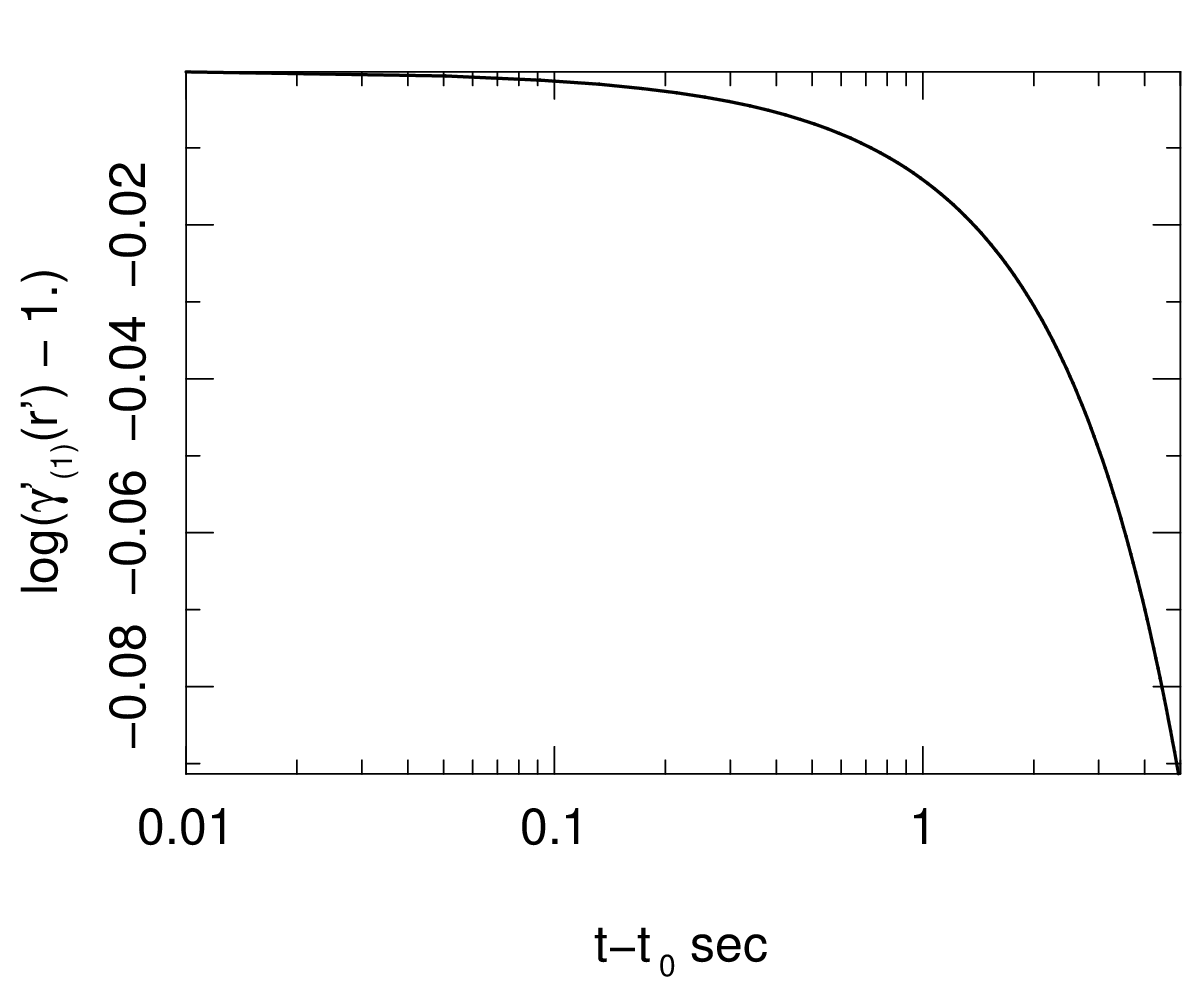} &
24)\includegraphics[height=4cm,angle=-90]{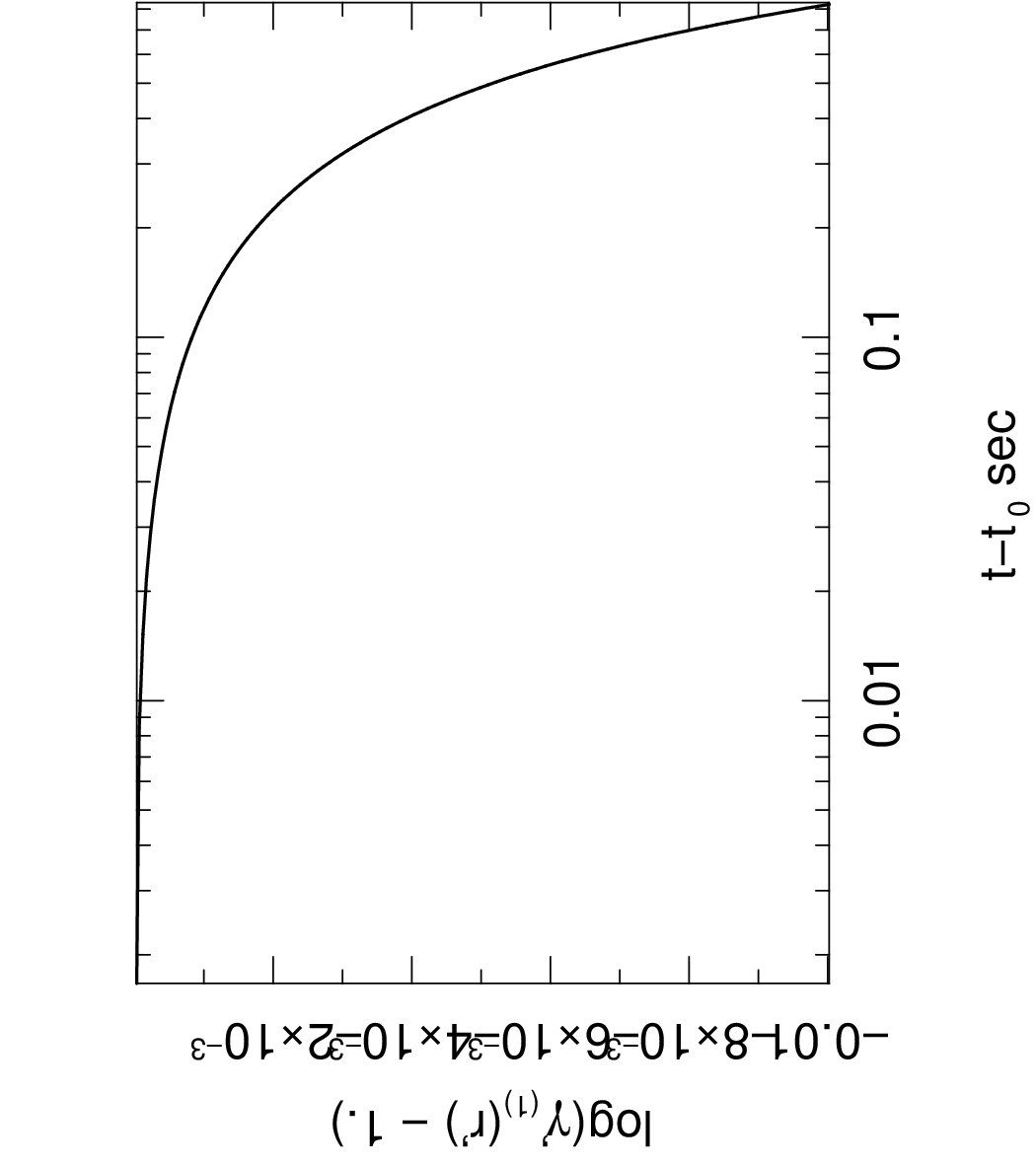} 
\end {tabular}
\end {center}
\caption{Evolution of the first order relative Lorentz factor $\gamma'_1$ with 
time/radius. In place of plotting $\gamma'_1$ itself we plot 
$\log(\gamma'_1 - 1)$ to see more easily when the shock becomes 
non-relativistic. Definitions of curves are the same as Fig. \ref{fig:betazero}.
\label{fig:gammaone}}
\end{figure*}

\begin{figure*}
\begin{center}
\begin {tabular}{cccc} 
1)\includegraphics[height=4cm,angle=-90]{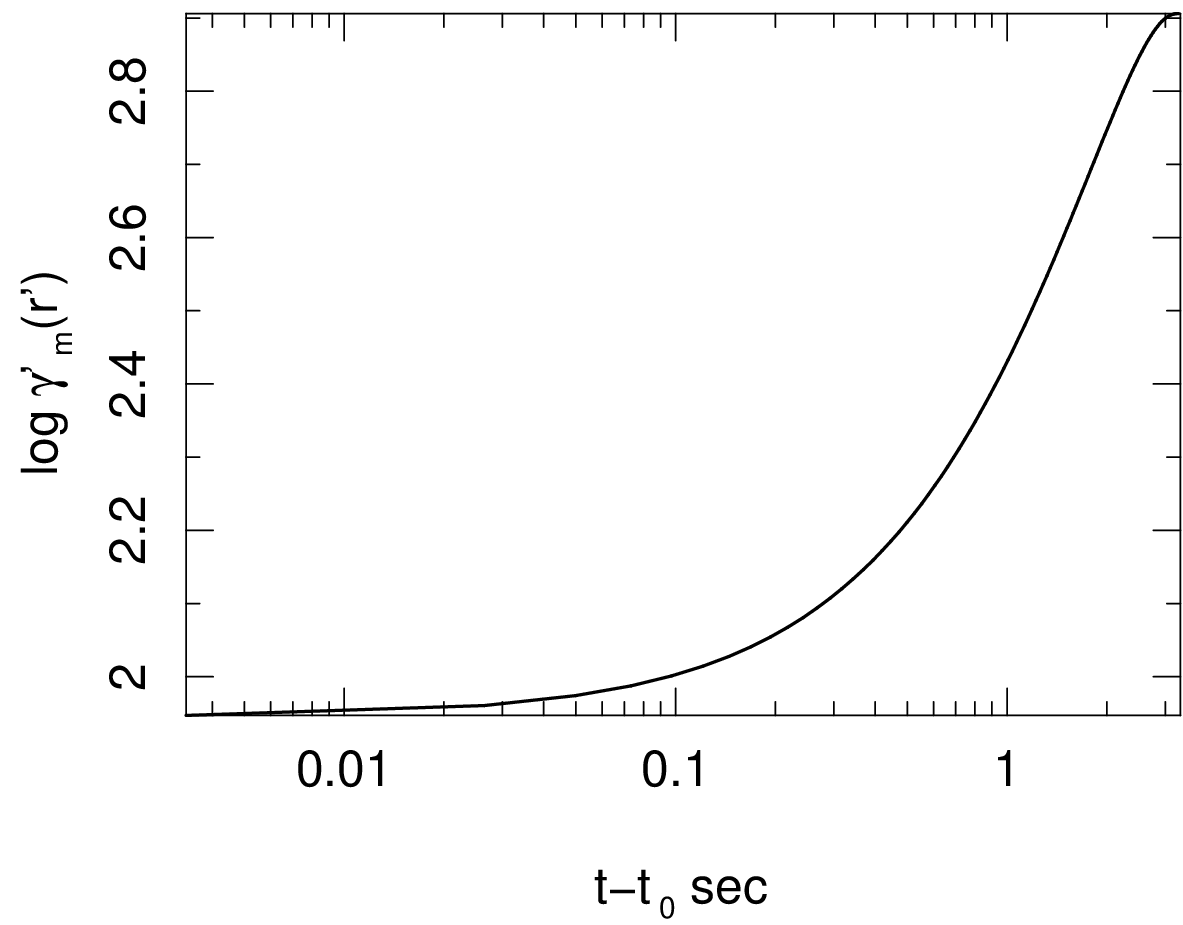} &
2)\includegraphics[height=4cm,angle=-90]{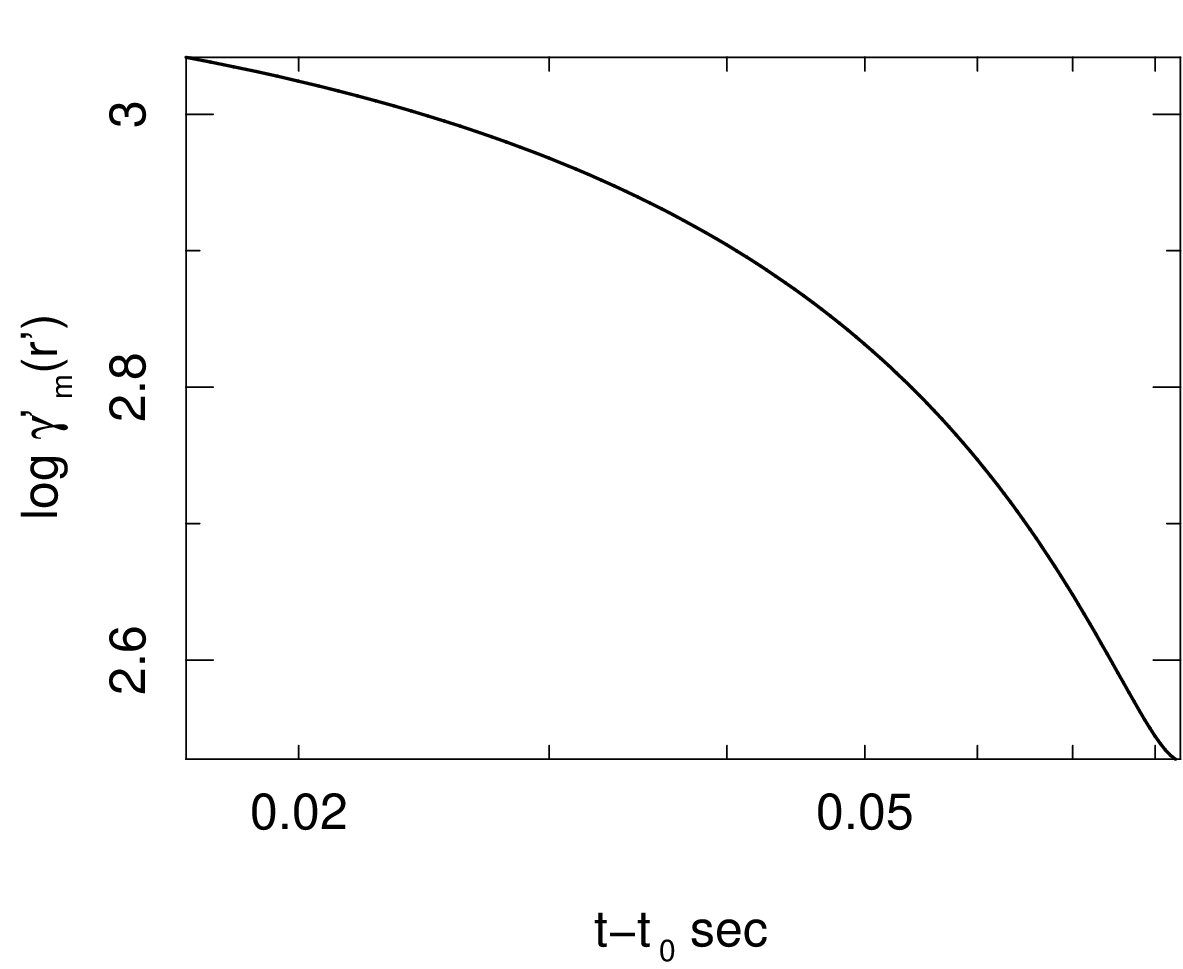} & 
3)\includegraphics[height=4cm,angle=-90]{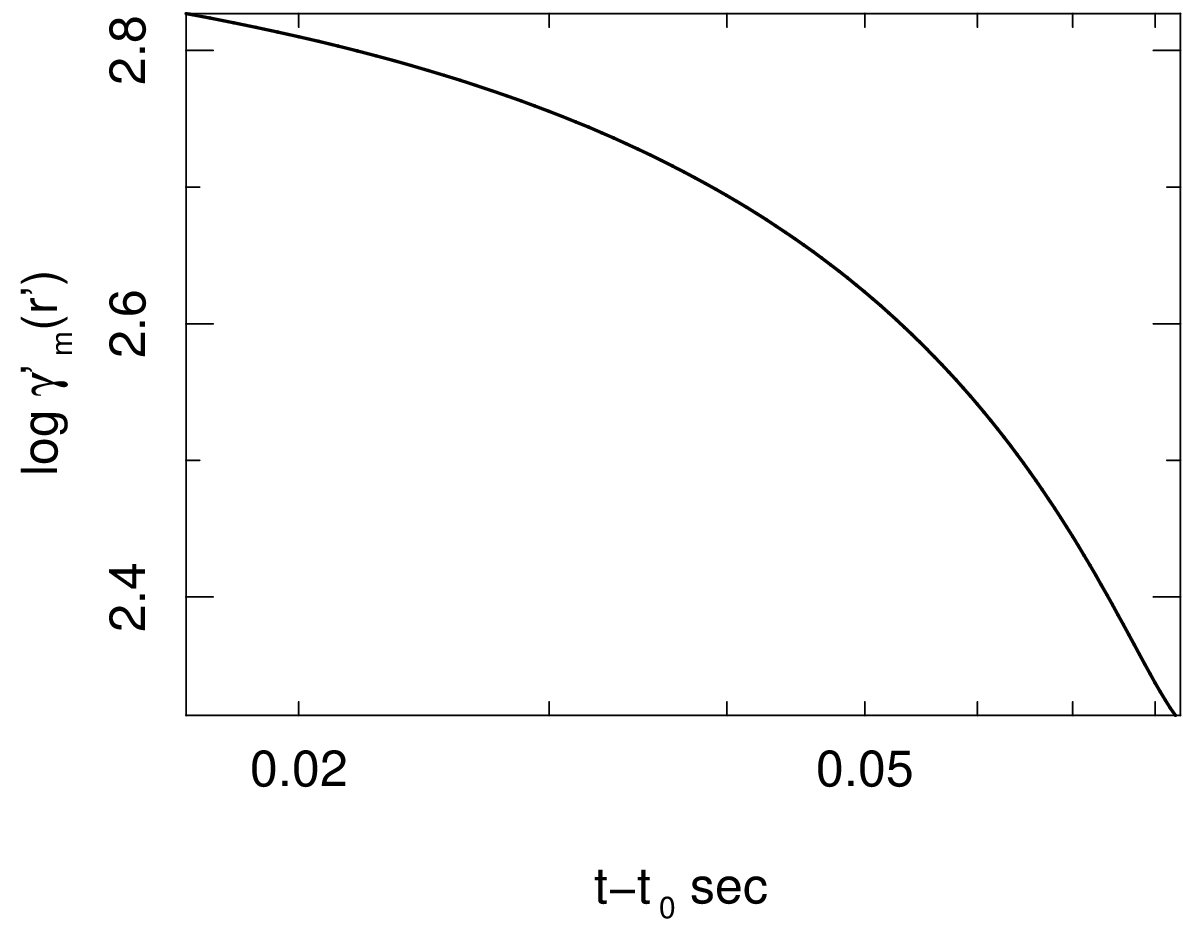} &
4)\includegraphics[height=4cm,angle=-90]{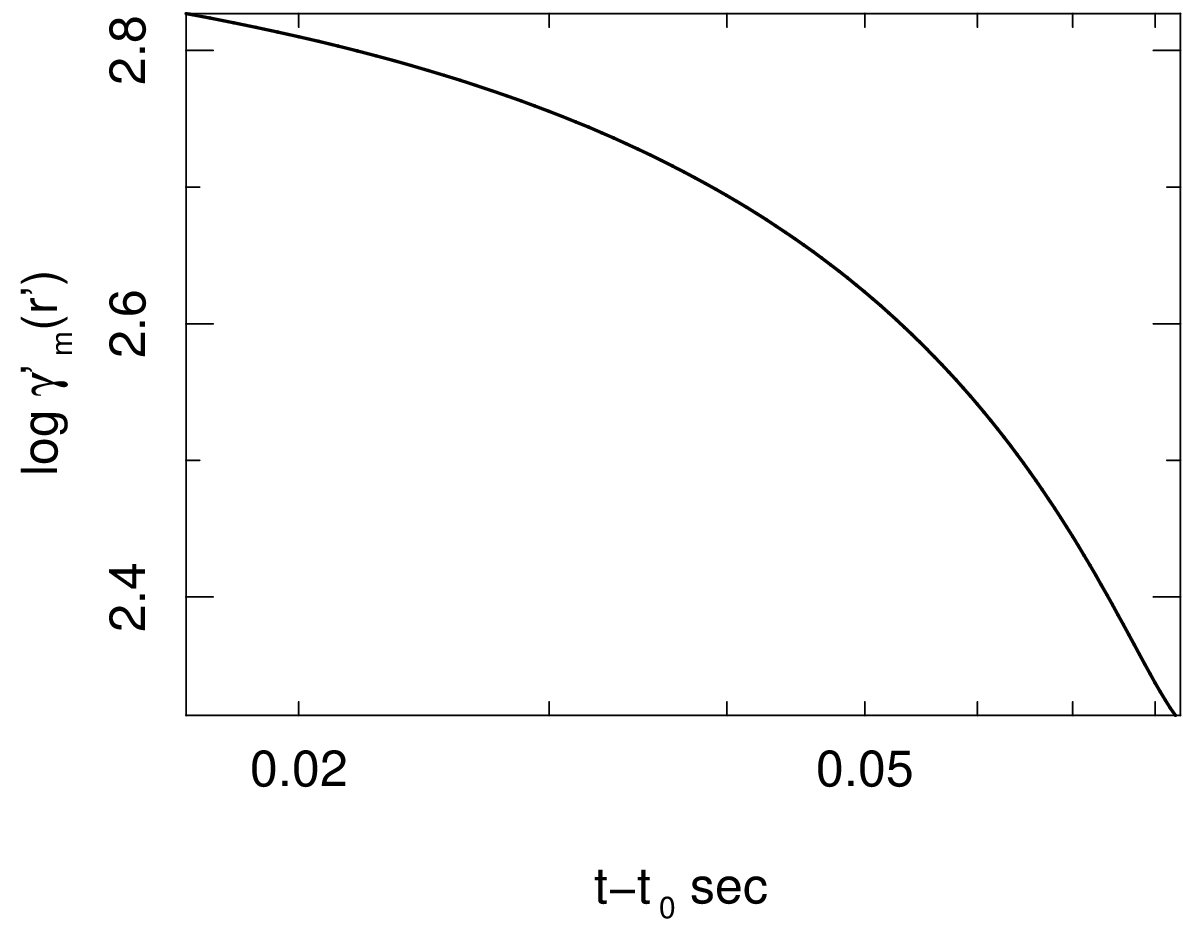} \\ 
5)\includegraphics[height=4cm,angle=-90]{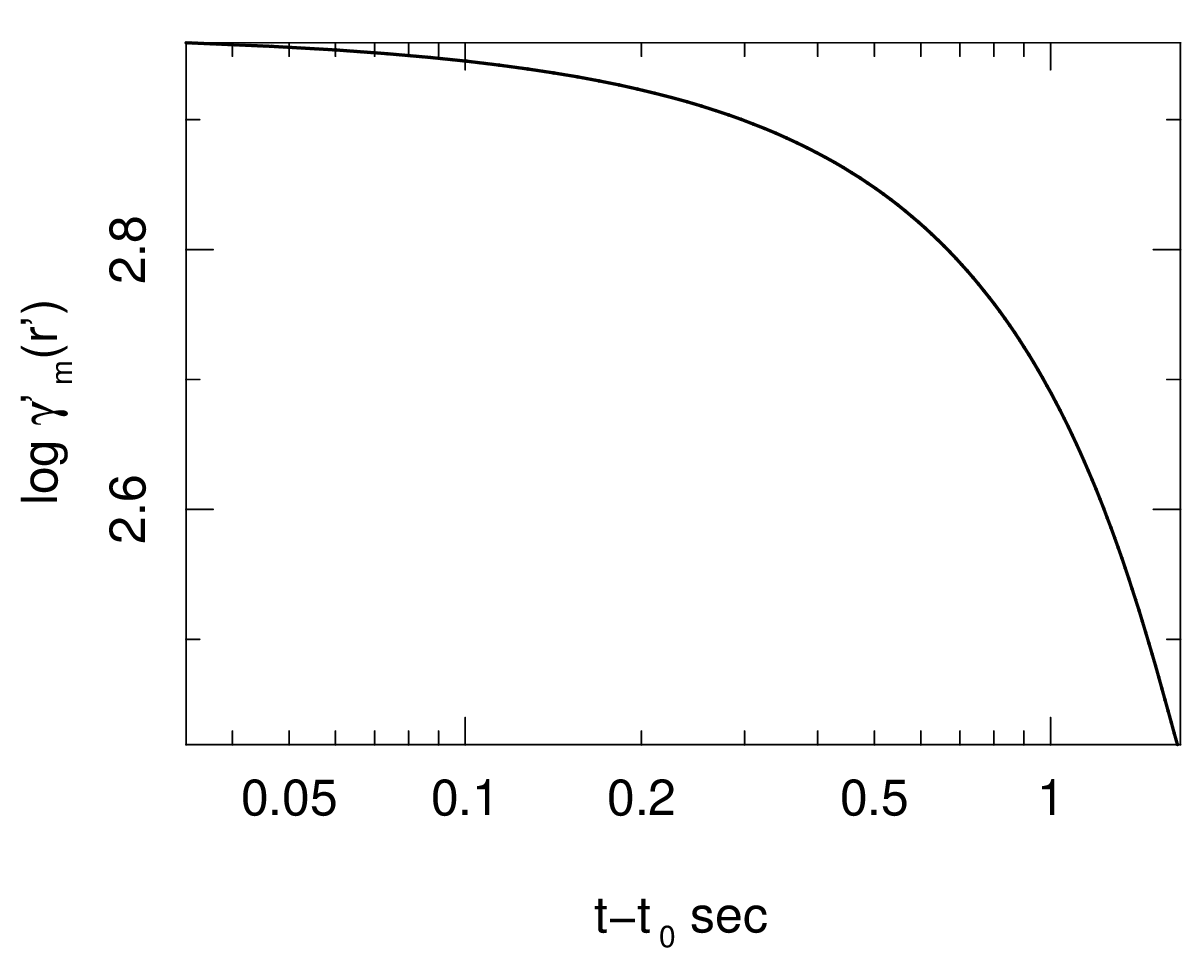} &
6)\includegraphics[height=4cm,angle=-90]{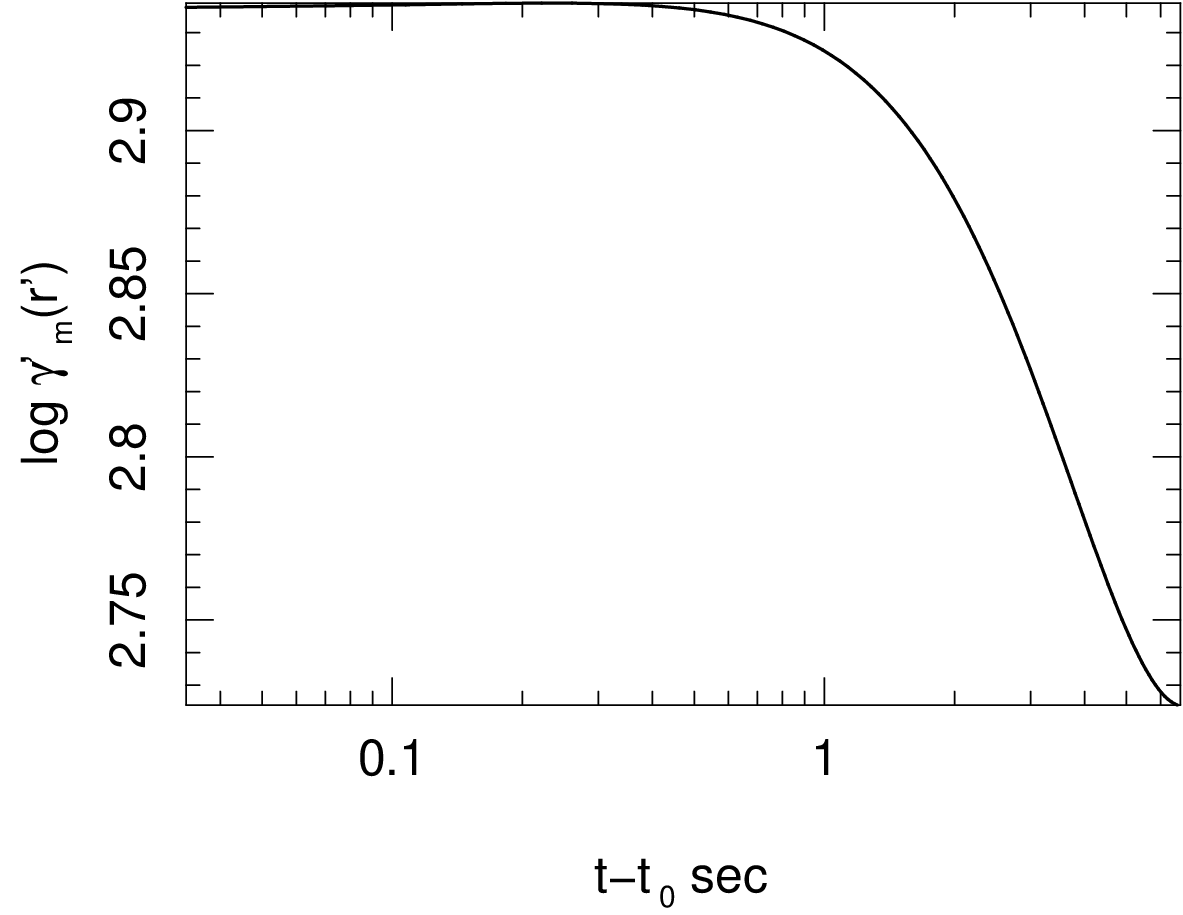} & 
7)\includegraphics[height=4cm,angle=-90]{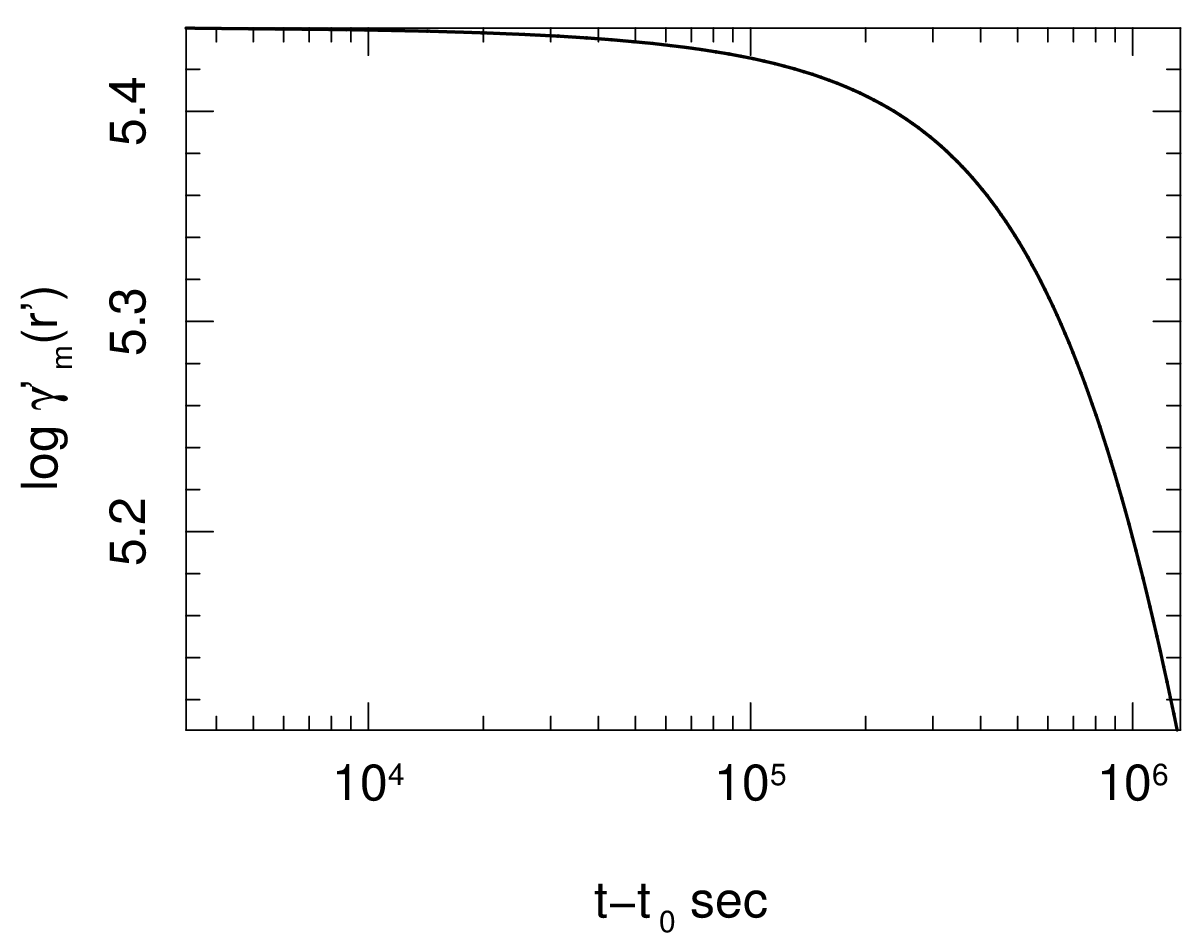} &
8)\includegraphics[height=4cm,angle=-90]{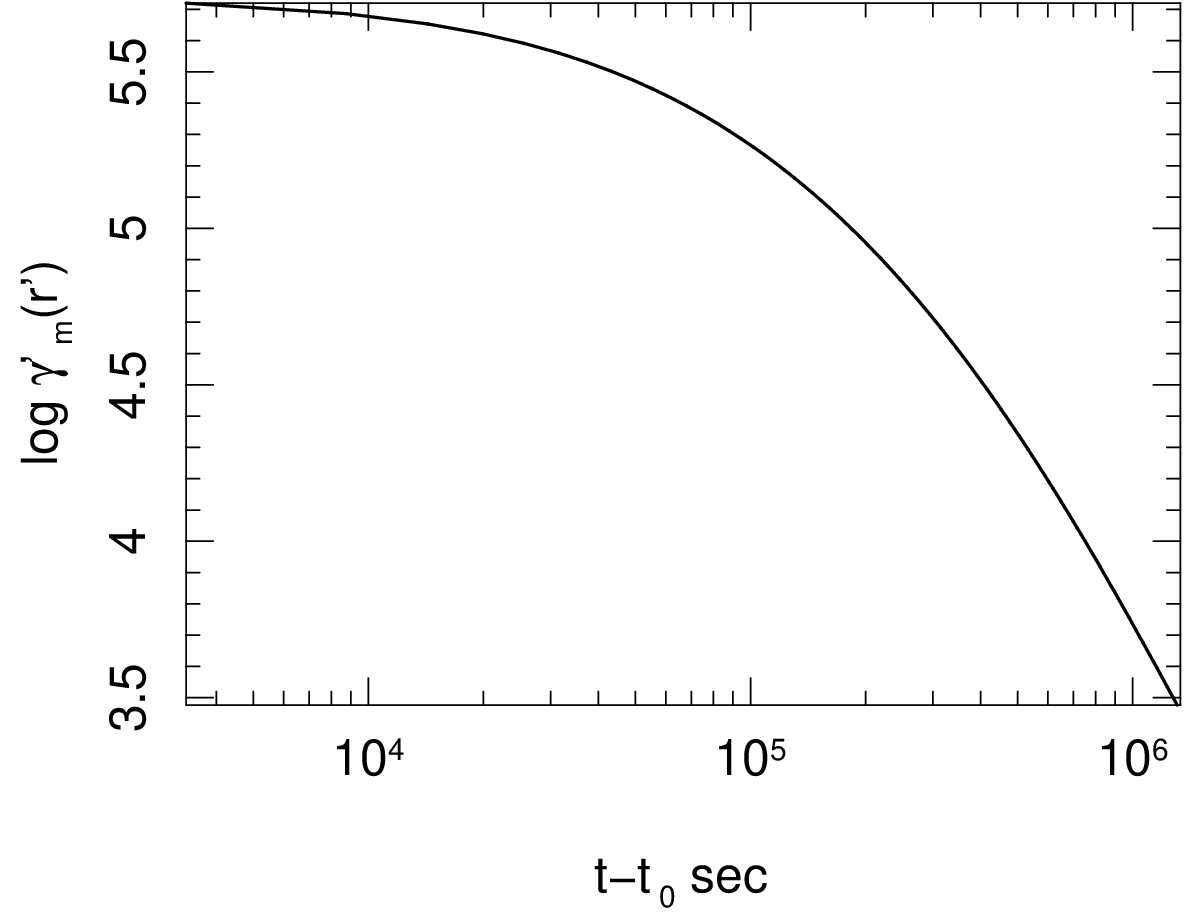} \\
9)\includegraphics[height=4cm,angle=-90]{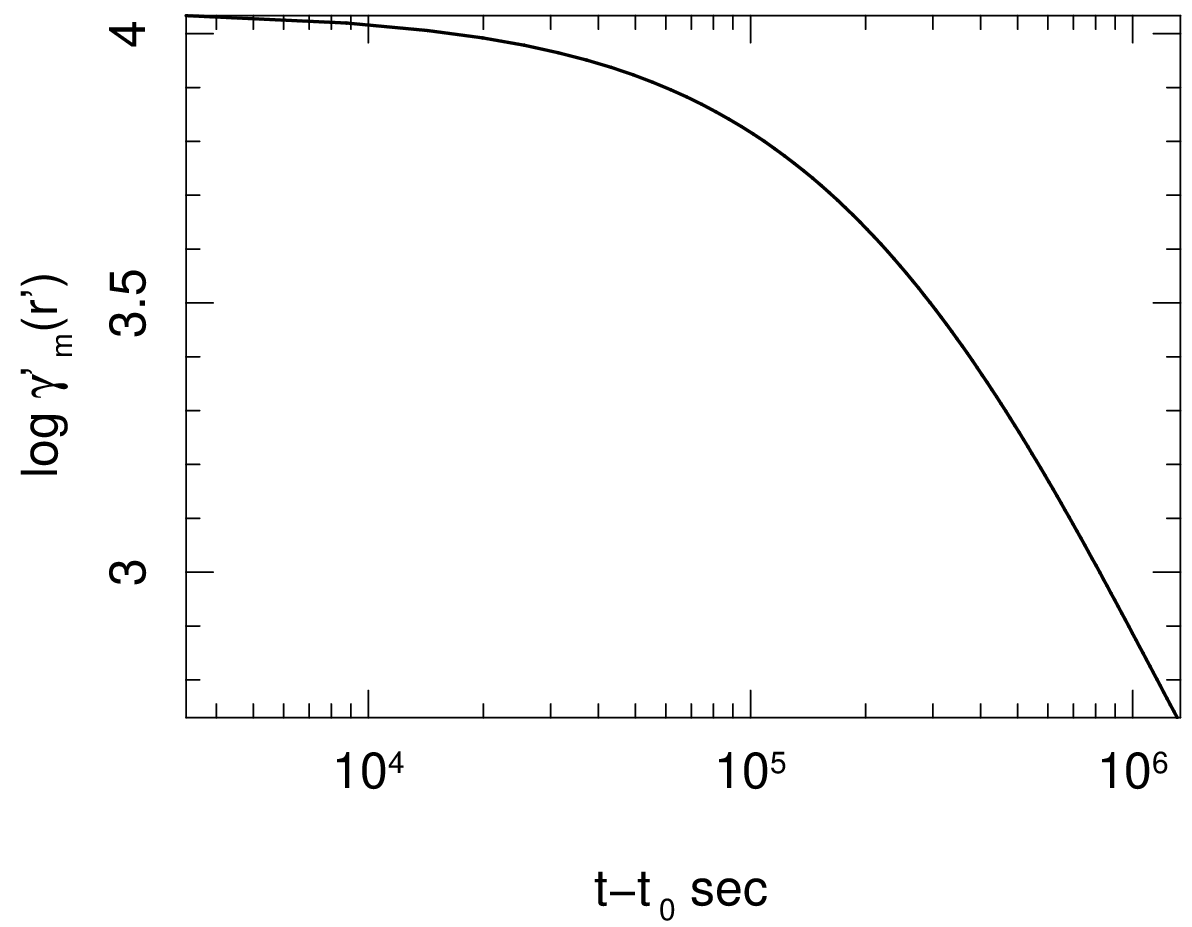} &
10)\includegraphics[height=4cm,angle=-90]{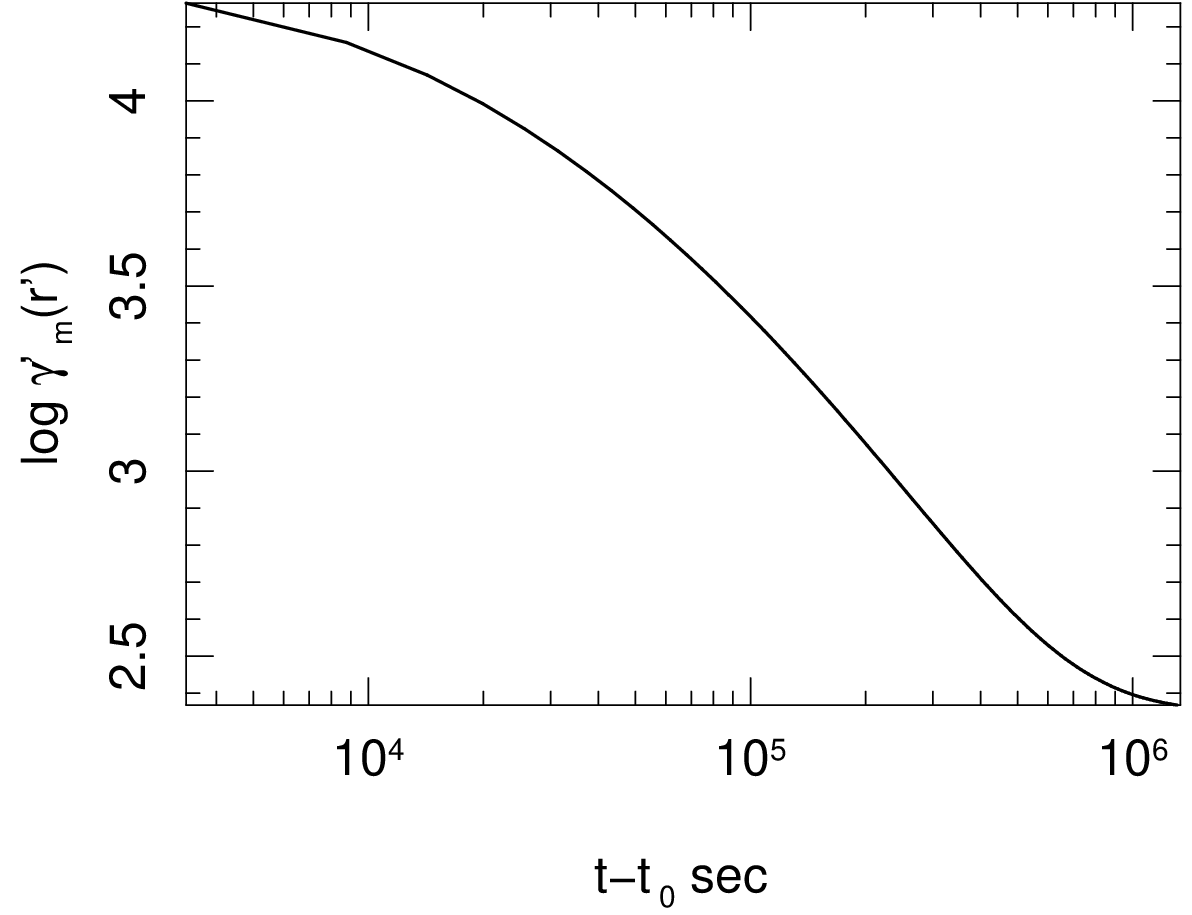} &
11)\includegraphics[height=4cm,angle=-90]{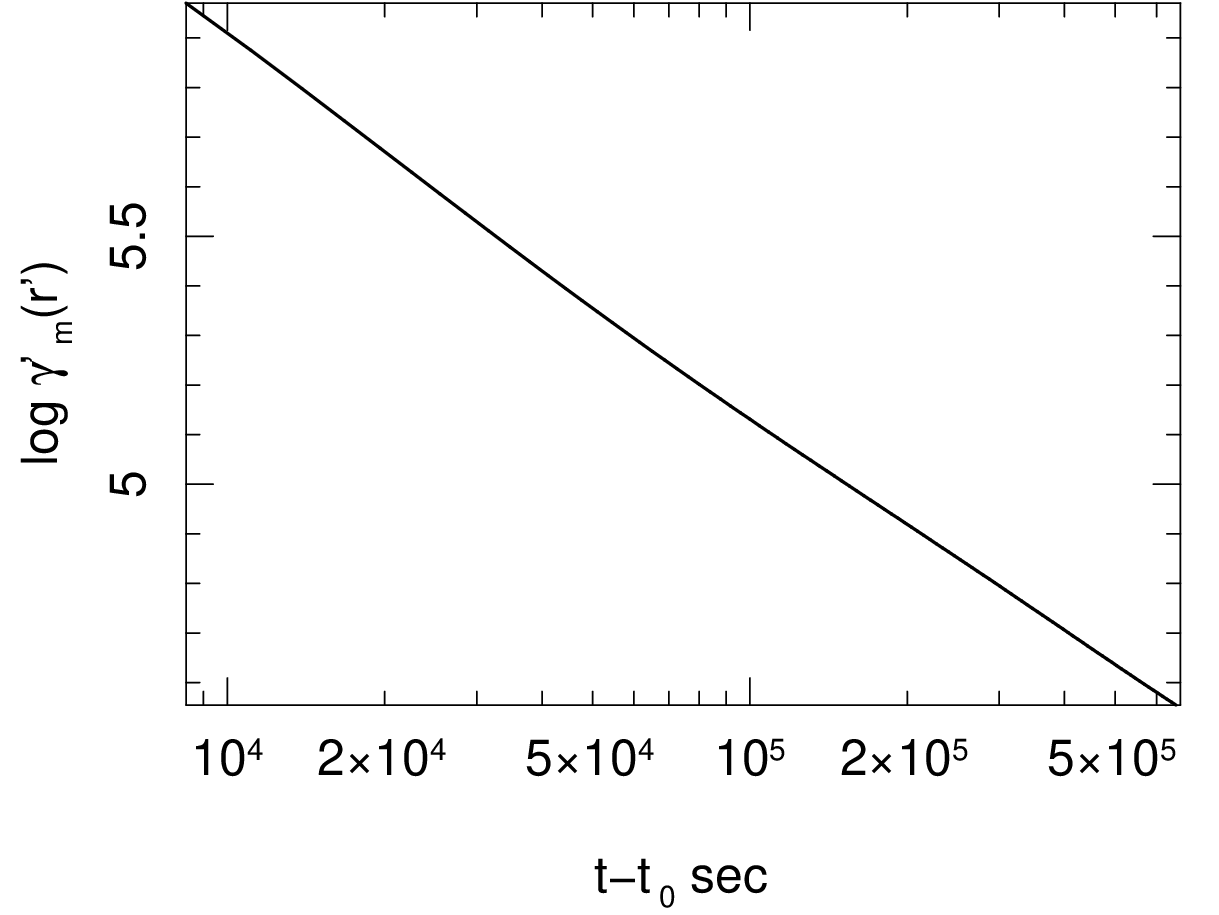} &
12)\includegraphics[height=4cm,angle=-90]{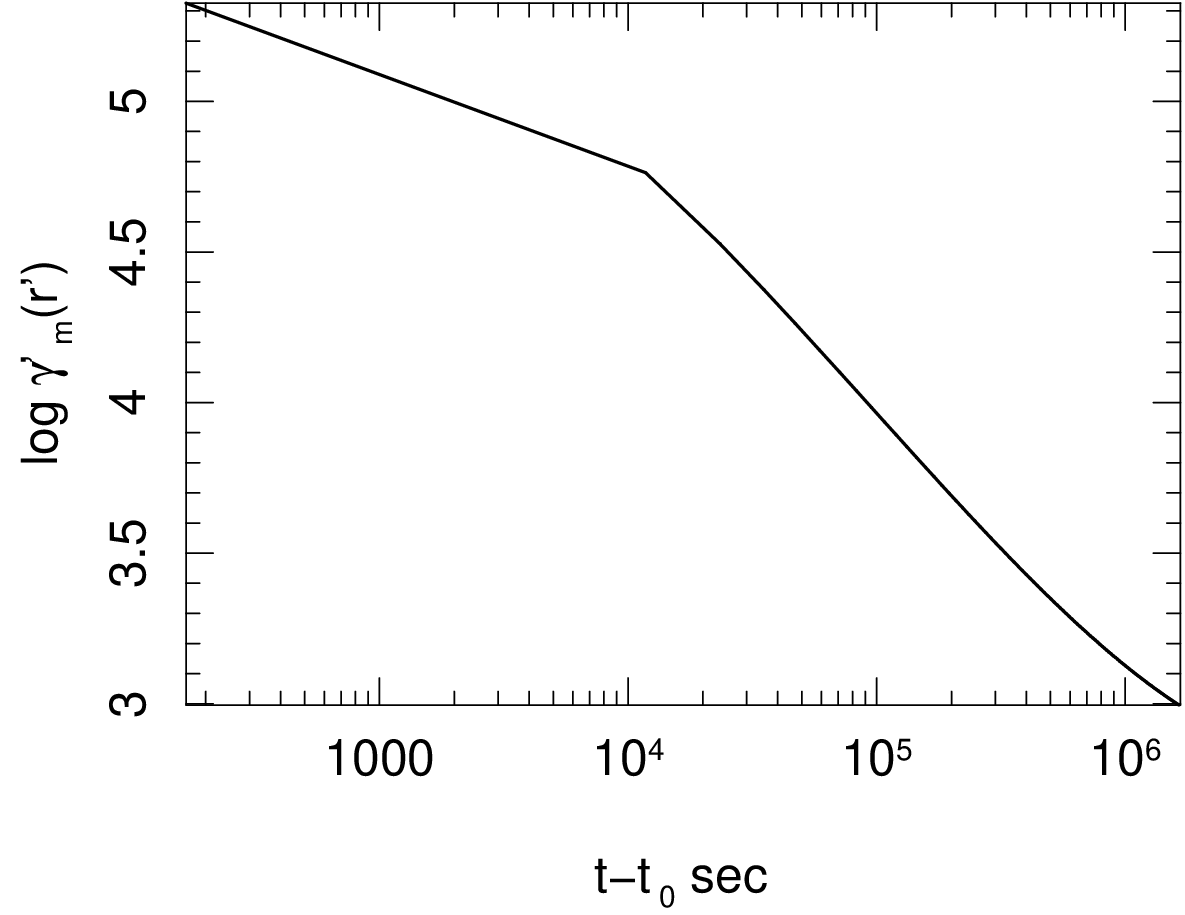} \\
13)\includegraphics[height=4cm,angle=-90]{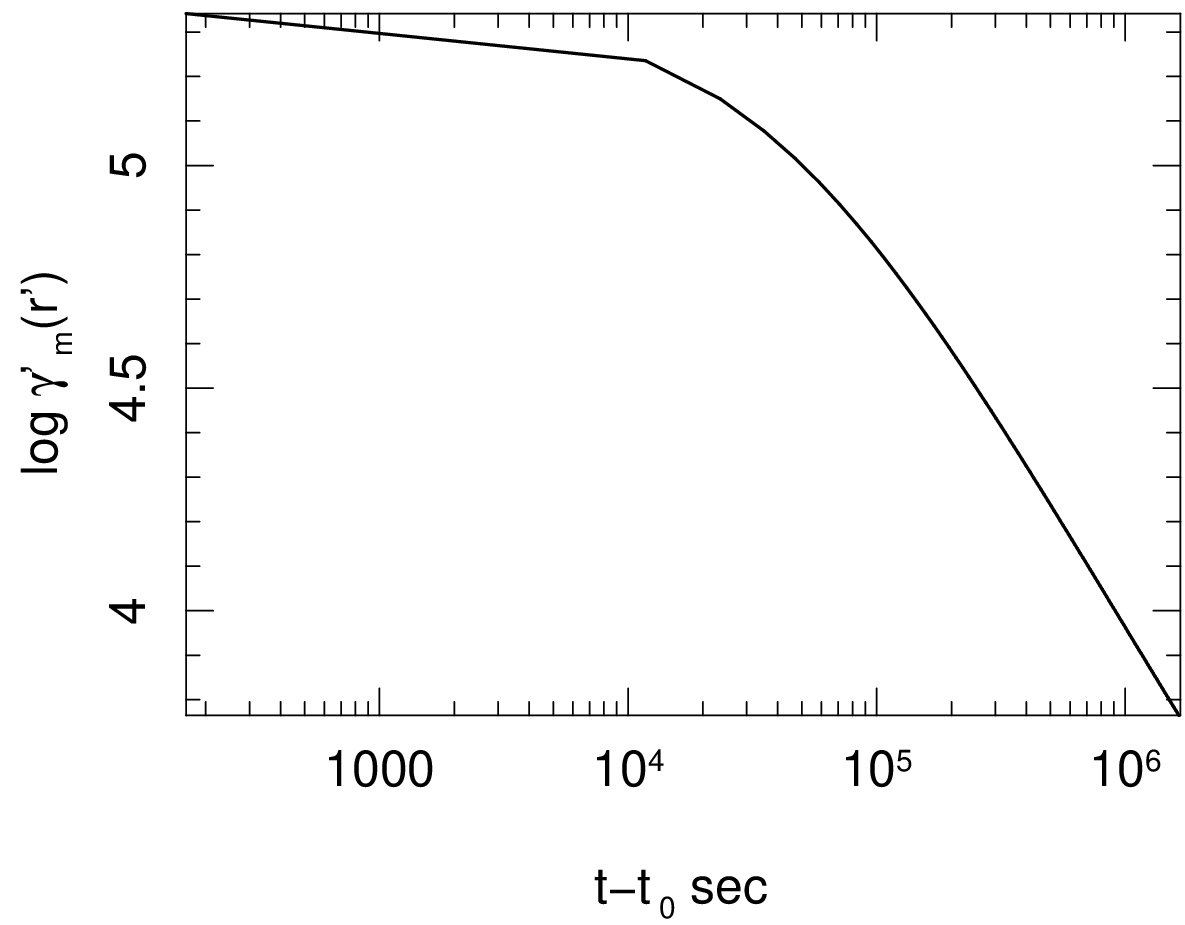} &
14)\includegraphics[height=4cm,angle=-90]{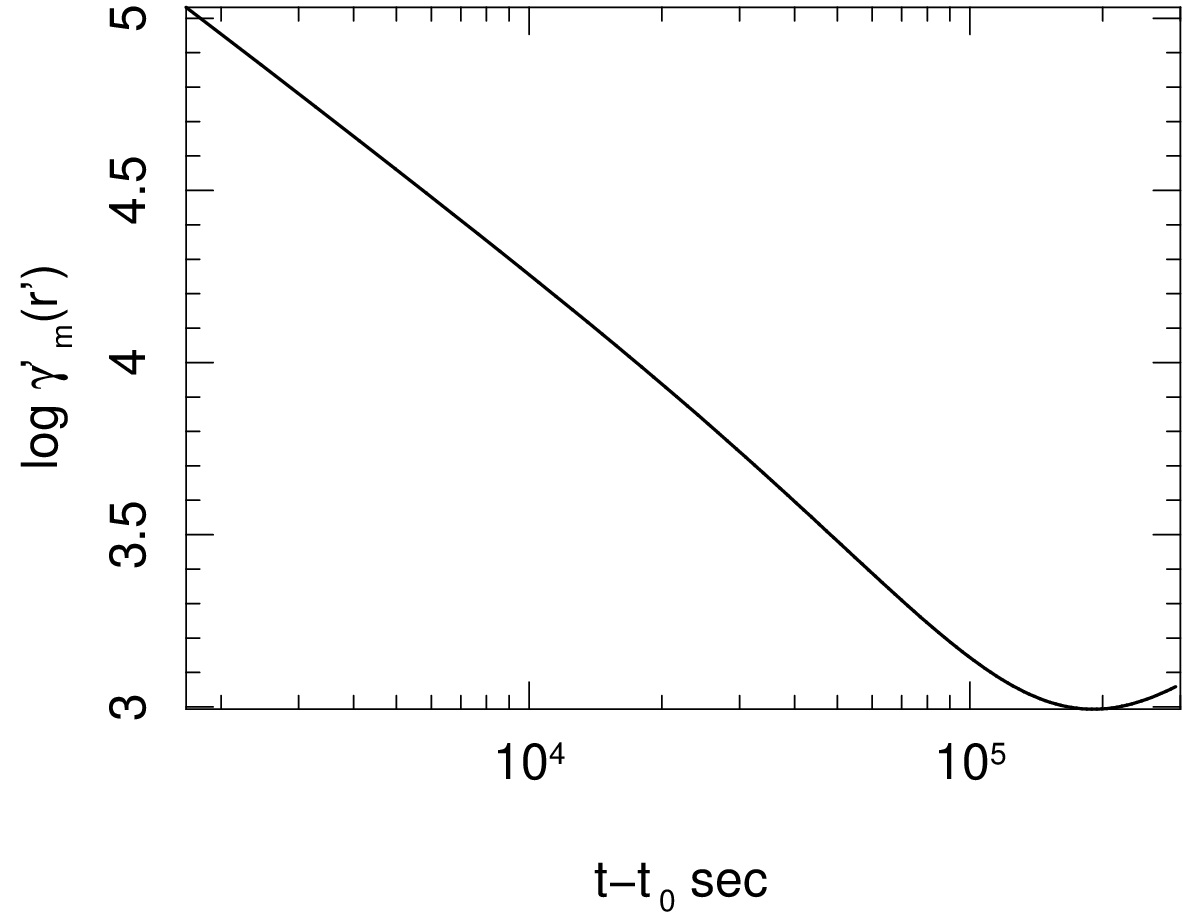} &
15)\includegraphics[height=4cm,angle=-90]{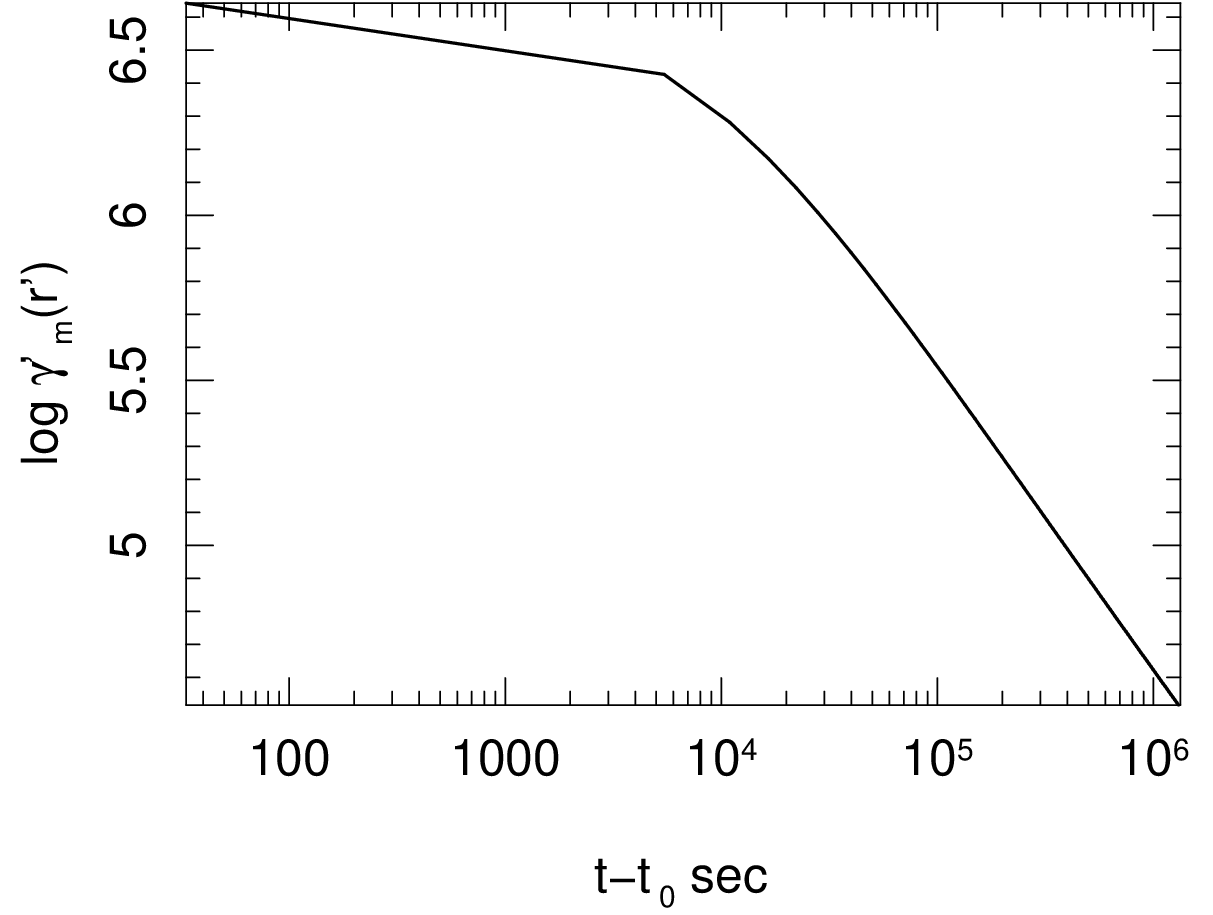} &
16)\includegraphics[height=4cm,angle=-90]{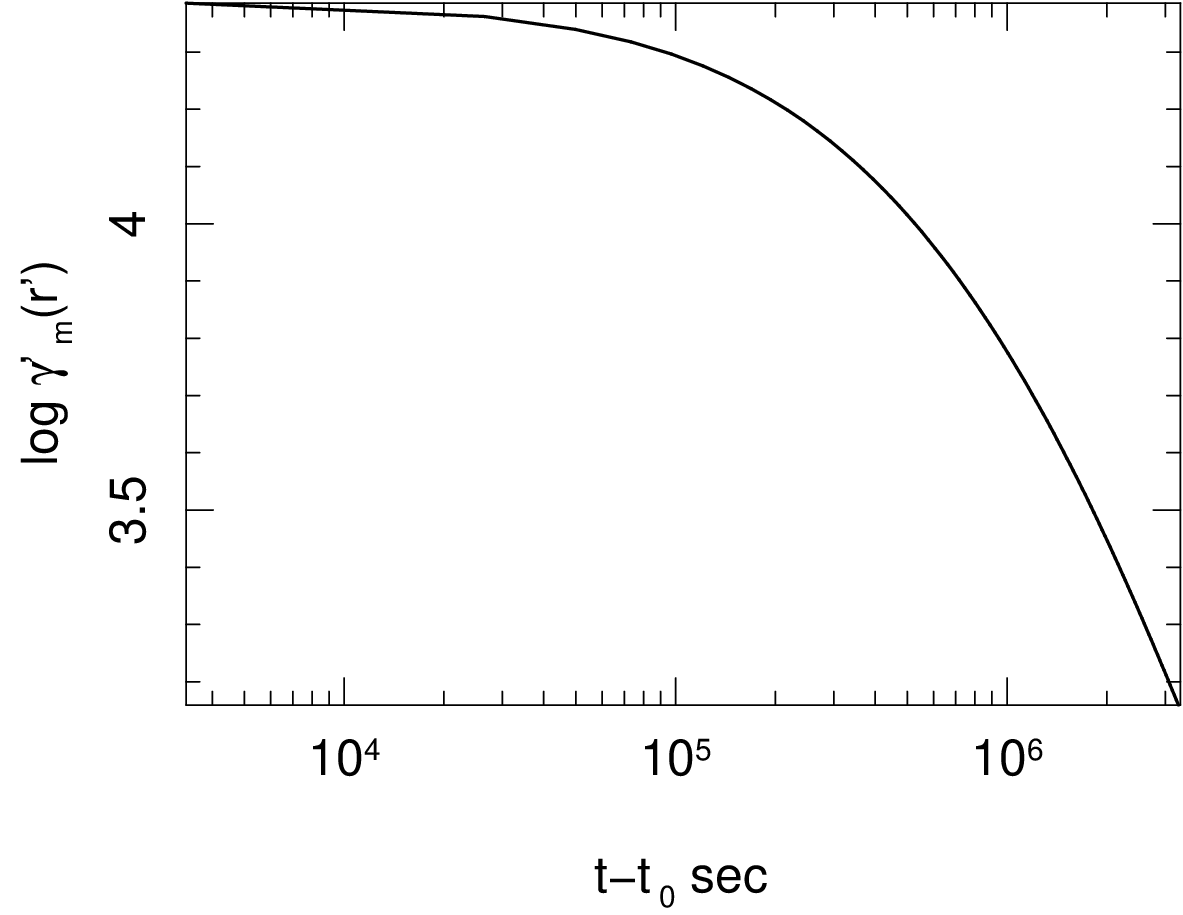} \\
17)\includegraphics[height=4cm,angle=-90]{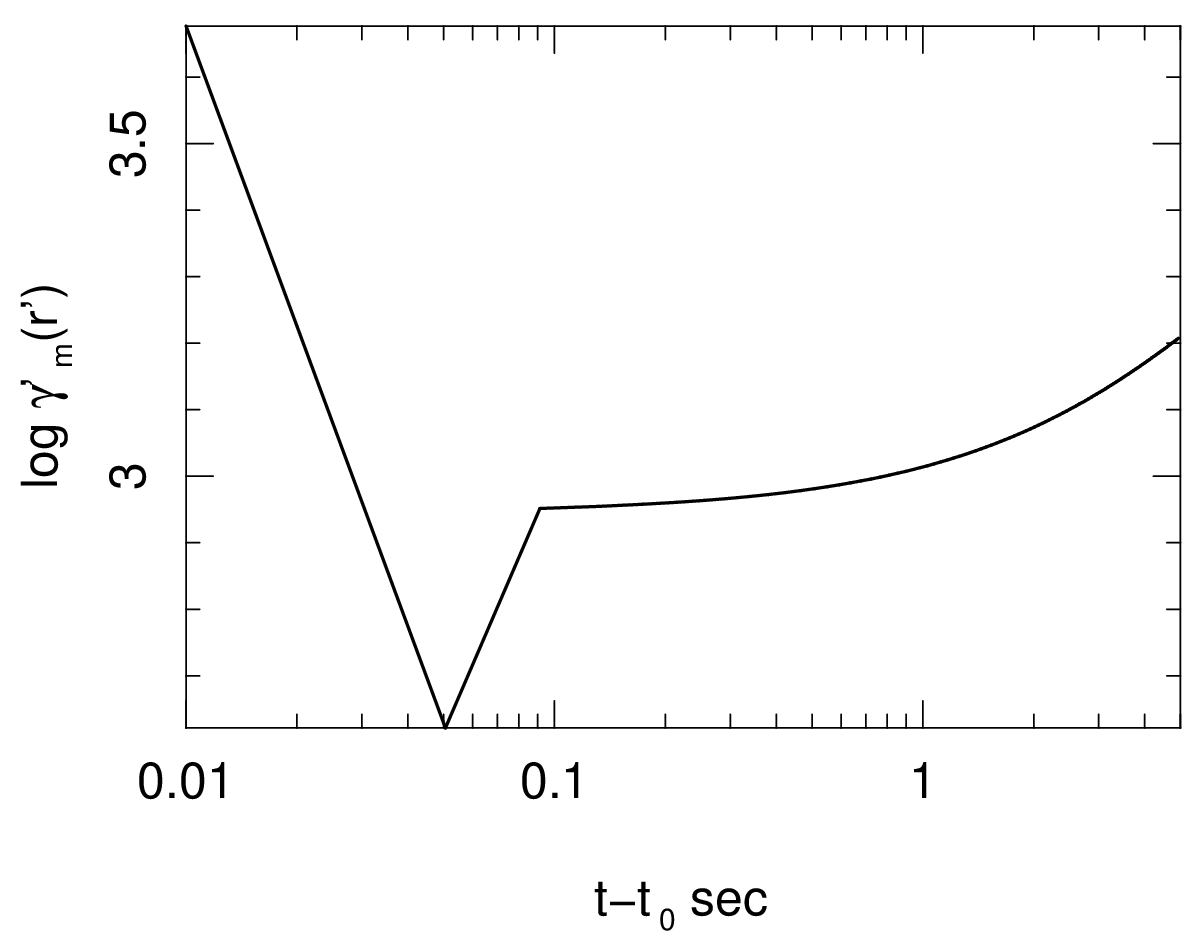} &
18)\includegraphics[height=4cm,angle=-90]{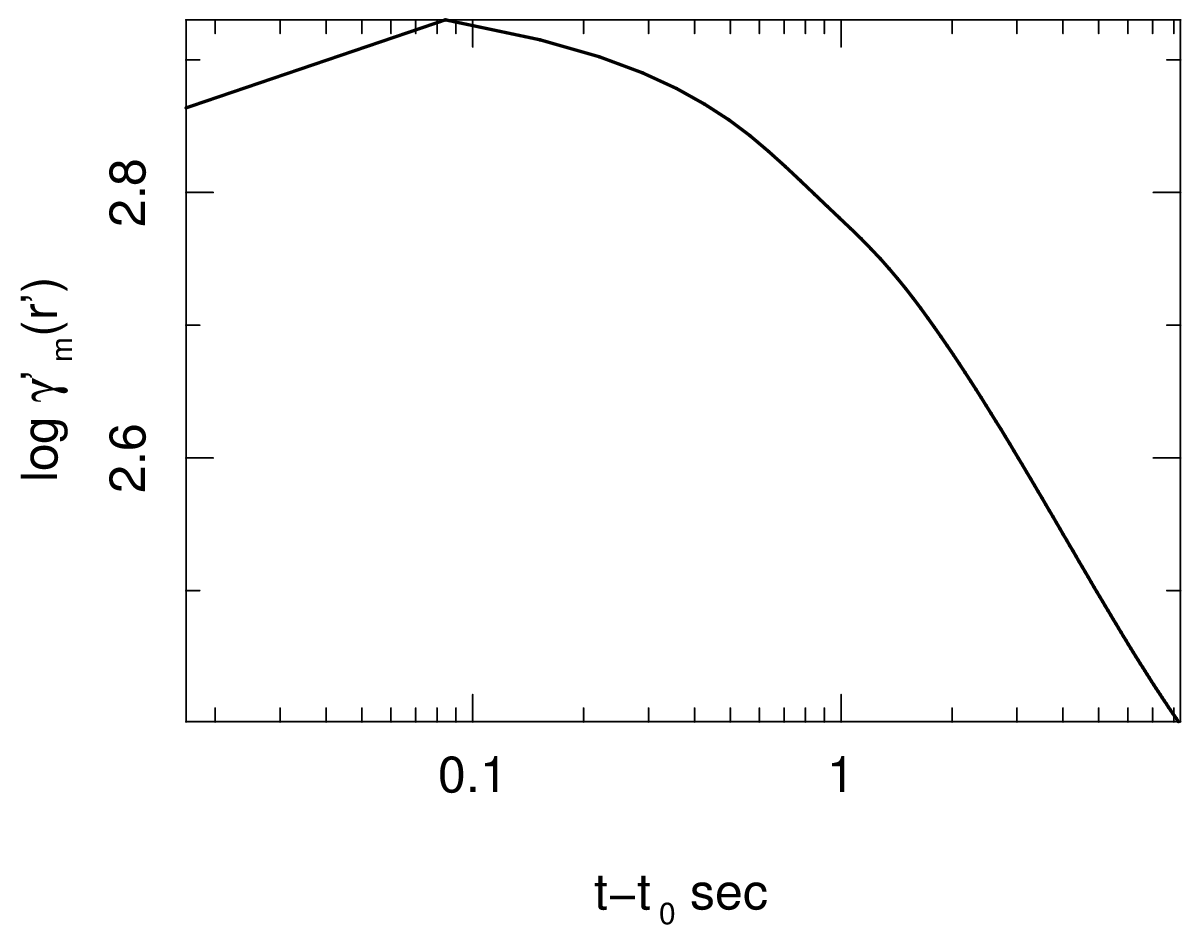} &
19)\includegraphics[height=4cm,angle=-90]{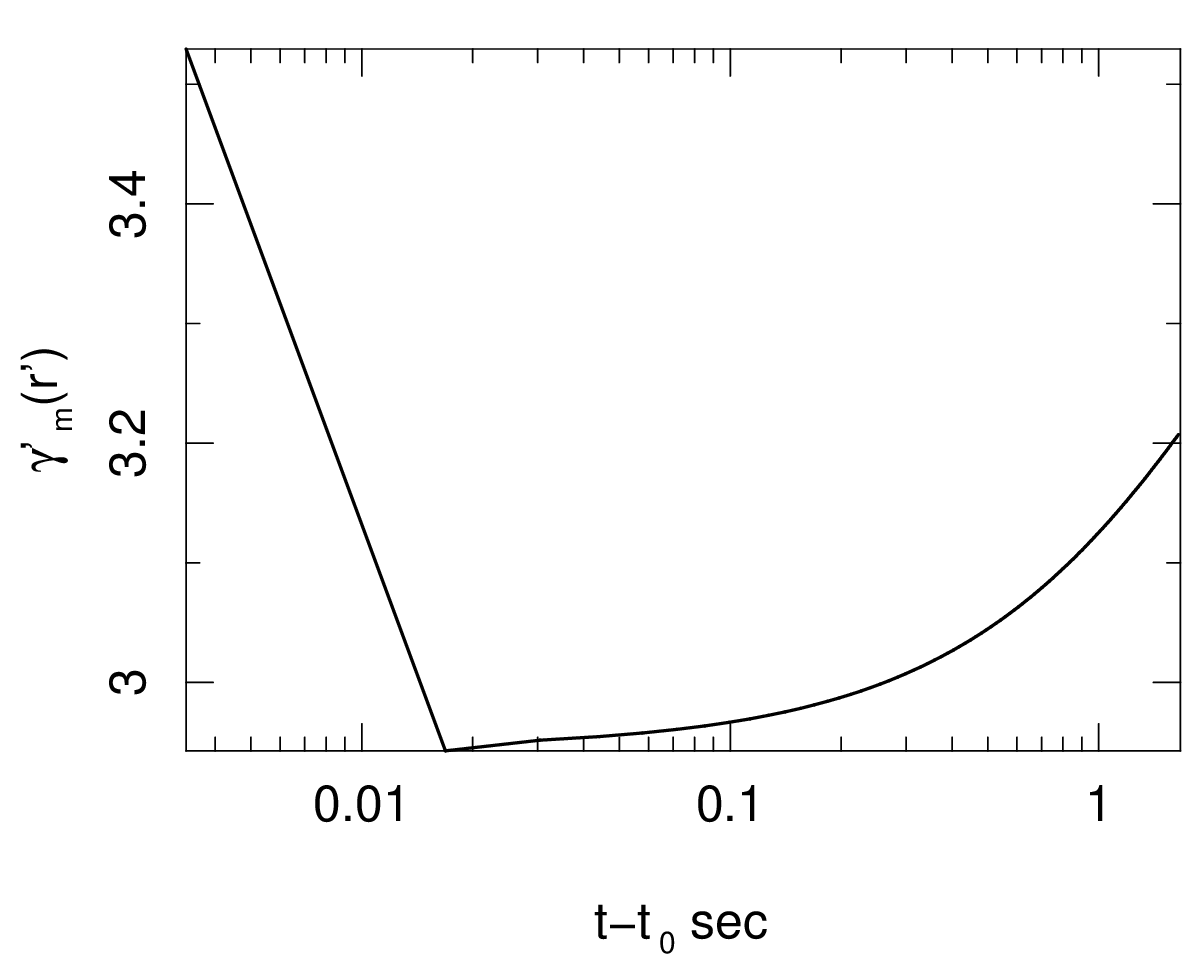} &
20)\includegraphics[height=4cm,angle=-90]{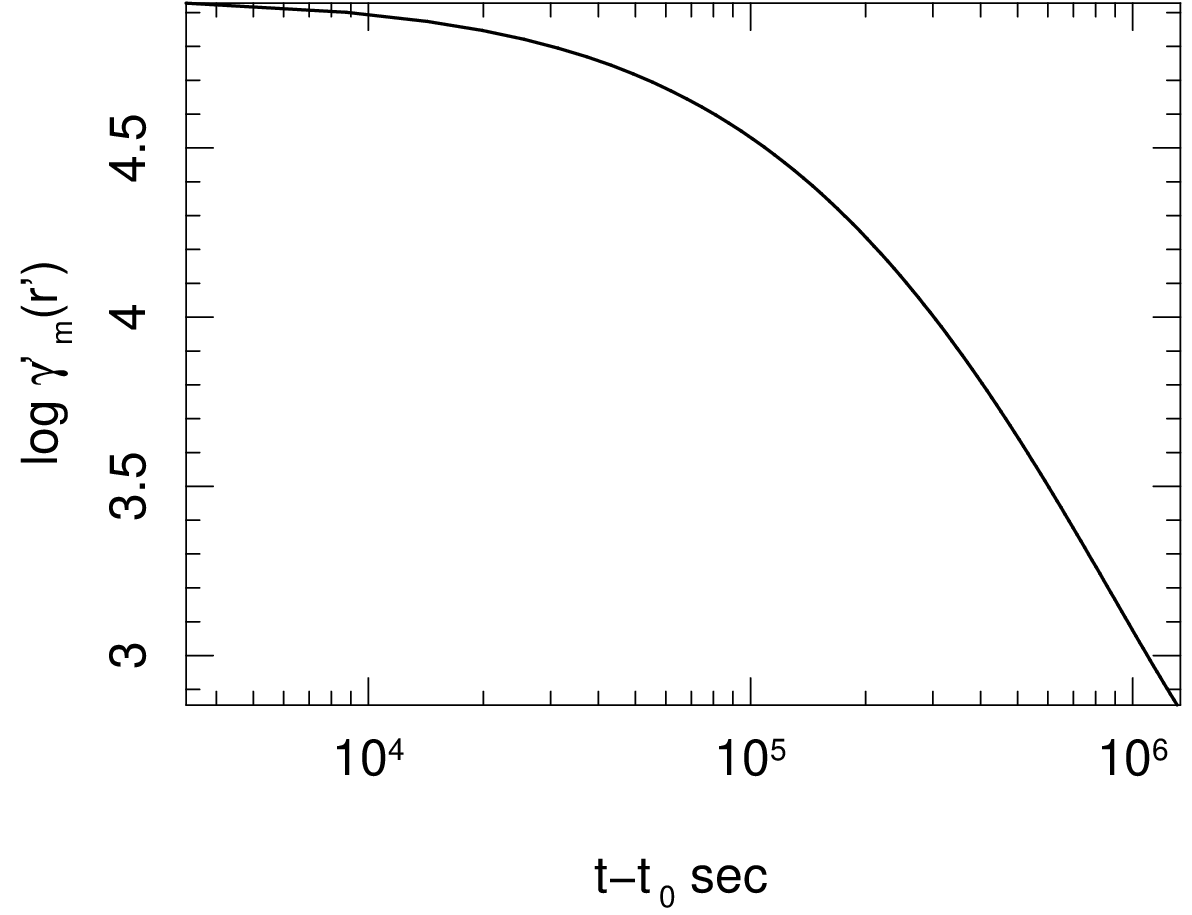} \\
21)\includegraphics[height=4cm,angle=-90]{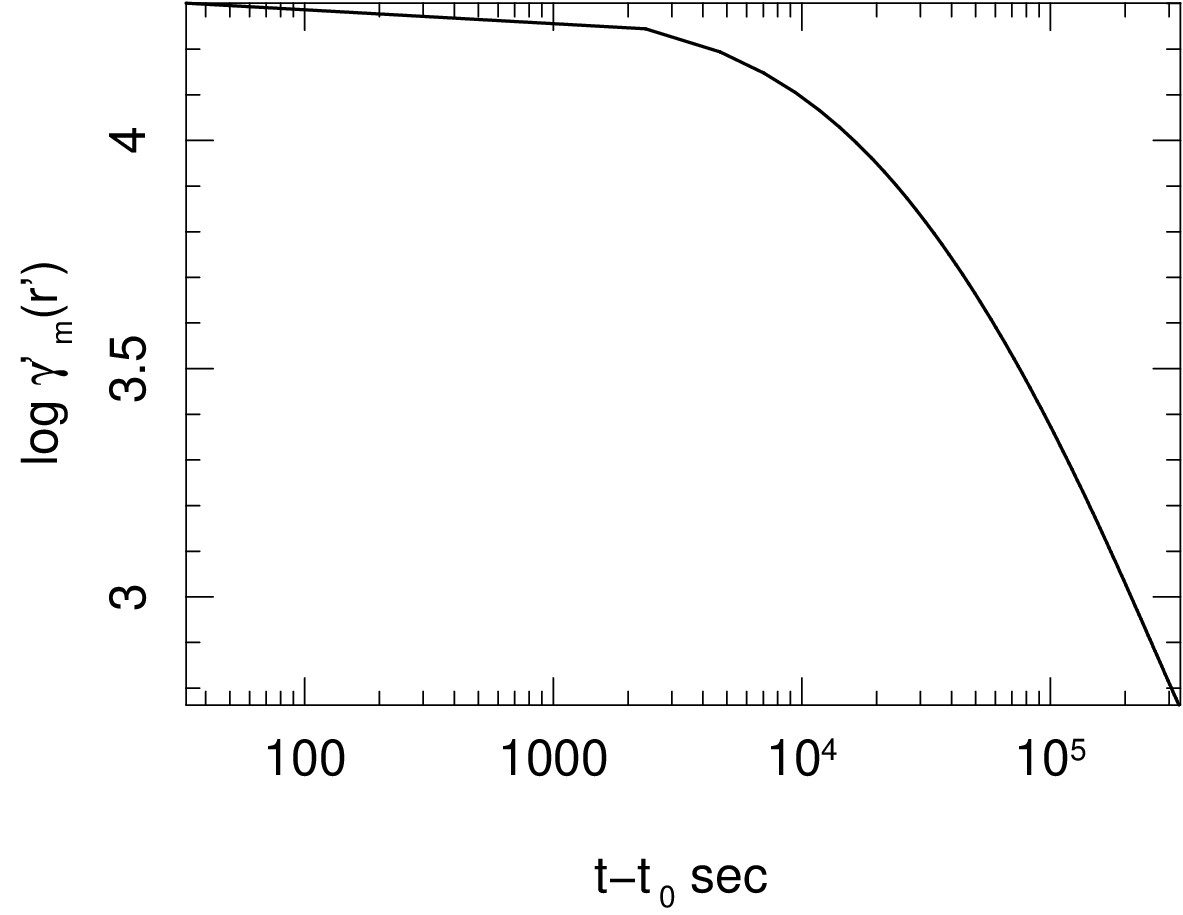} &
22)\includegraphics[height=4cm,angle=-90]{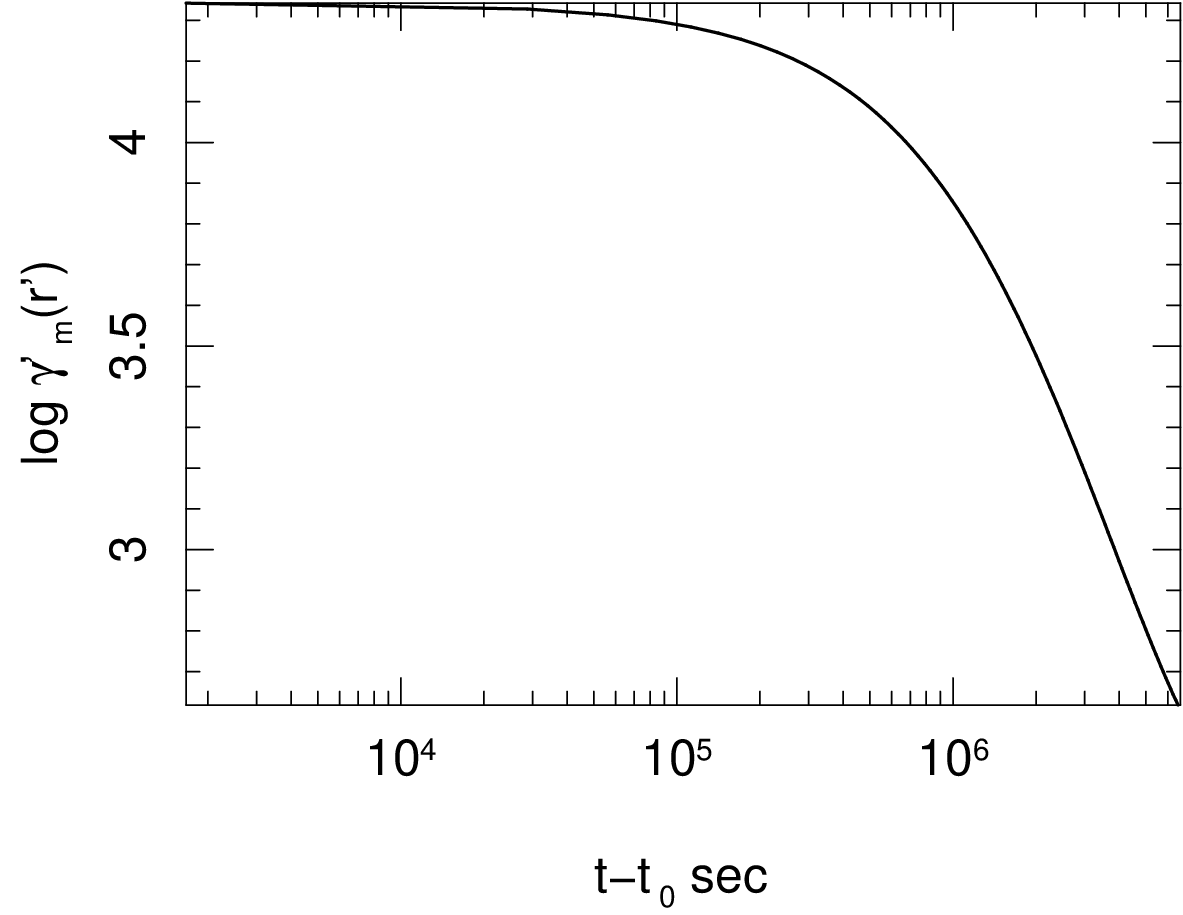} &
23)\includegraphics[height=4cm,angle=-90]{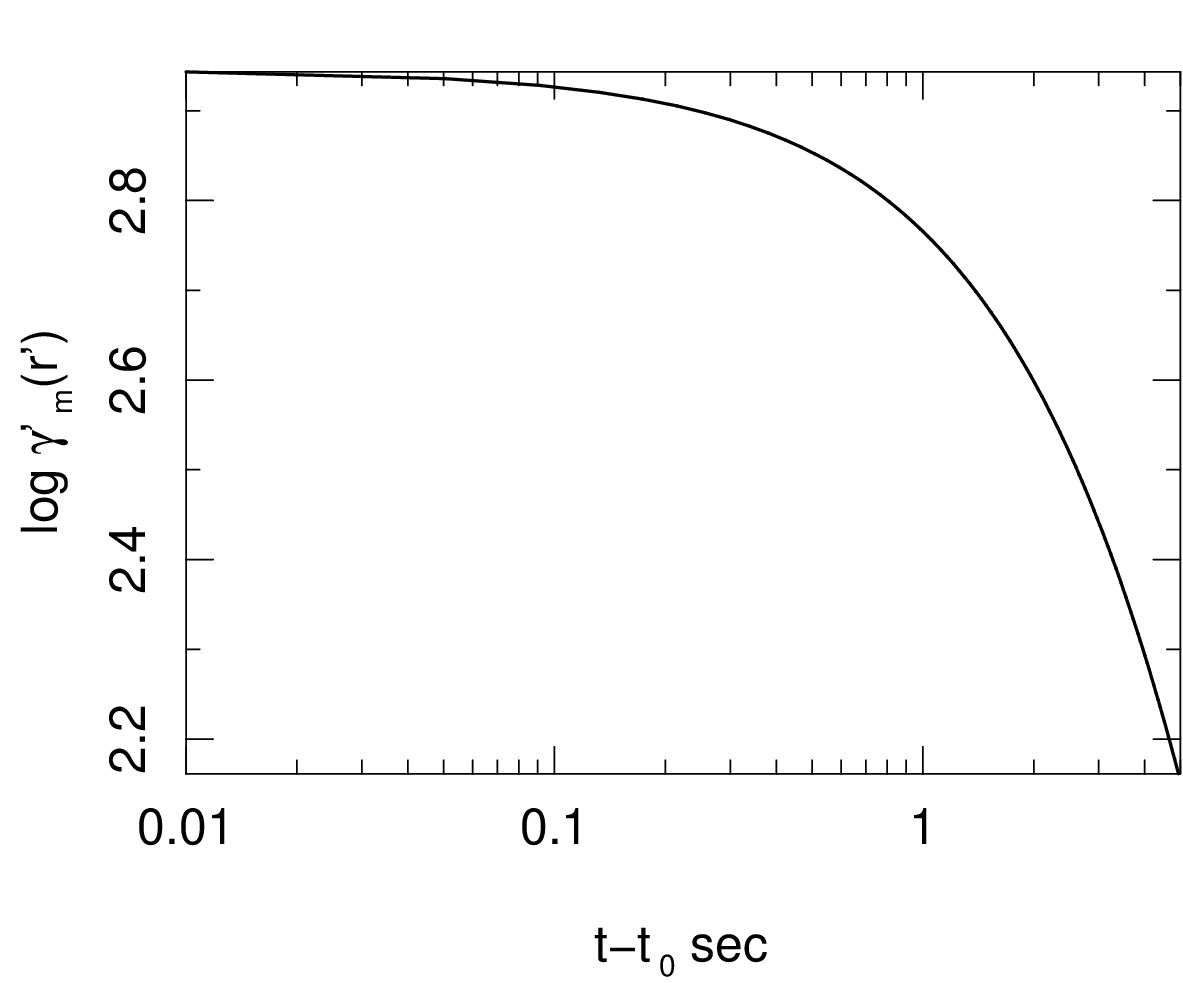} &
24)\includegraphics[height=4cm,angle=-90]{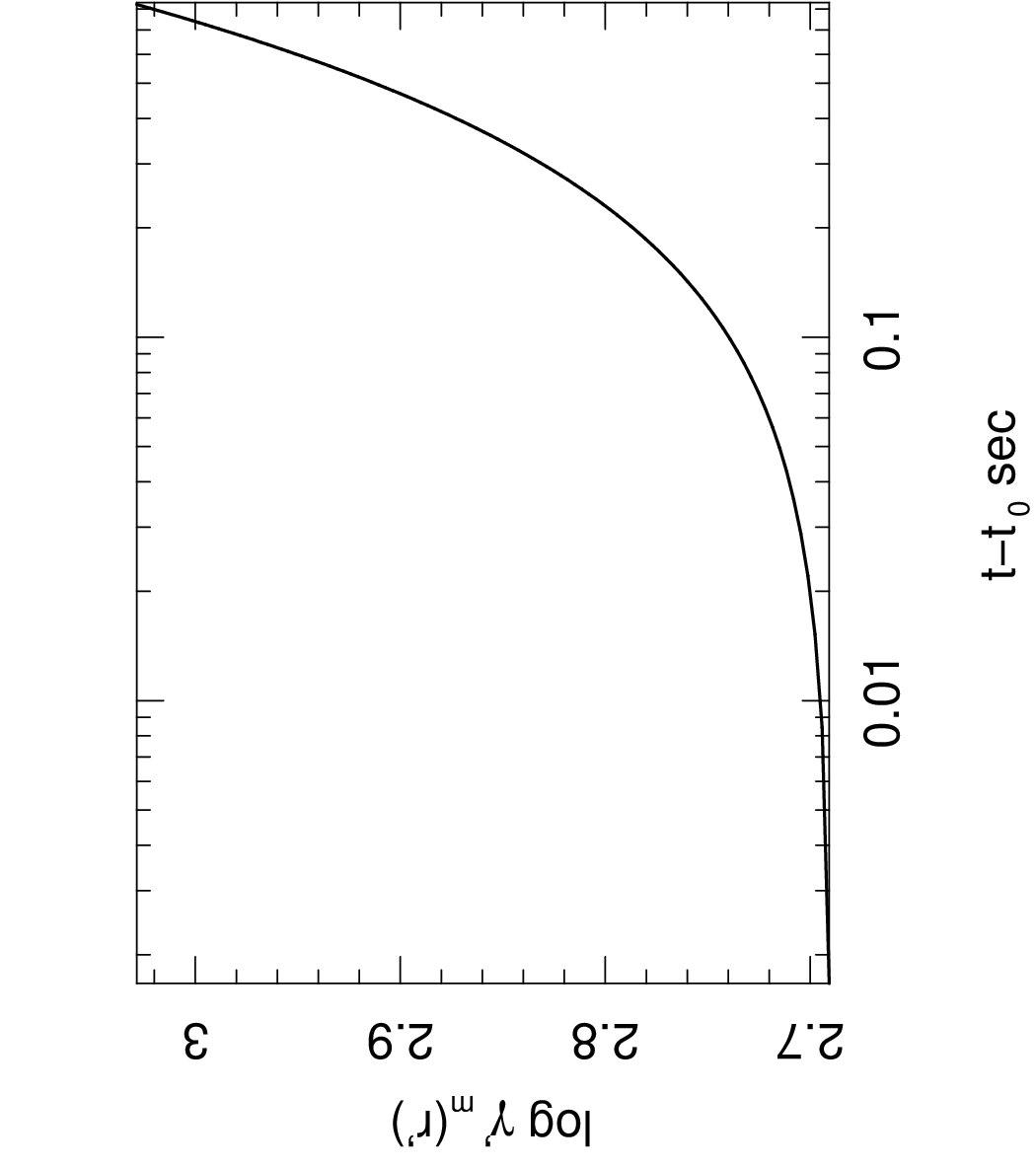} 
\end {tabular}
\end {center}
\caption{Evolution of $\gamma'_m$ with time/radius. Definitions of curves are 
the same as Fig. \ref{fig:betazero}. Note the hardening of emission with time for some 
of the models.
\label{fig:gammam}}
\end{figure*}

\begin{figure*}
\begin{center}
\begin {tabular}{cccc}
1)\includegraphics[height=4cm,angle=-90]{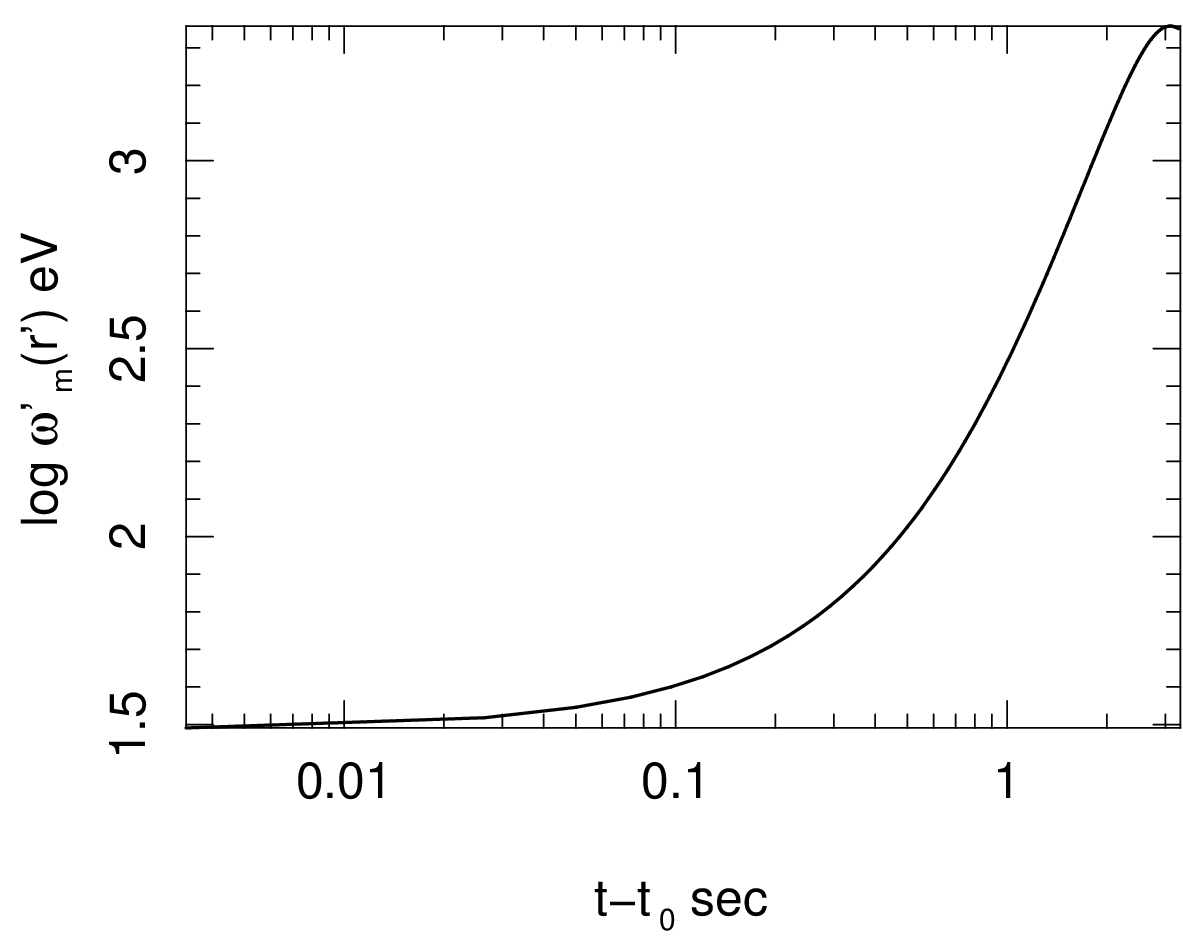} &
2)\includegraphics[height=4cm,angle=-90]{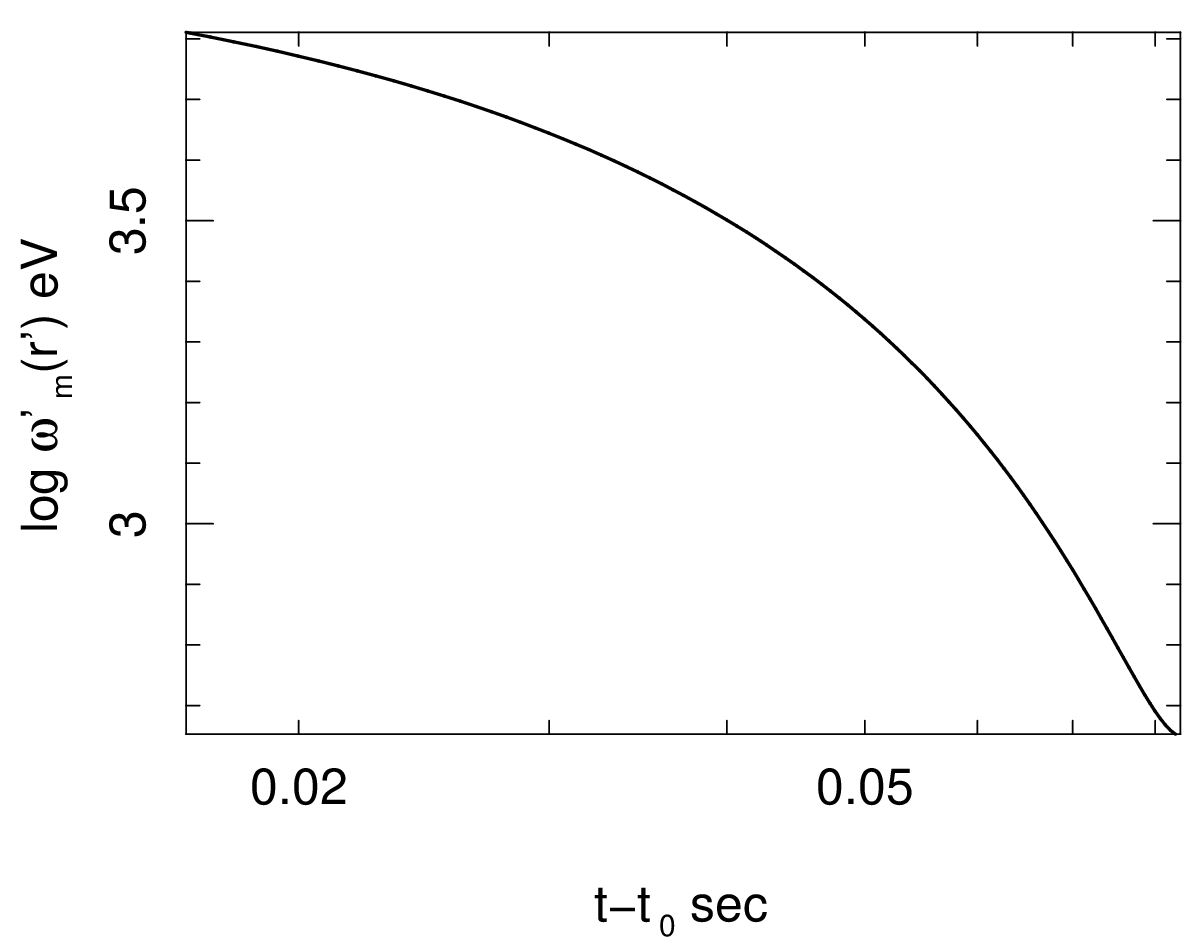} & 
3)\includegraphics[height=4cm,angle=-90]{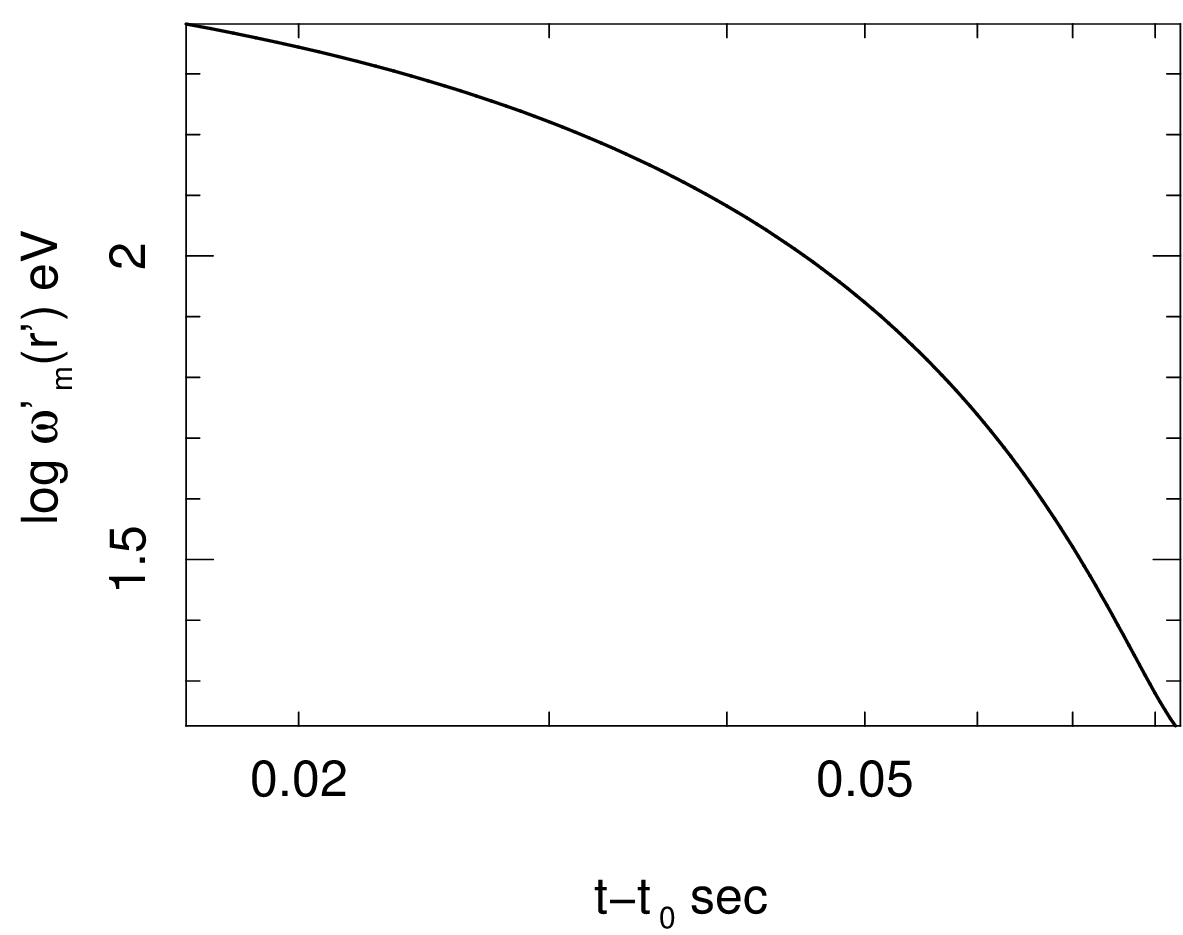} &
4)\includegraphics[height=4cm,angle=-90]{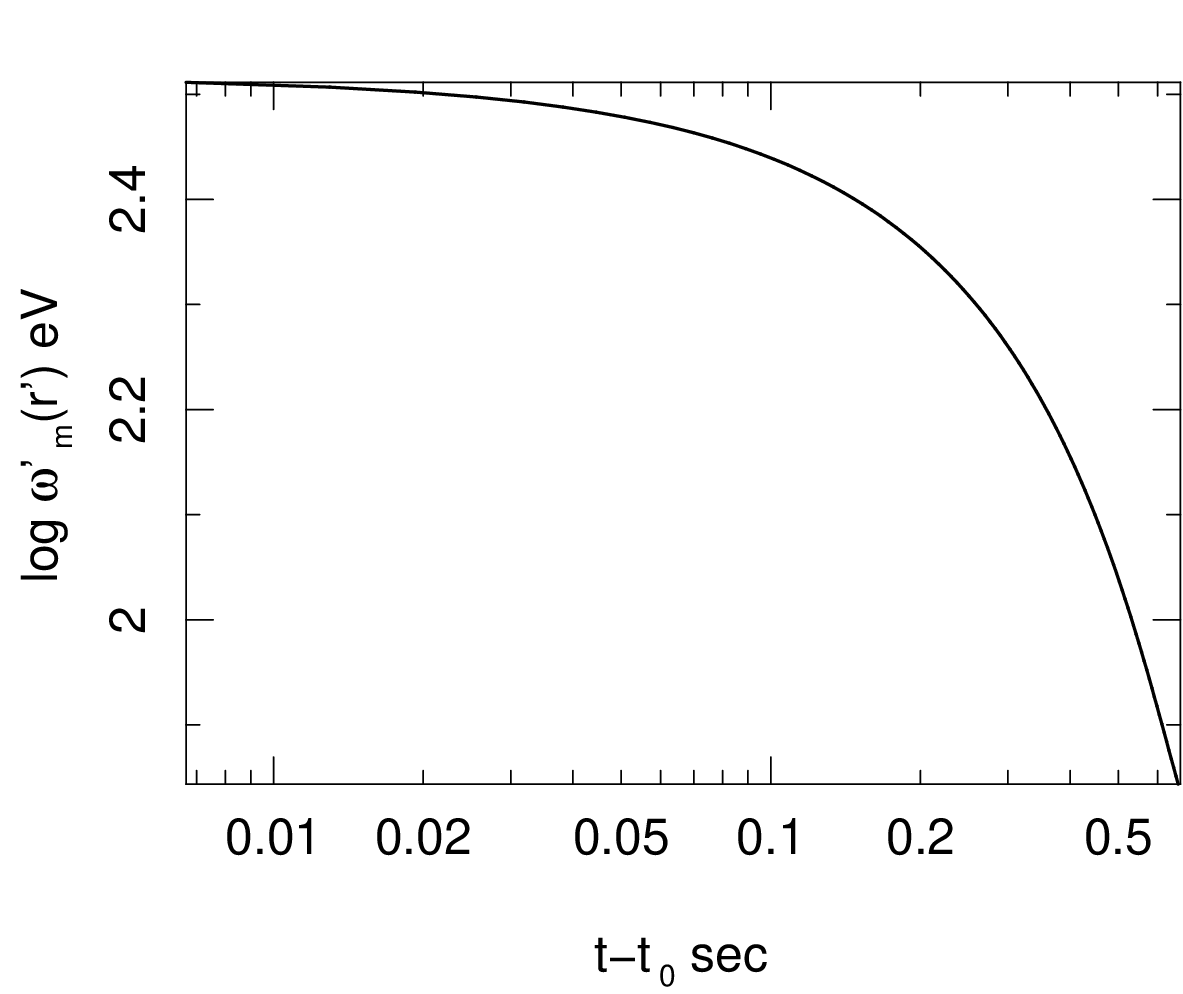} \\ 
5)\includegraphics[height=4cm,angle=-90]{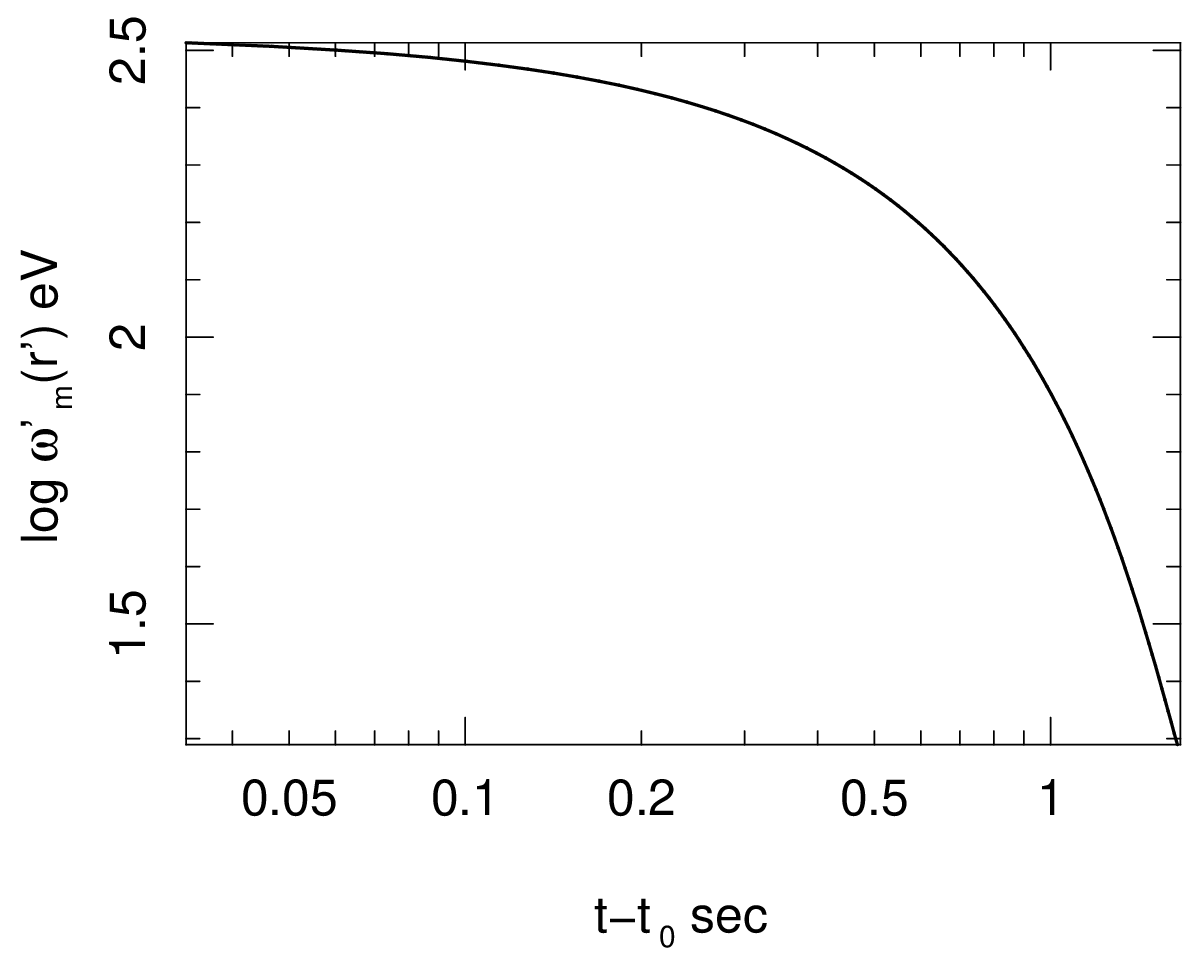} &
6)\includegraphics[height=4cm,angle=-90]{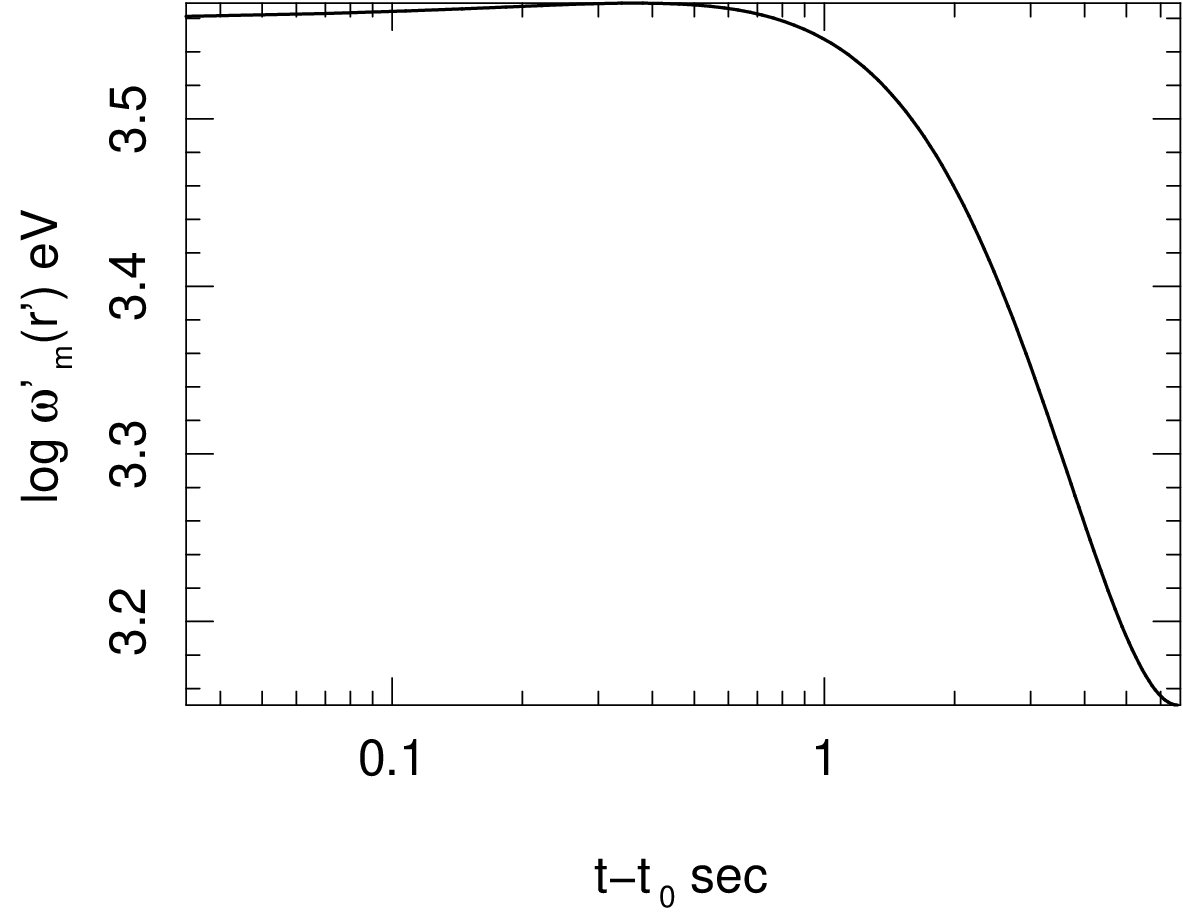} & 
7)\includegraphics[height=4cm,angle=-90]{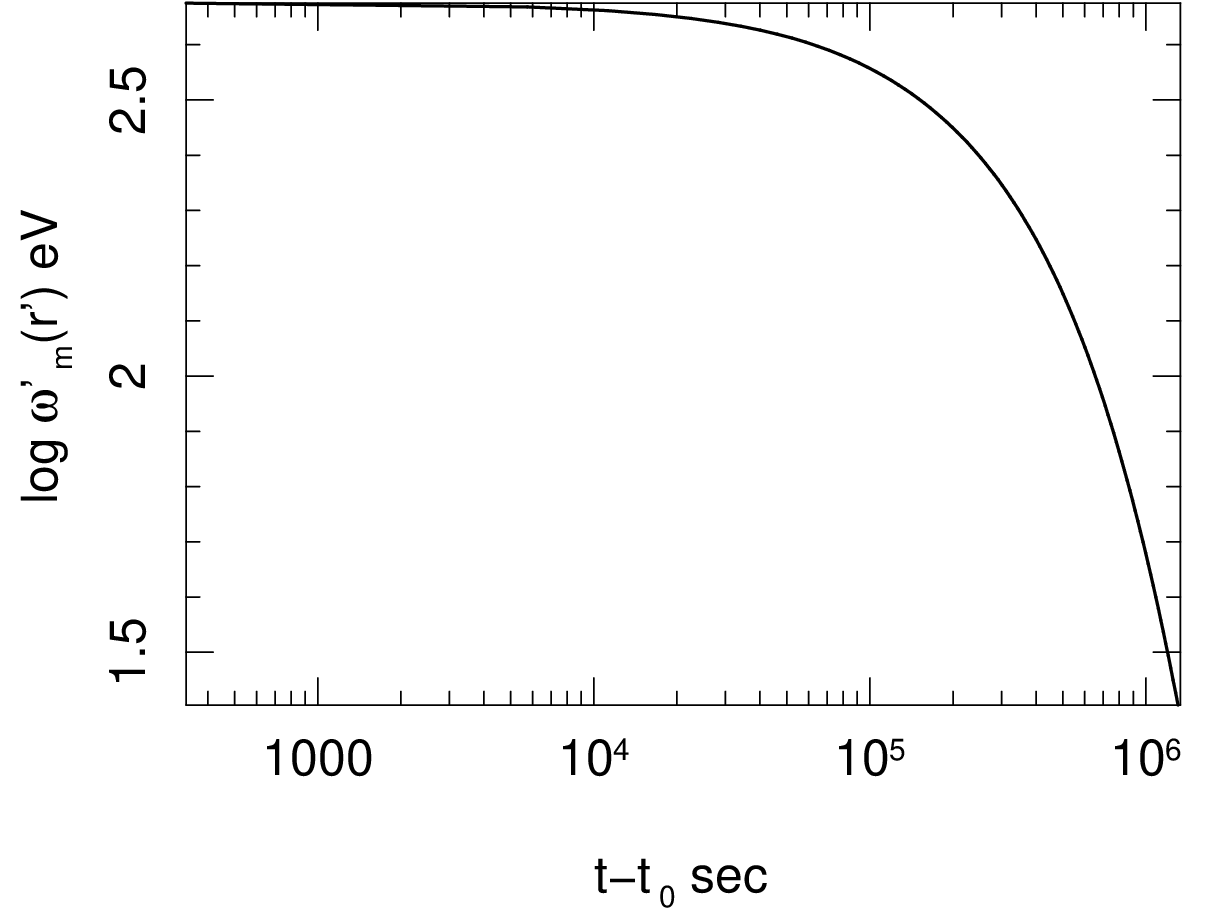} &
8)\includegraphics[height=4cm,angle=-90]{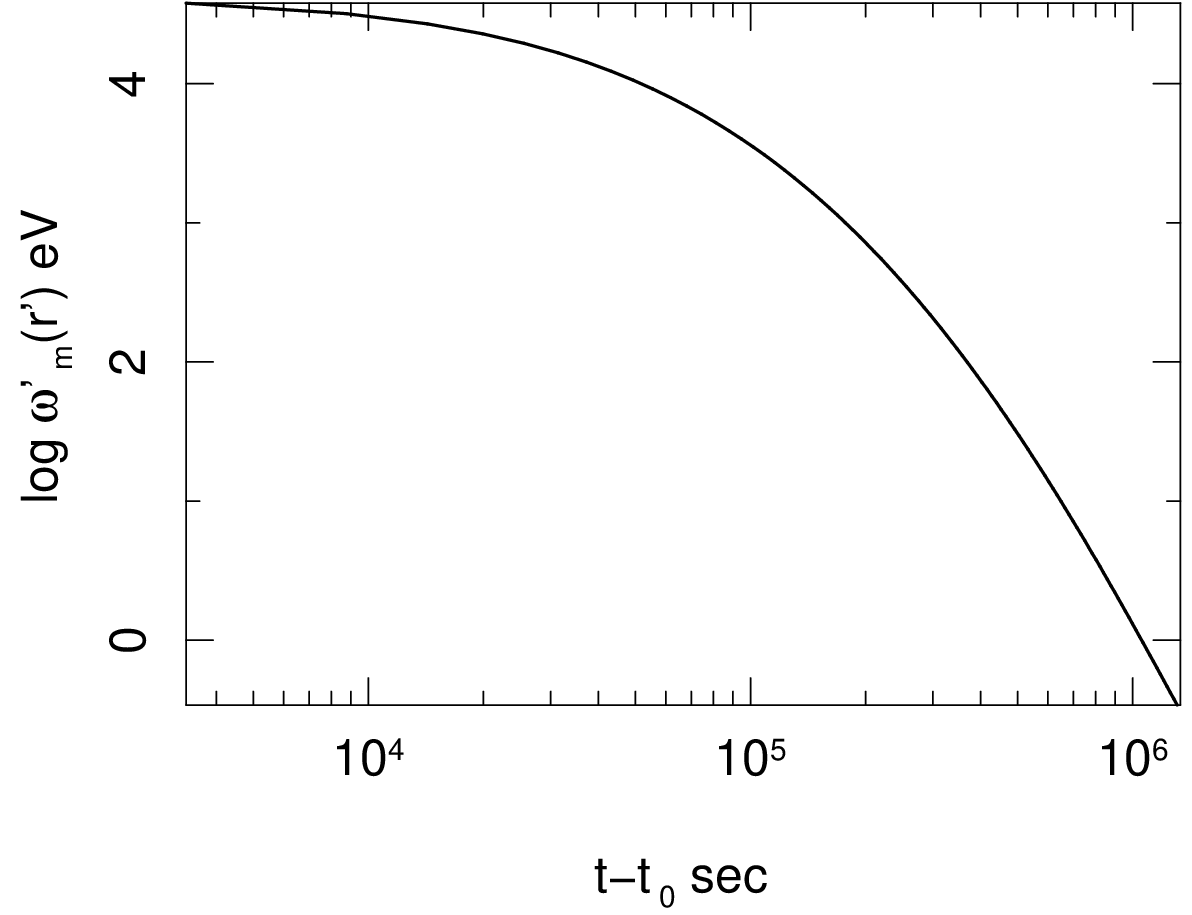} \\
9)\includegraphics[height=4cm,angle=-90]{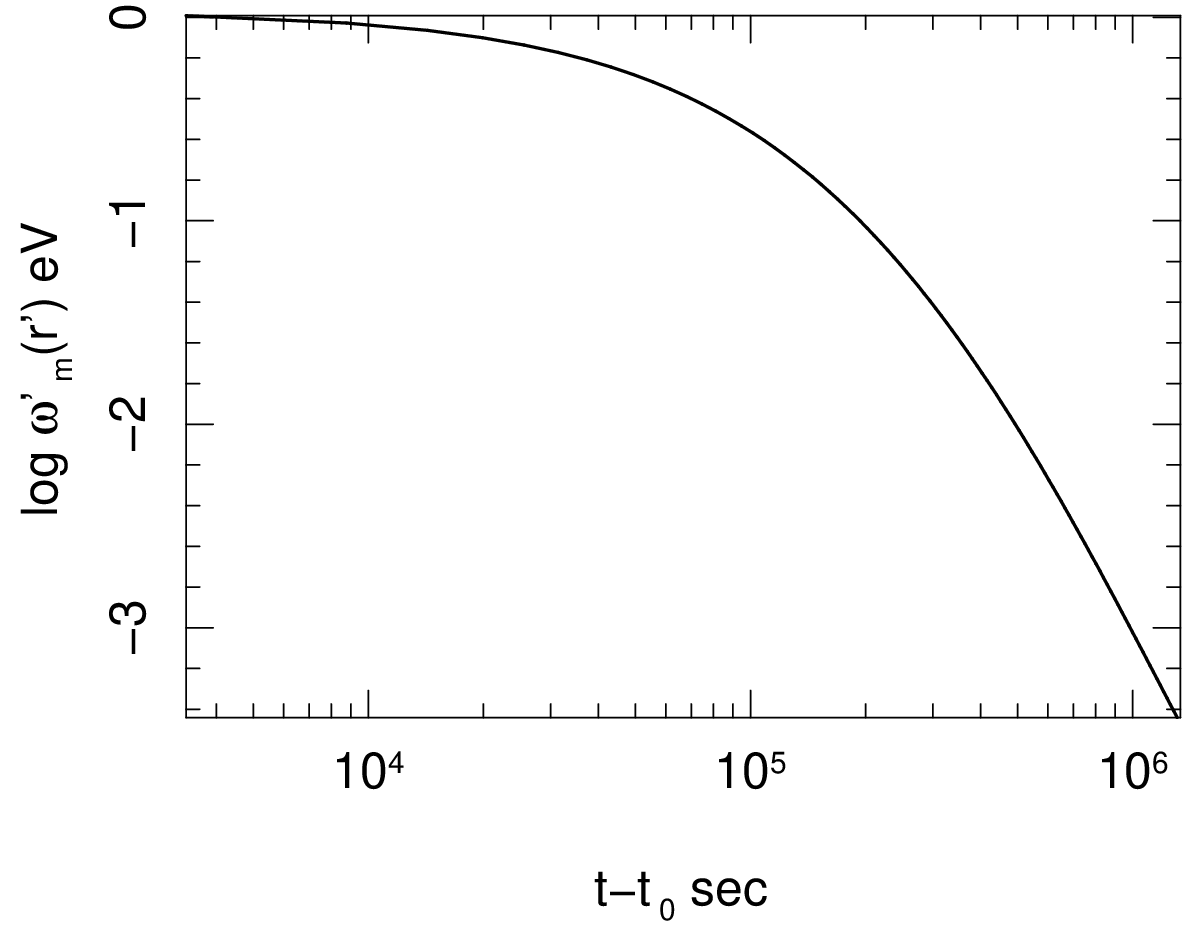} &
10)\includegraphics[height=4cm,angle=-90]{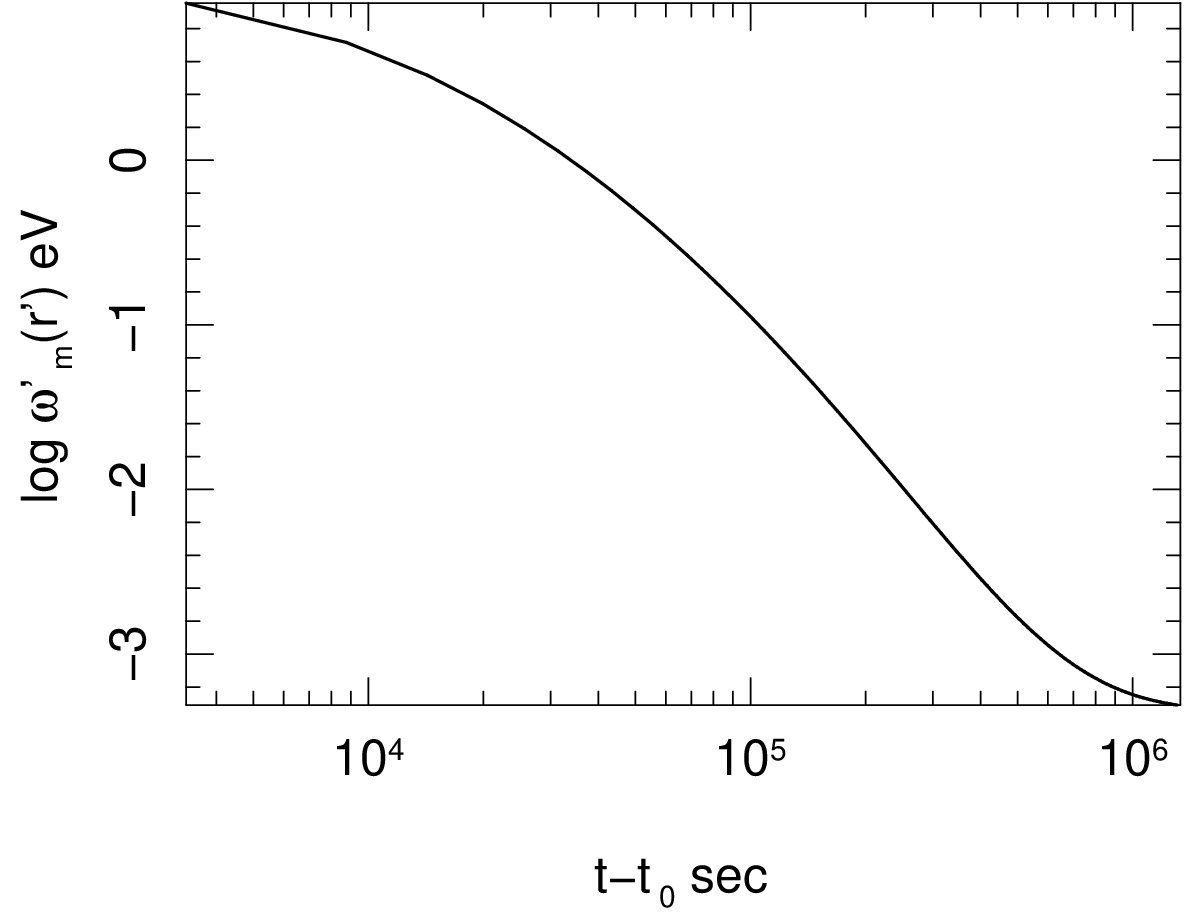} &
11)\includegraphics[height=4cm,angle=-90]{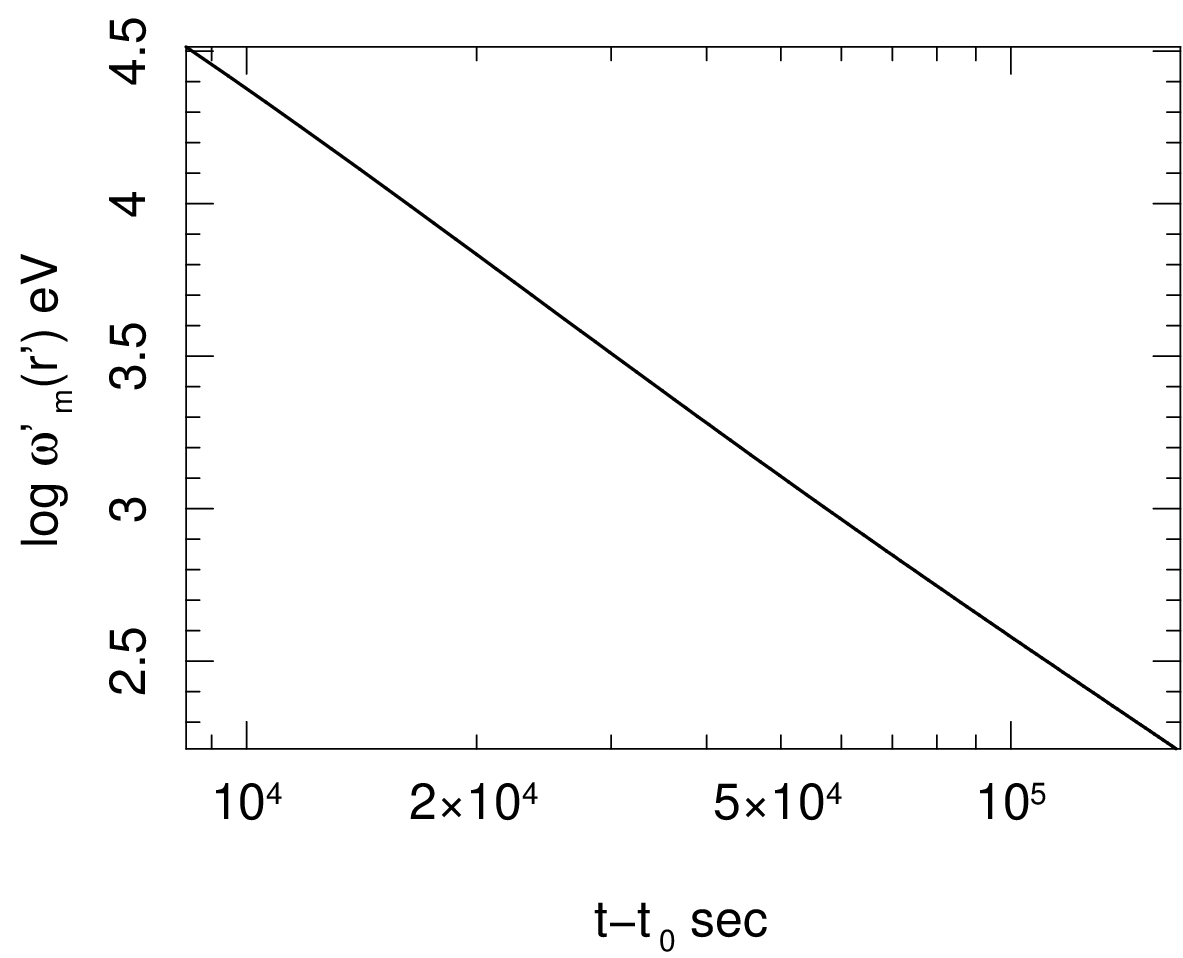} &
12)\includegraphics[height=4cm,angle=-90]{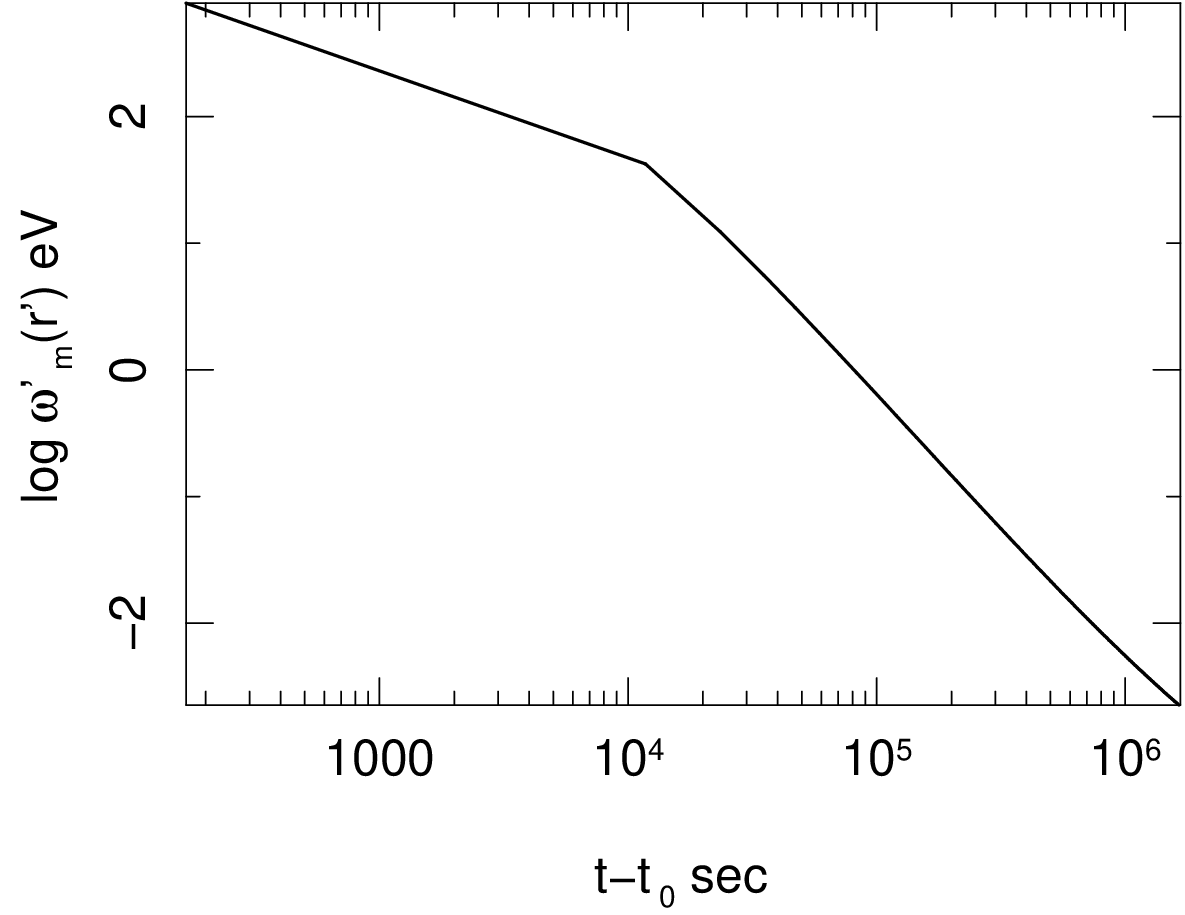} \\
13)\includegraphics[height=4cm,angle=-90]{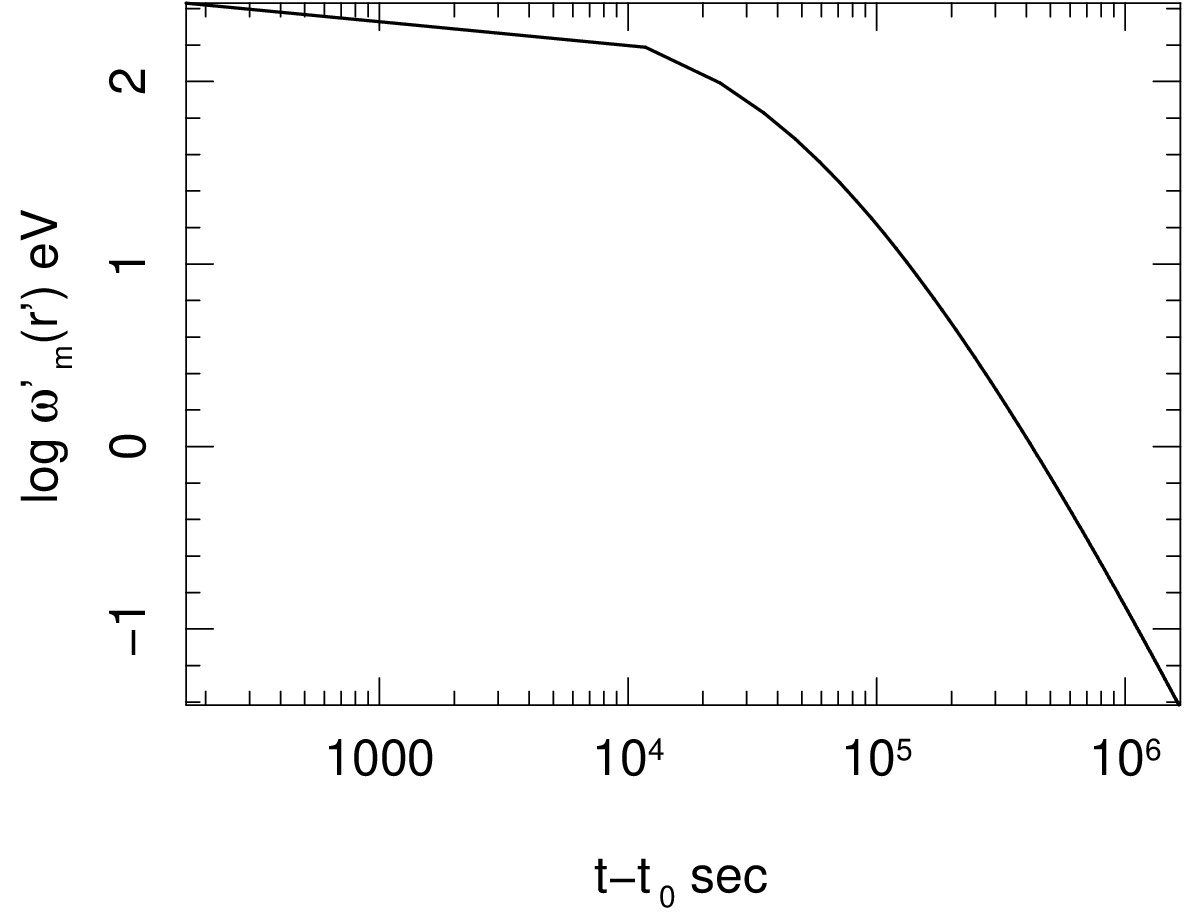} &
14)\includegraphics[height=4cm,angle=-90]{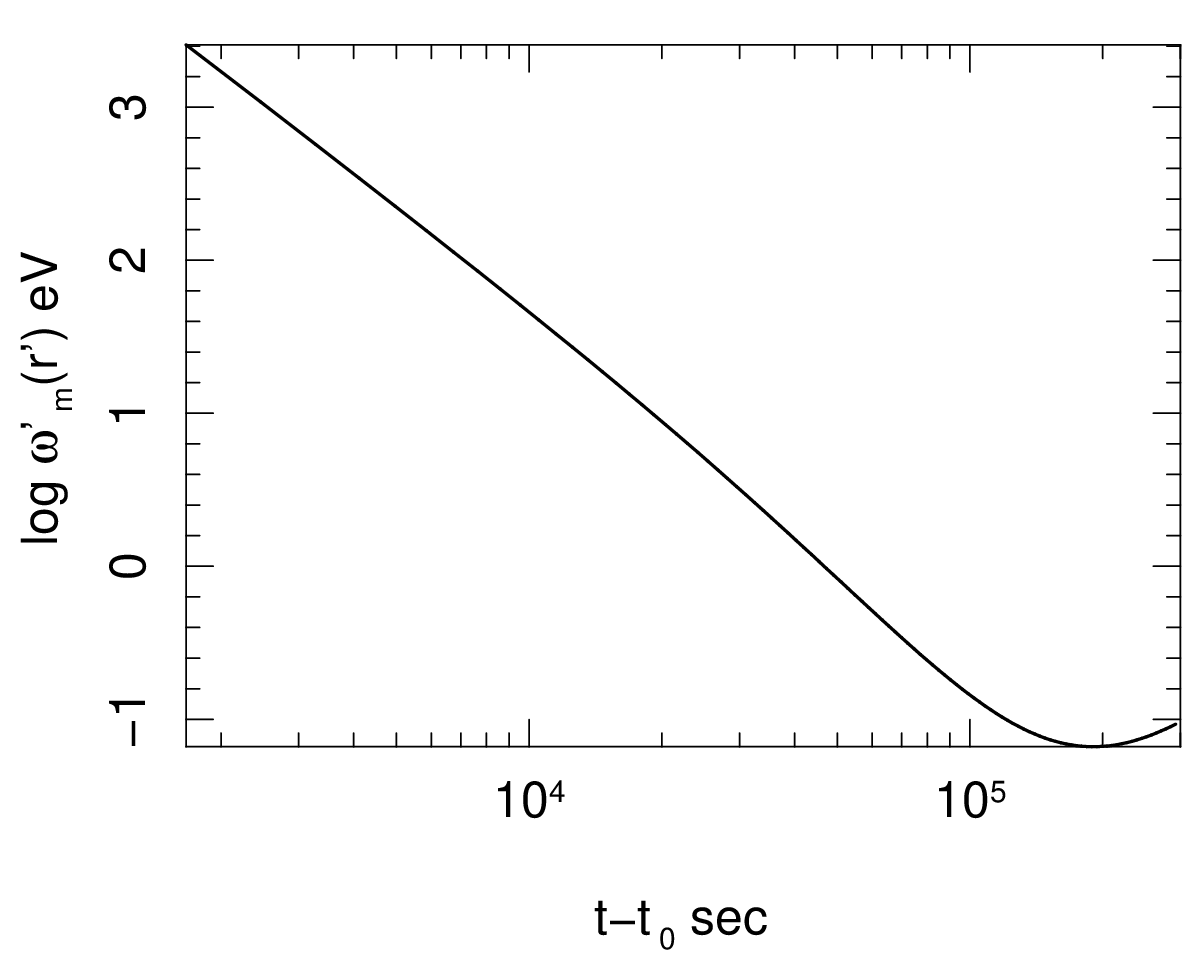} &
15)\includegraphics[height=4cm,angle=-90]{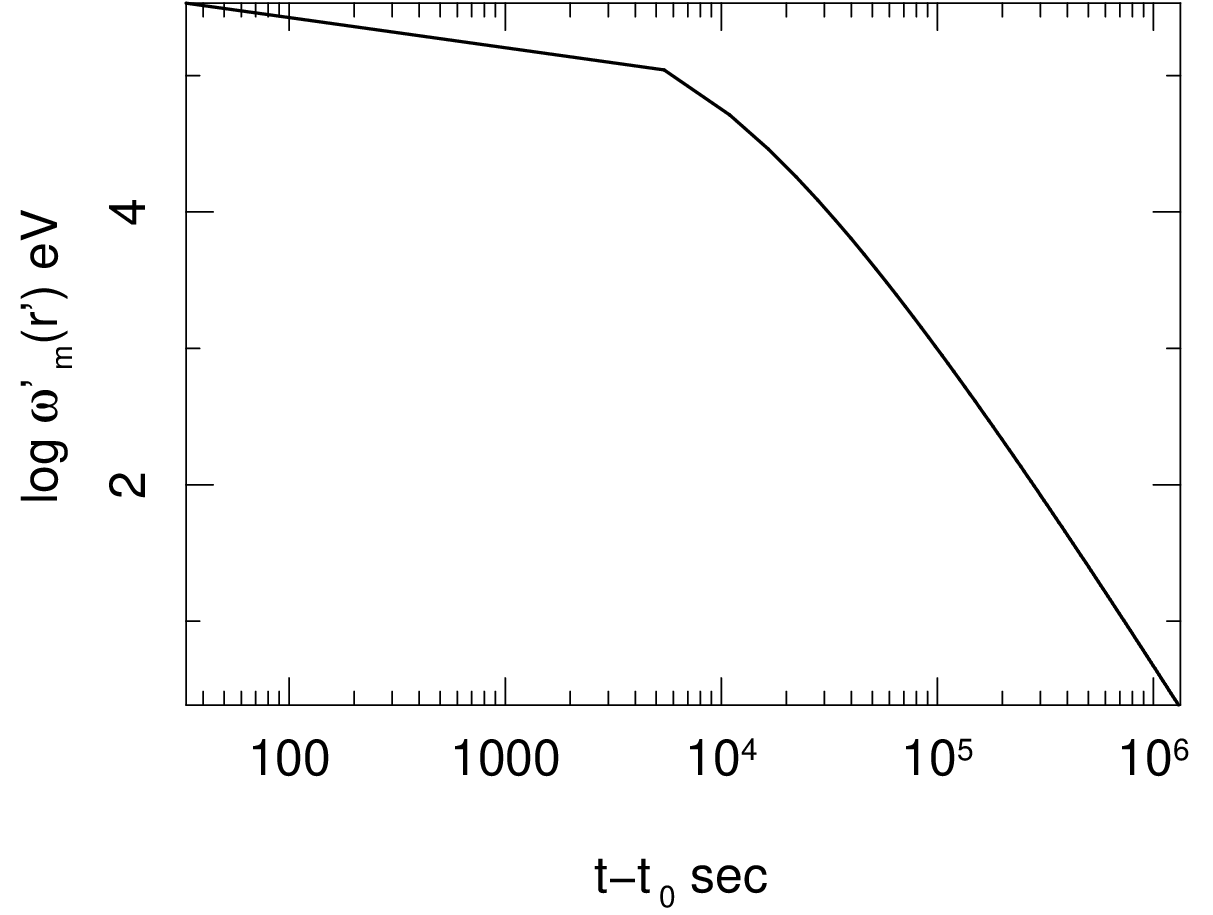} &
16)\includegraphics[height=4cm,angle=-90]{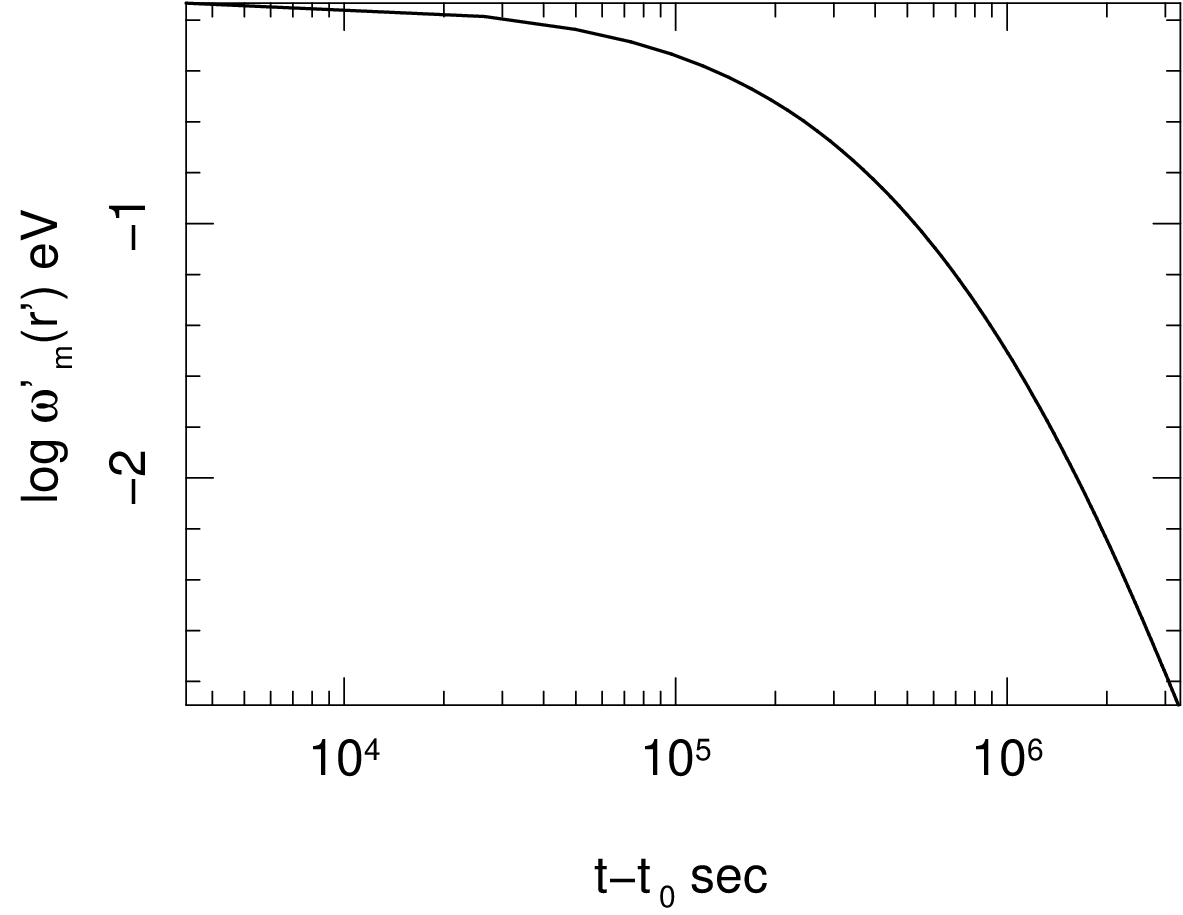} \\
17)\includegraphics[height=4cm,angle=-90]{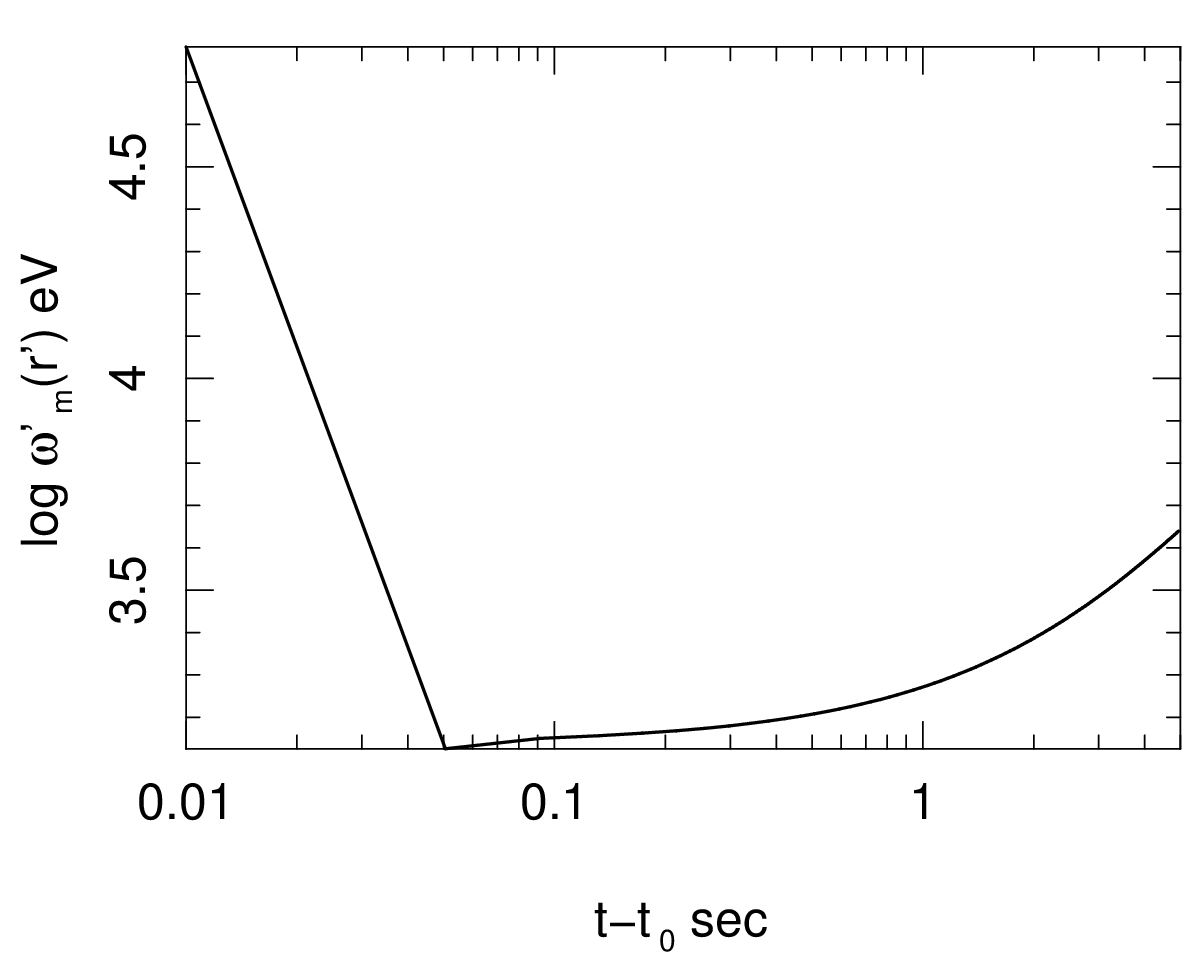} &
18)\includegraphics[height=4cm,angle=-90]{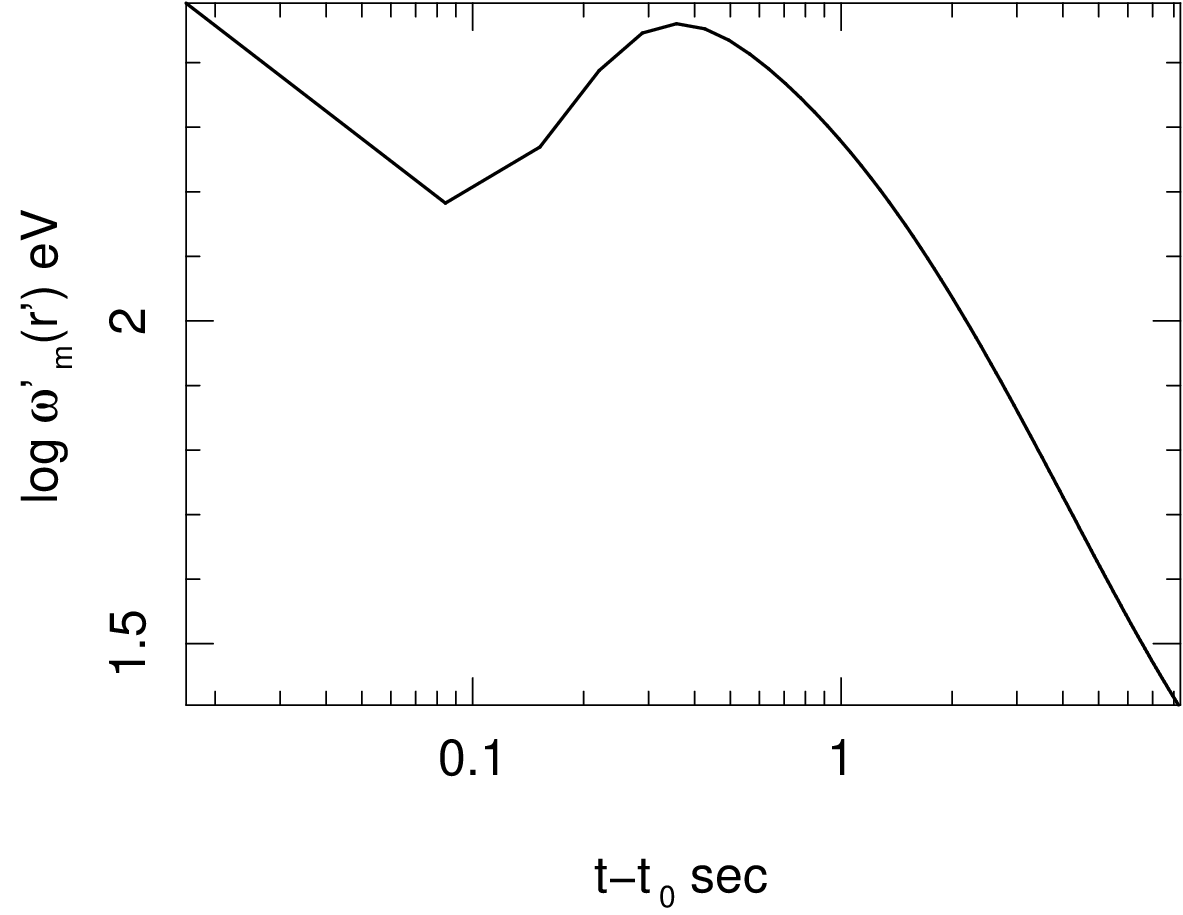} &
19)\includegraphics[height=4cm,angle=-90]{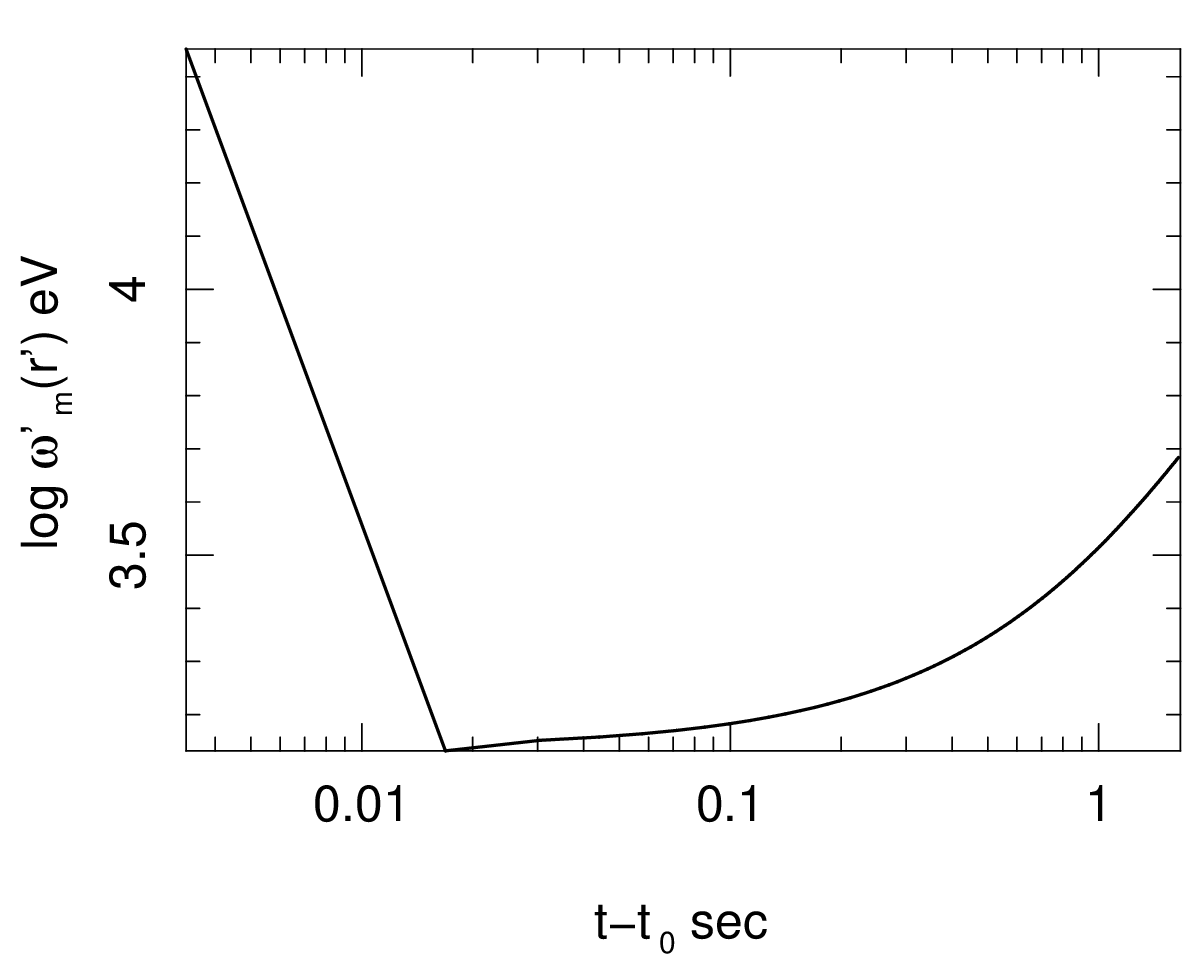} &
20)\includegraphics[height=4cm,angle=-90]{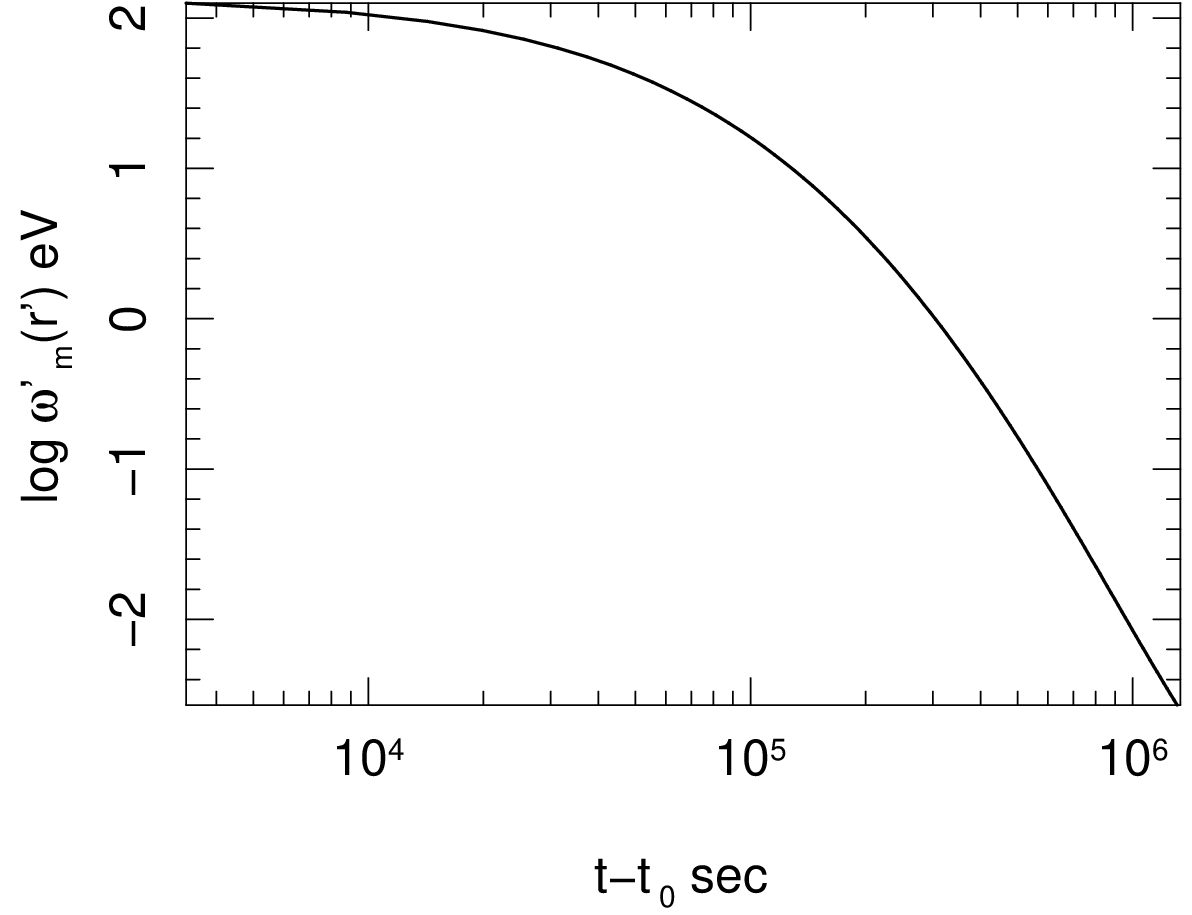} \\
21)\includegraphics[height=4cm,angle=-90]{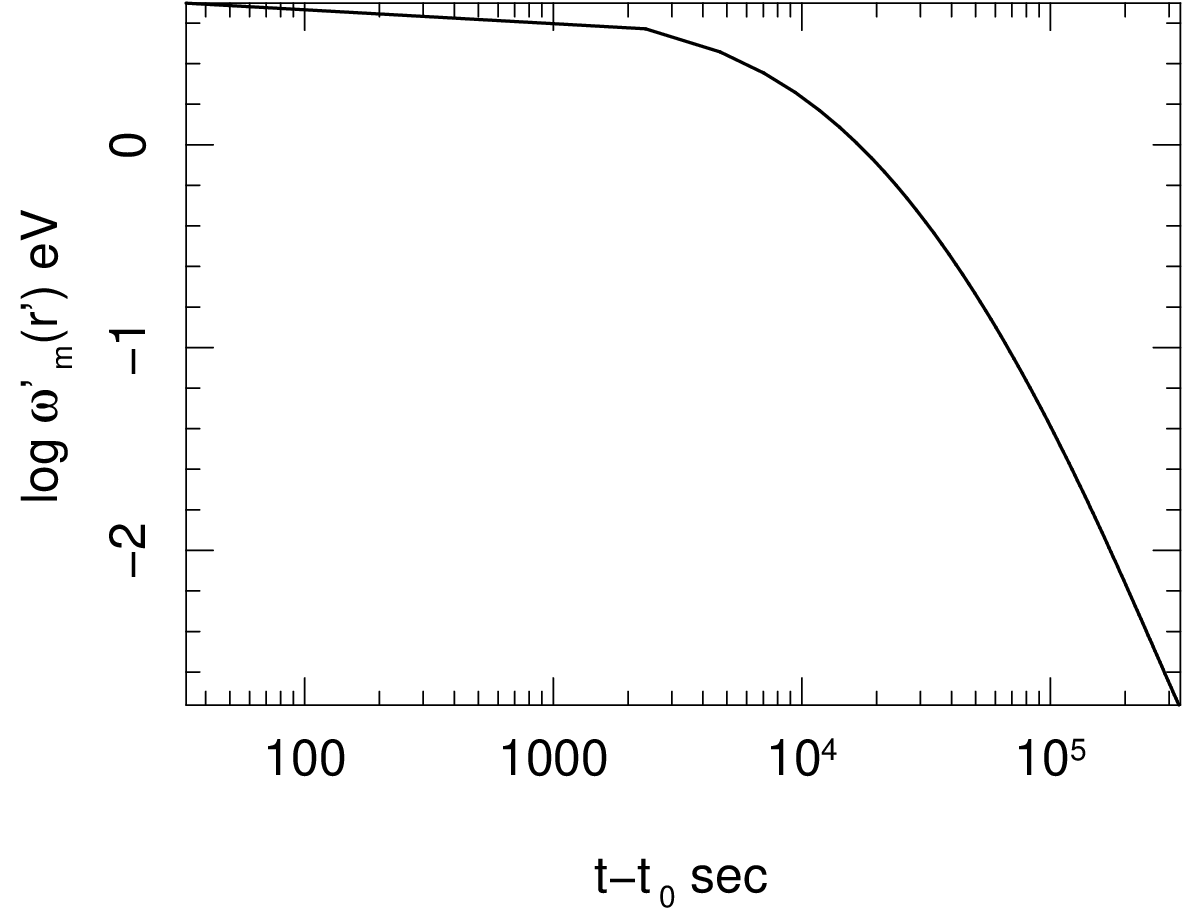} &
22)\includegraphics[height=4cm,angle=-90]{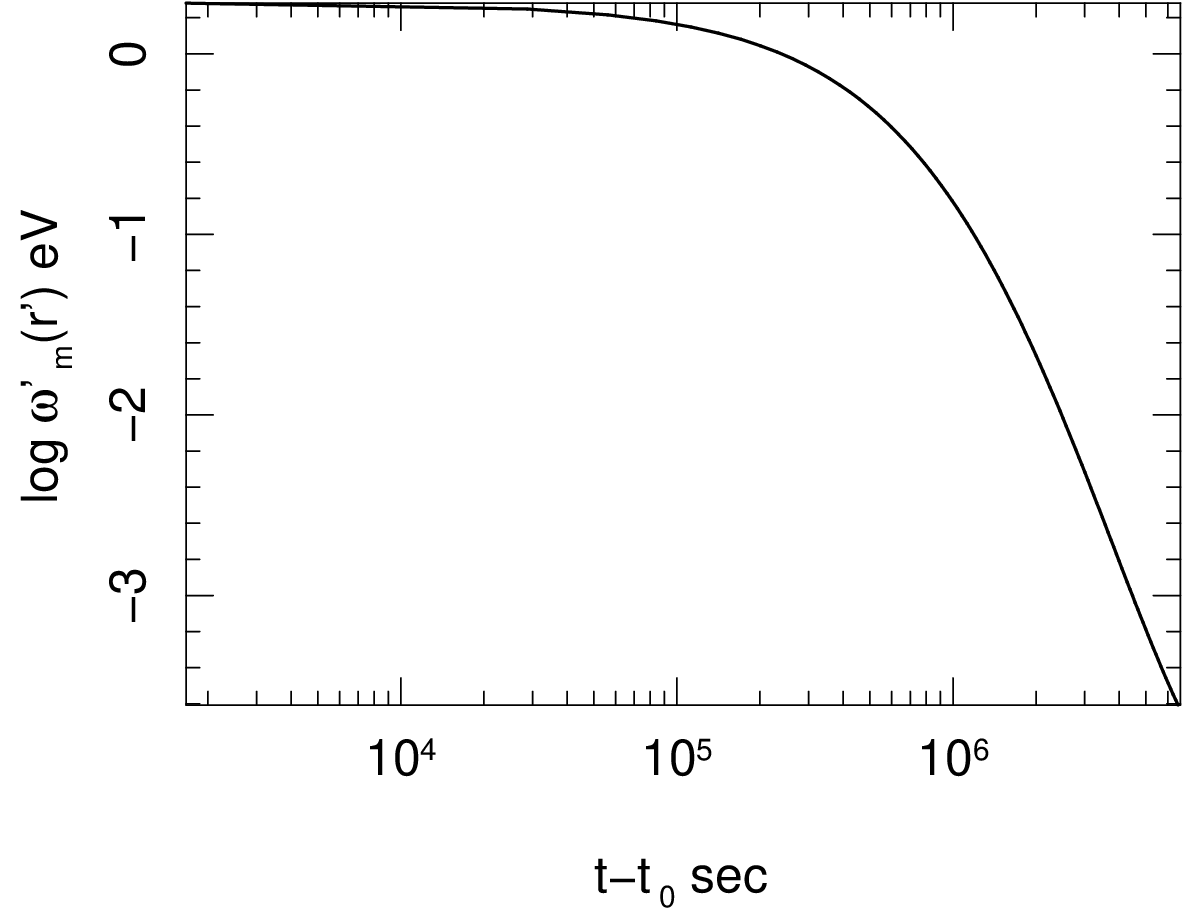} &
23)\includegraphics[height=4cm,angle=-90]{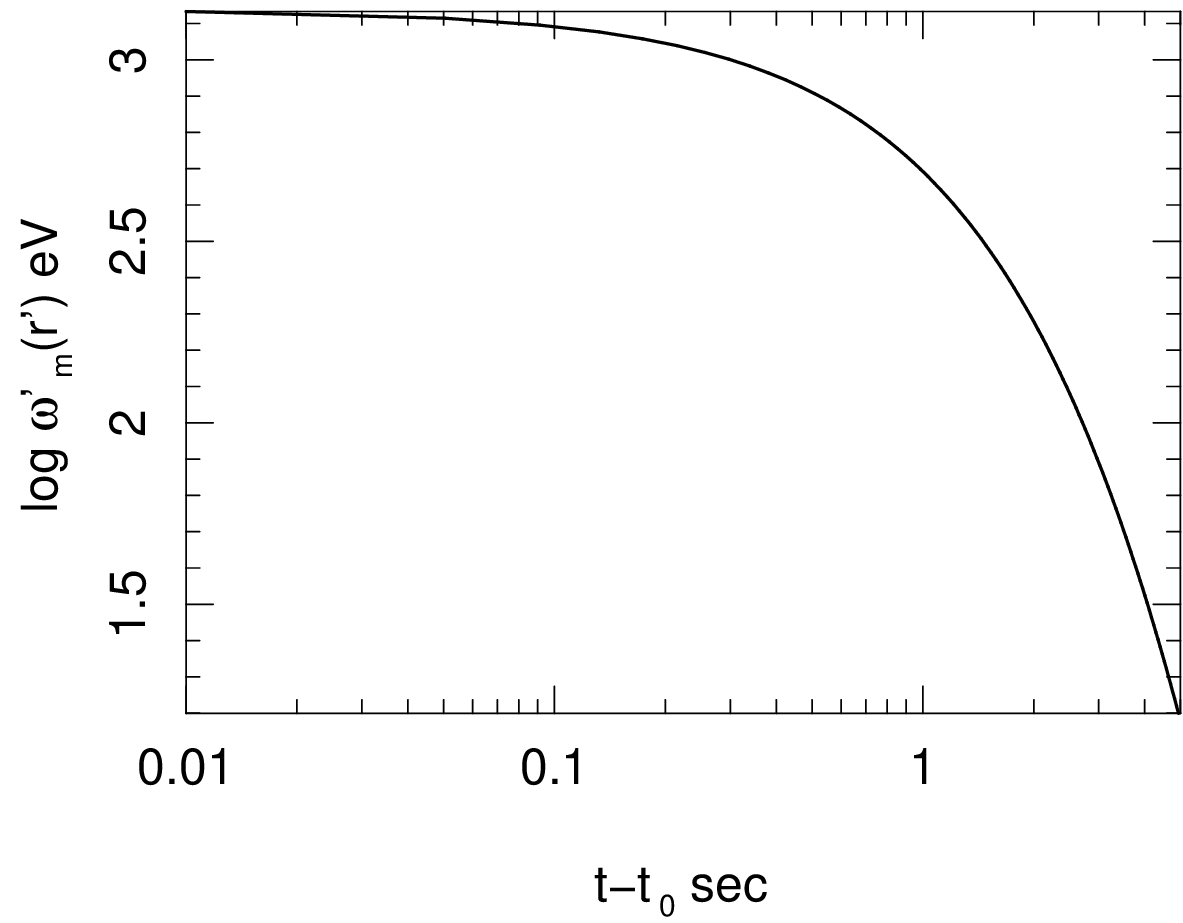} &
24)\includegraphics[height=4cm,angle=-90]{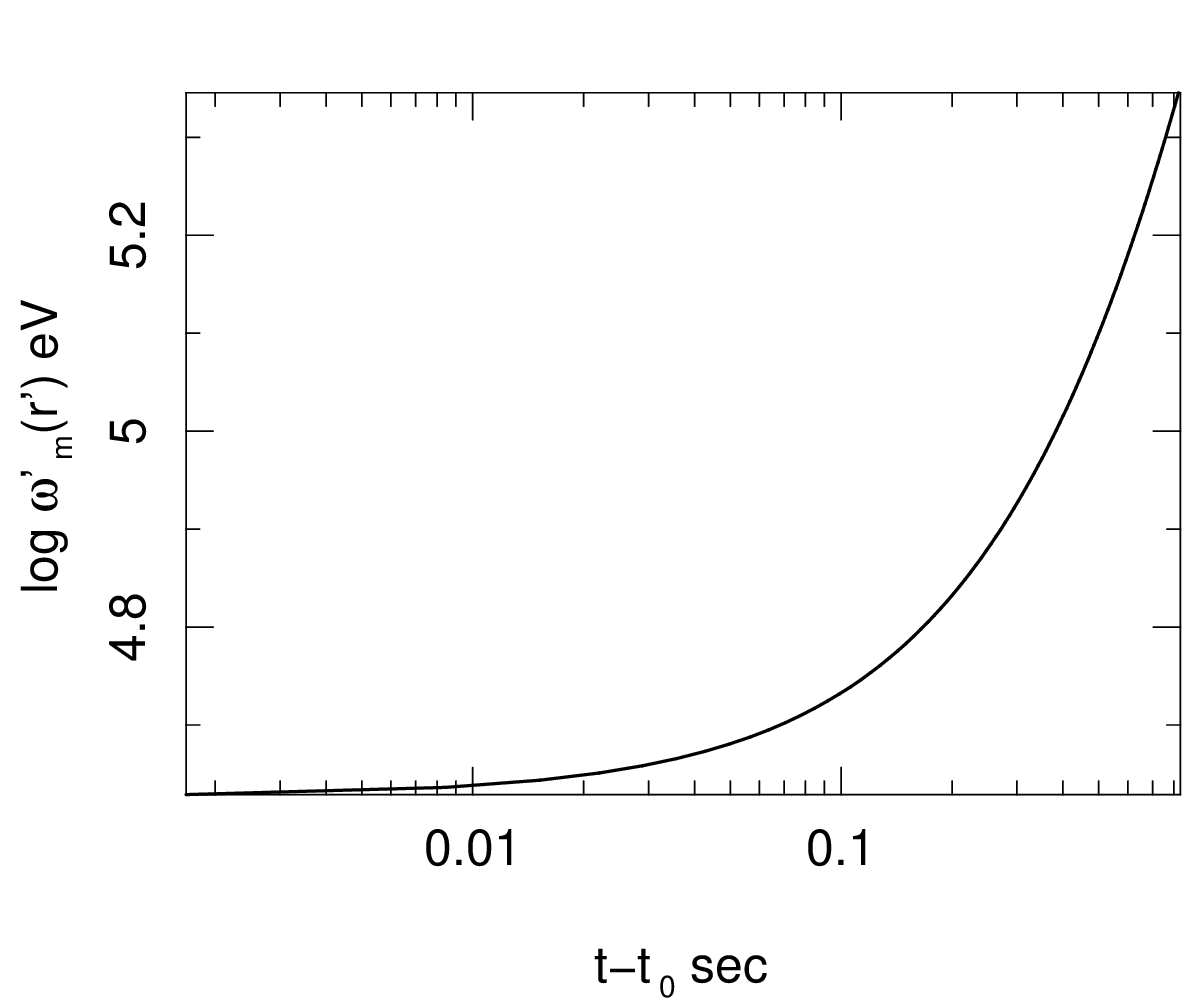} 
\end {tabular}
\end {center}
\caption{Evolution of $\omega'_m$ with time/radius. Definitions of curves are 
the same as Fig. \ref{fig:betazero}. These plots show that $\log\omega'_m$ and 
$\gamma'_m$ have a very similar evolution.
\label{fig:omegam}}
\end{figure*}

\begin{figure*}
\begin{center}
\begin {tabular}{cccc}
1)\includegraphics[height=4cm,angle=-90]{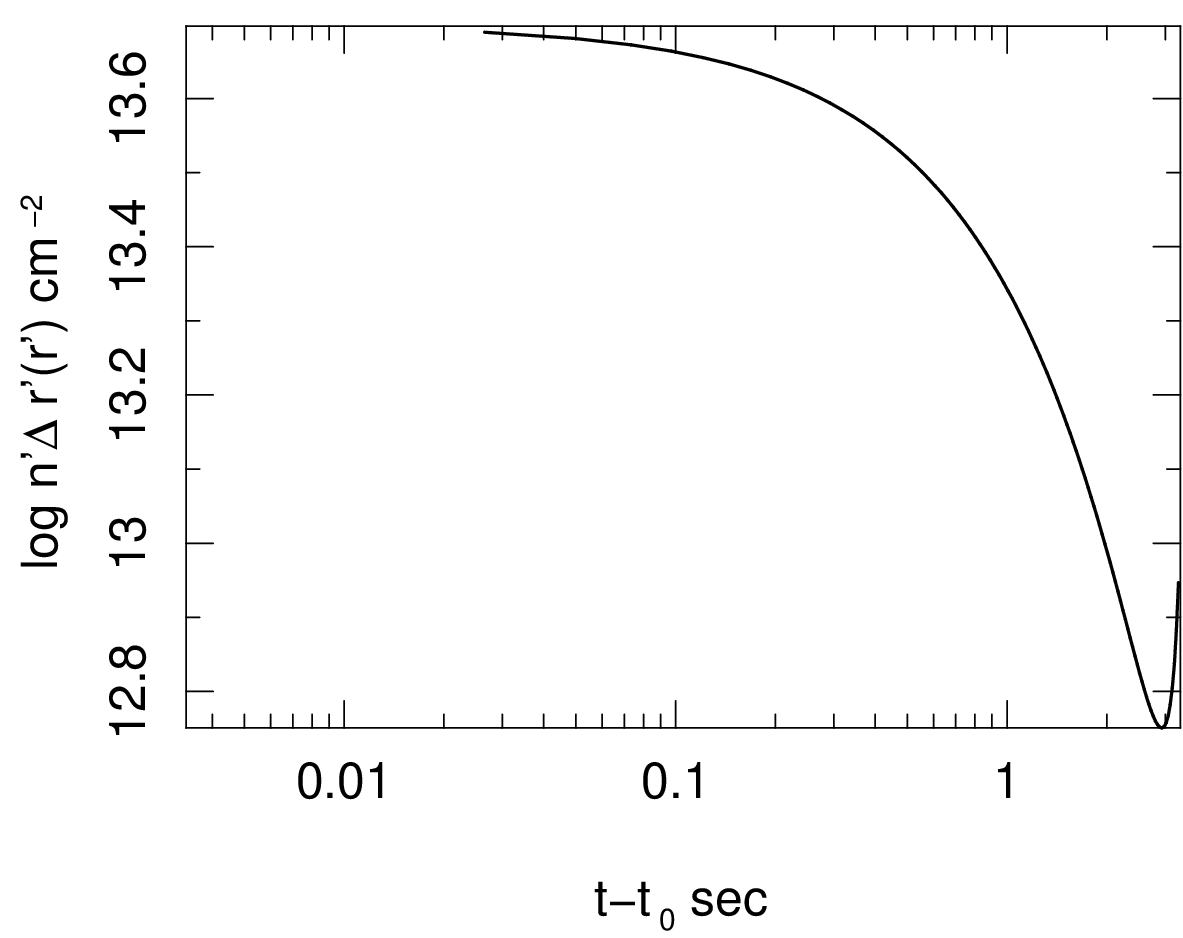} &
2)\includegraphics[height=4cm,angle=-90]{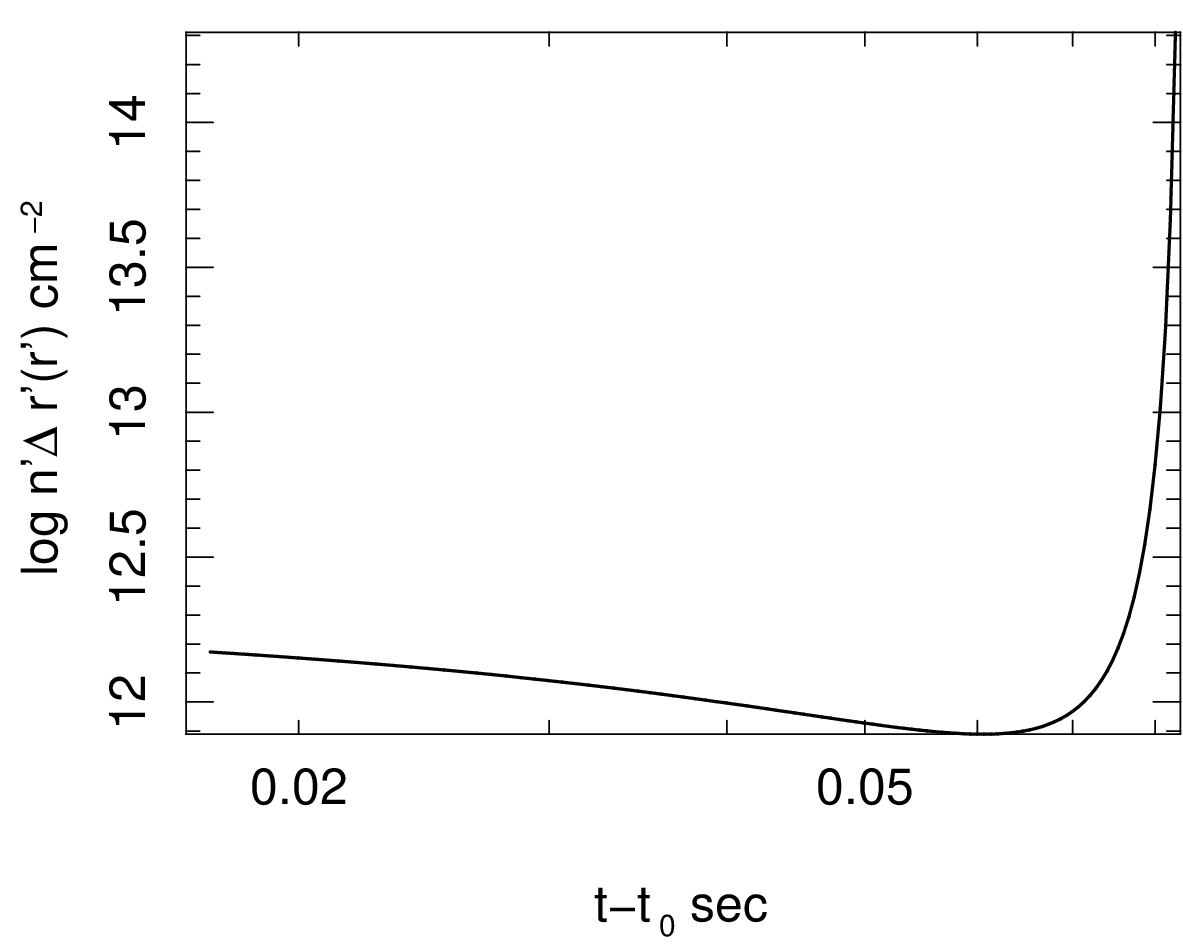} & 
3)\includegraphics[height=4cm,angle=-90]{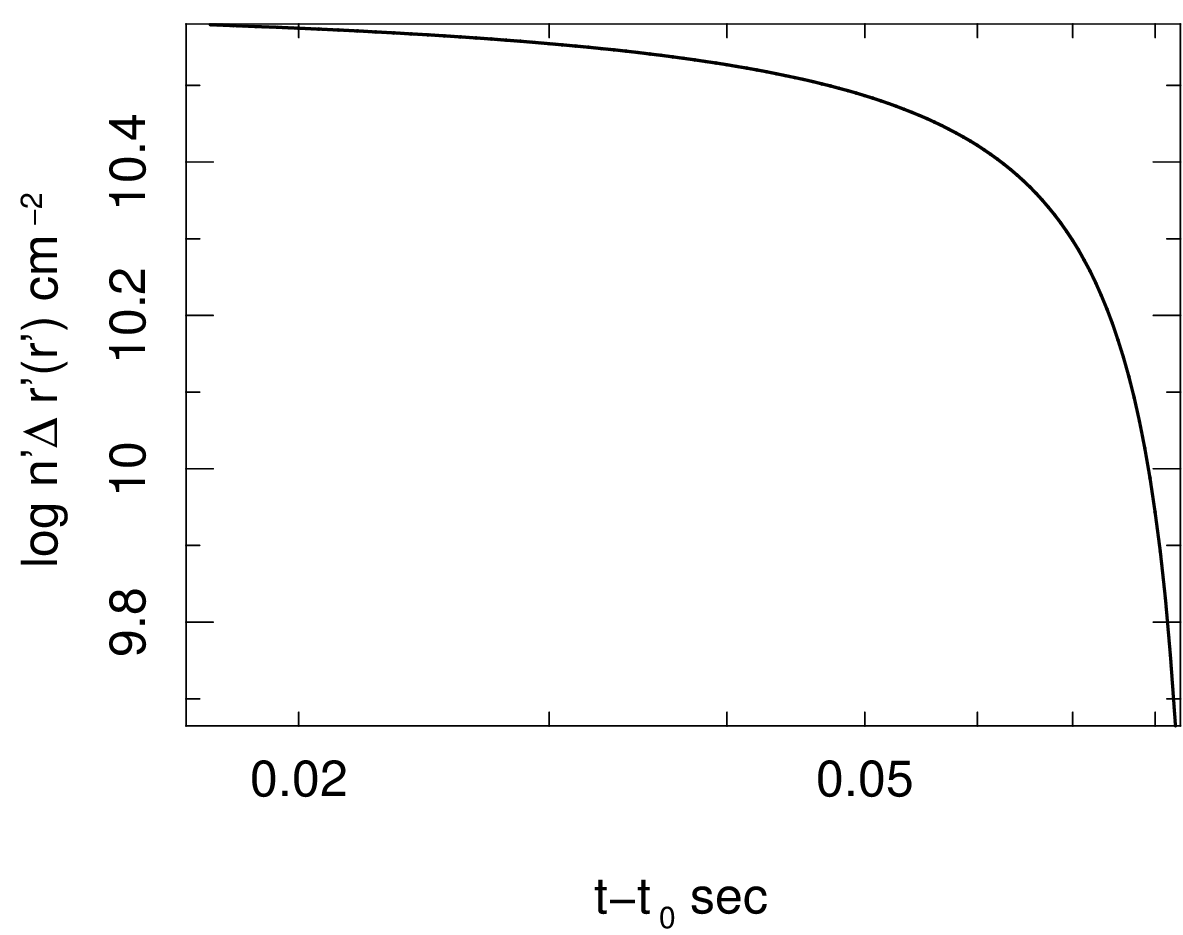} &
4)\includegraphics[height=4cm,angle=-90]{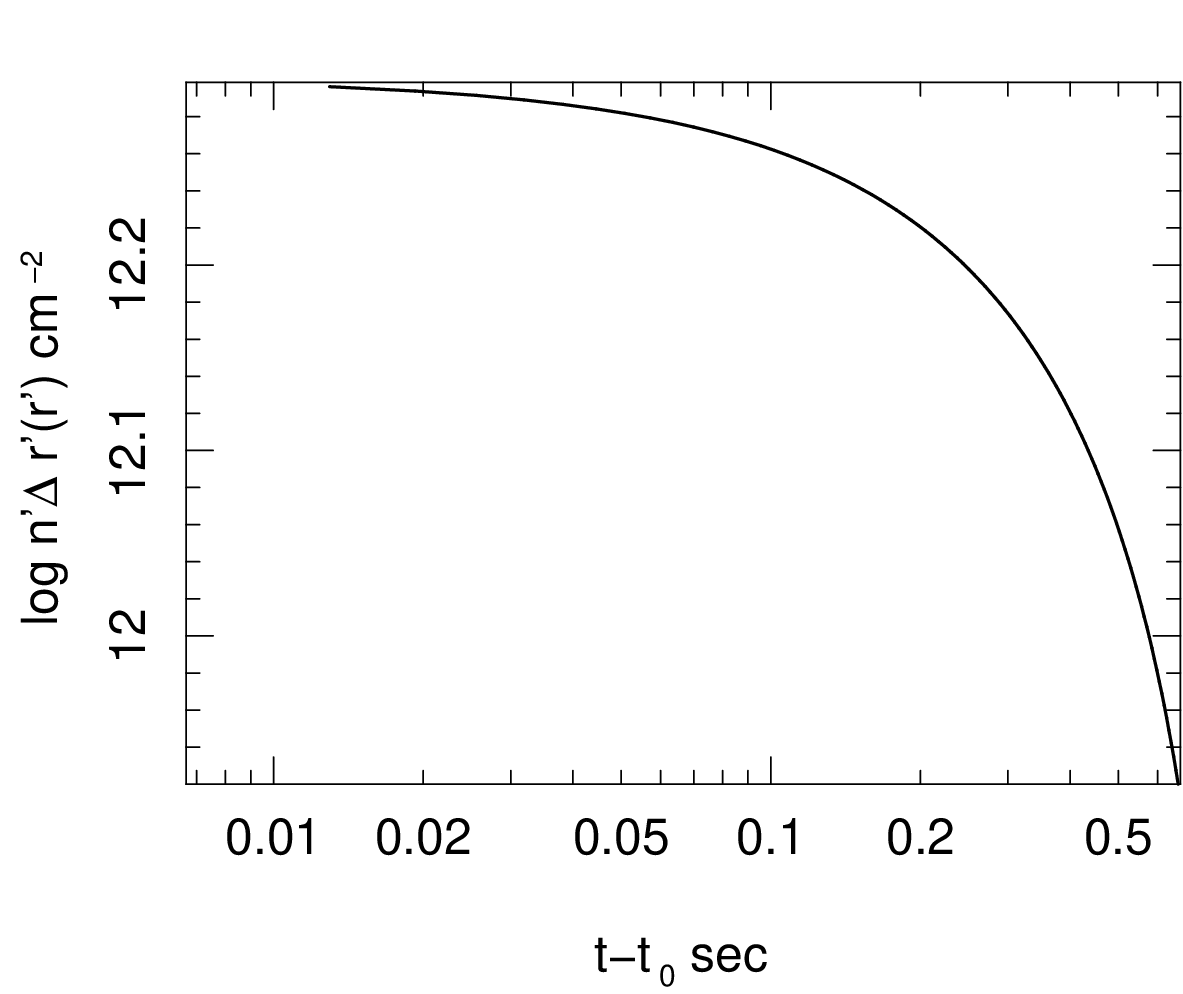} \\ 
5)\includegraphics[height=4cm,angle=-90]{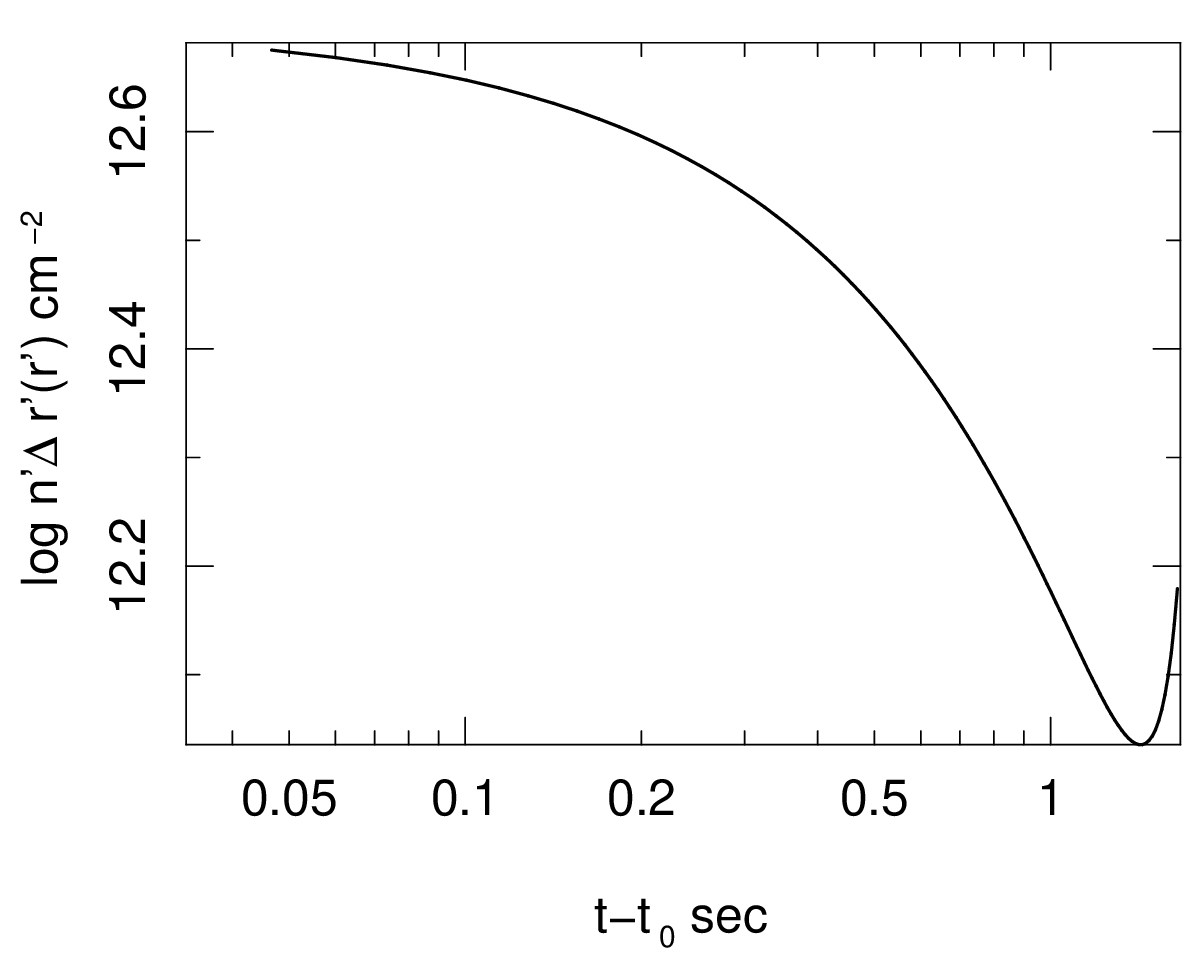} &
6)\includegraphics[height=4cm,angle=-90]{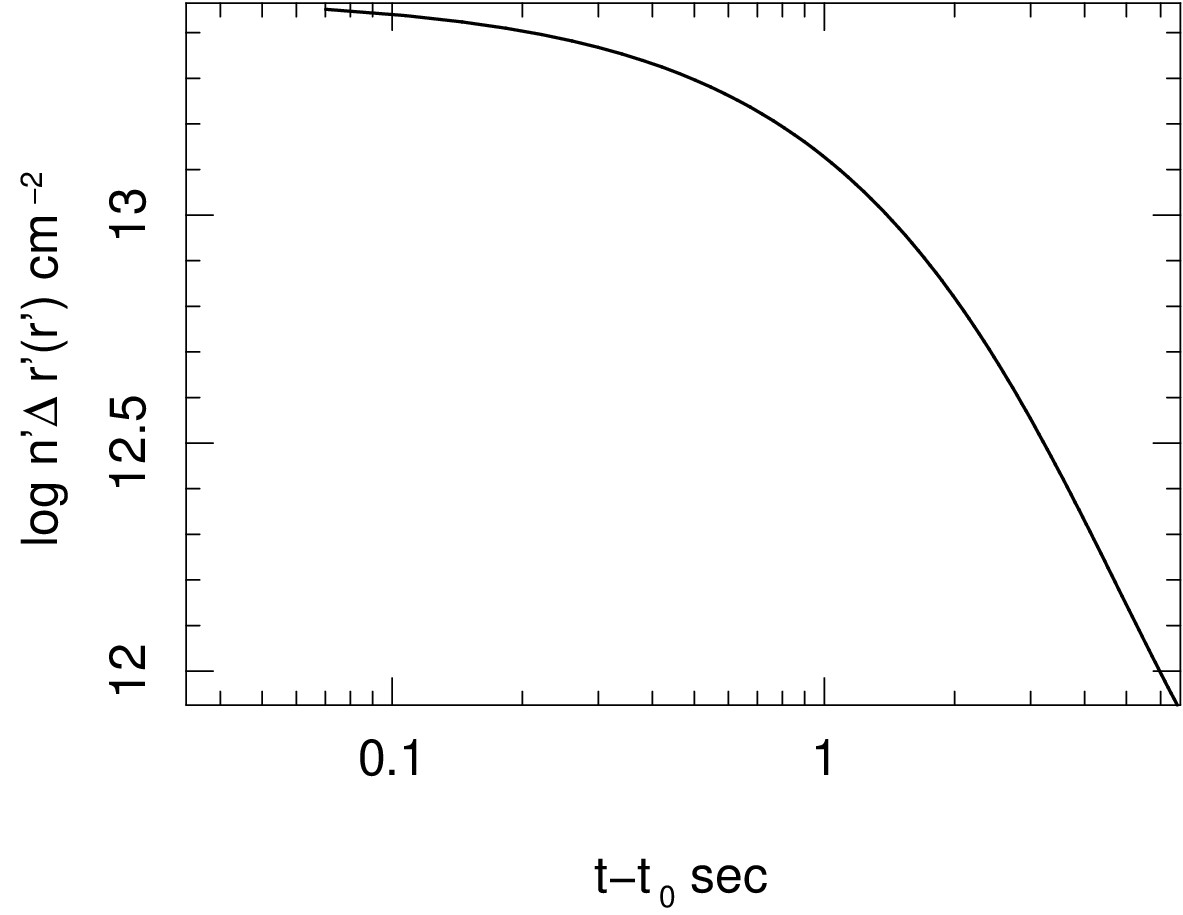} & 
7)\includegraphics[height=4cm,angle=-90]{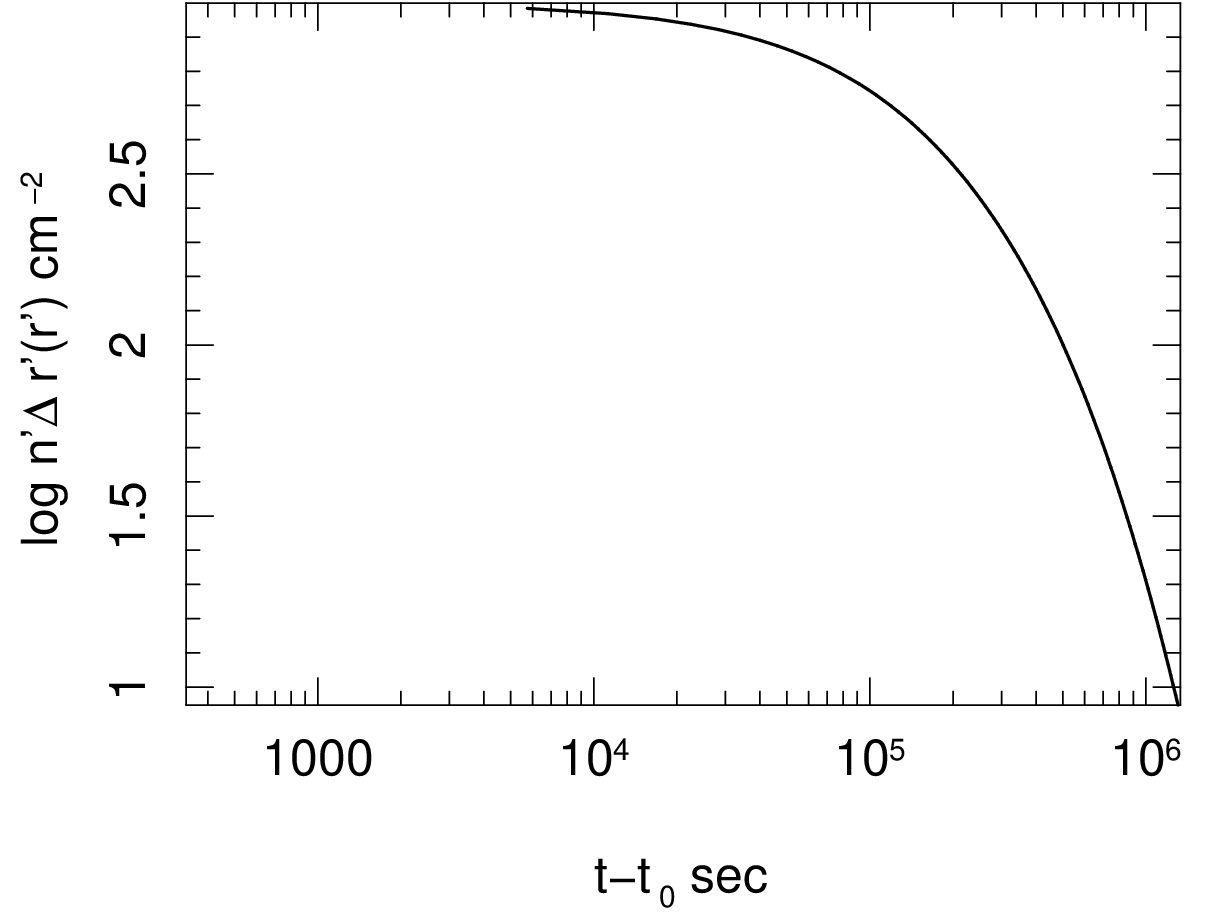} &
8)\includegraphics[height=4cm,angle=-90]{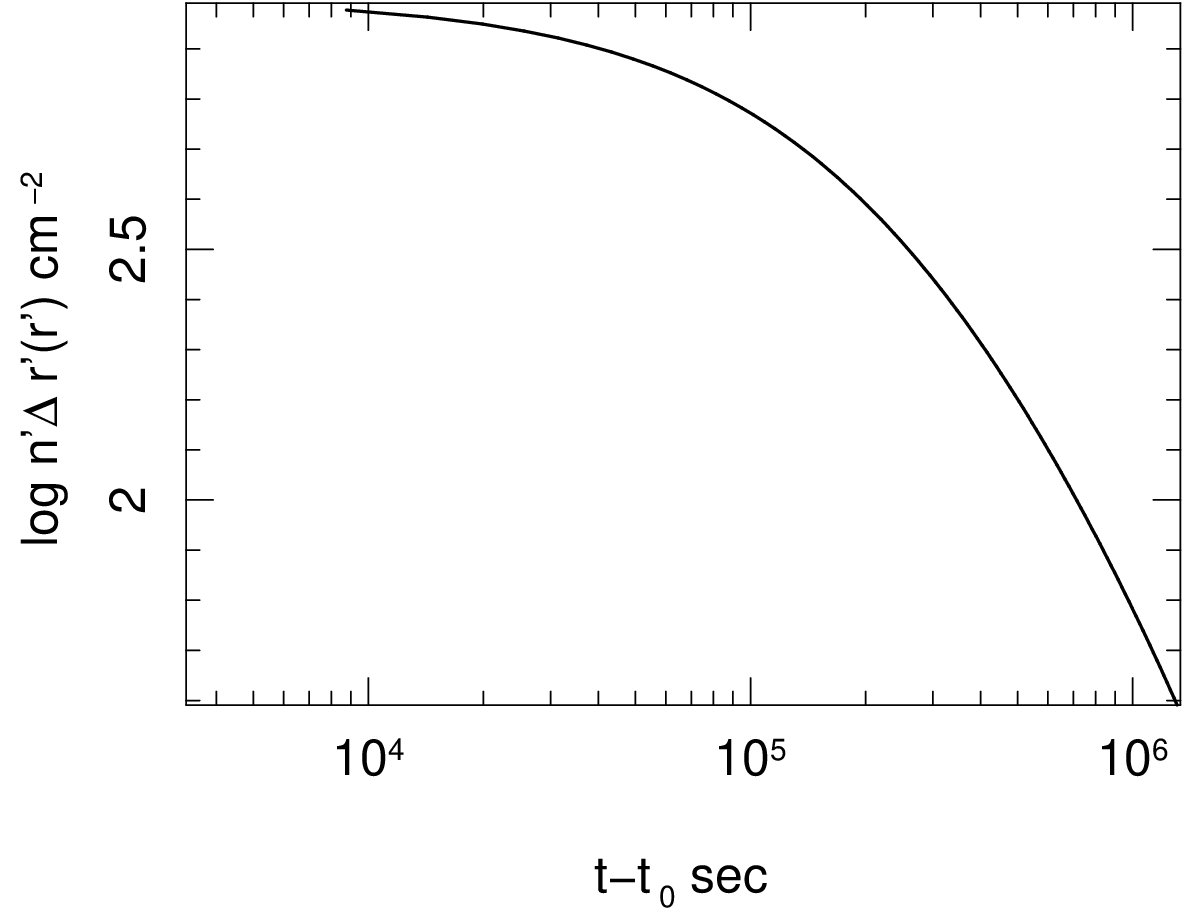} \\
9)\includegraphics[height=4cm,angle=-90]{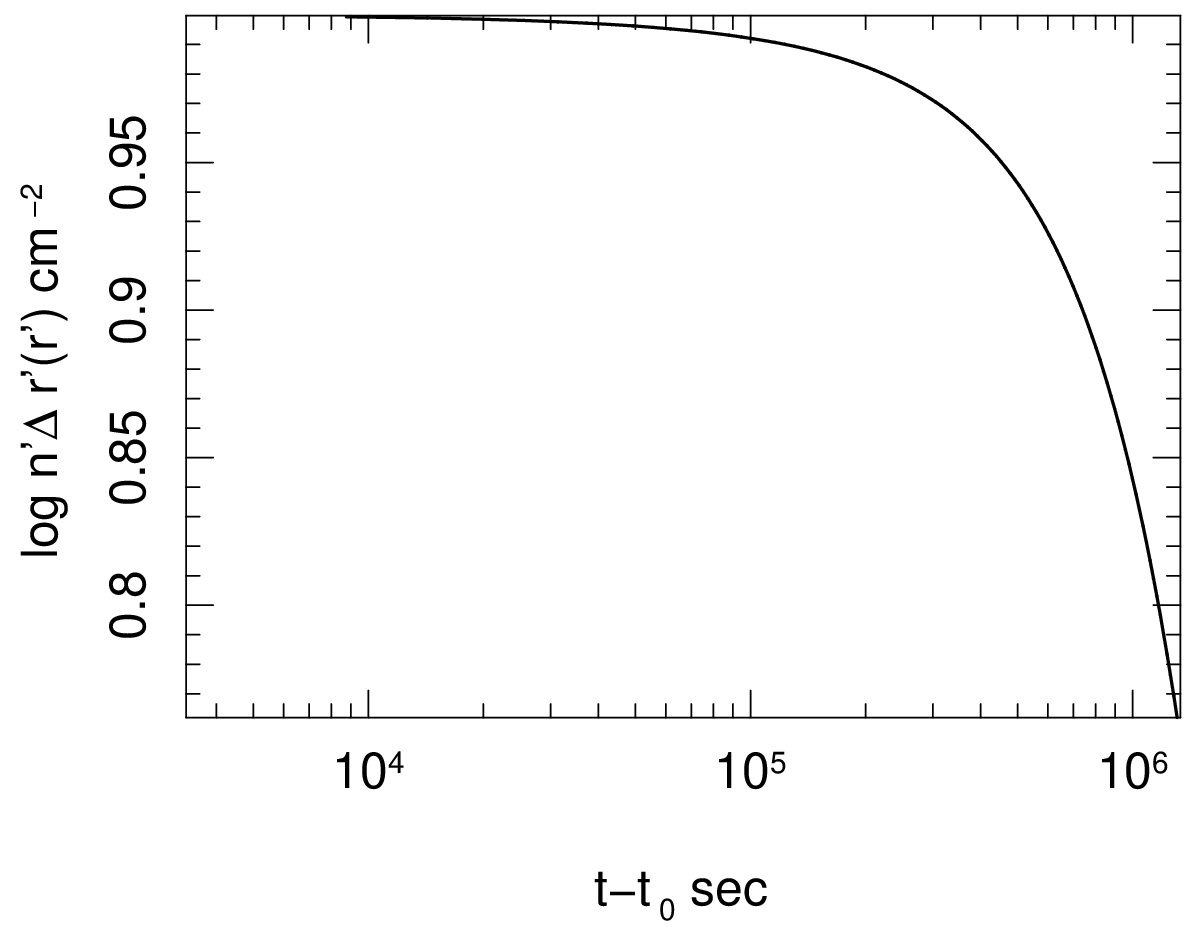} &
10)\includegraphics[height=4cm,angle=-90]{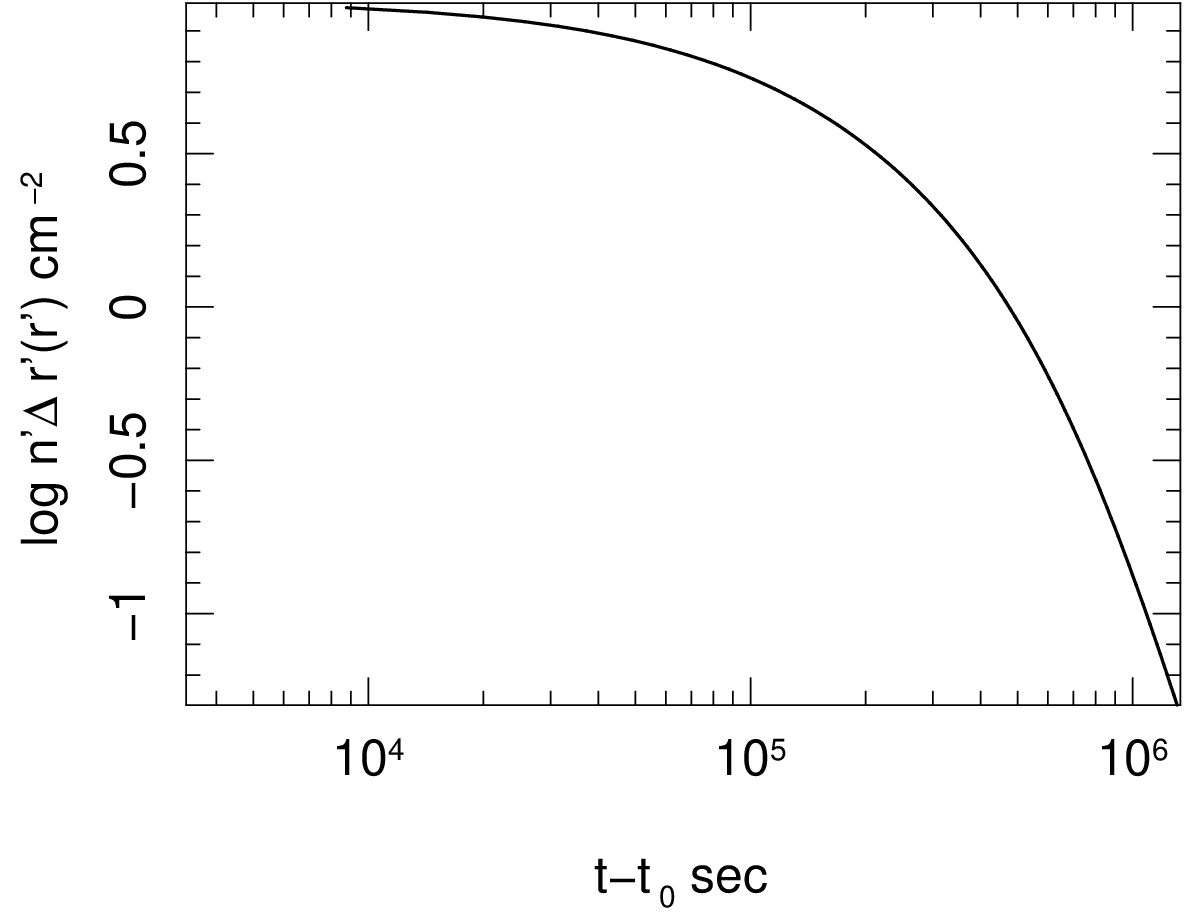} &
11)\includegraphics[height=4cm,angle=-90]{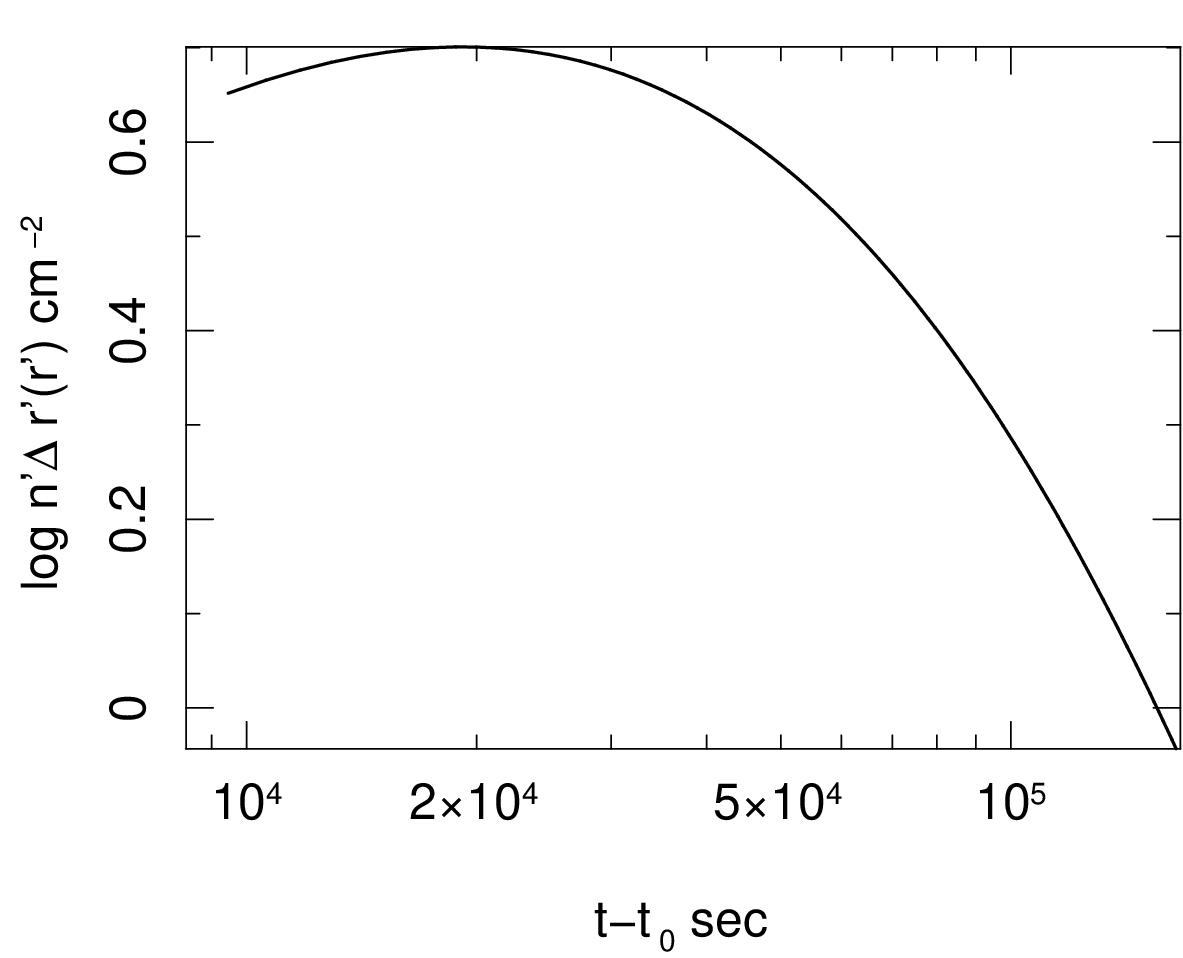} &
12)\includegraphics[height=4cm,angle=-90]{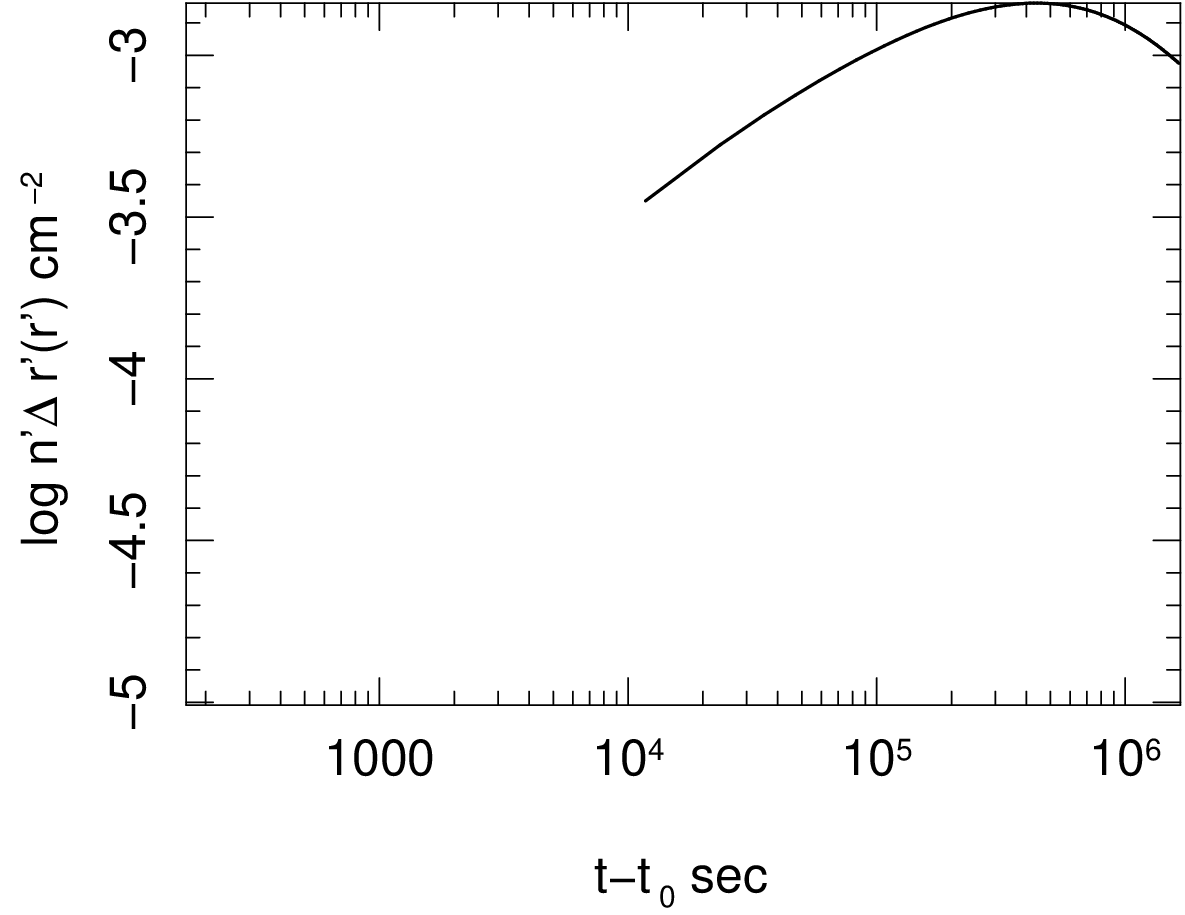} \\
13)\includegraphics[height=4cm,angle=-90]{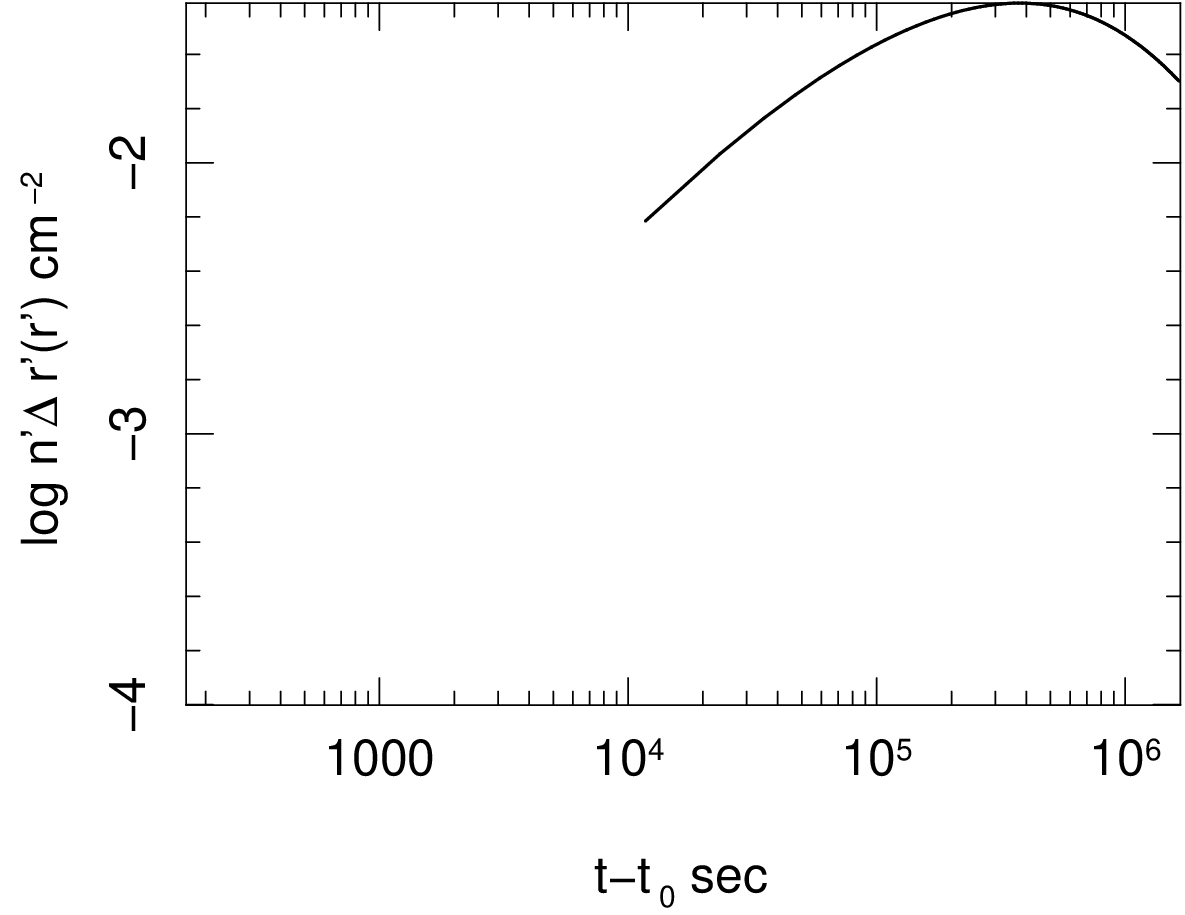} &
14)\includegraphics[height=4cm,angle=-90]{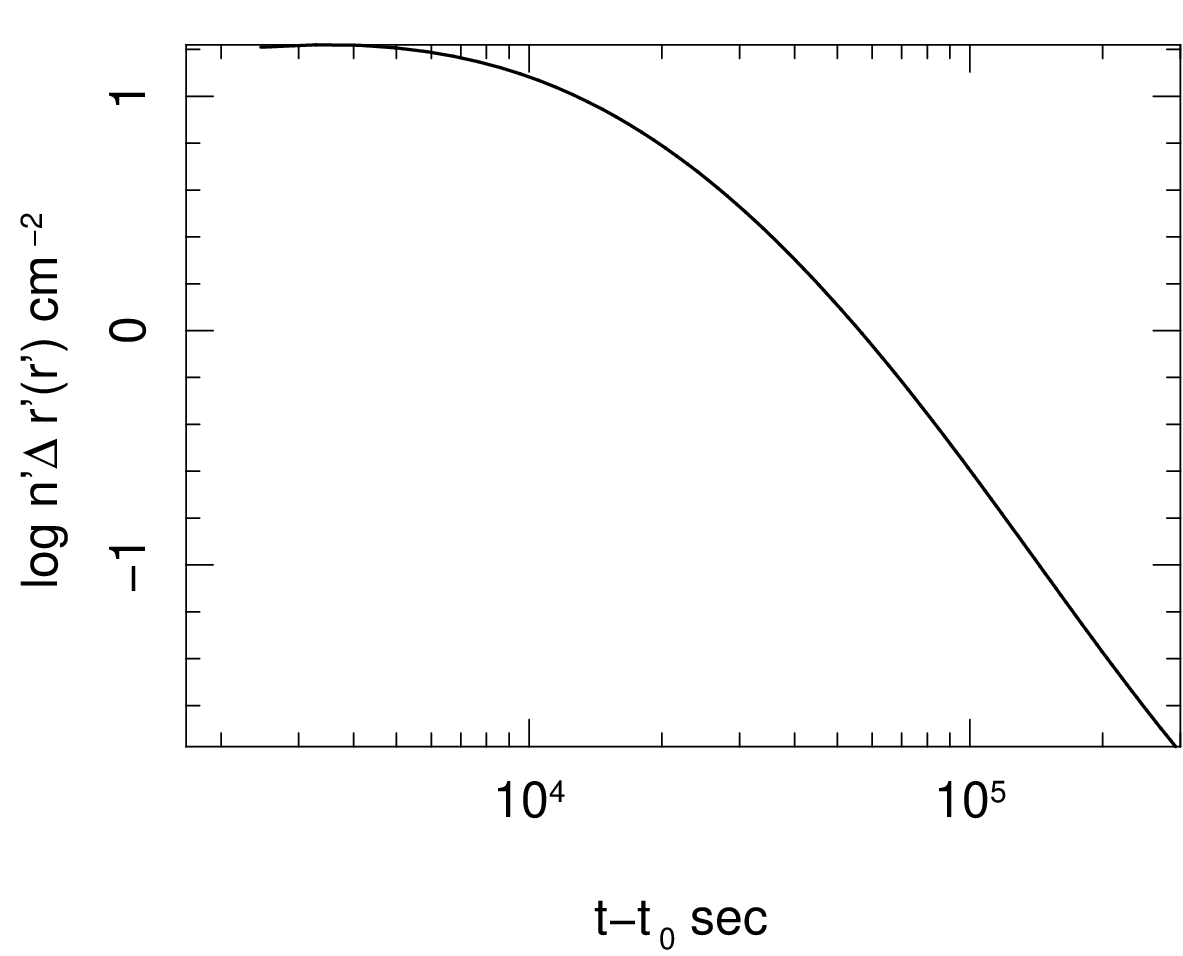} &
15)\includegraphics[height=4cm,angle=-90]{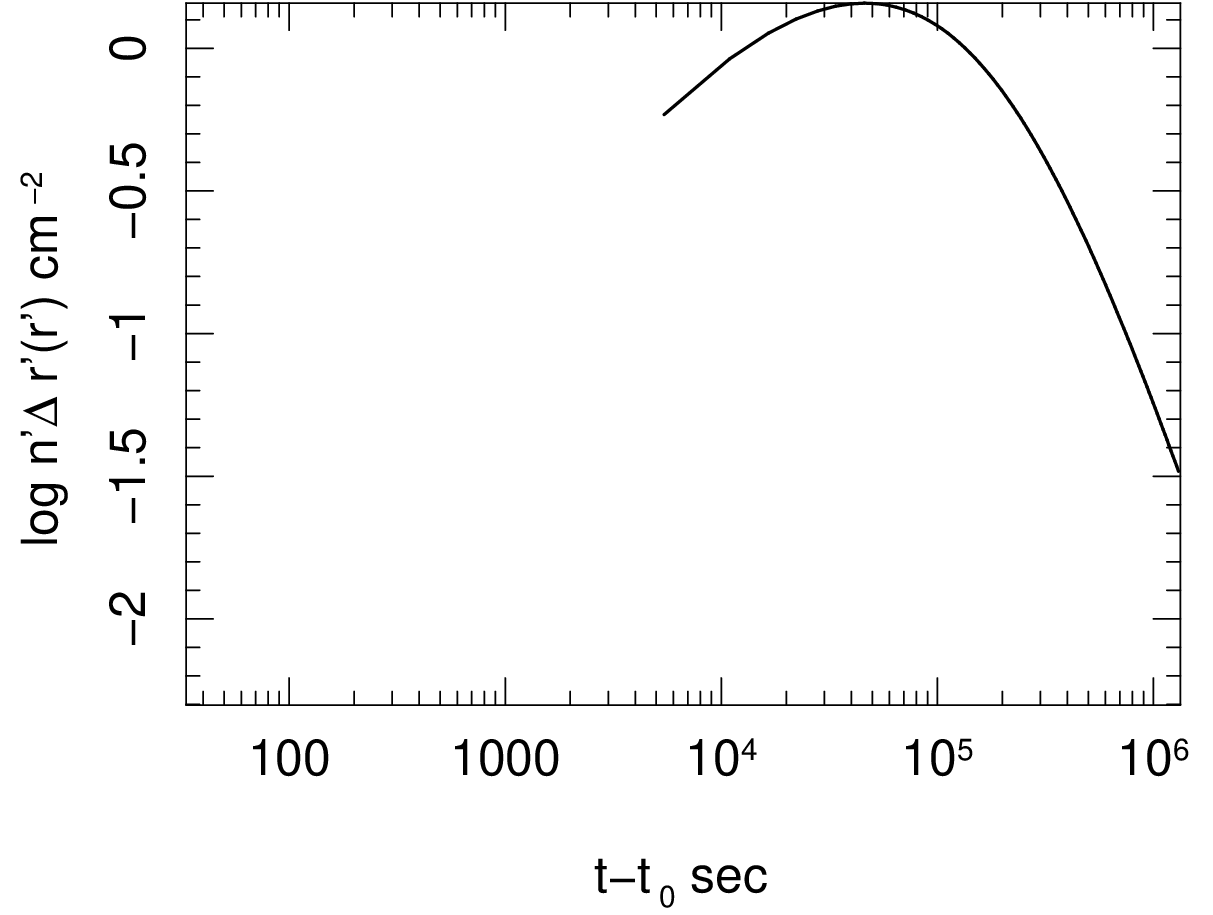} &
16)\includegraphics[height=4cm,angle=-90]{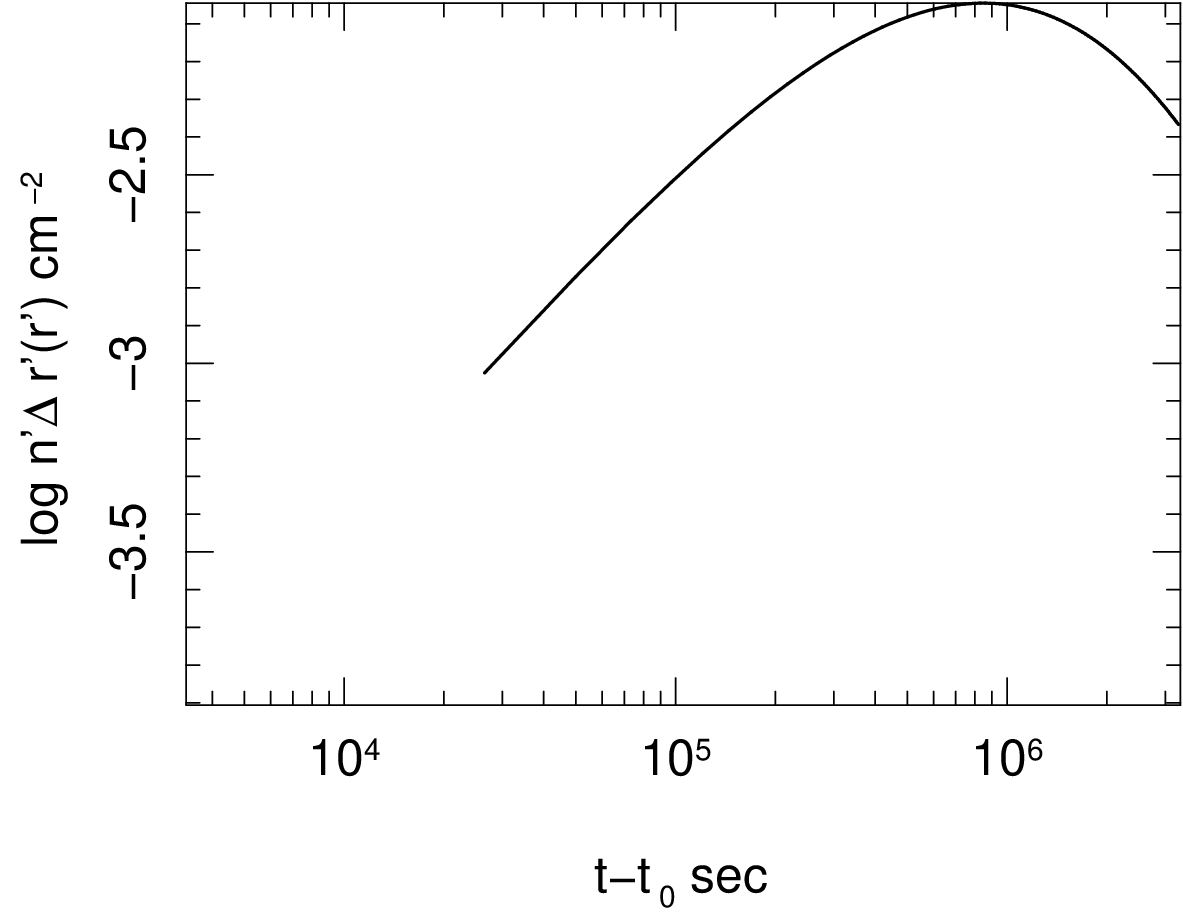} \\
17)\includegraphics[height=4cm,angle=-90]{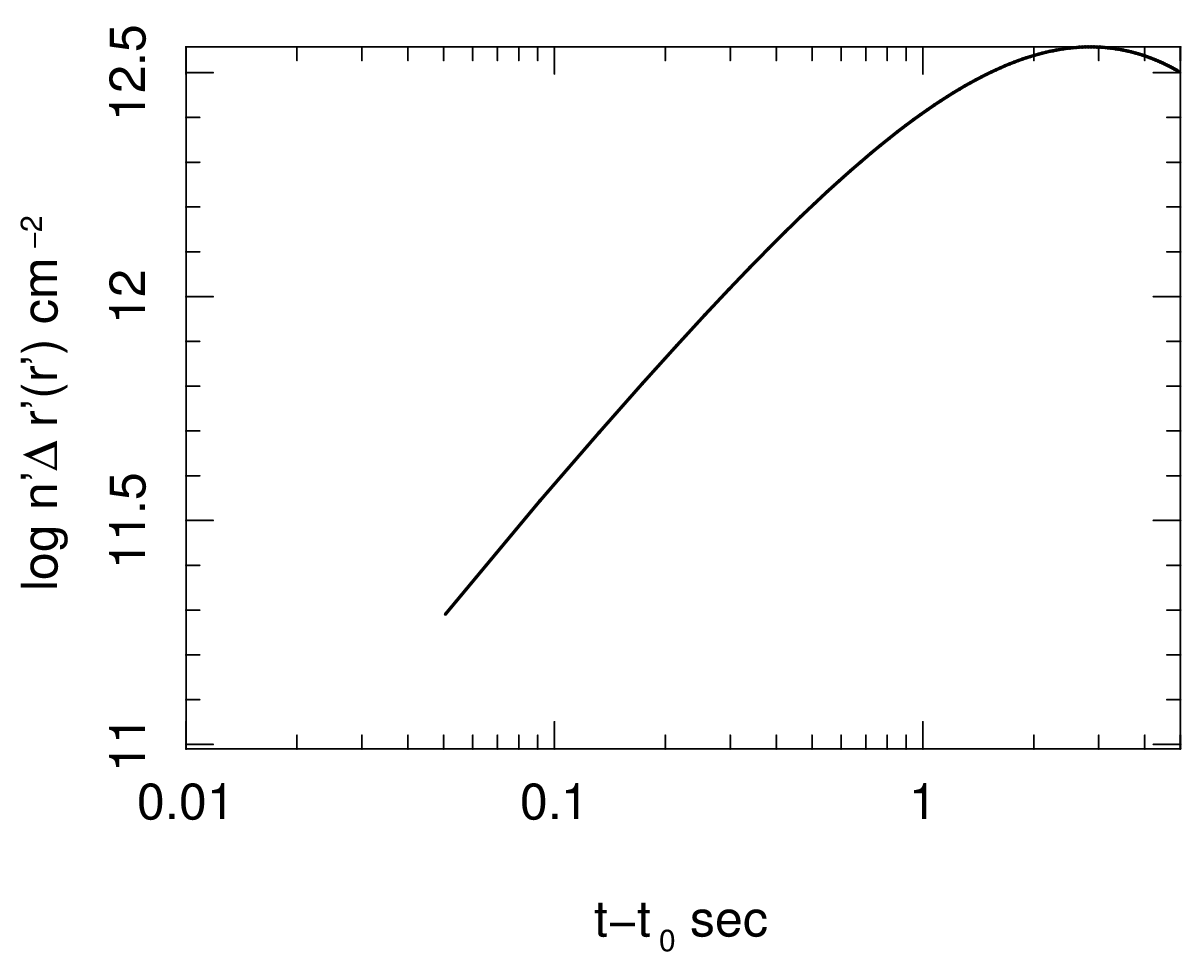} &
18)\includegraphics[height=4cm,angle=-90]{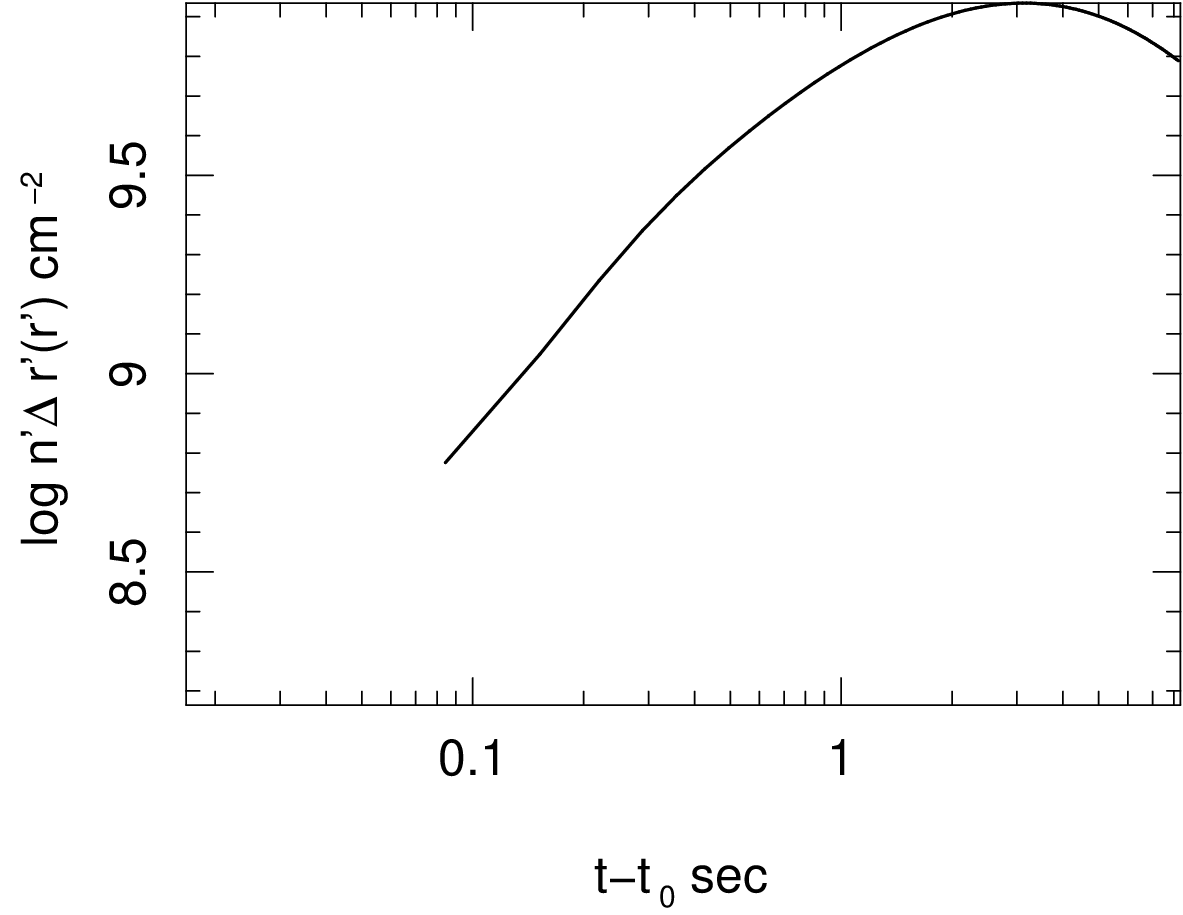} &
19)\includegraphics[height=4cm,angle=-90]{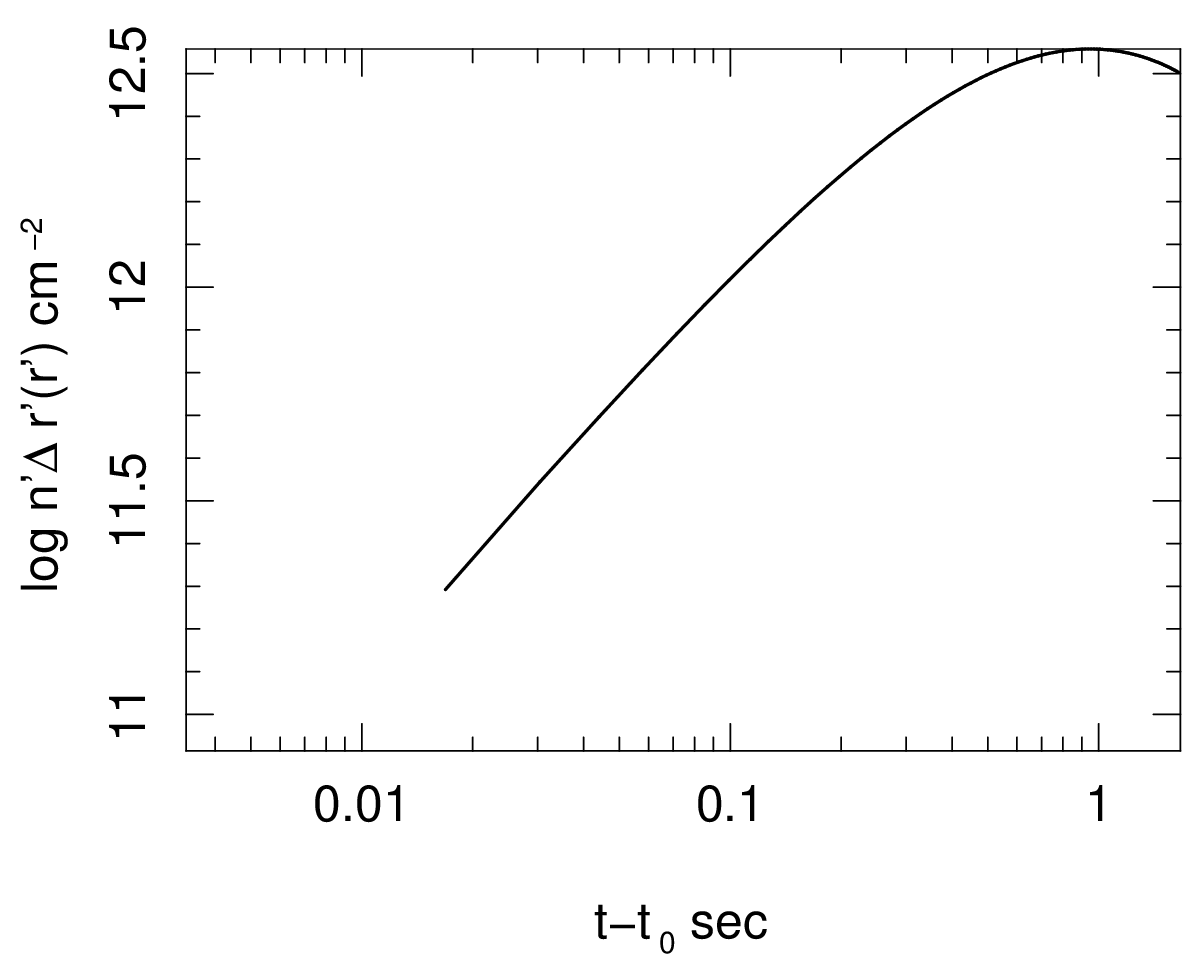} &
20)\includegraphics[height=4cm,angle=-90]{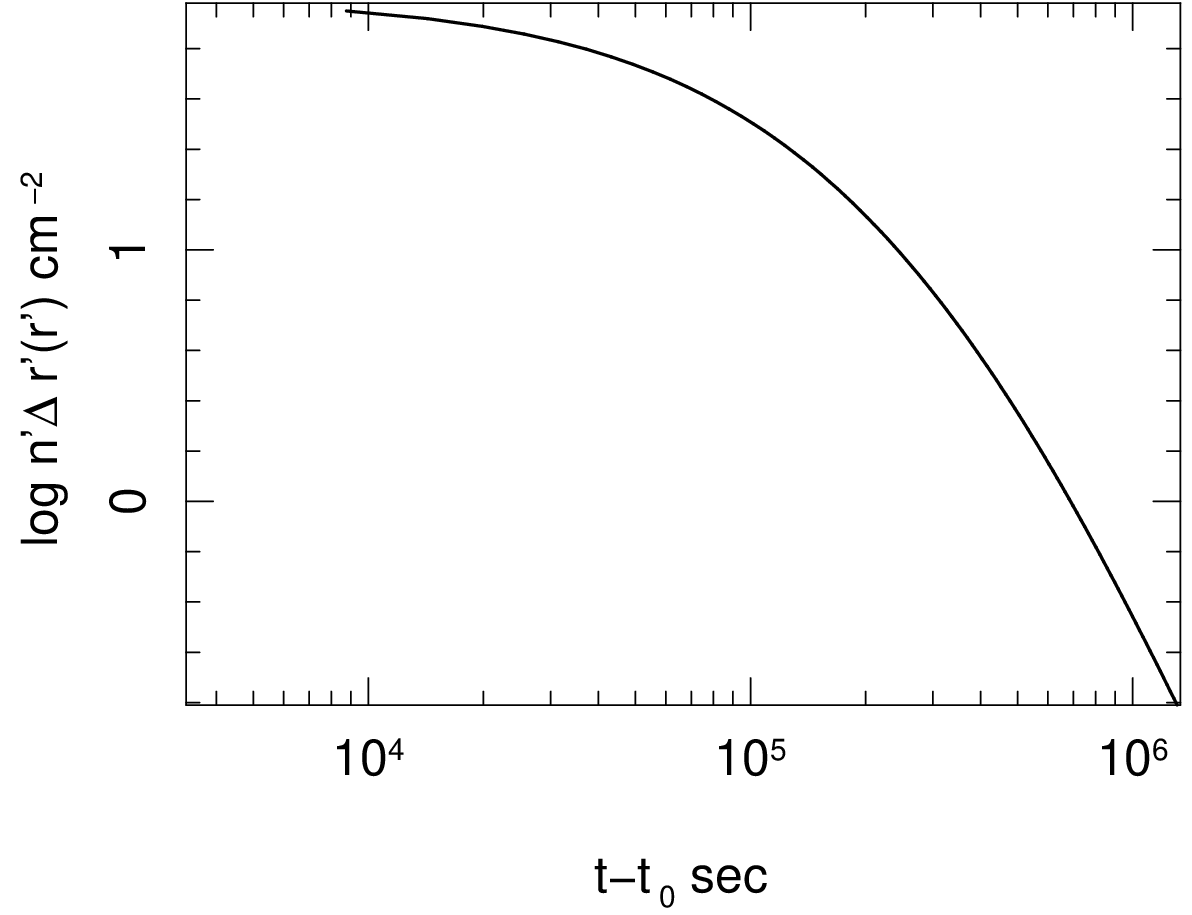} \\
21)\includegraphics[height=4cm,angle=-90]{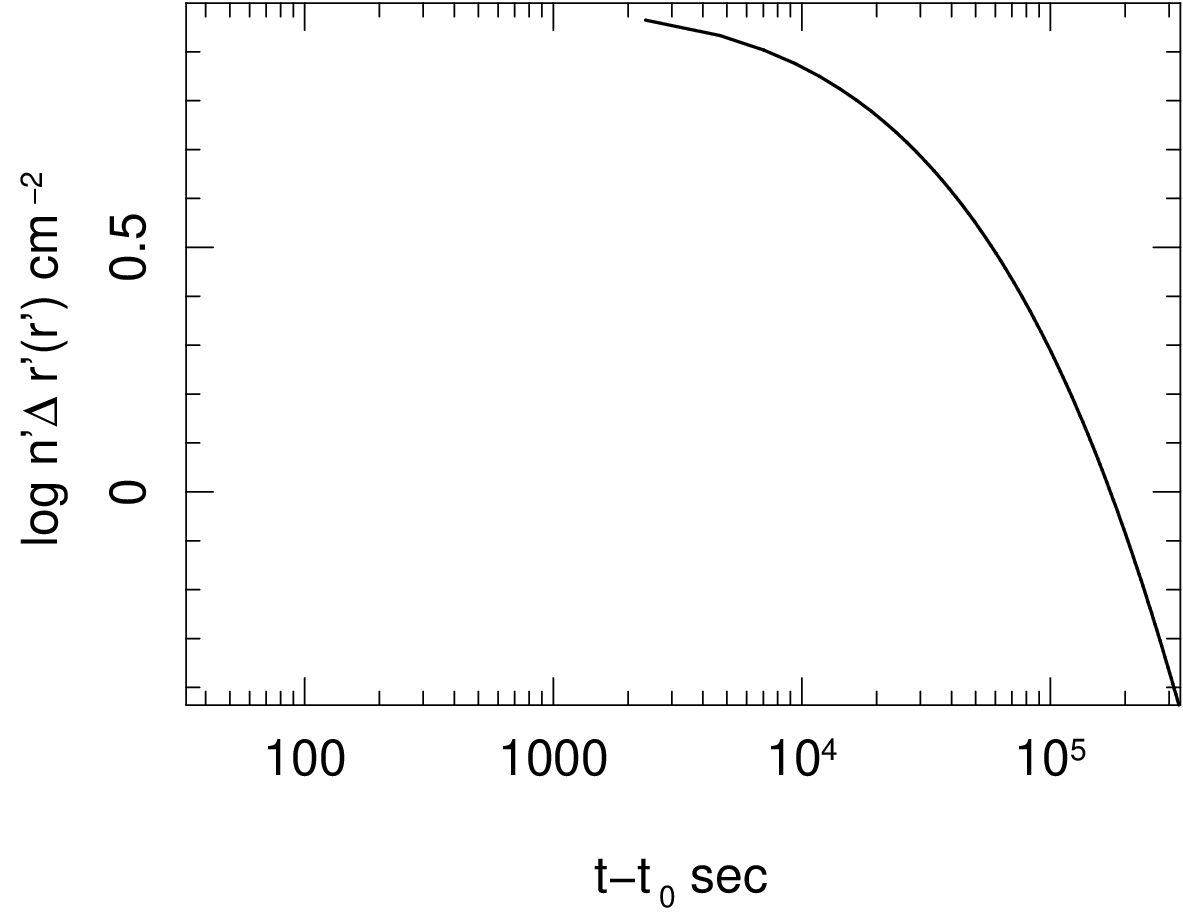} &
22)\includegraphics[height=4cm,angle=-90]{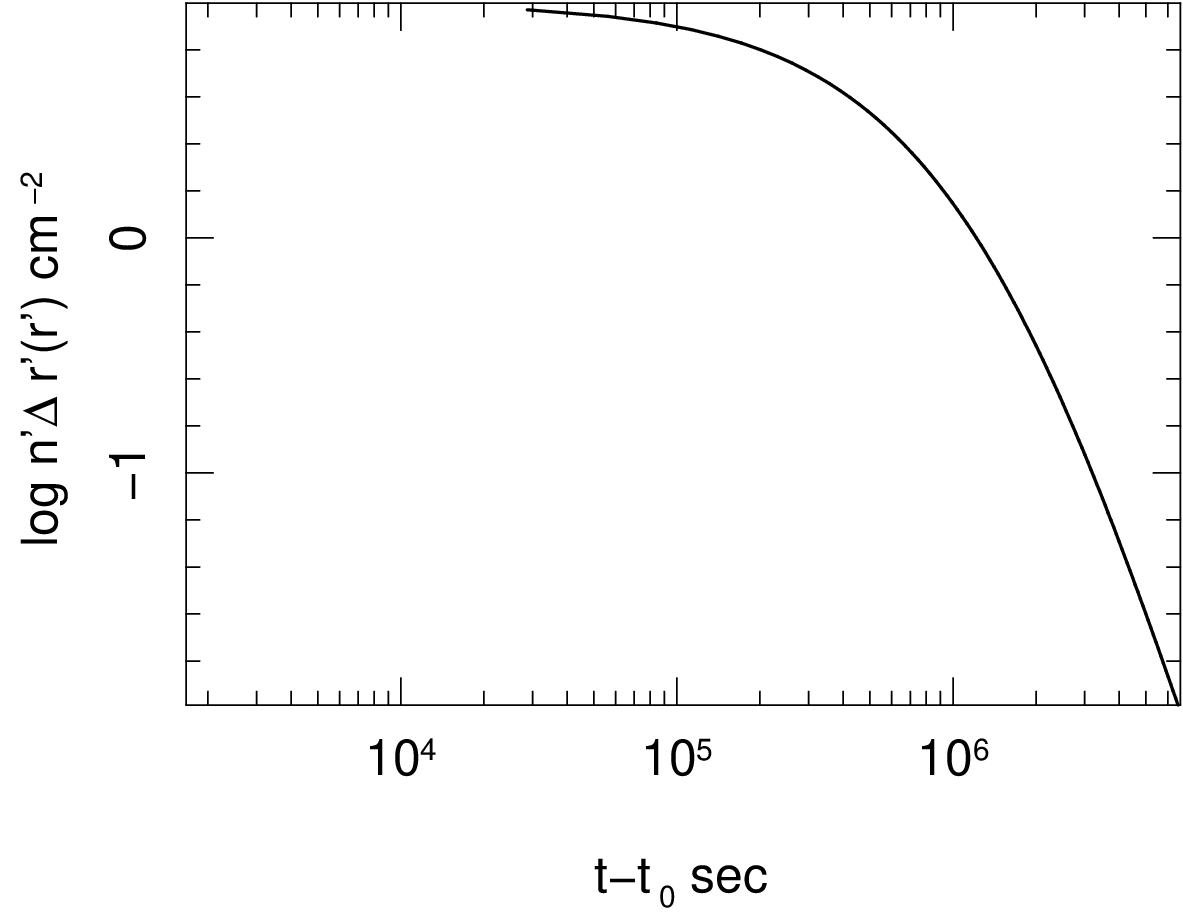} &
23)\includegraphics[height=4cm,angle=-90]{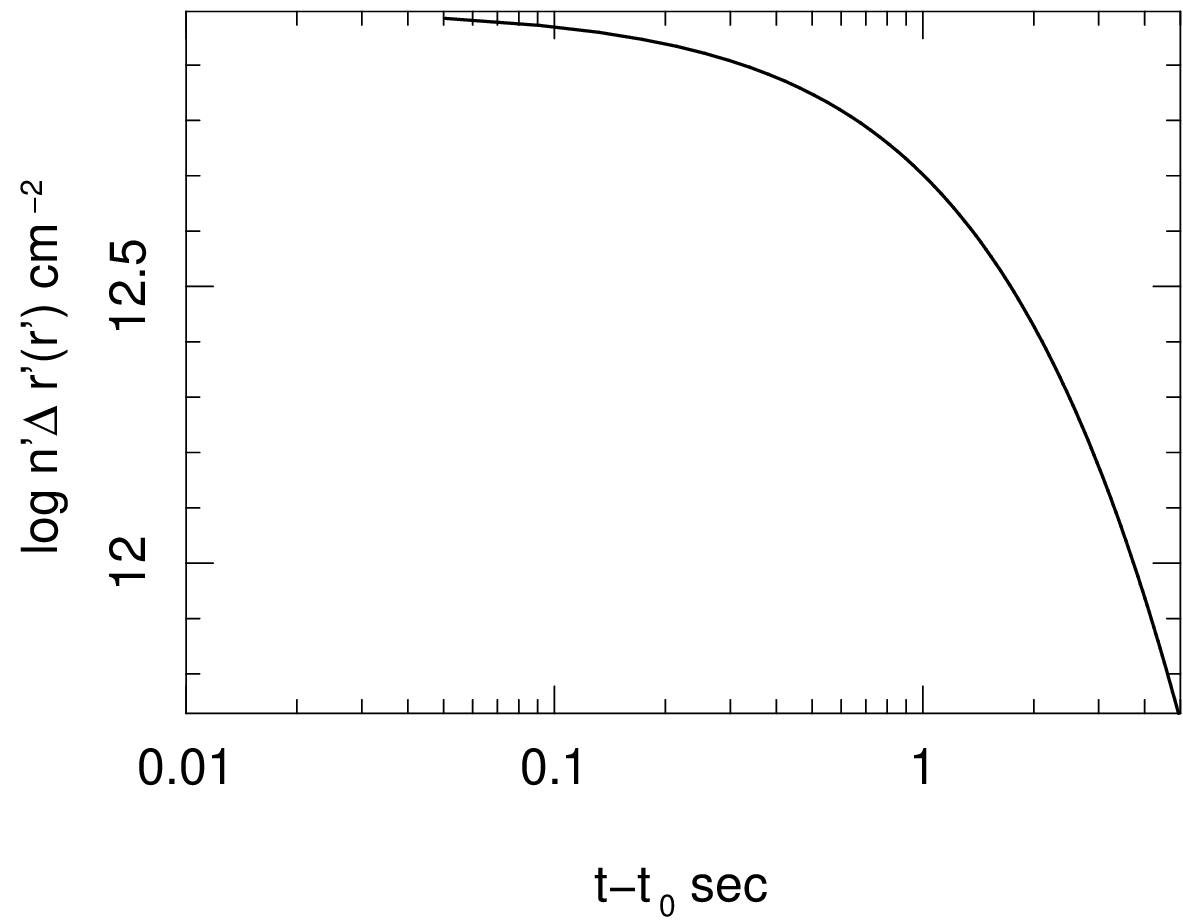} &
24)\includegraphics[height=4cm,angle=-90]{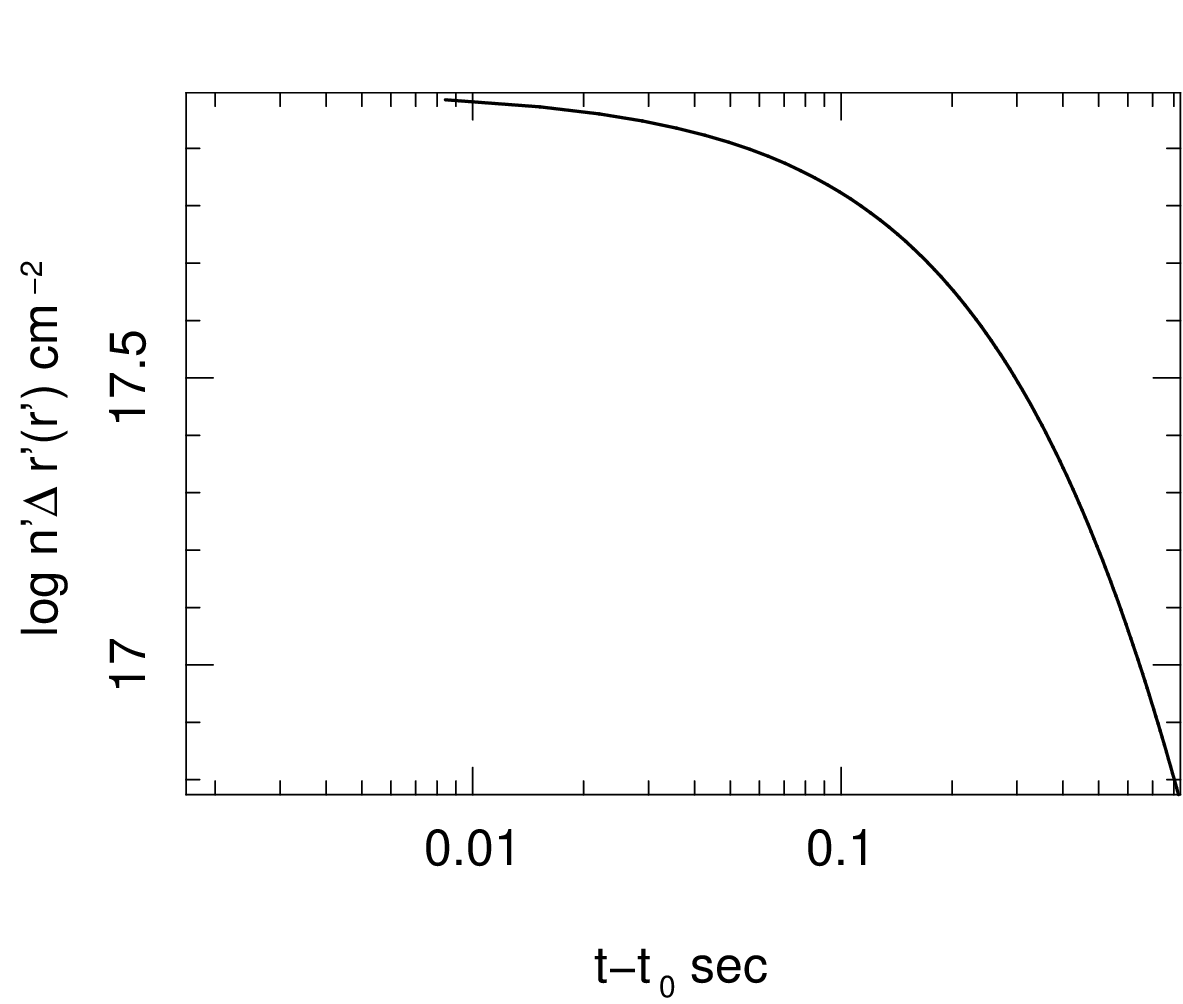} 
\end {tabular}
\end {center}
\caption{Evolution of the column density $n'\Delta r'$ with time/radius. 
Definitions of curves are the same as Fig. \ref{fig:betazero}. Depending on 
parameters and active region evolution model the column density shows a 
variety of behavior.
\label{fig:coldens}}
\end{figure*}

\begin{figure*}
\begin{center}
\begin {tabular}{cccc}
1)\includegraphics[height=4cm,angle=-90]{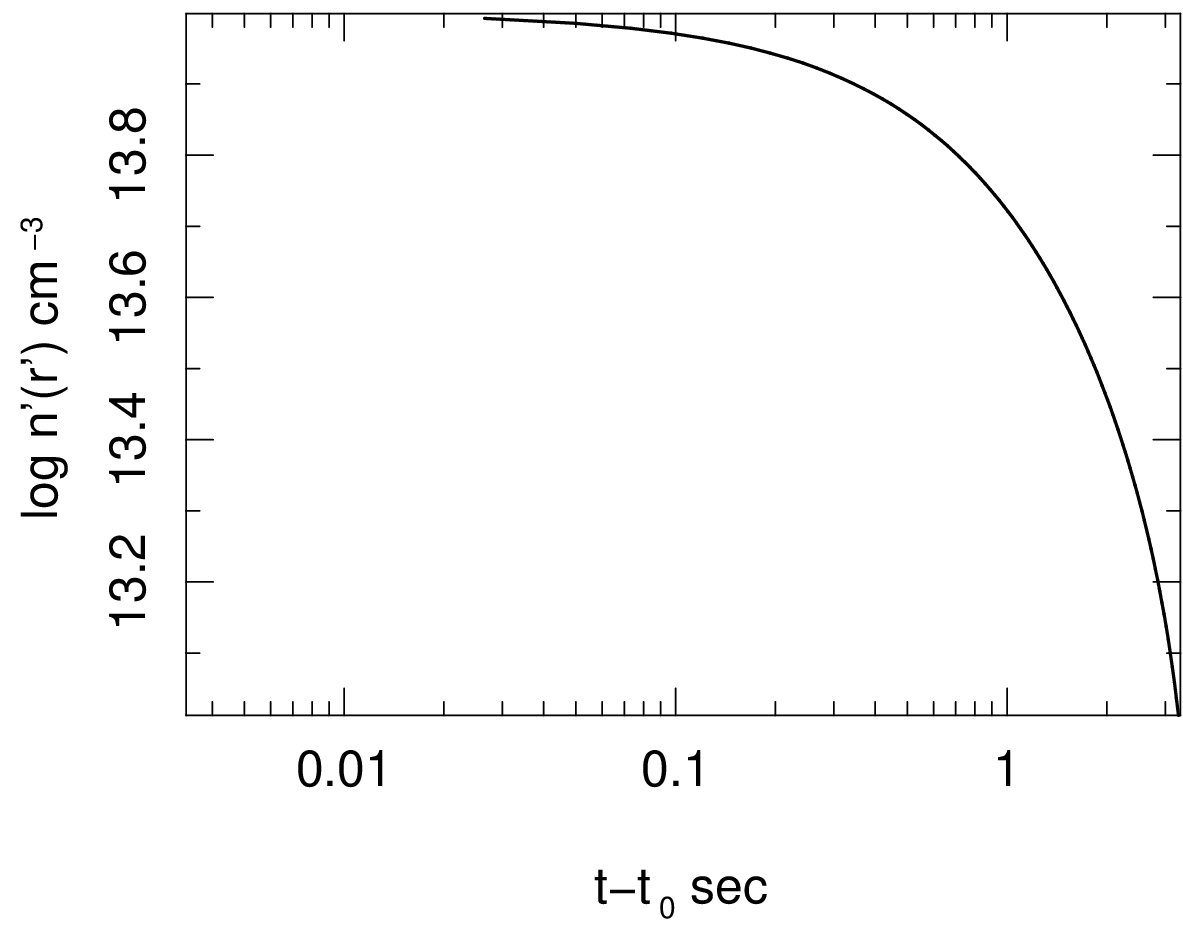} &
2)\includegraphics[height=4cm,angle=-90]{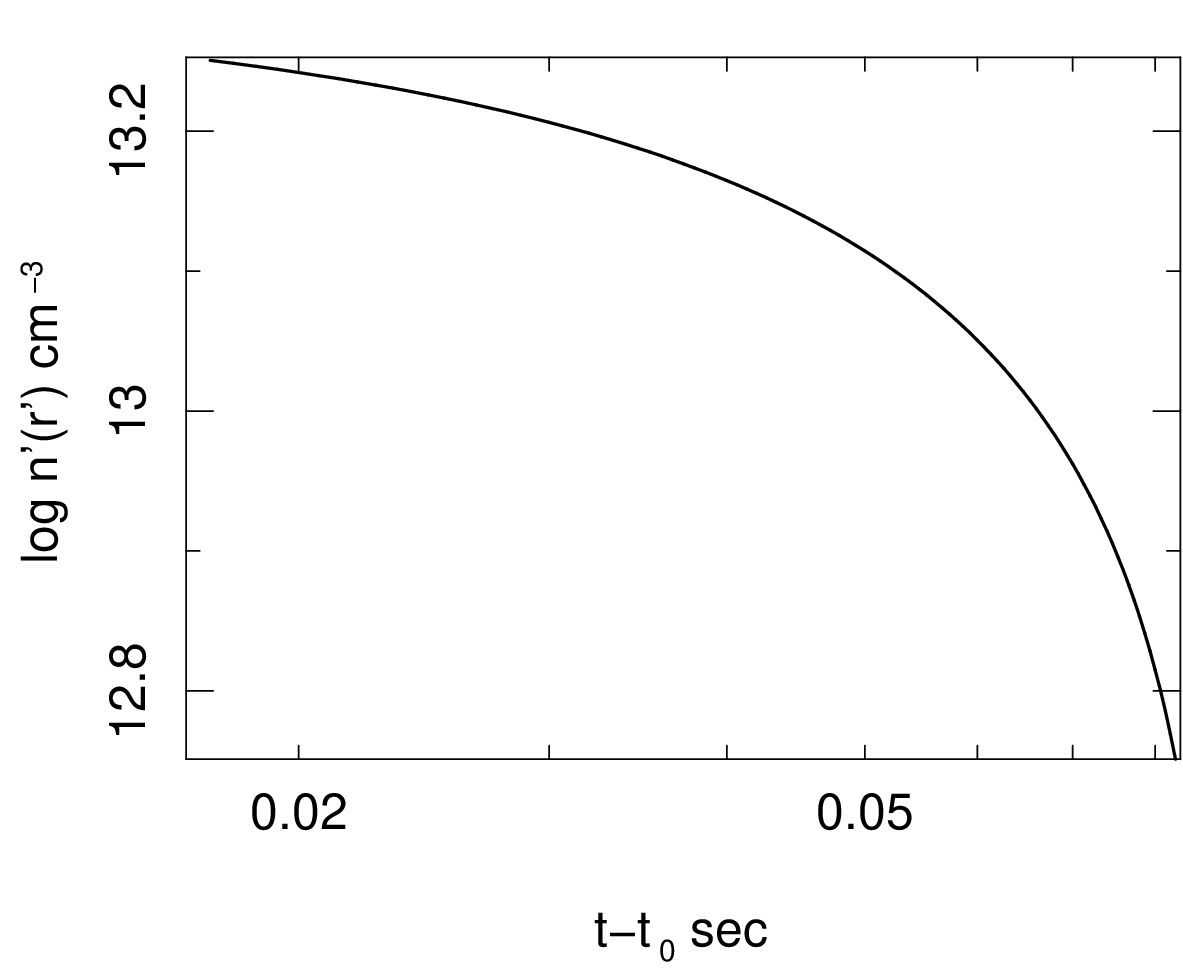} & 
3)\includegraphics[height=4cm,angle=-90]{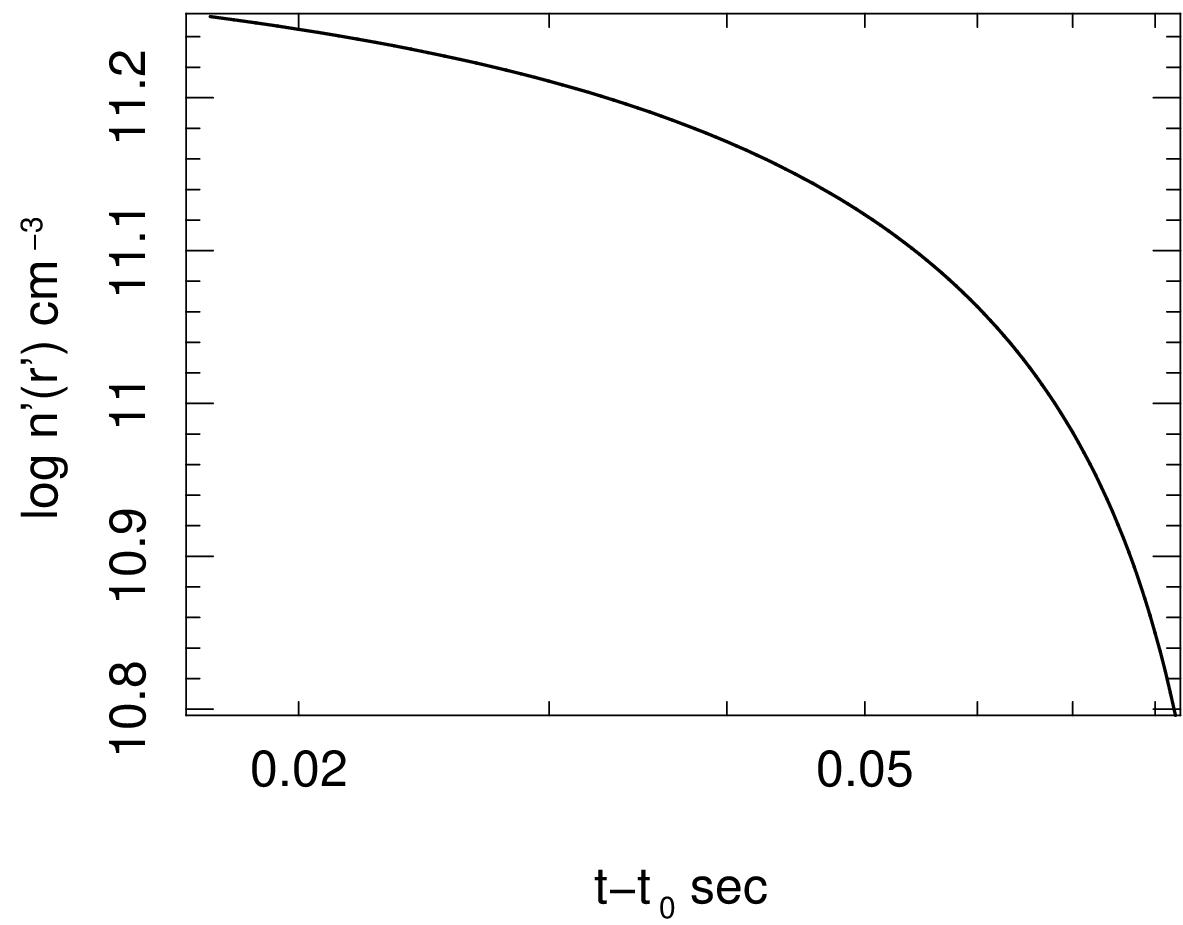} &
4)\includegraphics[height=4cm,angle=-90]{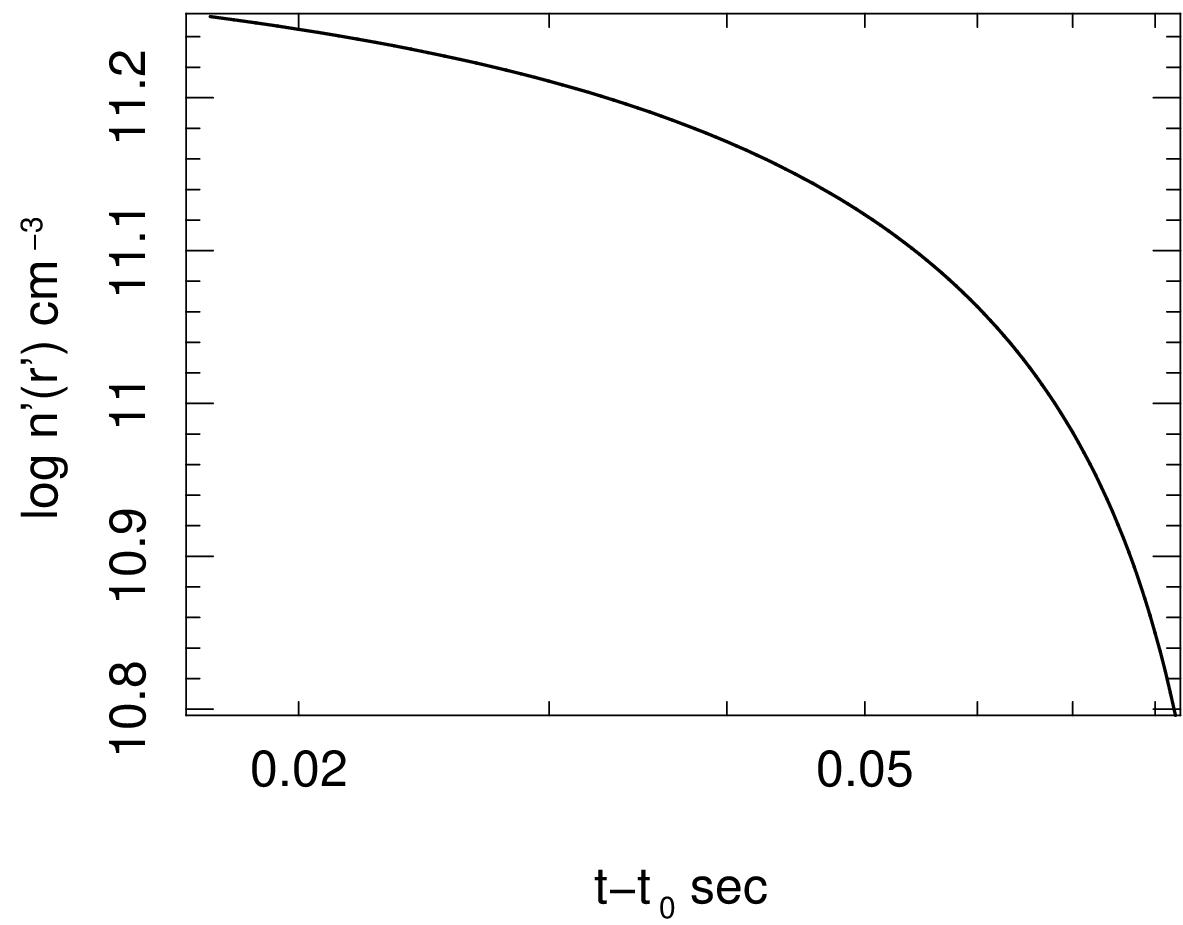} \\ 
5)\includegraphics[height=4cm,angle=-90]{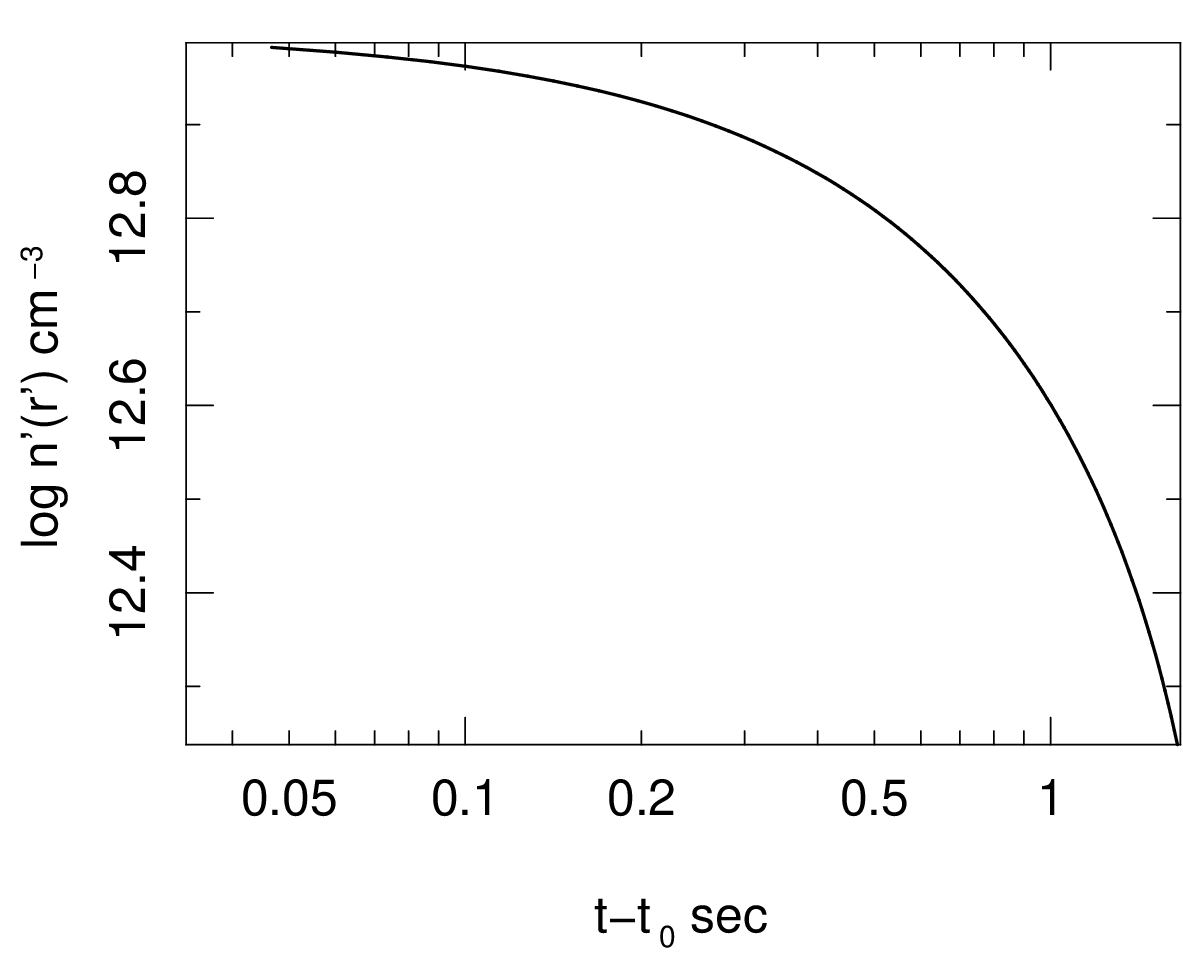} &
6)\includegraphics[height=4cm,angle=-90]{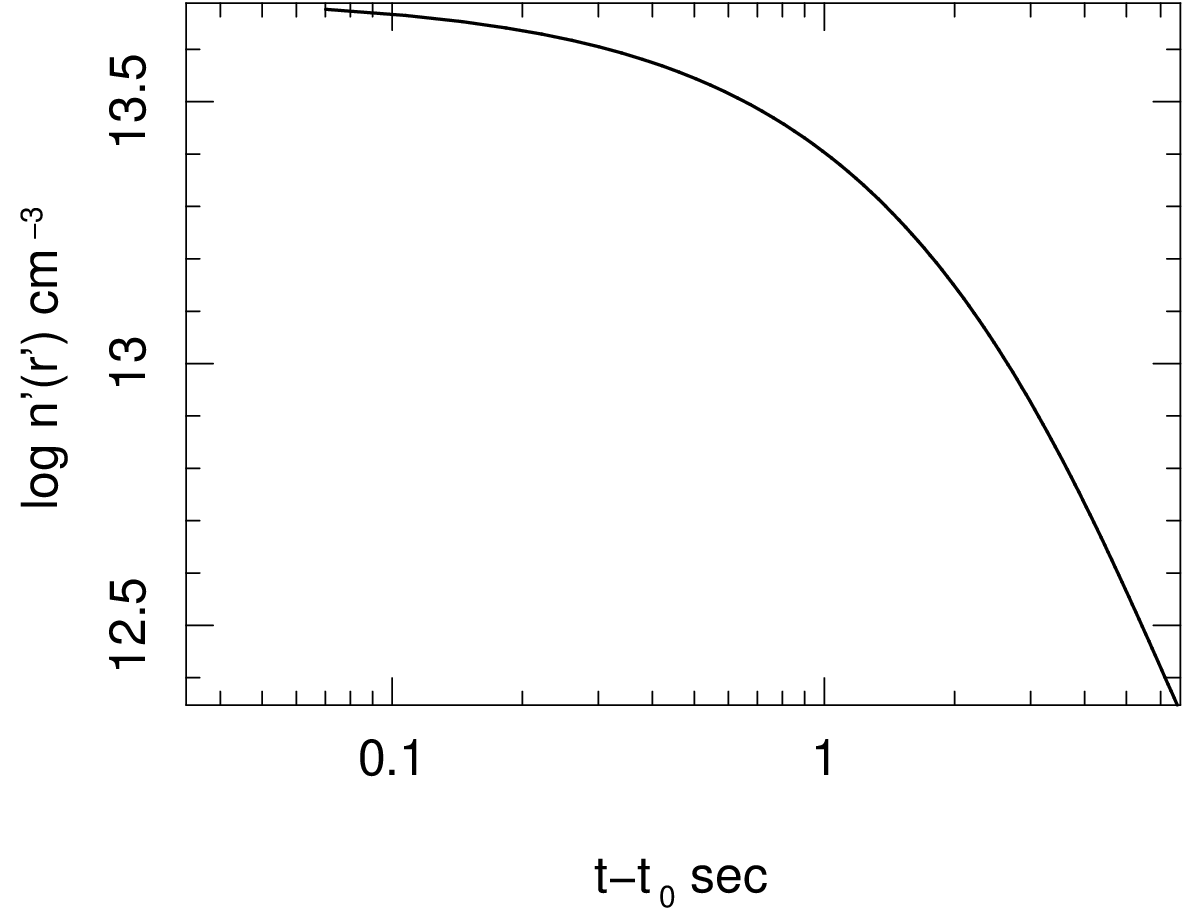} & 
7)\includegraphics[height=4cm,angle=-90]{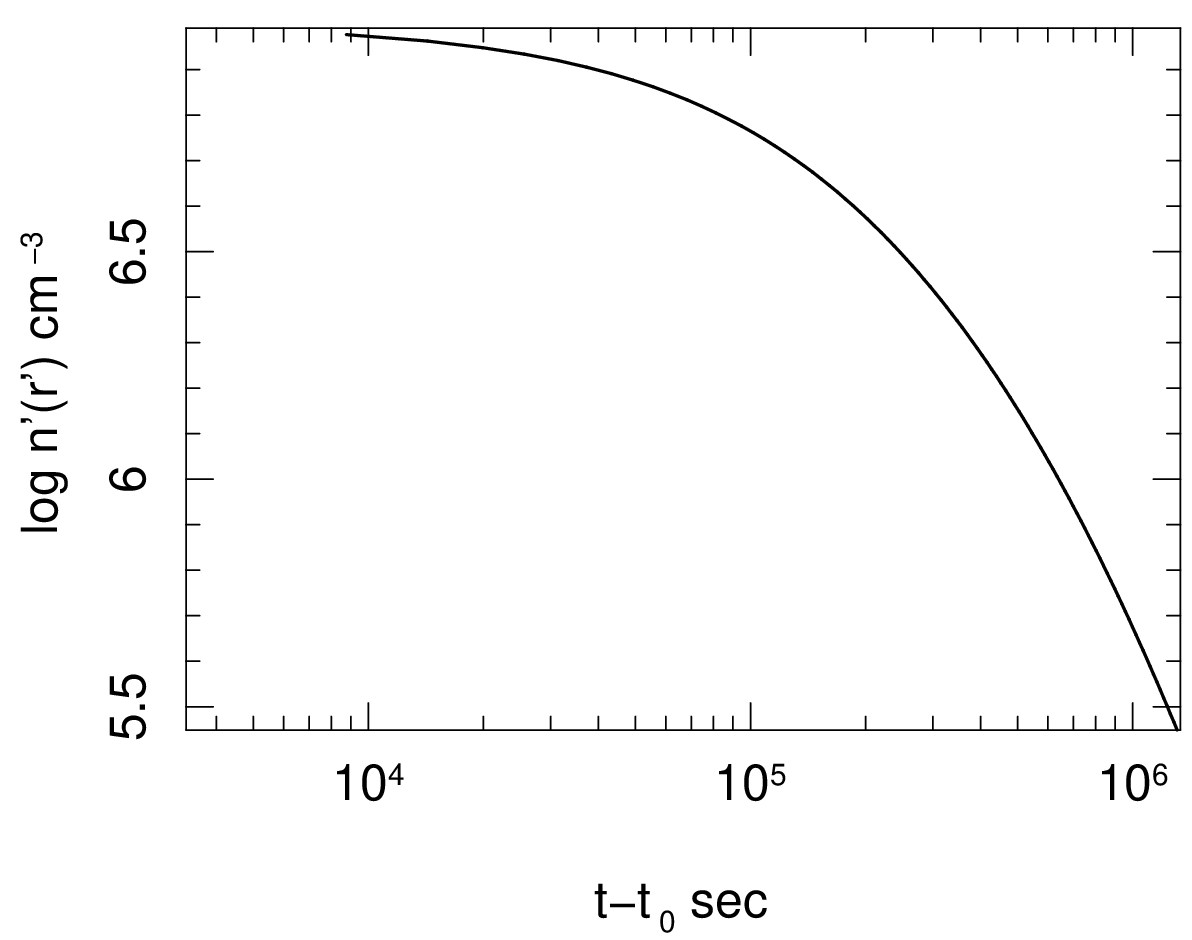} &
8)\includegraphics[height=4cm,angle=-90]{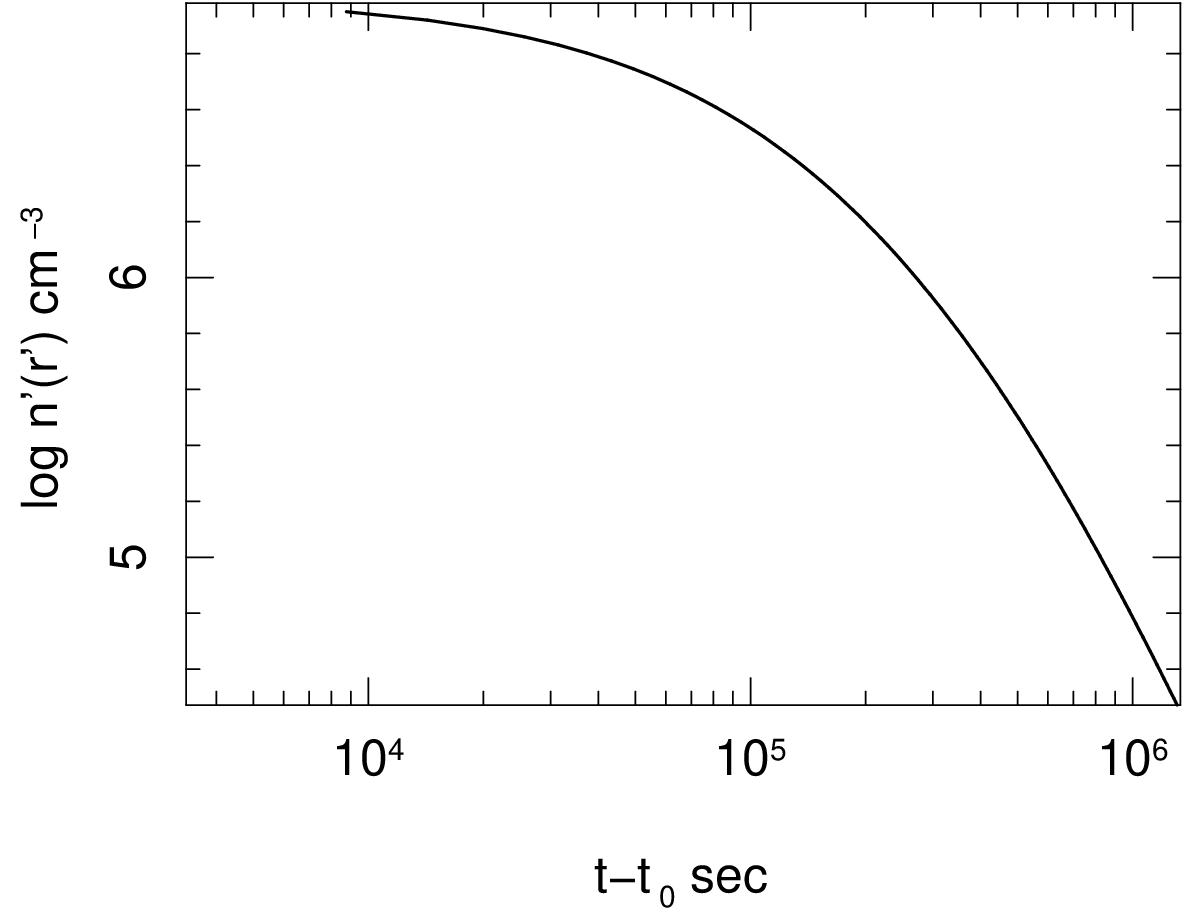} \\
9)\includegraphics[height=4cm,angle=-90]{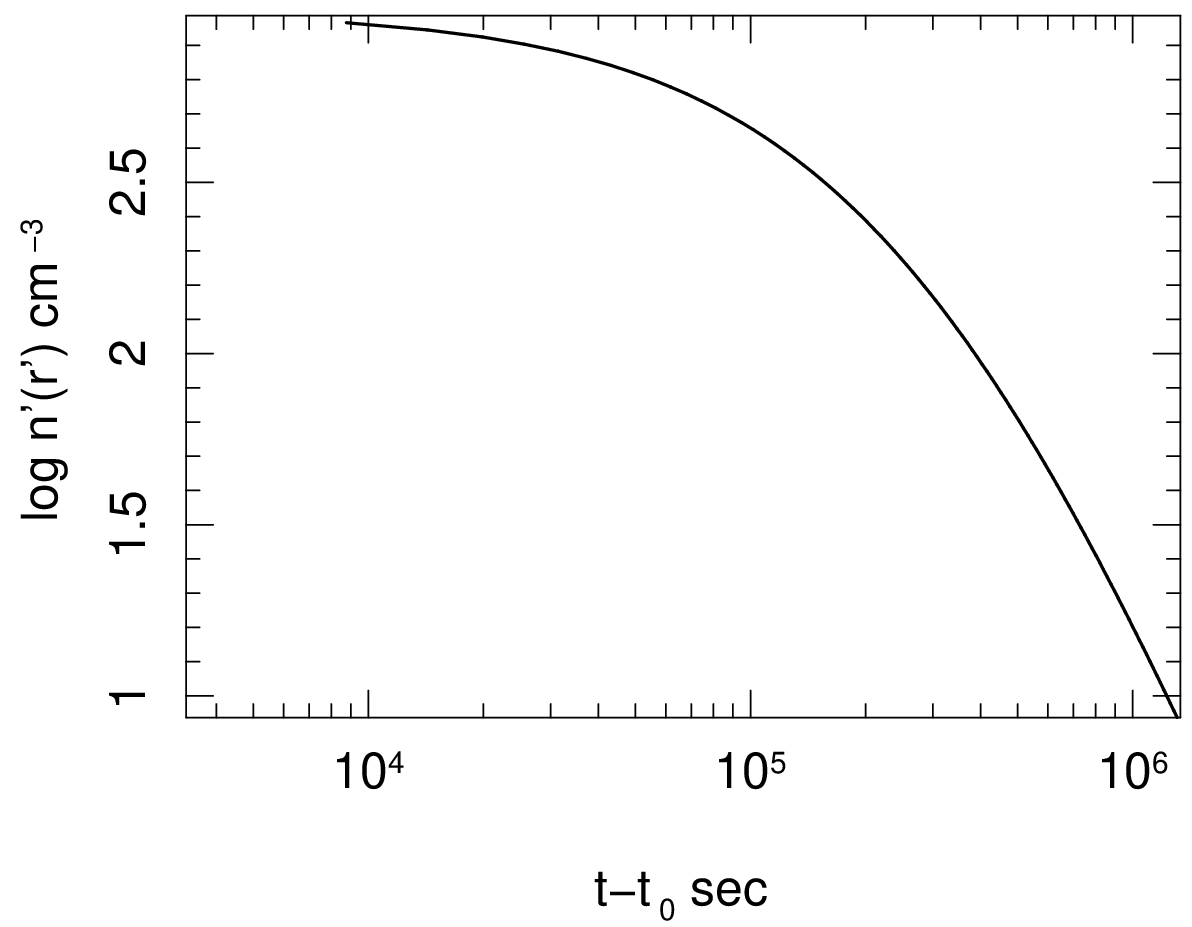} &
10)\includegraphics[height=4cm,angle=-90]{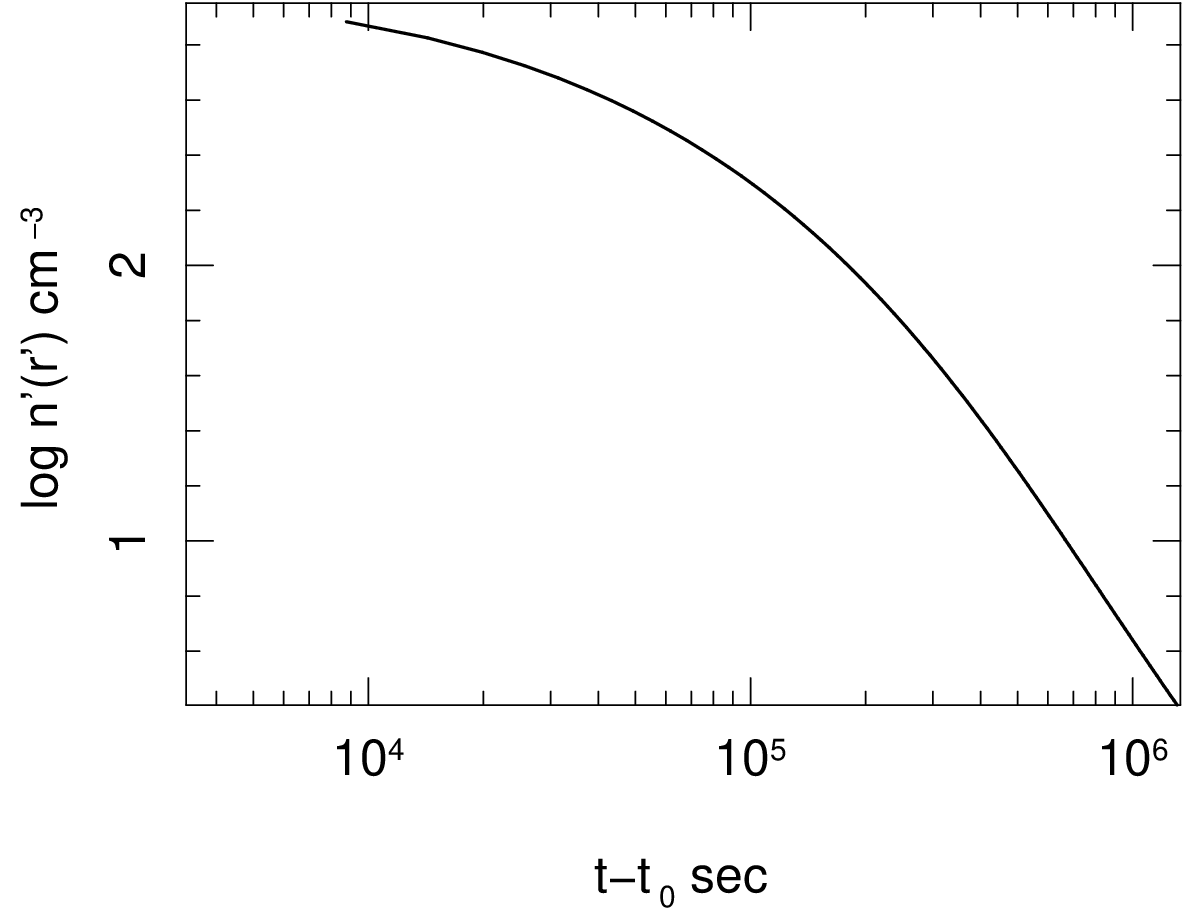} &
11)\includegraphics[height=4cm,angle=-90]{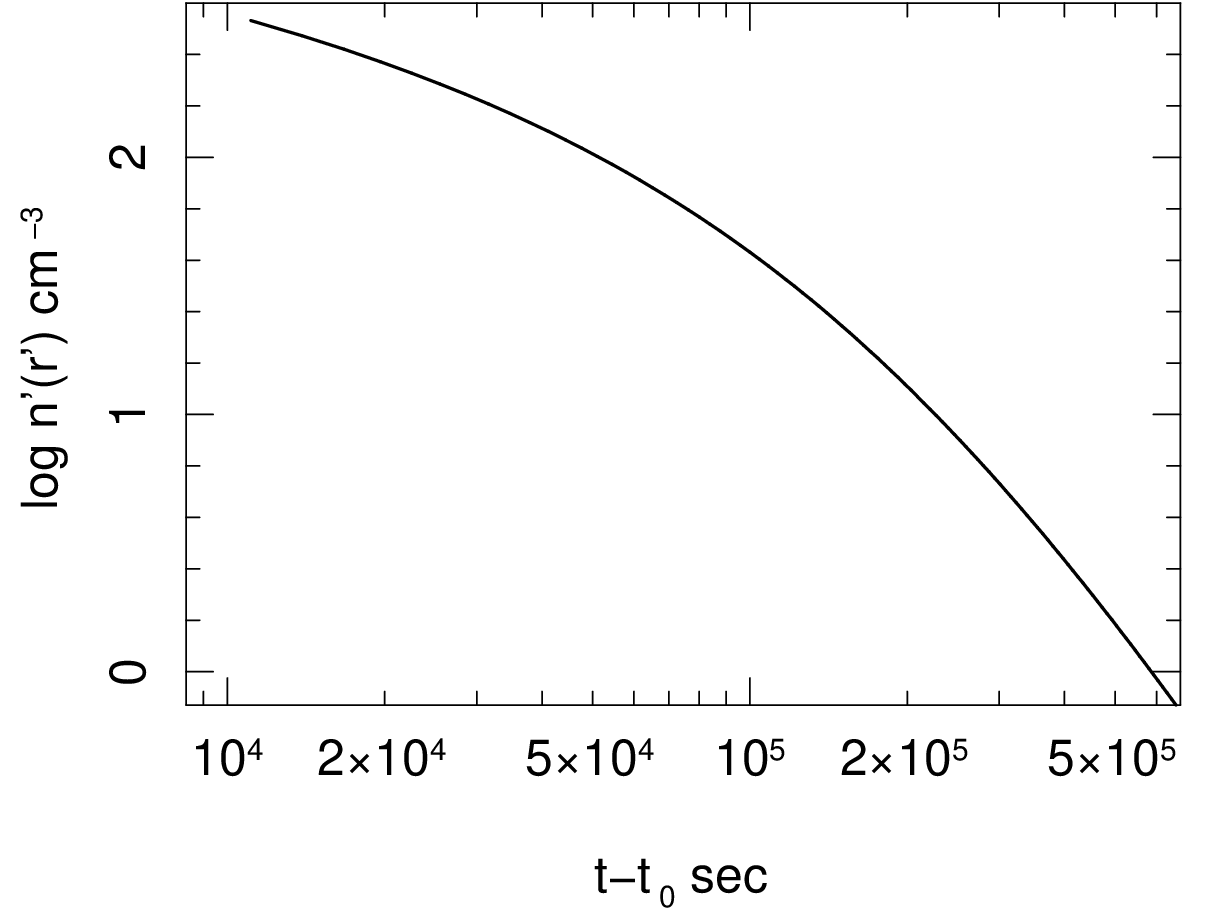} &
12)\includegraphics[height=4cm,angle=-90]{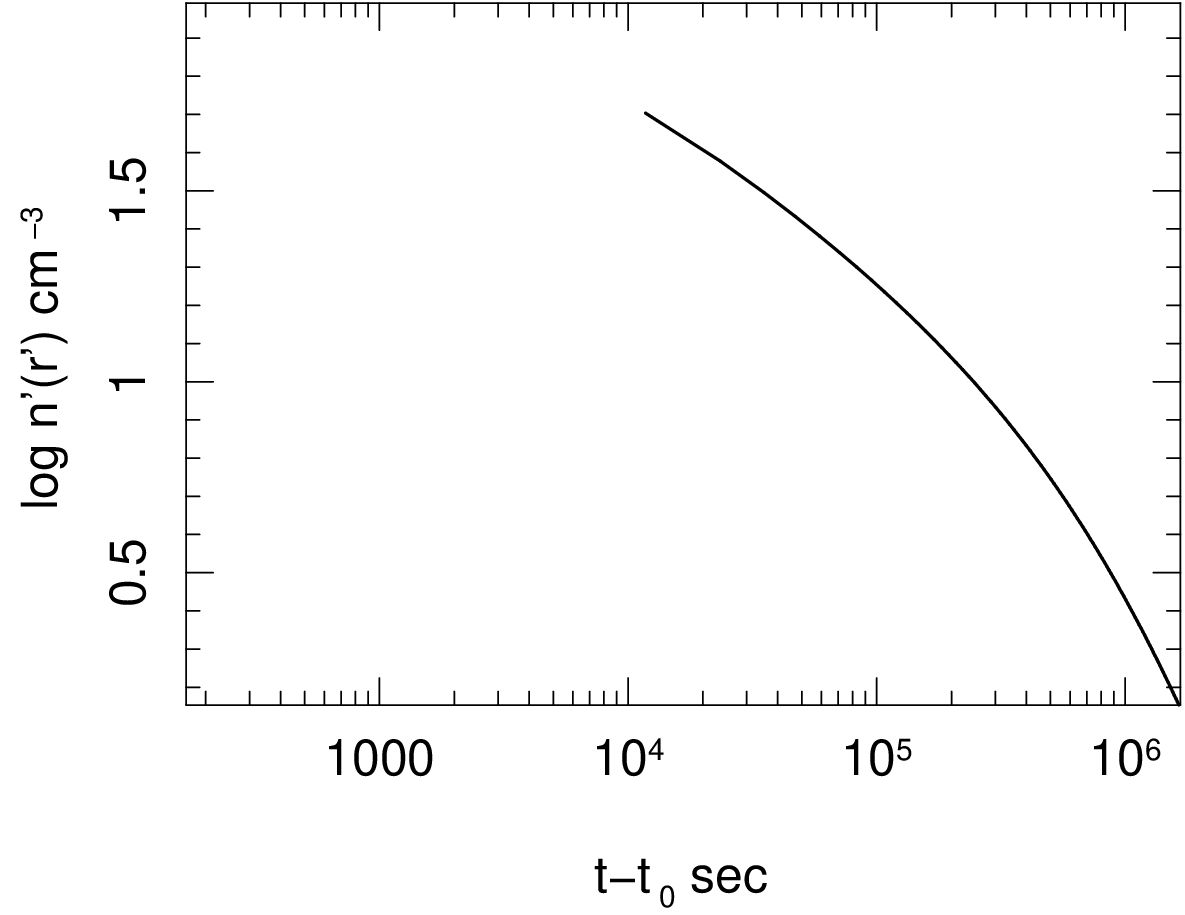} \\
13)\includegraphics[height=4cm,angle=-90]{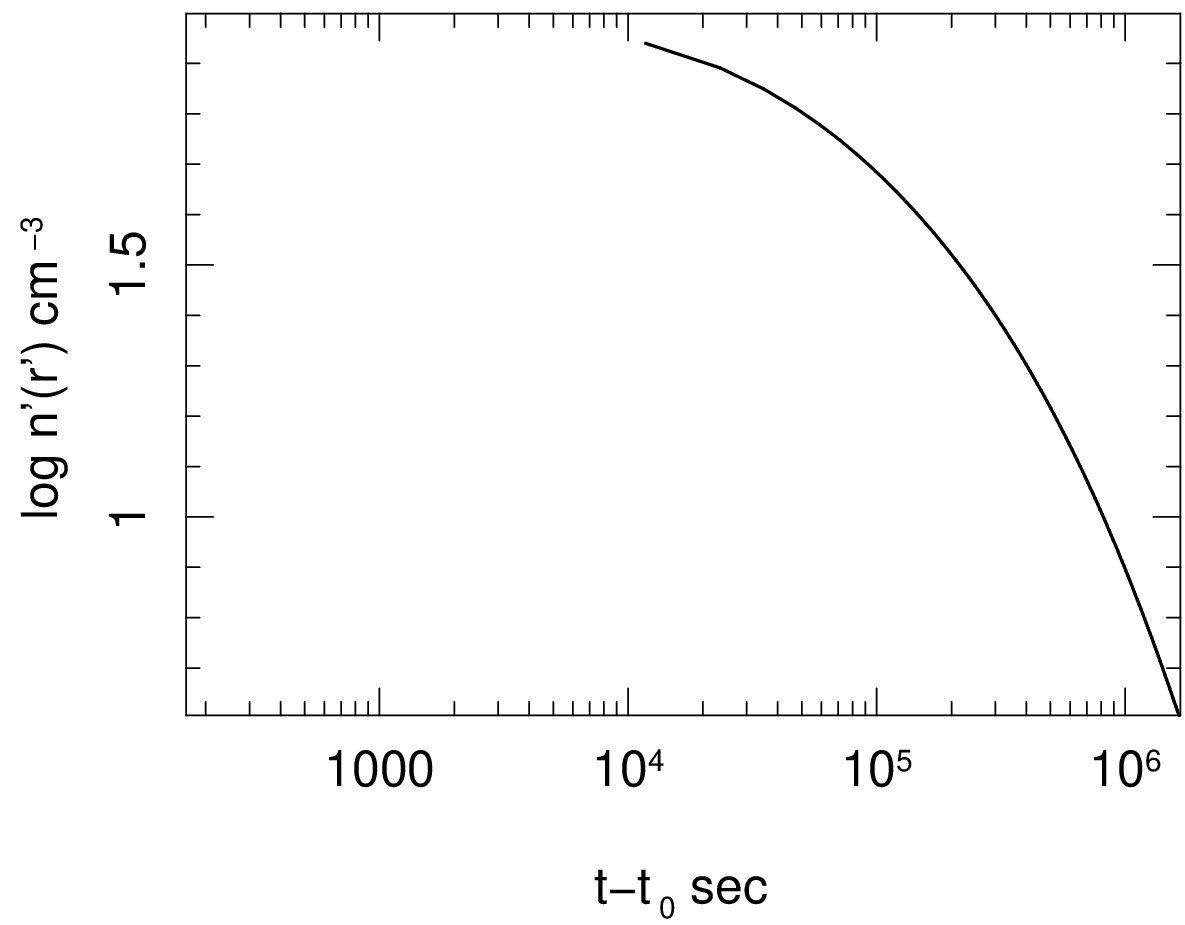} &
14)\includegraphics[height=4cm,angle=-90]{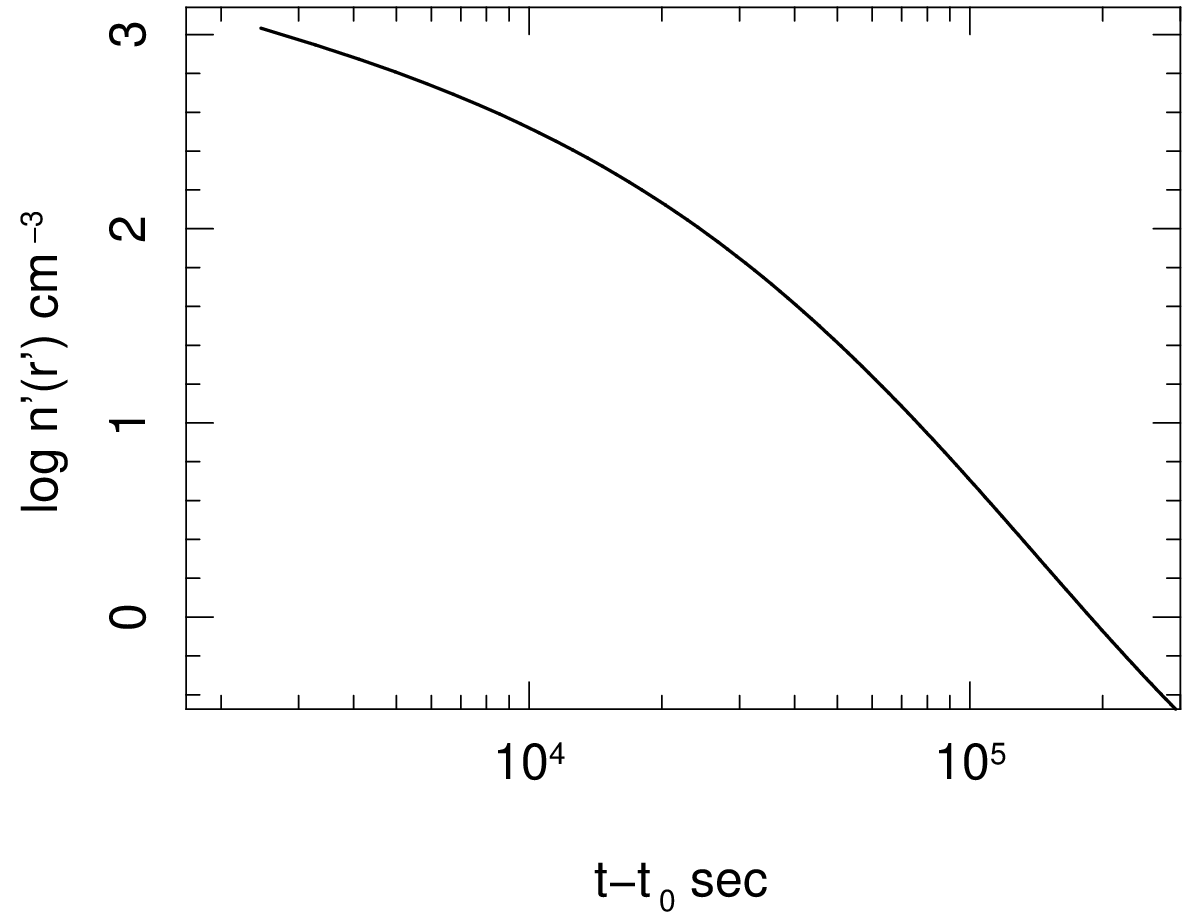} &
15)\includegraphics[height=4cm,angle=-90]{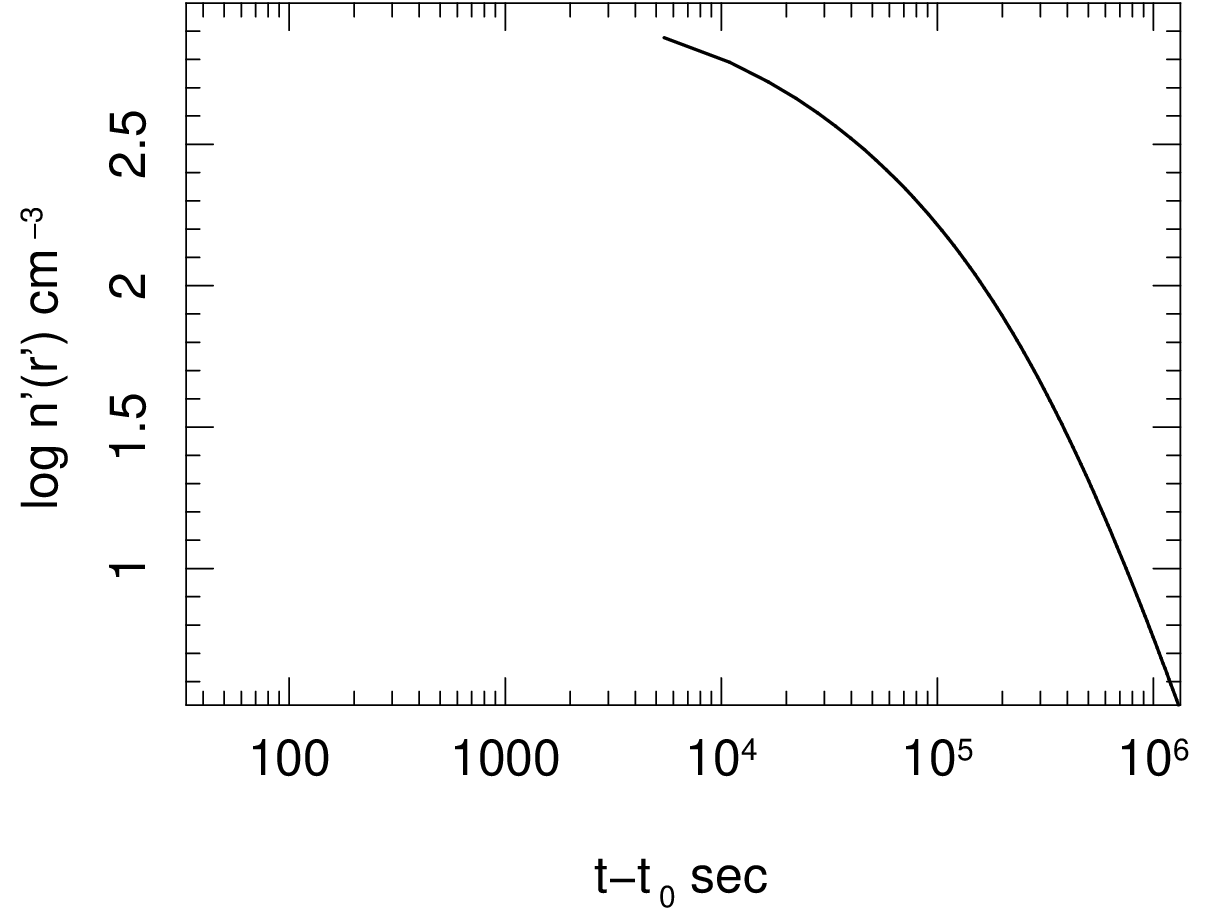} &
16)\includegraphics[height=4cm,angle=-90]{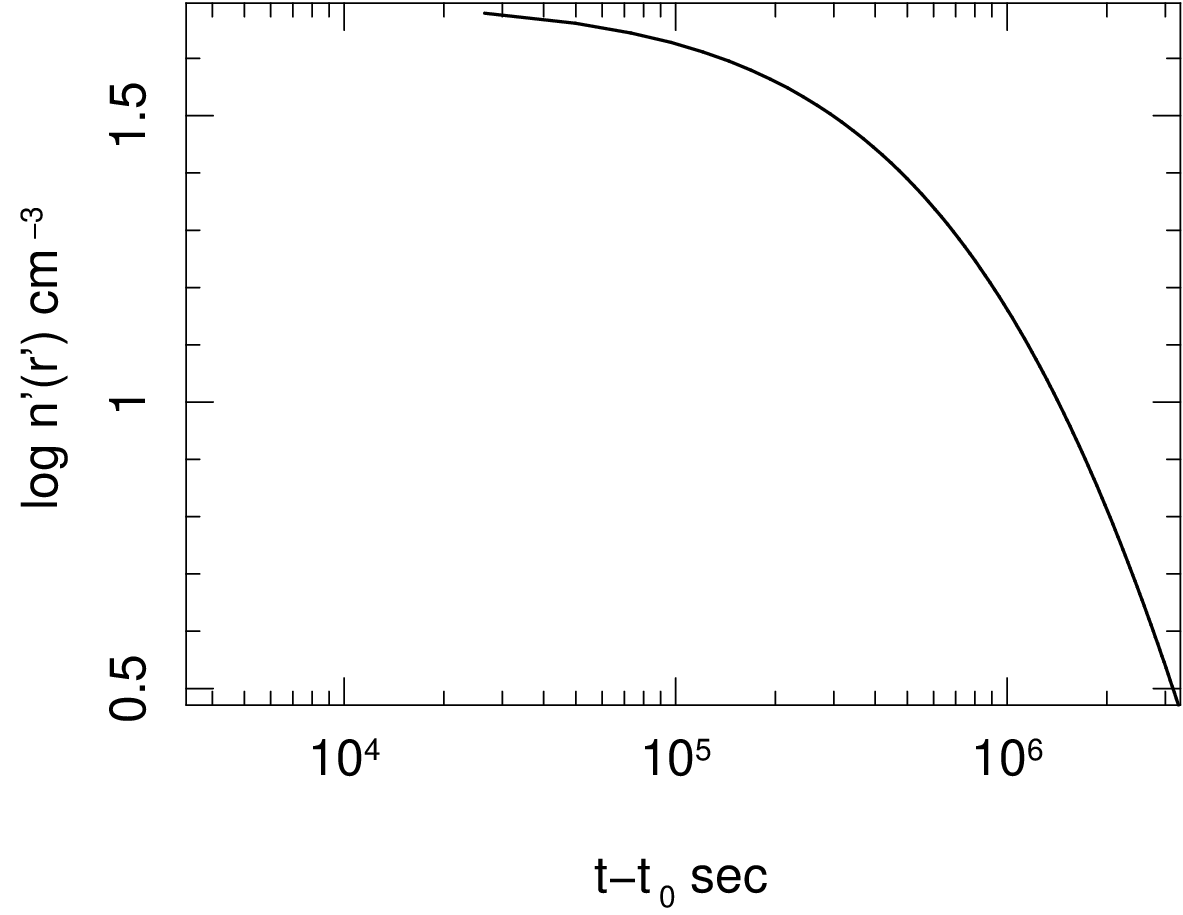} \\
17)\includegraphics[height=4cm,angle=-90]{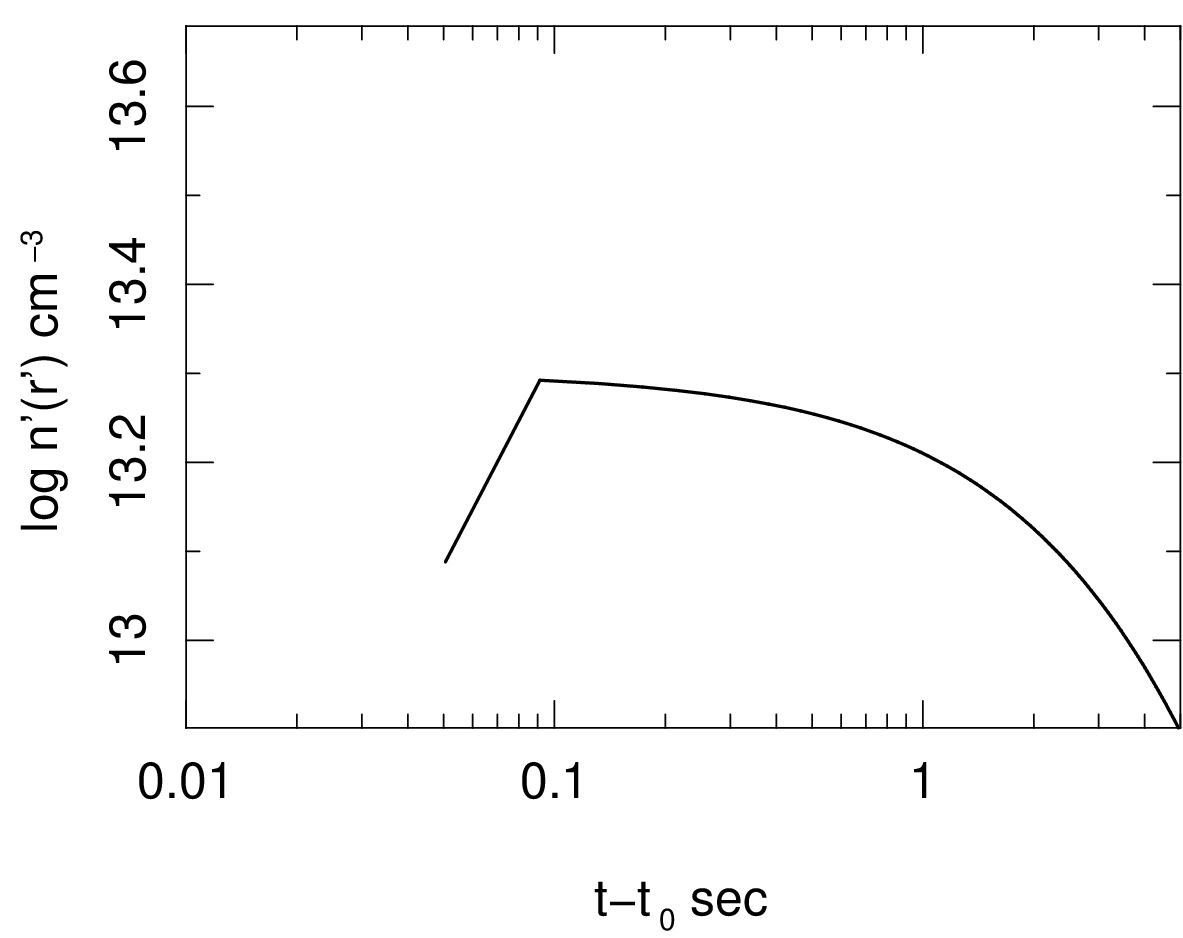} &
18)\includegraphics[height=4cm,angle=-90]{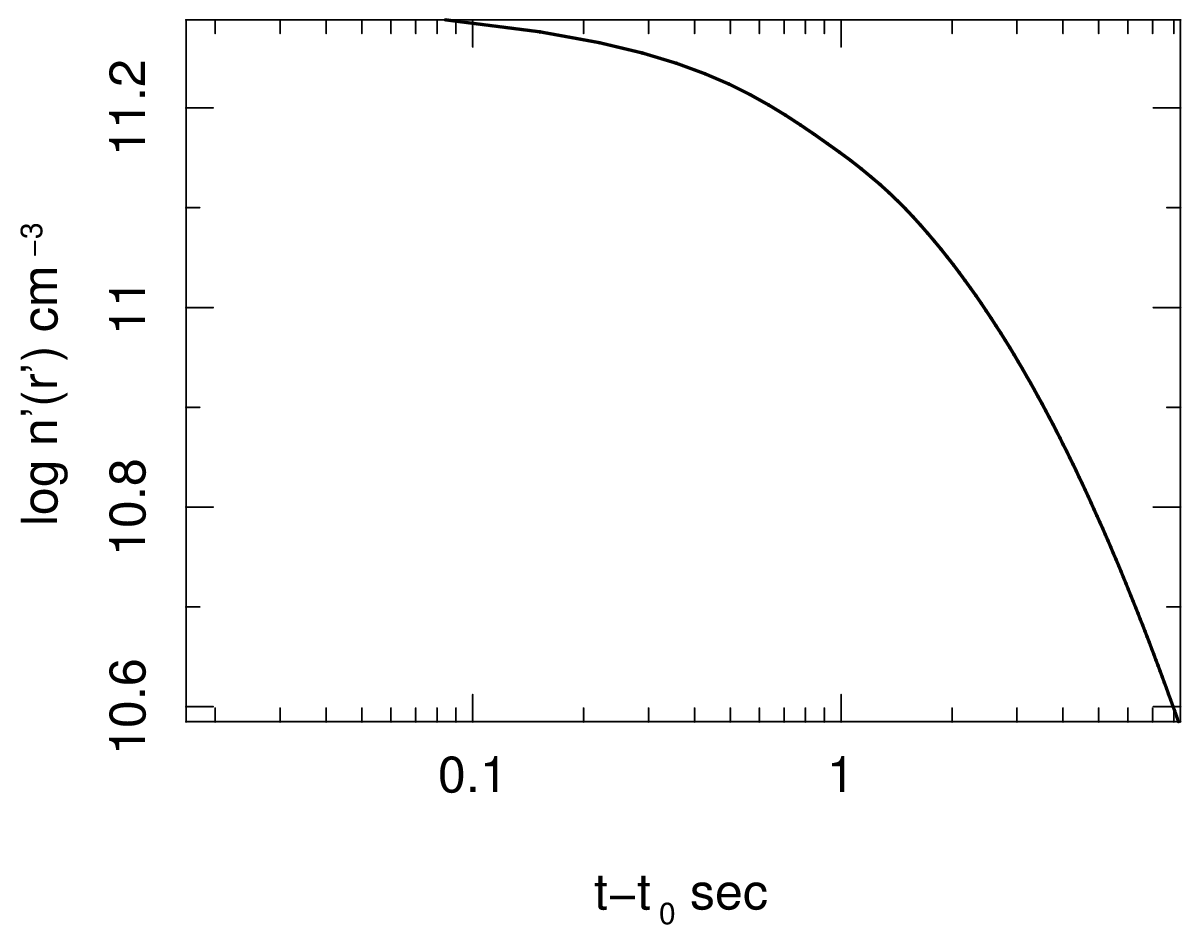} &
19)\includegraphics[height=4cm,angle=-90]{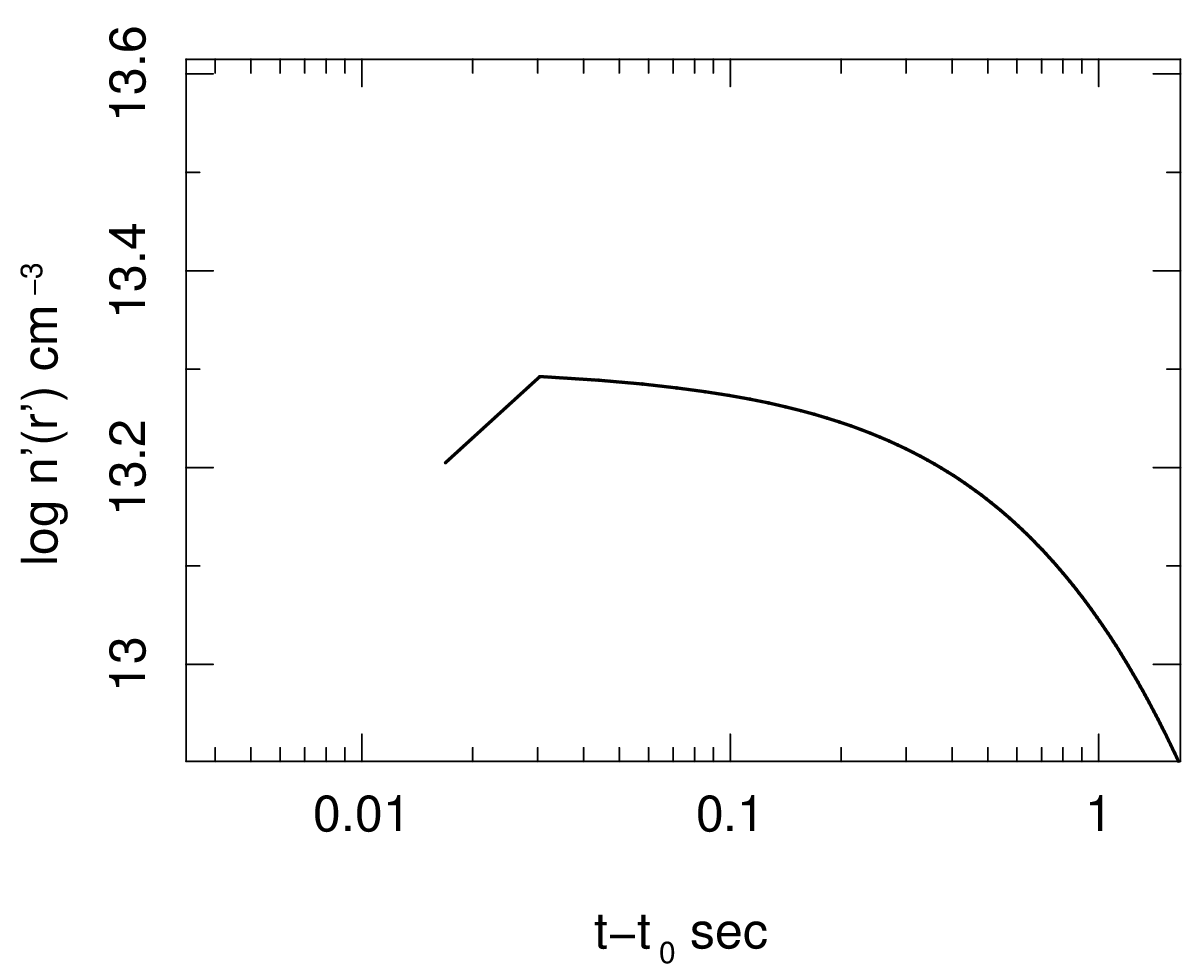} &
20)\includegraphics[height=4cm,angle=-90]{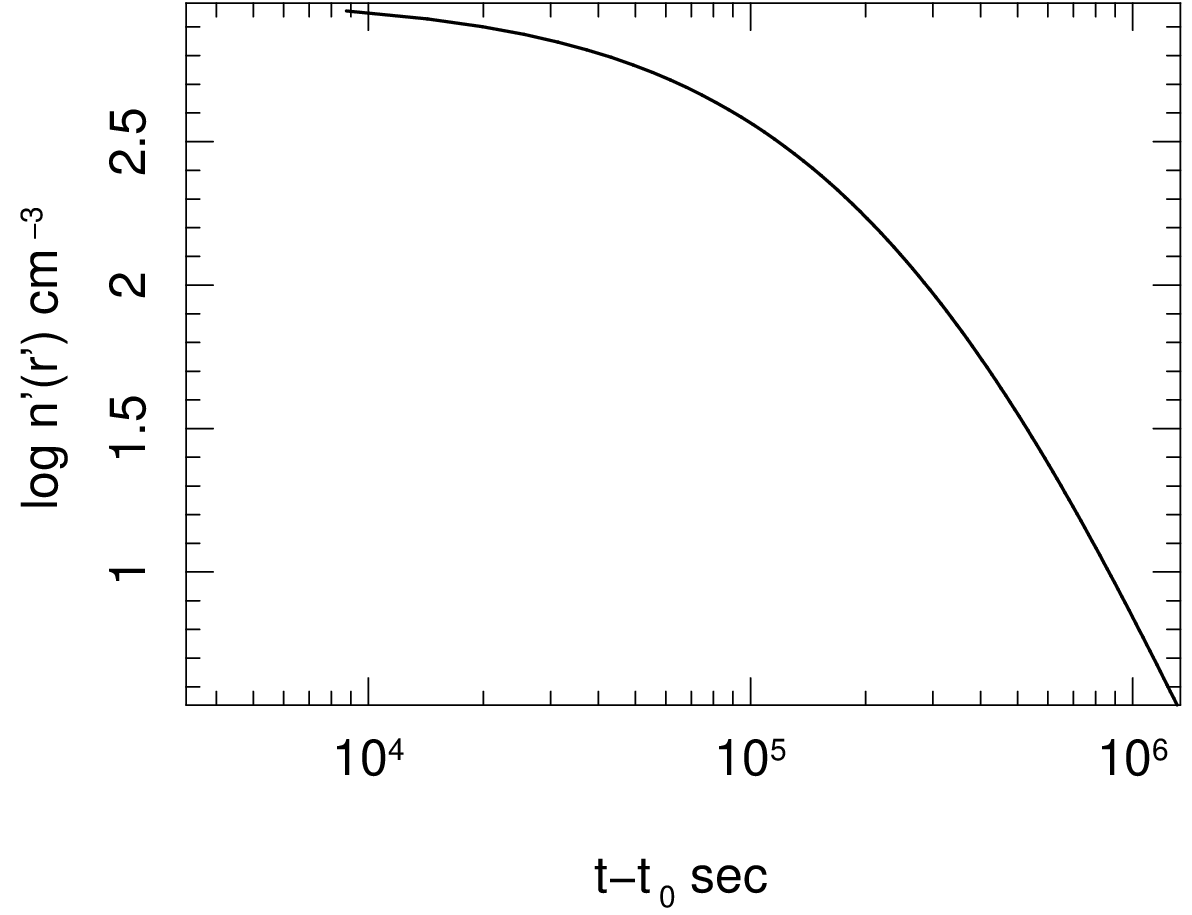} \\
21)\includegraphics[height=4cm,angle=-90]{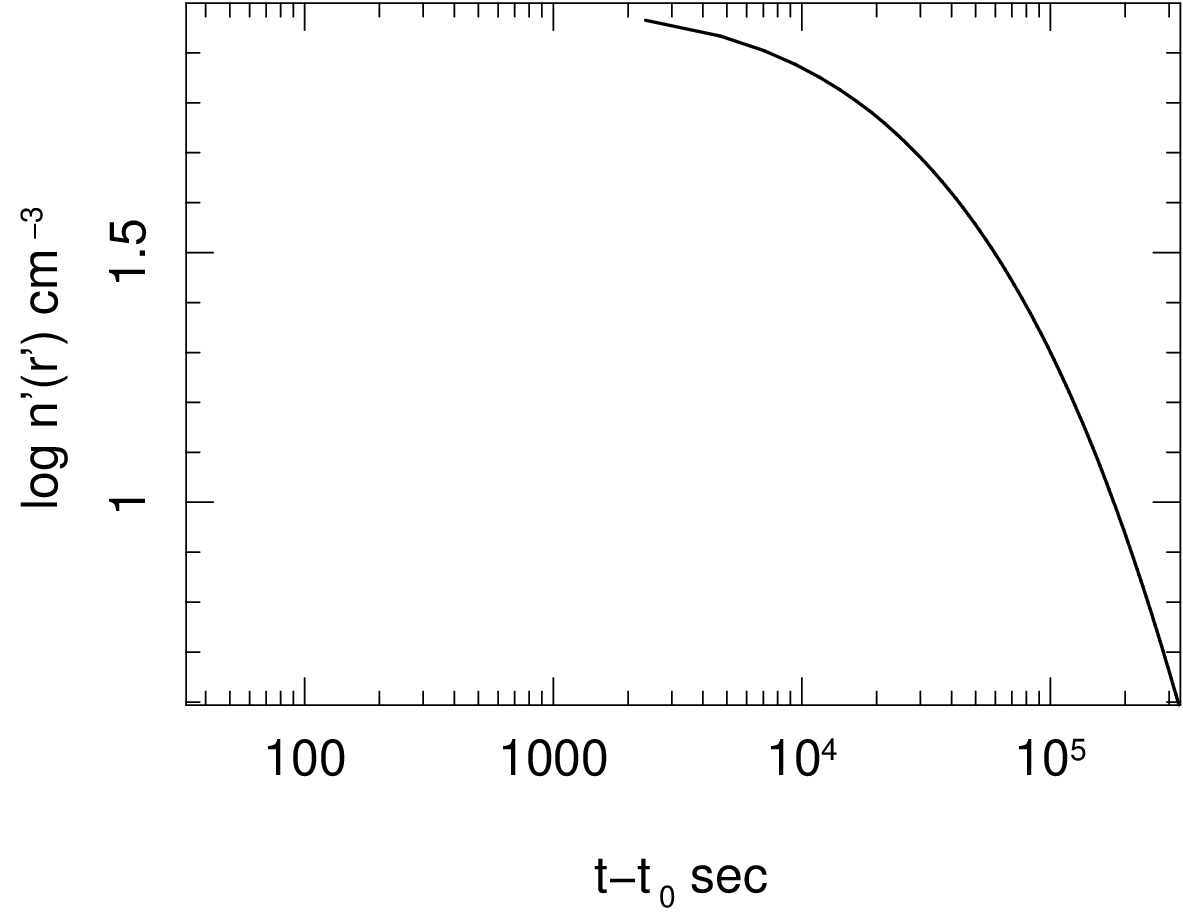} &
22)\includegraphics[height=4cm,angle=-90]{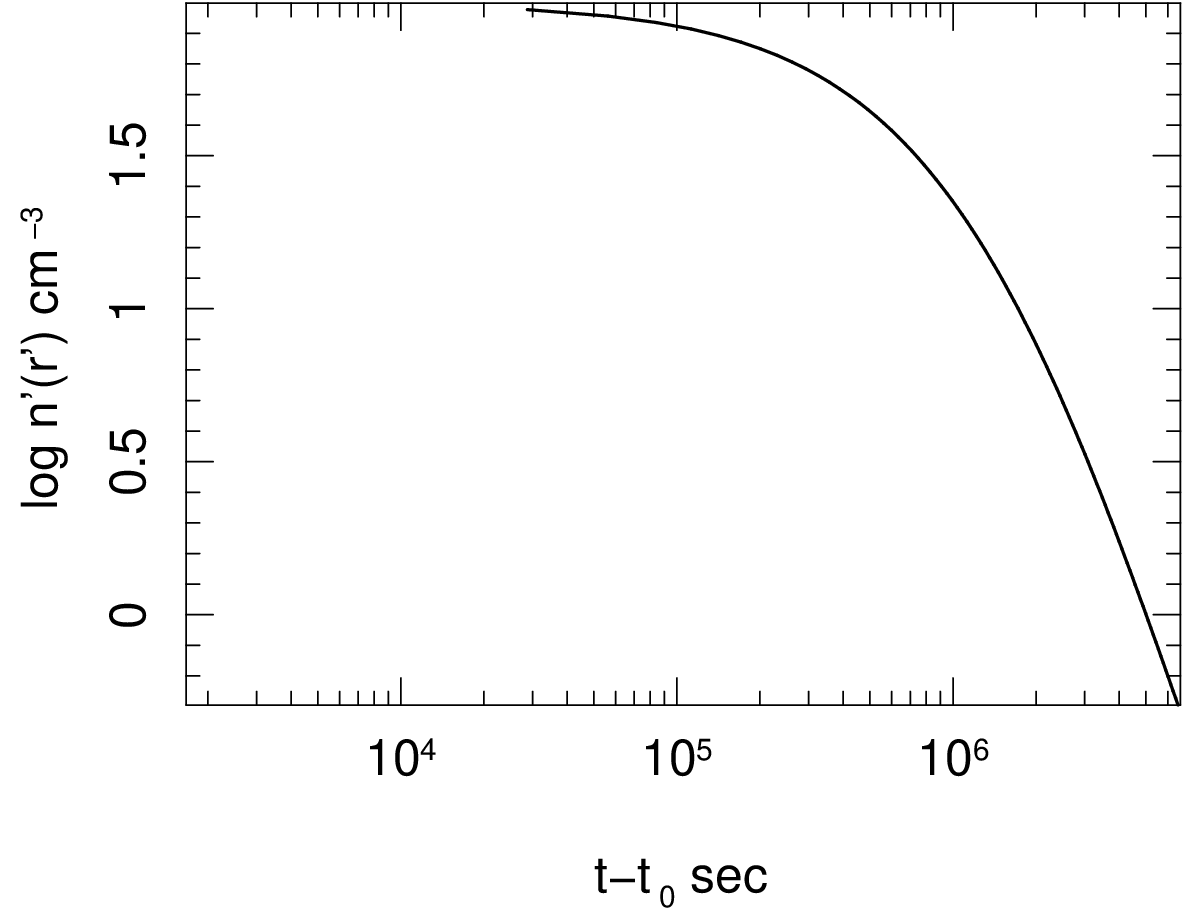} &
23)\includegraphics[height=4cm,angle=-90]{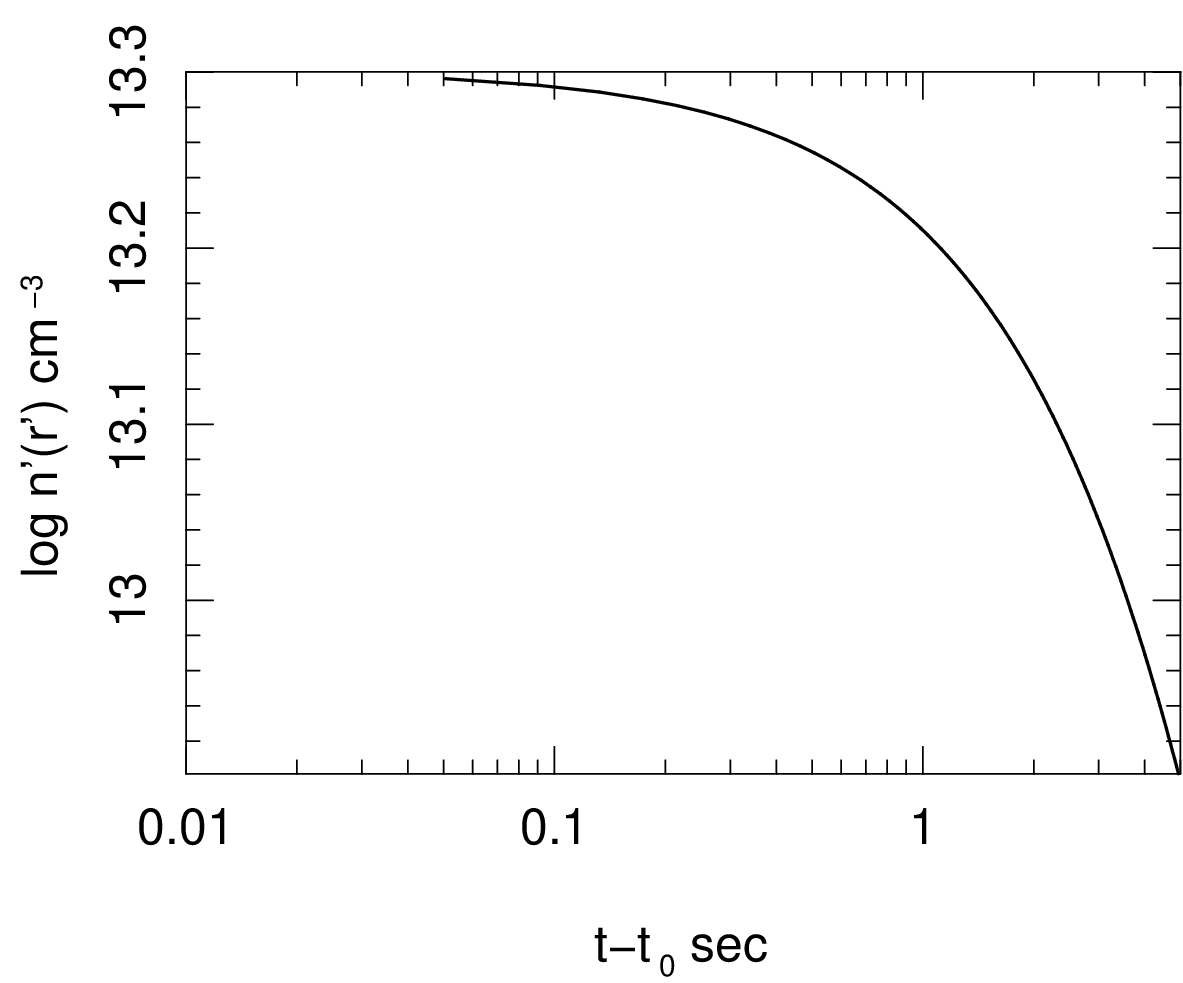} &
24)\includegraphics[height=4cm,angle=-90]{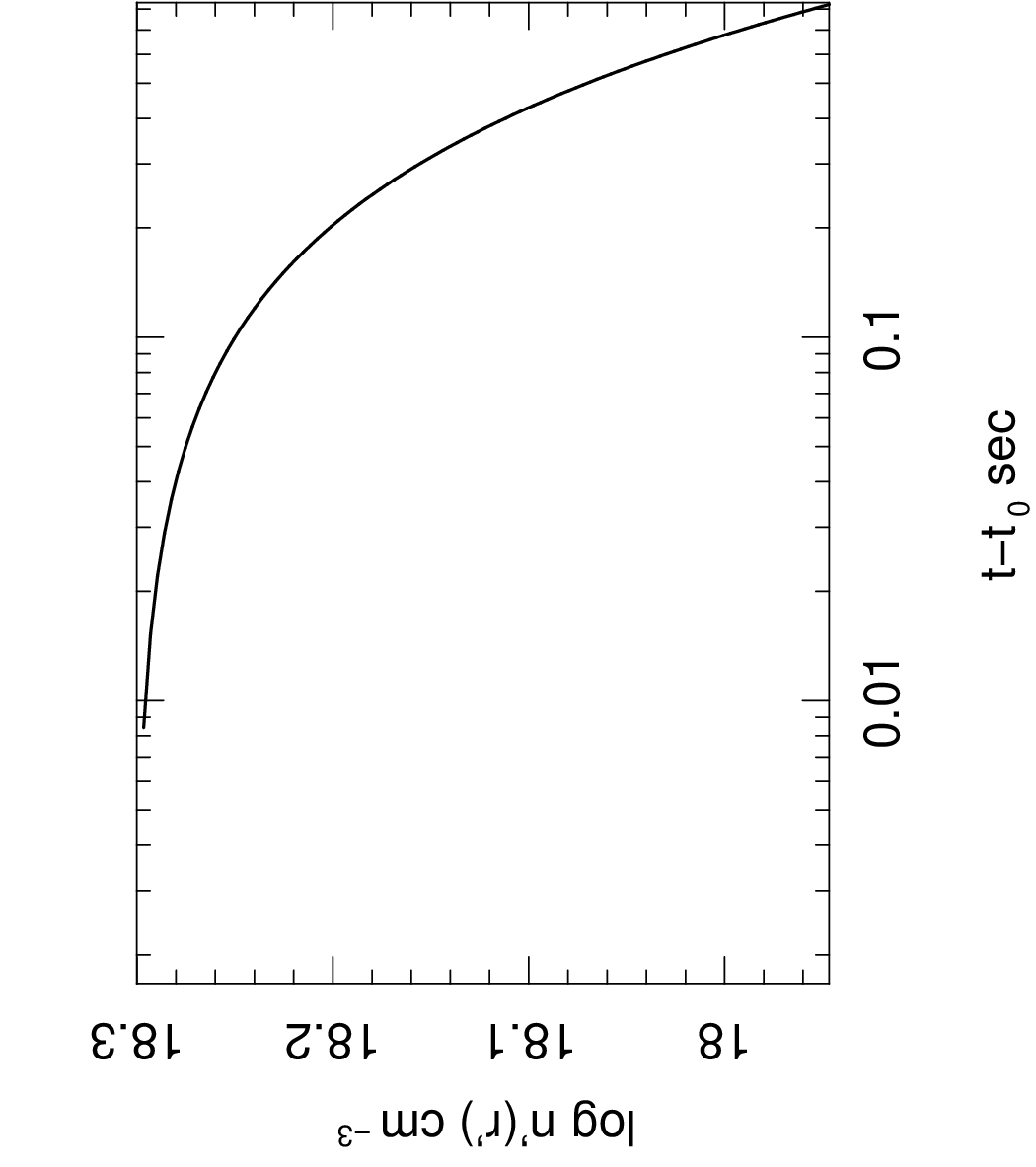} 
\end {tabular}
\end {center}
\caption{Evolution of the active region density $n'$ with time/radius. 
Definitions of curves are the same as Fig. \ref{fig:betazero}.
\label{fig:nprim}}
\end{figure*}

\begin{figure*}
\begin{center}
\begin {tabular}{cccc}
1)\includegraphics[height=4cm,angle=-90]{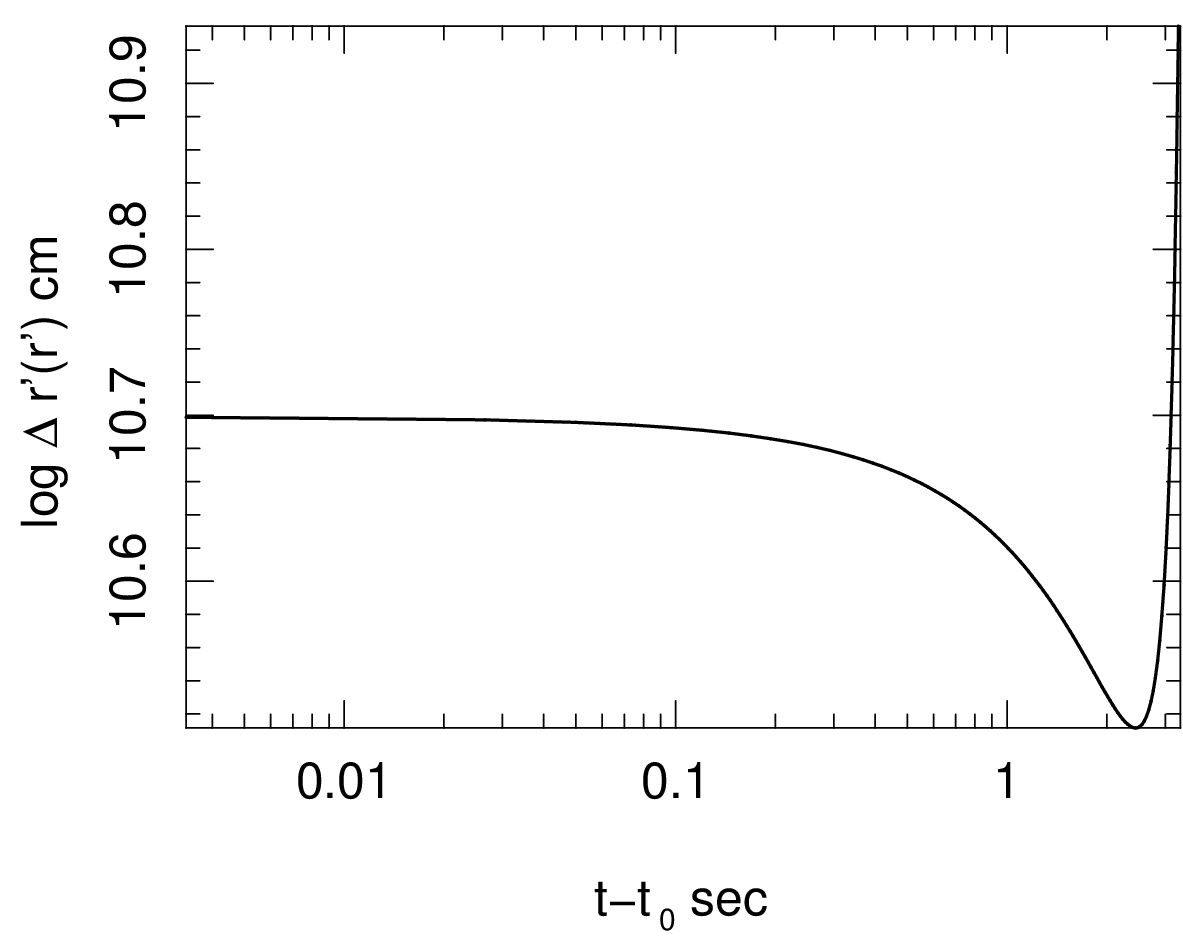} &
2)\includegraphics[height=4cm,angle=-90]{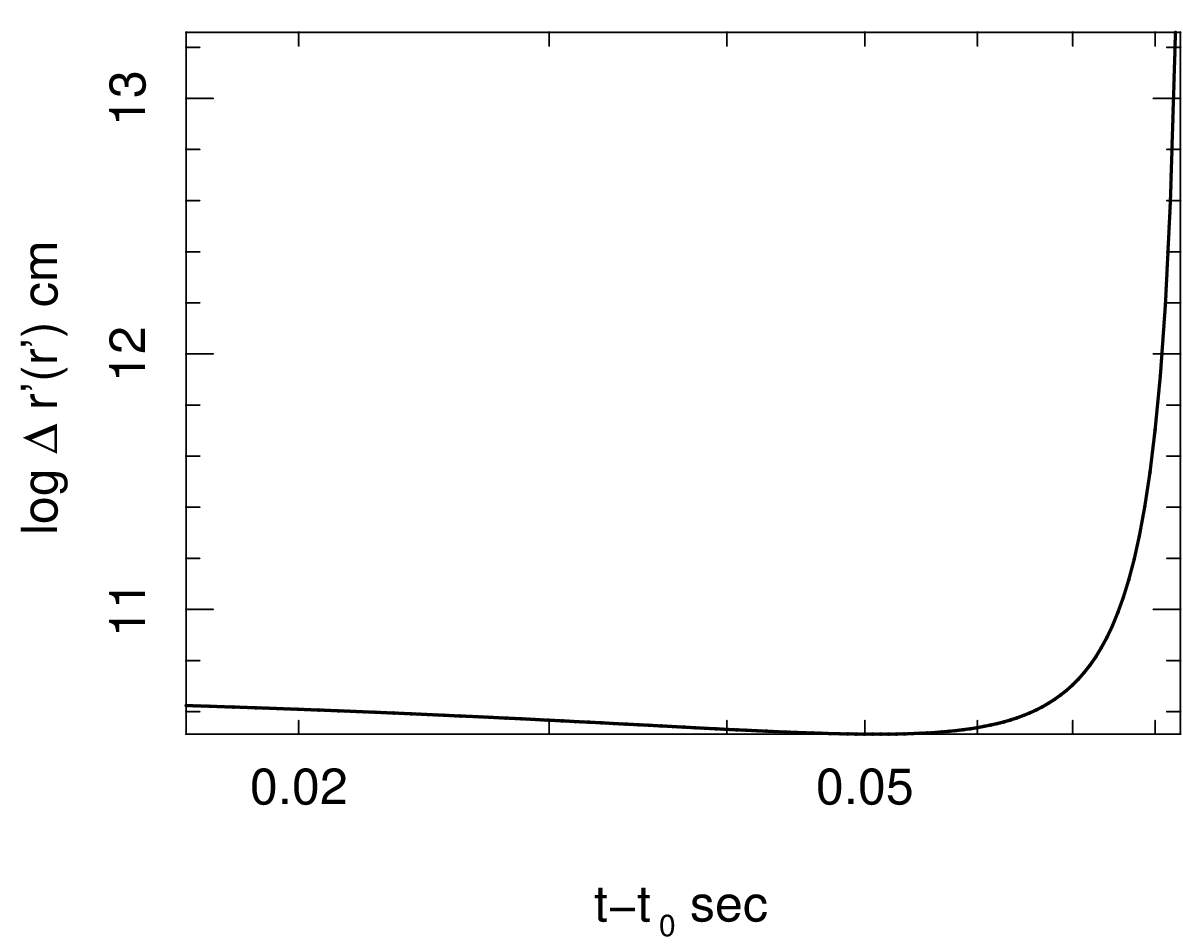} & 
3)\includegraphics[height=4cm,angle=-90]{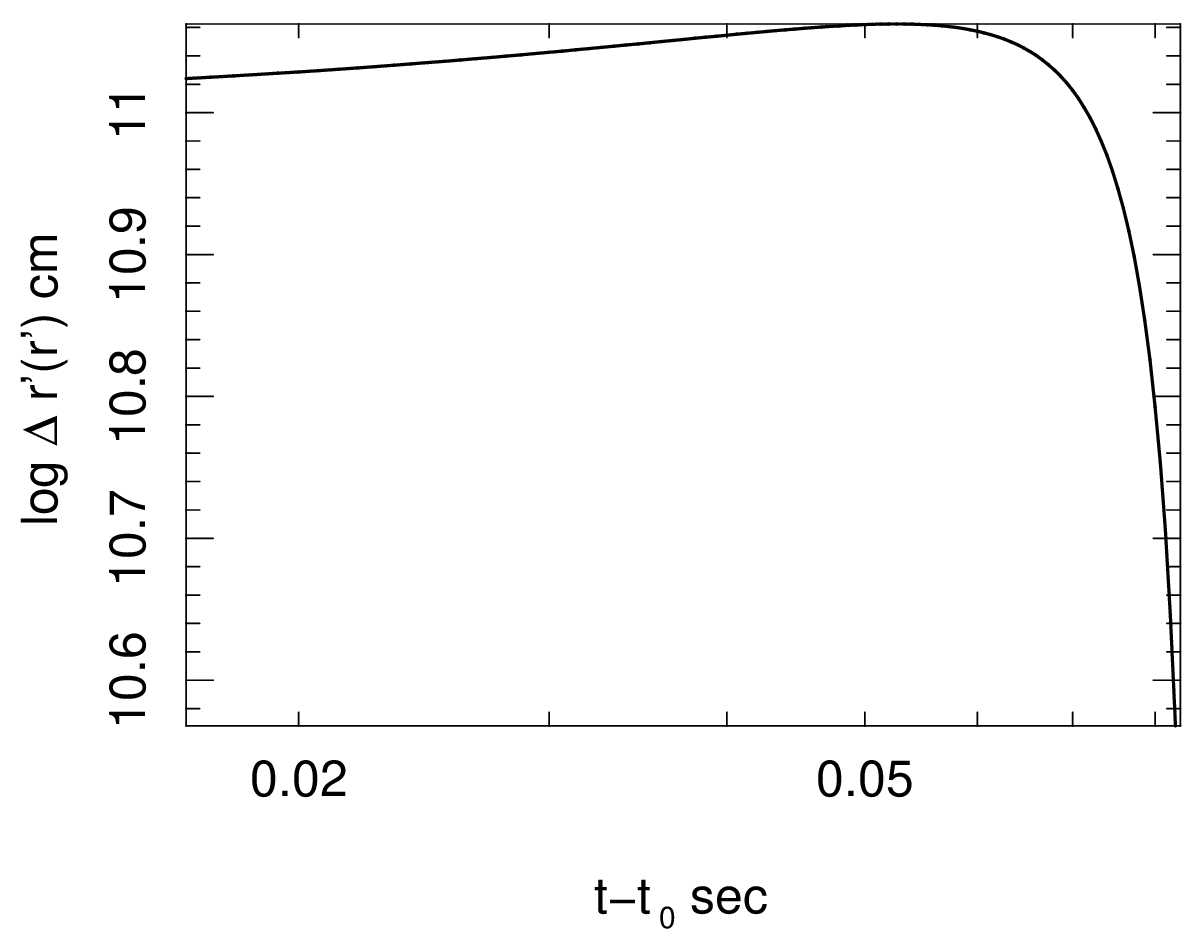} &
4)\includegraphics[height=4cm,angle=-90]{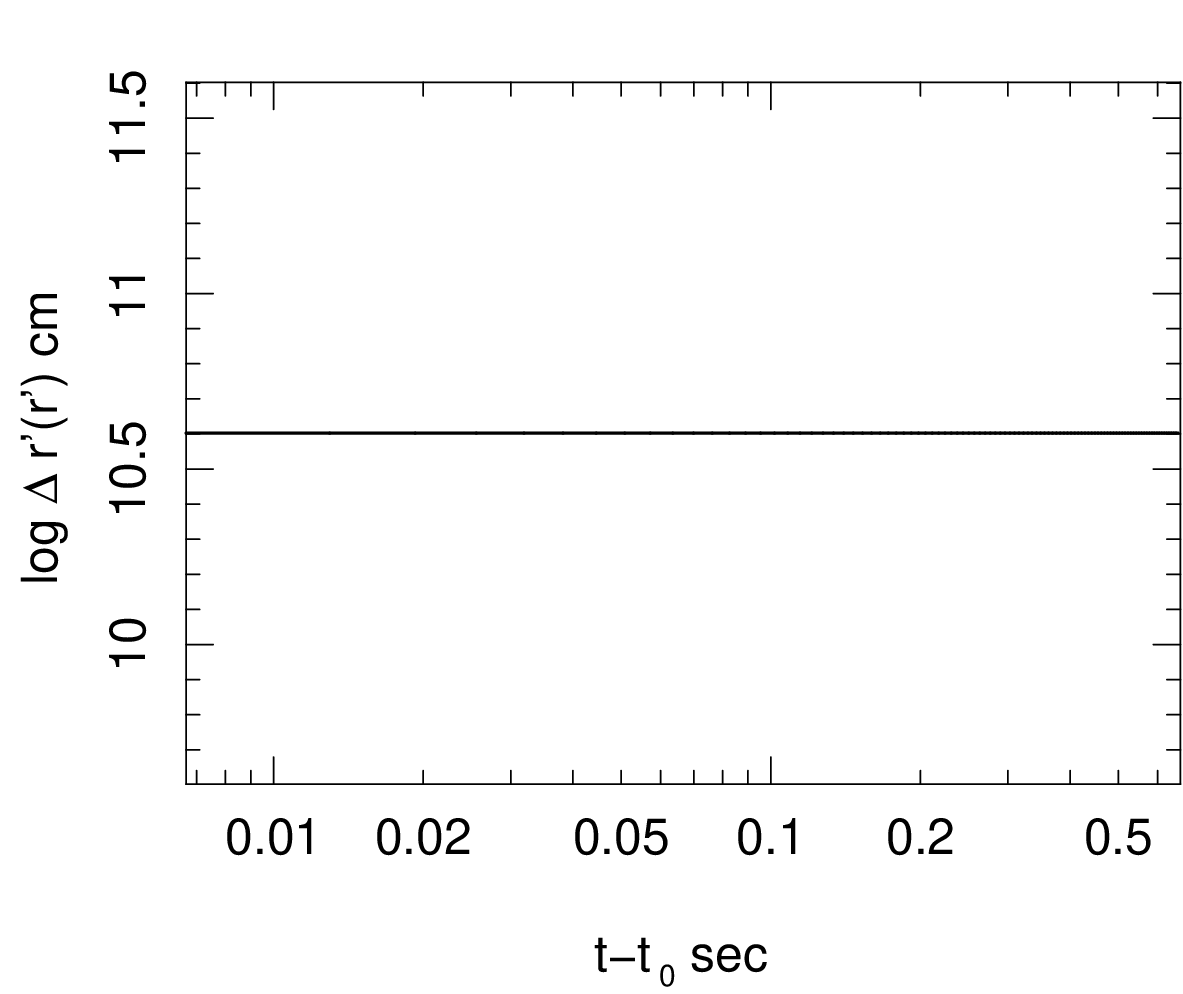} \\ 
5)\includegraphics[height=4cm,angle=-90]{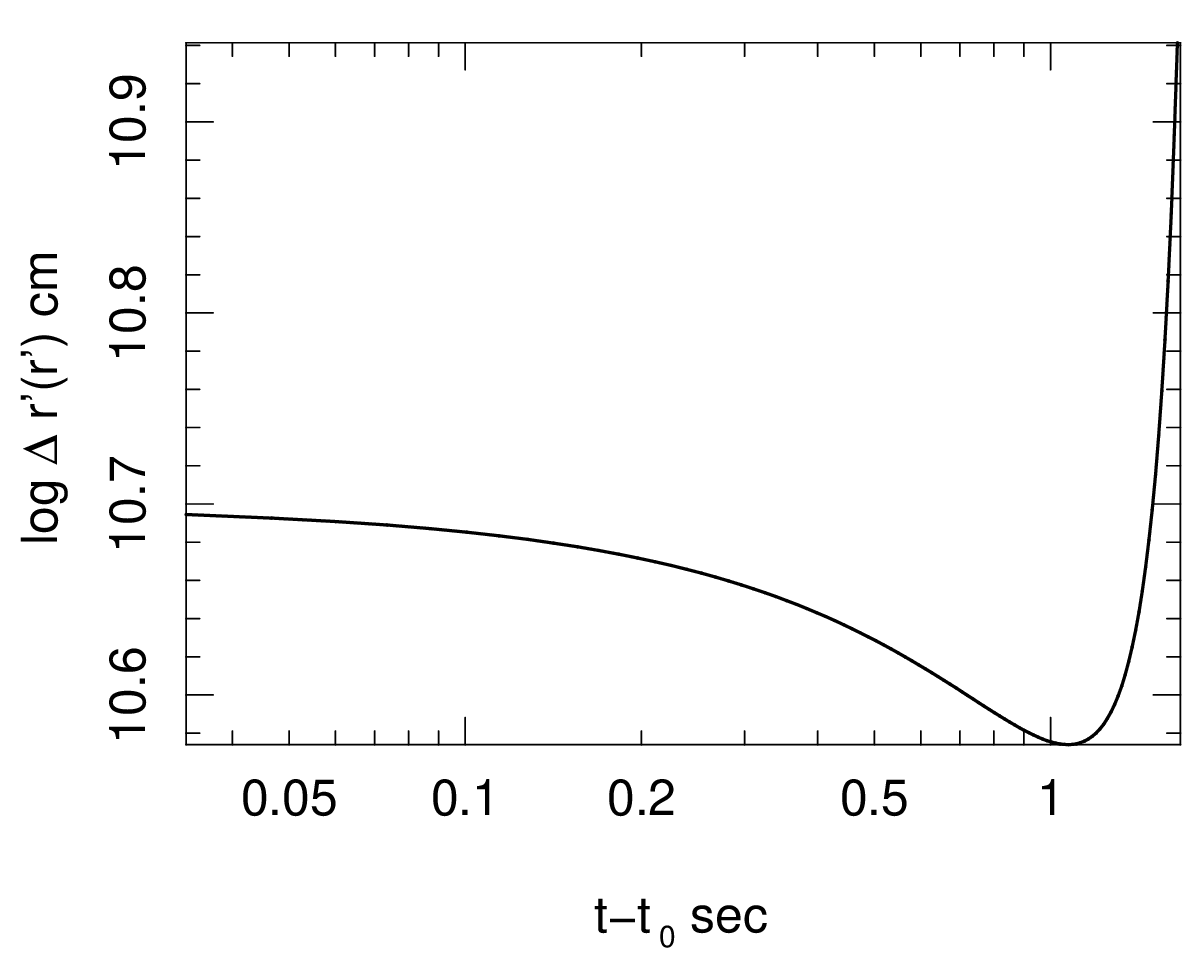} &
6)\includegraphics[height=4cm,angle=-90]{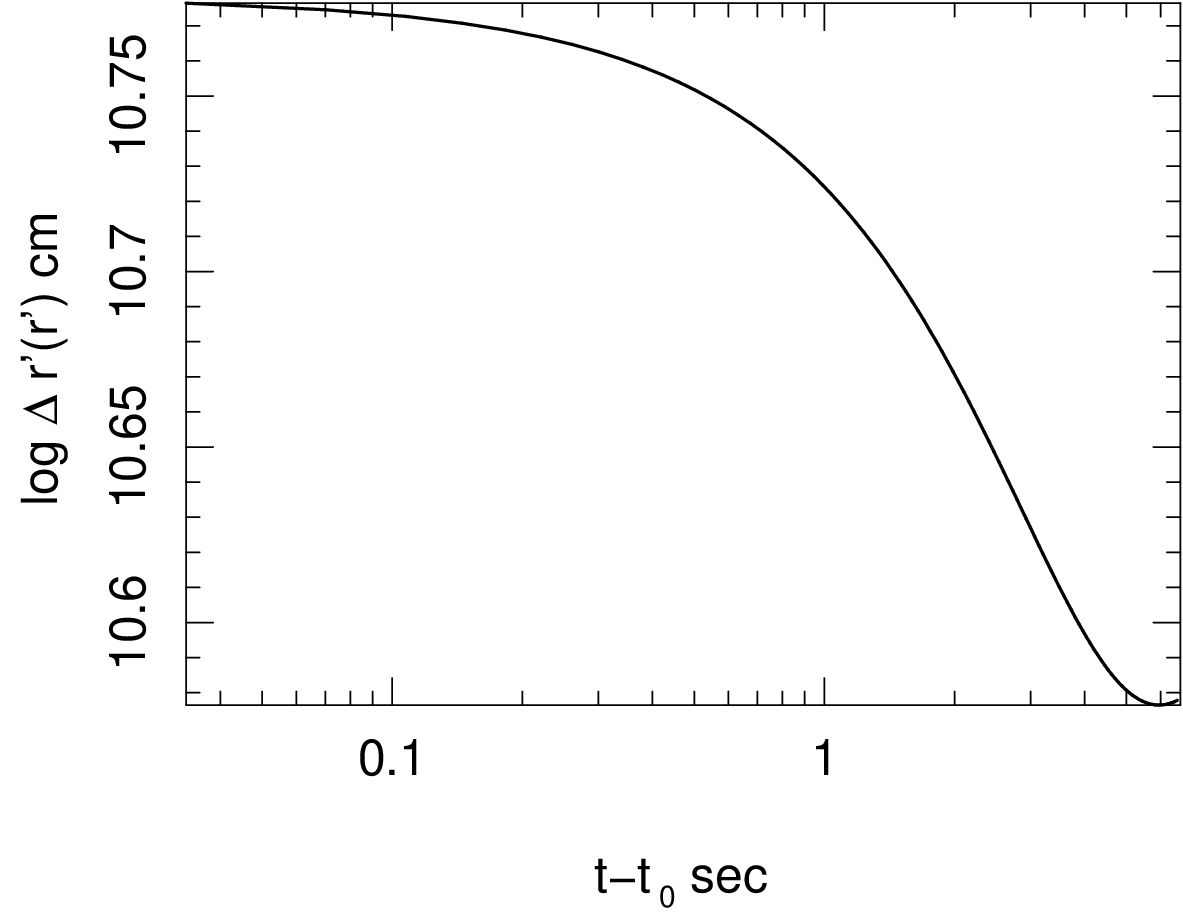} & 
7)\includegraphics[height=4cm,angle=-90]{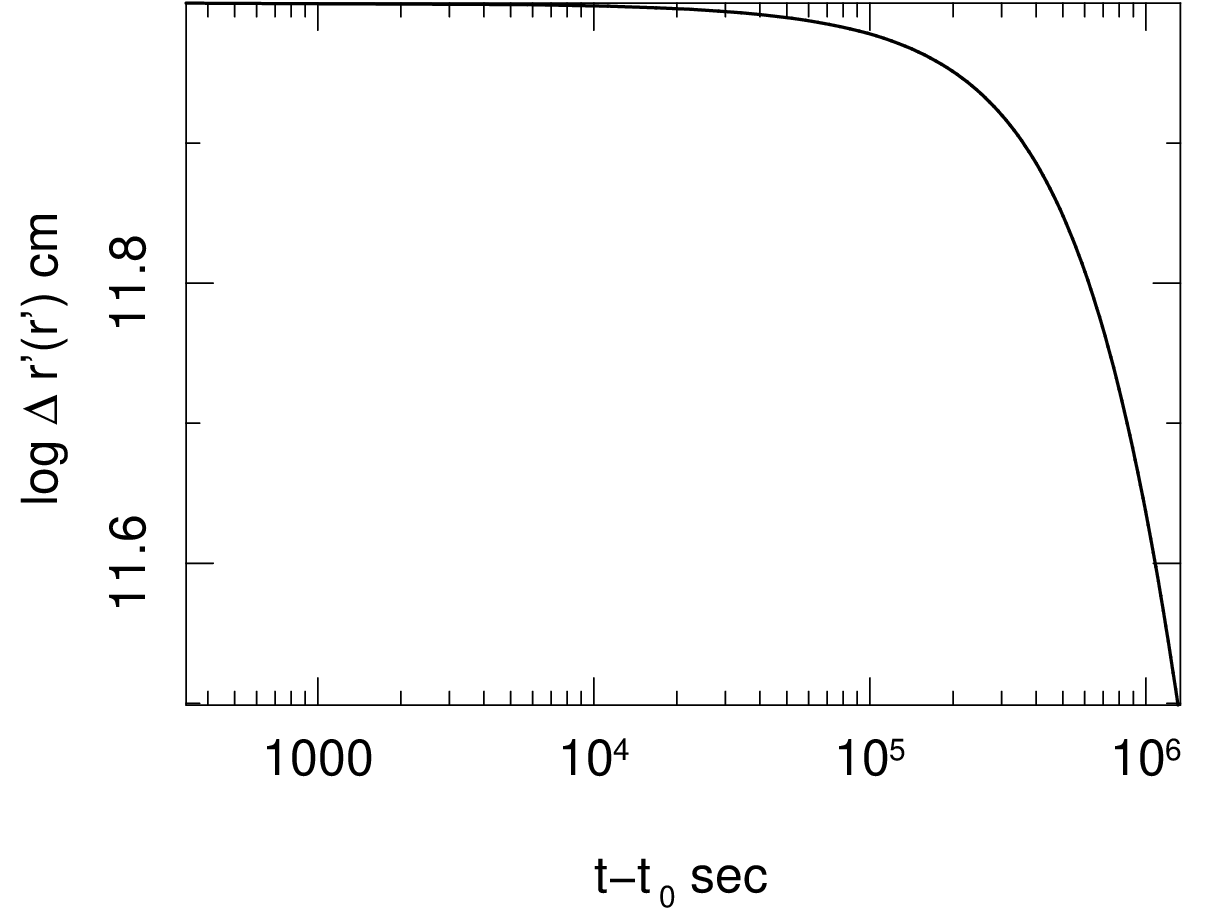} &
8)\includegraphics[height=4cm,angle=-90]{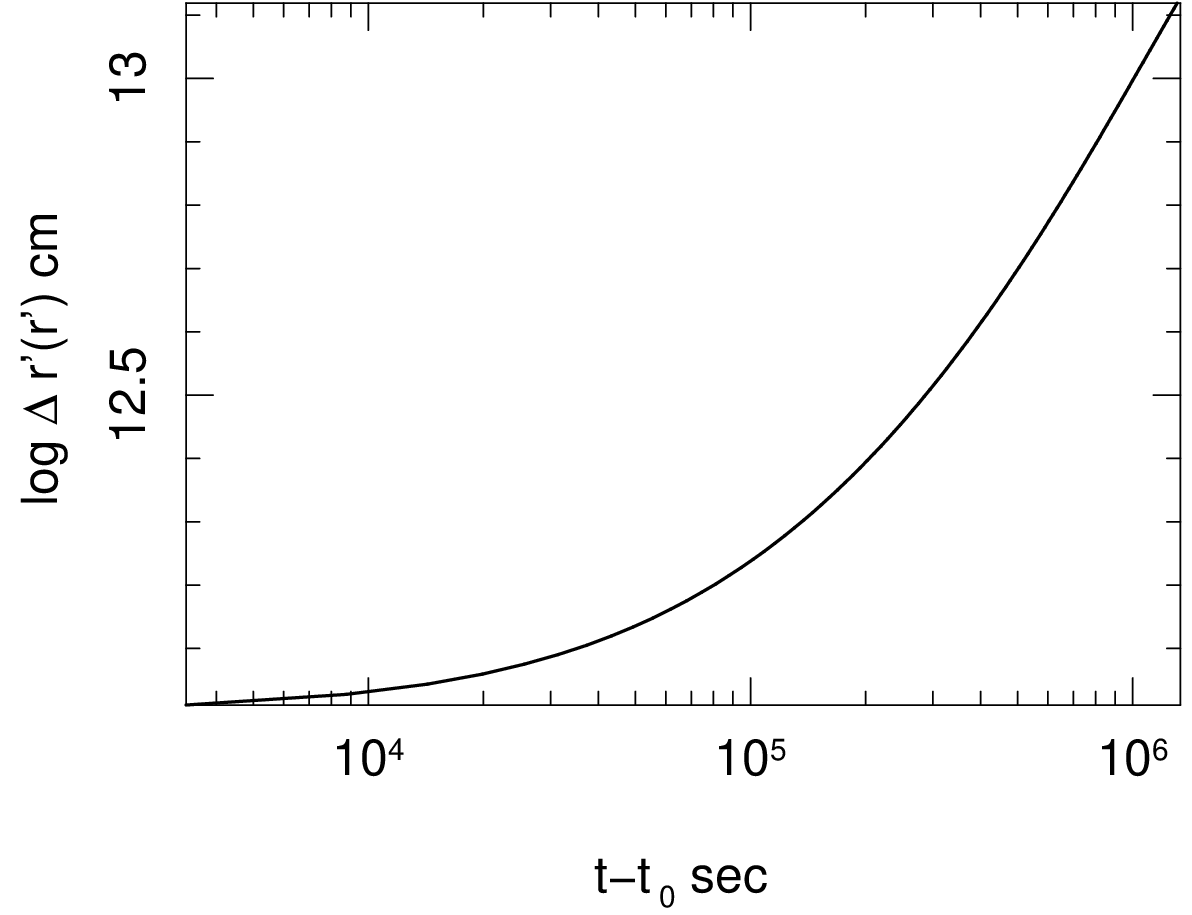} \\
9)\includegraphics[height=4cm,angle=-90]{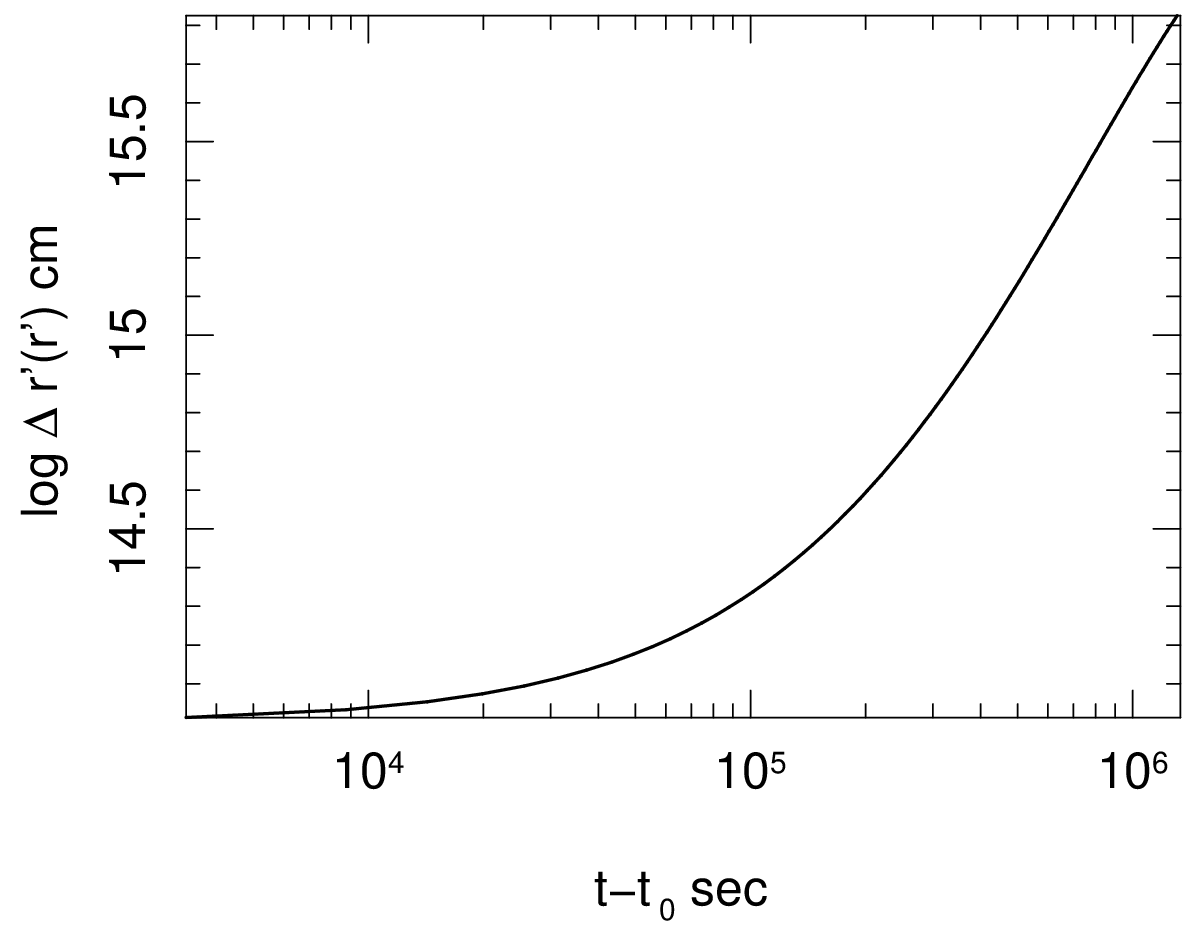} &
10)\includegraphics[height=4cm,angle=-90]{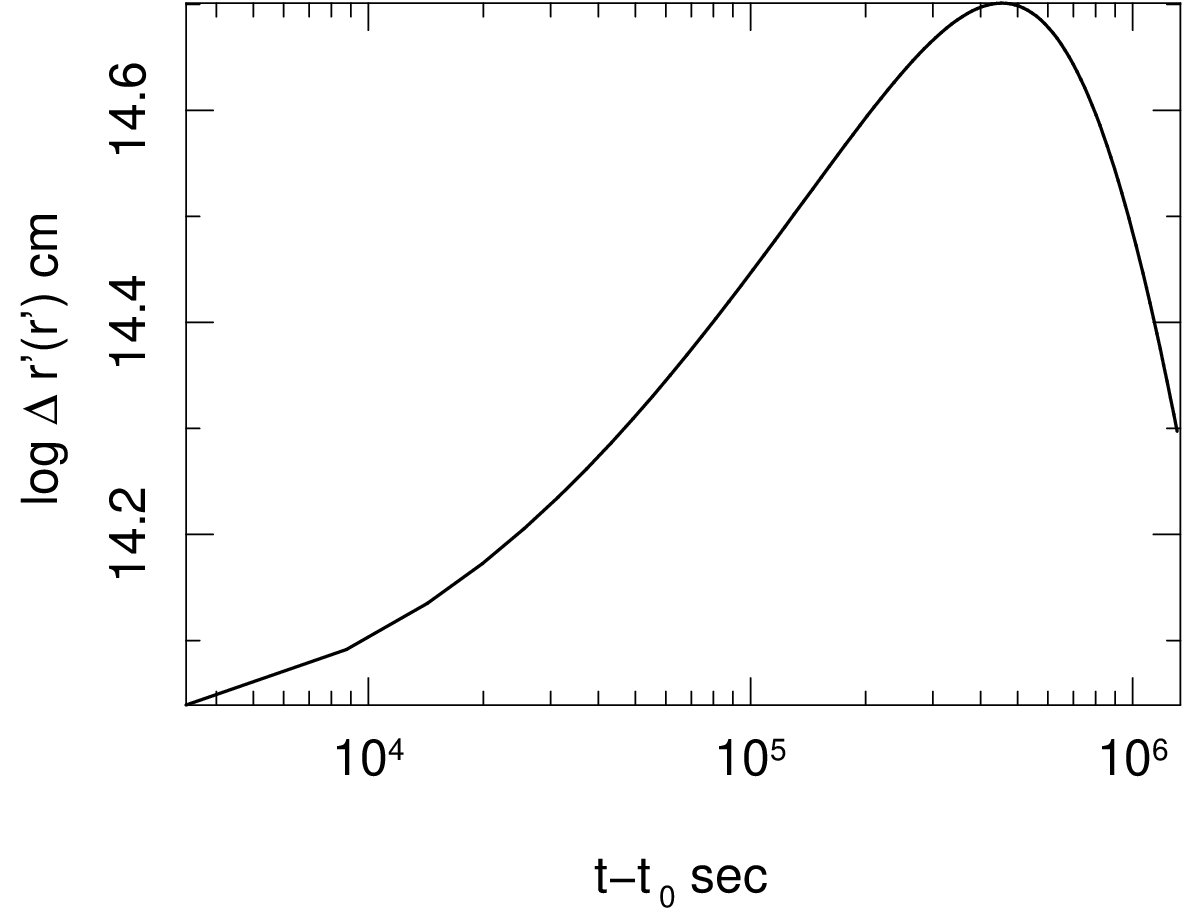} &
11)\includegraphics[height=4cm,angle=-90]{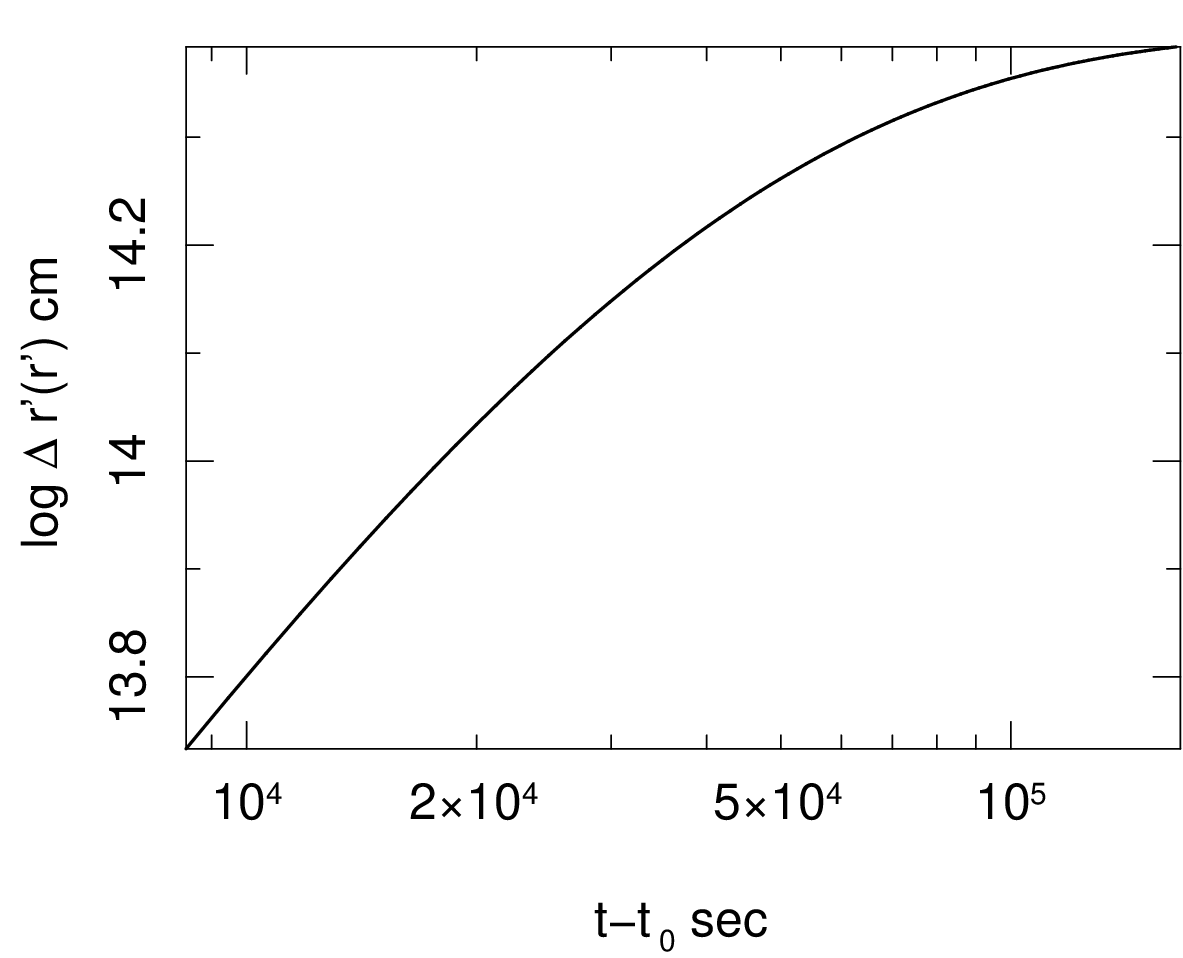} &
12)\includegraphics[height=4cm,angle=-90]{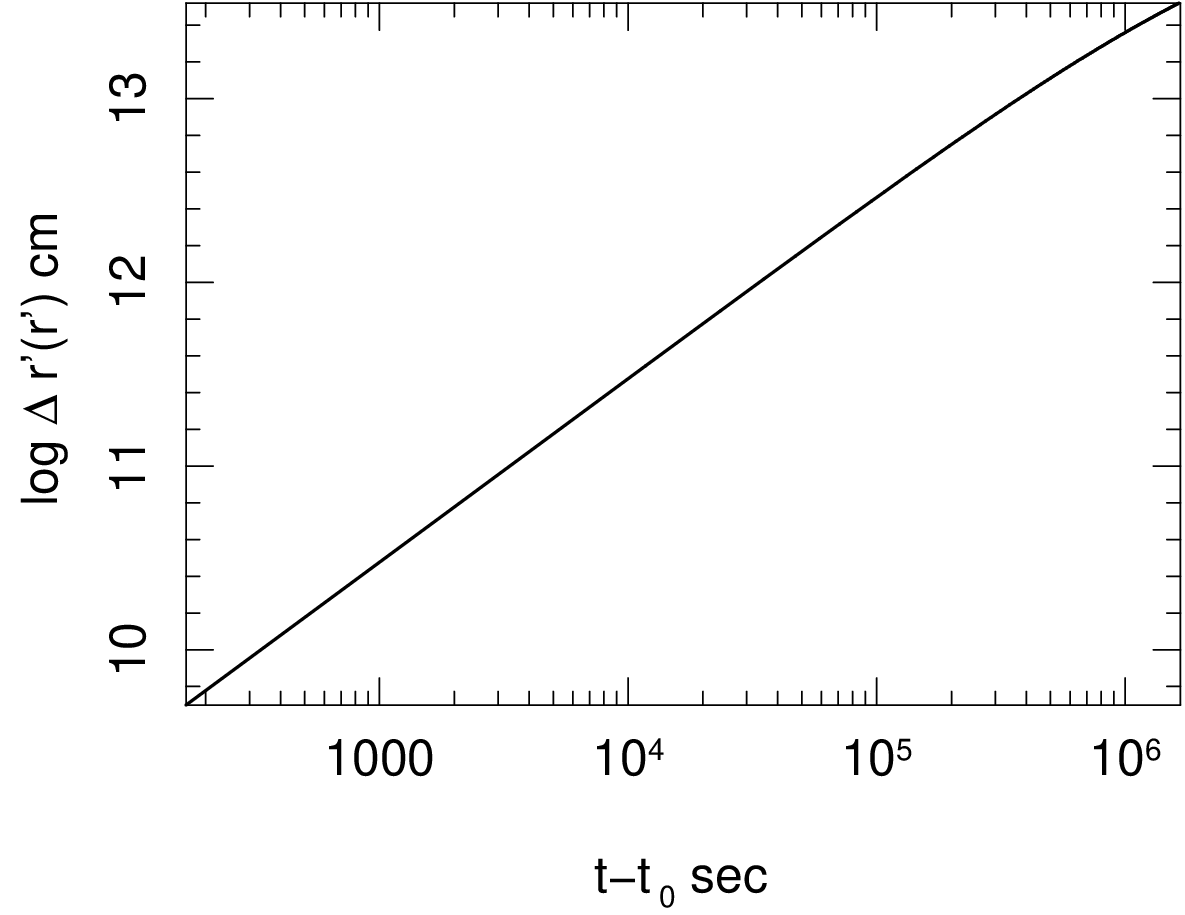} \\
13)\includegraphics[height=4cm,angle=-90]{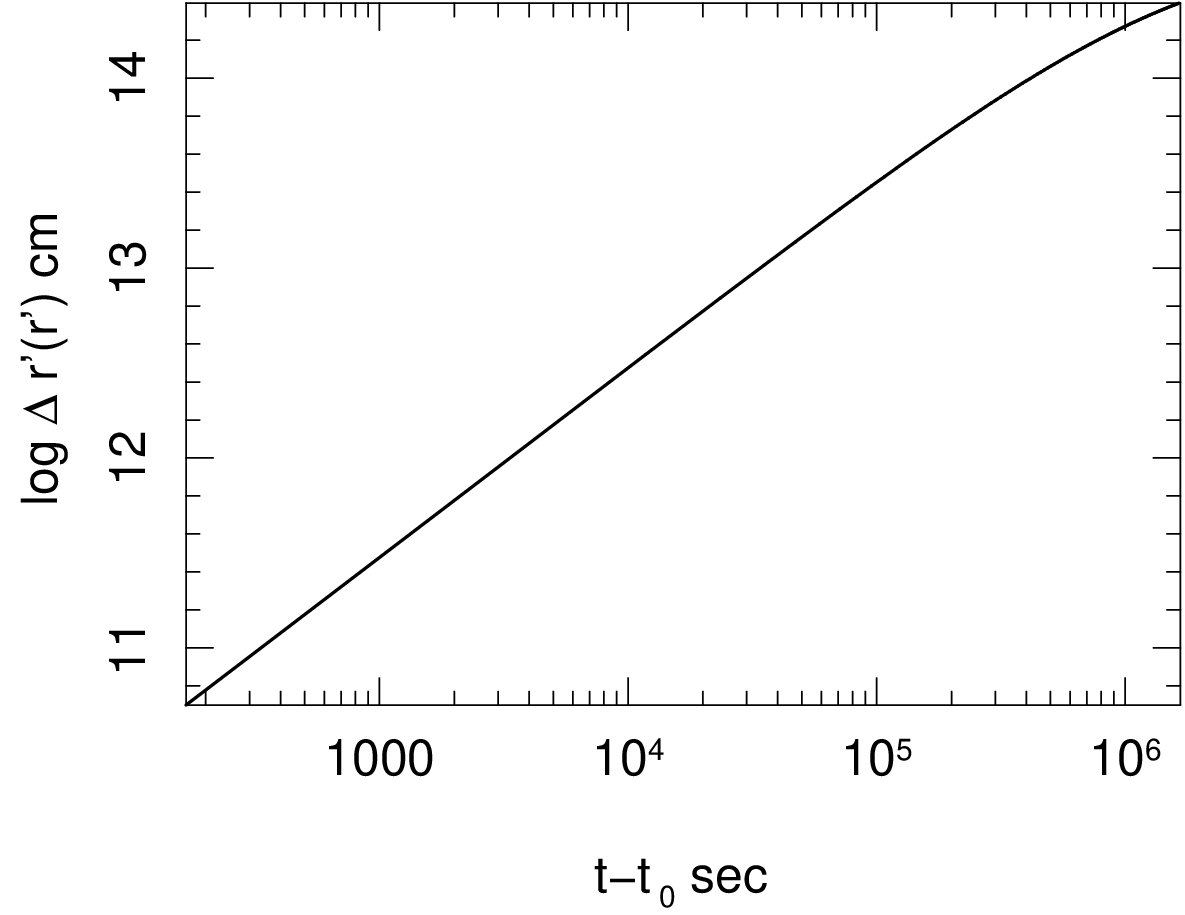} &
14)\includegraphics[height=4cm,angle=-90]{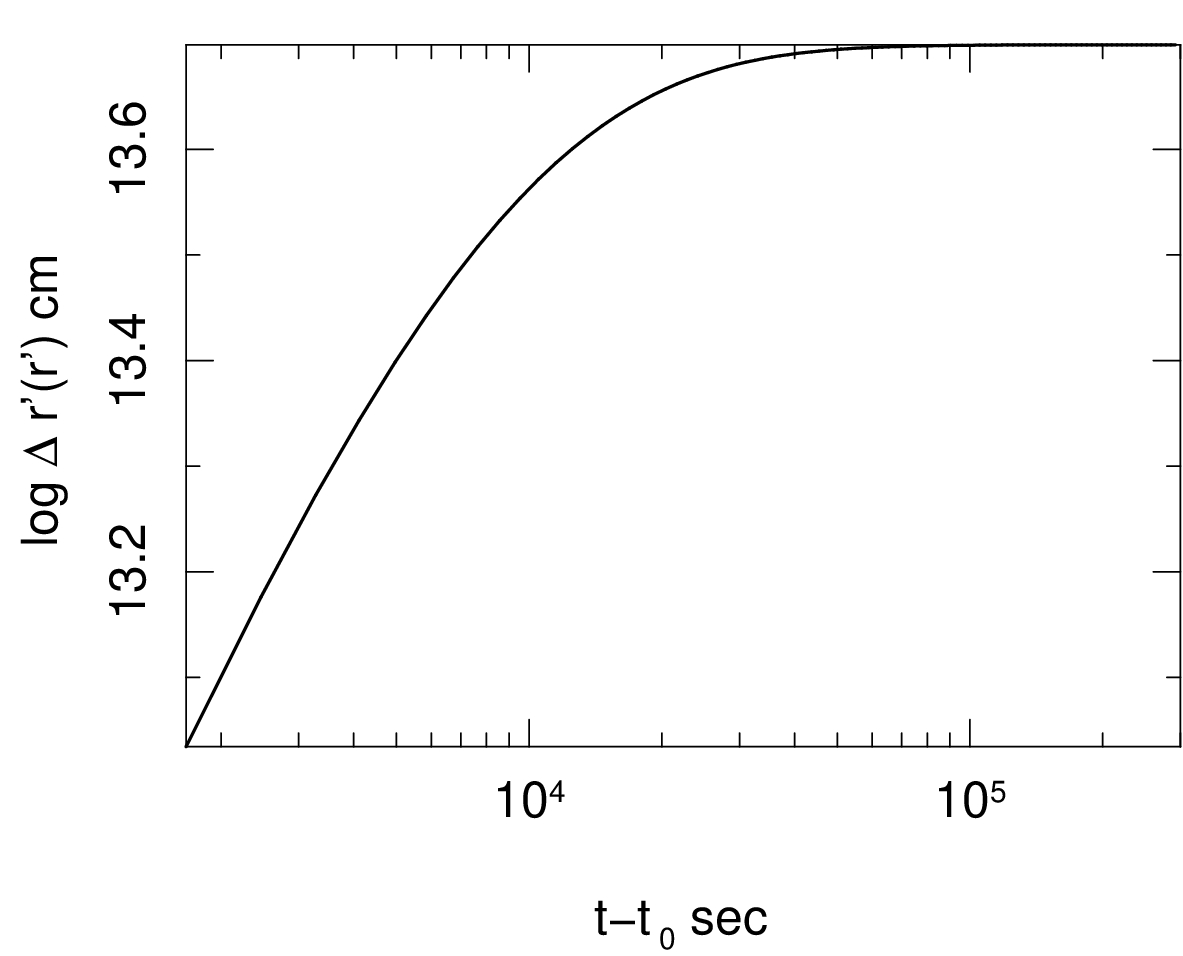} &
15)\includegraphics[height=4cm,angle=-90]{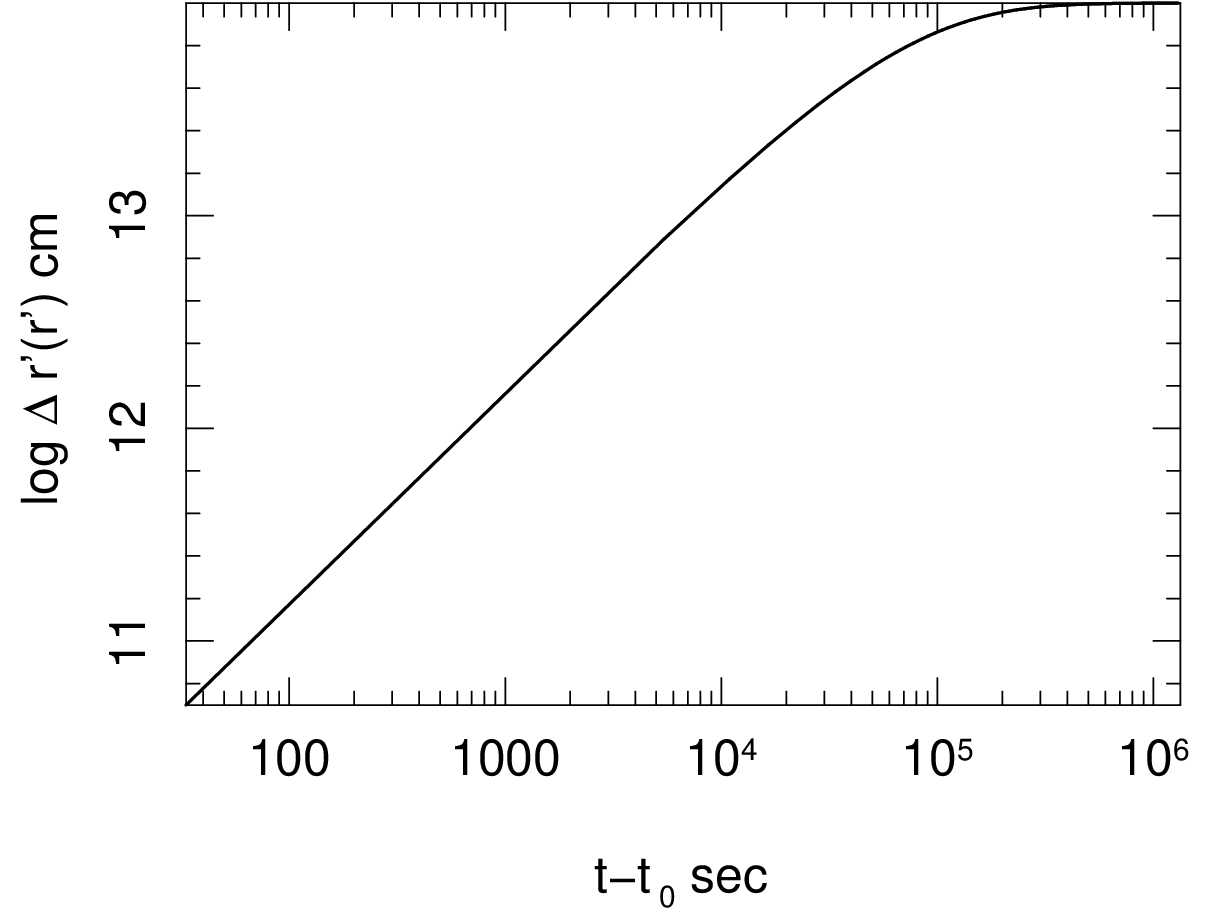} &
16)\includegraphics[height=4cm,angle=-90]{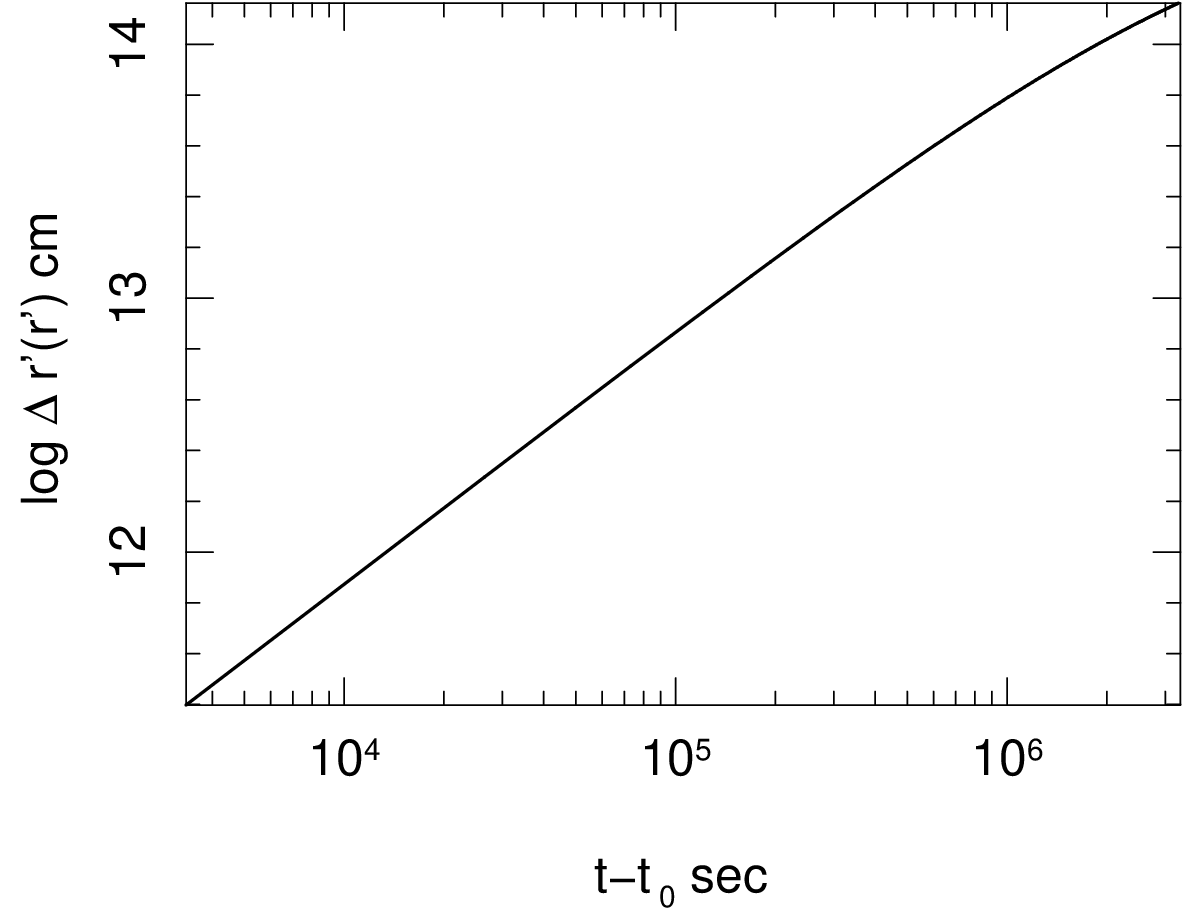} \\
17)\includegraphics[height=4cm,angle=-90]{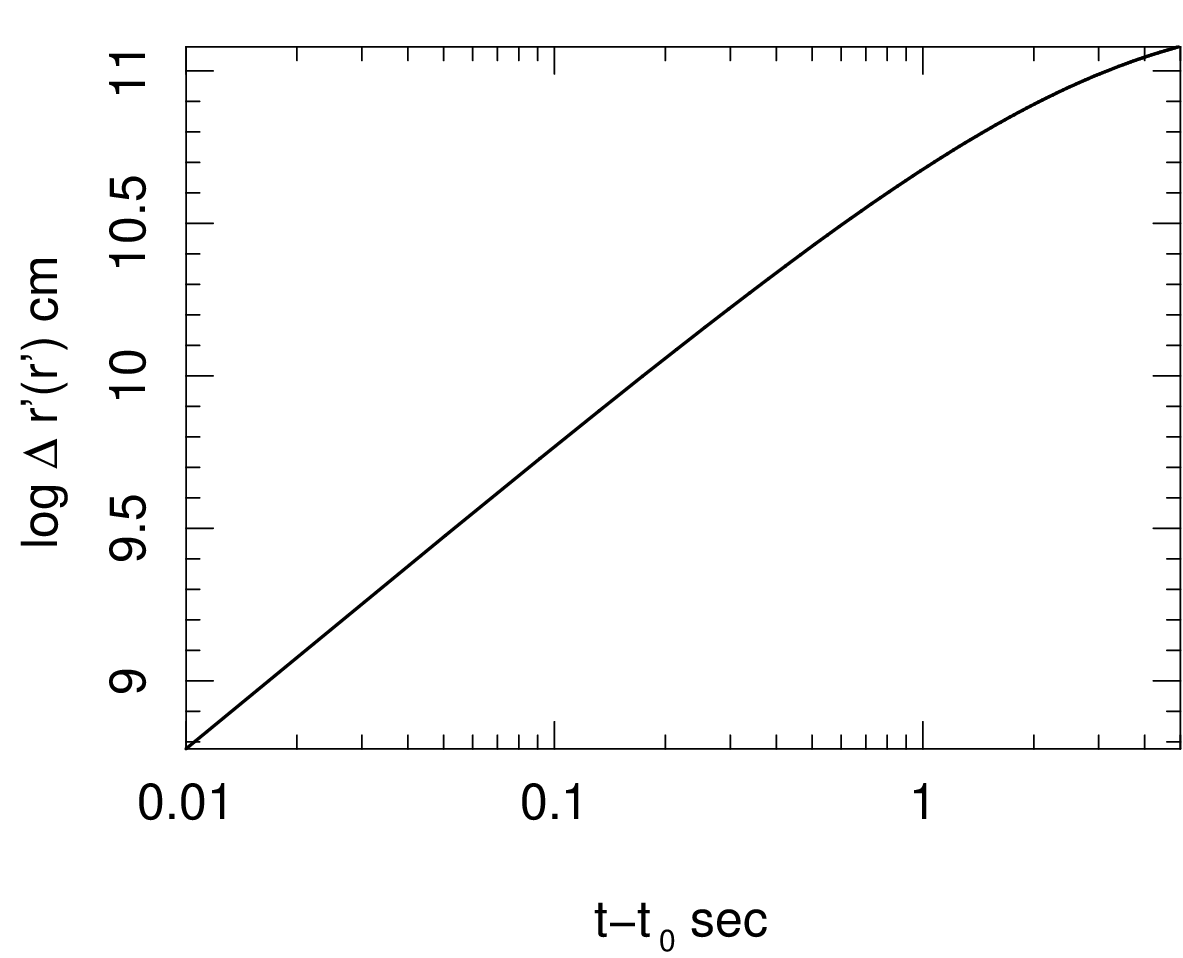} &
18)\includegraphics[height=4cm,angle=-90]{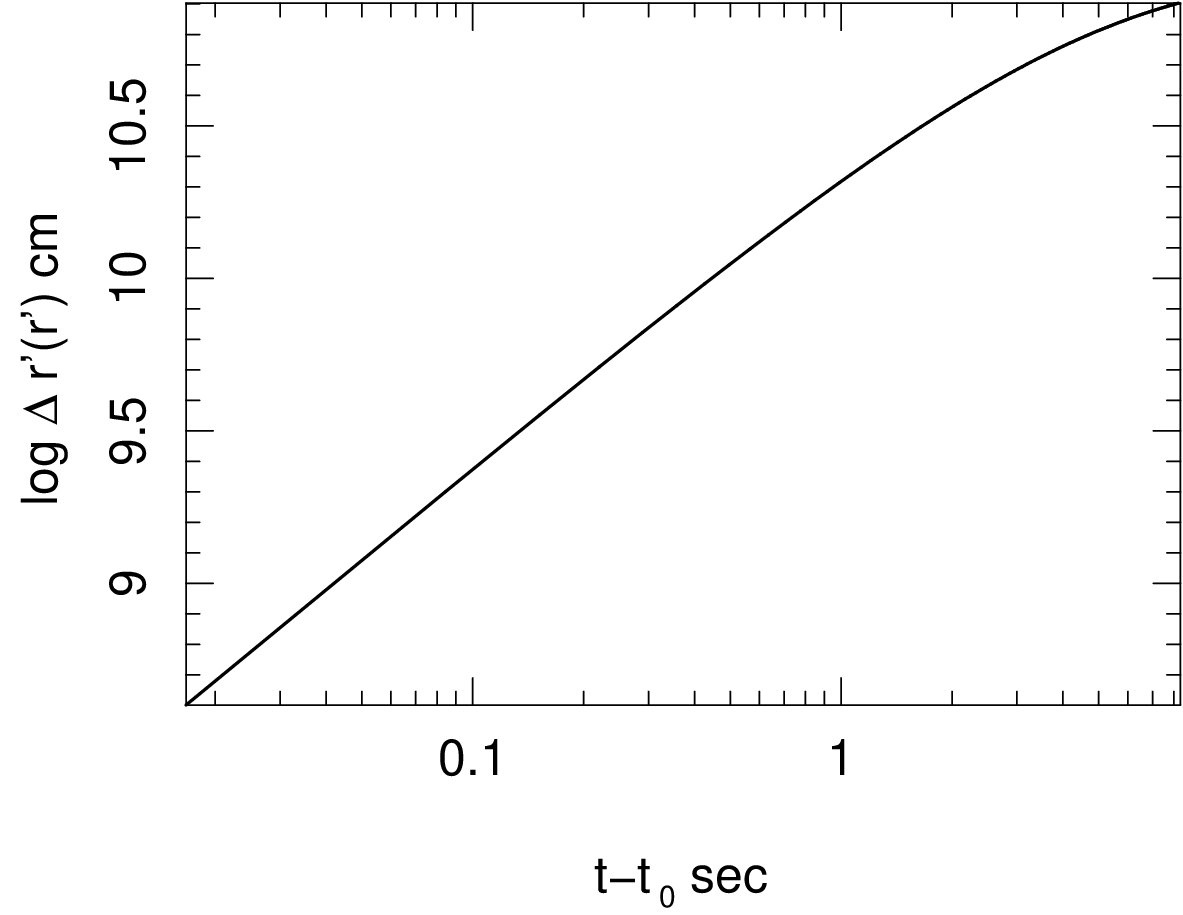} &
19)\includegraphics[height=4cm,angle=-90]{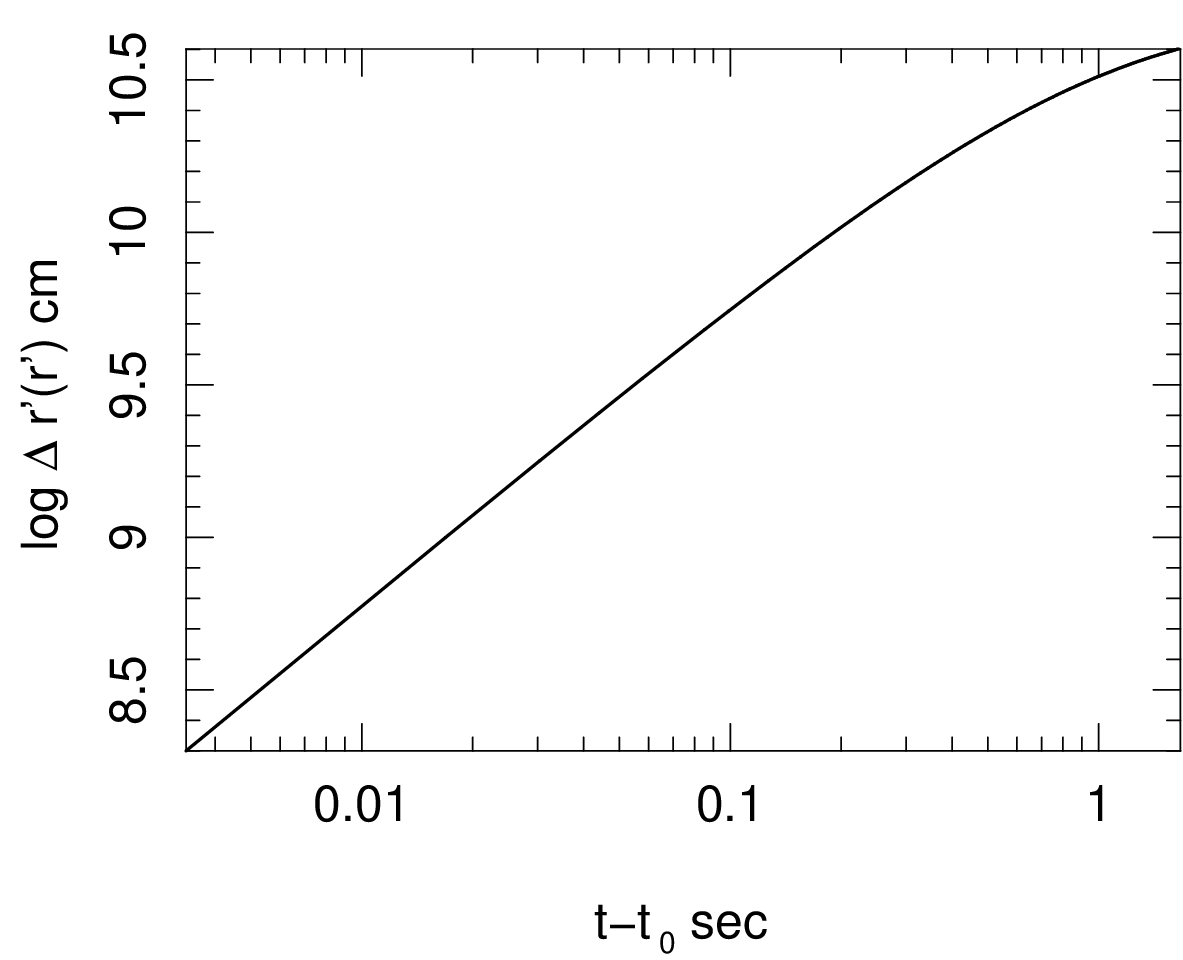} &
20)\includegraphics[height=4cm,angle=-90]{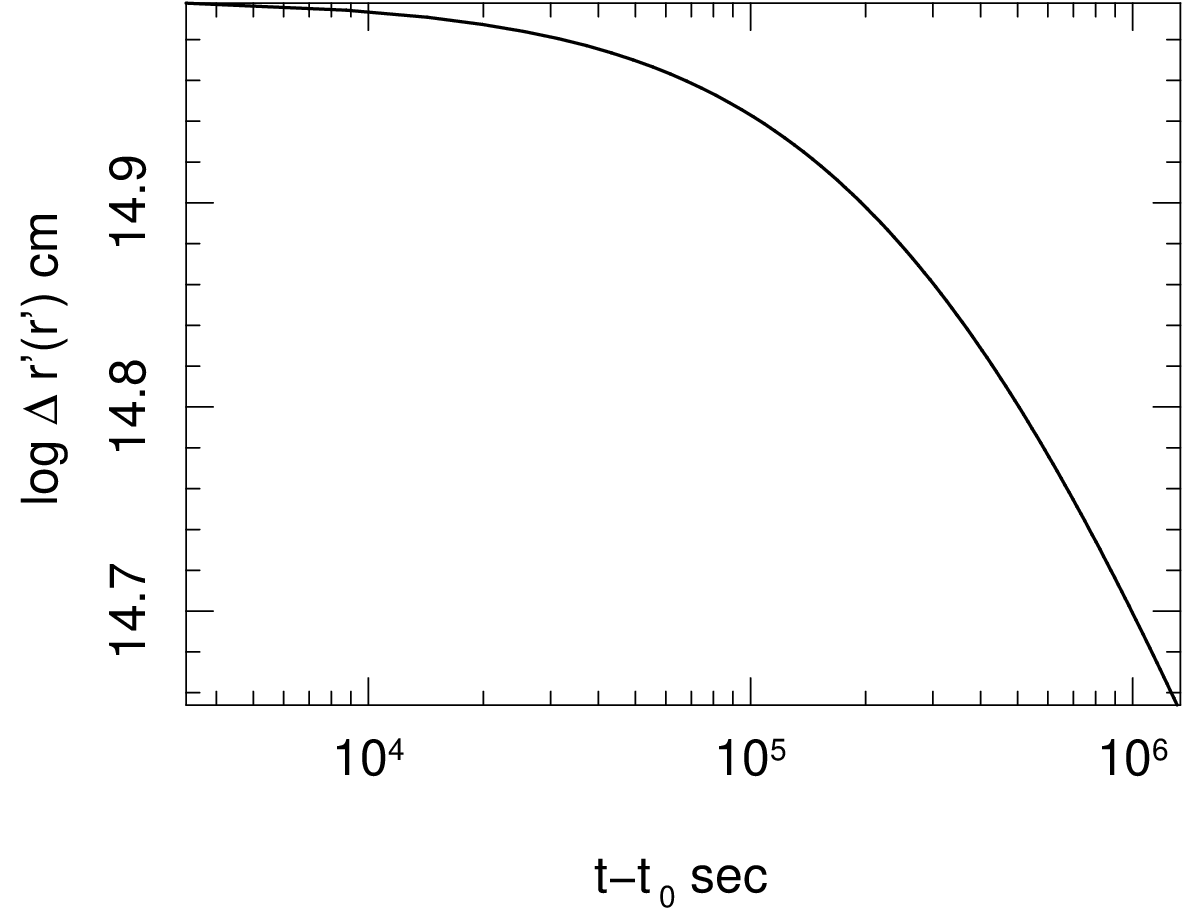} \\
21)\includegraphics[height=4cm,angle=-90]{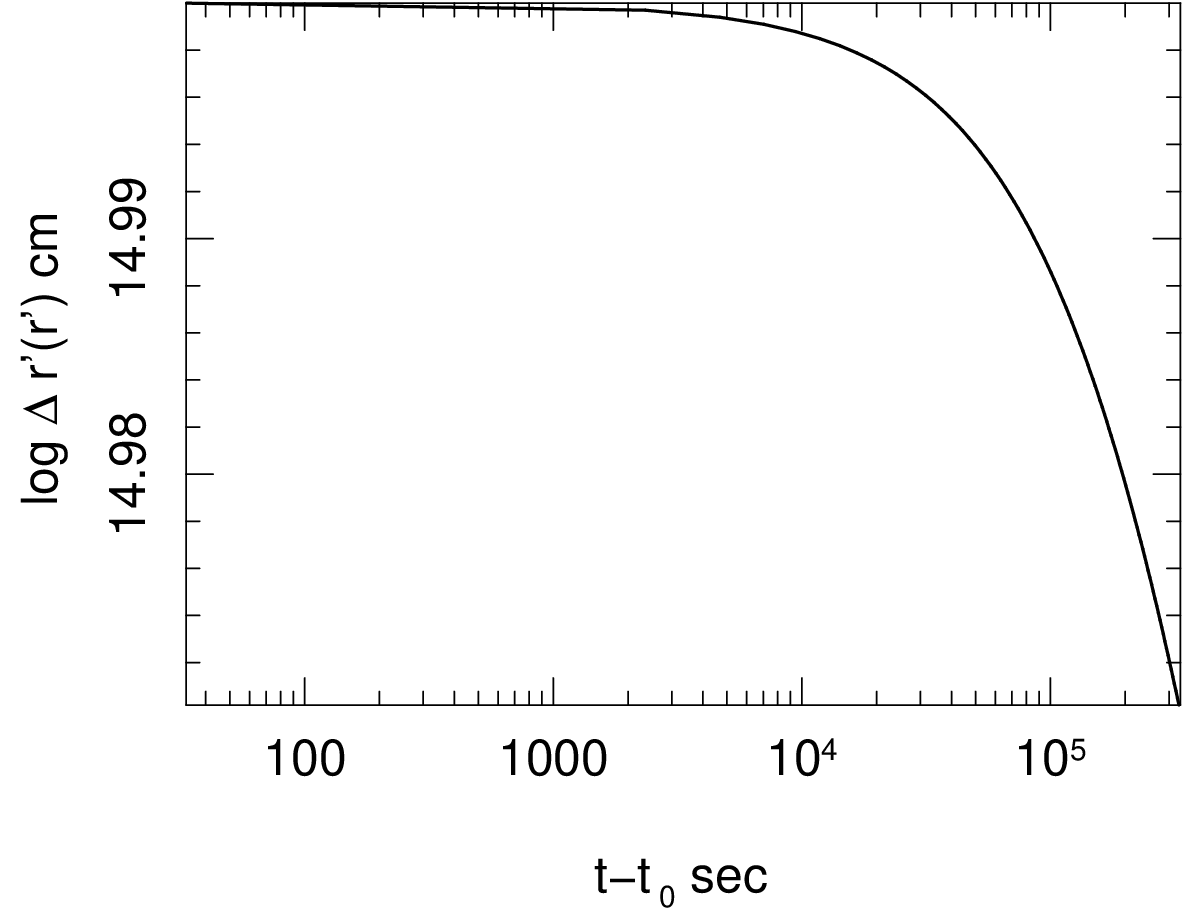} &
22)\includegraphics[height=4cm,angle=-90]{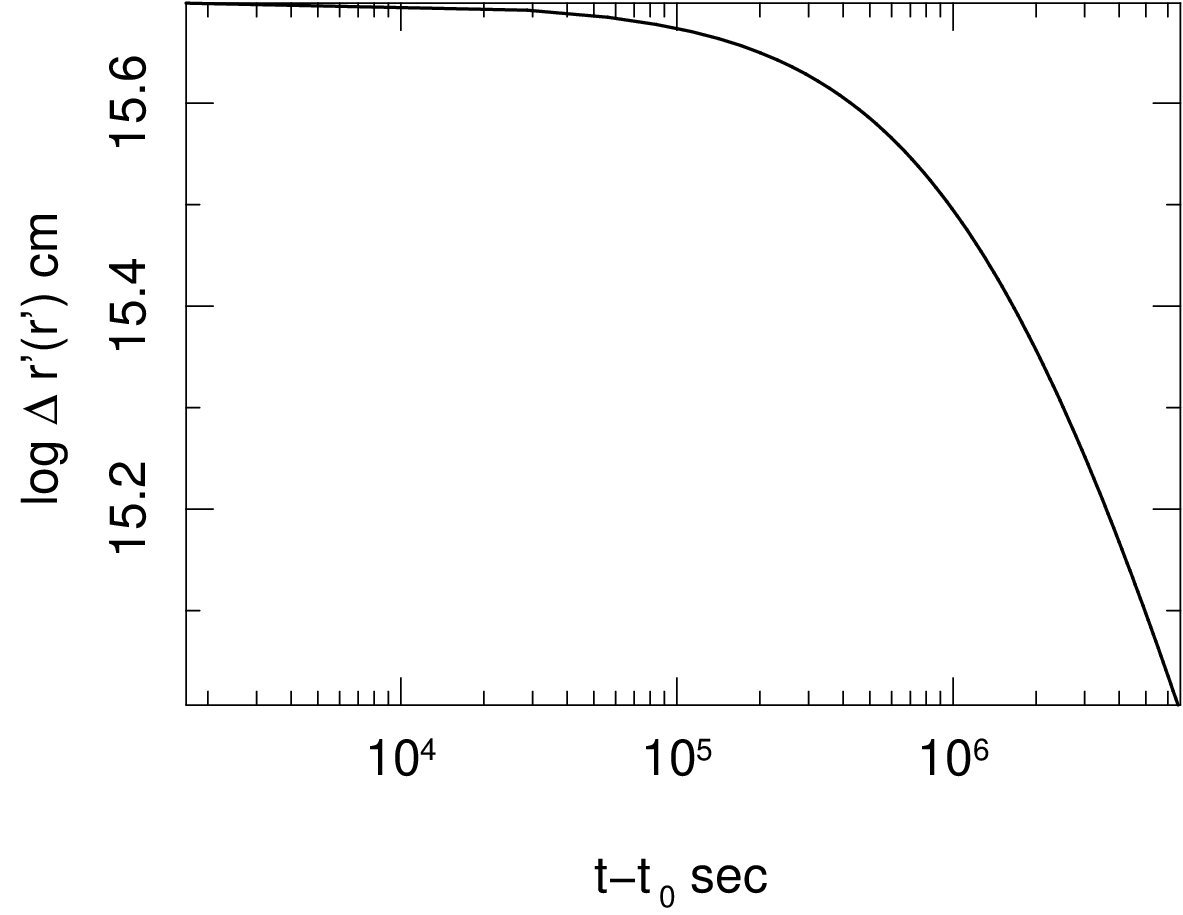} &
23)\includegraphics[height=4cm,angle=-90]{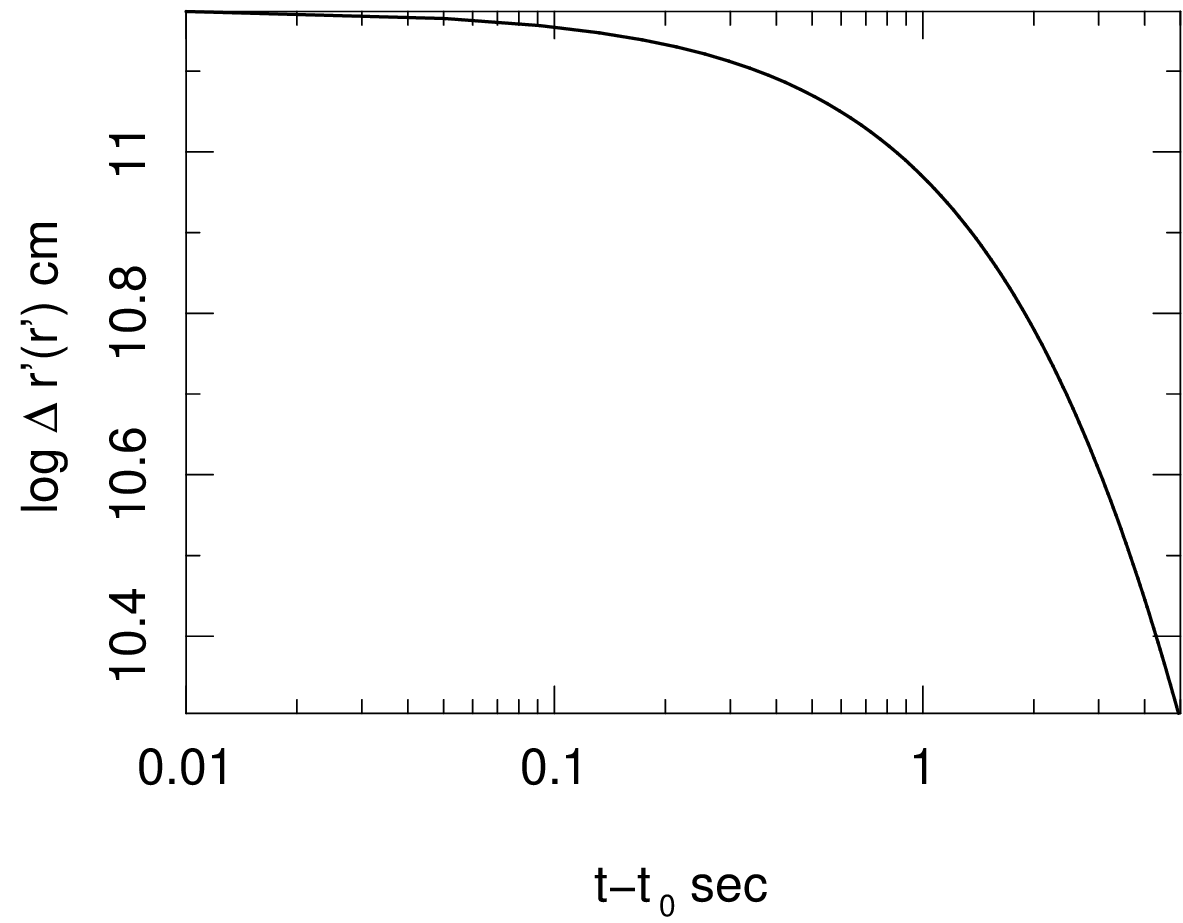} &
24)\includegraphics[height=4cm,angle=-90]{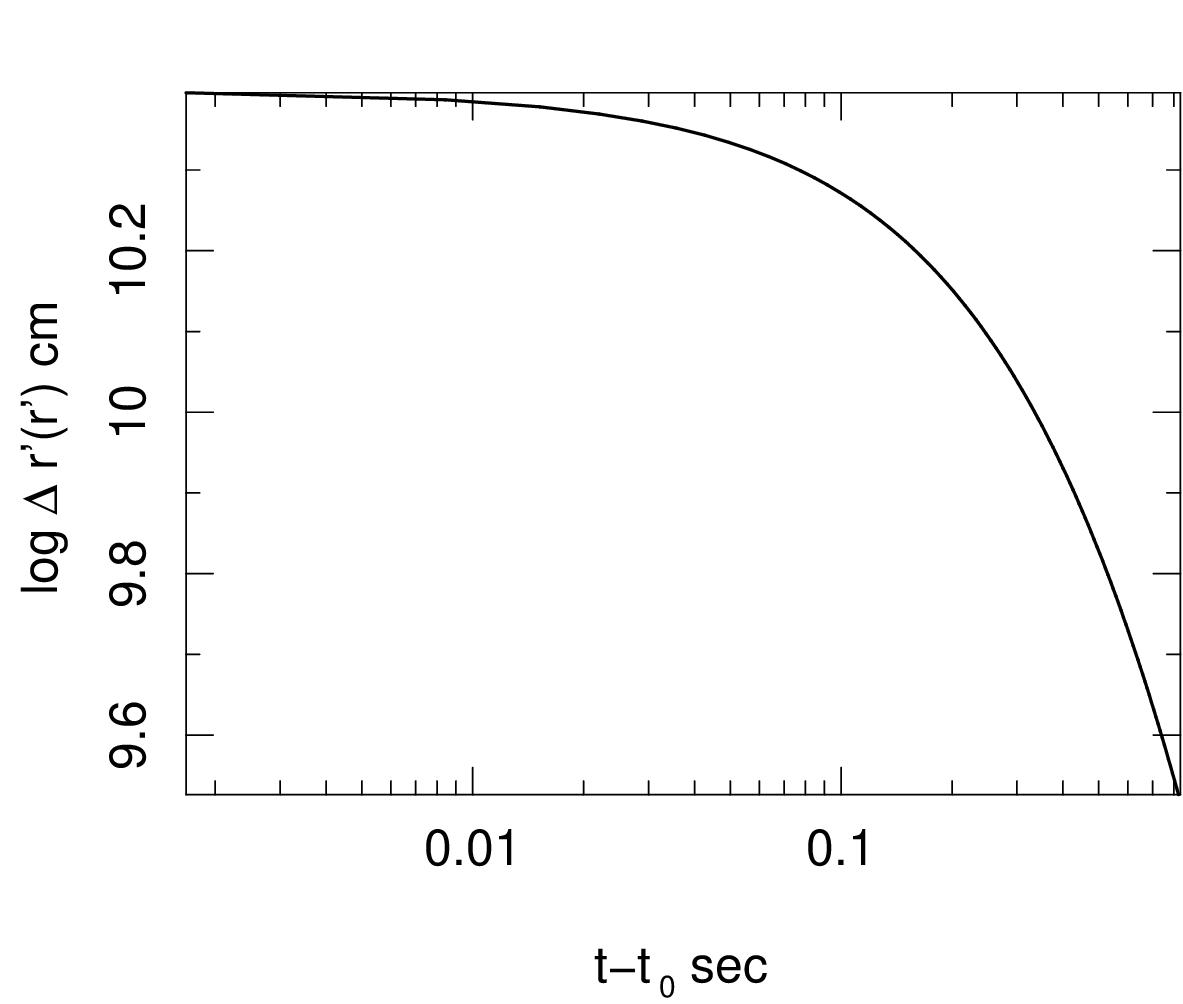} 
\end {tabular}
\end {center}
\caption{Evolution of the width of the active region $\Delta r'$ with 
time/radius. Definitions of curves are the same as Fig. \ref{fig:betazero}.
\label{fig:deltar}}
\end{figure*}

Figures \ref{fig:betazero} to \ref{fig:deltar} show the evolution of the 
quantities that determine the kinematics and dynamics of the shock and 
synchrotron emission. It is notable that they present a variety of behaviours 
despite the fact that in most cases parameters of the models are not too 
different 
from each others. Apriori some of these quantities can be extracted from 
observations (see Sec.\secextract). They give valuable information about the 
physical state of the ejecta and its surrounding. This information can be used 
to constrain engine models. 

The first application of these plots is the comparison and verification of 
the present model against previous works. For instance, comparing plots of 
$\log (\gamma' - 1)$ as a function of time in Fig. \ref{fig:gammaone} with the 
simulations of~\cite{shkdyn} for relativistic and with simulations of 
~\cite{shknewton} for 
non-relativistic regime shows that although the general behaviour of this 
quantity in all simulations is the same, the slope of the evolution is not. 
According to the self-similar solution in the ultra-relativistic regime the 
slope of $\gamma'(t') \sim -3/7$~\citep{kinematic,emission1,shkdyn}. Here we 
find roughly the same value at the beginning of the emission, see specially 
models of afterglows in Fig. \ref{fig:gammaone} where the relative Lorentz 
factor between two shells is highly relativistic. However, because in the 
models studied here the physical parameters such as the electric and magnetic 
fields are varying, the evolution of $\gamma'(t')$ at the beginning of the 
shock can be much flatter. On the other hand in non-relativistic regime, 
$(\gamma'(t') - 1) \rightarrow 0$ very quickly. This happens specially 
in the prompt shock where the relative Lorentz factor is small and approaches 
to zero when shells coalesce, see models 1 to 6 in Fig. \ref{fig:gammaone}. 
Therefore the transfer of kinematic energy to radiation is not really a simple 
power-law as usually assumed. This has important implications for the 
formation of various features of the light curves (We discuss the light 
curves in details in the next section). There are very few systematic 
simulation of prompt emission from first principles to compare with this 
results. There are however analytical estimations and simulations for the 
afterglow~\citep{shkdyn,shknewton}. Their slope of $\alpha \sim -1$ at very 
late times and somehow flatter at earlier earlier time is similar to models 9 
and 10 which meant to simulate the late time afterglow. Nonetheless, the 
differences between these models as well as models 20 to 22 which are also 
candidate models for late afterglow show that depending on the evolution of 
the active region and physical parameters, various kinematical and dynamical 
outcome can be expected from late afterglow. This is  consistent with the 
observation of very slow X-ray decay of e.g. 
GRB 050822~\citep{grb050822,grb050822-1} with final decay slope of 
$\alpha \sim -0.97$ and very fast decay of e.g. 
GRB 050730~\citep{grb050730,grb050730-1} and 
GRB 060607a~\citep{grb060607a,grb060607a-1} with final $\alpha \lesssim -3$

We notice that for models 17 to 19 which simulate the onset of prompt emission 
with an active region evolving according to quasi-steady state model, $\beta'$,
$\gamma'$, etc. have a pathological behaviour for one decade of time or so at 
the beginning of the simulation. We could not understand the reason because 
the same code has been used for the simulation of external shocks and they do 
not show any anormal behaviour. By changing the duration of the simulations 
and the same parameter sets we observed that the anormal interval varies, but 
the behaviour of normal section was stable and did not depend on the simulated 
duration. Therefore we conclude that the initial anormal behaviour is a 
purely numerical effect. Consequently, for these models we consider only the 
interval of time in which the simulation has good behaviour - from 
$t-t_0 \sim 0.1$ sec. for models 17 and 18, and $t-t_0 \sim 0.03$ sec. for 
model 19.

Another physically important quantity is $\omega'_m$. In~\citep{emission1} its 
time evolution slope in radiative external shocks is estimated to be 
$\sim -12/7 \approx -1.7$. From plots in Fig. \ref{fig:omegam} we can conclude 
that in most cases the late time slope of $\omega'_m$ is close to this value. 
However, we observe a more complex behaviour for this quantity. Notably, in 
prompt internal shocks it can increases at least during an interval of time. 
This is the case for models 1, 17, 19, and 24. For most of other models the 
initial decline of $\omega'_m$ is much slower than the predication of 
self-similar shocks. This means that the initial radiation in both internal 
and external shocks can be much harder than what is expected from self-similar 
shock models. This behaviour can be related to time variation of shocks 
micro-physics. We come back to this point in the next section.

The column density of the active region $n'\Delta r'$ is also an interesting 
quantity which apriori can be measured directly. Plots in 
Fig. \ref{fig:coldens} show that the intrinsic column density can evolve very 
differently from burst to burst, both in prompt internal and external 
shocks. Moreover its evolution depends on other physical parameters. In fact, 
from definition of this quantity it is clear that it directly depends on the 
evolution of density and the size of the active region see 
Figs. \ref{fig:nprim} and \ref{fig:deltar}. A remarkable property is 
that during the sudden rise of the prompt emission such as in model 2, the 
column density simultaneously increases. The same in true during the build-up 
of active region in external shocks, models 13 and 15 to 19. 

In the previous simulations~\citep{revshock,shockmag} the effect of 
accumulation of matter has been added to the simulations as a factor called 
{\it compression factor}. Because in these simulations it is usually 
considered that synchrotron emission happens just in the shock front, the 
only column density seen by the radiation is considered to be the material in 
the line of sight and downstream of the shock front. Therefore we 
can not compare the column density of active zone and its evolution in this 
model with previous simulations. At present the extraction column density of 
active region from observations is very difficult and depends on how precisely 
we can determine and remove the contribution of foreground matter on the line 
of sight of GRBs. We have discussed these issues in Sec.\secextract. 

\section {Light curves} \label {sec:lc}
The most prominent feature of the GRBs is the fast time variation in their 
emission reflected in the name of these transient sources. 
Any model of gamma ray bursts must reproduce their light curves and various 
observed features such as lags, Fast Rise Exponential Decay (FRED) behaviour 
of pulses, shallow slope regime in the early X-ray and sometimes optical 
afterglow emission, chromatic breaks, and the slow decay of optical and 
longer wavelength emissions.

To see if the model discussed in Paper I can reproduce the diversity of 
gamma-ray light curves and their afterglow in lower energy bands we calculate 
the time variation of the spectrum, equation (60-Paper I), assuming a 
power-law distribution
for the Lorentz factor of electrons. The parameter sets are the same as 
what we used for the calculation of the shell dynamics and emission 
characteristics summarized in Table \ref{tab:param}. We should remind that 
multiple peaks observed in many GRBs are most probably the result of multiple 
shell collisions. Our aim here is to reproduce the characteristics of a 
single peak interpreted as the collision between two shells.

We also remind that the parameters for the models with a dynamical active 
region have been chosen to simulate the prompt gamma-ray emission, except for 
models 7 to 10. For these models the parameter are chosen according to what we 
believe to correspond to the collision of the shells/ejecta with the low 
density slow material, with the wind around the central engine, or 
ISM. Such collisions are considered to be the origin of the late emissions 
that produces slowly varying emission at low energies, X-ray, optical, and 
radio. Although based on physical arguments the dynamical active region seems 
to be more suitable for the prompt emission, it can be also considered for 
the external shocks specially at late times. This is the aim of model 9 and 
10. The chosen parameters for the simulation of quasi steady models are most 
relevant to the afterglow, except models 17 to 19 for which they are similar 
to what we have chosen for the simulations of internal shocks with a dynamical 
active region. 
Similar to the previous case we want to see if the quasi steady model can 
reproduce the observed properties of the prompt emission. The light curves 
obtained from these simulations are shown in Figs. \ref{fig:lcgmzero} and 
\ref{fig:lcgmone}.

As mentioned in the introduction, in these simulations the high latitude 
emission - the 
terms proportional to ${\mathcal G} (r)$ defined in equation (61-Paper I) in 
the power spectrum, equation (60-Paper I) is neglected. This is a good 
approximation for the prompt and early afterglow emissions. However, at late 
times when the Lorentz factor is decreased, their contribution becomes 
important. The main effect of these retarded emission is a flattening of the 
lightcurve. Due to larger Doppler shift of higher latitudes with respect to 
the emission at zero latitude, it would be more significant in lower energies 
than higher ones. This effect is independent of the evolution of the active 
region and applies to all the models explained here. Therefore what ever the 
late slope of the simulations, we should assume that in full model they would 
be flatter.

\subsection{Light curve of simulated models with dynamical active region}
\label{sec:lcdar}
\begin{figure*}
\begin{center}
\begin {tabular}{ccc}
1)\includegraphics[height=5.5cm,angle=-90]{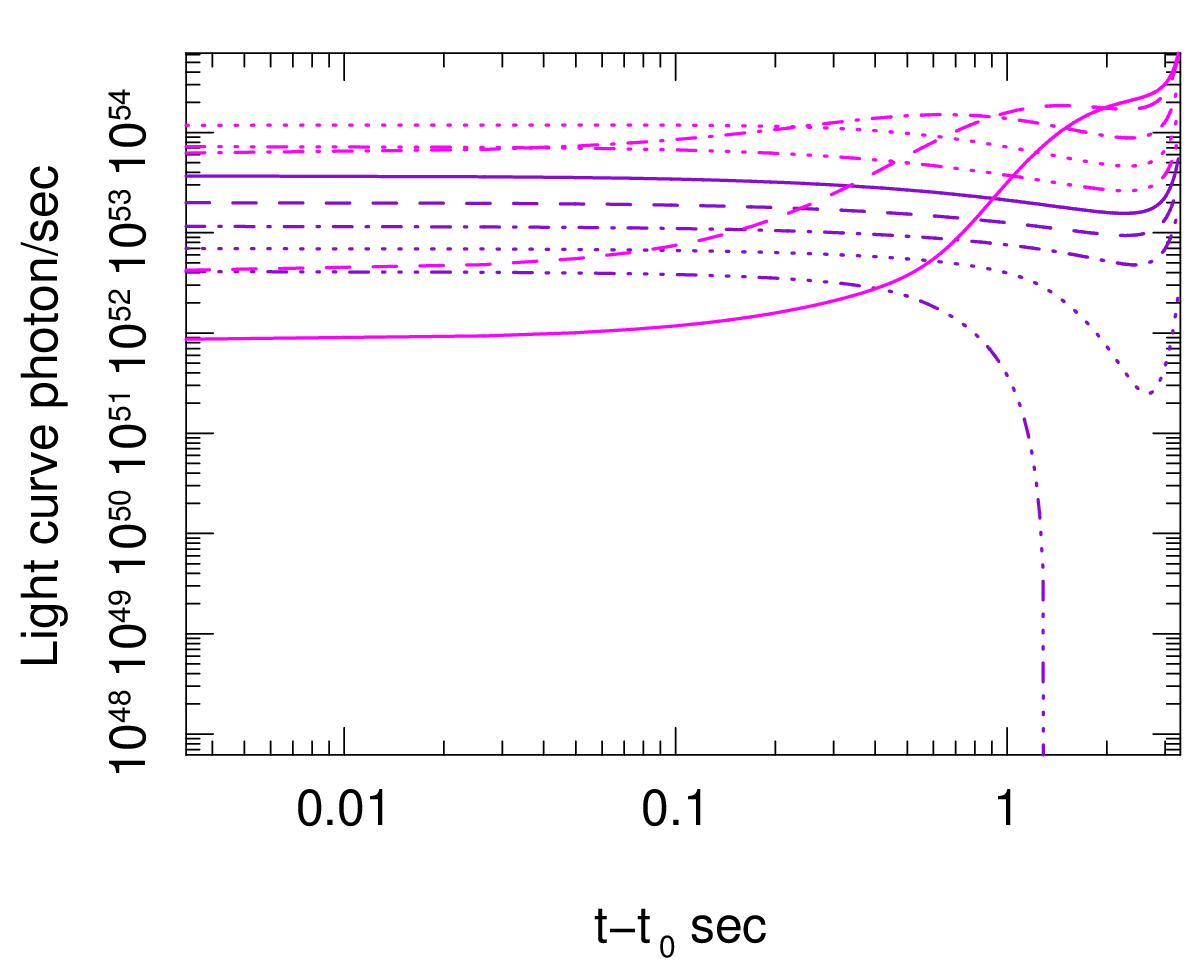} &
2)\includegraphics[height=5.5cm,angle=-90]{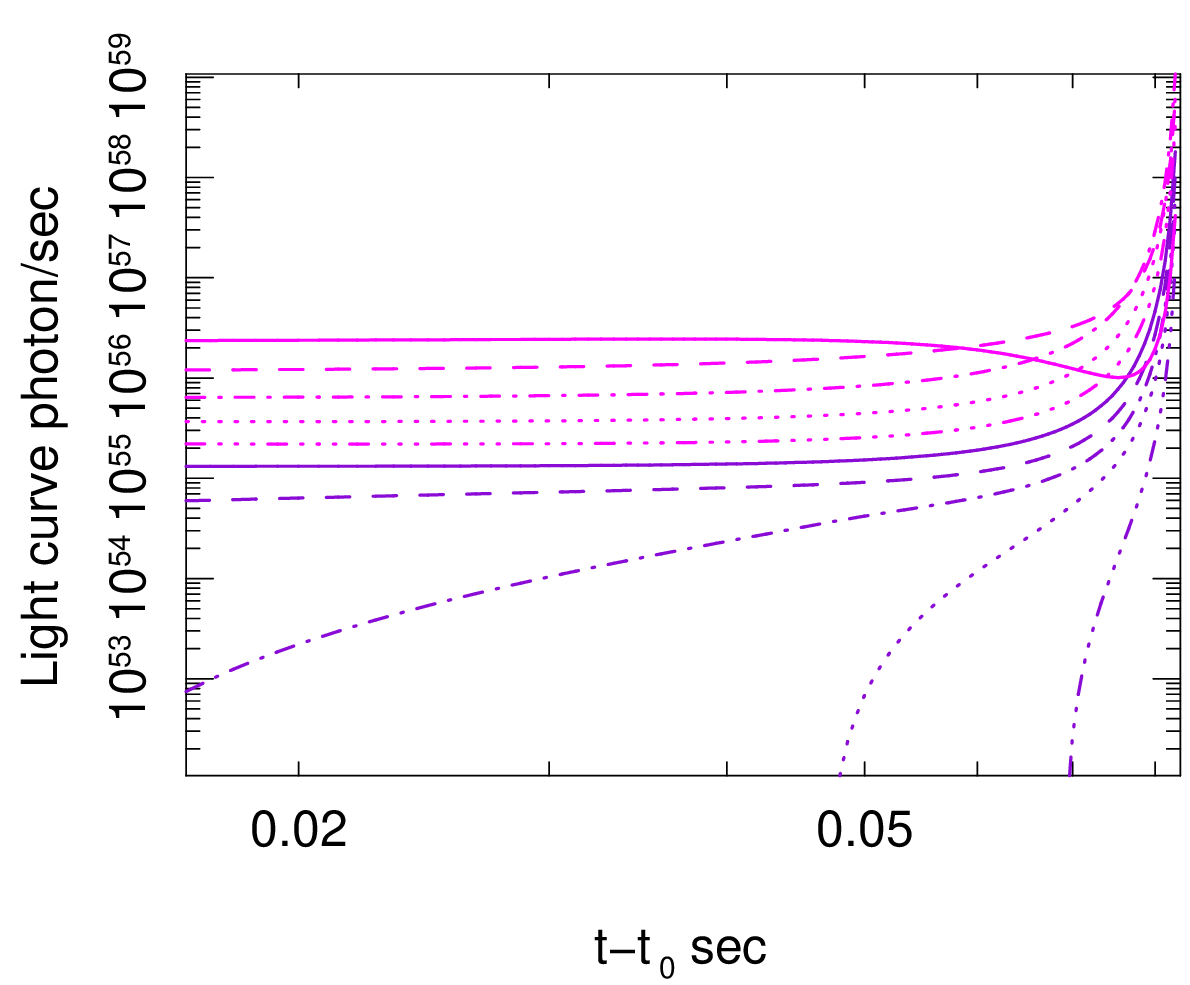} &
3)\includegraphics[height=5.5cm,angle=-90]{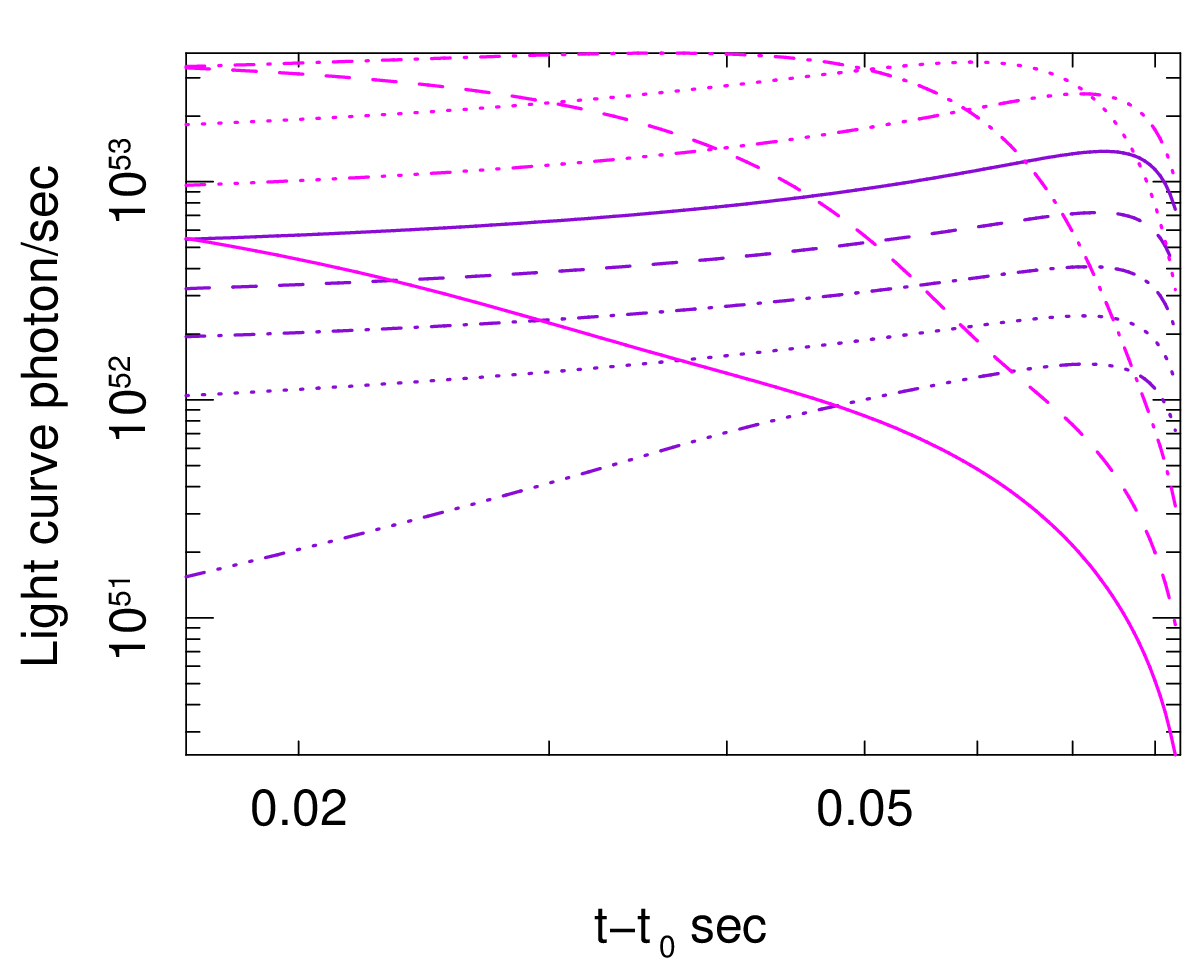} \\
4)\includegraphics[height=5.5cm,angle=-90]{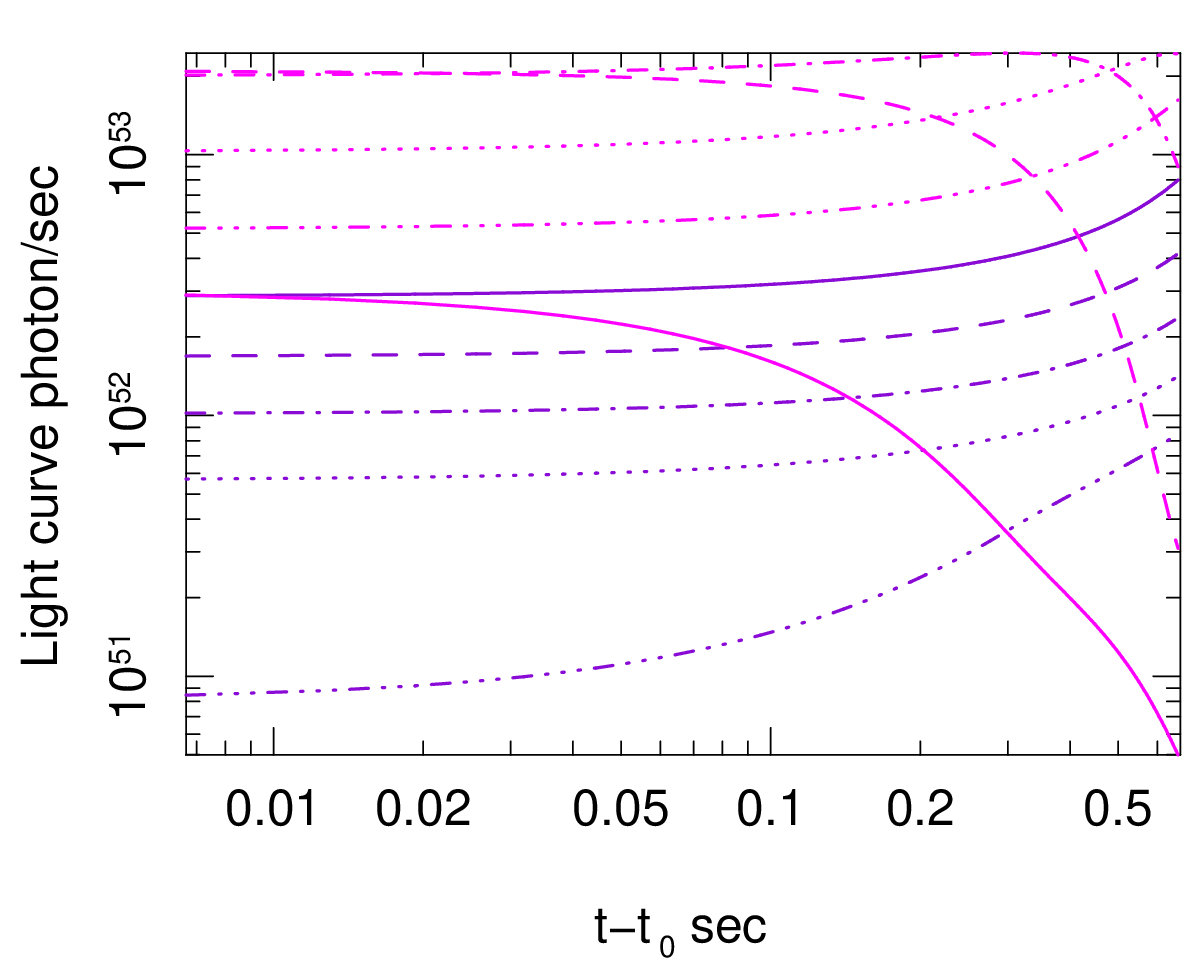} & 
5)\includegraphics[height=5.5cm,angle=-90]{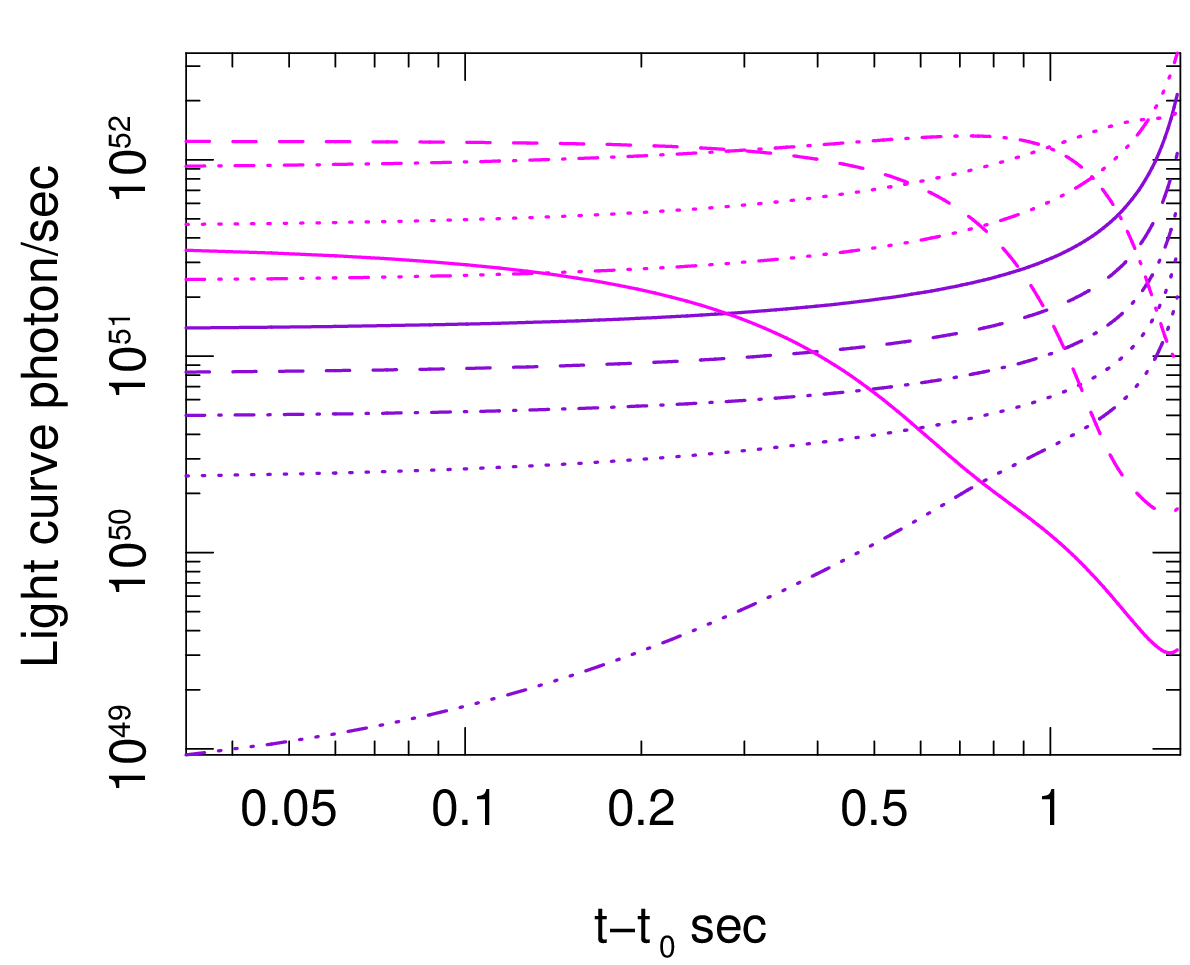} &
6)\includegraphics[height=5.5cm,angle=-90]{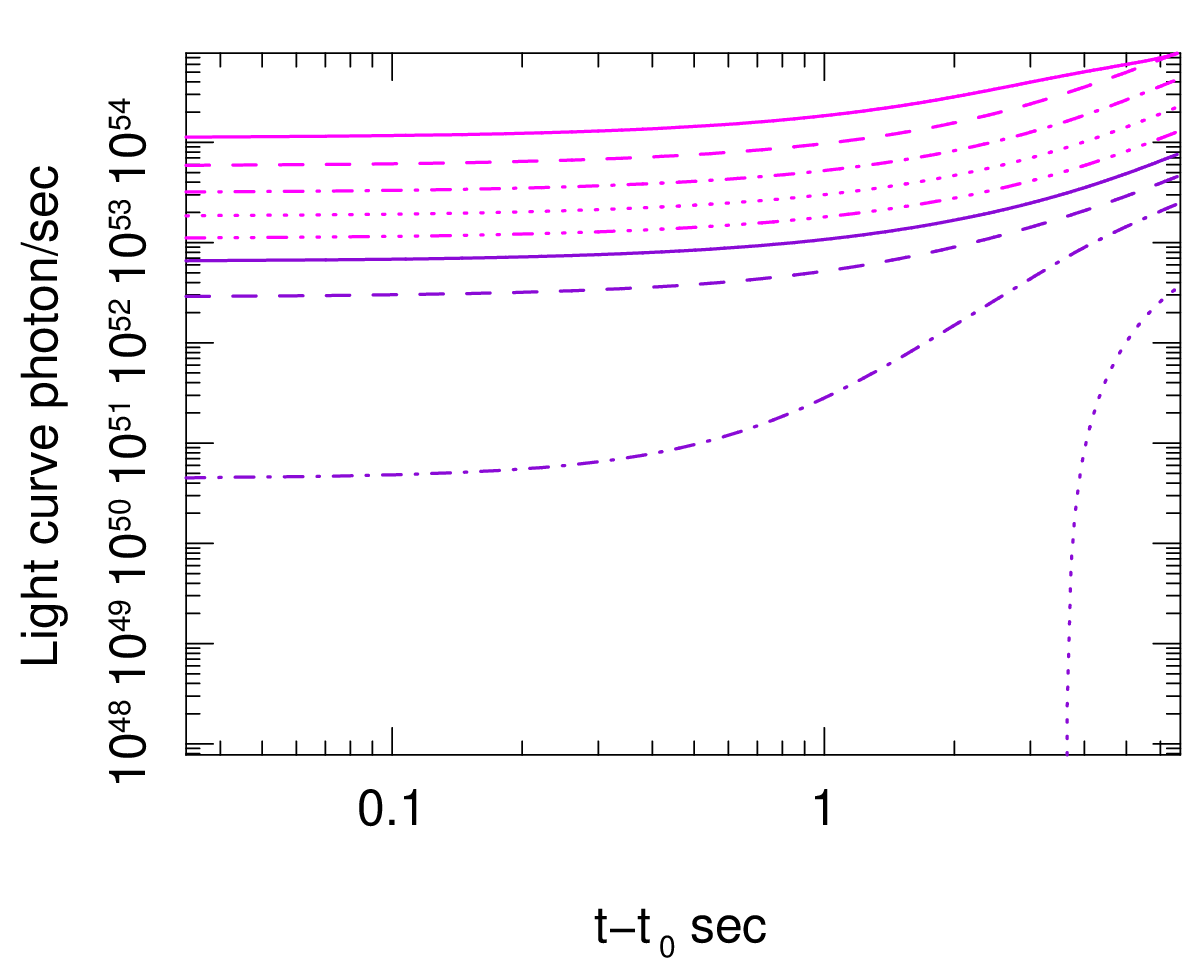} \\
7)\includegraphics[height=5.5cm,angle=-90]{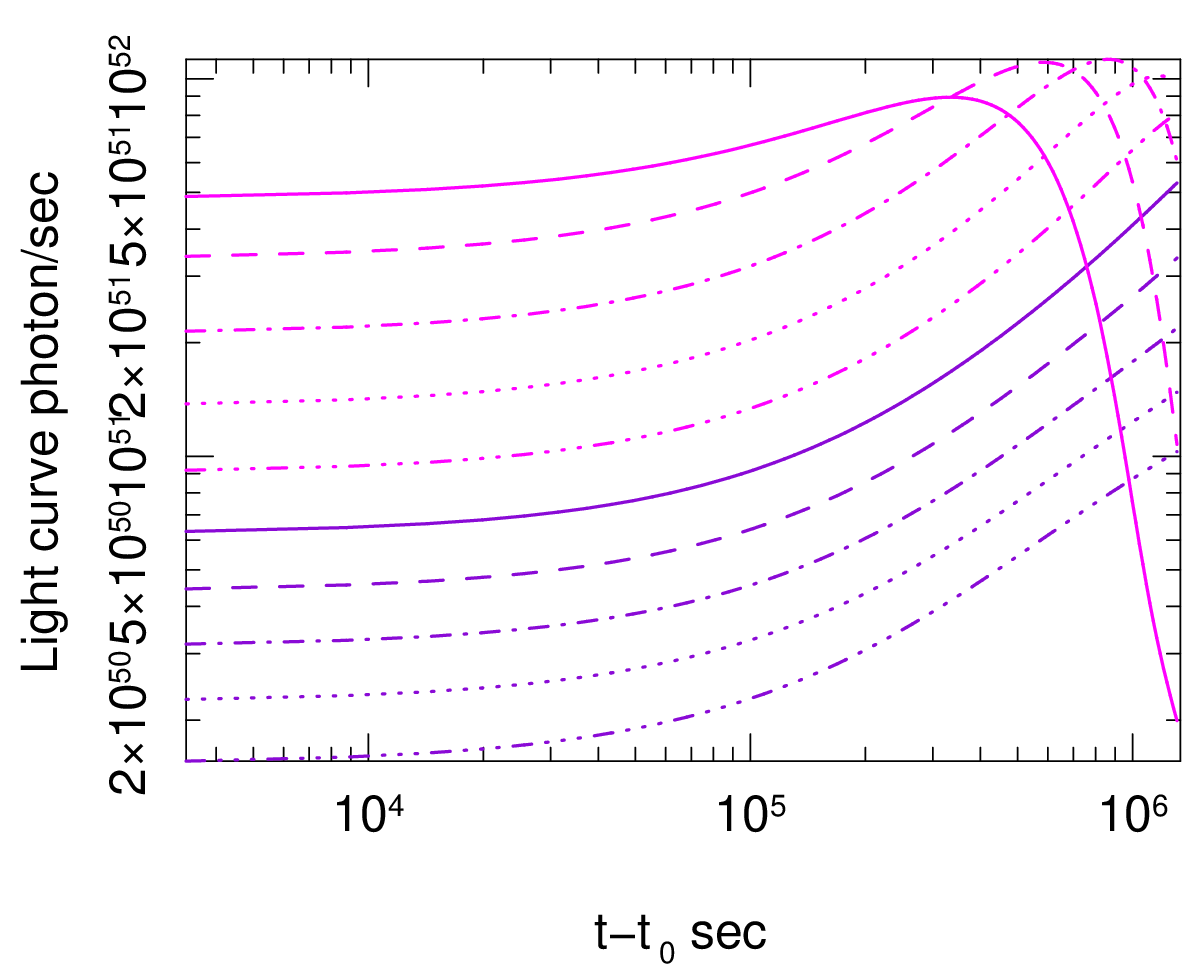} &
8)\includegraphics[height=5.5cm,angle=-90]{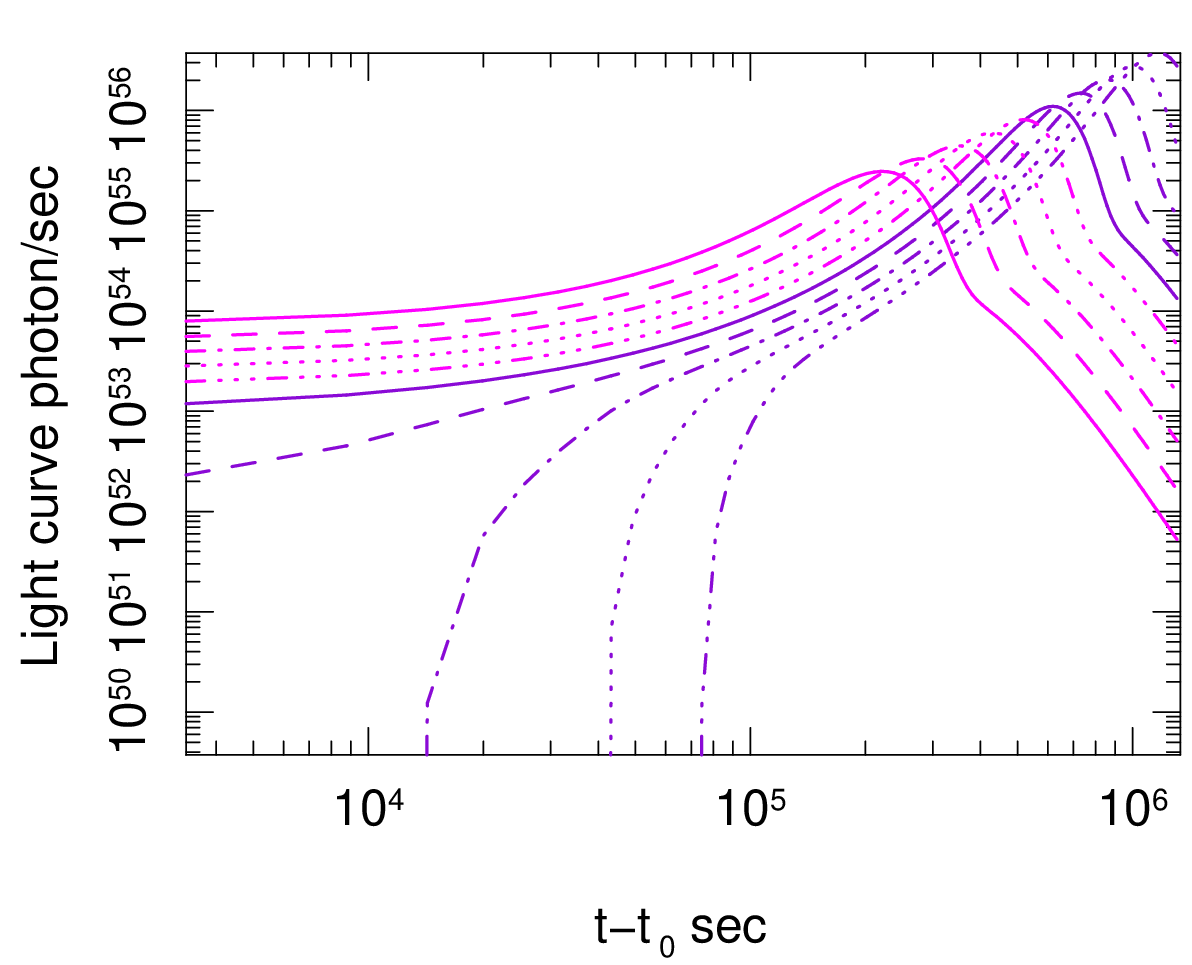} &
9)\includegraphics[height=5.5cm,angle=-90]{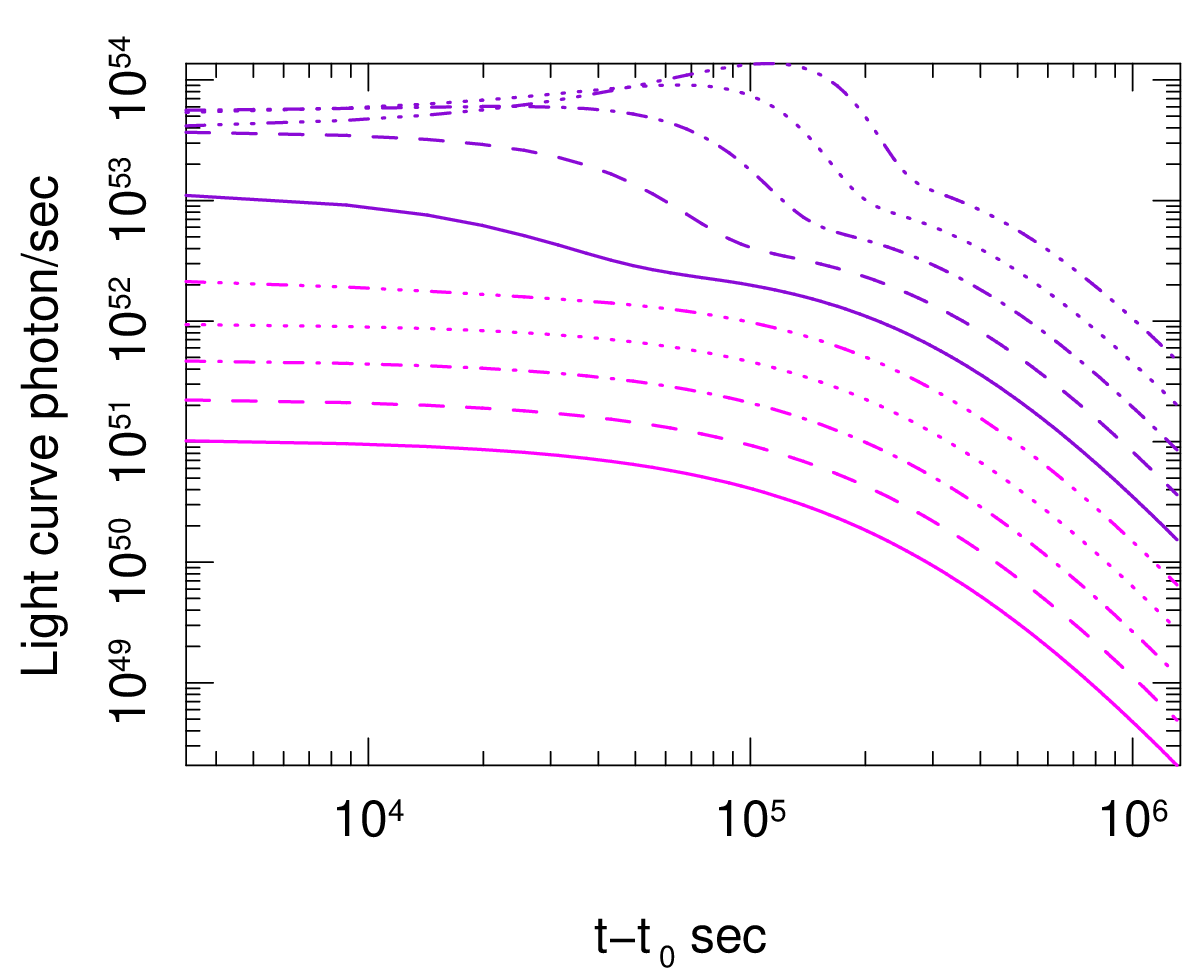} \\
10)\includegraphics[height=5.5cm,angle=-90]{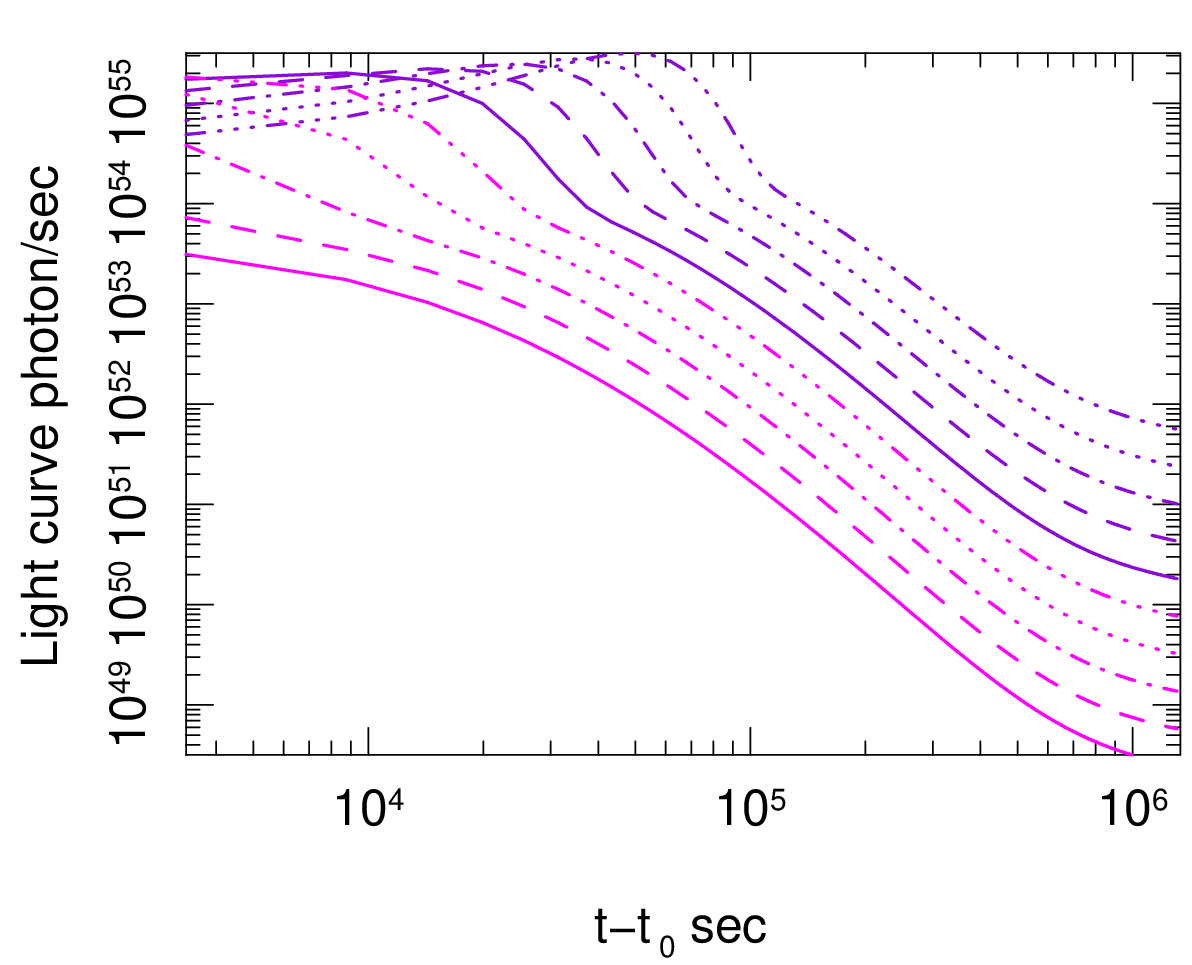} &
\end {tabular}
\end {center}
\caption{Light curves of models 1 to 10 in 10 energy bands from 
$2$ eV to $2$ MeV (models 1 to 6), $2$ eV to $20$ keV (models 7 and 8), and 
$2$ eV to $2$ keV (models 9 and 10), equally separated in logarithmic scale. 
The five highest energy bands (magenta/light grey), the five lowest energy 
bands (purple/dark grey). In each group the full line presents the highest 
energy band. Their active regions evolve according to dynamical model. Models 
1 to 6 are meant to correspond to various regimes during prompt gamma-ray 
emission. Models 7 and 8 presents the onset of afterglows, and models 9 and 
10 the end of afterglows (See the text for detailed interpretation).
\label{fig:lcgmzero}}
\end{figure*}

First we concentrate on the simulation of the models with a dynamical active 
region. Fig. \ref{fig:lcgmzero} shows a variety of behaviours both in the time 
evolution of the light curves and in the difference between the form of the 
light curve at different energy bands. We should remind that as the 
value of $\Delta r' (r'_0)$ for a dynamical active region model can not be 
zero, we should assume that these simulations show only the evolution of 
the collision after the formation of an active region that is used as the 
initial condition for these simulations. The plots in Fig. \ref{fig:lcgmzero} 
show that in some cases such as models 3, 4, and 5 the parameter set does not 
permit an increase of the emission and it decays. By contrast in models 1, 2, 
and 6 the initial emission evolves to a much harder and brighter emission. 

Model 1 which is hard and square-shape looks like some of the observed hard 
and very bright bursts such as GRB 060105~\citep{grb060105}, 
GRB 060813a~\citep{grb060813a}, GRB 061007~\citep{grb061007,grb061007-1}, and 
the super burst GRB 080319b~\citep{grb080319b,grb080319b1}. The hardening is 
also reflected in the increasing $\gamma'_m$ and $\omega'_m$ in this model see 
Figs. \ref{fig:gammam} and \ref{fig:omegam}. Comparing the 
parameters of this model with other models, it is clear that the main 
difference is the evolution of the electric and magnetic fields\footnote{When 
we talk about {\it the evolution of the electric and magnetic fields}, we 
actually mean the evolution of the fraction of kinetic energy of ejecta 
transferred to these fields. Fractions are scalar values in contrast to fields 
which are vectors and their variation is more complex.}. They both increase 
with radius, but the electric field rising slope is much sharper than 
magnetic field. Therefore it can maintain for relatively long time the 
acceleration of electrons. The growing magnetic field helps to have 
significant synchrotron emission but if the electric field was low, either 
all the synchrotron energy was emitted quickly - like model 2 - 
or the gross of emission was more gradual and made a triangle/Gaussian shape 
peak. The early decay of the emission in lowest energy bands is most probably 
due to numerical insufficiency. However, it is possible that the early fall 
of the optical and ultra-violet light curves are real and due to the 
hardening of the emission. 

In a real burst at some point the fields stop growing, the emission 
becomes softer and then breaks. Evidently, the evolution of fields according 
to a power-law with a constant index considered for each of the simulations 
here can not realistically reproduce the variation of parameters in a single 
simulation. Model 6, for instance, can be a prototype of long, few seconds, 
roughly constant emission following the rise of model 1. Model 2 shows a much 
faster rise similar to what happens in the short bursts. 
Fig. \ref{fig:gammaone} shows that at maximum emission the shock becomes very 
quickly non-relativistic, i.e. $\gamma' \rightarrow 1$. One can predict that 
this should stop the growing of fields and in fact leads to their fast decay 
and thereby fast decay of the mission. This situation is well presented by 
model 3 in which the light curve decay in less than 0.1 sec. In this model 
the lag between high energy bands is $\sim 20$ msec. It should be possible 
to reduce the lag and find values consistent with what is observed for short 
bursts. Most probably this needs an exponential decay of the fields which is 
not included in the present work. Models 4 and 5 are typical end of emission 
regime for long bursts in which a large lag, few hundreds of seconds, 
between energy bands appears. Another interesting aspect of this model is the 
very slow initial evolution of the optical and ultra-violet bands and 
relatively late rise of the optical bands. If the burst stays hard the peak of 
the emission in lower energy bands, X-ray and optical can have much larger 
lags. For instance, in models 4 and 5 soft X-ray and UV/optical bands are yet 
rising when the highest energy band in gamma-ray had decayed by more than one 
order of magnitude. Such large lag in low energy bands can be the origin of 
the peak observed in the X-ray - specially in short hard bursts see 
e.g.~\citep{grb070724a}- and optical light curve of some long 
bursts~\citep{sampaper}, although an external shock origin can not be ruled 
out. On the other hand the steep decay slope of these models specially at 
softer bands can explain the very steep decline of the X-ray and optical 
emissions - usually called the {\it tail emissions}- seen in the majority of 
the \swift early afterglows~\citep{enereff}. 
In fact multi-band simultaneous observations of late gamma-ray peaks in bursts 
such as GRB 060124~\citep{grb060124,grb060124-1}, 
GRB 061121~\citep{grb061121,grb061121-1}, and GRB 070724A~\citep{grb070724a} 
have already proved that this regime of X-ray light curves is in close 
relation with the prompt emission~\citep{enereff,earlyemi1,shallowx} 
consistent with the prediction of models presented here. In all these 
simulations the time duration of the emission (width of the peak) in higher 
energy bands are shorter than lower energies, consistent with the 
observations~\citep{peakwidth}

Models 7 to 10 are simulated with parameters for afterglows. From 
Fig. \ref{fig:deltar} it is clear that in model 7 the width of the active 
region 
is decreasing although it has a negative $\tau$. By contrast in model 8 which 
has a positive $\tau$ the $\Delta r'$ increases. Nonetheless in both cases 
the column density decreases rapidly, see Fig. \ref{fig:coldens}. These 
models have a relatively large Lorentz factor and hard emission in X-ray and 
their light curves are very flat or rising until hundreds of kilo seconds. 
Apriori this can explain the shallow regime of the X-ray light curves seen in 
many bursts~\citep{enereff,shallowx} if the shock stays only for a limited 
time of up to few thousands of seconds in this regime early after the onset 
of the external shock. The late decay of light curves of these models is 
too fast and inconsistent with observations. Therefore such a model can only 
be an example of early afterglows. We should also remember that 
the early emission in X-ray and optical bands is in part due to the tail 
emission of the prompt shock~\citep{xrtafterglow} and therefore a combination 
of the two component must be considered. We come back to this point in the 
next section.

In model 9 and 10 all the 
parameters have values similar to the previous two models but their Lorentz 
factor is much smaller. In this case the shallow regime is declining, in 
contrast to the previous models and the late decay has a slope of $\sim -1.5$ 
to $-2.5$ consistent with observations~\citep{enereff}. The form of the 
light curve also is grossly consistent with previous 
simulations~\citep{emission1} although the present models include more 
parameters. In fact if 
a regime similar to model 9 or 10 happens at late times $\gtrsim 10^5$ sec 
after trigger, in logarithmic scale the shallow part will be a very short and 
insignificant interval. Thus the observer sees a break from a previous 
regime, presumably a shallow regime similar to models 7 and 8, to a declining 
regime with a slope of $\sim -1.5$ to $-2$ that then breaks to a steeper 
slope of $\sim -2.5$ to $\sim -3$ at very late times. Such a behaviour has 
been 
detected in many of \swift bursts~\citep{xlcuni}. Our tests show that slopes 
somehow depend on $p$, larger $p$ steeper the decline slope. This is also 
consistent with previous simulations~\citep{emission1}. Nonetheless, as we 
explained in Paper I, flatter slopes can be obtained at low energies if there 
is an exponential cutoff in the electron distribution at high energies. This 
option is not included in present simulations and we leave its exploration 
to a future work. As mentioned before, the inclusion of high latitude 
emissions also make the decline flatter.

\subsection{Light curve of simulated models with quasi-steady active region}
\label{sec:lcqsar}
\begin{figure*}
\begin{center}
\begin {tabular}{ccc}
11)\includegraphics[height=5.5cm,angle=-90]{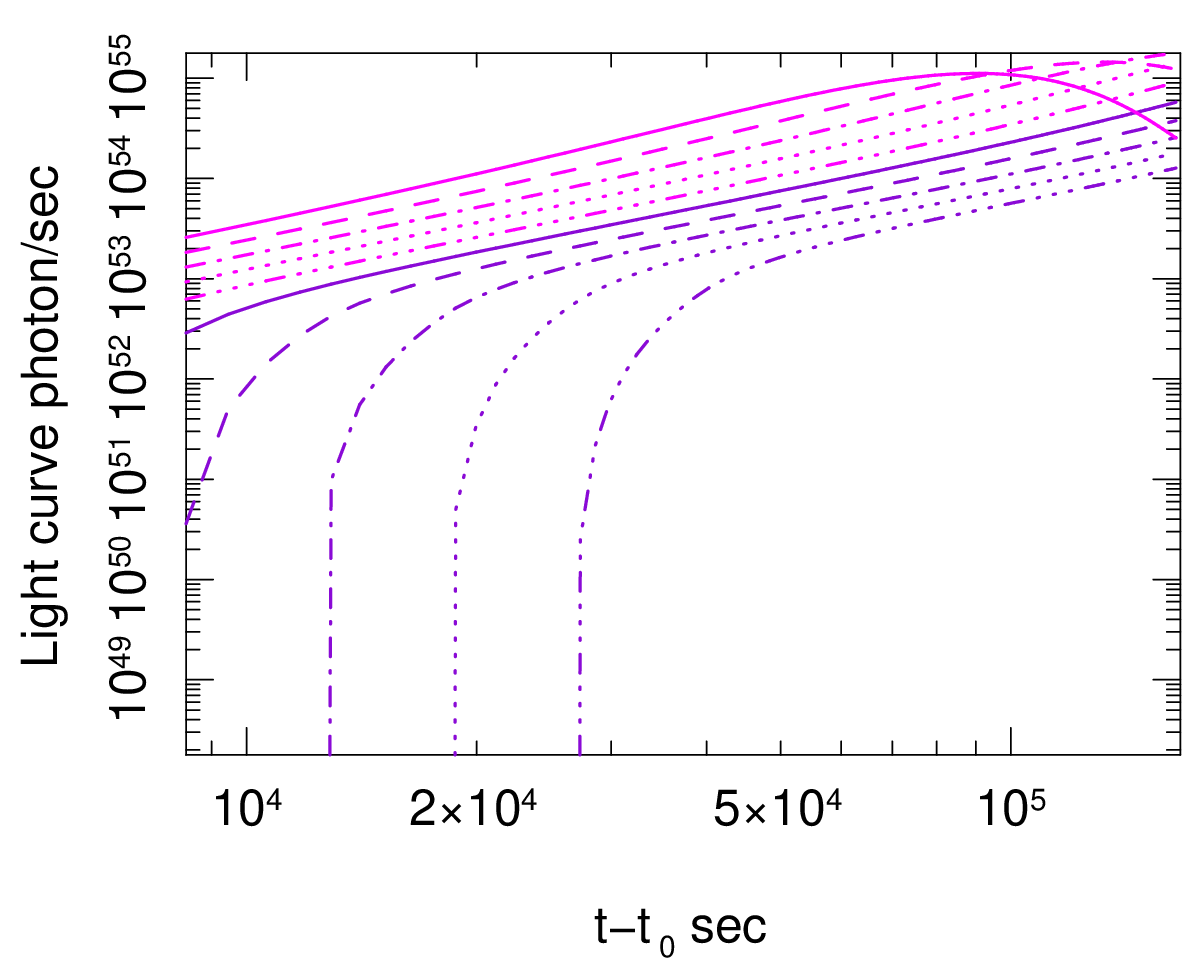} &
12)\includegraphics[height=5.5cm,angle=-90]{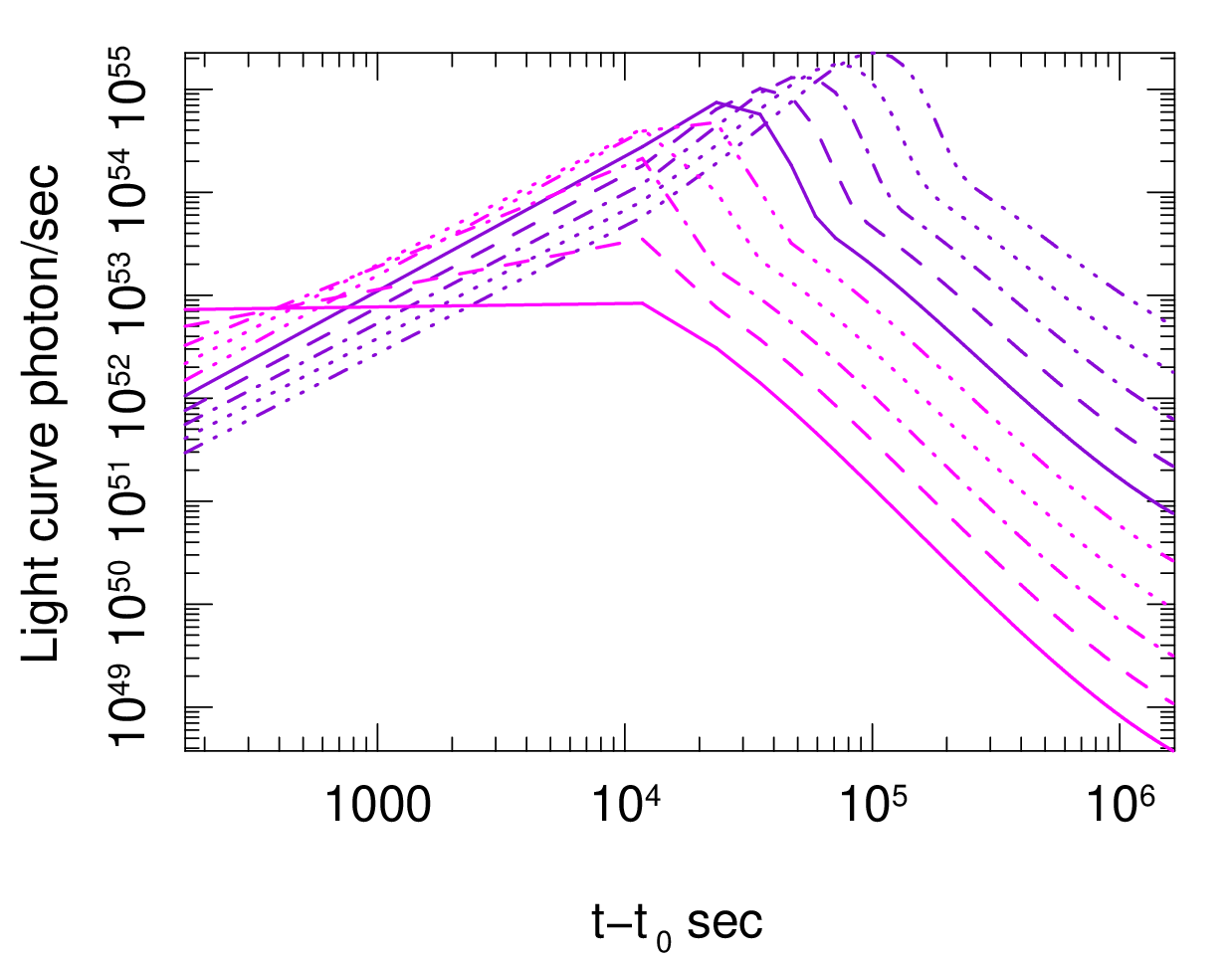} &
13)\includegraphics[height=5.5cm,angle=-90]{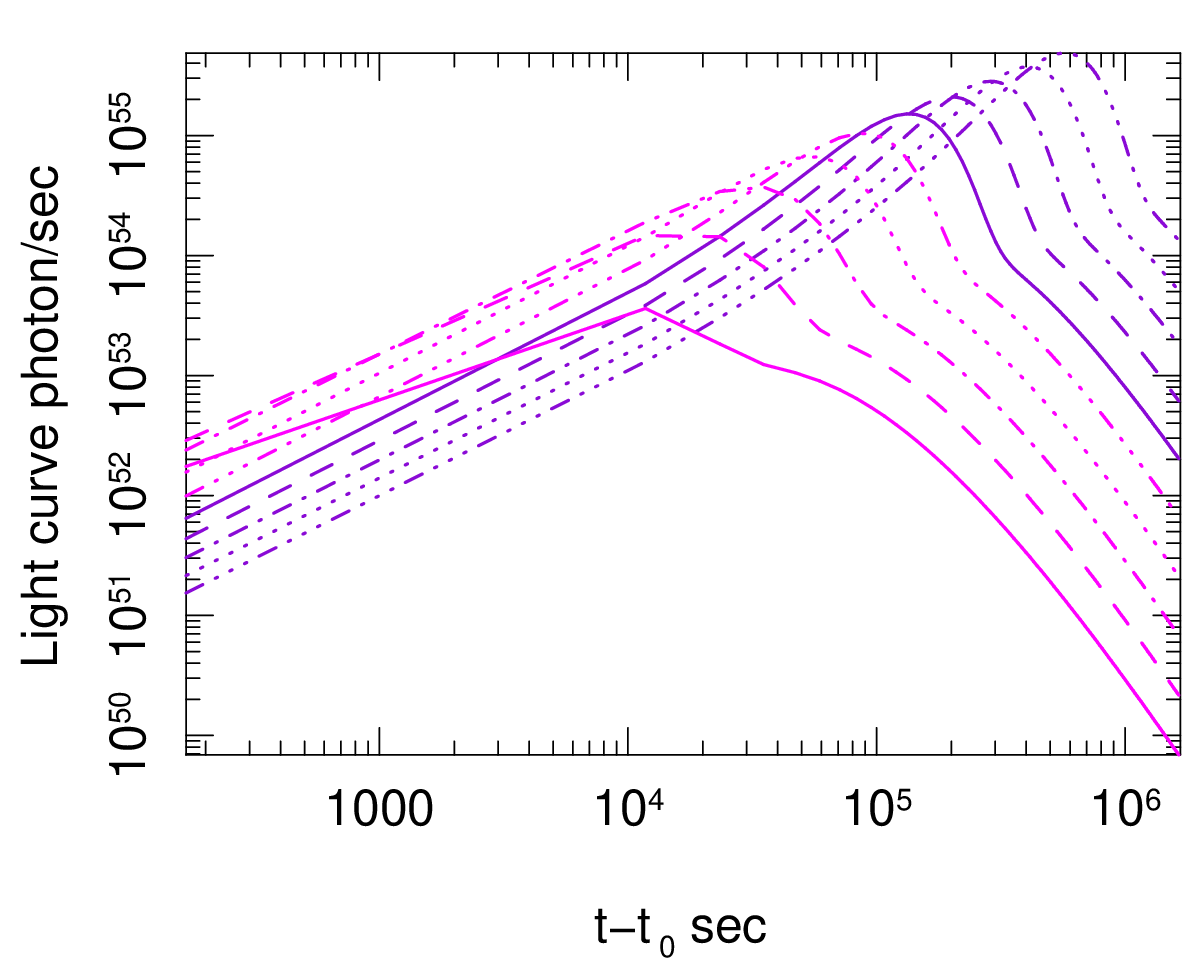} \\
14)\includegraphics[height=5.5cm,angle=-90]{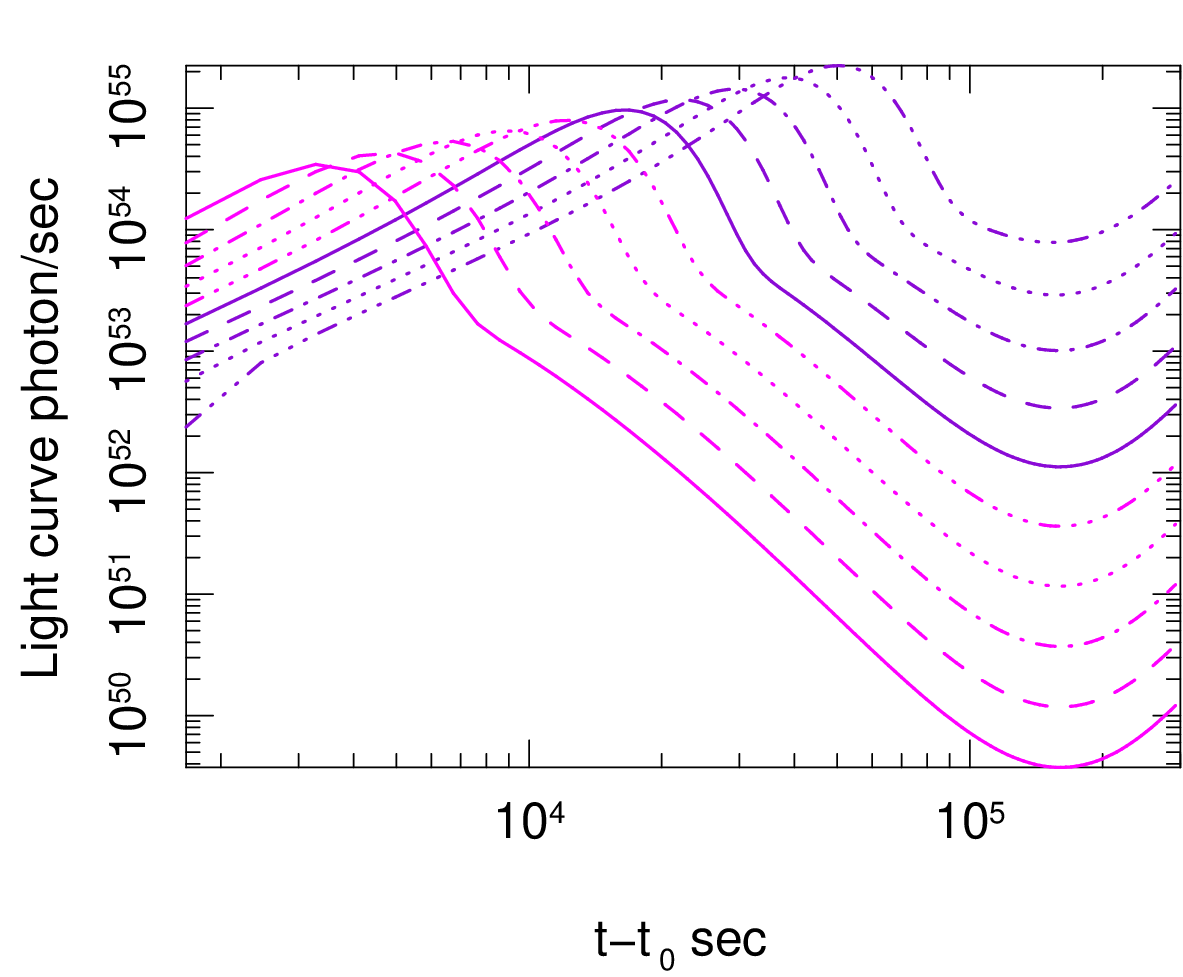} &
15)\includegraphics[height=5.5cm,angle=-90]{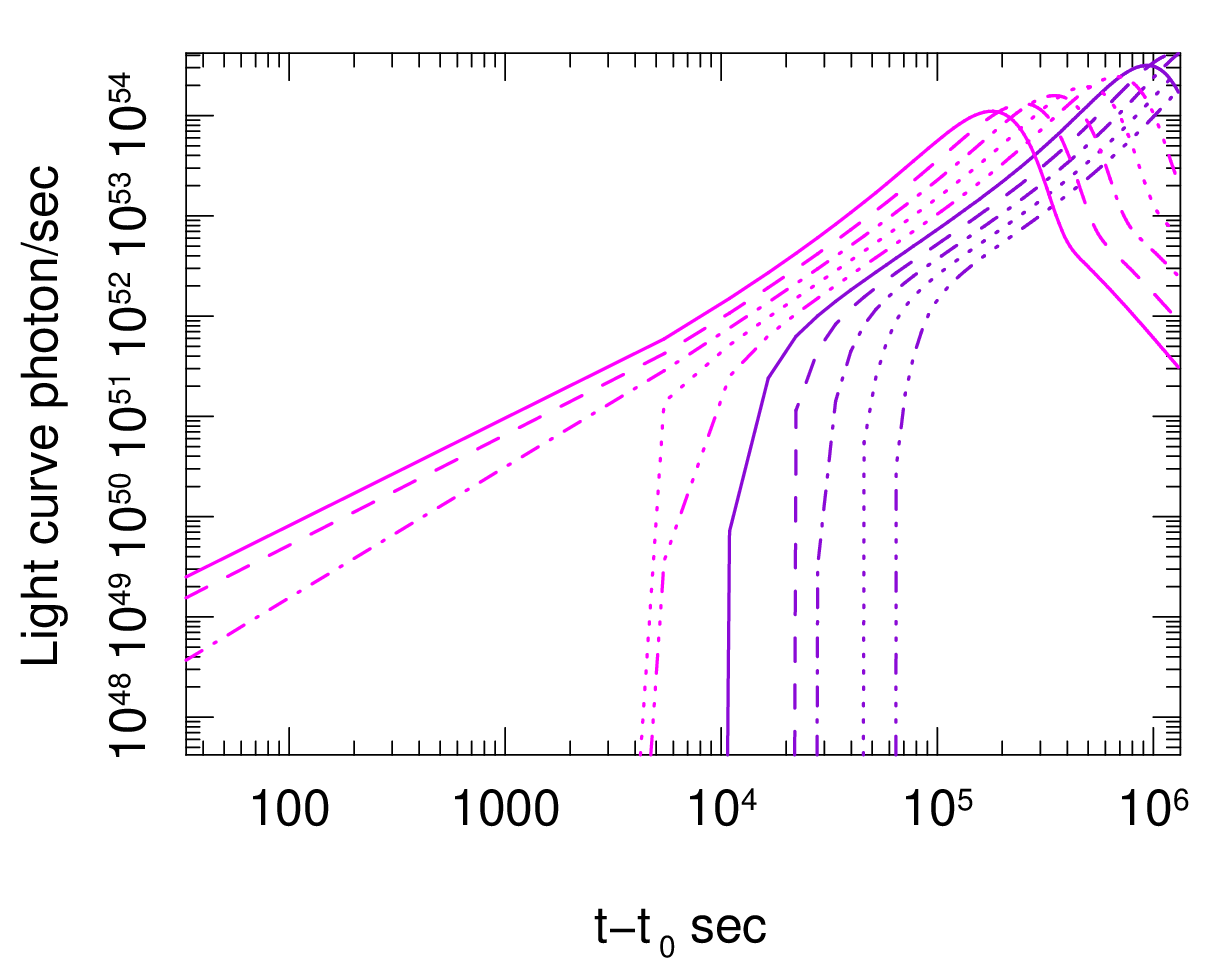} &
16)\includegraphics[height=5.5cm,angle=-90]{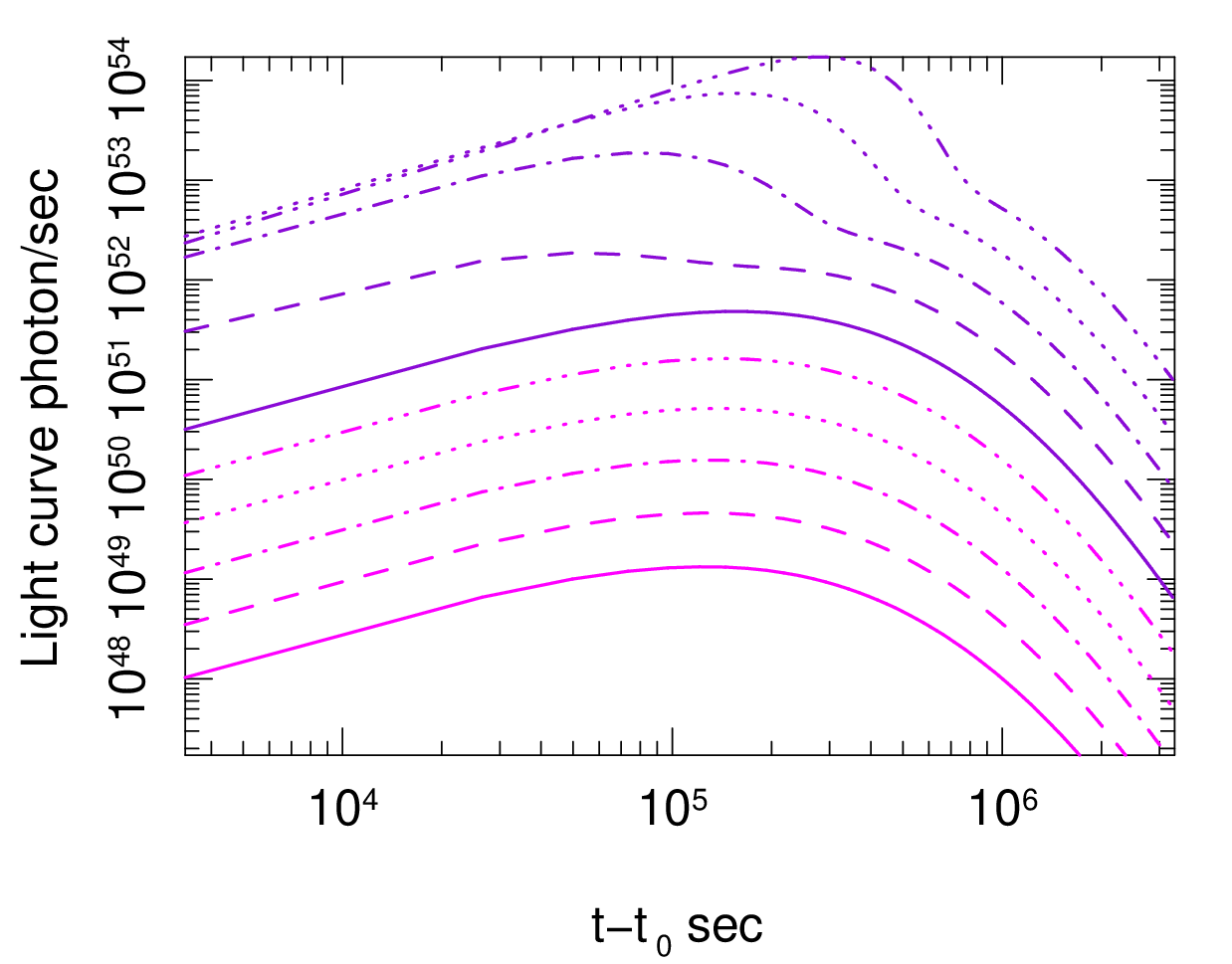} \\
17)\includegraphics[height=5.5cm,angle=-90]{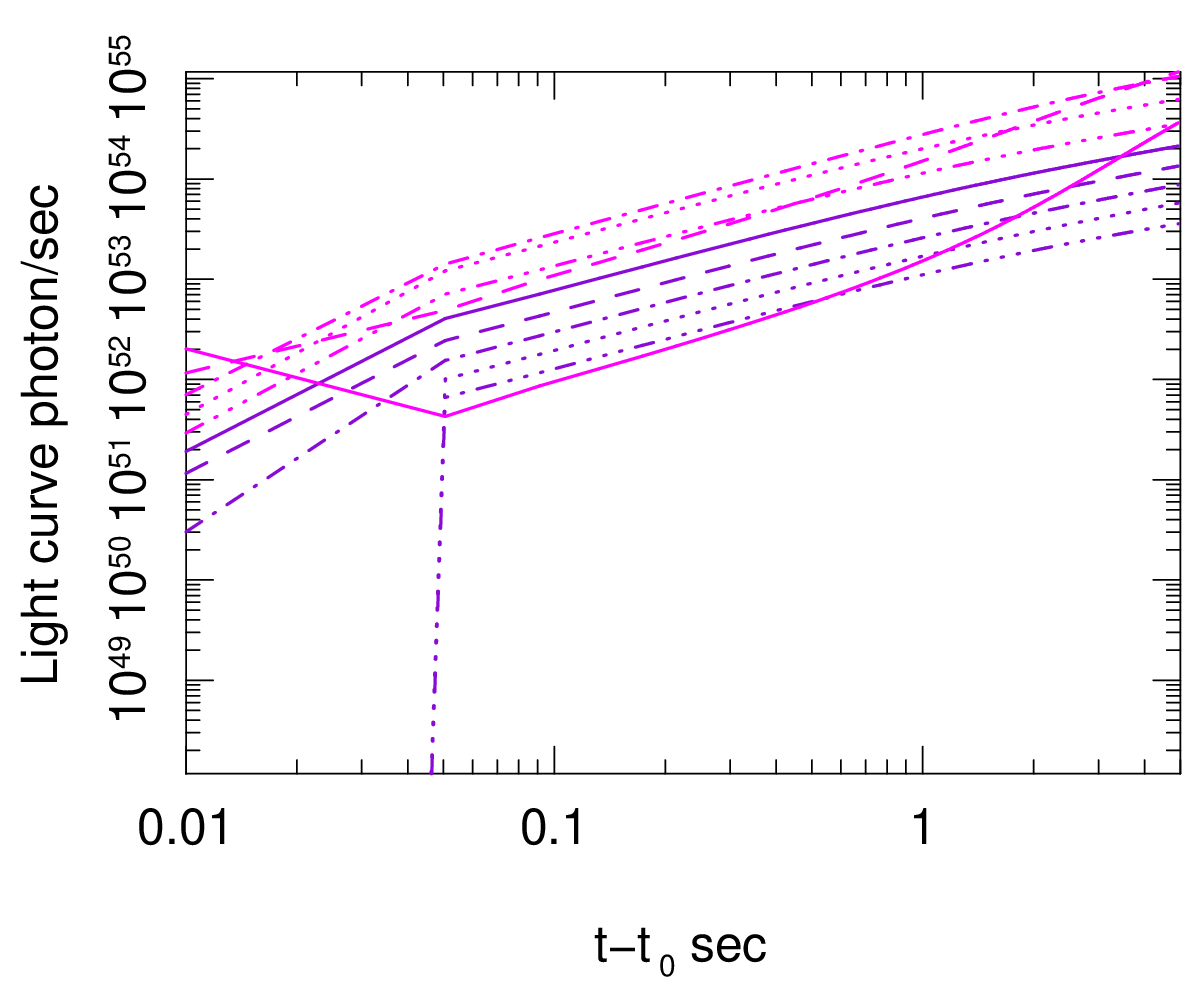} &
18)\includegraphics[height=5.5cm,angle=-90]{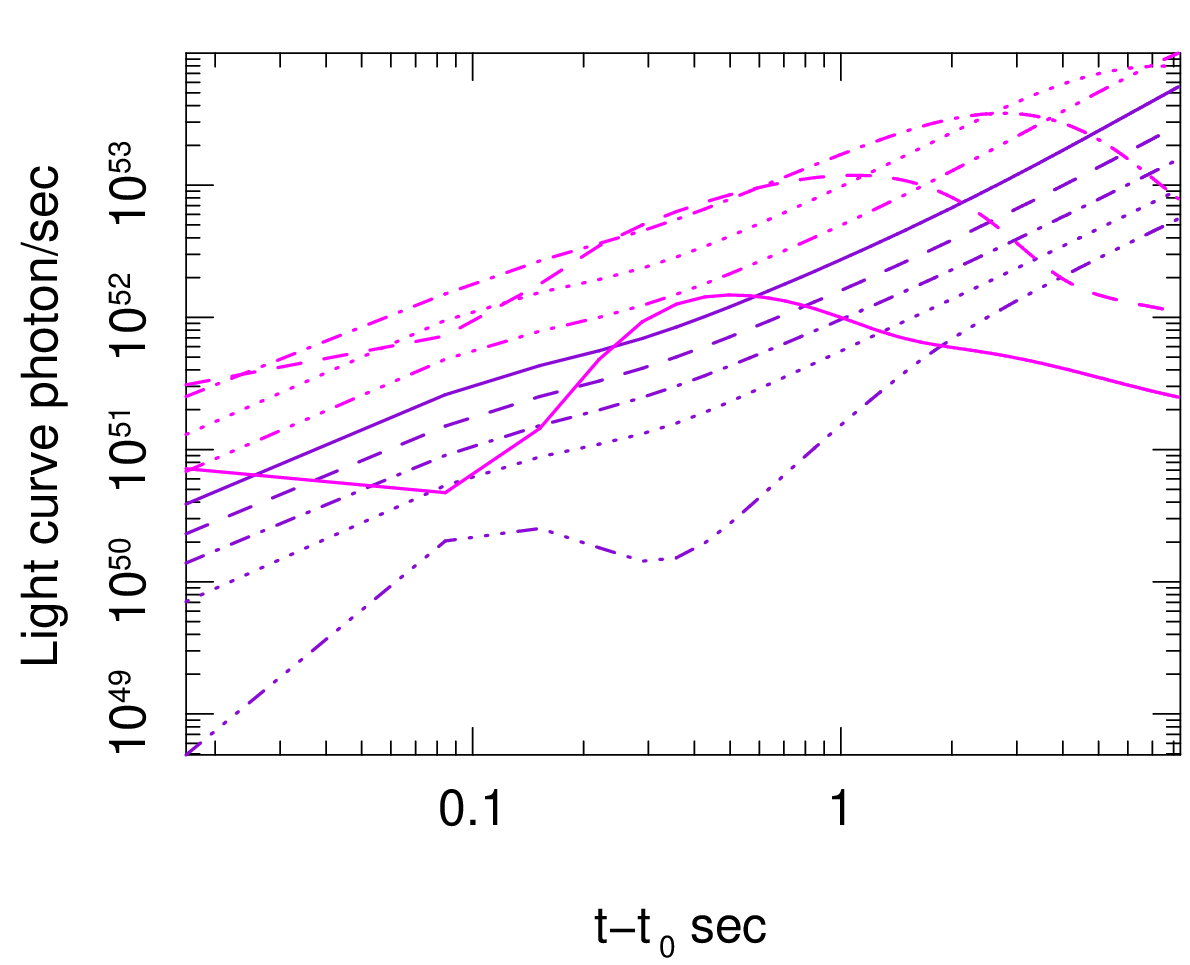} &
19)\includegraphics[height=5.5cm,angle=-90]{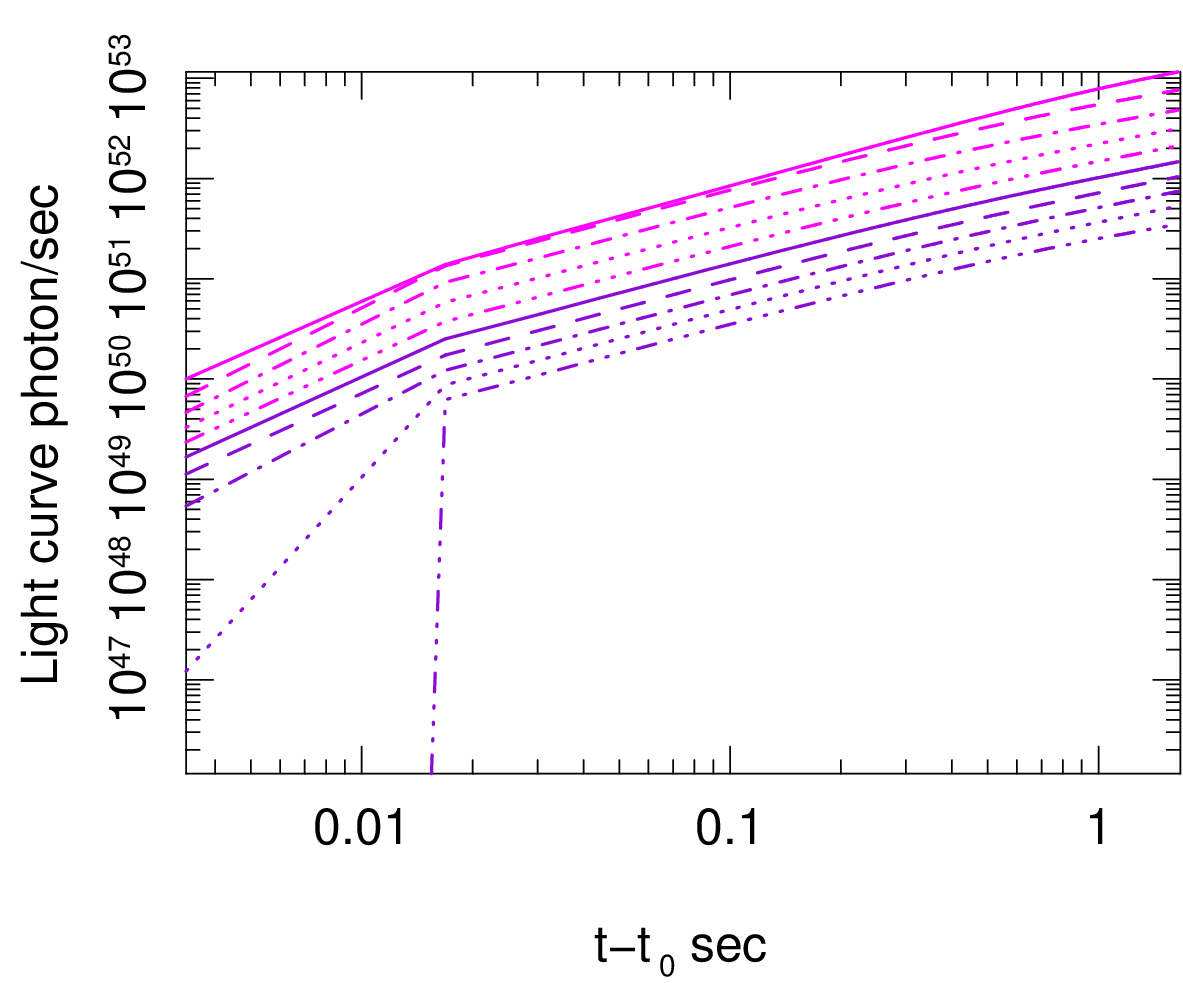} 
\end {tabular}
\end {center}
\caption{Light curves of models 10 to 19 in 10 energy bands. For models 11 to 
16 the energy bands are from $2$ eV to $10$ or $20$ keV, equally separated in 
logarithmic scale. They are meant to correspond to onset of an afterglow. For 
models 17 to 19 the energy bands are from $2$ eV to $2$ MeV corresponding a 
prompt emission. Their active regions evolve according to quasi-steady state 
model. Descriptions of the curves are the same as Fig. \ref{fig:lcgmzero}.
\label{fig:lcgmone}}
\end{figure*}

Models 11 to 18 in Fig. \ref{fig:lcgmone} are simulated with a quasi steady 
active region, the first six models as external shocks and the last three  
as internal shocks. The first noticeable feature is the similarity between 
the global trends of their light curves, in contrast to the variety of 
behaviours we have seen in models with dynamical active region. This is 
consistent with the \swift~\citep{swift} observations of X-ray and UV/optical 
late emissions\footnote{We discussed that the observed emission in lower 
energies after the prompt peak most probably consists of a combination of 
emission from the prompt collision and the {\it afterglow} defined as the 
emission from the collision of ejecta/jet with the surrounding material or 
ISM~\citep{xrtafterglow}. Therefore we refer to this emission as 
{\it late emission} rather than the {\it afterglow} which usually is 
considered to be the emission from external shocks}~\citep{xrtafterglow}. 
Both X-ray and optical light curves look like the light curves of 
simulated external shocks in~\cite{emission1}. Their similarly to previous 
simulations is more visible than models 7 and 8 with dynamical active 
region. 

In general these models show a large lag of few thousands of seconds between 
the time of peak emission at hard and soft X-ray bands, versus UV and optical 
bands. This is consistent with the previous calculations~\citep{emission1}. 
But the general aspects of the light curves are not similar to what is 
observed in real bursts. Therefore if the early emission was exclusively 
from external shocks these models were rules out. On the other 
hand, it has been shown~\citep{xrtafterglow} that the X-ray light curves are 
composed of two overlapping components, one related to the prompt emission, 
and the other can be so called the afterglow. Moreover, the general form of 
the light curves in Fig. \ref{fig:lcgmone} is similar and consistent with the 
fit performed in~\cite{xrtafterglow} for the afterglow component. In this 
case the shallow regime observed in many bursts presents the interval in 
which the prompt component is decaying and the afterglow is gradually taking 
over~\citep{varmicro,shallowregime}. This can be also the reason for the 
hardening of the emission, because the decaying prompt emission can be at 
this time softer than growing emission from the external shock. As explained 
in the previous section, another possibility for explaining the shallow 
regime is the change in the evolution trend of the active region from 
quasi-steady state to dynamical after the initial rise of emission in the 
previous regime. The slower decay rate of the low energy light curves in 
internal shocks and the large lag of the break in these bands with respect 
to X-ray break in the external shock emissions is consistent with the lack or 
very late break seen in the optical light curves of the bursts detected by 
\swift/UVOT~\citep{sampaper}. A shallower 
electron distribution at low energies and an exponential break at high energy 
can also be responsible for shallow optical lightcurves~\citep{jetbreakexp0}.
As for the decay slope after the break in models 11 to 13 it is about $-2$ and 
in model 14 about $-2.5$. Similar to dynamical models for late external shock 
emission a shallower electron distribution leads to a shallower break slope. 
In model 15 the break slope is very steep and inconsistent with 
observations. Therefore this model can simulate the rise of the afterglow 
and not its decay. Model 16 also has a steep slope but its Lorentz factor is 
not very large and high latitude emissions should significantly flatten the 
decay slope.

\begin{figure*}
\begin{center}
\begin {tabular}{ccc}
11)\includegraphics[height=5.5cm,angle=-90]{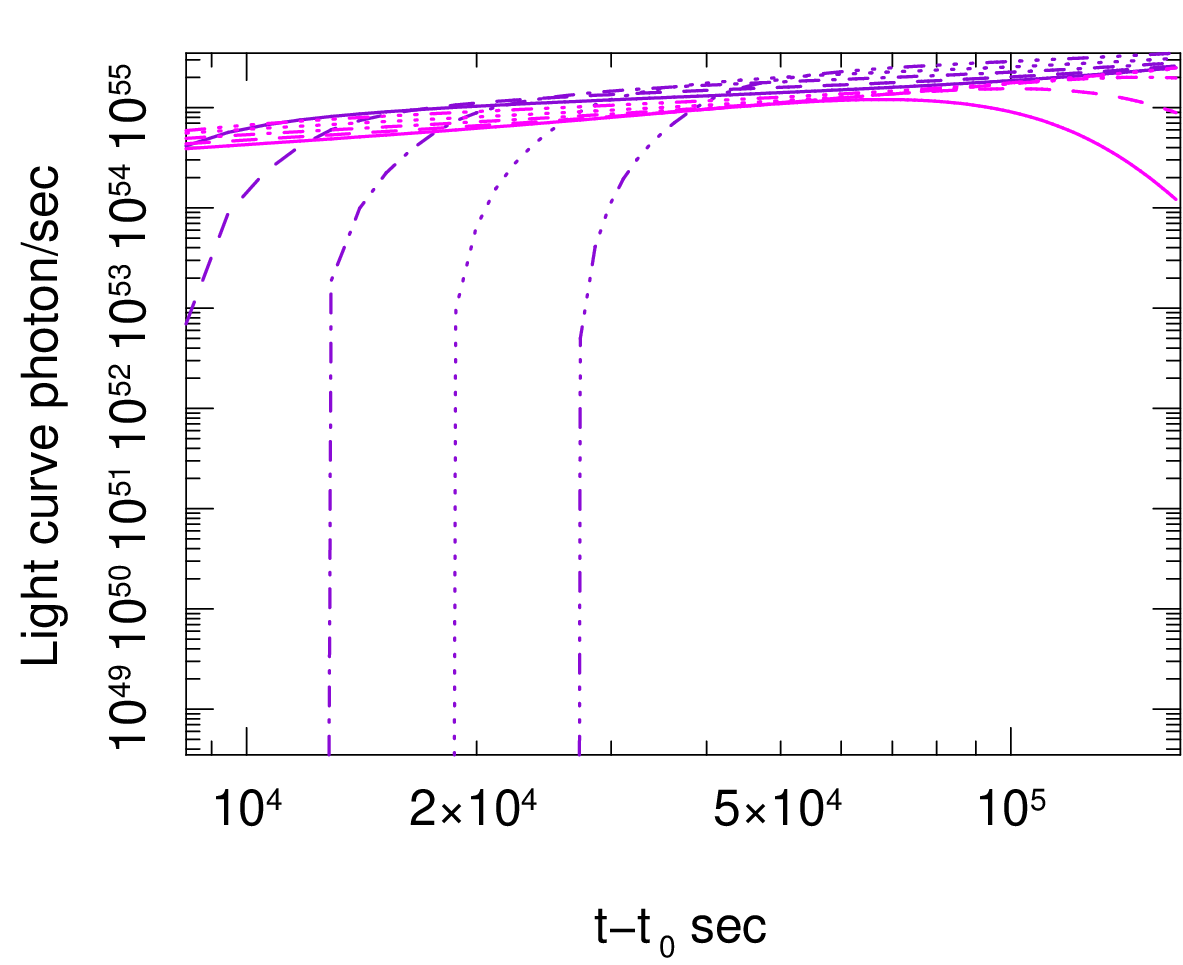} &
12)\includegraphics[height=5.5cm,angle=-90]{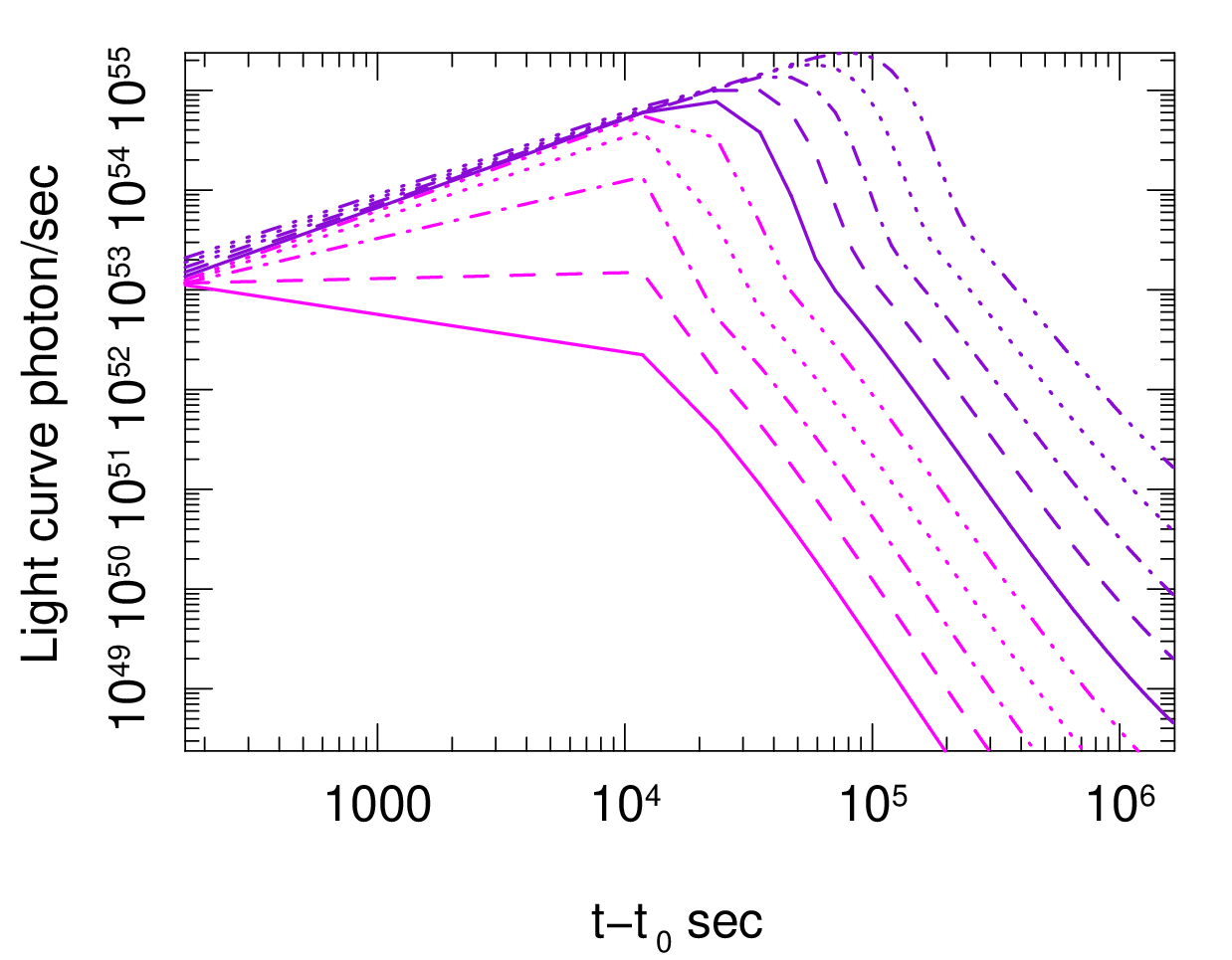} &
13)\includegraphics[height=5.5cm,angle=-90]{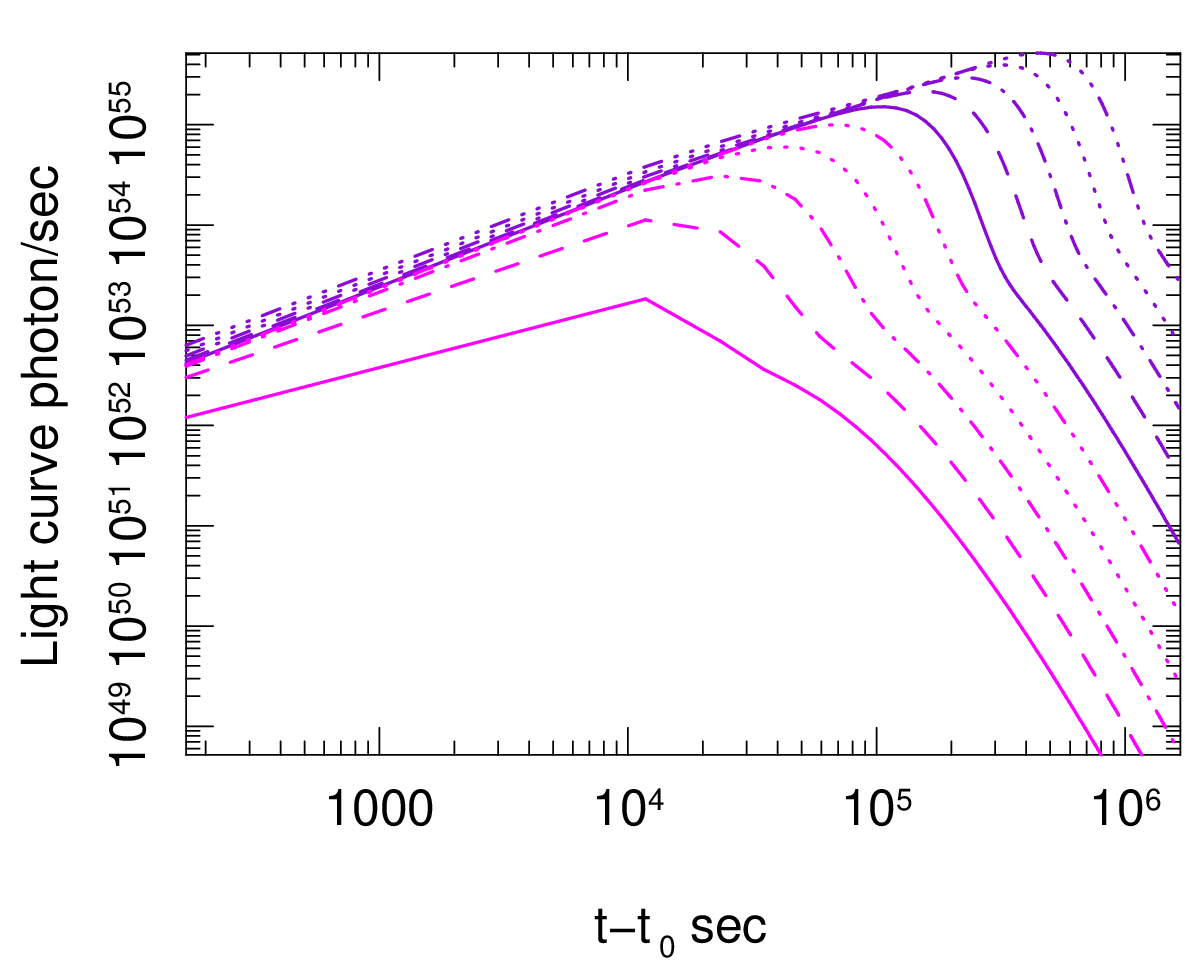} \\
14)\includegraphics[height=5.5cm,angle=-90]{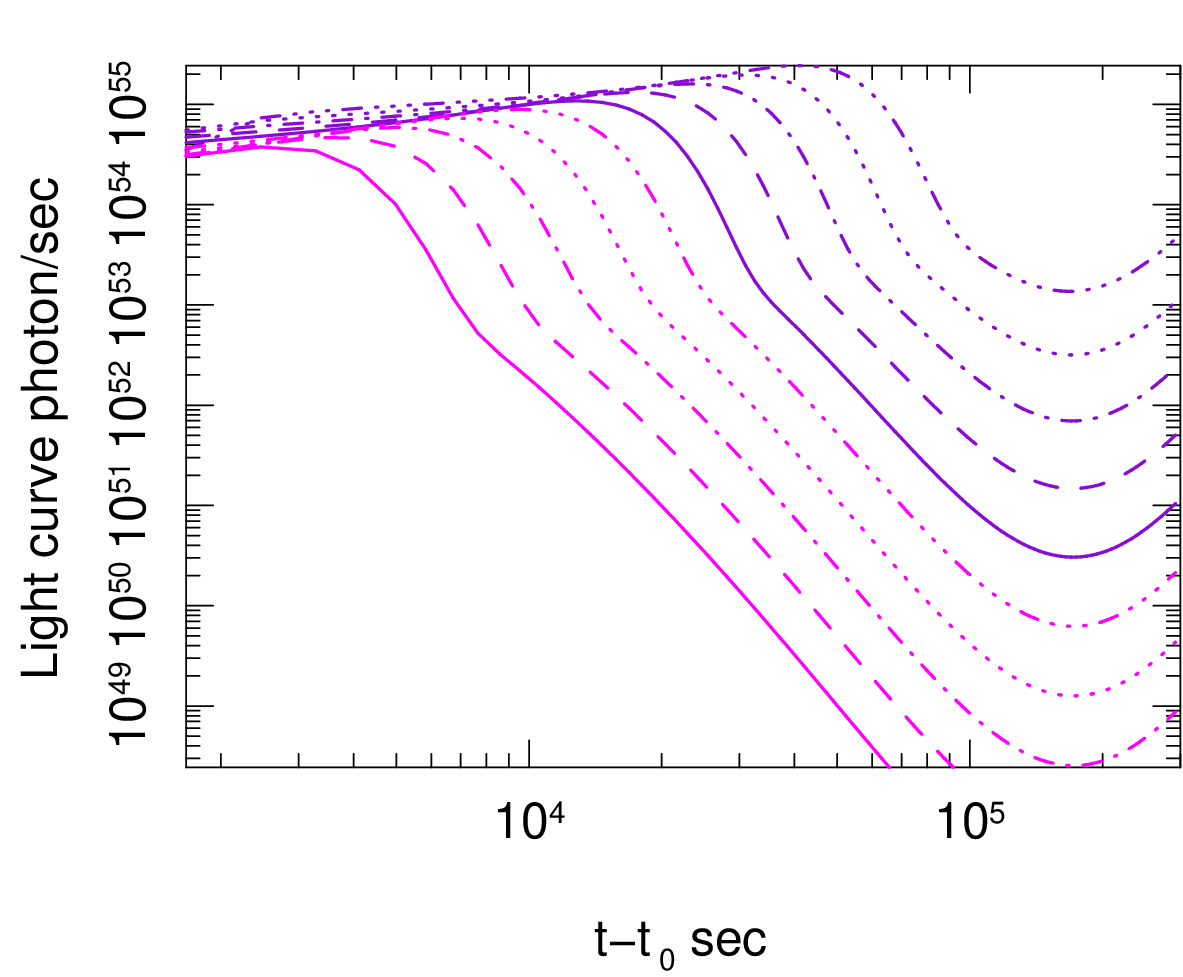} &
15)\includegraphics[height=5.5cm,angle=-90]{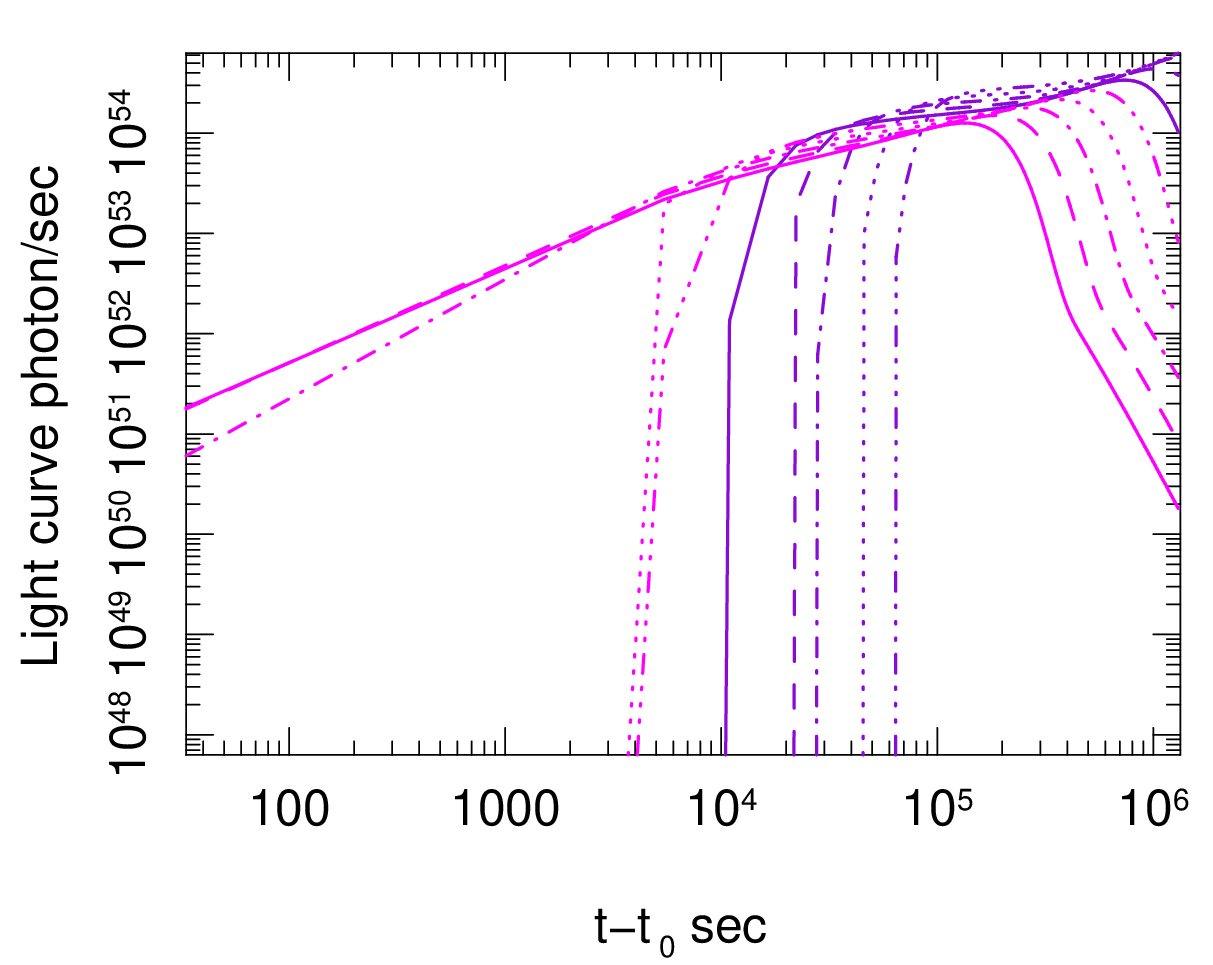} &
16)\includegraphics[height=5.5cm,angle=-90]{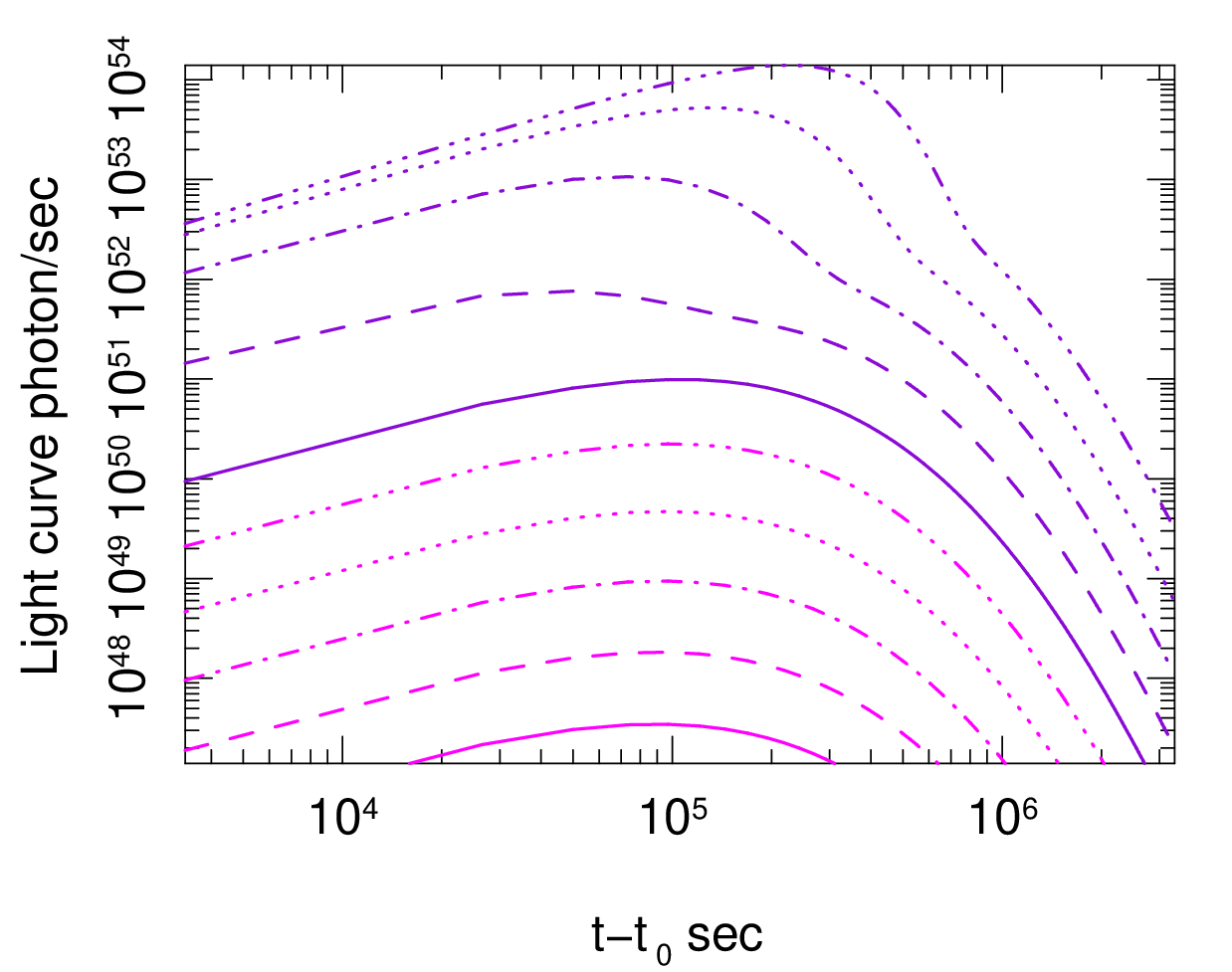} 
\end {tabular}
\end {center}
\caption{Light curves of models 11 to 16 in 10 energy bands with 
$\Delta r \propto (\omega' / \omega'_m)^{-1/2}$. Descriptions of the energy 
bands and curves are the same as Fig. \ref{fig:lcgmone}. 
\label{fig:lcgmonemod}}
\end{figure*}

In these simulations we have neglected the energy dependence of $\Delta r$, 
the width of the active region. The initial ejecta responsible for the 
internal shock is expected to be compact and dense. Thus, its emission in all 
bands come roughly from the same region and the assumption of an energy 
independent $\Delta r$ is justified. Following the prompt collision(s) and 
scattering of the particles, one expects the formation of a low energy 
tail which nonetheless can have enough strong electric and magnetic fields to 
accelerate electrons and produce low energy synchrotron emission. To consider 
such a possibility we have also simulated the same models with the same 
parameters listed in Table \ref{tab:param} but with $\Delta r_0 \rightarrow 
\Delta r_0 (\omega/\omega_m)^{-1/2}$. The selected index of $-1/2$ for energy 
dependence is arbitrary, and we leave a more precise estimation of this 
parameter to a future work in which we try to reconstruct real bursts 
according to the models presented here. The results of these simulations are 
presented in Fig. \ref{fig:lcgmonemod}. The general properties of the light 
curves are preserved but lower energy bands have relatively higher fluxes and 
later breaks consistent with the theoretical explanation in Paper I.

Models 17 to 19 are simulated with parameters relevant to internal shocks. 
They are generally hard and suitable for the rising regime of the prompt 
emission. In the dynamical active region model the initial 
$\Delta r_0 \neq 0$. Therefore somehow an initial active region must form 
before it evolves according to dynamical model. The quasi steady model is a 
suitable approximation for this regime. The hardness of the simulations is 
consistent with the rapid hardening of the emission at the beginning of the 
collision. At high energies model 18 has a FRED shape light curve 
but the lag between gamma-ray energy bands is $\sim 2$ sec. much longer than 
what is observed in long bursts, although very long faint bursts with 
supernova counterpart can accommodate such long lags.

\begin{figure*}
\begin{center}
\begin {tabular}{ccc}
20)\includegraphics[height=5.5cm,angle=-90]{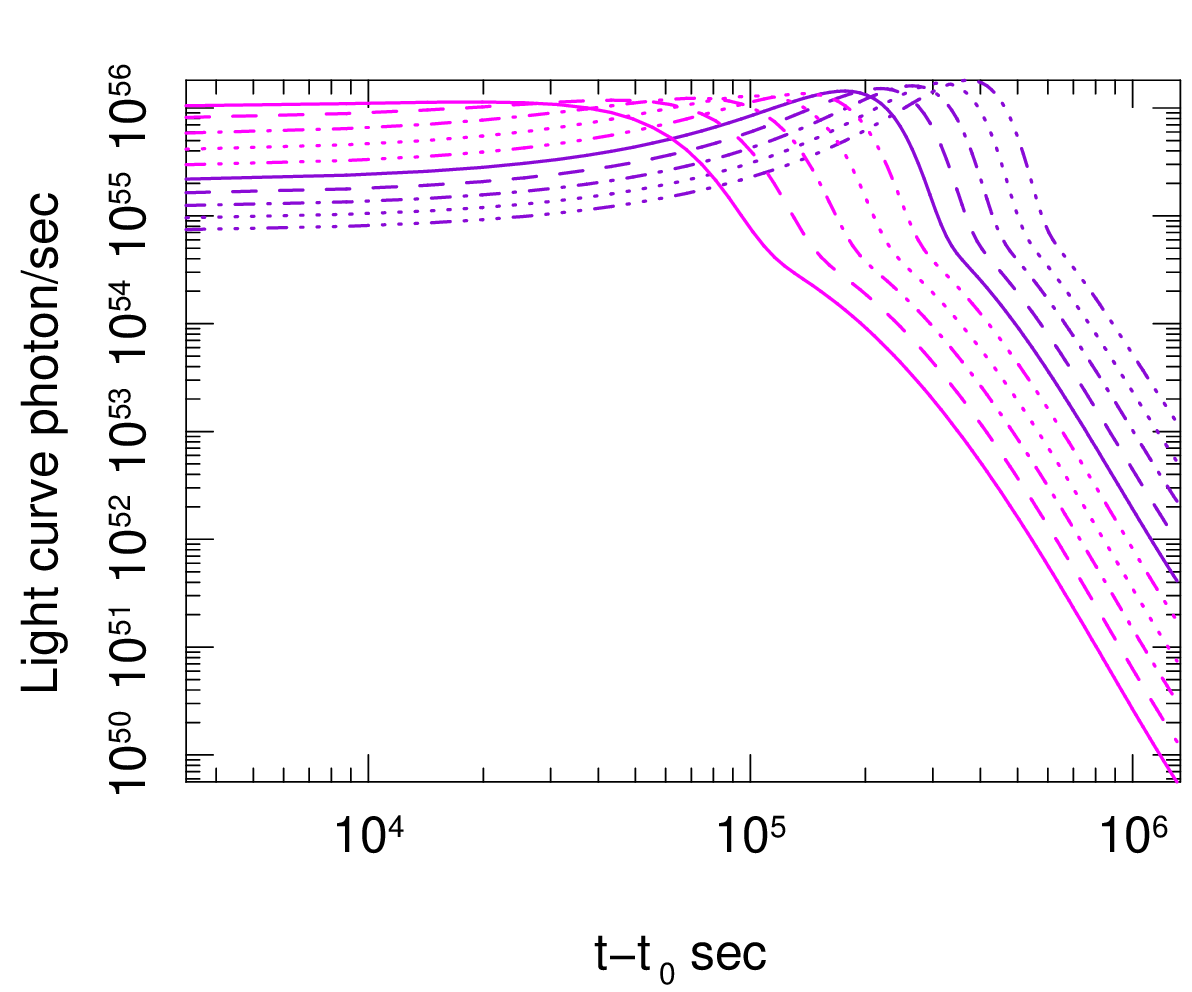} &
21)\includegraphics[height=5.5cm,angle=-90]{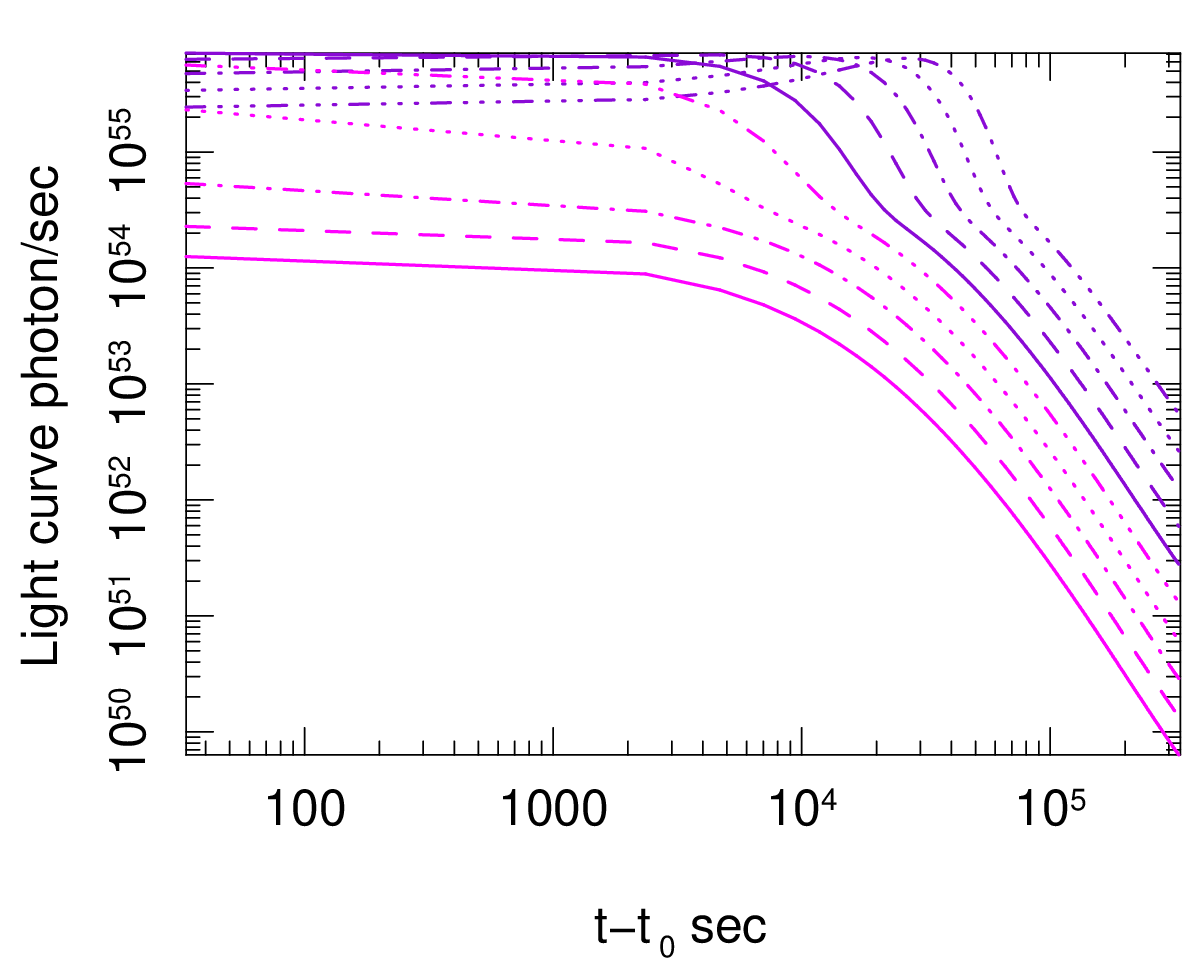} &
22)\includegraphics[height=5.5cm,angle=-90]{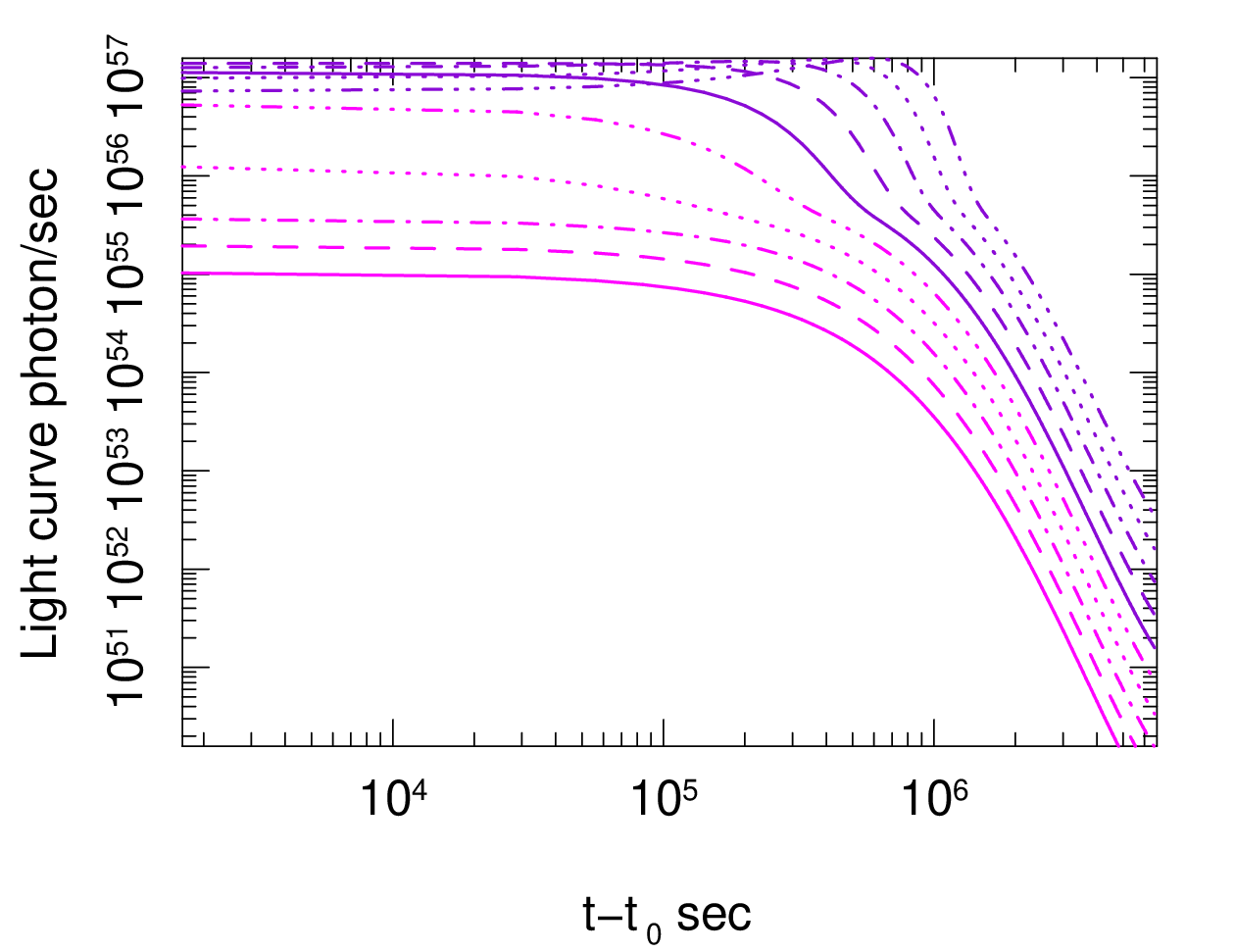} \\
23)\includegraphics[height=5.5cm,angle=-90]{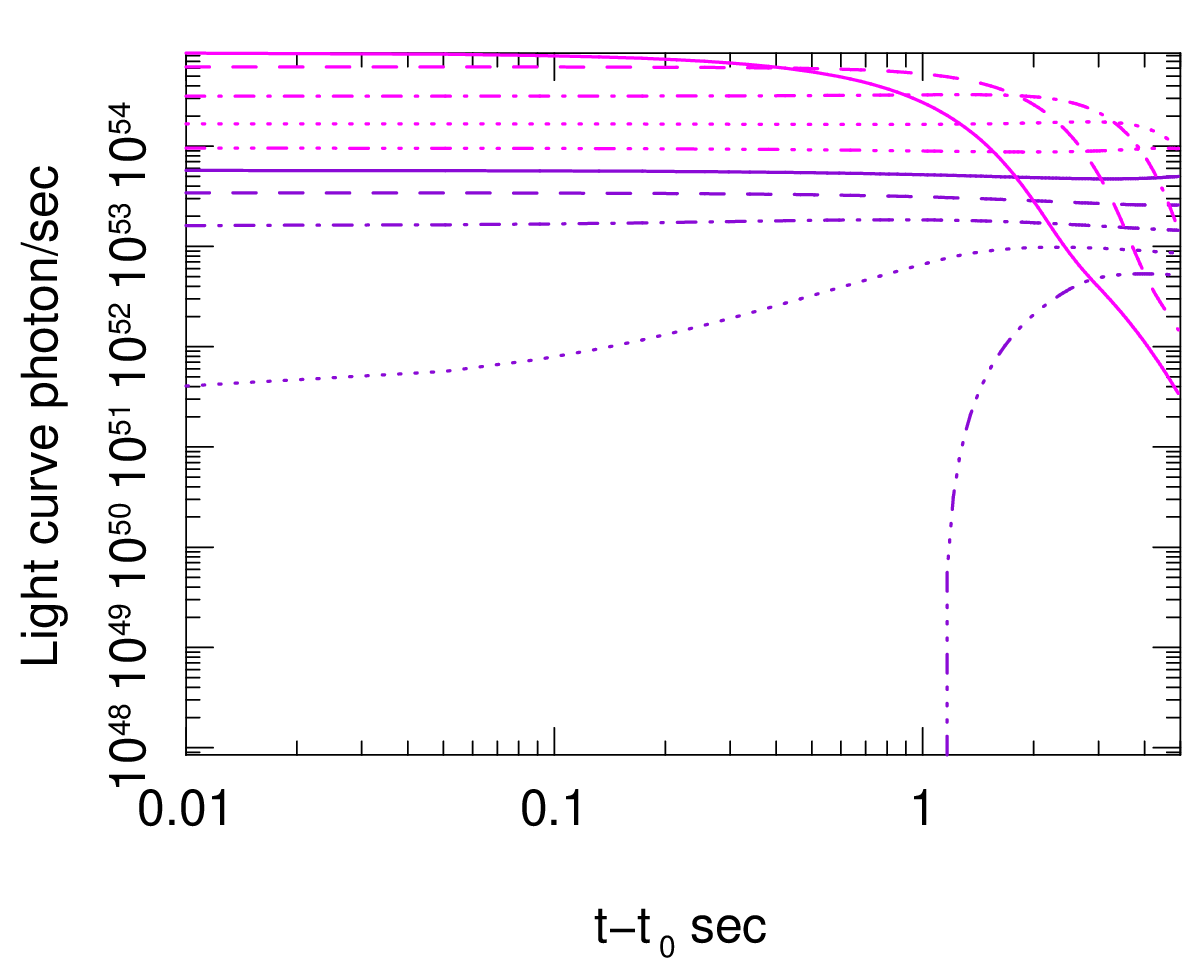} &
24)\includegraphics[height=5.5cm,angle=-90]{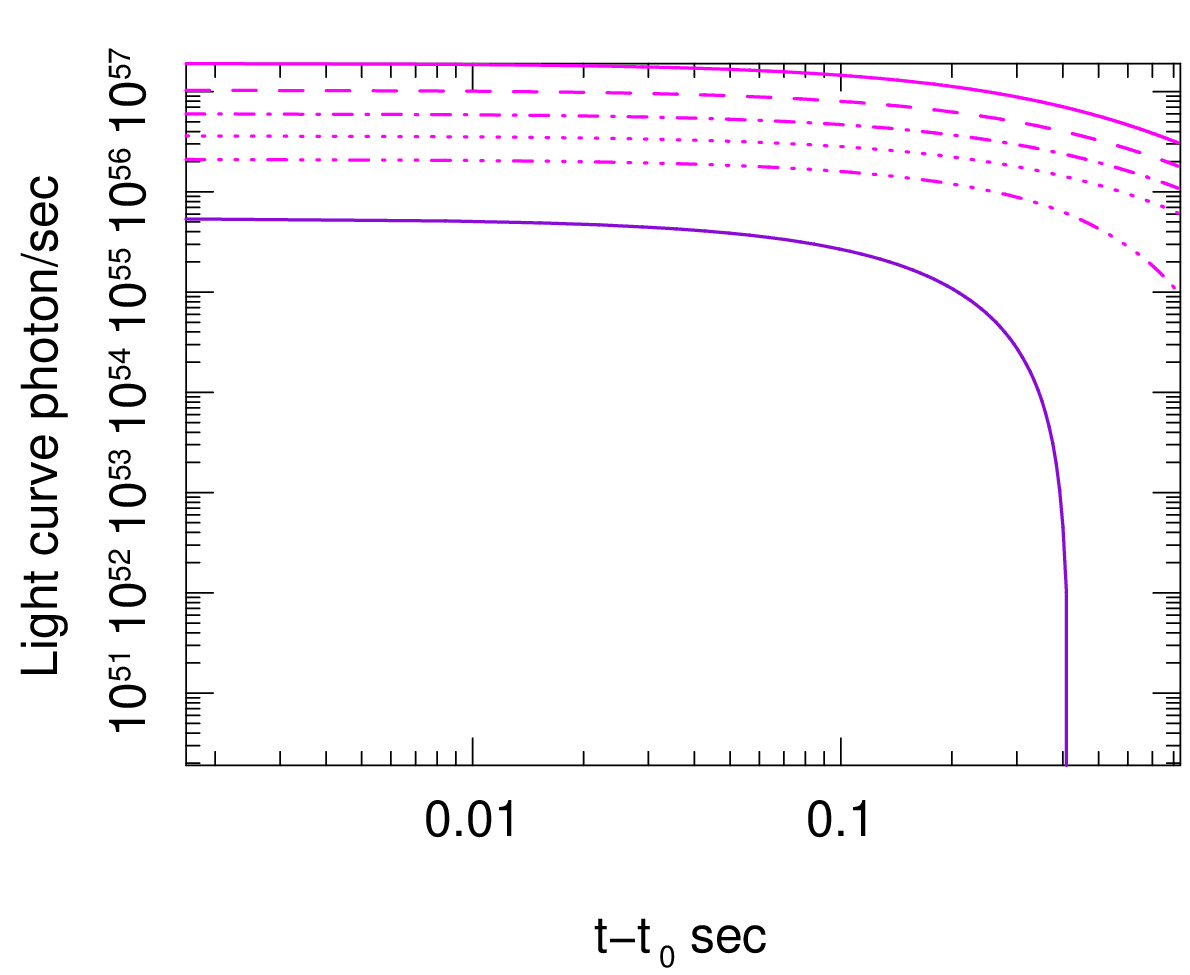} 
\end {tabular}
\end {center}
\caption{Light curves of models 20 to model 24 in 10 energy bands, from $2$ eV 
to $1$ or $2$ keV for models 20 to 22 and from $2$ eV to $2$ MeV in models 23 
and 24. Descriptions of the energy bands and curves are the same as 
Fig. \ref{fig:lcgmone}. Their active regions evolve according to equation 
(29-Paper I). Models 20 to 22 presents the end of of afterglow emission and 
models 23 and 24 the end of prompt emission.
\label{fig:lcgmtwo}}
\end{figure*}

Finally Fig. \ref{fig:lcgmtwo} show the light curves for models 20 to 24 in 
which the active region width evolves from an initial non-zero value at $r'_0$ 
according to a power-law, see equation (29-Paper I). The expression for 
$\Delta r'$ evolution in this model apparently looks like quasi-steady state 
model, but in reality it is more similar to the dynamical model because in 
short intervals of time $\beta'$ and $\gamma'$ on which the dynamical model 
depends, have a close to power-law evolution, see Figs. \ref{fig:betaone} and 
\ref {fig:gammaone}. Therefore, the evolution of $\Delta r'$ is also a close 
to a power-law in a short time intervals. This fact is reflected in the 
form of the light curves of models 20 to 22 that present external shocks. 
They are very similar to models 7 to 10 generated by the dynamical model and 
their lightcurves have very shallow initial slopes and steep break. Their 
slopes after the break seem to be too steep. However, the contribution of 
the high latitude emissions as explained at the beginning of the 
Sec. \ref{sec:lc} should flatten their decay slopes. 
Moreover, our tests and also the difference between the after-break slopes of 
models 20 and 21 show that a flatter electron distribution reduces the 
steepness of the break. In particular, a power-law with exponential break 
distribution can make flatter late time breaks at lower energies and faster 
break at high energies. This feature has been also considered to explain 
the behaviour of late X-ray emission of some 
bursts~\citep{jetbreakexp0,ecutoff}. Models 23 and 24 can be good candidate 
for the decay of emission during the prompt internal shock.

In summary, the general behaviour of light curves in the simulations of active 
region models studied here are consistent with observations if we admit 
switching between models during the evolution of both internal and external 
shocks. Many of unexplained features of both prompt and afterglows are 
explained in these simulations such as: the energy dependence of the prompt 
emission peak width, the steep decay of the early afterglow and its connection 
to prompt emission, and the shallow regime which can be in part due to the way 
the active region evolves - in another word to the microphysics of the 
shock - and in part due to the combination of tail emission from prompt shock 
and the onset of external shock. Anther notable conclusion from the 
simulations of external shocks is that in all cases there is a chromatic 
break in the light curves due to the accumulation of in-falling matter into 
the shock. Therefore, changes of physical parameters and thereby the way 
the active region evolves, can completely smear an achromatic break due to 
the kinematical jet break in a collimated ejecta. We observed a jet break in 
all the simulations although the model considered here has a spherical 
symmetry and therefore no kinematical jet break. Multiple jet breaks observed 
in a significant fraction of late emissions of the gamma-ray bursts also can 
be explained by switching between simple evolution models for the active 
region presented here as well as the change in the physical quantities such as 
the energy distribution of electrons. Evidently, it is up to 
magnetohydrodynamical simulations of relativistic and non-relativistic shocks 
and Fermi processes to verify the validity and conditions for the presence of 
the simple models presented here, the possibility of switching between models, 
different regimes, etc.

\section{Spectral behaviour} \label{sec:spectbehave}
\begin{figure*}
\begin{center}
\begin {tabular}{ccc}
1)\includegraphics[height=5.5cm,angle=-90]{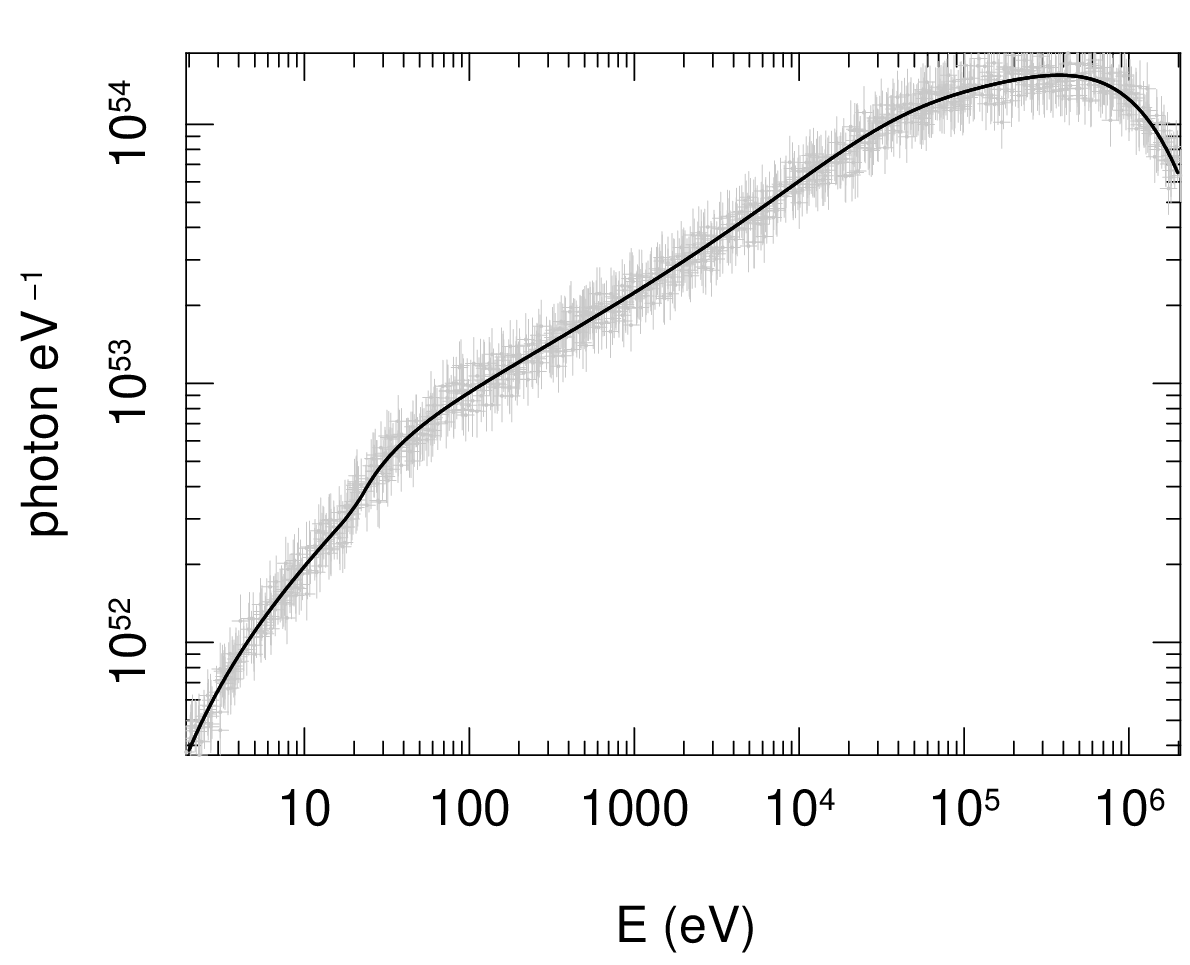} &
2)\includegraphics[height=5.5cm,angle=-90]{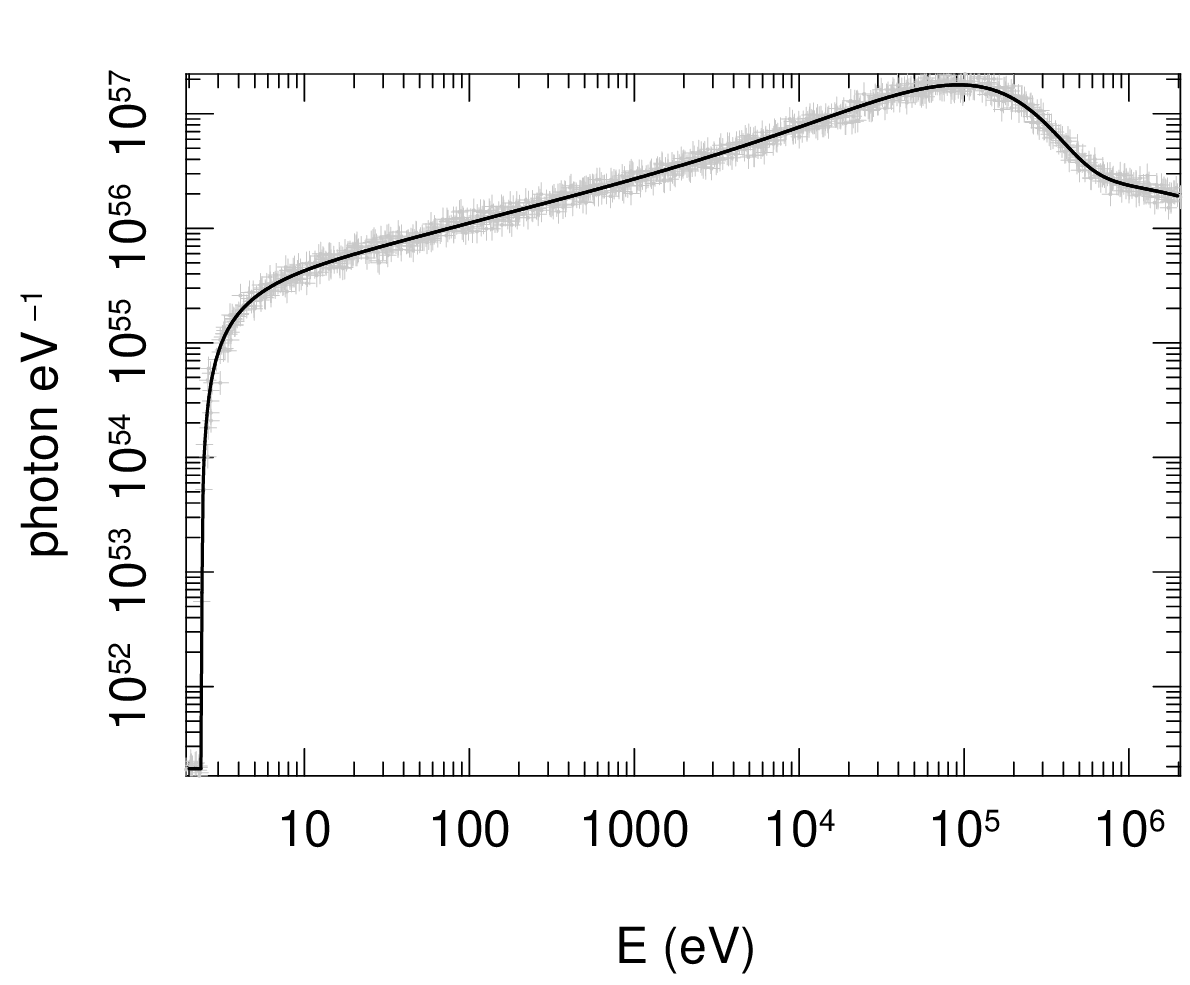} &
3)\includegraphics[height=5.5cm,angle=-90]{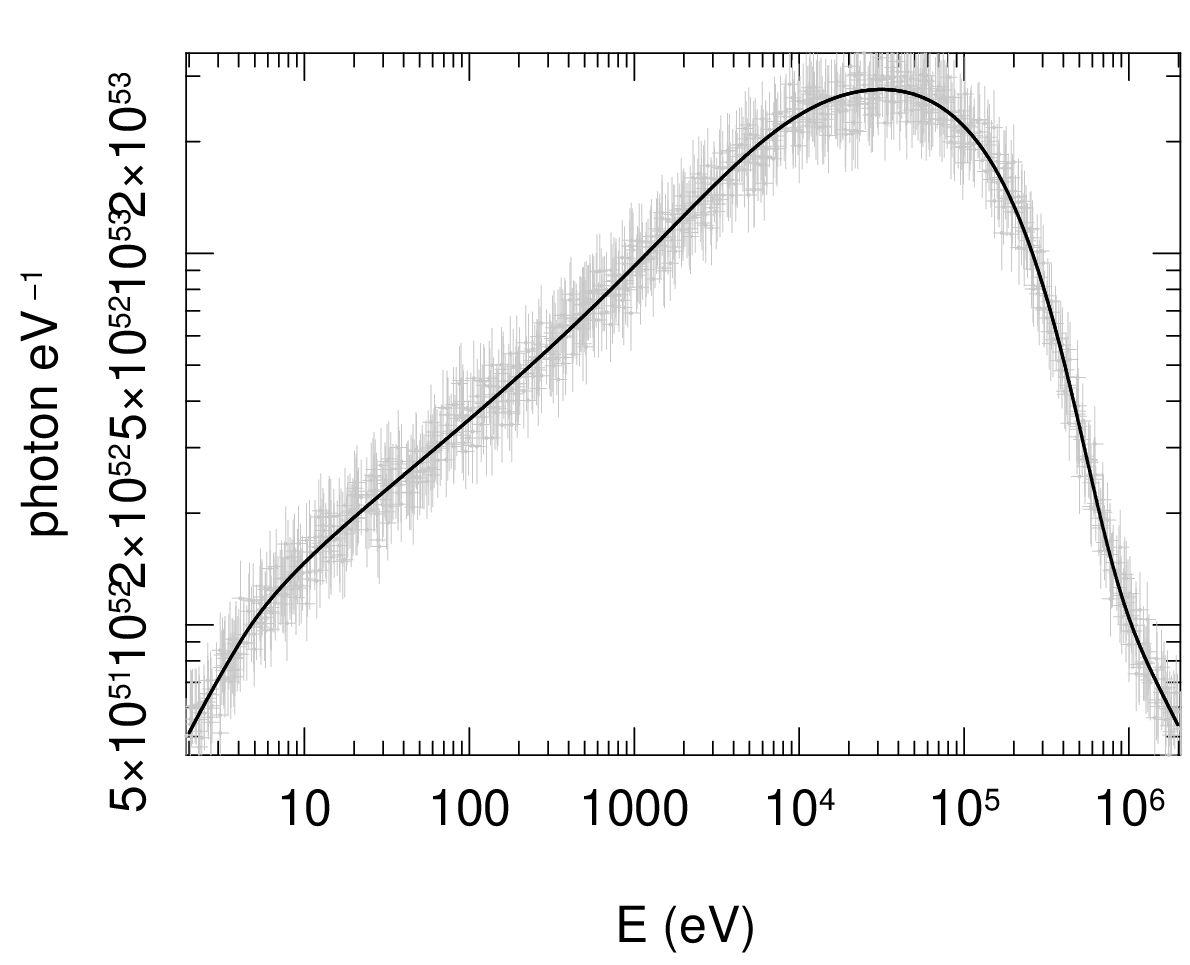} \\
4)\includegraphics[height=5.5cm,angle=-90]{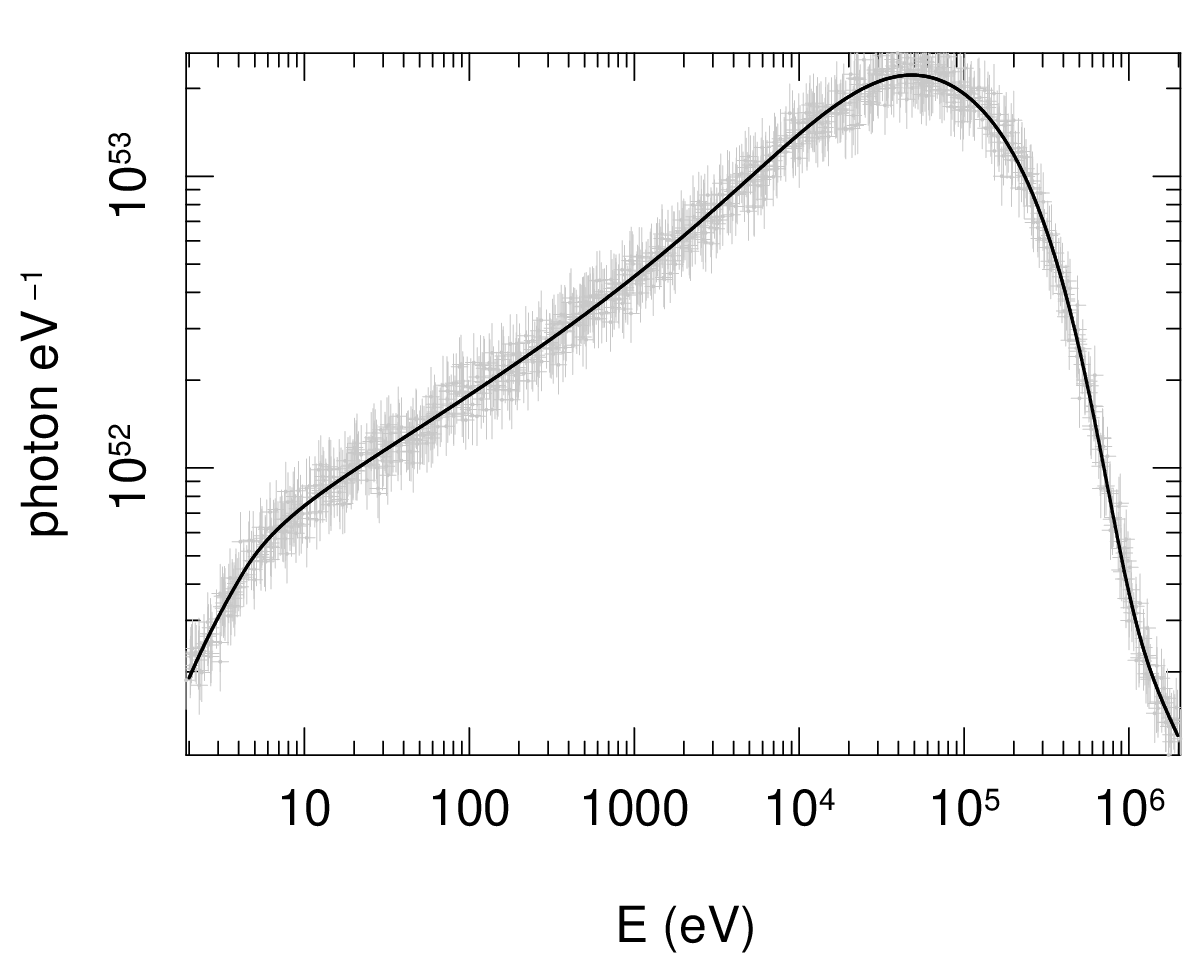} & 
5)\includegraphics[height=5.5cm,angle=-90]{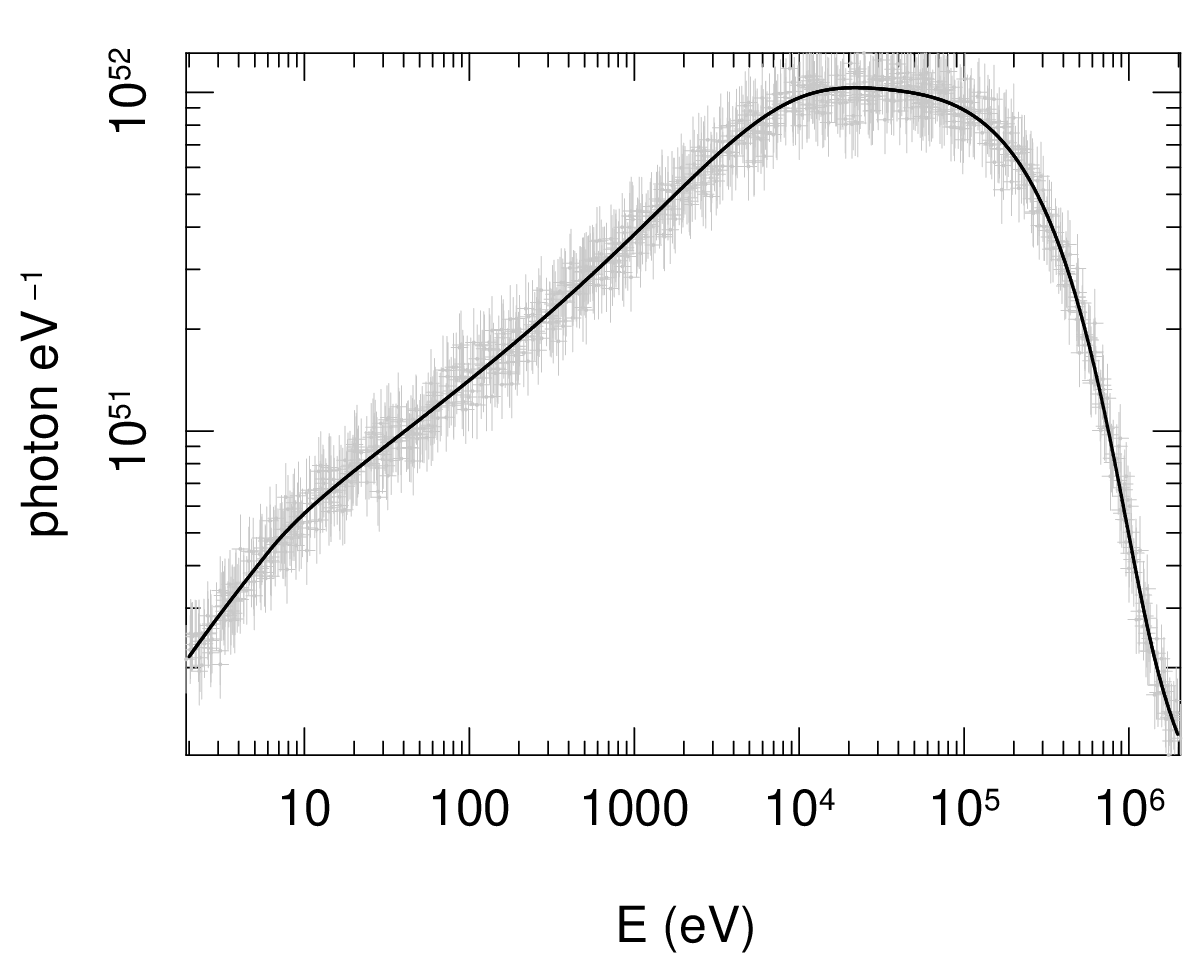} &
6)\includegraphics[height=5.5cm,angle=-90]{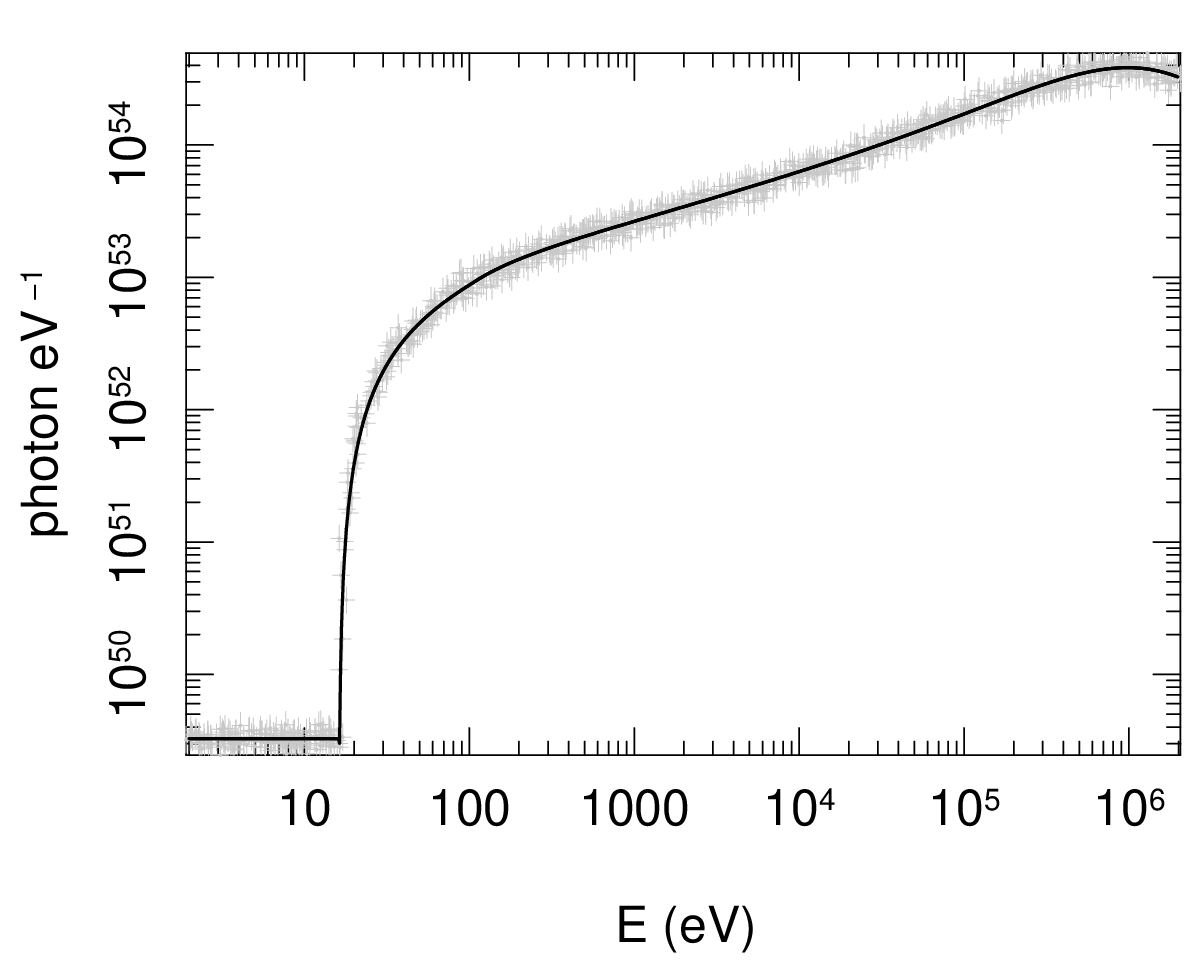} \\
7)\includegraphics[height=5.5cm,angle=-90]{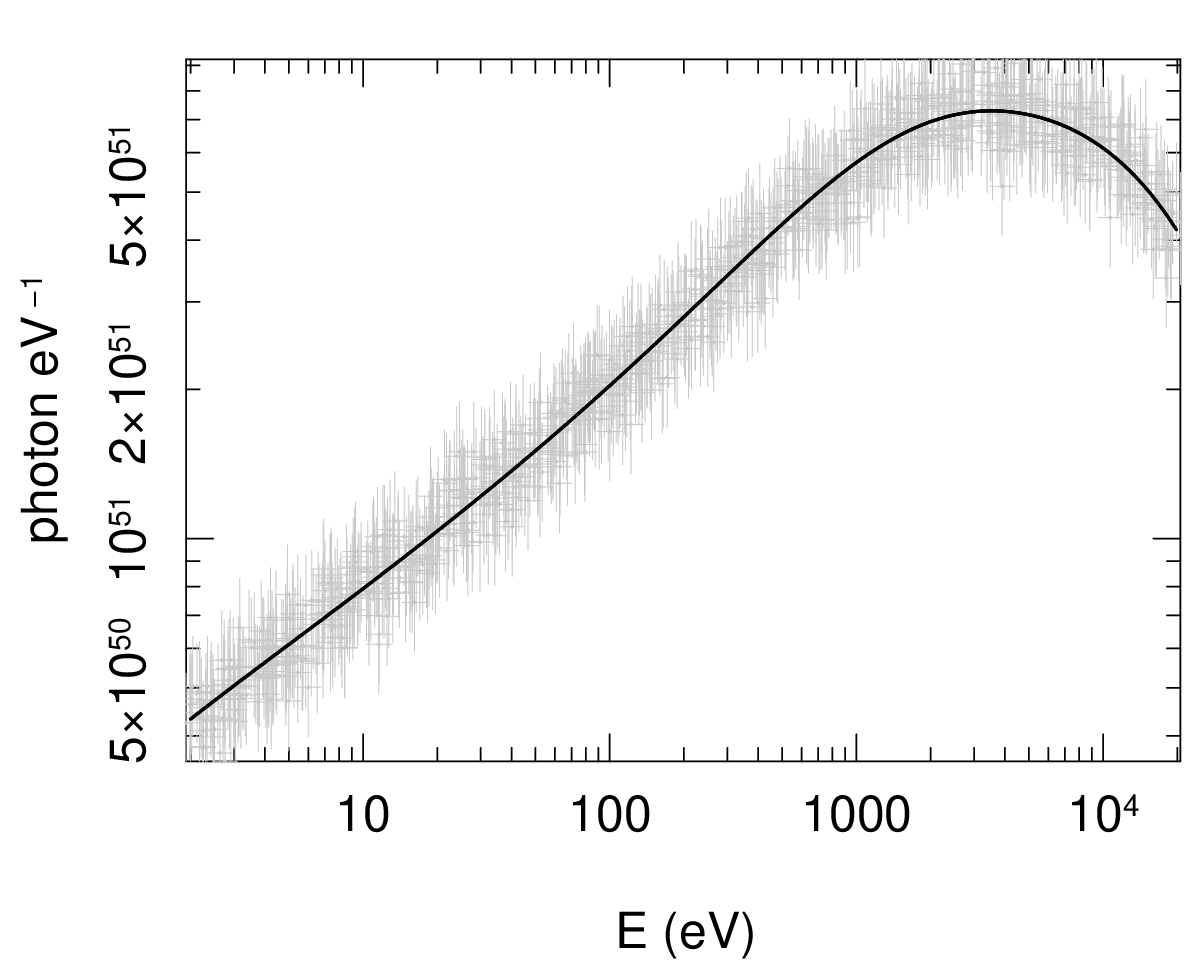} &
8)\includegraphics[height=5.5cm,angle=-90]{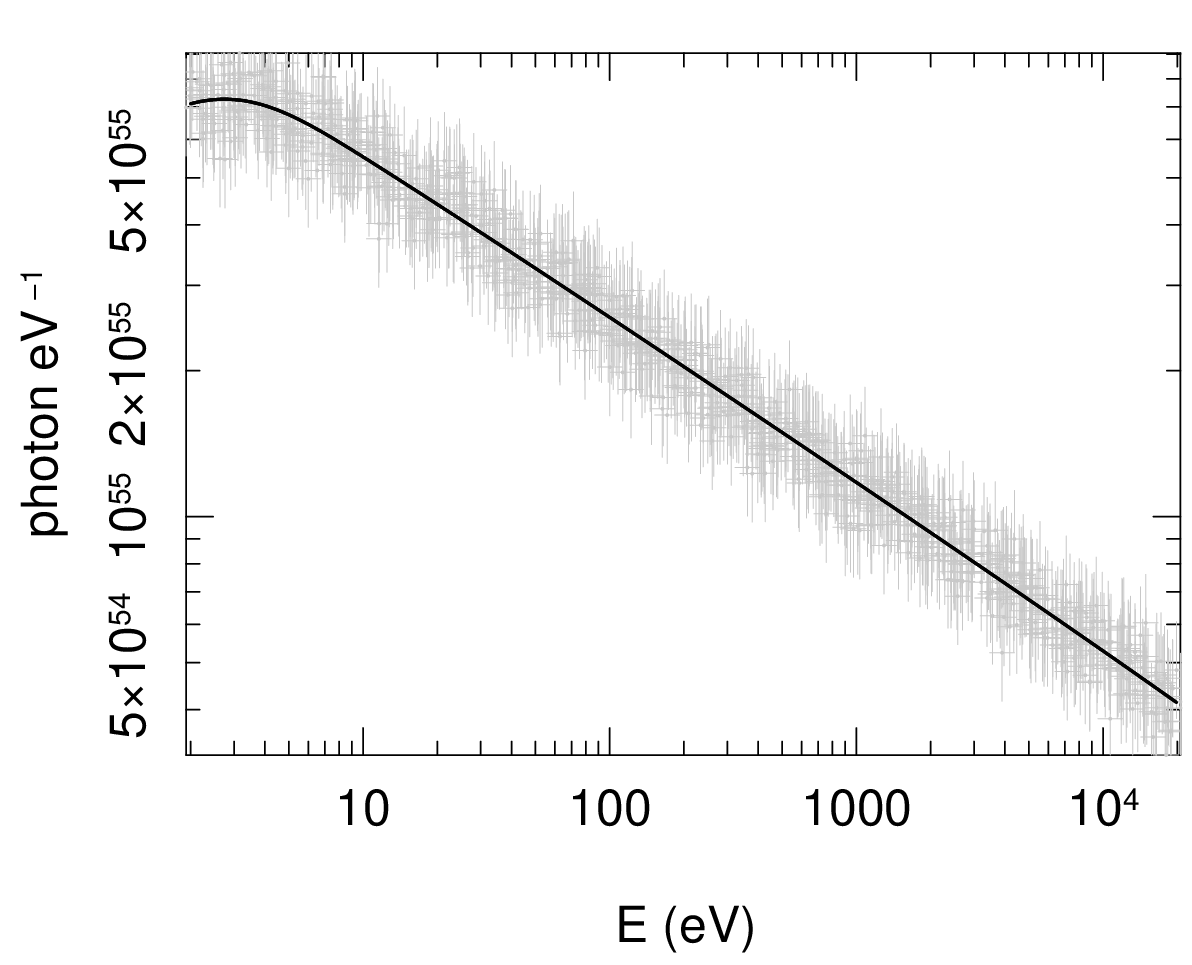} &
9)\includegraphics[height=5.5cm,angle=-90]{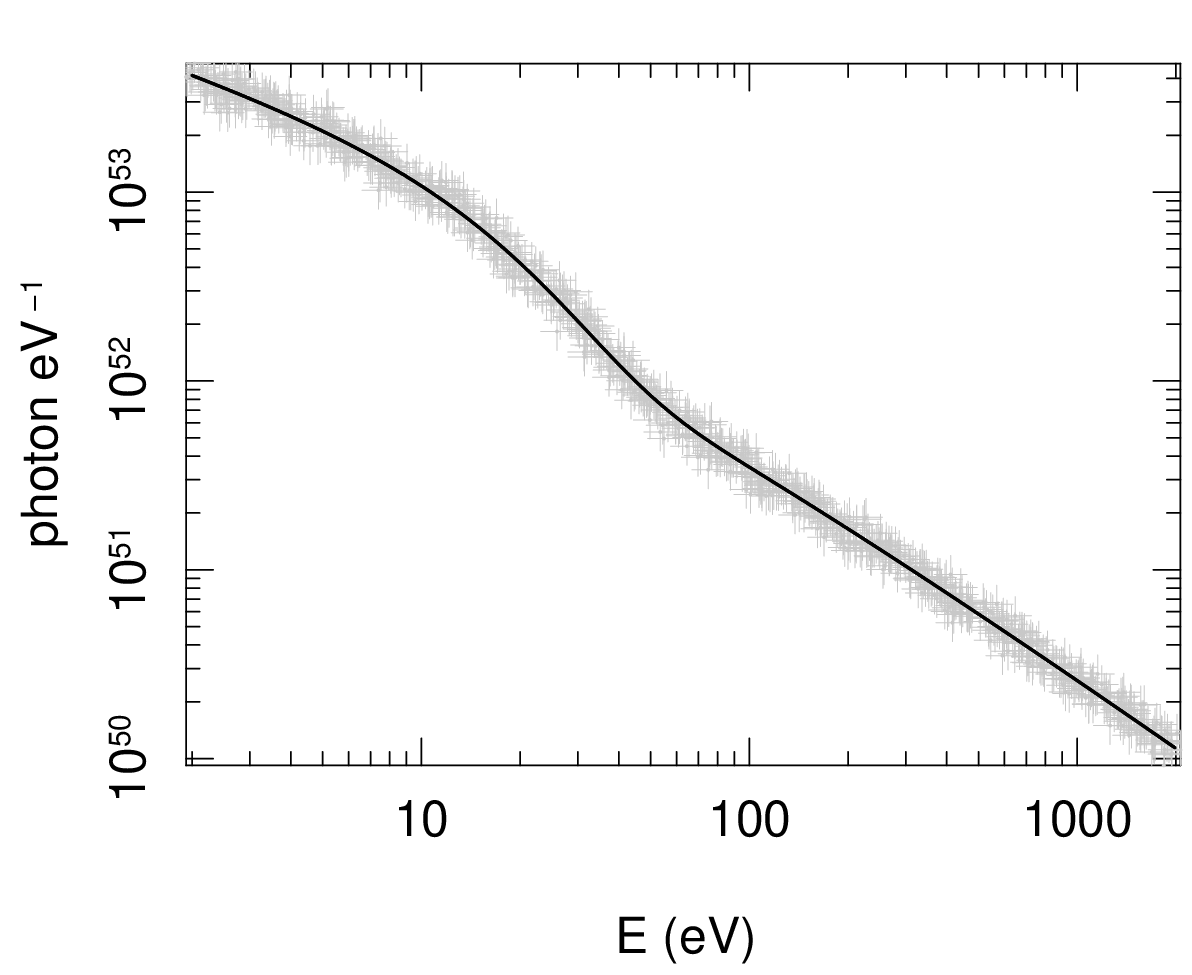} \\
10)\includegraphics[height=5.5cm,angle=-90]{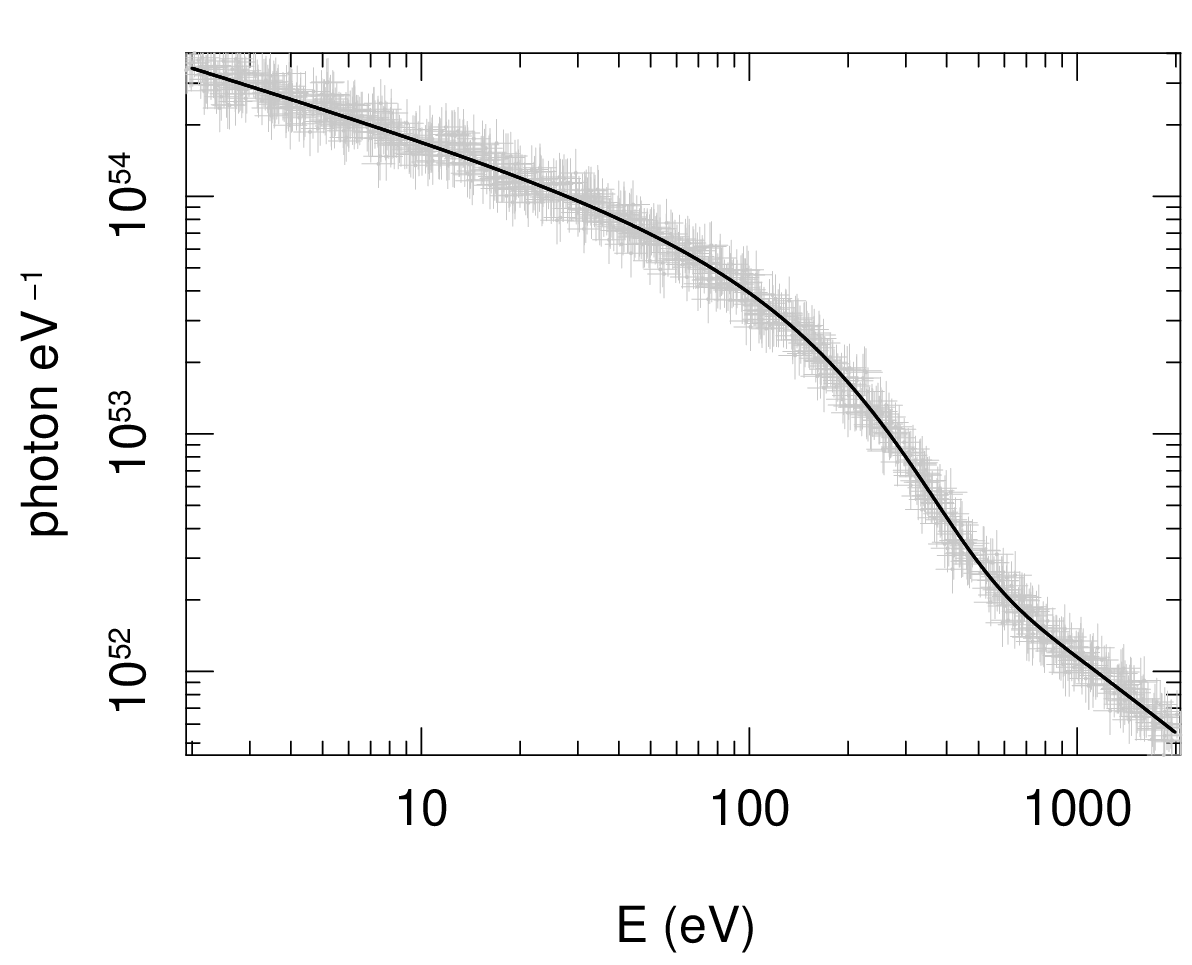} 
\end {tabular}
\end {center}
\caption{Total broad band ($2$ eV - $2$ MeV)/($2$ eV - $20$ keV) spectrum of 
the simulated models with dynamical active region. To show the level of 
ambiguity in the observational data due to systematic errors, we have added a 
random error with a normalized sigma of 5\% for energy and 10\% for spectrum 
amplitude (crosses).
\label{fig:specdar}}
\end{figure*}

In the GRB data analysis usually all the photons detected during a given 
interval of time, e.g. the duration of the observation by the gamma-ray 
telescope - $T_{100}$ - is used for the determination of an averaged 
spectrum. Therefore for comparing the theoretical spectrum, equation 
(60-Paper I), with data we must integrate it with respect to $r$ for a given 
radius/time interval and divide it by the duration length. However, here we 
do not consider a lower limit for the detectable flux in the simulated 
models, and therefore $T_{100}$ is the same as the simulation time. For this 
reason we simply determine the total spectrum for the duration of the 
simulations without dividing it by time. The results are shown in 
Fig. \ref{fig:specdar} for dynamical and in Fig. \ref{fig:specqsar} 
for quasi steady active region models explained in Table \ref{tab:param}.

\subsection{Spectrum of simulated models with dynamical active region}
\label{sec:specdar}
From variety of behaviours we have seen in the light curve of dynamical models 
we must expect the same level of variation in the spectrum, and it is exactly 
what we find in Fig. \ref{fig:specdar}. We should remind that the spectrum 
of the observed bursts are usually time averaged on the total duration of the 
burst, or belongs to the total duration of a peak. Thus, their behaviour can 
be very different from the spectrum in a shorter interval of time. Each of The 
simulations presented in this work correspond to a fraction of a burst, and 
therefore we should not expect that it has a spectrum similar to time averaged 
observed spectrum of complete bursts.

The spectra of models 1, 2, and 6 are much harder than observed time averaged 
spectra. However, considering their light 
curves, they correspond to a short time at the beginning of the burst, and 
therefor their impact on the integrated spectrum would be small. In fact 
some bright hard bursts such as GRB 060105~\citep{grb060105}, 
GRB 080319B~\citep{grb080319b}, and GRB 080706~\citep{grb080607} 
(long bursts) and GRB 051221~\citep{grb051221, grb051221-1}, 
GRB 060313~\citep{grb060313}, and GRB 080123~\citep{grb080123} (short bursts) 
seem to have a very flat or even rising spectrum at the onset of the 
burst.

At highest energies, $E \gtrsim$ few times $10^4$ eV, models 3, 4, and 5 which 
present the end of a peak, see Fig. \ref{fig:lcgmzero} have a power-law with 
an exponential cutoff spectrum at $E \gtrsim 10^4$ eV and at 
lower energies the slope is positive. This is consistent with 
magnetohydrodynamics simulations~\citep{simulspec,simulspec0} and shock 
emission models~\citep{spectheory}. At $E \sim 10^4 - 10^5$ eV 
where their spectra are roughly flat, with an error at the level or higher 
than what is shown in the plots ($10\%$ of the simulated values), they can be 
fit with a power-law consistent with observation of long \swift 
GRBs~\citep{batcat}. Although these models present only the end of a peak, the 
general behaviour of the broad band spectra is consistent the broad band 
spectrum of GRB 080916C which has been observed simultaneously by a large 
number of space and ground based instruments such as 
{\it Fermi}~\citep{grb080916c}, \swift~\citep{grb080916c-1,grb080916c-3}, 
{\it Konus-Wind}~\citep{grb080916c-2}, etc. This means that the effect of 
hard initial rise and soft tail emission after the decay of emission is small. 
Moreover, from similarity of these spectra with the spectrum of the only 
burst for which we have broad spectrum up to GeV bands we can conclude that 
these models present {\it typical prompt emissions}.

A common feature between all these spectra is their low luminosity at optical 
energies. This is consistent with the lack of very bright peak in optical 
frequencies when multiple-band observations of late gamma-ray peaks were
possible. Examples are GRB 060526~\citep{grb060526,grb060526-1} and 
GRB 061121~\citep{grb061121,grb061121-1}. In GRB 060526 the X-ray 
emission increased about $100$ times during a relatively soft but long peak 
observed also in gamma-ray. The optical emission during this peak increased 
only by a factor of $\lesssim 2$ in V band from the continuum. In GRB 061121
also the X-ray increased about $100$ times but V band peak was only $\sim 5$ 
times larger than the continuum and in B band only $\sim 3$ times. Most 
authors consider the null assumption of a single power-law spectrum at low 
energies and try to explain the low luminosity of bursts at low energies by 
dust absorption. In the case of GRB 061121 apriori a dust to gas model 
similar to the value for Milky Way can explain observations~\citep{grb061121}. 
However, we ignore the exact amount of dust in the host galaxy and we can not 
rule out a break at low energies consistent with magnetohydrodynamics 
simulations~\citep{simulspec,simulspec0}. 

Models 7 to 10 simulate external shocks and have a much steeper spectrum at 
high energies. The spectrum of model 7 which can correspond to the rise of 
the external shock is hard with a break at few keV. The spectrum of model 8 
and the soft X-ray spectra of models 9 and 10 can be fit by a single power-law 
with negative index. At very soft X and UV bands the spectrum index of 
models 9 and 10 becomes steeper but at optical bands it becomes again flatter. 
This should correspond to the transition region from $E > \bar{\omega}_m$ to 
$E < \bar{\omega}_m$ where $\bar{\omega}_m$ is the average minimum synchrotron 
characteristic energy in the observer frame. The spectrum indexes in X-ray 
band varies from $\sim -0.4$ in model 8 to $\sim -1.3$ for model 9 and roughly 
flat for $10^3 < E < 10^4$ eV for model 7. The large scatter of the X-ray 
spectrum index in these simulations is consistent with the observations of the 
\swift-XRT~\citep{xrt}.

\subsection{Spectrum of simulated models with quasi-steady active region}
\label{sec:specqsar}
\begin{figure*}
\begin{center}
\begin {tabular}{ccc}
11)\includegraphics[height=5.5cm,angle=-90]{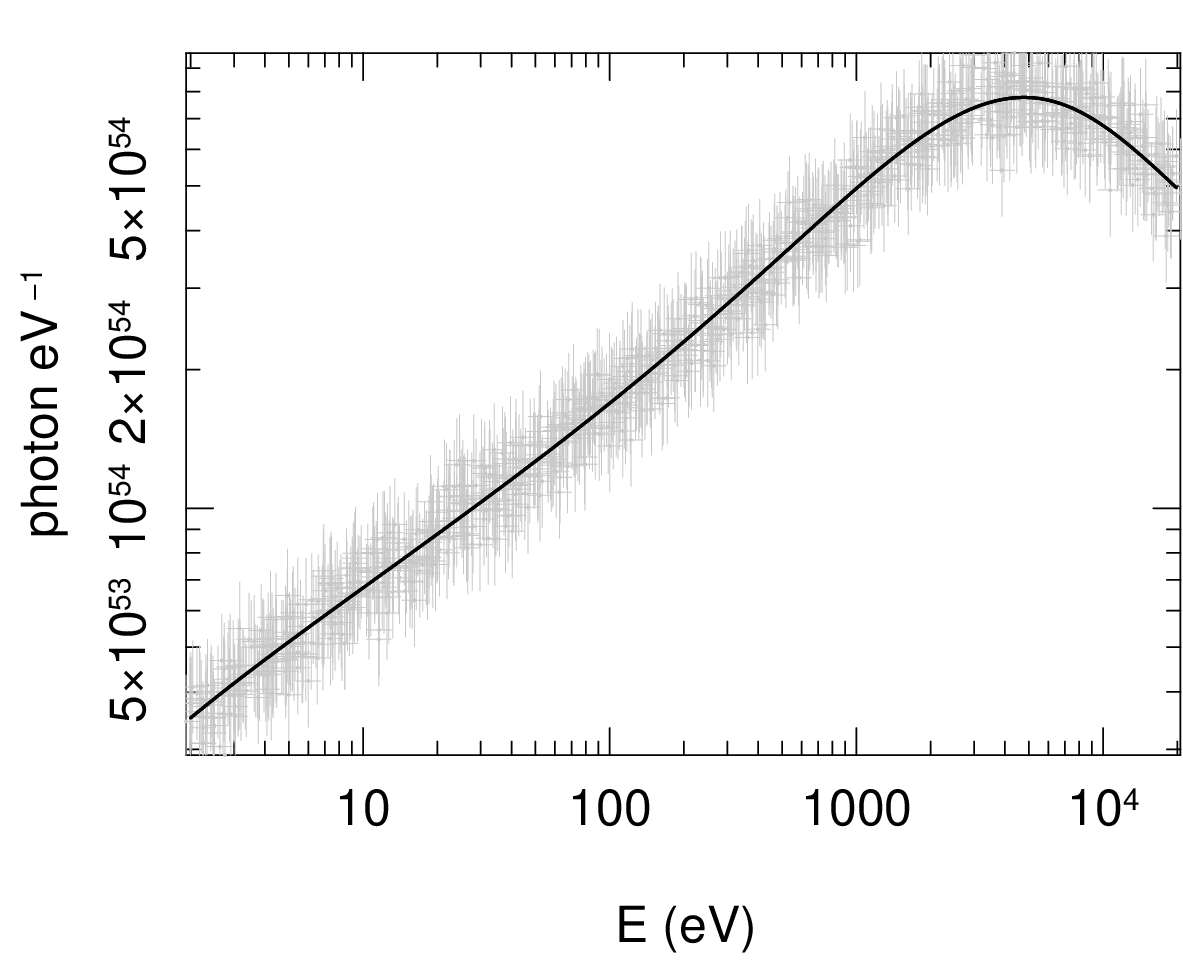} &
12)\includegraphics[height=5.5cm,angle=-90]{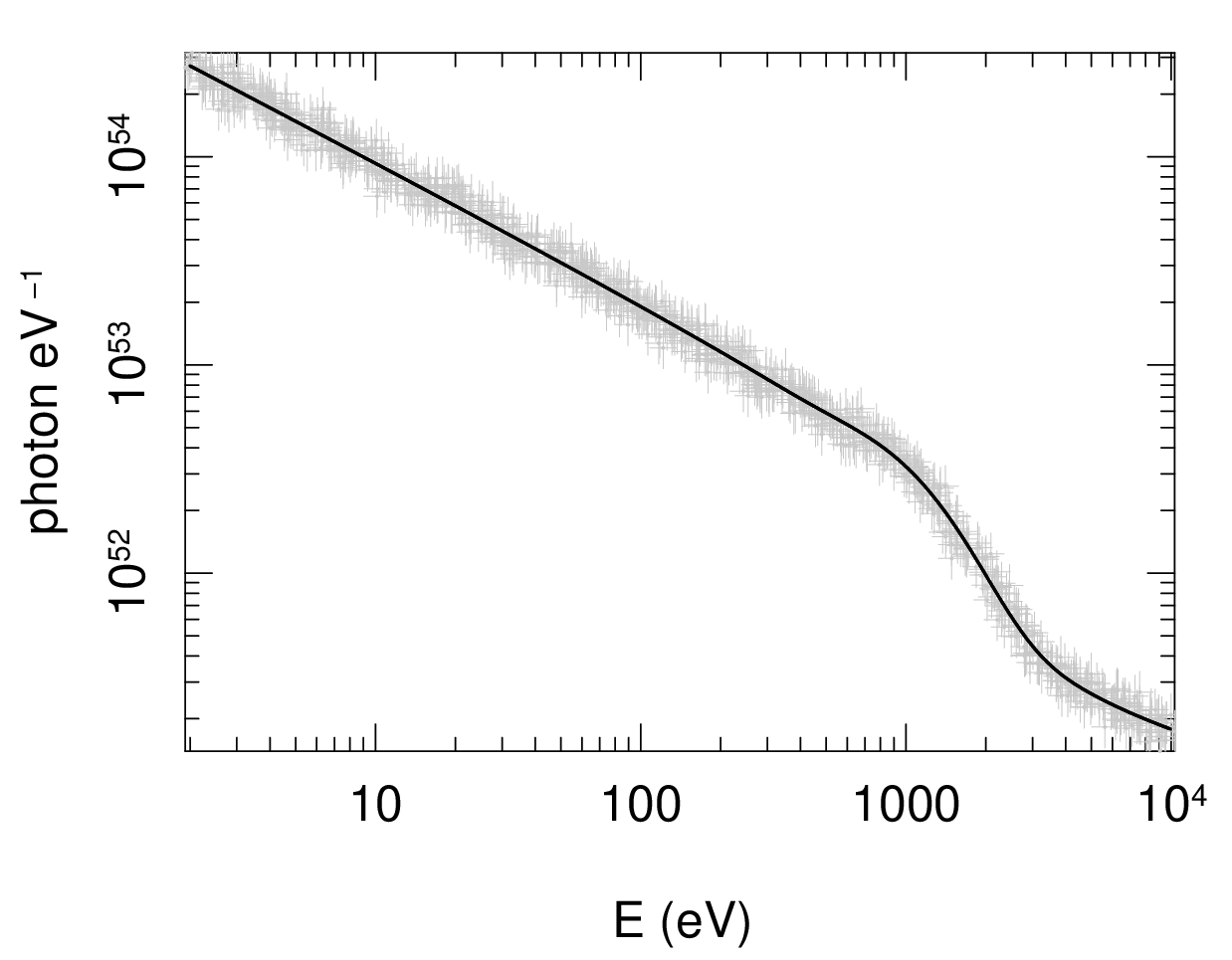} &
13)\includegraphics[height=5.5cm,angle=-90]{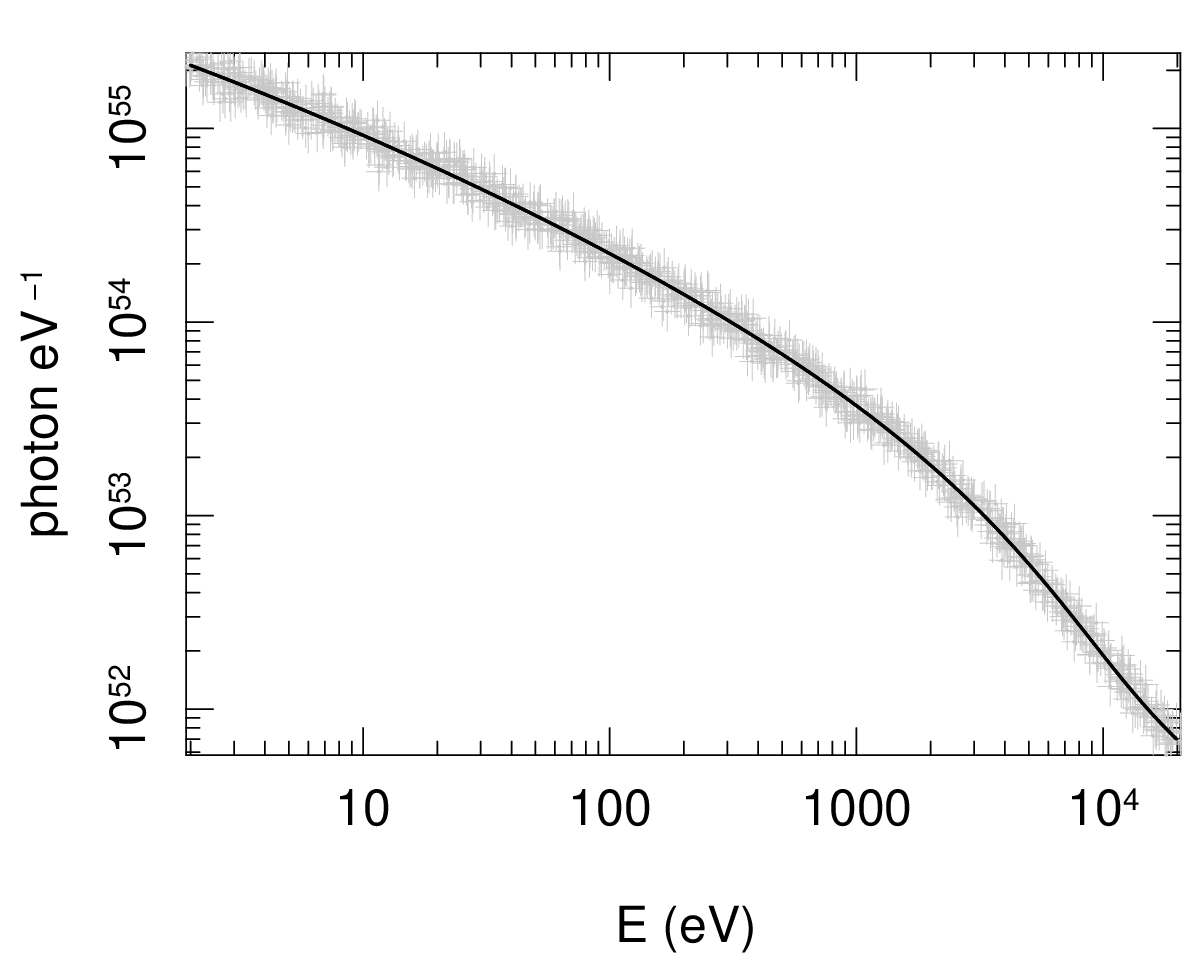} \\
14)\includegraphics[height=5.5cm,angle=-90]{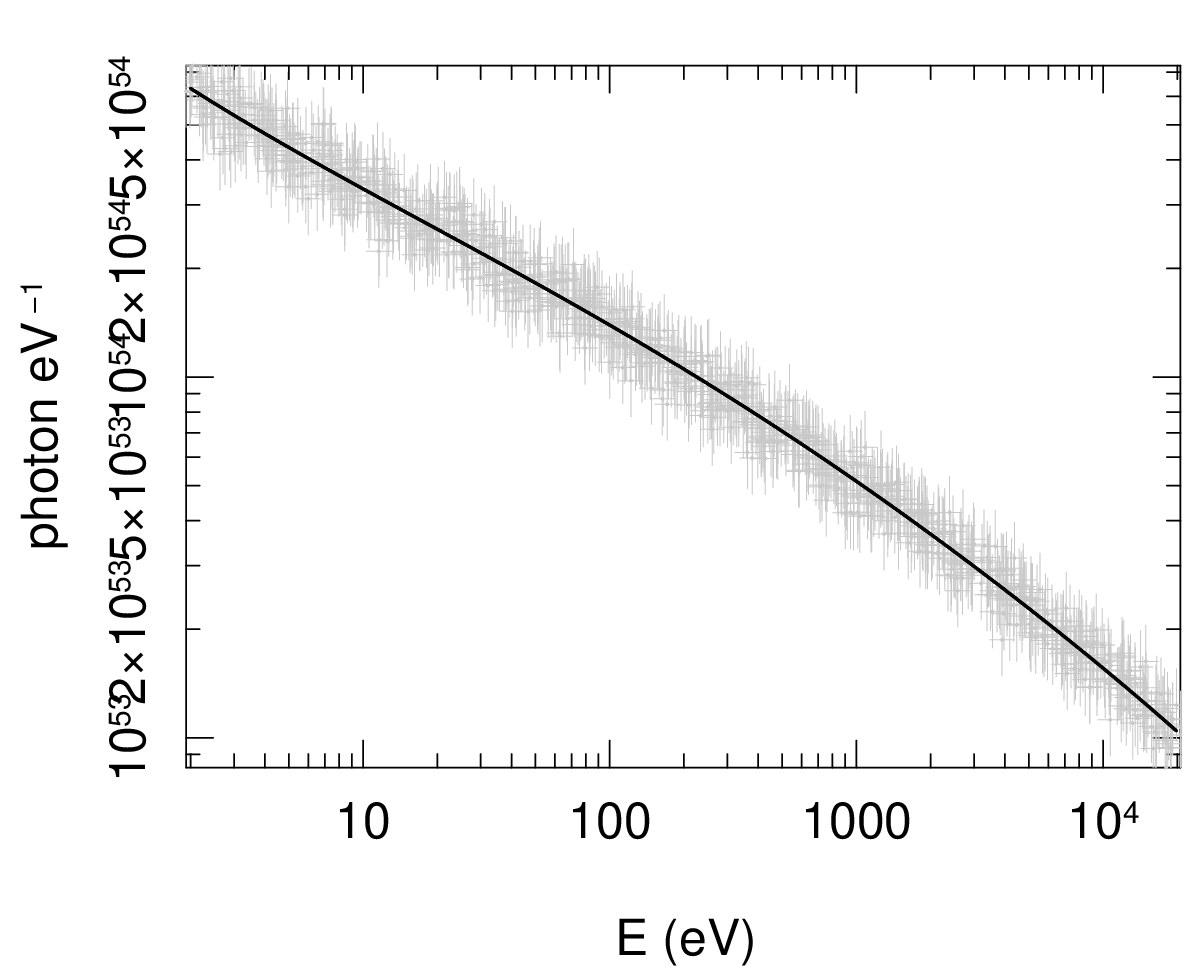} &
15)\includegraphics[height=5.5cm,angle=-90]{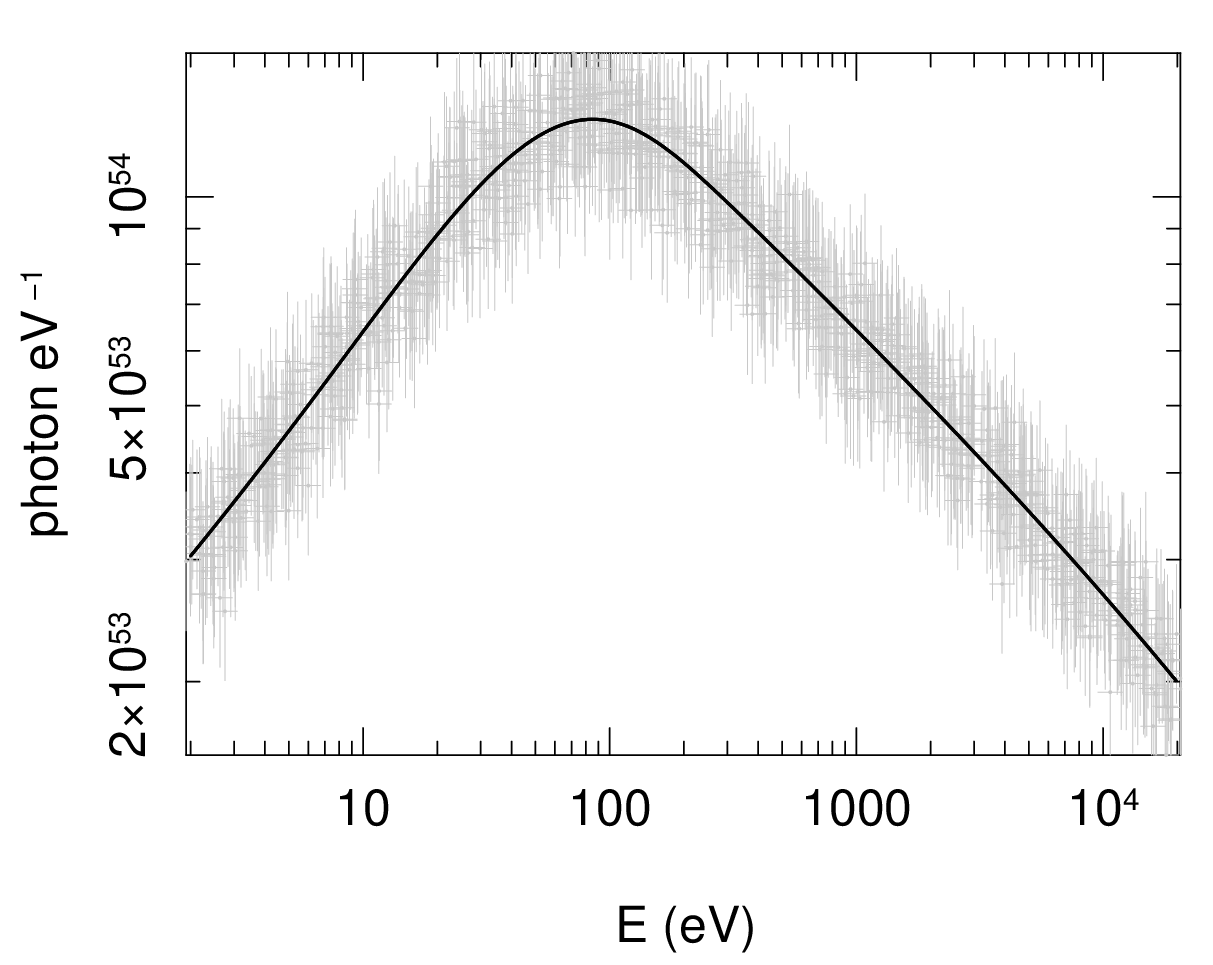} &
16)\includegraphics[height=5.5cm,angle=-90]{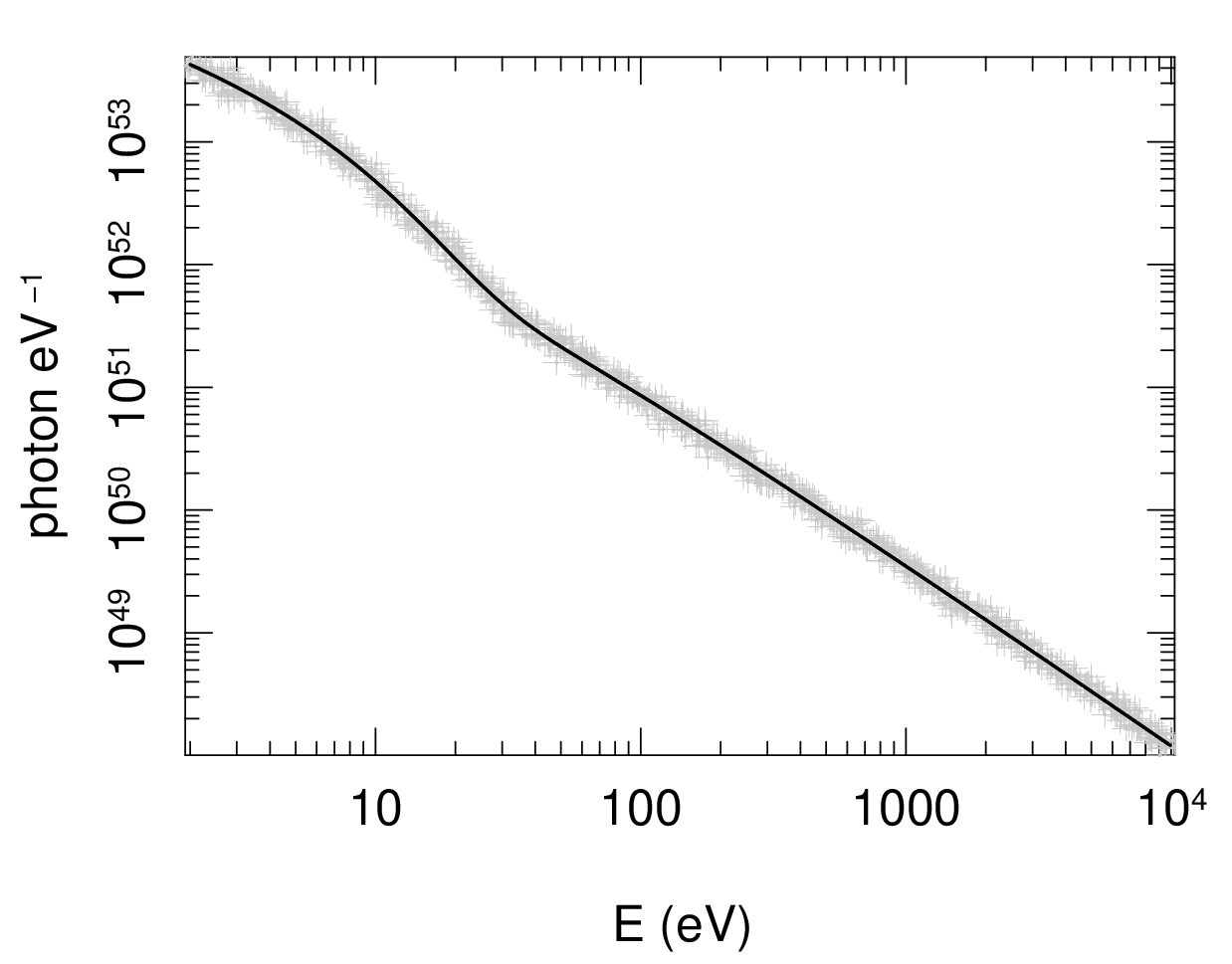} \\
17)\includegraphics[height=5.5cm,angle=-90]{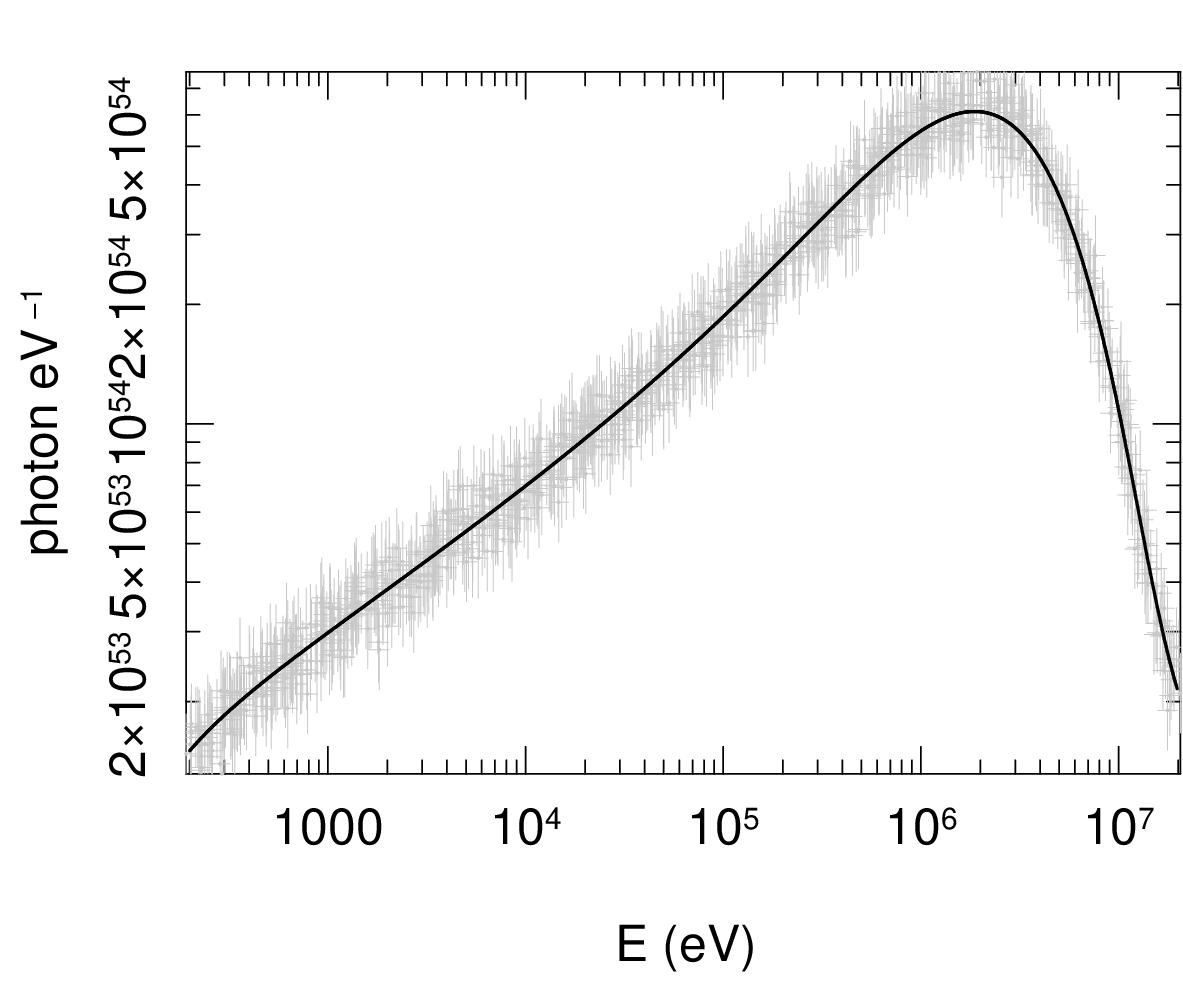} &
18)\includegraphics[height=5.5cm,angle=-90]{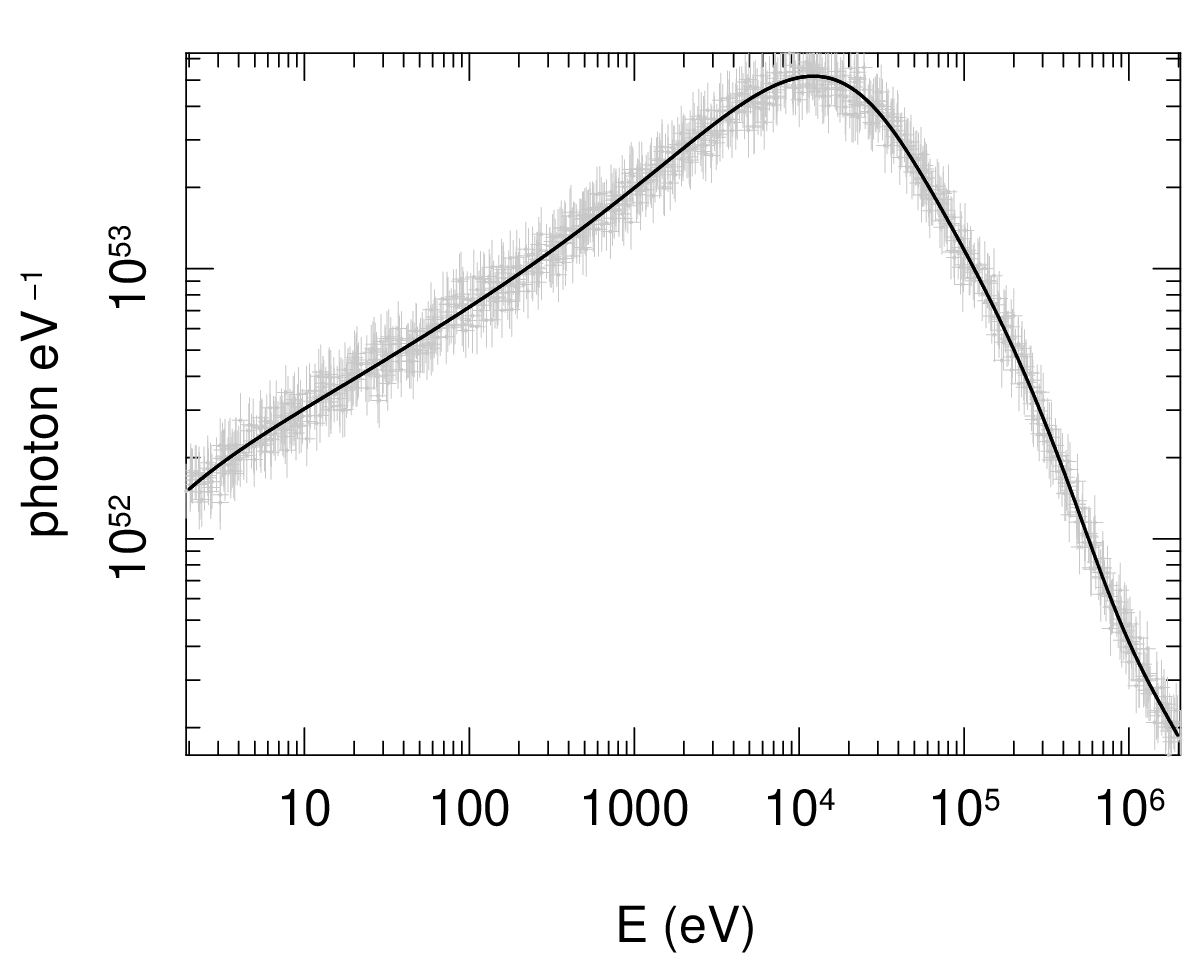} &
19)\includegraphics[height=5.5cm,angle=-90]{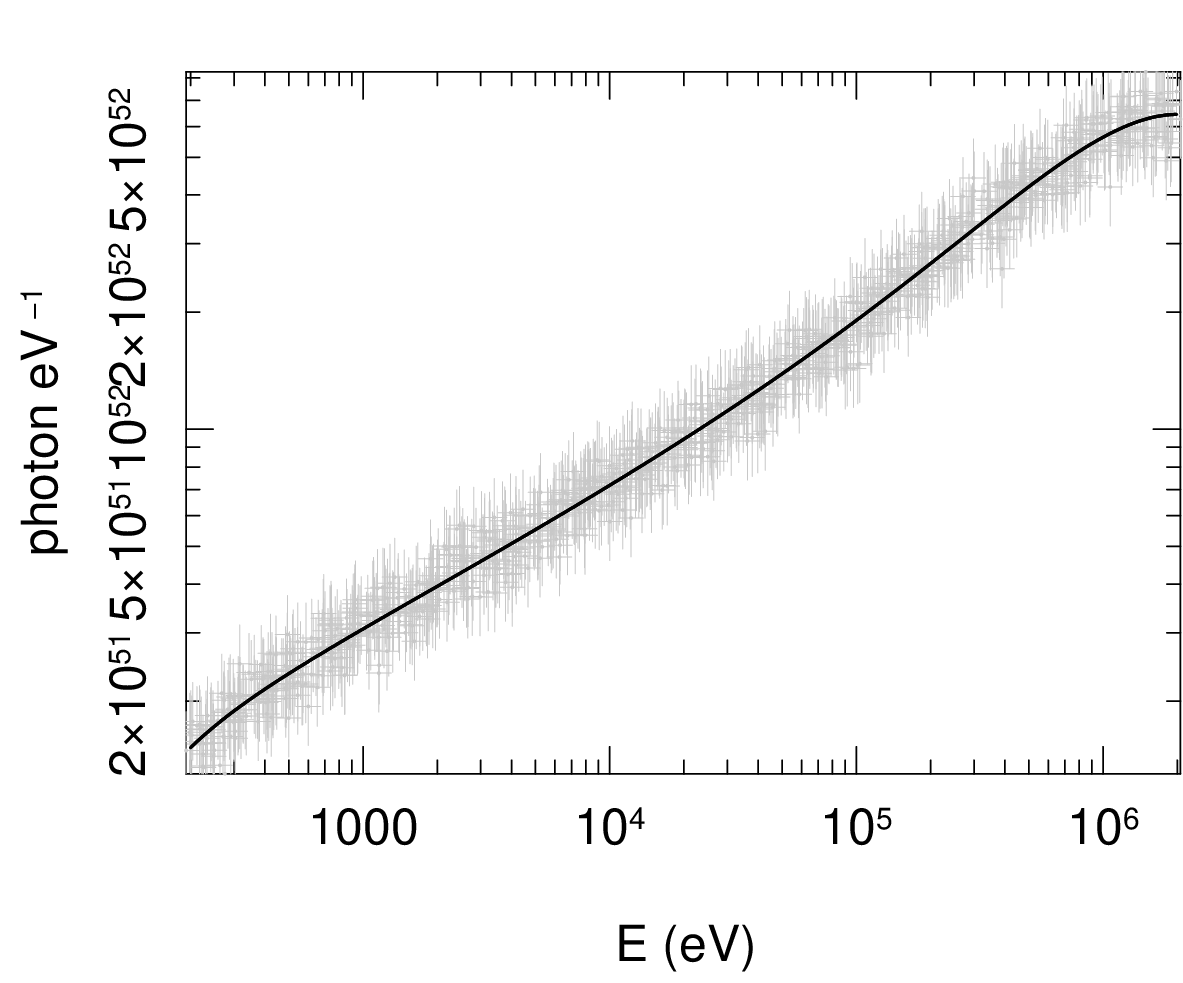}
\end {tabular}
\end {center}
\caption{Total broad band ($2$ eV - $20$ keV)/($2$ eV - $2$ MeV) spectrum of 
the simulated models with quasi steady active region. Definitions of the 
curves are similar to Fig. \ref{fig:specdar}.
\label{fig:specqsar}}
\end{figure*}

Fig. \ref{fig:specqsar} shows that the spectrum of external shock simulations 
according to this model is very similar to what have been obtained for the 
dynamical models. Except model 11 which is hard and corresponds to the rising 
of the emission, others can be fit by a simple power-law or power-law with 
exponential cutoff at high and/or low energies. Like dynamical models, their 
spectrum index in X-ray band are mostly shallow with $\alpha > -1$ and get 
even 
shallower in optical bands. But when the averaged minimum characteristic 
energy is in the plotted bands, as in model 12, there is a transient steep 
region similar to what we saw in dynamical models. Models 11 and 15 are hard 
with a positive index at low energies. As these models can only be relevant 
for the rising of the external shock emission which is usually 
overlapped with the tail emission of the prompt, it is very difficult to 
compare this results directly with observations. 

\begin{figure*}
\begin{center}
\begin {tabular}{ccc}
11)\includegraphics[height=5.5cm,angle=-90]{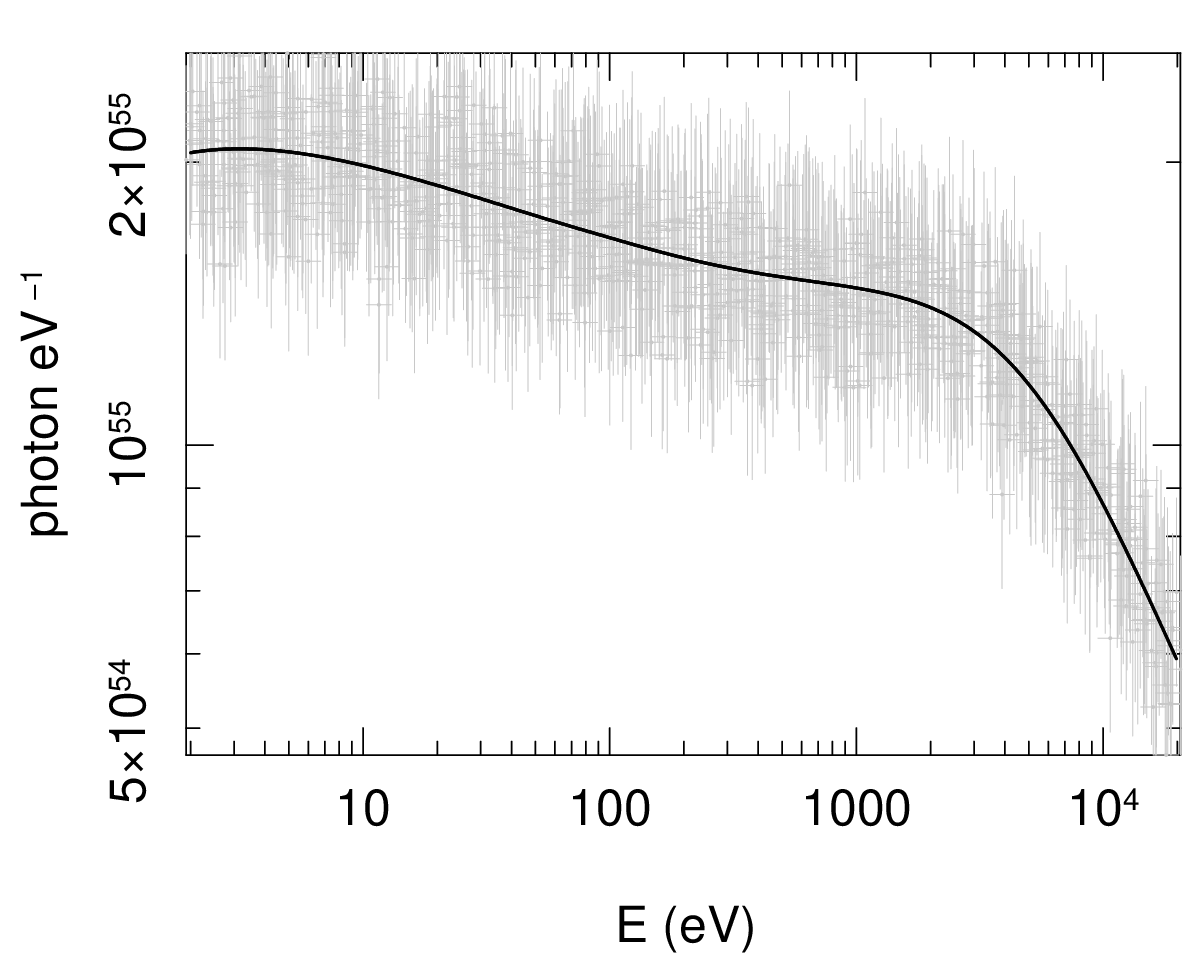} &
12)\includegraphics[height=5.5cm,angle=-90]{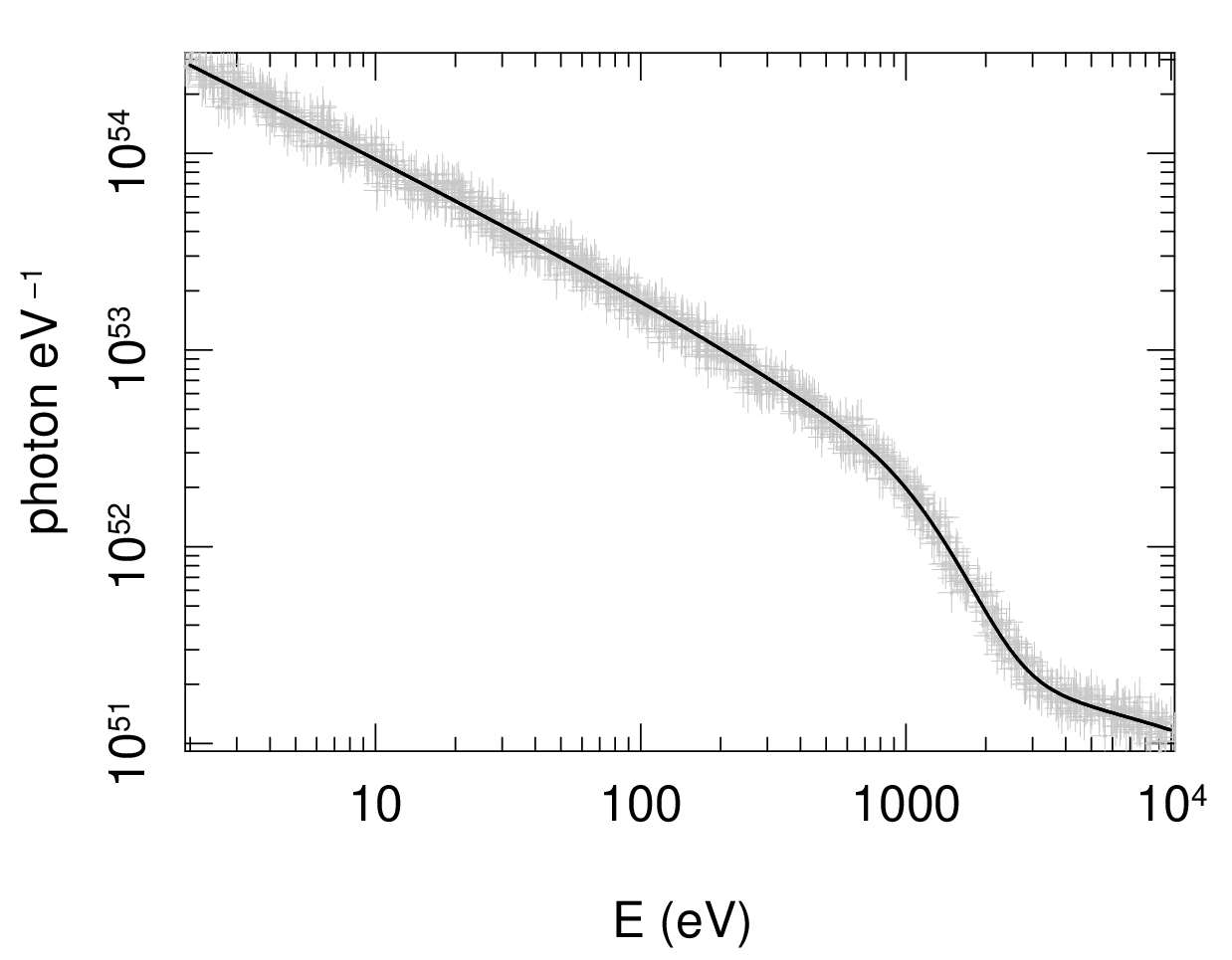} &
13)\includegraphics[height=5.5cm,angle=-90]{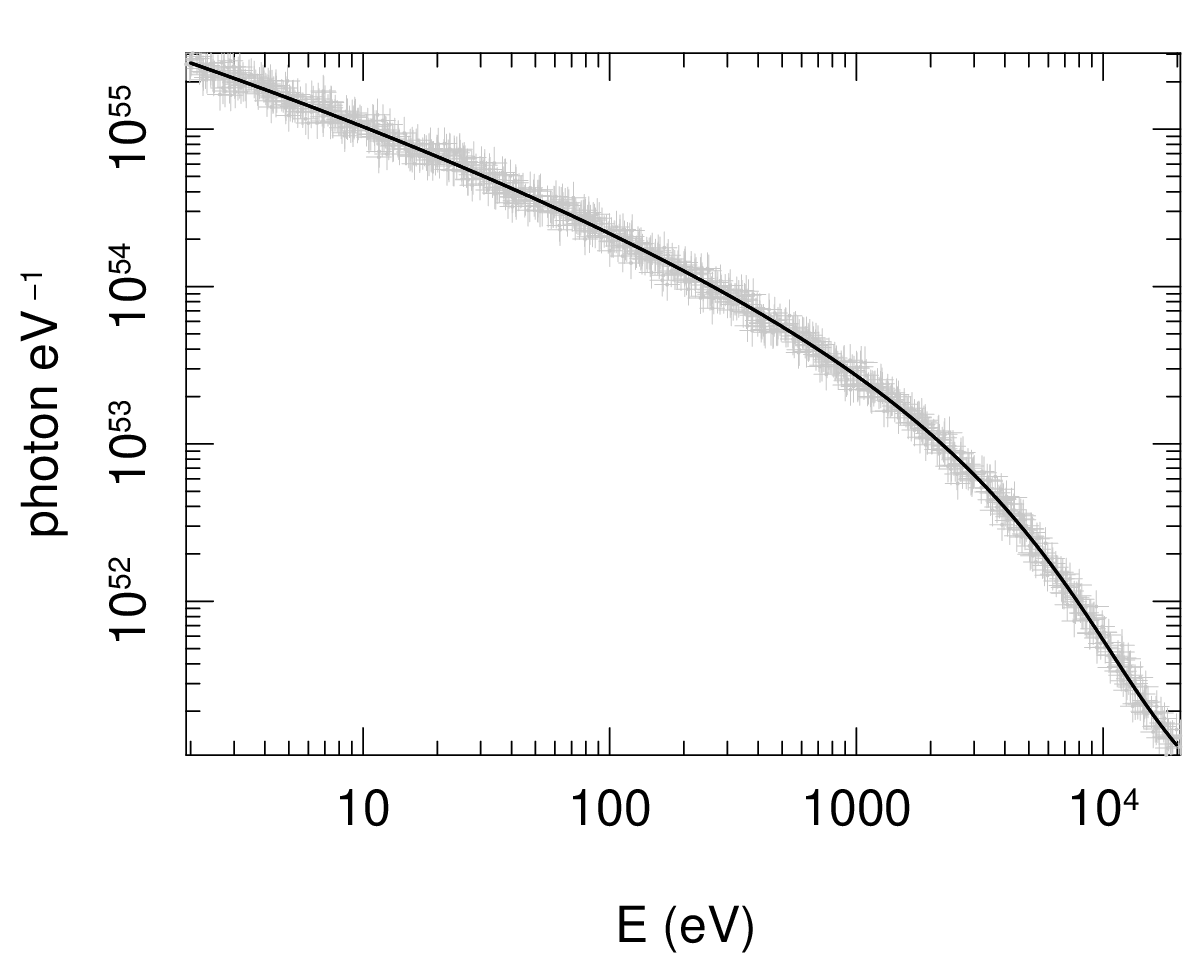} \\
14)\includegraphics[height=5.5cm,angle=-90]{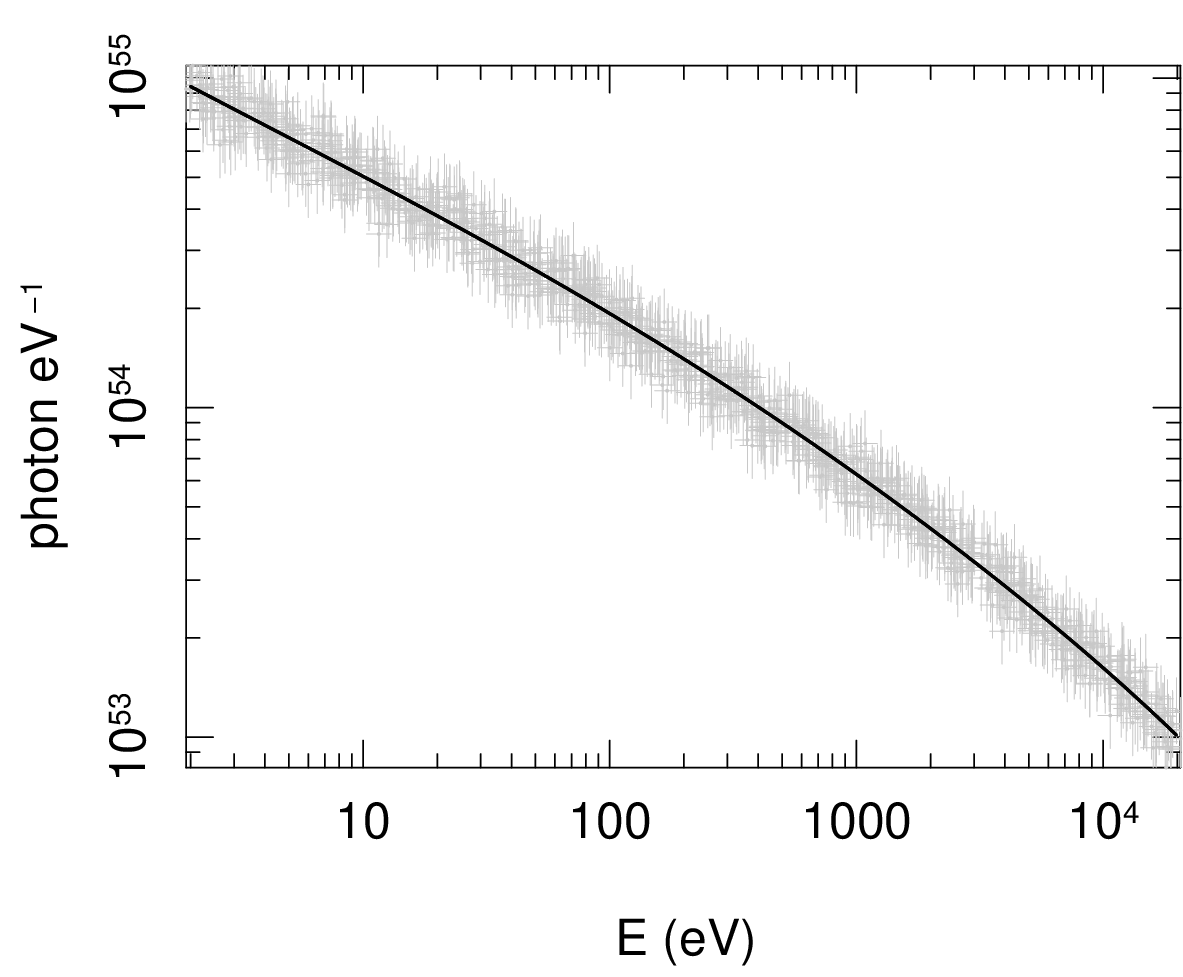} &
15)\includegraphics[height=5.5cm,angle=-90]{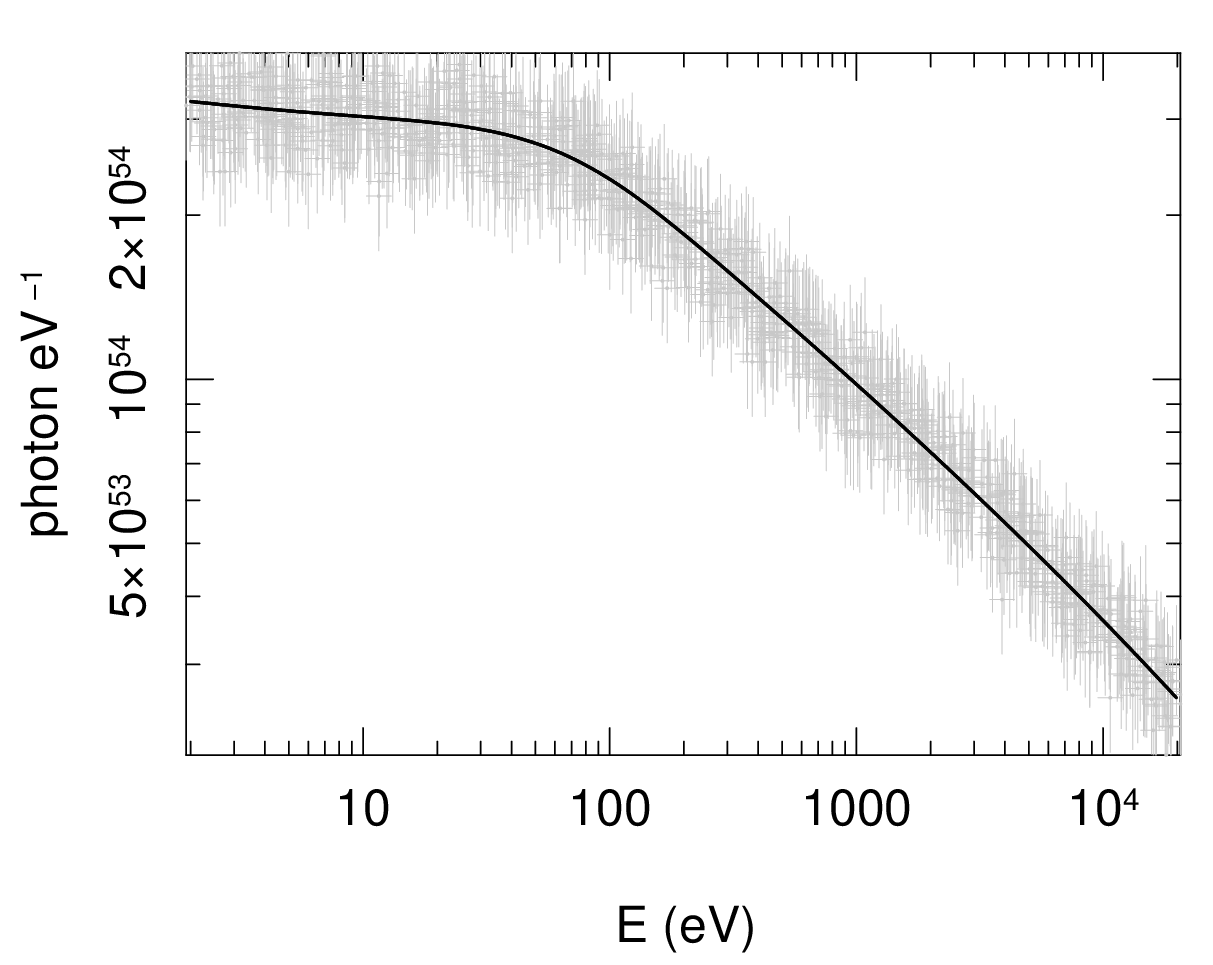} &
16)\includegraphics[height=5.5cm,angle=-90]{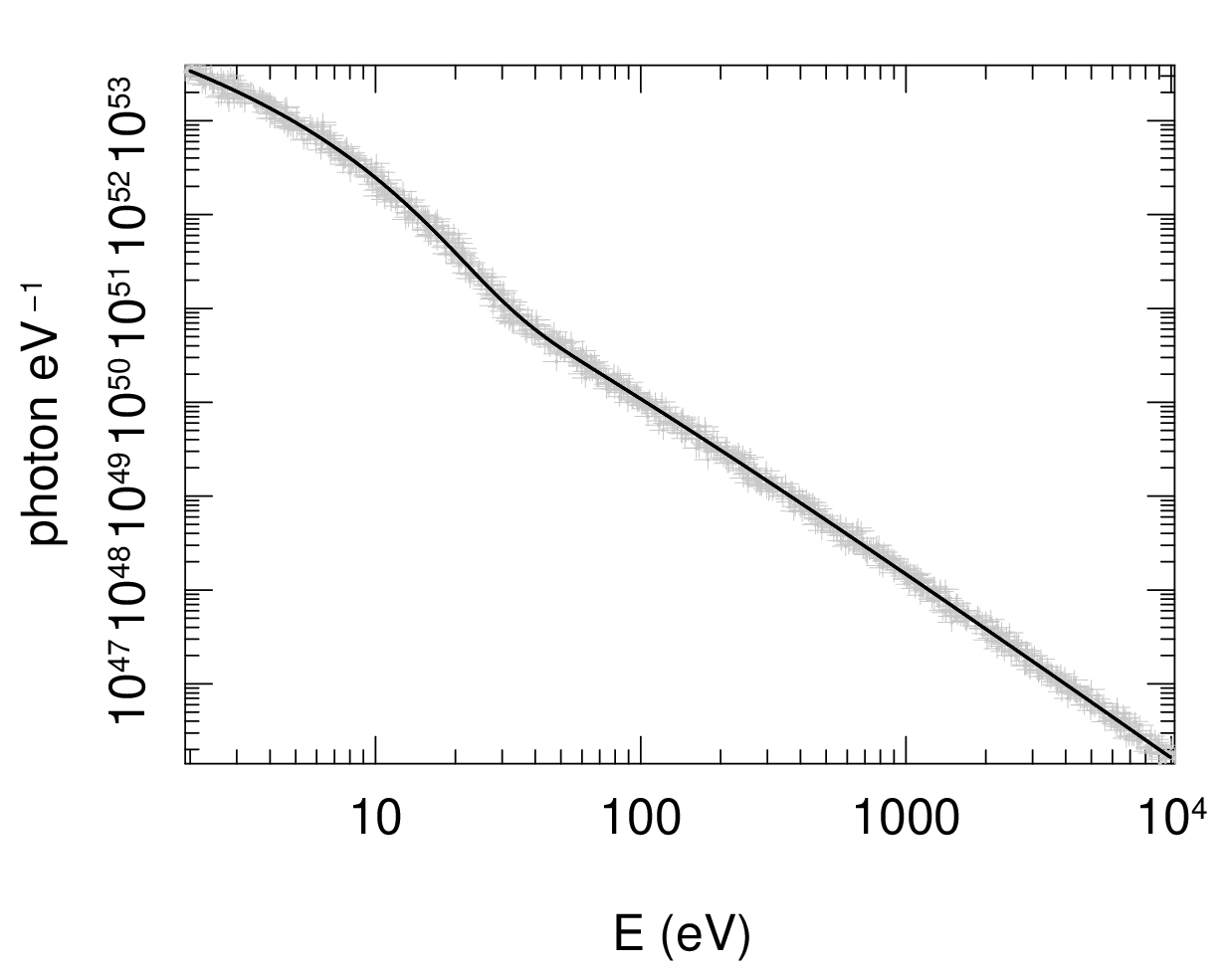} 
\end {tabular}
\end {center}
\caption{Total broad band ($2$ eV - $20$ keV) spectrum of the simulated 
models 11 to 16 with $\Delta r \propto (\omega' / \omega'_m)^{-1/2}$. 
Definitions of the curves are similar to Fig. \ref{fig:specdar}.
\label{fig:specqsardr}}
\end{figure*}

Fig. \ref{fig:specqsardr} shows the spectrum of the same models but with an 
energy dependent $\Delta r$ as explained in Sec.\ref{sec:lcqsar}. As expected, 
they are softer. Models 11 and 15 have negative slopes at low energies and a 
cutoff at high energies. Other models have the same form of the spectrum as 
in Fig. \ref{fig:specqsar} but softer.

Models 17 to 19 in Fig. \ref{fig:specqsar} which present internal shocks with 
quasi-steady state active region are hard and has spectra similar to prompt 
models with dynamical active region.

For various fundamental and observational reasons it is very difficult to 
compare these models with observations and with previous simulations or 
theoretical estimations. As we mentioned at the beginning of this section, 
these simulations does not correspond to a complete burst but just to part 
of a burst that can be simulated with a single and simple evolution model. 
Moreover, the emission after the prompt gamma-ray peak is most 
probably a combination of emission from the prompt internal shock and the
afterglow / external shocks~\citep{xrtafterglow}. It is not always easy to 
distinguish these components and in any case the separation would be model 
dependent. Nonetheless, if we assume that the missing sectors do not change 
the spectrum significantly, we can compare these simulations with 
observations and previous theoretical estimations. For instance models 
7, 11, and 15 which present the early emission of the afterglow with what is 
called {\it fast cooling regime} in~\citep{emission1}, the general form of 
the spectrum is similar. Notably, their slope at high energies is 
$\lesssim -1$ when $p = 2.5$ for all models except model 16. This value is 
consistent with the value of $-p/2$ predicted in~\cite{emission1}. In model 
16 the index $p = 3$ and the value of slope at high energies is $\sim -1.5$ 
consistent with estimation of~\cite{emission1}.

As for comparison with observations, it strongly depends on the time interval 
of observations. For instant, the available spectrum fit for the \swift bursts 
belong mainly to the first couple of thousands of seconds because at later 
times the emission is faint and it is very difficult to determine the 
spectrum. However, as we mentioned before the early X-ray and optical emission 
are most probably due to the tail emission from the remnant of the prompt 
shock and therefore they can not be compared with what is expected from 
external shock. This mixing can be the reason for the large scatter in the 
spectral index of the X-ray spectrum~\citep{xrtafterglow}. For instance 
GRB 070724A~\citep{grb070724a}, GRB 070721B~\citep{grb070721b}, 
GRB 061004~\citep{grb061004} have an average spectrum index of $\sim -1.4$ 
when others such as GRB 050128~\citep{grb050128}, 
GRB 070208~\citep{grb070208}, and GRB 060804~\citep{grb060804} 
have an index of $\sim -2.3$.

On the other hand, one of the principle differences between the 
prediction of these models and what is usually considered in the literature 
for the broad X-ray and optical spectrum is a single power law with or 
without cutoff at high or low energies. The simulated models here show 
the possibility of variant slopes and presence of cutoffs even in the energy 
range accessible to one instrument. This can be the reason for the absence of 
an optical afterglow specially when the burst is hard. Nonetheless, the 
under-luminous optical emission can be also explained by the presence of 
dust in the host galaxy or even in the circumburst environment~\citep{dust}. 
One argument against the presence of a large amount of dust in the host 
galaxy of GRBs is the fact that most of them have relatively low 
metalicity~\citep{grbhostmetal}. On the other hand, if the absorbers are 
limited to the star-forming clouds and their surroundings, then 
the local quantity of dust can be significant although the total fraction of 
dust in the host galaxy can be low. Present observations and reduction 
techniques are not yet sufficient to clarify this issue. In conclusion, these 
simulations show that the spectrum of emission itself can be in part 
responsible for the lack of observation of optical and lower energy 
afterglows in the observed GRBs.

\section{Summary}
We discussed the results of a few numerical simulations of GRBs based on the 
model explained in Paper I. We showed that the behaviour of these models is 
in general consistent with some regimes in the observed GRBs. It is difficult  
to verify the reality of some of the features seen in these simulations. For 
instance, a long lag between the peak of the X-ray and optical afterglows is 
predicted by some of these simulations. We need long multi-wavelength 
observations to be able to see such a feature in real bursts, and usually 
such data do not exist. Nonetheless, there are exceptions. Long follow-up 
of some bursts such as GRB 081029~\citep{grb080129} and 
GRB 081007~\citep{grb081007} show a wide late peak in optical wavelengths 
undetected in the X-ray. It can be from the afterglow, although other 
possibility such as the appearance of a supernova can not be ruled out.
These observations increase the confidence on this model and analytical 
results presented here. Moreover, they encourage long term monitoring of 
GRBs to see if similar behaviour can be found in other bursts. 

To complete 
these simulations we need to put together the simulation of various regimes 
and make the light curves and spectra corresponding to the whole burst. This 
should help to better understand the contribution of internal and external 
shocks in the light curves and spectrum as well as features such as the shallow
regime, breaks, etc. We leave these investigations for future works.

\section*{Acknowledgments}
I would like to thank Keith Mason for encouraging me to work on GRB science. 
The ideas presented in this work couldn't be developed without long 
discussions with the past and present members of the \swift science team at 
MSSL: A. Blustin, A. Breeveld, M. De Pasquale, P. Kuin, S. Oates, S. Rosen, 
M. Page, P. Schady, M. Still, and S. Zane, as well as the other members of the 
\swift team in particular: S. Barthelmy, Ph. Evens, E.E. Fenimore, N. Gehrels, 
P. M\'es\'zaros, J. Osborne, and K. Page. I thank all of them.

\begin{table*}
\caption{Definition of symbols \label{tab:def}}
{\small
\begin{tabular}{lp{7cm}}
\hline
Symbol & Definition \\
\hline\hline
 & All quantities without, except constants, are defined with respect to 
the rest frame of far but at the same redshift observer. \\
$'$ & All quantities with a prime are defined with respect to the rest frame 
of slower shell. \\\hline
$\beta' (r')$ & Relative $\beta$ of two colliding shell in the rest frame of 
slower shell. \\\hline
$\omega'_m (r')$ & Characteristic synchrotron frequency for slowest electrons, 
Eq. (78-Paper I)
\\\hline
$\gamma'_1$ & Relative Lorentz factor of two shells in the rest frame of 
slower shell calculated with first order radiative correction. \\\hline
$n'\Delta r'$ & intrinsic column density of the active region. \\\hline
$p$ & the index of electrons Lorentz factor distribution. \\\hline
\end{tabular}
}
\end{table*}

\begin{table*}
\caption{Parameter set of the simulated models. \label{tab:param}}
{\small
\begin{tabular}{p{2mm}|p{2mm}llllllllllllp{11mm}p{5mm}l}
\hline
& No. & $r'_0(cm)$ & $\frac{\Delta r'_0}{r'_0} $ & $p$ & $\kappa$ & $ \gamma'_0$ & $ \tau$ & $ \delta$ & $\epsilon_B$ & $\alpha_B$ & $\epsilon_e$ & $\alpha_e$ & $N'_0$ (cm$^{-3}$) & $n'(r'_0)$ cm$^{-3}$ & $(\frac{r'}{r'_0})$ max & $\Gamma_f$ \\
\hline\hline 
\multirow{10}{2mm}{D. A. R.} 
& 1 & $10^{11}$ & 0.5 & 2.5 & 0 & 2 & -3 & 0 & $10^{-3}$ & -0.5 & 0.02 & -5 & $5 \times 10^{12}$ & $10^{14}$ & 2 & 200 \\  
& 2 & $5 \times 10^{11}$ & 0.1 & 2.5 & 0 & 2 & -3 & 0 & $10^{-3}$ & -1 & 0.3 & -3 & $10^{13}$ & $2 \times 10^{13}$ & 1.005 & 200 \\  
& 3 & $5 \times 10^{11}$ & 0.2 & 2.2 & 0 & 2 & 1 & 0 & $2 \times 10^{-3}$ & 0.1 & 0.2 & 0 & $5 \times 10^{10}$ & $2 \times 10^{11}$ & 1.005 & 200 \\  
& 4 & $2 \times 10^{11}$ & 0.2 & 3 & 0 & 1.5 & 0 &  0 & $10^{-3}$ & -0.3 & 0.2 & -0.5 & $5 \times 10^{11}$ & $10^{13}$ & 1.1 & 200 \\ 
& 5 & $10^{11}$ & 0.5 & 2.7 & 0 & 2 & -2 & 0 & $10^{-4}$ & 0.1 & 0.2 & $10^{-2}$ & $5 \times 10^{11}$ & $10^{13}$ & 1.5 & 200 \\ 
& 6 & $10^{11}$ & 0.6 & 2.1 & 0 & 3 & -1 & 0 & $10^{-3}$ & -0.8 & 0.1 & -1 & 5 $\times 10^{12}$ & $5 \times 10^{13}$ & 3 & 200 \\ 
\cline{2-17}
& 7 & $10^{16}$ & $10^{-4}$ & 2.5 & 1 & 50 & -3 & 0 & $5 \times 10^{-4}$ & 0.5 & 0.1 & 0 & 1 & $10^{7}$ & 5 & 1 \\ 
& 8 & $10^{16}$ & $10^{-4}$ & 2.5 & 0 & 50 & 1 & 0 & $5 \times 10^{-3}$ & -0.01 & 0.2 & -0.01 & 50 & $10^{7}$ & 5 & 1 \\ 
& 9 & $10^{16}$ & 0.01 & 2.5 & 1 & 10 & 3 & 0 & $5 \times 10^{-3}$ & 0.5 & 0.1 & 0 & 1 & $10^3$ & 5 & 1 \\ 
& 10 & $10^{16}$ & 0.01 & 2.5 & 0 & 10 & 1 & 0 & $5 \times 10^{-3}$ & -0.01 & 0.2 & -0.01 & 10 & $10^3$ & 5 & 1 \\ 
\hline\hline
\multirow{9}{2mm}{Q. S. A. R.}  
& 11 & $5 \times 10^{15}$ & 0.05 & 2.5 & 0 & 200 & 0 & 5 & $10^{-3}$ & -0.5 & 0.1 & -2 & 10 & $10^3$ & 5 & 1 \\ 
& 12 & $5 \times 10^{16}$ & 0.01 & 2.5 & 2 & 20 & 0 & 0.1 & $10^{-3}$ & 0.01 & 0.5 & 0.01 & 10 & $10^2$ & 2 & 1 \\ 
& 13 & $5 \times 10^{16}$ & 0.01 & 2.5 & 2 & 20 & 0 & 1 & $10^{-3}$ & 0.01 & 0.5 & 0.01 & 1 & $10^2$ & 2 & 1 \\ 
& 14 & $10^{15}$ & 0.05 & 2.5 & 0 & 200 & 0 & 5 & $10^{-3}$ & -0.5 & 0.1 & -1 & $10^3$ & $10^4$ & 10 & 1 \\ 
& 15 & $10^{16}$ & 0.01 & 2.5 & 2 & 200 & 0 & 5 & $10^{-3}$ & -1 & 0.1 & -0.1 & 1 & $10^3$ & 5 & 1 \\ 
& 16 & $10^{17}$ & 5 $\times 10^{-3}$ & 3 & 0 & 10 & 0 & 0.5 & $10^{-3}$ & 0.01 & 0.2 & 0.01 & $10^{-1}$ & 50 & 2 & 1 \\ 
\cline{2-17}
& 17 & $3 \times 10^{11}$ & 0.5 & 2.5 & 0 & 2 & 0 & 4 & $10^{-3}$ & 0.01 & 0.2 & -2 & $10^{12}$ & $2 \times 10^{13}$ & 1.5 & 200 \\ 
& 18 & $5 \times 10^{11}$ & 0.2 & 2.5 & 0 & 2 & 0 & 4 & $10^{-3}$ & 0.2 & 0.2 & 2 & $10^{10}$ & $2 \times 10^{11}$ & 1.5 & 200 \\ 
& 19 & $10^{11}$ & 0.5 & 2.5 & 0 & 2 & 0 & 4 & $10^{-3}$ & -0.5 & 0.2 & -2 & $10^{12}$ & $2 \times 10^{13}$ & 1.5 & 200 \\ 
\hline\hline
\multirow{5}{2mm}{Eq. (29) in Paper I} 
& 20 & $10^{16}$ & 0.1 & 2.5 & 0 & 20 & 0 & 0.5 & $10^{-3}$ & 0.01 & 0.2 & 0.01 & 10 & $10^3$ & 5 & 1 \\ 
& 21 & $10^{16}$ & 0.1 & 2.2 & 0 & 10 & 0 & 0.1 & $10^{-3}$ & 0.01 & 0.2 & 0.01 & 10 & $10^2$ & 2 & 1 \\ 
& 22 & $5 \times 10^{16}$ & 0.1 & 2.5 & 0 & 10 & 0 & 1 & $10^{-3}$ & 0.05 & 0.2 & 0.05 & 1 & $10^2$ & 5 & 1 \\ 
\cline{2-17}
& 23 & $3 \times 10^{11}$ & 0.5 & 2.5 & 0 & 2 & 0 & 5 & $10^{-3}$ & 4 & 0.2 & 4 & $10^{12}$ & $2 \times 10^{13}$ & 1.5 & 200 \\ 
& 24 & $5 \times 10^{10}$ & 0.5 & 3 & 0 & 2 & 0 & 5 & $10^{-3}$ & -0.5 & 0.1 & -2 & $10^{16}$ & $2 \times 10^{18}$ & 1.5 & 200 \\ 
\hline
\end{tabular}
\footnote*{Dynamical Active Region (D.A.R.), Quasi-Steady Active Region 
(Q.S.A.R.).}\\
\footnote*{The aim for using large negative values for $\alpha_e$ in the first 
and second parameter sets were simulating an exponential rise of the electric 
field.}
}
\end{table*}

\begin{figure*}
\begin{center}
\begin {tabular}{ccc}
20)\includegraphics[height=5.5cm,angle=-90]{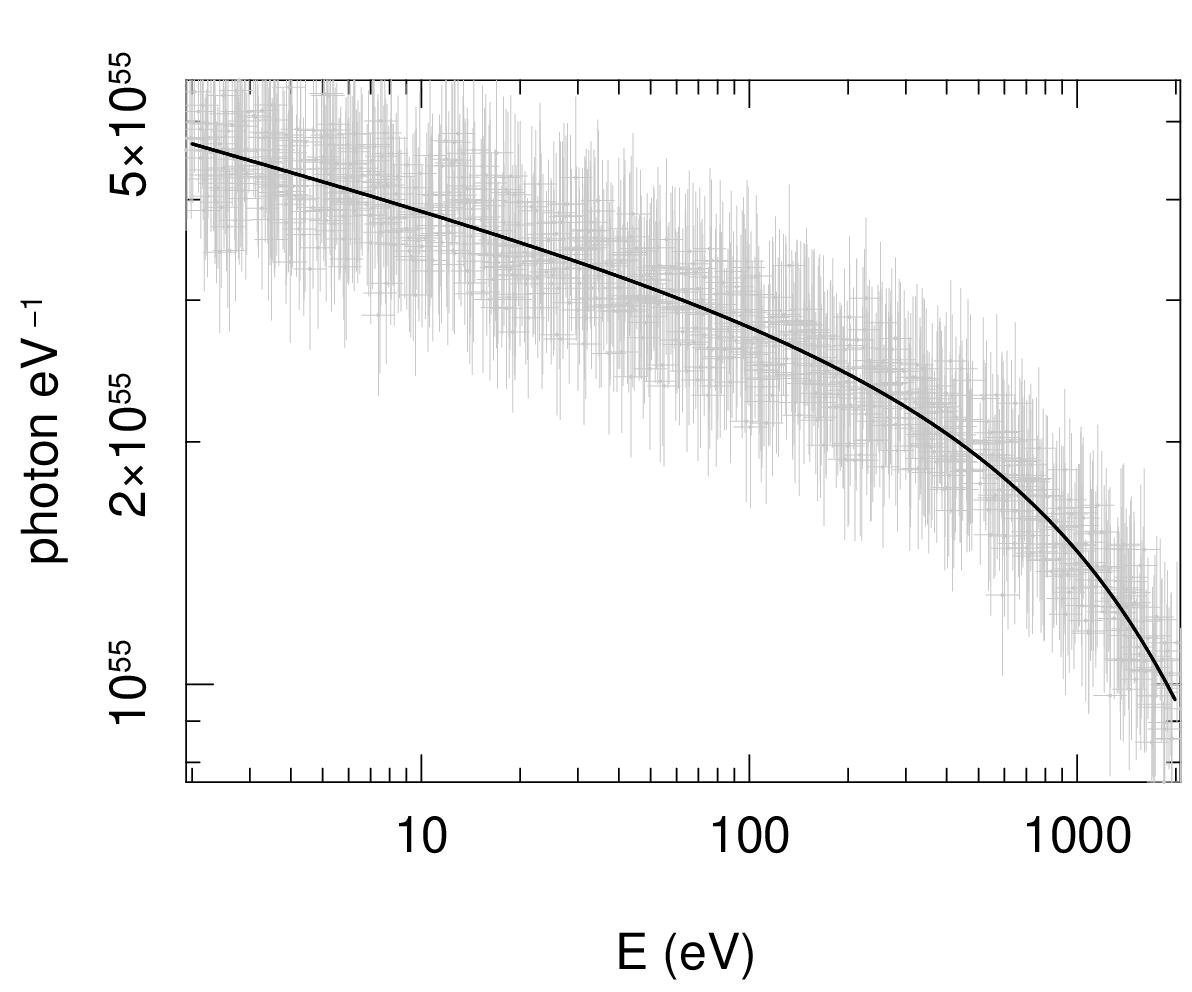} &
21)\includegraphics[height=5.5cm,angle=-90]{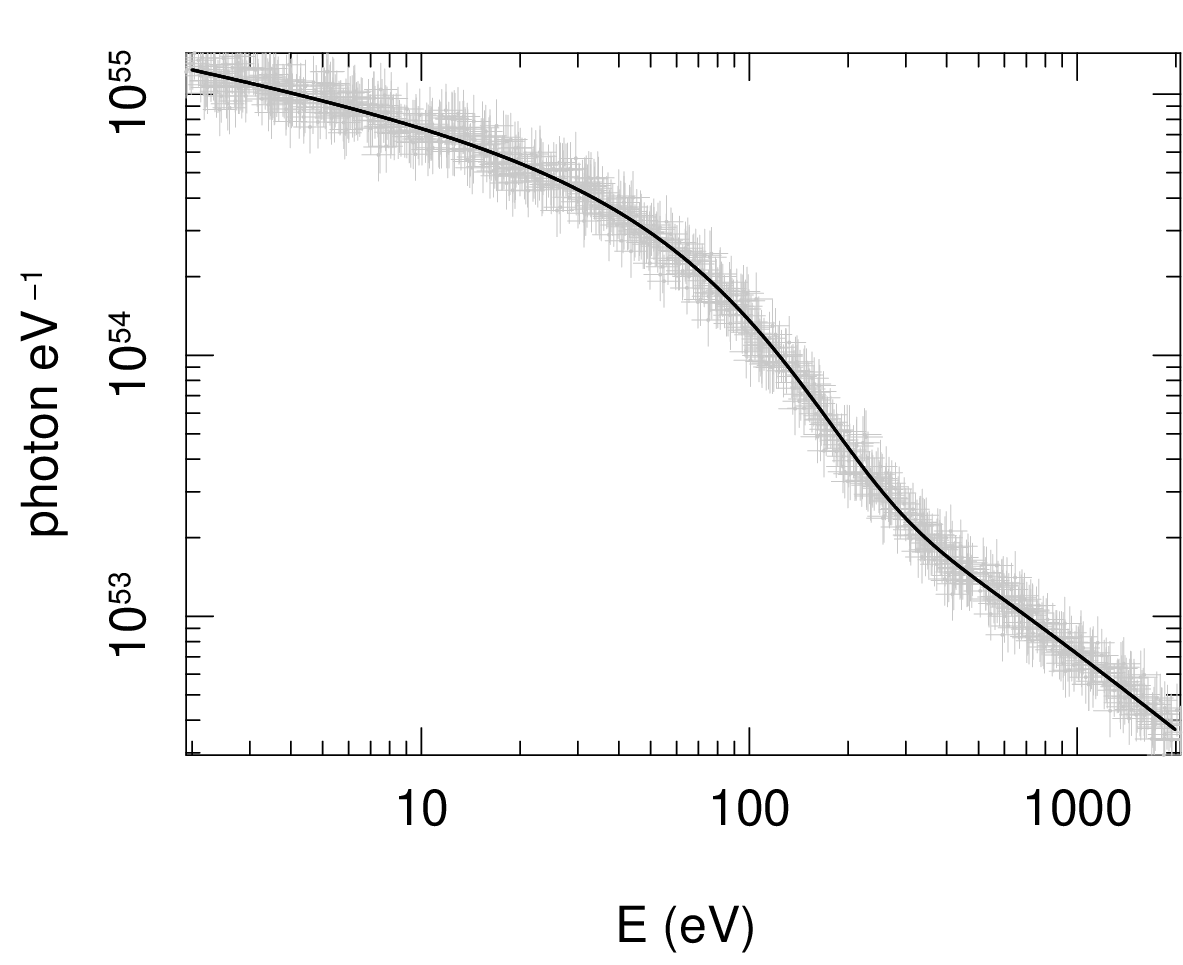} &
22)\includegraphics[height=5.5cm,angle=-90]{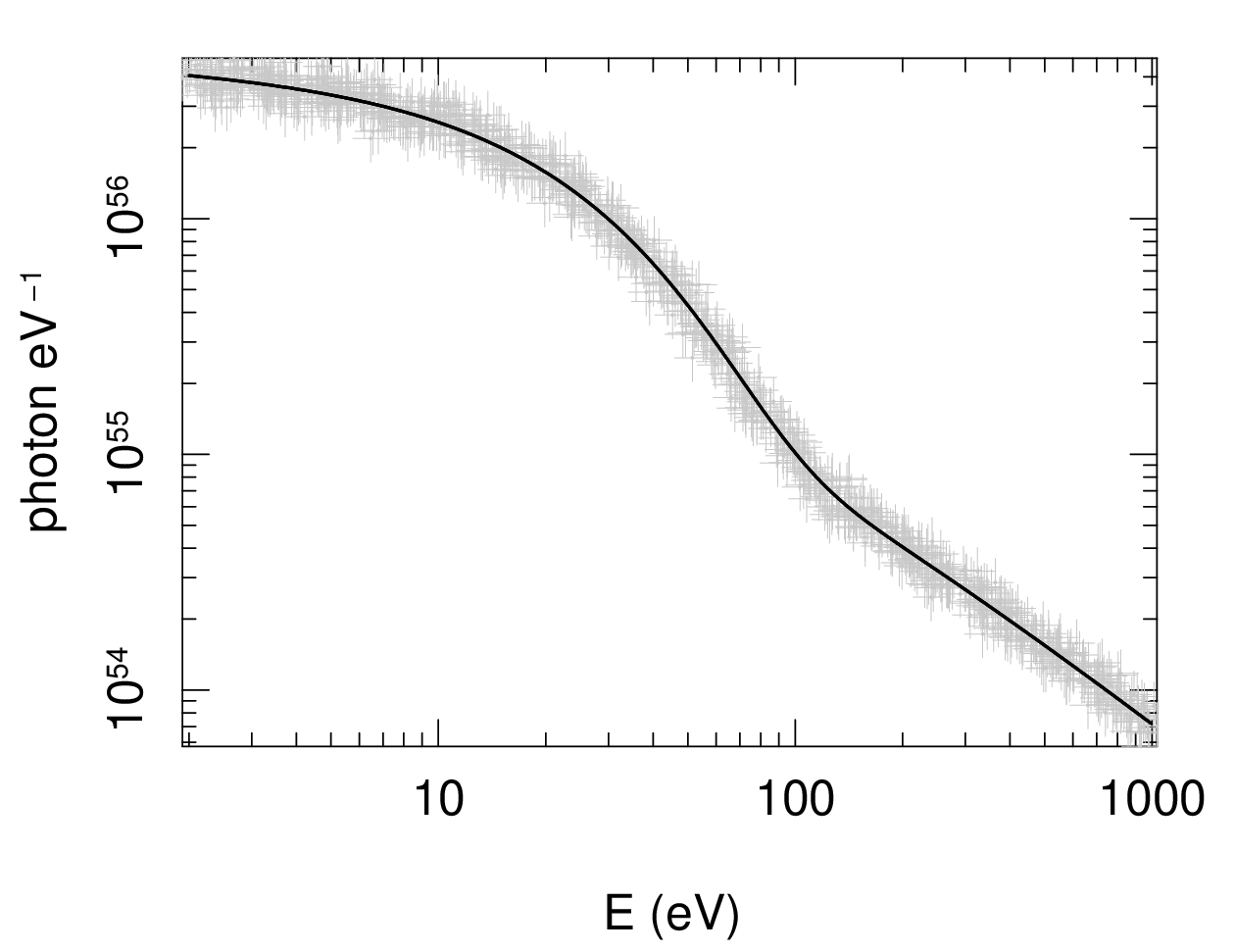} \\
23)\includegraphics[height=5.5cm,angle=-90]{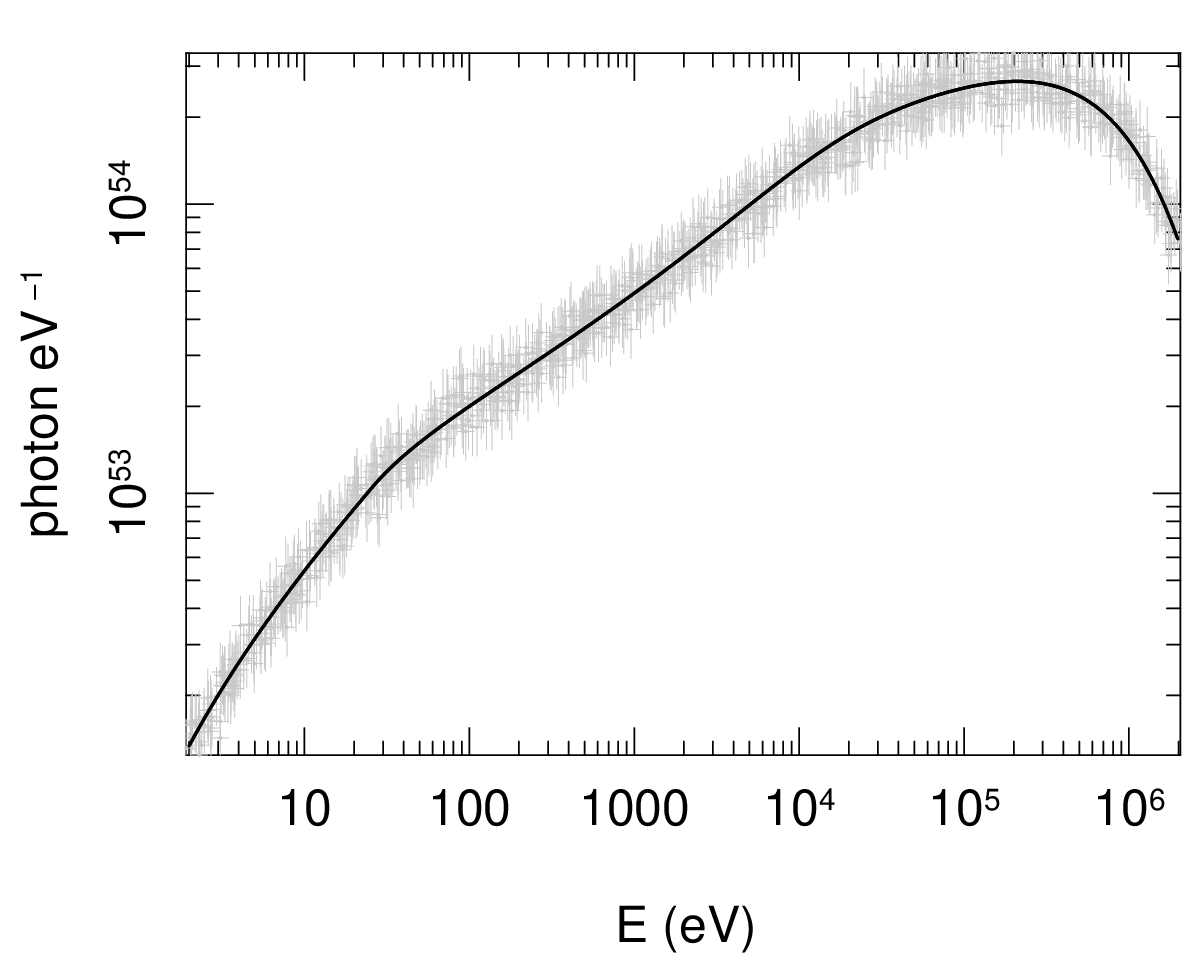} &
24)\includegraphics[height=5.5cm,angle=-90]{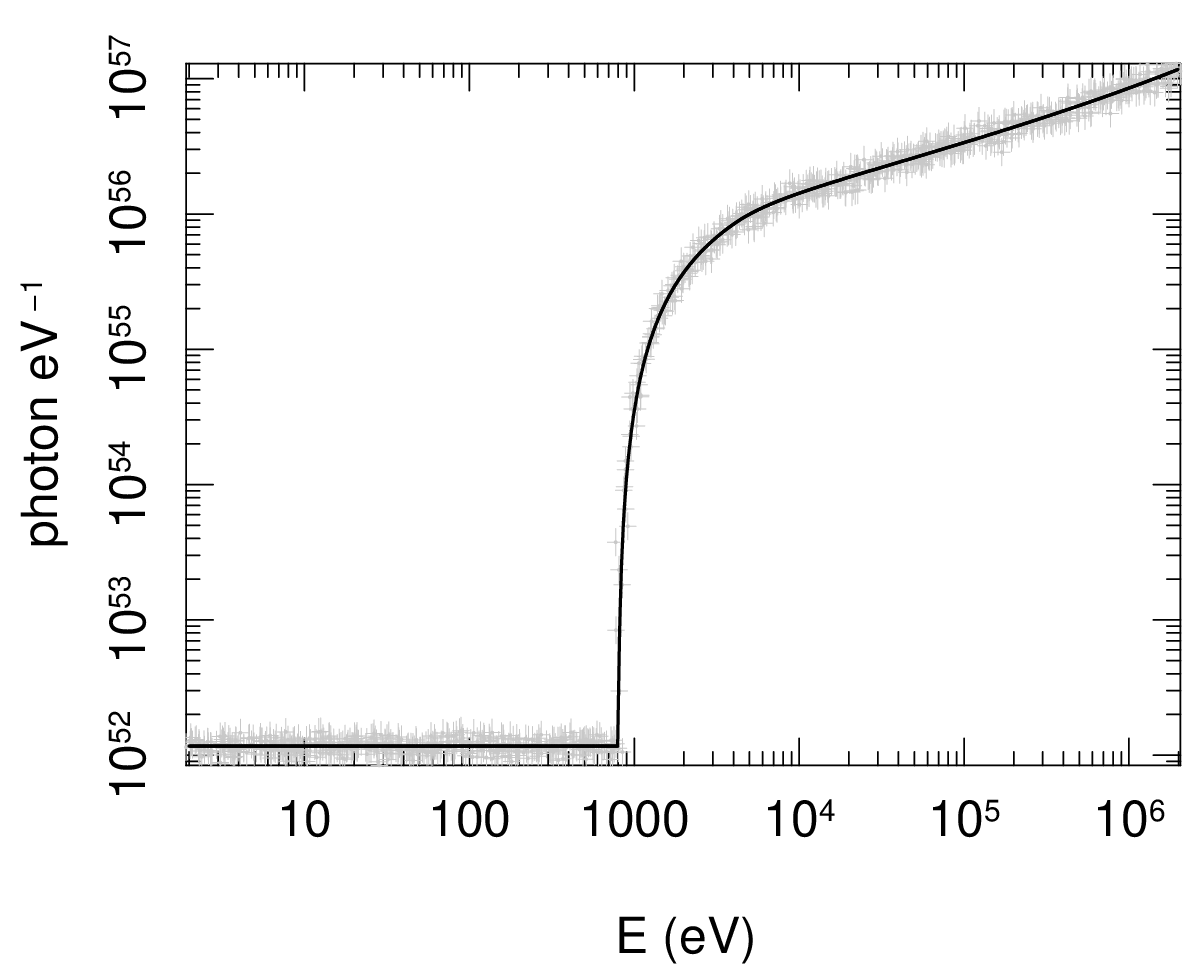} 
\end {tabular}
\end {center}
\caption{Total broad band ($2$ eV - $20$ keV)/($2$ eV - $2$ MeV) spectrum of 
models 20 to 24. Definitions of the curves are similar to 
Fig. \ref{fig:specdar}.
\label{fig:specqsarend}}
\end{figure*}

\end{document}